\documentclass[doctor, english, final]{postech-ucs}

\usepackage{amsmath}
\usepackage{enumitem}
\usepackage{comment}

\usepackage{array,enumitem,multirow,caption,graphicx,subcaption,multicol,lipsum,float, adjustbox, rotating}
\usepackage{amsmath, amsfonts, amsthm, booktabs}
\usepackage[ruled,vlined,linesnumbered]{algorithm2e}
\usepackage{kotex} 
\usepackage{makecell}
\usepackage{color, colortbl}
\usepackage{array,enumitem,multirow,caption,graphicx,subcaption,multicol,lipsum, float}
\usepackage{amsmath, amsfonts, amsthm, balance}
\newcolumntype{C}[1]{>{\centering\let\newline\\\arraybackslash\hspace{0pt}}m{#1}}
\newcolumntype{Z}{>{\centering\arraybackslash}m{0.062\linewidth}}
\newcommand{\RowStretch}[1]{\renewcommand{\arraystretch}{#1}}
\let\oldnl\nl
\newcommand{\nonl}{\renewcommand{\nl}{\let\nl\oldnl}}
\usepackage{float}
\usepackage{setspace}
\usepackage{threeparttable}
\usepackage{pdfpages}

\title[korean] {정확성 및 효율성을 갖춘 추천 시스템을 위한 지식 증류 기술}
\title[english]{Knowledge Distillation Approaches for Accurate and Efficient Recommender System}

\author[korean] {강}{성 구}
\author[english]{Kang}{SeongKu}

\advisor[major]{유환조}{Hwanjo Yu}{nosign}

\department{CSE}{engineering}

\studentid{20182821}

\referee[1]{Hwanjo Yu}
\referee[2]{Sang-Wook Kim}
\referee[3]{Minsu Cho}
\referee[4]{Dongwoo Kim}
\referee[5]{Chanyoung Park}

\approvaldate{2022}{11}{21}

\refereedate{2022}{11}{21}

\gradyear{2023}

\begin{document}


\begin{abstract}
In this era of information explosion, recommender systems are widely used in various industries to provide personalized user experience, playing a key role in promoting corporate profits.
Recent recommender systems tend to adopt increasingly complex and large models to better understand the complex nature of user-item interactions.
Large models with numerous parameters have high recommendation accuracy due to their excellent expressive power.
However, they also incur correspondingly high computational costs as well as high latency for inference, which has become one of the major obstacles to deployment.

To reduce the model size while maintaining accuracy, we focus on knowledge distillation (KD), a model-independent strategy that transfers knowledge from a well-trained large model to a compact model.
The compact model trained with KD has an accuracy comparable to that of the large model and can make more efficient online inferences due to its small model size.
Despite its breakthrough in classification problems, KD to recommendation models and ranking problems has not been studied well in the previous literature.
This dissertation is devoted to developing knowledge distillation methods for recommender systems to fully improve the performance of a compact model.

We propose novel distillation methods designed for recommender systems.
The proposed methods are categorized according to their knowledge sources as follows:
\textbf{(1) Latent knowledge:} 
we propose two methods that transfer latent knowledge of user/item representation.
They effectively transfer knowledge of niche tastes with a balanced distillation strategy that prevents the KD process from being biased towards a small number of large preference groups.
Also, we propose a new method that transfers user/item relations in the representation space.
The proposed method selectively transfers essential relations considering the limited capacity of the compact model.
\textbf{(2) Ranking knowledge:} 
we propose three methods that transfer ranking knowledge from the recommendation results.
They formulate the KD process as a ranking matching problem and transfer the knowledge via a listwise learning strategy.
Further, we present a new learning framework that compresses the ranking knowledge of heterogeneous recommendation models.
The proposed framework is developed to ease the computational burdens of model ensemble which is a dominant solution for many recommendation applications.

We validate the benefit of our proposed methods and frameworks through extensive experiments.     
To summarize, this dissertation sheds light on knowledge distillation approaches for a better accuracy-efficiency trade-off of the recommendation models.

    \end{abstract}

    \tableofcontents





\chapter{Introduction}

\section{Research Motivation}
In this era of information explosion, recommendation system (RS) is being used as a core technology in web page search services (e.g., Naver, Google) and personalized item recommendation services (e.g., Amazon, Netflix). 
In particular, RS has established itself as an essential technology for E-commerce and Over-the-top (OTT) media services in that it supports users' decision-making process and maximizes corporate profits.

Recent RS tends to adopt increasingly complex and large models to better understand the complex nature of user-item interactions.
A large model with numerous parameters can accurately capture the users' complex preferences through excellent expression ability, and thus has a high recommendation accuracy. 
However, such a large model incurs high computation costs and high inference latency, which has become one of the major obstacles to real-time service.
To reduce the model size while maintaining accuracy, we focus on knowledge distillation (KD), a model-independent strategy that transfers knowledge from a well-trained large model to a compact model.
The compact model trained with KD has an accuracy comparable to that of the large model and can make more efficient online inferences due to its small model size.
Despite its breakthrough in classification problems, KD to recommendation models and ranking problems has not been studied well in the previous literature.

This dissertation is devoted to developing KD methods for the recommendation model for a better accuracy-efficiency trade-off.
We aim to effectively transfer two knowledge sources of the recommendation model; \textbf{latent knowledge} and \textbf{ranking knowledge}.
Latent knowledge refers to all information about users, items, and their relationships discovered and stored in the model, providing detailed explanations of the model's final prediction.
Ranking knowledge refers to the relative preference order among items, providing direct guidance on the model prediction.

\section{Contributions}
\begin{figure}[t]
\centering
\begin{subfigure}[t]{1.\linewidth}
    \includegraphics[width=\linewidth]{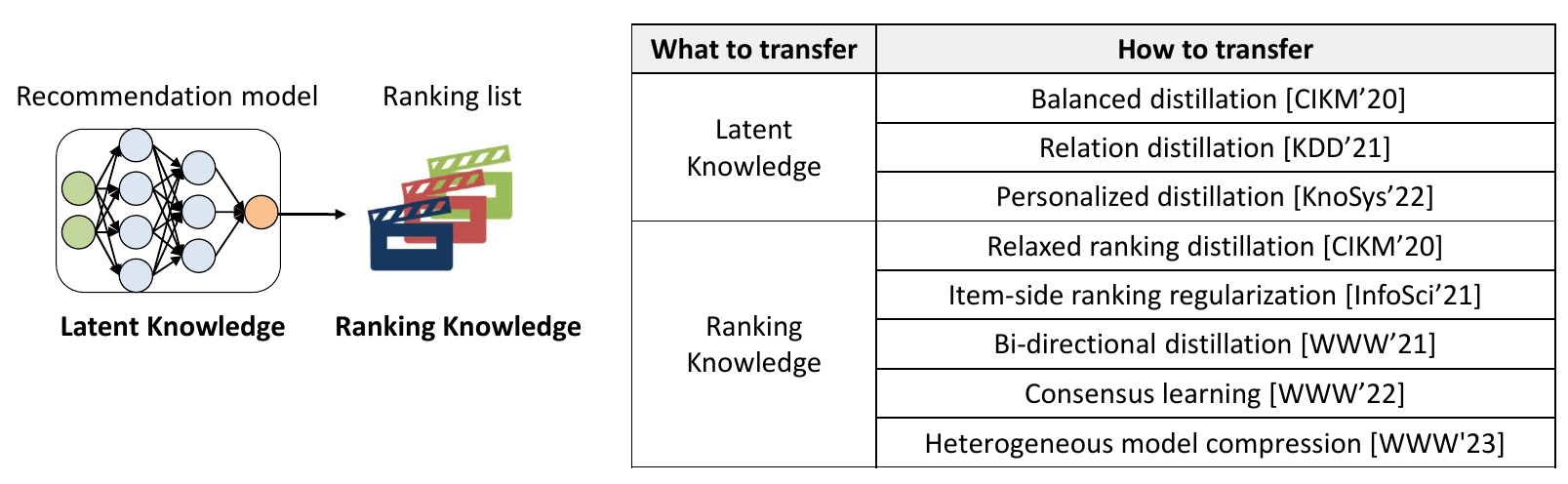}
\end{subfigure}
\caption{An overview of the proposed approaches.}
\label{fig:thesis_overview}
\end{figure}

Figure \ref{fig:thesis_overview} shows an overview of our works included in this thesis.
We provide a summary of our contributions as follows:
\begin{itemize}
    \item Latent knowledge distillation (Chapter 2 to 4): 
    We propose Distillation Experts (DE), Personalized Hint Regression (PHR), and Topology Distillation (TD) to effectively transfer the latent knowledge of the recommendation model.
    They are designed to identify and selectively transfer only essential information considering the limited capacity of our target model.
    Our extensive experiments demonstrate that they significantly improve the recommendation quality of the compact target model.

    \item Ranking knowledge distillation (Chapter 2, 5 to 7): 
    We propose Relaxed Ranking Distillation (RRD), Item-side Ranking Regularized Distillation (IR-RRD), Consensus Learning for Collaborative Filtering (ConCF), and Heterogeneous Model Compression (HetComp) to transfer ranking knowledge from the prediction of the recommendation model.
    They formulate the distillation process as a ranking matching problem among the ranking list of each model and introduce new strategies to fully improve the distillation quality.
    We provide extensive experiment results showing the superiority and rationality of the proposed methods.
    
\end{itemize}

\section{Thesis Organization}
The rest of this thesis is organized as follows.
In Chapter \ref{chapt:de-rrd}, we propose DE for transferring the latent knowledge in a balanced way and RRD for transferring the ranking knowledge based on a relaxed permutation probability.
In Chapter \ref{chapt:phr}, we present PHR, which is a follow-up study of DE, removing all hyperparameters of DE.
In Chapter \ref{chapt:TD}, we propose FTD and HTD for transferring the topological structure built upon the relations in the representation space.
In Chapter \ref{chapt:IR-RRD}, we present IR-RRD, which is a follow-up study of RRD, for improving the distillation quality based on item-side ranking regularization.
In Chapter \ref{chapt:ConCF}, we present ConCF for generating a more generalizable model by using online knowledge distillation among heterogeneous learning objectives.
In Chapter \ref{chapt:HetComp}, we propose HetComp for compressing knowledge of heterogeneous recommendation models into a compact model.
In Chapter \ref{chapt:conclusion}, we make a conclusion and present future research directions.

\chapter{A Knowledge Distillation Framework for Recommender System}
\label{chapt:de-rrd}
Recent recommender systems have started to employ knowledge distillation, which is a model compression technique distilling knowledge from a cumbersome model (teacher) to a compact model (student), to reduce inference latency while maintaining performance.
The state-of-the-art methods have only focused on making the student model accurately imitate the predictions of the teacher model.
They have a limitation in that the prediction results incompletely reveal the teacher’s knowledge.
In this paper, we propose a novel knowledge distillation framework for recommender system, called \textit{DE-RRD}, which enables the student model to learn from the latent knowledge encoded in the teacher model as well as from the teacher's predictions.
Concretely, \textit{DE-RRD} consists of two methods:
1) \textit{Distillation Experts (DE)} that directly transfers the latent knowledge from the teacher model.
DE exploits ``experts'' and a novel expert selection strategy for effectively distilling the vast teacher's knowledge to the student with limited capacity.
2) \textit{Relaxed Ranking Distillation (RRD)}
that transfers the knowledge revealed from the teacher's prediction with consideration of the relaxed ranking orders among items.
Our extensive experiments show that \textit{DE-RRD} outperforms the state-of-the-art competitors and achieves comparable or even better performance to that of the teacher model with faster inference~time.

\section{Introduction}
\label{sec:DE-RRD_introduction_derrd}

In recent years, recommender system (RS) has been broadly adopted in various industries, helping users’ decisions in the era of information explosion, and playing a key role in promoting corporate profits.
However, a growing scale of users (and items) and sophisticated model architecture to capture complex patterns make the size of the model continuously increasing \cite{RD, CD, GCN_distill, DCF}.
A large model with numerous parameters has a high capacity, and thus usually has better recommendation performance.
On the other hand, it requires a large computational time and memory costs, and thus incurs a high latency during the inference phase, which makes it difficult to apply such large model to real-time platform.

Motivated by the significant success of \textit{knowledge distillation} (KD) in the computer vision field, a few work \cite{RD, CD} have employed KD for RS to reduce the size of models while maintaining the performance.
KD is a model-agnostic strategy to accelerate the learning of a new compact model (student) by transferring knowledge from a previously trained large model (teacher) \cite{KD}.
The knowledge transfer is conducted as follows:
First, the teacher model is trained with the user-item interactions in the training set which has binary labels -- `1' for observed interactions, and `0' for unobserved interactions.
Then, the student model is trained with the ``soft'' labels generated by the teacher model (i.e., teacher's predictions) along with the available binary labels.
The student model trained with KD has comparable performance to that of the teacher, and also has a lower latency due to its small size \cite{RD, CD}.

The core idea behind this process is that the soft labels predicted by the teacher model reveal hidden relations among entities (i.e., users and items) not explicitly included in the training set, so that they accelerate and improve the learning of the student model.
Specifically, the items ranked near the top of a user’s recommendation list would have strong correlations to the items that the user has interacted before \cite{RD}.
Also, the soft labels provide guidance for distinguishing the items that each user would like and the items that each user would not be interested in among numerous unobserved items only labeled as `0' \cite{CD}.
By using the additional supervisions from the teacher model, the state-of-the-art methods \cite{RD, CD} have achieved comparable or even better performance to the teacher models with faster inference time.

\begin{figure}[t]
\centering
  \includegraphics[height=5cm]{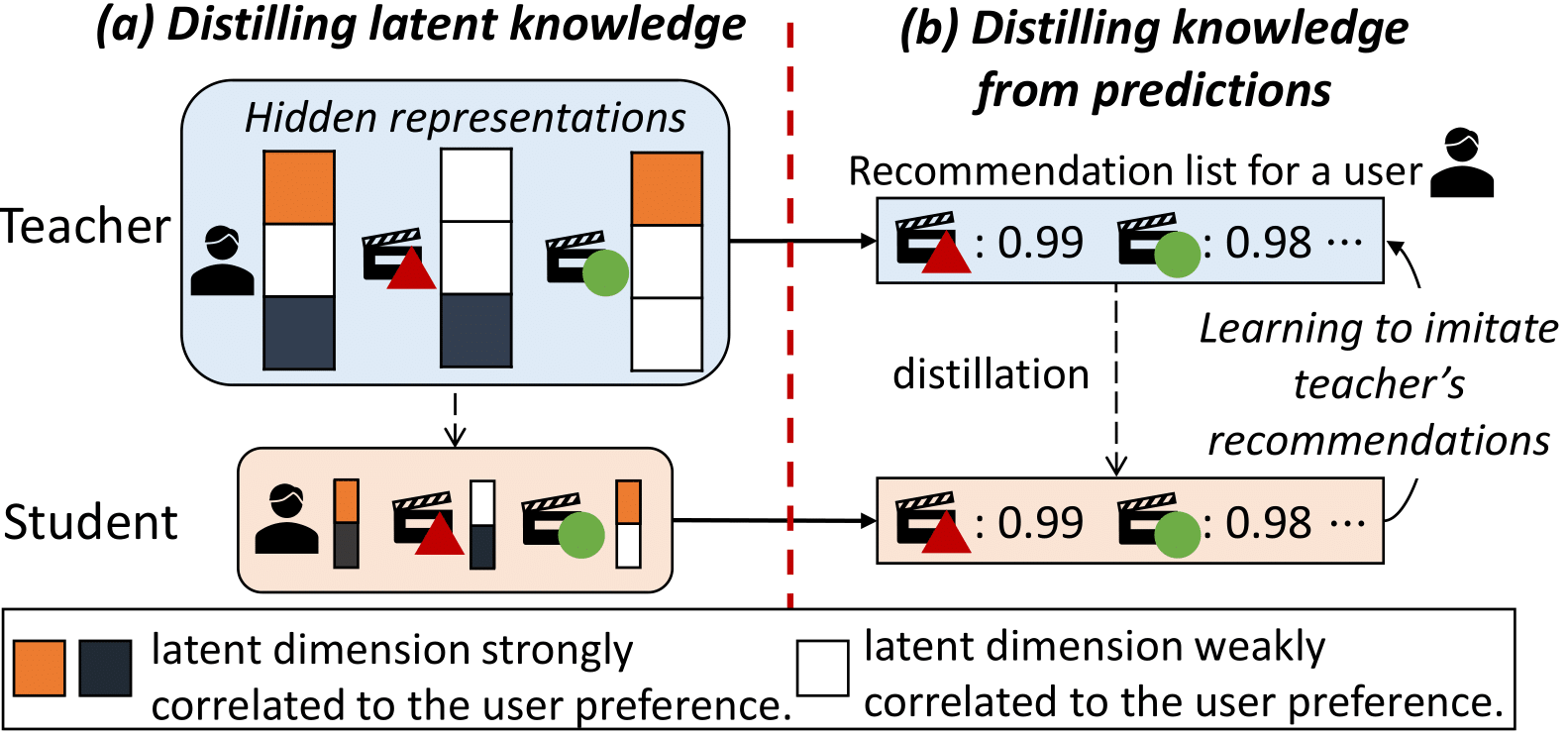}
  \caption{The existing methods \cite{RD, CD} distill the knowledge only based on the teacher’s predictions (b). The proposed framework directly distills the latent knowledge stored in the teacher (a) along with the knowledge revealed from the predictions (b).}
  \label{fig:de_rrd_overview}
\end{figure}

However, there are still limitations in existing methods \cite{RD, CD}.
First, the learning of the student is only guided by the teacher's prediction results, which is not sufficient to fully take advantage of the knowledge stored in the teacher.
This is because the prediction results incompletely reveal the teacher’s knowledge.
As illustrated in Figure \ref{fig:de_rrd_overview}, the recommendation list from the teacher only shows that a user has a similar degree of preference on the two items (0.99 and 0.98).
However, \textit{latent knowledge} in the teacher, which is used to make such predictions, contains more detailed information that the user likes different aspects of the two items (marked as navy blue and orange, respectively).
In this regard, we argue that the training process and the performance of the student can be further improved by directly distilling such latent knowledge stored in the teacher model.
Second, they distill the knowledge from the teacher's predictions in a point-wise manner that considers a single item at a time.
Because the point-wise approach does not consider multiple items simultaneously, it has a limitation in accurately maintaining the ranking orders predicted by the teacher model \cite{NCR}, which leads to degraded recommendation performance.

In this paper, we propose a novel knowledge distillation framework for RS, named DE-RRD, which distills both the latent knowledge stored in the teacher model (Fig. \ref{fig:de_rrd_overview}a) and the knowledge revealed from teacher's predictions (Fig. \ref{fig:de_rrd_overview}b).
By learning both the teacher's final predictions and the detailed knowledge that provides the bases for such predictions, the student model can be further improved.
The proposed framework consists of two methods: 1) Distillation Experts (DE) and 2) Relaxed Ranking Distillation (RRD).
The main contributions of this paper lie in the following aspects:

\vspace{3pt} \noindent
\textbf{Distilling latent knowledge in the teacher model.}
We propose a novel method---\textbf{DE}---for directly distilling latent knowledge stored in the teacher model.
Specifically, DE transfers the knowledge from hidden representation space (i.e., the output of the intermediate layer) of the teacher to the representation space of the student.
Due to the limited capacity, the student model cannot learn all the knowledge in the teacher representation space.
DE first introduces an ``expert'', which is a small feed-forward network, to distill the summarized knowledge that can restore the detailed knowledge of each entity in the teacher.
However, distilling the knowledge of all entities with a single expert intermingles the information of weakly correlated entities and further hinders the entire distillation process.
To tackle this problem, DE adopts the multiple experts and a novel expert selection strategy that clearly distinguishes the knowledge that each expert distills based on the correlations among the entities in the teacher representation space.
To the best of our knowledge, our approach is the first attempt to directly distill the latent knowledge in the teacher model for RS.
We demonstrate its rationality and superiority through extensive experiments and comprehensive analyses.

\vspace{1.5pt} \noindent
\textbf{Relaxed Ranking Distillation from the teacher’s predictions.}
We propose a new method---\textbf{RRD}---that transfers the knowledge from the teacher's predictions with direct consideration of ranking orders among items.
Unlike the existing methods \cite{RD, CD} that distill the knowledge of an item at a time, RRD formulates this as a ranking matching problem between the recommendation list of the teacher and that of the student.
To this end, RRD adopts the list-wise learning-to-rank approach \cite{xia2008list-wise} and learns to ensure the student to preserve the ranking orders predicted by the teacher.
However, directly applying the list-wise approach can have adverse effects on the recommendation performance.
Since a user is interested in only a few items among the numerous total items \cite{candidategeneration}, learning the detailed ranking orders of all items is not only daunting but also ineffective.
To tackle this challenge, RRD reformulates the daunting task to a \textit{relaxed ranking matching} problem.
Concretely, RRD matches the recommendation list from the teacher and that from the student, \textit{ignoring} the detailed ranking orders among the uninteresting items that the user would not be interested in.
RRD achieves superior recommendation performance compared to the state-of-the-art methods \cite{RD, CD}.



\vspace{2pt} \noindent
\textbf{An unified framework.}
We propose a novel framework---\textbf{DE-RRD}---which enables the student model to learn both from the teacher’s predictions and from the latent knowledge stored in the teacher model.
Our extensive experiments on real-world datasets show that DE-RRD considerably outperforms the state-of-the-art competitors. 
DE-RRD achieves comparable performance to that of the teacher with a smaller number of learning parameters than all the competitors.
Also, DE-RRD shows the largest performance gain when the student has the identical structure to the teacher model (i.e., self-distillation \cite{self_distill1}).
Furthermore, we provide both qualitative and quantitative analyses to further investigate the superiority of each proposed component.
The source code of DE-RRD is publicly available\footnote{\url{https://github.com/SeongKu-Kang/DE-RRD_CIKM20}}.

\section{Related Work}
\label{sec:DE-RRD_relatedwork}
\label{reference}
Balancing \textit{effectiveness} and \textit{efficiency} is a key requirement for real-time recommender system (RS);
the system should provide \textit{accurate recommendations} with \textit{fast inference time}.
Recently, the size of the recommender model is continuously increasing, and the computational time and memory cost required for the inference are also increasing accordingly \cite{RD, CD, GCN_distill, DCF}.
Due to the high latency, it becomes difficult to apply such large recommender to the real-time large-scale platform.
In this section, we review several approaches to alleviate this problem.

\vspace{2pt} \noindent
\textbf{Balancing Effectiveness and Efficiency.}
Several methods have adopted hash techniques to reduce the inference cost \cite{hash1, hash2, DCF, candidategeneration}.
They first learn binary representations of users and items, then construct the hash table.
Although exploiting the binary representation can significantly reduce the inference costs, due to the constrained capability, their recommendation performance is limited compared to models that use real-values representations.
In addition, several work has focused on accelerating the inference of the existing recommenders \cite{pruning_RS2_inner_only, tree_RS, compression1}.
Specifically, tree-based data structures \cite{ KDtree}, data compression techniques \cite{compression1}, and approximated nearest neighbor search techniques \cite{LSH, LSH_inner_product} have been successfully adopted to reduce the  search costs.
However, they still have problems such as applicable only to specific models (e.g., k-d tree for metric learning-based models \cite{METAS}), or easily falling into a local optimum due to the local search.

\vspace{2pt} \noindent
\textbf{Knowledge Distillation.}
Knowledge distillation (KD) is a model-agnostic strategy to improve the learning and the performance of a new “compact” model (student) by transferring knowledge from a previously trained “large” model (teacher) \cite{KD, FitNet, chen2017learning, self_distill1}.
The student model trained with KD has comparable performance to that of the teacher model, and also has lower inference latency due to its ~small size.
Most KD methods have focused on the image classification problem.
An early work \cite{KD} matches the softmax distribution of the teacher and the student.
The predicted label distribution contains more rich information (e.g., inter-class correlation) than the one-hot class label, which leads to improved learning of the student model.
Subsequent methods \cite{FitNet, chen2017learning} have focused on distilling knowledge from intermediate layers.
Because teacher's intermediate layers are generally bigger than that of the student, they \cite{FitNet, chen2017learning} utilize additional layers to bridge the different dimensions.
Interestingly, KD has turned out to be effective in improving the teacher model itself by self-distillation \cite{self_distill1}.

\vspace{2pt} \noindent
\textbf{Knowledge Distillation in Recommender System.}
Recently, inspired by the huge success of KD in the computer vision field, a few work \cite{RD, CD} have adopted KD to RS.
A pioneer work is Ranking Distillation (RD) \cite{RD} which applies KD for the ranking problem; Providing recommendations of top-$N$ unobserved items that have not interacted with a user.
RD jointly optimizes a base recommender's loss function with a \textit{distillation loss}.
\begin{equation}
    \begin{aligned}
\min_{\theta_{s}} \mathcal{L}_{B a s e} + \lambda \mathcal{L}_{R D}
    \end{aligned}
\end{equation}
where $\theta_{s}$ is the learning parameters of the student model, $\lambda$ is a hyperparameter that controls the effects of RD.
The base recommender can be any existing RS model such as BPR \cite{BPR}, NeuMF \cite{NeuMF}, and $\mathcal{L}_{B a s e}$ is its loss function (e.g., binary cross-entropy).
The distillation loss of RD for user $u$ is defined as follows:
\begin{equation}
    \begin{aligned}
    \mathcal{L}_{R D} = - \sum_{\pi_k \in \boldsymbol{\pi}} w_{\pi_k} \log \bigl( P\left(rel=1 | u, \pi_k \right)  \bigr)
    \end{aligned}
\end{equation}
where $\boldsymbol{\pi}$ is a ranked list of top-$K$ unobserved items for user $u$ predicted by the teacher, $\pi_k$ is the $k$-th item in this ranking, and $P\left(rel=1 | u, \pi_k \right)$ is the relevance probability of user $u$ to $\pi_k$ predicted by the student model. 
$w_{\pi_k}$ is the weight, which is computed based on each item's ranking from the student and the teacher, for reflecting relative importance among top-$K$ items.

A subsequent work
Collaborative Distillation (CD) \cite{CD} first samples unobserved items from the teacher's recommendation list according to their ranking; high-ranked items are more frequently sampled, then trains the student to mimic the teacher's prediction score (e.g., relevance probability) on the sampled items.
The distillation loss of CD for user $u$ is defined as follows:
\begin{equation}
    \begin{aligned}
\mathcal{L}_{C D} = - & \Bigl( \sum_{\pi_k \in \boldsymbol{\pi}} q_{\pi_k} \log \bigl( P\left(rel=1 | u, \pi_k \right)  \bigr)  \\ &+ (1-q_{\pi_k}) \log \bigl( 1-P\left(rel=1 | u, \pi_k \right)  \bigr) \Bigr)
    \end{aligned}
\end{equation}
where $\boldsymbol{\pi}$ is a ranked list of $K$ unobserved items sampled from teacher's recommendations for user $u$, $q_{\pi_k}$ is the weight, which is computed based on teacher's prediction score on each item, for reflecting relative importance among the sampled items.

In summary, the distillation loss of the existing methods makes the student model follow the teacher's predictions on unobserved items with particular emphasis on the high-ranked items.
In RS, only high-ranked items in the recommendation list are matter.
Also, such high-ranked items reveal hidden patterns among entities (i.e., users and items); the high-ranked items in the recommendation list would have strong correlations to the user \cite{RD}.
By using such additional supervisions from the teacher, they have achieved the comparable performance to the teacher with faster inference time.

However, the existing methods still have room for improvement by the following reasons:
First, the student can be further improved by directly distilling the \textit{latent knowledge} stored in the teacher model.
Latent knowledge refers to all information of users, items, and relationships among them that is discovered and stored in the teacher model.
Such knowledge is valuable for the student because it provides detailed explanations on the final prediction of the teacher.
Second, they transfer the knowledge from the teacher's predictions with a point-wise approach that considers a single item at a time.
Since the point-wise approach does not take into account multiple items simultaneously, it has a limitation in accurately maintaining the ranking orders in the teacher's ranking list \cite{NCR}.
This can lead to limited recommendation performance.

\section{Problem Formulation}
\label{sec:DE-RRD_problem}
In this work, we focus on top-$N$ recommendations for implicit feedback.
Let $\mathcal{U}$ and $\mathcal{I}$ denote the set of users and items, respectively.
Given collaborative filtering (CF) information (i.e., implicit interactions between users and items), we build a binary matrix $\boldsymbol { R } \in \{0,1\}^{| \mathcal { U } | \times | \mathcal { I } |}$. 
Each element of $\boldsymbol { R }$ has a binary value indicating whether a user has interacted with an item (1) or not (0).
Note that an unobserved interaction does not necessarily mean a user's negative preference on an item, it can be that the user is not aware of the item.
For each user, a recommender model ranks all items that have not interacted with the user (i.e., unobserved items) and provides a ranked list of top-$N$ unobserved items.

The knowledge distillation is conducted as follows:
First, a teacher model with a large number of learning parameters is trained with the training set which has binary labels.
Then, a student model with a smaller number of learning parameters is trained with the help from the teacher model in addition to the binary labels.
The goal of KD is to fully improve the inference efficiency without compromising the effectiveness;
We aim to design a KD framework that enables the student model to maintain the recommendation performance of the teacher with a small number of learning parameters.

\section{Proposed Framework---DE-RRD}
\label{sec:DE-RRD_method}
\begin{figure}[t]
  \includegraphics[width=0.99\textwidth]{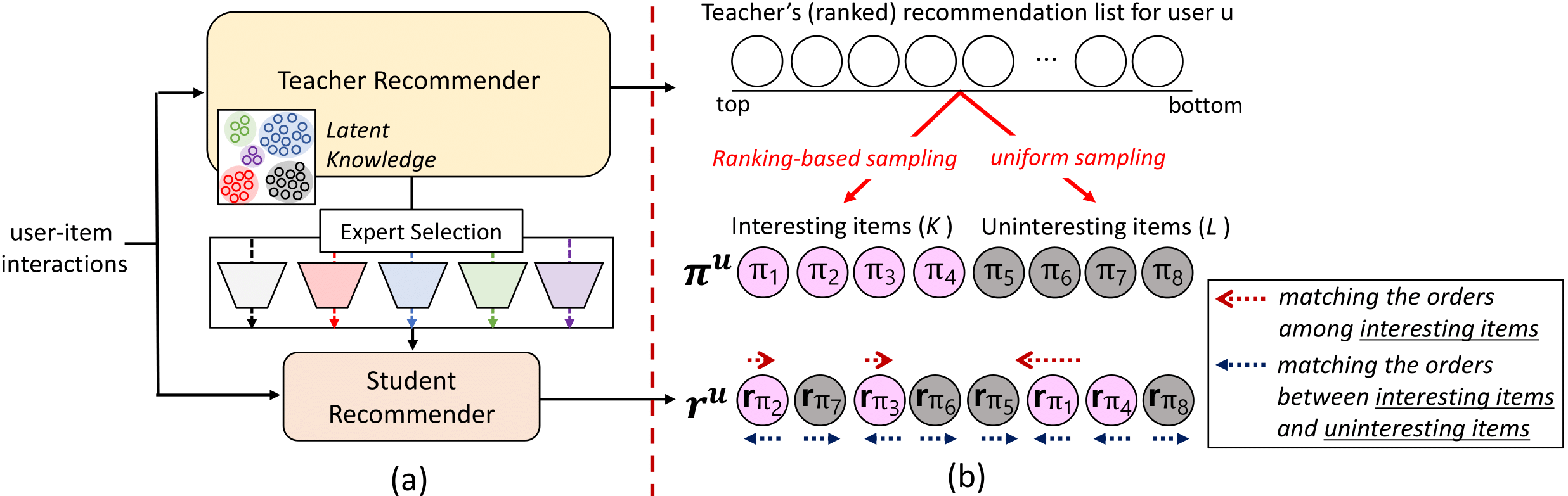}
  \caption{Illustration of DE-RRD framework.  (a) \textit{Distillation Experts} (DE) directly distills the teacher's latent knowledge with the experts and the selection strategy. (b) \textit{Relaxed Ranking Distillation} (RRD) distills the knowledge from the teacher's prediction based on the relaxed ranking approach that ignores  orders among the uninteresting items. Best viewed in color.}
  \label{fig:de-rrd_method}
\end{figure}

We propose DE-RRD framework which enables the student model to learn both from the teacher’s predictions and from the latent knowledge encoded in the teacher model.
DE-RRD consists of two methods: 1) \textit{Distillation Experts} (DE) that directly transfers the latent knowledge from the teacher, 2) \textit{Relaxed Ranking Distillation} (RRD) that transfers the knowledge revealed from the teacher’s predictions with direct consideration of ranking orders among items. 
This section is organized as follows.
We first describe each component of the proposed framework: DE in Section \ref{sec:DE-RRD_DE}, RRD in Section \ref{sec:DE-RRD_RRD}.
Then, we explain the end-to-end optimization process in Section \ref{sec:DE-RRD_opt}. 
The overview of DE-RRD is provided in Figure \ref{fig:de-rrd_method}.

\subsection{Distillation Experts (DE)}
\label{sec:DE-RRD_DE}
In this section, we provide the details of DE which distills the latent knowledge from the hidden representation space (i.e., the output of the intermediate layer) of the teacher to the corresponding representation space of the student.
We first introduce ``expert'' to distill the summarized knowledge that can restore the detailed teacher's knowledge of each entity.
Then, we introduce a novel expert selection strategy for effectively distilling CF knowledge that contains information of all the entities having diverse preferences and characteristics.

\subsubsection{\textbf{Expert for distillation}}\noindent
DE exploits “expert” to distill knowledge from the teacher's hidden representation space.
An expert, which is a small feed-forward network, is trained to \textit{reconstruct} the representation on a selected intermediate layer of the teacher from the representation on the corresponding intermediate layer of the student.
Let $h_t(\cdot)$ denote a mapping function to the representation space ($\in \mathbb{R}^{d_t}$) of the teacher model (i.e., a nested function up to the intermediate layer of the teacher).
Similarly, let $h_s(\cdot)$ denote a mapping function to the student’s representation space ($\in \mathbb{R}^{d_s}$).
The output of the mapping function can be a separate representation of a user, an item (e.g., BPR \cite{BPR}) or their combined representation (e.g., NeuMF \cite{NeuMF}) based on the base model’s structure and the type of selected layer.
Here, we use user $u$ as an example for convenience.
An expert $E$ is trained to reconstruct $h_{t}\left(u\right)$ from $h_{s}\left(u\right)$ as follows:
\begin{equation}
\begin{aligned}
\mathcal{L}(u)=\| h_{t}\left(u\right) - E \bigl(h_{s}\left(u\right)\bigr)\|_2
\end{aligned}
\end{equation}
Note that in the KD process, the teacher model is already trained and frozen.
By minimizing the above equation, parameters in the student model (i.e., $h_s(\cdot)$) and the expert are updated.

The student model has smaller capacity compared to the teacher ($d_s << d_t$). 
By minimizing the equation 4, the student learns compressed information on the user's preference that can restore more detailed knowledge in the teacher as accurate as possible.
This approach provides a kind of filtering effect and improves the learning of the student model.

\subsubsection{\textbf{Expert selection strategy}}\noindent
Training a single expert to distill all the CF knowledge in the teacher is not sufficient to achieve satisfactory performance.
The CF knowledge contains vast information of user groups with various preferences and item groups with diverse characteristics.
When a single expert is trained to distill the knowledge of all the diverse entities, the information of the weakly correlated entities (e.g., users that have dissimilar preferences) is mixed and reflected in the expert's weights.
This leads to the adulterated distillation that hinders the student model from discovering some users' preferences.

To alleviate the problem, DE puts multiple experts in parallel and clearly distinguishes the knowledge that each expert distills.
The key idea is to divide the representation space into exclusive divisions based on the teacher's knowledge and make each expert to be specialized in distilling the knowledge in a division (Fig. \ref{fig:de-rrd_method}a).
The representations belonging to the same division has strong correlations with each other, and they are distilled by the same expert without being mixed with weakly correlated representations belonging to the different divisions.
The knowledge transfer of DE is conducted in the two steps:
(1) a selection network first computes each expert's degree of specialization for the knowledge to be distilled.
(2) DE selects an expert based on the computed distribution, then distills the knowledge through the selected expert.

Concretely, DE has $M$ experts ($E_1, E_2, ... , E_M$) and a selection network $S$ whose output is $M$-dimensional vector.
To distill user $u$'s knowledge from the teacher, the selection network $S$ first computes the normalized specialization score vector $\boldsymbol{\alpha}^{u} \in \mathbb{R}^{M}$ as follows:
\begin{equation}
\begin{aligned}
\mathbf{e}^{u} &= S\bigl(h_{t}\left(u\right)\bigr),\\
\alpha^{u}_{m} &= \frac{\exp \left(e^{u}_m\right)}{\sum_{i=1}^{M} \exp(e^{u}_i)} \quad \text{for} \quad m = 1, ..., M
\end{aligned}
\end{equation}

\noindent
Then, DE selects an expert based on the computed distribution.
We represent the selection variable $\mathbf{s}^{u}$ that determines which expert to be selected for distilling $h_t(u)$.
$\mathbf{s}^{u}$ is a $M$-dimensional one-hot vector where an element is set to 1 if the corresponding expert is selected for distillation.
DE samples this selection variable $\mathbf{s}^u$ from a multinoulli distribution parameterized by $\{\alpha^{u}_m\}$ i.e., $p\left(s^{u}_{m}=1 | S, h_{t}\left(u\right)\right) = \alpha^{u}_{m}$, then reconstructs teacher's representation as follows:
\begin{equation}
\begin{aligned}
\mathbf{s}^{u} &\sim \text{Multinoulli}_M \left(\{\alpha^{u}_{m}\}\right)\\
\mathcal{L}(u) &=\|h_{t}\left(u\right)-\sum_{m=1}^{M} s^{u}_m \cdot E_{m}\bigl(h_{s}\left(u\right)\bigr)\|_{2}
\end{aligned}
\end{equation}
\noindent
However, the sampling process is non-differentiable, which would block the gradient flows and disable the end-to-end training. 
As a workaround, we adopt a continuous relaxation of the discrete distribution by using Gumbel-Softmax \cite{GumbelSoftmax}. 
The Gumbel-Softmax is a continuous distribution on the simplex that can approximate samples from a categorical distribution; it uses the Gumbel-Max trick \cite{GumbelMax} to draw samples from the categorical distribution, then uses the softmax function as a continuous approximation of argmax operation to get the approximated one-hot representation.
With the relaxation, the selection network can be trained by the backpropagation.

\begin{figure}[t]
\centering
  \includegraphics[height=3.5cm]{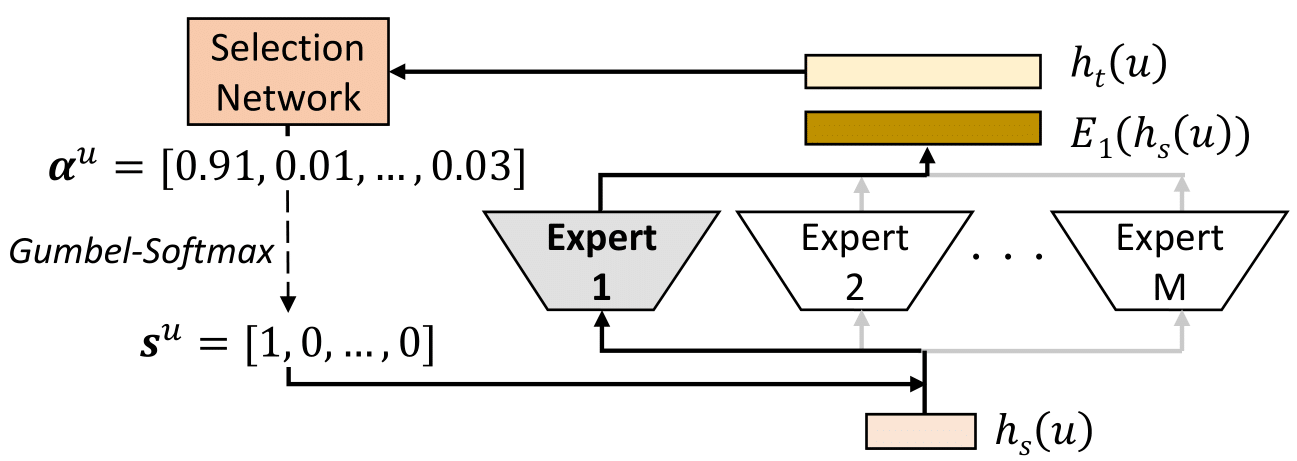}
  \caption{Illustration of the expert selection process of DE. During the training, $\mathbf{s}^{u}$ becomes a one-hot vector and selects the most specialized expert in the knowledge to be distilled.}
  \label{fig:de-rrd_selection}
\end{figure}

DE gets the approximated one-hot selection variable $\mathbf{s}^{u}$ by using the Gumbel-Softmax and reconstructs the teacher's representation as~follows:
\begin{equation}
\begin{aligned}
s^{u}_{m} &= \frac{\exp \Bigl(\left(\log \alpha^{u}_m +g_{m} \right) / \tau \Bigr)}{\sum_{i=1}^{M} \exp \Bigl(\bigl(\log \alpha^{u}_i +g_{i}\bigr)/ \tau\Bigr)}  \quad \text{for} \quad m = 1, ..., M\\
&\mathcal{L}(u) =\|h_{t}\left(u\right)-\sum_{m=1}^{M} s^{u}_m \cdot E_{m}\bigl(h_{s}\left(u\right)\bigr)\|_{2}
\end{aligned}
\end{equation}
where $g_i$ is i.i.d drawn from Gumbel$(0, 1)$ distribution\footnote{$g_i=-\text{log}(-\text{log}(r))$, where $r$ is sampled from $Uniform(0,1)$.}.
The extent of relaxation is controlled by a temperature parameter $\tau$.
In the beginning of the training, we set a large value on $\tau$, and gradually decreases its value during the training.
As $\tau$ is decreased to $0$, 
$\mathbf{s}^{u}$ smoothly becomes one-hot vector where $s^{u}_m = 1$ with probability $\alpha^{u}_m$.
In other words, during the training, each expert gradually gets specialized on certain information that has strong correlations.
This process is illustrated in Figure \ref{fig:de-rrd_selection}.

\vspace{1pt} \noindent
\textbf{Discussion: Effects of expert selection.}
As the expert selection is based on the teacher's knowledge, correlations among the entities in the teacher representation space are naturally reflected in the expert selection;
the user representations with very similar preferences (i.e., located closely in the space) would be distilled by the same expert with a high probability.
This allows each expert to be trained to distill only the knowledge of strongly correlated entities, and thus each expert can provide better guidance that does not include the information of weakly correlated entities.

\vspace{1pt} \noindent
\textbf{Discussion: selection vs. attention.}
Instead of selecting one expert, the attention mechanism (i.e., the softmax function) can be adopted.
However, we think the selection is a more appropriate choice to distill the CF knowledge containing all the entities having diverse preferences and characteristics.
This is because the attention makes every expert involved in distilling the knowledge of each entity.
In other words, like in the case of a single expert, all the experts and attention network are trained to minimize the overall reconstruction errors of all the diverse entities. 
By doing so, information of weakly relevant entities gets mixed together, and this leads to performance degrade in some user groups.
We provide experiment results to support our claims. Please refer to Section \ref{sec:DE-RRD_experiments}.

\subsubsection{\textbf{Optimization of DE}}\noindent
DE is jointly optimized with the base model's loss function in the end-to-end manner as follows:
\begin{equation}
\begin{aligned}
\min_{\theta_{s}, \theta_{DE}} \mathcal{L}_{B a s e} + \lambda_{D E} \cdot \mathcal{L}_{D E}
\end{aligned}
\end{equation}
where $\theta_{s}$ is the learning parameters of the student model, $\theta_{DE}$ is the learning parameters of DE (i.e., the selection network and the experts), and $\lambda_{D E}$ is a hyperparameter that controls the effects of DE.
The base model can be any existing recommender (e.g., BPR, NeuMF), and $\mathcal{L}_{B a s e}$ corresponds to its loss function.
Note that the experts are not used in the inference phase.

The loss function of DE can be flexibly defined based on the base model's structure and the types of hidden layer chosen for the distillation.
Concretely, for NeuMF \cite{NeuMF}, which is a state-of-the-art deep recommender, the loss function can be defined to 1) separately distill knowledge of users and items in a mini-batch (i.e., $\sum_{u \in B} \mathcal{L}(u) +  \sum_{i \in B} \mathcal{L}(i)$) 
or 2) distill the combined knowledge (i.e., $\sum_{(u, i) \in B} \mathcal{L}(u, i)$). 
Also, we adopt a simple temperature annealing schedule, which gradually decays the temperature from $\tau_0$ to $\tau_P$ as done in \cite{DRE}:
$\tau (p)=\tau_0(\tau_P / \tau_0)^{p/P}$ where $\tau(p)$ is the temperature at epoch $p$, and $P$ is the total training epochs.

\subsection{Relaxed Ranking Distillation (RRD)}
\label{sec:DE-RRD_RRD}
We propose RRD, a new method to distill the knowledge revealed from the teacher's predictions with direct consideration of ranking orders among items.
RRD formulates this as a ranking matching problem between the recommendation list of the teacher model and that of the student model.
To this end, RRD adopts the classical list-wise learning-to-rank approach \cite{xia2008list-wise}.
Its core idea is to define a probability of a permutation (i.e., a ranking order) based on the ranking score predicted by a model, and train the model to maximize the likelihood of the ground-truth ranking order.
For more details about the list-wise approach, please refer to \cite{xia2008list-wise}.

However, merely adopting the list-wise loss can have adverse effects on the ranking performance.
Because a user is interested in only a few items among the numerous total items \cite{candidategeneration}, learning the detailed ranking orders of all the unobserved items is not only daunting but also ineffective.
The recommendation list from the teacher model contains information about a user’s potential preference on each unobserved item; A few items that the user would be interested in (i.e., interesting items) are located near the top of the list, whereas the majority of items that the user would not be interested in (i.e., uninteresting items) are located far from the top.

Based on this information, RRD reformulates the daunting task of learning all the precise ranking orders to a \textit{relaxed ranking matching} problem.
In other words, RRD aims to match the recommendation list from the teacher and that from the student, \textit{ignoring} the detailed ranking orders among the uninteresting items.
Concretely, RRD distills the information of (1) the detailed ranking orders among the interesting items, (2) the relative ranking orders between the interesting items and the uninteresting items.
The overview of RRD is provided in Figure \ref{fig:de-rrd_method}b.

\subsubsection{\textbf{Sampling interesting/uninteresting items}}\noindent
The first step of RRD is to sample items from the teacher's recommendation list.
In specific, RRD samples $K$ interesting items and $L$ uninteresting items for each user.
As a user would not be interested in the vast majority of items, the interesting items should be sampled from a very narrow range near the top of the list, whereas the uninteresting items should be sampled from the wide range of the rest.
To sample the interesting items, we adopt a ranking position importance scheme \cite{rendle2014improving, RD} that places more emphasis on the higher positions in the ranking list.
In the scheme, the probability of the $k$-th ranked item to be sampled is defined as: $p_k \propto e^{-k/T}$
where $T$ is the hyperparameter that controls emphasis on top positions.
With the scheme, RRD samples $K$ interesting items according to the user’s potential preference on each item (i.e., item's ranking) predicted by the teacher. 
To sample the uninteresting items that corresponds the majority of items, we use a simple uniform sampling.
Concretely, RRD uniformly samples $L$ uninteresting items from a set of items that have lower rankings than the previously sampled interesting~items.

\subsubsection{\textbf{Relaxed permutation probability}}\noindent
Then, RRD defines a relaxed permutation probability motivated by \cite{xia2008list-wise}.
For user $u$, $\boldsymbol{\pi^{u}}$ denotes a ranked list of all the sampled items ($K+L$) sorted by the original order in the teacher's recommendation list.
$\mathbf{r}^{u}$ denotes ranking scores on the sampled items predicted by the student model.
The relaxed permutation probability is formulated as follows:
\begin{equation}
\begin{aligned}
    p\left(\boldsymbol{\pi}^u_{1:K} | \mathbf{r}^u\right)=\prod_{k=1}^{K} \frac{\exp ({r}^u_{\pi_{k}})}{\sum_{i=k}^{K} \exp ({r}^u_{\pi_{i}})+\sum_{j=K}^{K+L} \exp ({r}^u_{\pi_{j}})}
\end{aligned}
\end{equation}
where ${r}^{u}_{\boldsymbol{\pi}_k}$ denotes a ranking score predicted by the student for the $k$-th item in $\boldsymbol{\pi}^u$, $\boldsymbol{\pi}^u_{1:K}$ denotes the partial list that contains the interesting items.
RRD learns to maximize the log-likelihood $\log p\left(\boldsymbol{\pi}_{1:K} | \mathbf{r}\right)$ for all users.
The proposed permutation probability is not affected by the detailed ranking orders among the uninteresting items ($L$).
By maximizing the log-likelihood, the student model is trained to locate all the interesting items ($K$) higher than all the uninteresting items ($L$) in the recommendation list, while maintaining the detailed ranking orders (from the teacher's recommendation list) among the interesting items.

\subsubsection{\textbf{Optimization of RRD}}\noindent
RRD is jointly optimized with the base model's loss function in the end-to-end manner as follows:
\begin{equation}
\begin{aligned}
\min_{\theta_{s}} \mathcal{L}_{B a s e} + \lambda_{R R D} \cdot \mathcal{L}_{R R D}
\end{aligned}
\end{equation}
where $\theta_{s}$ is the learning parameters of the student model and $\lambda_{R R D}$ is a hyperparameter that controls the effects of RRD.
The base model can be any existing recommender, and $\mathcal{L}_{B a s e}$ corresponds to its loss function.
The sampling process is conducted at every epoch.
The loss function of RRD is defined to distill the knowledge of users in the mini-batch: $-\frac{1}{|B|}\sum_{u \in B} \log p(\boldsymbol{\pi}^{u}_{1:K}|\mathbf{r}^{u})$.

\subsection{Optimization of DE-RRD}
\label{sec:DE-RRD_opt}
The proposed DE-RRD framework is optimized in the end-to-end manner as follows:
\begin{equation}
\begin{aligned}
\min_{\theta_{s}, \theta_{DE}} \mathcal{L}_{B a s e} + \lambda_{D E} \cdot \mathcal{L}_{D E} + \lambda_{R R D} \cdot \mathcal{L}_{R R D}
\end{aligned}
\end{equation}
where $\theta_{s}$ is the learning parameters of the student model, $\theta_{DE}$ is the learning parameters of DE (i.e., the selection network and the experts).
The base model can be any existing recommender, and $\mathcal{L}_{B a s e}$ corresponds to its loss function.

\section{Experiments}
\label{sec:DE-RRD_experiments}
We validate the superiority of DE-RRD on 12 experiment settings (2 real-world datasets $\times$ 2 base models $\times$ 3 different student model sizes).
We first provide extensive experiment results supporting that DE-RRD outperforms the state-of-the-art competitors.
We also provide both quantitative and qualitative analyses to verify the rationality and superiority of each proposed component.
Lastly, we provide hyperparameter study.

\subsection{Experimental Setup}
\noindent
\textbf{Datasets.}
We use two public real-world datasets: CiteULike \cite{wang2013collaborative}, Foursquare \cite{liu2017experimental}.
We remove users and items having fewer than five ratings for CiteULike, twenty ratings for Foursquare as done in \cite{BPR, NeuMF, SSCDR}.
Data statistics are summarized in Table \ref{tbl:de-rrd_statistic}.
\begin{table}[h]
\centering
\renewcommand{\arraystretch}{0.7}
  \caption{Data Statistics (after preprocessing)}
  \begin{tabular}{ccccc}
    \toprule
    Dataset & \#Users & \#Items & \#Interactions & Sparsity \\
    \midrule
    CiteULike & 5,220 & 25,182 & 115,142 & 99.91\% \\
    Foursquare & 19,466 & 28,594 & 609,655 & 99.89\% \\
    \bottomrule
  \end{tabular}
    \label{tbl:de-rrd_statistic}
\end{table}

\vspace{2pt} \noindent
\textbf{Base Models.}
We validate the proposed framework on base models that have different architectures and optimization strategies.
We choose a latent factor model and a deep learning model that are broadly used for top-$N$ recommendation with implicit feedback.
\begin{itemize}[leftmargin=*]
    \item \textbf{BPR \cite{BPR}}: 
    A learning-to-rank model for implicit feedback.
    It assumes that observed items are more preferred than unobserved items and optimizes Matrix Factorization (MF) with the pair-wise ranking loss function.
    \item \textbf{NeuMF \cite{NeuMF}}: The state-of-the-art deep model for implicit feedback. 
    NeuMF combines MF and Multi-Layer Perceptron (MLP) to learn the user-item interaction, and optimizes it with the point-wise objective function (i.e., binary cross-entropy).
\end{itemize}

\vspace{2pt} \noindent
\textbf{Teacher/Student.}
For each base model and dataset, we increase the number of learning parameters until the recommendation performance is no longer increased, and use the model with the best performance as Teacher model.
For each base model, we build three student models by limiting the number of learning parameters.
We adjust the number of parameters based on the size of the last hidden layer.
The limiting ratios ($\phi$) are \{0.1, 0.5, 1.0\}.
Following the notation of the previous work \cite{RD, CD}, we call the student model trained without the help of the teacher model (i.e., no distillation) as ``Student'' in this experiment sections.

\vspace{2pt} \noindent
\textbf{Comparison Methods.}
The proposed framework is compared with the following methods:
\begin{itemize}[leftmargin=*]
    \item \textbf{Ranking Distillation (RD) \cite{RD}}: A KD method for recommender system that uses items with the highest ranking from the teacher's predictions for distilling the knowledge.
    \item \textbf{Collaborative Distillation (CD) \cite{CD}}: The state-of-the-art KD method for recommender system.
    CD samples items from teacher's predictions based on their ranking, then uses them for distillation.
    As suggested in the paper, we use unobserved items only for distilling the knowledge.
\end{itemize}
Finally, \textbf{DE-RRD} framework consists of the following two methods:
\begin{itemize}[leftmargin=*]
    \item \textbf{Distillation Experts (DE)}: A KD method that directly distills the latent knowledge stored in the teacher model.
    It can be combined with any \textit{prediction-based} KD methods (e.g., RD, CD, RRD).
    \item \textbf{Relaxed Ranking Distillation (RRD)}: A KD method that distills the knowledge revealed from the teacher’s predictions with consideration of relaxed ranking orders among items. 
\end{itemize}

\noindent
\textbf{Evaluation Protocol.}
We follow the widely used \textit{leave-one-out} evaluation protocol \cite{NeuMF, transCF, SSCDR}.
For each user, we leave out a single interacted item for testing, and use the rest for training.
In our experiments, we leave out an additional interacted item for the validation.
To address the time-consuming issue of ranking all the items, we randomly sample 499 
items from a set of unobserved items of the user, then evaluate how well each method can rank the test item higher than these sampled unobserved items. 
We repeat this process of sampling a test/validation item and unobserved items five times and report the average results.

As we focus on the top-$N$ recommendation task based on implicit feedback, we evaluate the performance of each method with widely used three ranking metrics \cite{NeuMF, SSCDR, candidategeneration}: 
hit ratio (H@$N$), normalized discounted cumulative gain (N@$N$), and mean reciprocal rank (M@$N$). 
H@$N$ measures whether the test item is present in the top-$N$ list, while N@$N$ and M@$N$ are position-aware ranking metrics that assign higher scores to the hits at upper ranks.

\vspace{2pt} \noindent
\textbf{Implementation Details for Reproducibility.}
We use PyTorch to implement the proposed framework and all the baselines, and use Adam optimizer to train all the methods.
For RD, we use the public implementation provided by the authors.
For each dataset, hyperparameters are tuned by using grid searches on the validation set. 
The learning rate for the Adam optimizer is chosen from \{0.1, 0.05, 0.01, 0.005, 0.001, 0.0005, 0.0001\}, the model regularizer is chosen from $\{10^{-1}, 10^{-2}, 10^{-3}, 10^{-4}, 10^{-5}\}$.
We set the total number of epochs as 1000, and adopt early stopping strategy; stopping if H@$5$ on the validation set does not increase for 30 successive epochs.
For all base models (i.e., BPR, NeuMF), the number of negative sample is set to 1, and no pre-trained technique is used.
For NeuMF, the number of the hidden layers is chosen from \{1, 2, 3, 4\}.

For all the distillation methods (i.e., RD, CD, DE, RRD), weight for KD loss ($\lambda$) is chosen from $\{1, 10^{-1}, 10^{-2}, 10^{-3}, 10^{-4}, 10^{-5}\}$.
For DE, the number of experts ($M$) is chosen from \{5, 10, 20, 30\}, MLP is employed for the experts and the selection network. 
The shape of the layers of an expert is [$d_s \rightarrow (d_s + d_t)/2 \rightarrow d_t$] with \textit{relu} activation, and that of the selection network is [$d_t \rightarrow M$].
We select the last hidden layer of all the base models to distill latent knowledge.
We put the experts according to the structure of the selected layer;
For the layer where user and item are separately encoded (i.e., BPR), we put $M$ user-side experts and $M$ item-side experts, and for the layer where user and items are jointly encoded (i.e., NeuMF), we put $M$ experts to distill the combined information.
$\tau_0$ and $\tau_{P}$ are set to $1, 10^{-10}$, respectively.
For prediction-based KD methods (i.e., RD, CD, RRD), the number of high-raked (or interesting) items ($K$) for distillation
is chosen from \{10, 20, 30, 40, 50\}, weight for controlling the importance of top position ($T$) is chosen from \{1, 5, 10, 20\}.
For RRD, the number of uninteresting items ($L$) is set to the same with $K$, but it can be further tuned.
For RD, the number of the warm-up epoch is chosen from \{30, 50, 100\}, the number of negative items in the dynamic weight is chosen from \{50, 100\}.
Also, RD and CD have additional hyperparameters for reflecting the relative importance of the items used for distillation.
We follow the recommended values from the public implementation and from the original papers.

\begin{sidewaystable}[!htbp]
\setlength\tabcolsep{7pt}
\small
\RowStretch{0.2}
  \caption{Recommendation performances ($\phi=0.1$). \textit{Improv.b} and \textit{Improv.s} denote the improvement of DE-RRD over the best baseline and student respectively.
  We conduct the paired t-test with 0.05 level on H@5 and all \textit{Improv.b} are statistically significant.
  }
  
  \begin{tabular}{cclccc ccc ccc}
    \toprule 
     Dataset & Base Model & KD Method & H@5 & M@5 & N@5 & H@10 & M@10 & N@10 & H@20 & M@20 & N@20 \\
    \midrule
     &&Teacher&0.5135&0.3583&0.3970&0.6185&0.3724&0.4310&0.7099&0.3788&0.4541\\
     &&Student&0.4441&0.2949&0.3319&0.5541&0.3102&0.3691&0.6557&0.3133&0.3906\\
     &&RD&0.4533&0.3019&0.3395&0.5601&0.3161&0.3740&0.6633&0.3232&0.3993\\
     &\multirow{6}{*}{BPR}&CD&0.4550&0.3025&0.3404&0.5607&0.3167&0.3746&0.6650&0.3240&0.4011\\
     \cmidrule{3-12}
     &&DE&0.4817&0.3230&0.3625&0.5916&0.3372&0.3977&0.6917&0.3441&0.4229\\
     &&RRD &0.4622&0.3076&0.3461&0.5703&0.3220&0.3809&0.6746&0.3293&0.4074\\
     &&DE-RRD &\textbf{0.4843}&\textbf{0.3231}&\textbf{0.3632}&\textbf{0.5966}&\textbf{0.3373}&\textbf{0.3989}&\textbf{0.6991}&\textbf{0.3447}&\textbf{0.4251}\\
     \cmidrule{3-12}
     &&\textit{Improv.b}&6.44\%&6.81\%&6.7\%&6.4\%&6.47\%&6.47\%&5.12\%&6.4\%&5.98\%\\
    \multirow{9}{*}{\rotatebox[origin=c]{0}{CiteULike}}&&\textit{Improv.s}&9.06\%&9.57\%&9.44\%&7.66\%&8.7\%&8.06\%&6.62\%&10.02\%&8.83\%\\
    \cmidrule{2-12}
    &&Teacher&0.4790&0.3318&0.3684&0.5827&0.3457&0.4020&0.6748&0.3521&0.4254\\
    &&Student&0.3867&0.2531&0.2865&0.4909&0.2670&0.3202&0.5833&0.2738&0.3436\\
    &&RD&0.4179&0.2760&0.3113&0.5211&0.2896&0.3444&0.6227&0.2958&0.3696\\
    &\multirow{6}{*}{NeuMF }&CD&0.4025&0.2633&0.2979&0.5030&0.2769&0.3306&0.6053&0.2822&0.3550\\
    \cmidrule{3-12}
    &&DE &0.4079&0.2625&0.2986&0.5139&0.2766&0.3328&0.6238&0.2843&0.3607\\
    &&RRD &0.4737&0.3086&0.3497&0.5800&0.3236&0.3847&0.6765&0.3305&0.4094\\
    &&DE-RRD &\textbf{0.4758}&\textbf{0.3108}&\textbf{0.3518}&\textbf{0.5805}&\textbf{0.3246}&\textbf{0.3856}&\textbf{0.6770}&\textbf{0.3312}&\textbf{0.4099}\\
    \cmidrule{3-12}
    &&\textit{Improv.b}&13.83\%&12.6\%&13.03\%&11.42\%&12.09\%&11.95\%&8.72\%&11.95\%&10.9\%\\
    &&\textit{Improv.s}&23.03\%&22.79\%&22.8\%&18.26\%&21.58\%&20.42\%&16.07\%&20.95\%&19.28\%\\
    \midrule
     &&Teacher&0.5598&0.3607&0.4101&0.7046&0.3802&0.4571&0.8175&0.3882&0.4859\\
    &&Student&0.4869&0.3033&0.3489&0.6397&0.3239&0.3984&0.7746&0.3338&0.4333\\
    &&RD&0.4932&0.3102&0.3555&0.6453&0.3302&0.4045&0.7771&0.3391&0.4377\\
    &\multirow{6}{*}{BPR}&CD&0.5006&0.3147&0.3608&0.6519&0.3354&0.3237&0.7789&0.3440&0.4421\\
    \cmidrule{3-12}
    &&DE &0.5283&0.3344&0.3824&0.6810&0.3544&0.4316&0.8032&0.3631&0.4627\\
    &&RRD &0.5132&0.3258&0.3722&0.6616&0.3455&0.4202&0.7862&0.3540&0.4516\\
    &&DE-RRD &\textbf{0.5308}&\textbf{0.3359}&\textbf{0.3843}&\textbf{0.6829}&\textbf{0.3565}&\textbf{0.4336}&\textbf{0.8063}&\textbf{0.3647}&\textbf{0.4647}\\
    \cmidrule{3-12}
     &&\textit{Improv.b}&6.03\%&6.74\%&6.51\%&4.76\%&6.29\%&7.19\%&3.52\%&6.02\%&5.11\%\\
    \multirow{9}{*}{\rotatebox[origin=c]{0}{Foursquare}}&&\textit{Improv.s}&9.02\%&10.75\%&10.15\%&6.75\%&10.06\%&8.84\%&4.09\%&9.26\%&7.25\%\\
    \cmidrule{2-12}
    &&Teacher&0.5436&0.3464&0.3954&0.6906&0.3662&0.4430&0.8085&0.3746&0.4731\\
    &&Student&0.4754&0.2847&0.3319&0.6343&0.3060&0.3833&0.7724&0.3157&0.4185\\
    &&RD&0.4789&0.2918&0.3380&0.6368&0.3110&0.3878&0.7761&0.3173&0.4205\\
    &\multirow{6}{*}{NeuMF}&CD&0.4904&0.2979&0.3456&0.6477&0.3156&0.3940&0.7845&0.3260&0.4293\\
    \cmidrule{3-12}
    &&DE &0.4862&0.2977&0.3444&0.6413&0.3174&0.3938&0.7742&0.3278&0.4284\\
    &&RRD &0.5172&0.3110&0.3621&0.6739&0.3321&0.4132&0.7982&0.3409&0.4450\\
    &&DE-RRD &\textbf{0.5193}&\textbf{0.3130}&\textbf{0.3641}&\textbf{0.6741}&\textbf{0.3332}&\textbf{0.4139}&\textbf{0.7983}&\textbf{0.3421}&\textbf{0.4454}\\
    \cmidrule{3-12}
     &&\textit{Improv.b}&5.89\%&5.07\%&5.35\%&4.08\%&5.58\%&5.05\%&1.76\%&4.94\%&3.75\%\\
    &&\textit{Improv.s}&9.23\%&9.94\%&9.7\%&6.27\%&8.89\%&7.98\%&3.35\%&8.36\%&6.43\%\\
    \bottomrule
  \end{tabular}
  \label{tab:de-rrd_main}
\end{sidewaystable}

\begin{figure*}[!htbp]
\centering
\begin{subfigure}[t]{0.8\linewidth}
    \includegraphics[width=\linewidth]{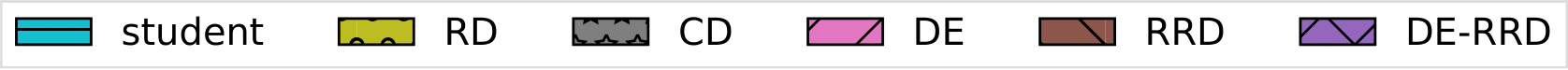}
\end{subfigure}

\begin{subfigure}[t]{0.4\linewidth}
    \includegraphics[width=\linewidth]{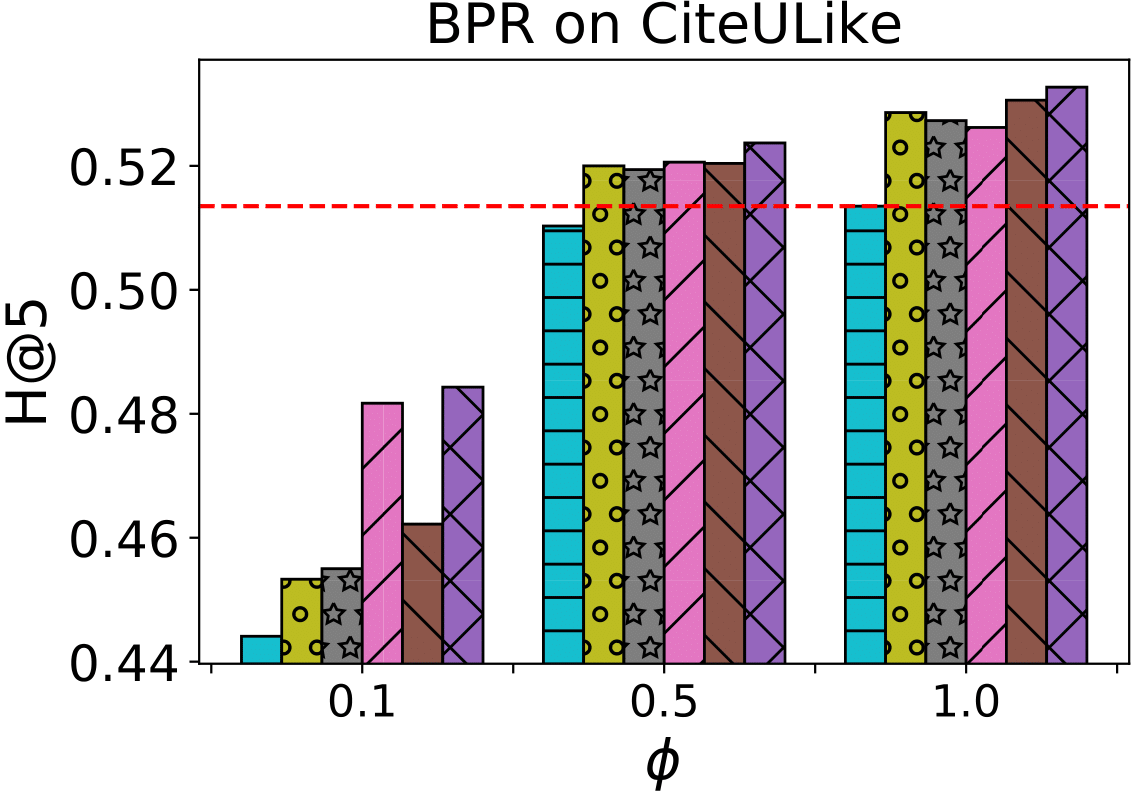}
\end{subfigure}
\begin{subfigure}[t]{0.4\linewidth}
    \includegraphics[width=\linewidth]{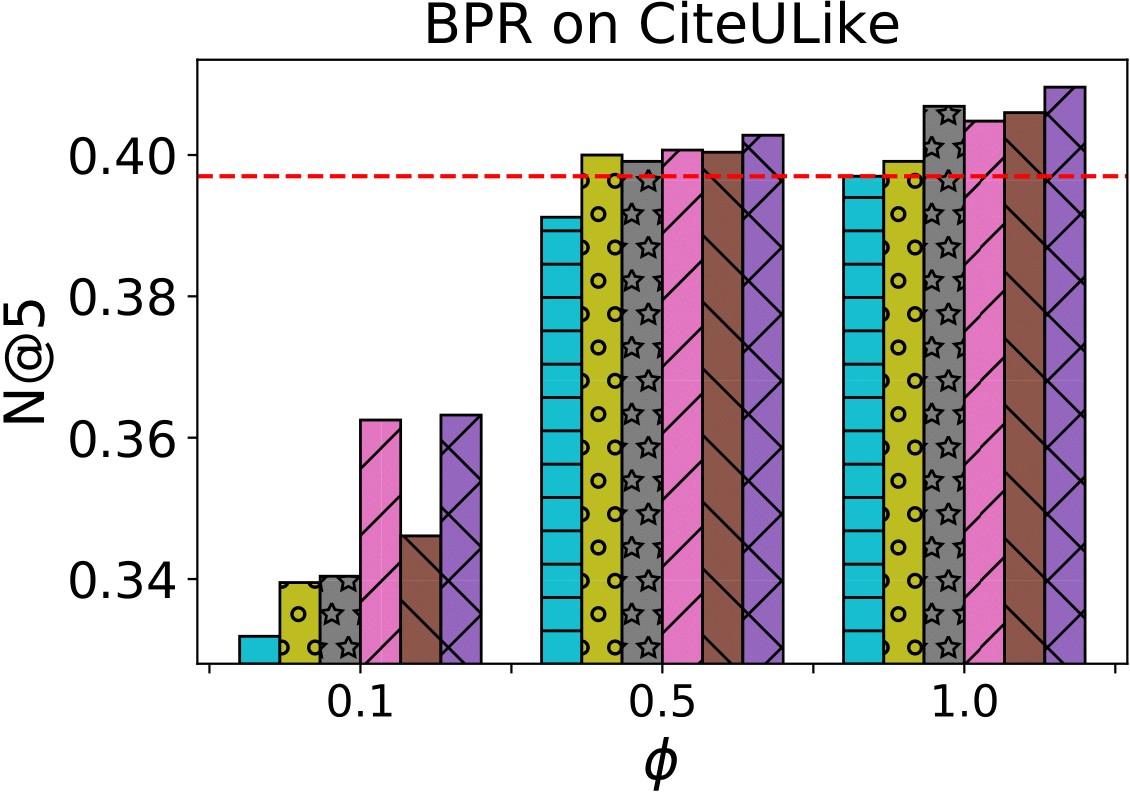}
\end{subfigure} 
\begin{subfigure}[t]{0.4\linewidth}
    \includegraphics[width=\linewidth]{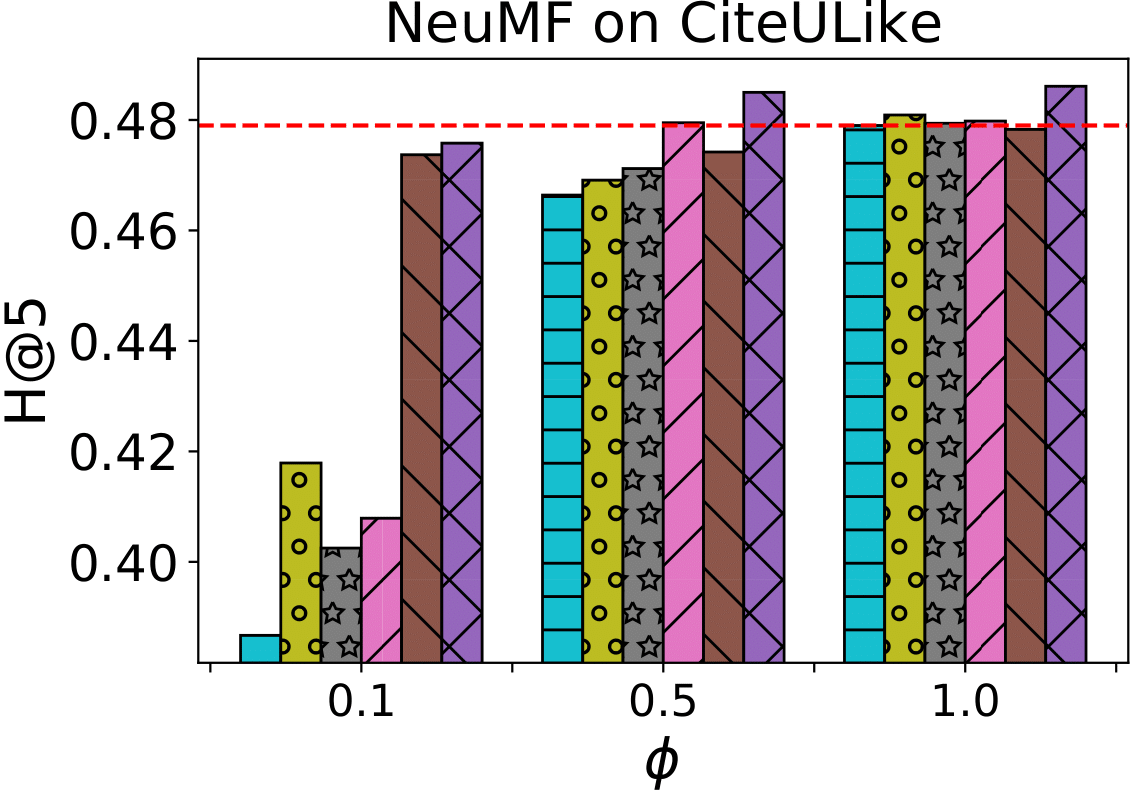}
\end{subfigure} 
\begin{subfigure}[t]{0.4\linewidth}
    \includegraphics[width=\linewidth]{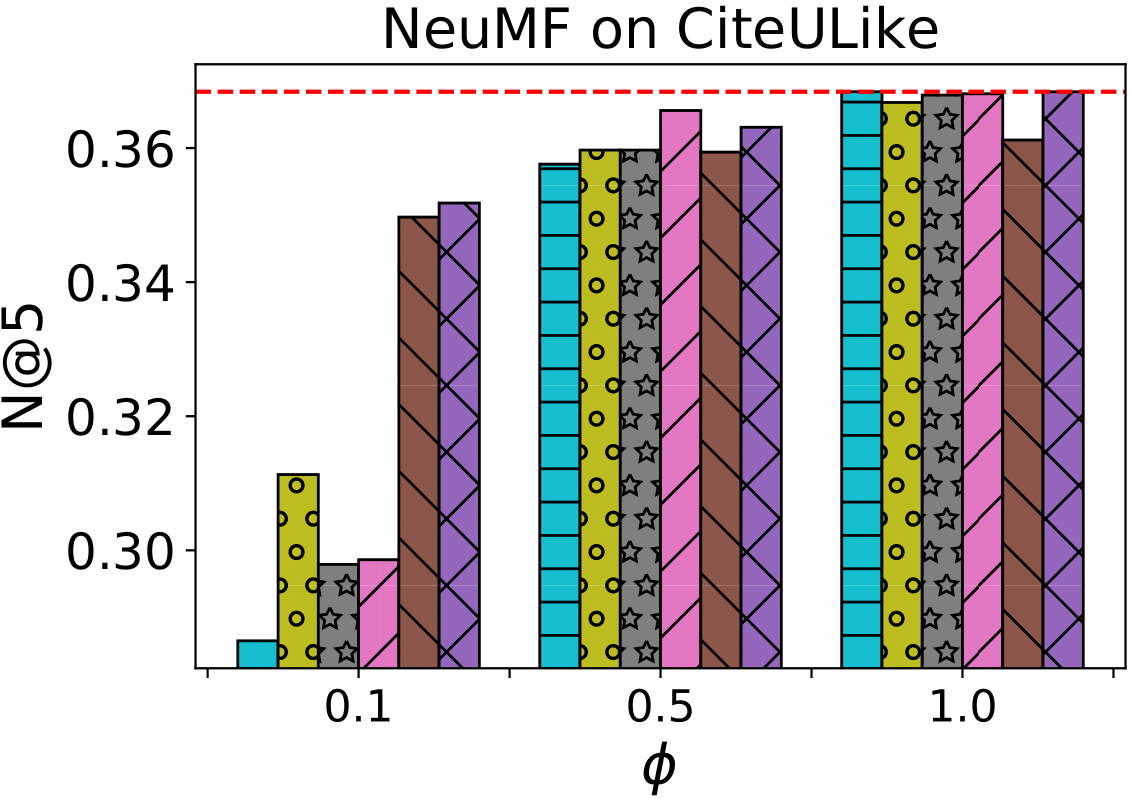}
\end{subfigure} 
\begin{subfigure}[t]{0.4\linewidth}
    \includegraphics[width=\linewidth]{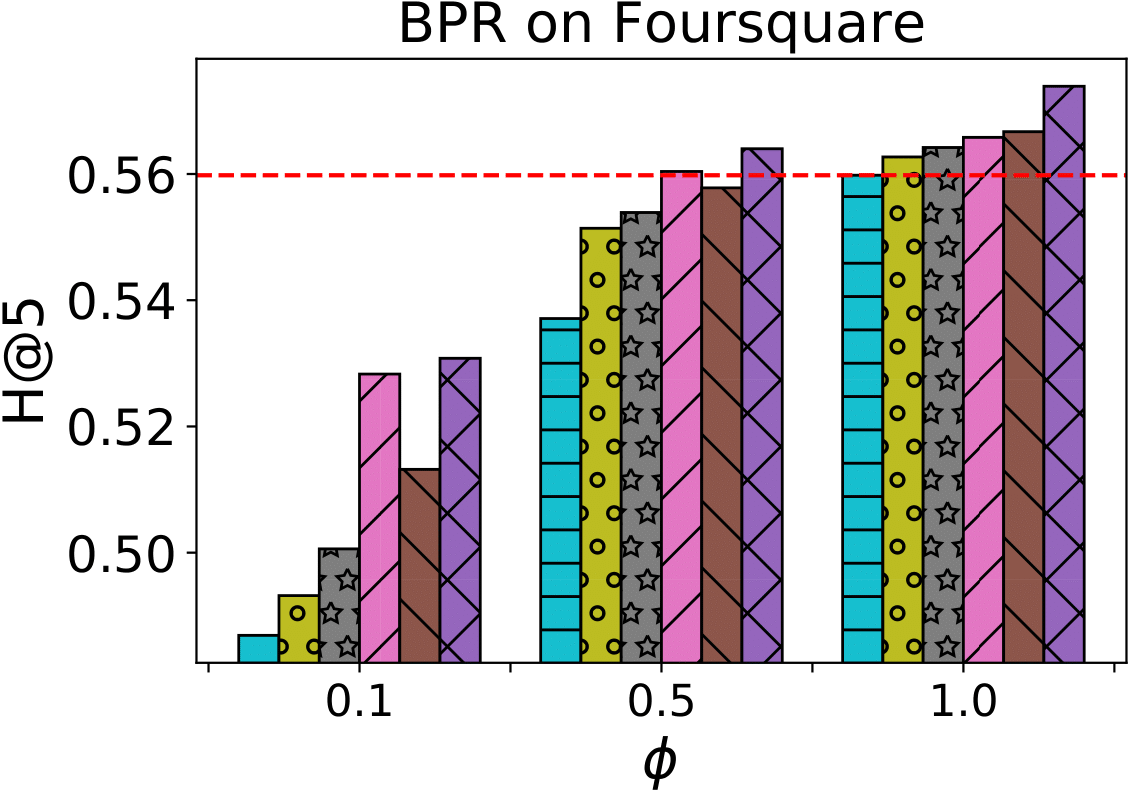}
\end{subfigure}
\begin{subfigure}[t]{0.4\linewidth}
    \includegraphics[width=\linewidth]{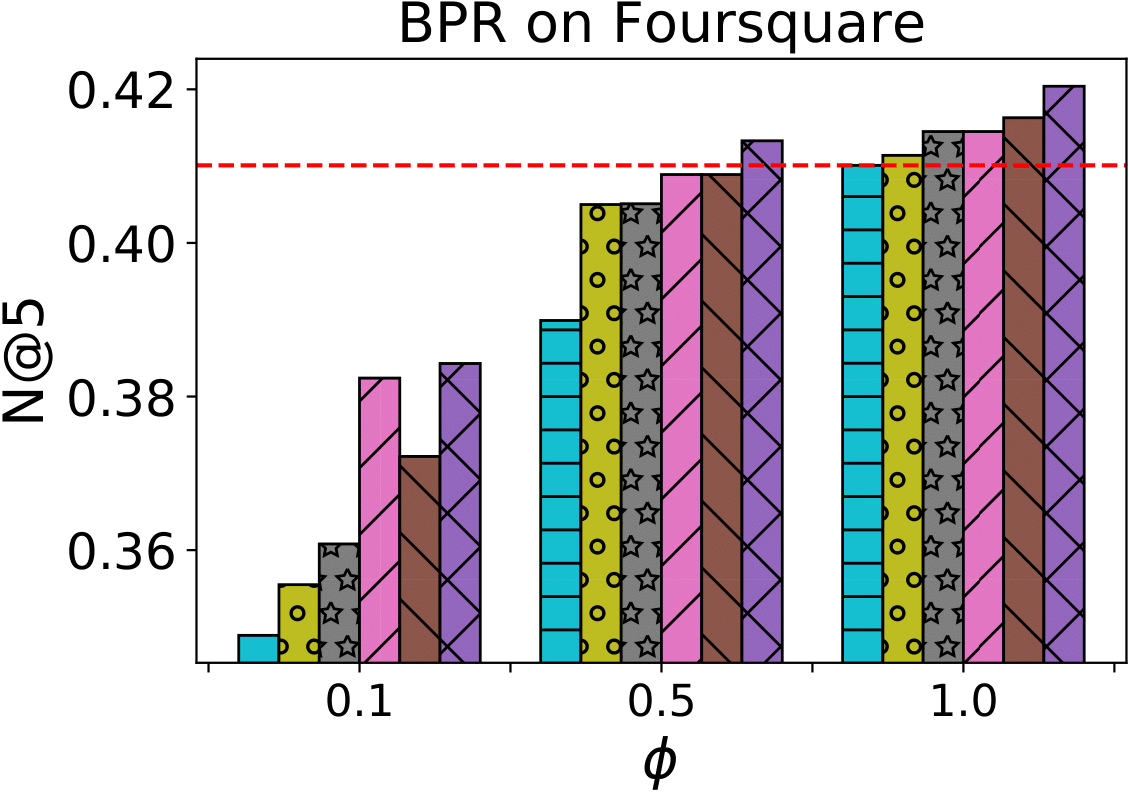}
\end{subfigure} 
\begin{subfigure}[t]{0.4\linewidth}
    \includegraphics[width=\linewidth]{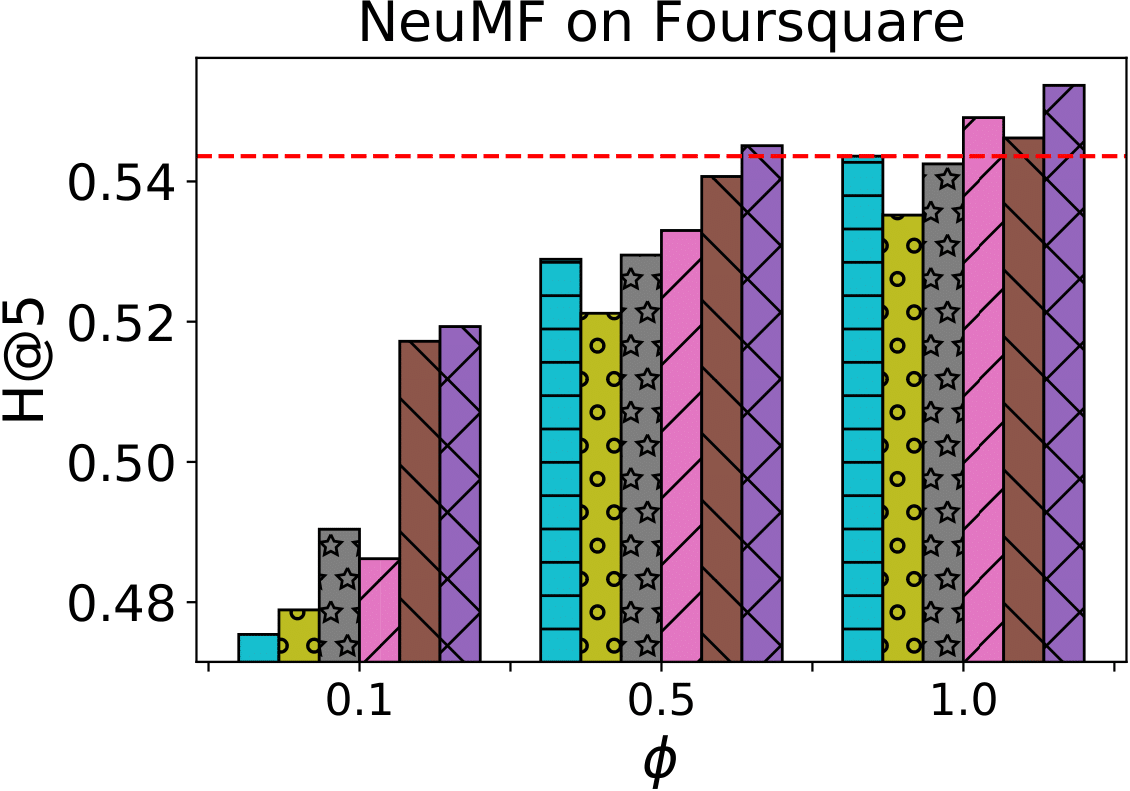}
\end{subfigure} 
\begin{subfigure}[t]{0.4\linewidth}
    \includegraphics[width=\linewidth]{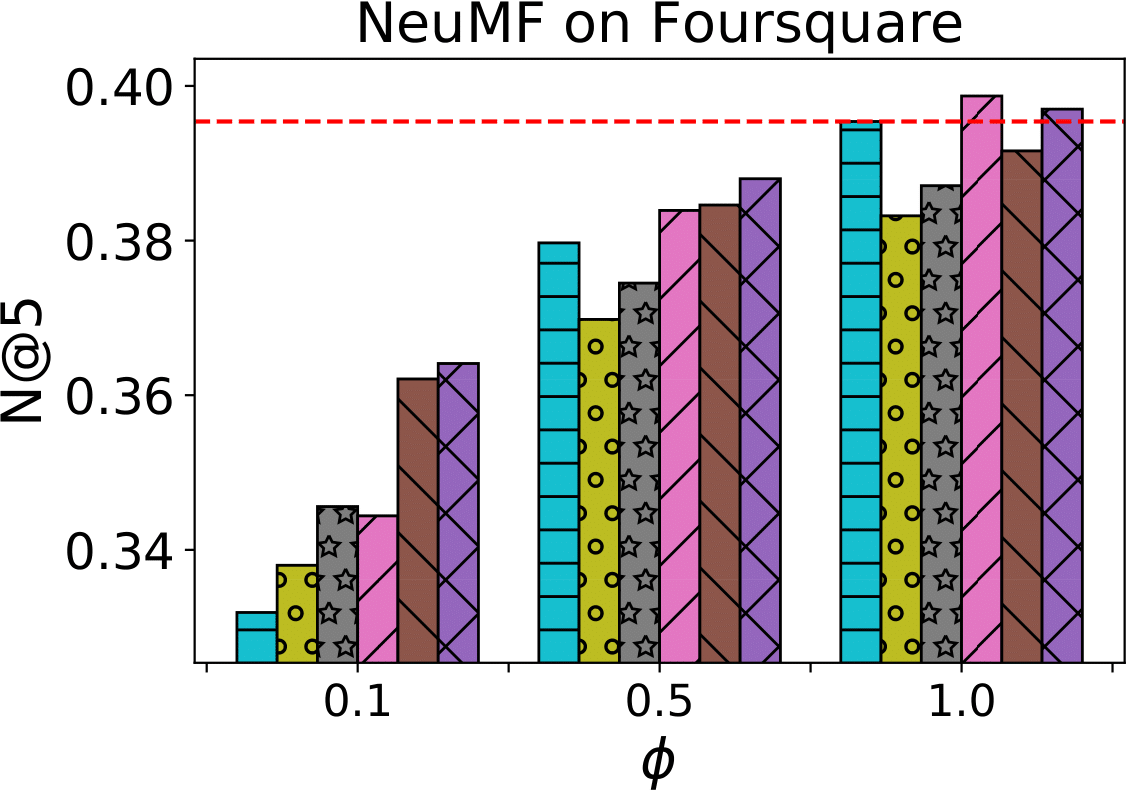}
\end{subfigure} 
\caption{Recommendation Performance on across three different student model sizes. (Red dotted line: Teacher)}
\label{fig:de-rrd_sizes}
\end{figure*}

\subsection{Performance Comparison}
Table \ref{tab:de-rrd_main} shows top-$N$ recommendation accuracy of different methods in terms of various ranking metrics.
In summary, DE-RRD shows the significant improvement compared to the state-of-the-art KD methods on two base models that have different architectures and optimization strategies.
Also, DE-RRD consistently outperforms the existing methods on three different sizes of the student model in Figure \ref{fig:de-rrd_sizes}.
We analyze the results from various perspectives.

We first observe that the two methods of the proposed framework (i.e., DE, RRD) improve the performance of the student model. 
DE directly distills the teacher's latent knowledge that includes detailed information on users, items, and the relationships among them.
This enables the student to be more effectively trained than finding such information from scratch with a limited capacity.
RRD distills the knowledge from the teacher's predictions based on the relaxed ranking approach which makes the student to effectively maintain the ranking orders of interesting items predicted by the teacher. 
Unlike the existing methods (i.e., RD, CD), it directly handles the ranking violations among the sampled items, which can lead to better ranking performance.

Also, we observe that RRD achieves large performance gain particularly in NeuMF ($\phi=0.1$).
One possible reason is that NeuMF is trained with the point-wise loss function (i.e., binary cross-entropy) which considers only one item at a time.
In general, it is known that the approaches considering the preference orders between items (e.g., pair-wise, list-wise) can achieve better ranking performance than the point-wise approach \cite{NCR}.
RRD enables the model to capture the ranking orders among the unobserved items, so that it can lead to the large performance gain.
Interestingly, we observe that the prediction-based KD methods (i.e., RD, CD, RRD) can have an adverse effect when the model size is large (NeuMF with $\phi = 0.5, 1.0$ in Figure \ref{fig:de-rrd_sizes}).
We conjecture that this is because when a model has sufficient capacity to achieve comparable performance to the teacher, enforcing it to exactly mimic the teacher's prediction results can act as a strong constraint that rather hinders~its~learning.

In addition, we observe that DE-RRD achieves the best performance among all the methods in general.
DE-RRD enables the student to learn both from the teacher's prediction and from the latent knowledge that provides the bases for such predictions.
Interestingly, DE-RRD also shows a large performance gain when the student model has the identical structure to the teacher model (i.e., self-distillation with $\phi=1.0$ in Figure \ref{fig:de-rrd_sizes}).
This result shows that it can be also used to maximize the performance of the existing recommender.

Lastly, we provide the result of the online inference efficiency test in Table \ref{table:de-rrd_latency}.
All inferences are made using PyTorch with CUDA from Tesla P40 GPU and Xeon on Gold 6148 CPU.
The student model trained with DE-RRD achieves comparable performance with only 10-50\% of learning parameters compared to the teacher.
The smaller model requires less computations and memory costs, so it can achieve lower latency.
In particular, deep recommender (i.e., NeuMF) which has a large number of learning parameters and complex structures takes more benefits from the smaller model size.
On real-time RS application that has larger numbers of users (and items) and has a more complex model structure, DE-RRD can lead to a larger improvement in online inference efficiency.

\begin{table}[t]
\centering
\small
\renewcommand{\arraystretch}{0.9}
\renewcommand{\tabcolsep}{2.2mm}
  \caption{Model compactness and online inference efficiency. Time (seconds) indicates the wall time used for generating recommendation list for every user. 
  H@5 Ratio denotes the ratio of H@5 from DE-RRD over that from Teacher.}
  \begin{tabular}{cccccc}
  \toprule
    Dataset & Base Model & $\phi$ & Time (s) & \#Params.& H@5 Ratio\\
   \midrule
   &\multirow{3}*{BPR}&1.0&59.27s&6.08M& 1.03\\
   &&0.5&57.53s&3.04M& 1.01\\
   \multirow{2}*{\rotatebox[origin=c]{0}{CiteULike}}&&0.1&55.39s&0.61M& 0.94\\
   \cmidrule{2-6}
   &\multirow{3}*{NeuMF}&1.0&79.27s&15.33M&1.01\\
   &&0.5&68.37s&7.63M& 1.01\\
   &&0.1&58.27s&1.52M& 0.99\\
   \midrule
&\multirow{3}*{BPR}&1.0&257.28s&9.61M& 1.03\\
   &&0.5&249.19s&4.81M& 1.01\\
   \multirow{2}*{\rotatebox[origin=c]{0}{Foursquare}}&&0.1&244.23s&0.96M& 0.95\\
   \cmidrule{2-6}
   &\multirow{3}*{NeuMF}&1.0&342.84s&24.16M& 1.02\\
   &&0.5&297.34s&12.05M&1.01\\
   &&0.1&255.24s&2.40M&0.95\\
   \bottomrule
  \end{tabular}
  \label{table:de-rrd_latency}
\end{table}

\subsection{Design Choice Analysis}
We provide both quantitative and qualitative analyses on the proposed methods and alternative design choices (i.e., ablations) to verify the superiority of our design choice.
The performance comparisons with the ablations are summarized in Table \ref{table:de-rrd_ablations}.

For DE, we consider three ablations: (a) Attention (b) One expert (large) (c) One expert (small).
As discussed in Section \ref{sec:DE-RRD_DE}, instead of the selection strategy, attention mechanism can be adopted.
We also compare the performance of one large expert\footnote{We make one large expert by adopting the average pooling.}  
and one small expert.
Note that DE, attention, and one expert (large) has the exact same number of learning parameters for experts.
We observe that the increased numbers of learning parameters do not necessarily contribute to performance improvement ((a) vs. (c) in BPR).

\begin{figure}[!htbp]
\centering
\vspace{-0.3cm}
\begin{subfigure}[t]{0.5\linewidth}
    \includegraphics[width=\linewidth]{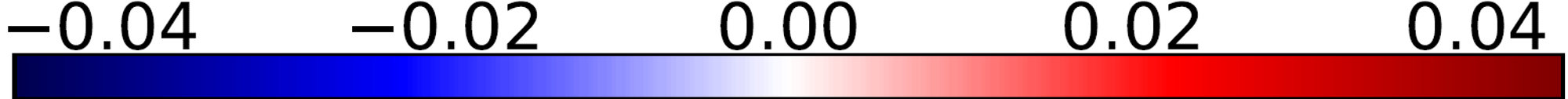}
\end{subfigure} 
\\
\begin{subfigure}[t]{0.33\linewidth}
    \includegraphics[width=\linewidth]{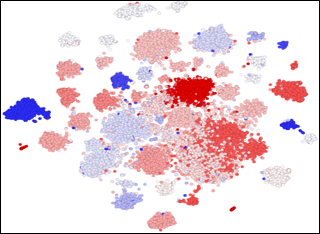}
    \caption*{Attention}
\end{subfigure} 
\hspace*{-0.07in}
\begin{subfigure}[t]{0.33\linewidth}
    \includegraphics[width=\linewidth]{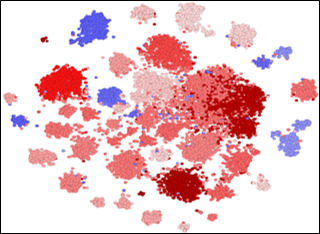}
    \caption*{One expert (large)}
\end{subfigure}
\hspace*{-0.07in}
\begin{subfigure}[t]{0.33\linewidth}
    \includegraphics[width=\linewidth]{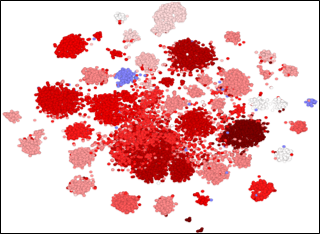}
    \caption*{DE (selection)}
\end{subfigure} 
\caption{Performance (N@20) gain map (BPR with $\phi=0.1$ on Foursquare).}
\label{fig:de-rrd_pgain}

\begin{subfigure}[t]{0.33\linewidth}
    \includegraphics[width=\linewidth]{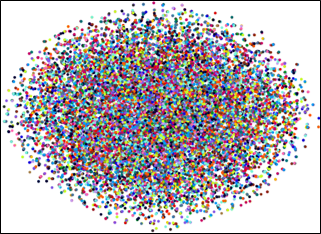}
    \caption{Epoch 0}
\end{subfigure} 
\hspace*{-0.07in}
\begin{subfigure}[t]{0.33\linewidth}
    \includegraphics[width=\linewidth]{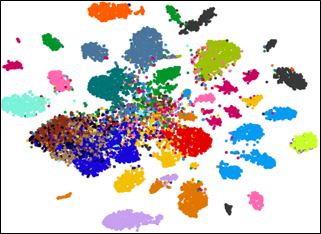}
    \caption{Epoch 20}
\end{subfigure}
\hspace*{-0.07in}
\begin{subfigure}[t]{0.33\linewidth}
    \includegraphics[width=\linewidth]{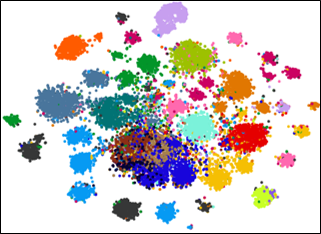}
    \caption{Train Done}
\end{subfigure} 
\caption{Expert selection map of DE. Each color corresponds to an expert (BPR with $\phi=0.1$ on Foursquare).}
\label{fig:de-rrd_selection_map}
\end{figure}

We also observe that the selection shows the best performance among all the ablations.
To further investigate this result, we conduct qualitative analysis on user representation spaces induced by each design choice.
Specifically, we first perform clustering\footnote{We use $k$-Means clustering in Scikit-learn. $k$ is set to 20.} on user representation space from the teacher model to find user groups that have strong correlations (or similar preferences).
Then, we visualize the average performance gain (per group) map in Figure \ref{fig:de-rrd_pgain}.
We observe that distilling the knowledge by the attention, one large expert can cause performance decreases in many user groups (blue clusters), whereas the selection improves the performance in more numbers of user groups (red clusters).
In the ablations (a)-(c), the experts are trained to minimize the overall reconstruction errors on all the diverse entities.
This makes the information of weakly correlated entities to be mixed together and further hinders discovering the preference of a particular user group.
Unlike the ablations, DE clearly distinguishes the knowledge that each expert distills, and makes each expert
to be trained to distill only the knowledge of strongly correlated entities.
So, it can alleviate such problem.
The expert selection map of DE is visualized in Figure \ref{fig:de-rrd_selection_map}.
We can observe that each expert gets gradually specialized in certain user groups that share similar preferences during the training.
\begin{table}[t]
\centering
 \caption{Performance comparison for alternative design choices on Foursquare ($\phi=0.1$).}
\small
\renewcommand{\arraystretch}{0.9}
\renewcommand{\tabcolsep}{2.2mm}
\resizebox{\columnwidth}{!}{%

  \begin{tabular}{clcccc}
   
  \toprule
    Base Model & Design choices & H@5 & N@5 & H@10 & N@10\\
  \midrule
  &DE&\textbf{0.5283}&\textbf{0.3824}&\textbf{0.6810}&\textbf{0.4316}\\
  &(a) Attention&0.5019&0.3625&0.6575&0.4131\\
  \multirow{3}*{BPR}&(b) One expert (large)&0.5151&0.3716&0.6733&0.4230\\
  &(c) One expert (small)&0.5136&0.3717&0.6683&0.4213\\
  \cmidrule{2-6}
  &RRD&\textbf{0.5132}&\textbf{0.3722}&\textbf{0.6616}&\textbf{0.4202}\\
  &(d) Full ranking &0.4983&0.3595&0.6474&0.4080\\
  &(e) \textit{Interesting} ranking & 0.4814&0.3479&0.6416&0.3999\\
  \midrule
  &DE&\textbf{0.4862}&\textbf{0.3444}&\textbf{0.6413}&\textbf{0.3938}\\
  &(a) Attention&0.4770&0.3364&0.6364&0.3903\\
  \multirow{3}*{NeuMF}&(b) One expert (large)&0.4741&0.3367&0.6341&0.3885\\
  &(c) One expert (small)&0.4740&0.3339&0.6316&0.3860\\
  \cmidrule{2-6}
  &RRD&\textbf{0.5172}&\textbf{0.3621}&\textbf{0.6739}&\textbf{0.4132}\\
  &(d) Full ranking &0.4799&0.3457&0.6324&0.3949\\
  &(e) \textit{Interesting} ranking & 0.4641&0.3294&0.6228&0.3809\\
  \bottomrule
  \end{tabular}%
  }
  \label{table:de-rrd_ablations}
\end{table}

For RRD, we consider two ablations: (d) and (e).
The ablations are intended to show the effects of the proposed relaxed ranking.
Concretely, we apply the list-wise loss (i.e., no relaxation) on all the sampled items (interesting and uninteresting items) for (d), on the top-ranked items (interesting items) for (e).
Note that all the methods use the same number of items for distillation.
We observe that merely adopting the list-wise loss has adverse effects on the ranking performance.
First, (d) learns to match the full ranking order among all the sampled items.
Learning the detailed order among the uninteresting items is not necessarily helpful to improve the ranking performance, and may further interfere with focusing on the interesting items.
Also, (e), which only considers the interesting items, shows even worse performance than Student.
The list-wise loss does not take into account the absolute ranking positions of the items; a ranking order can be satisfied regardless of the items' absolute positions.
Since (e) does not consider the relative orders between the interesting items and the uninteresting items, it may push such interesting items far from the top of the ranking list.
Unlike the ablations, RRD adopts the relaxed ranking approach, which enables the student to better focus on the interesting items while considering the relative orders with the uninteresting items.
\begin{figure}[t]
\begin{subfigure}[t]{0.24\linewidth}
    \includegraphics[width=\linewidth, height=0.9\linewidth]{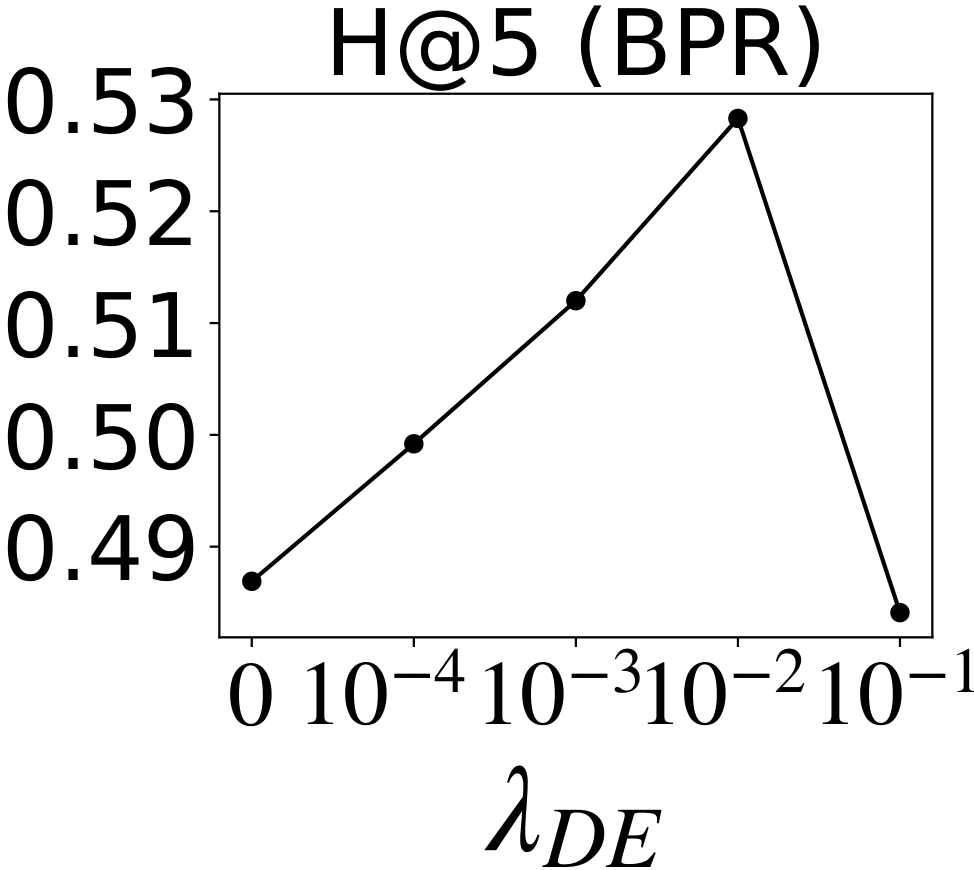}
\end{subfigure}
\hspace*{-0.05in}
\begin{subfigure}[t]{0.24\linewidth}
    \includegraphics[width=\linewidth, height=0.9\linewidth]{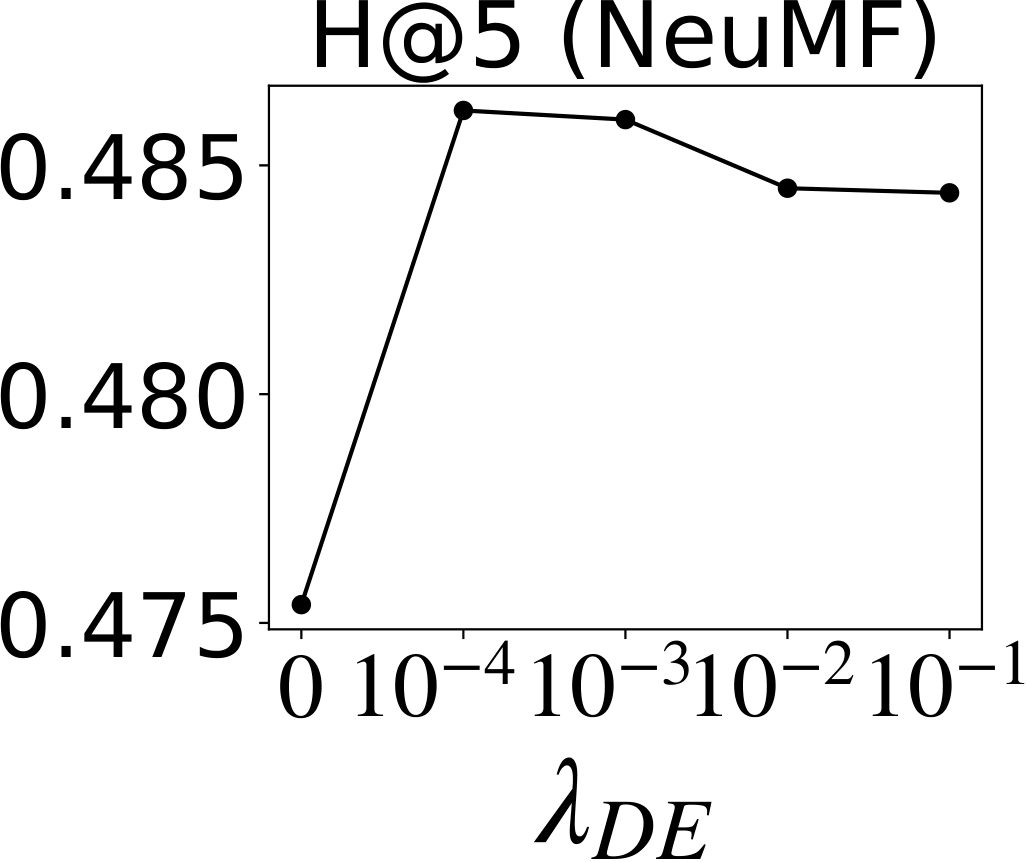}
\end{subfigure} 
\hspace*{-0.05in}
\begin{subfigure}[t]{0.24\linewidth}
    \includegraphics[width=0.98\linewidth, height=0.9\linewidth]{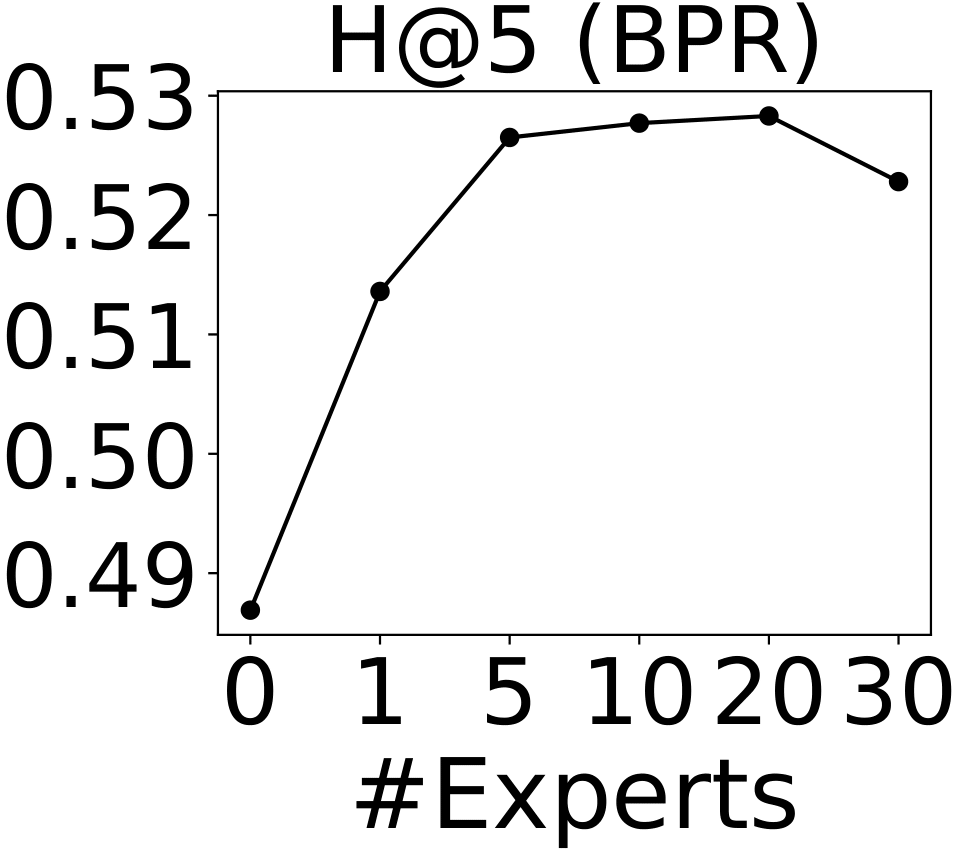}
\end{subfigure} 
\hspace*{-0.07in}
\begin{subfigure}[t]{0.24\linewidth}
    \includegraphics[width=\linewidth, height=0.9\linewidth]{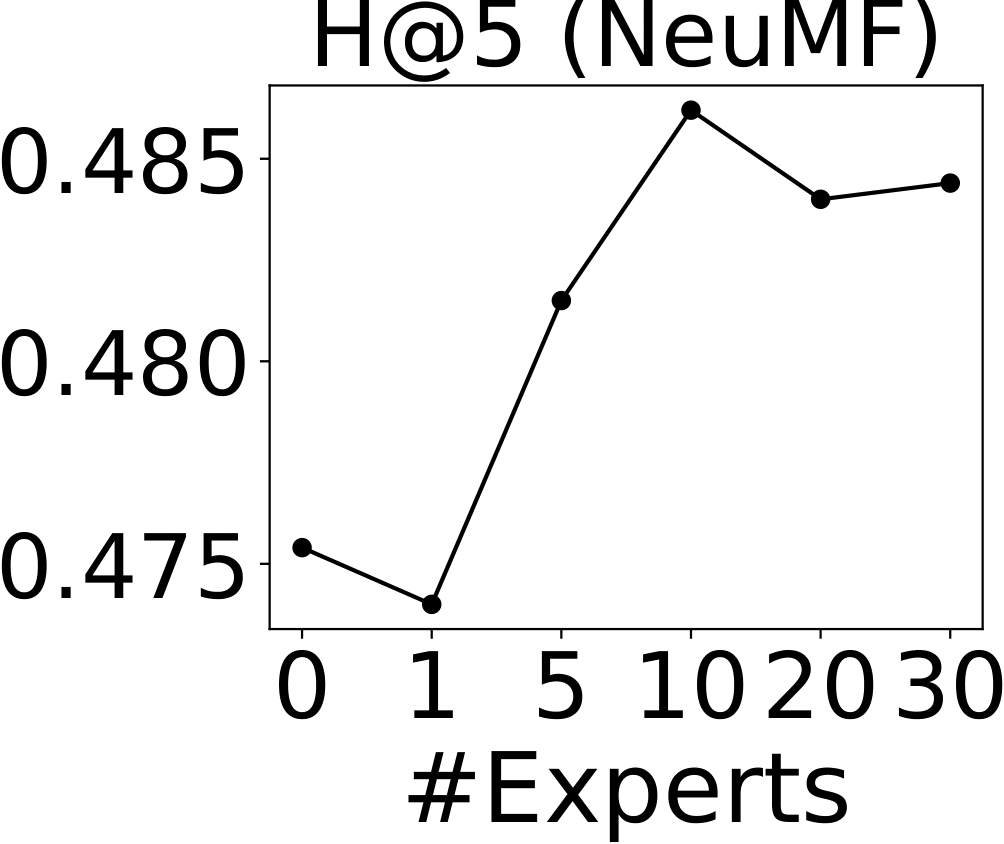}
\end{subfigure} 
\caption*{(a) Effects of $\lambda_{DE}$ and the number of experts\quad}
\vspace*{0.05in}
\begin{subfigure}[t]{0.24\linewidth}
    \includegraphics[width=\linewidth, height=0.9\linewidth]{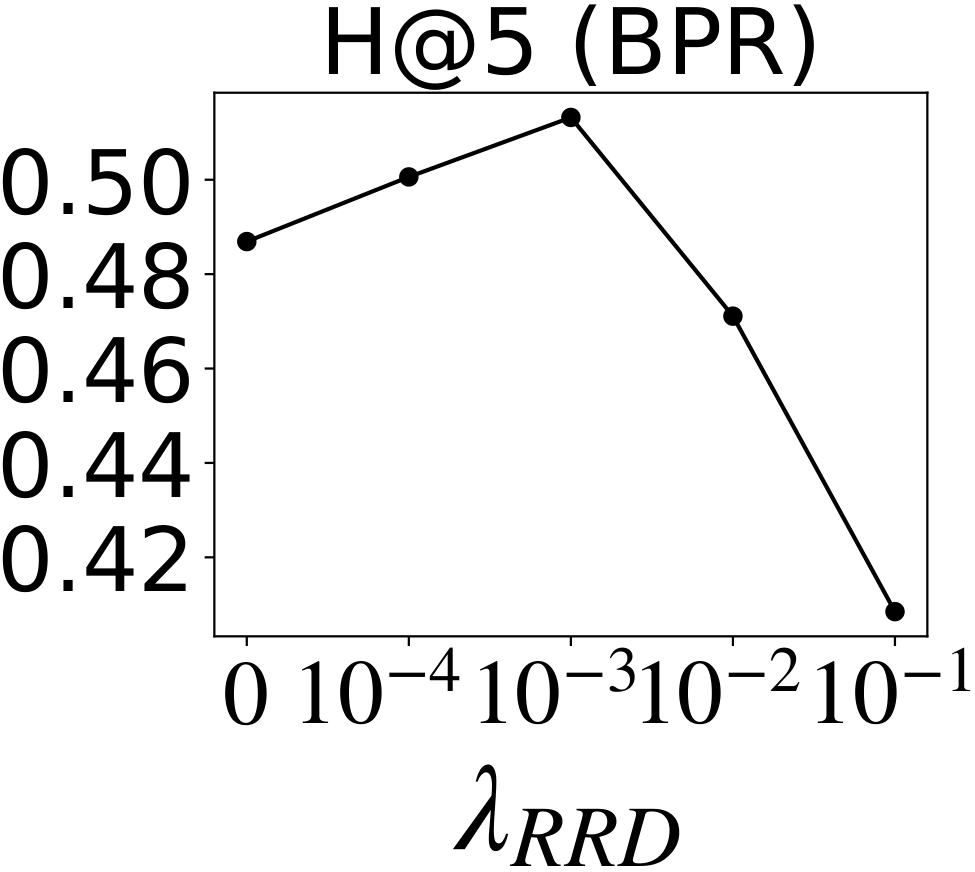}
\end{subfigure}
\hspace*{-0.05in}
\begin{subfigure}[t]{0.24\linewidth}
    \includegraphics[width=\linewidth, height=0.9\linewidth]{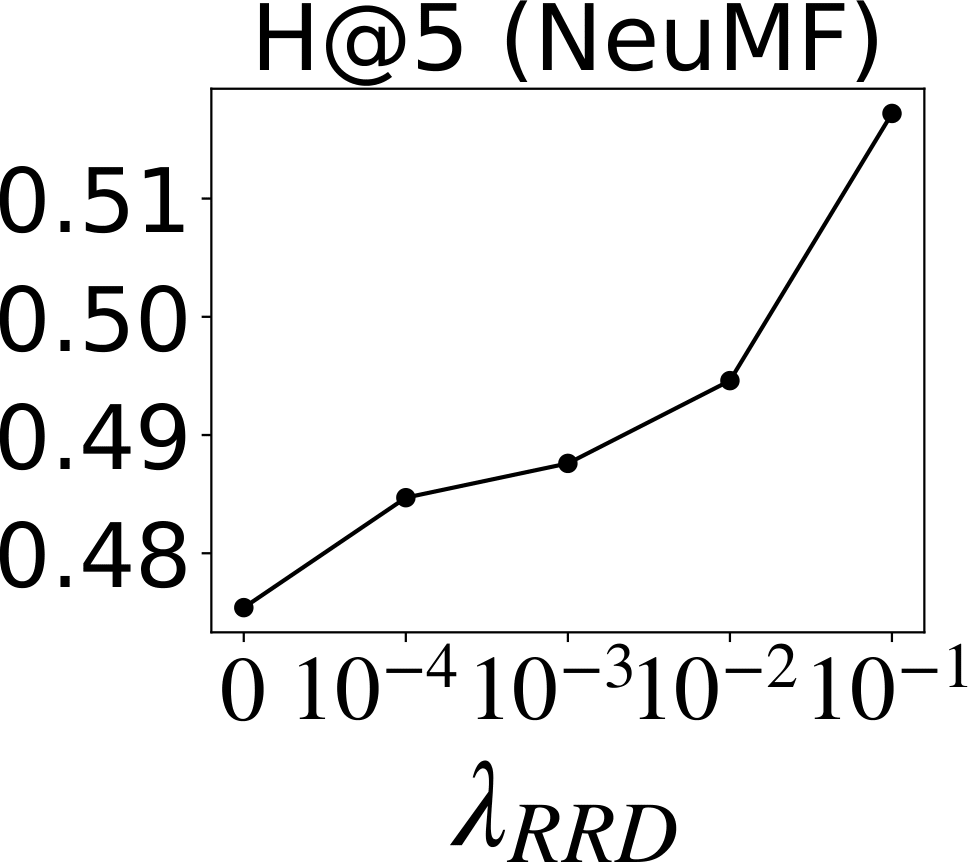}
\end{subfigure} 
\hspace*{-0.05in}
\begin{subfigure}[t]{0.24\linewidth}
    \includegraphics[width=0.98\linewidth, height=0.9\linewidth]{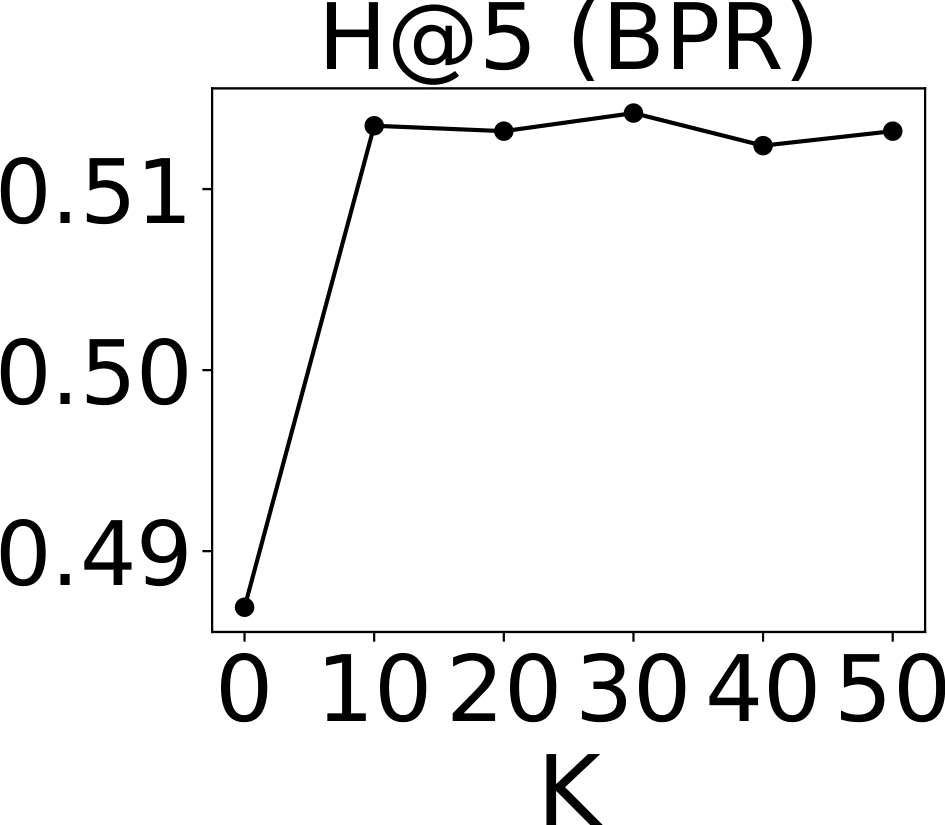}
\end{subfigure} 
\hspace*{-0.07in}
\begin{subfigure}[t]{0.24\linewidth}
    \includegraphics[width=\linewidth, height=0.9\linewidth]{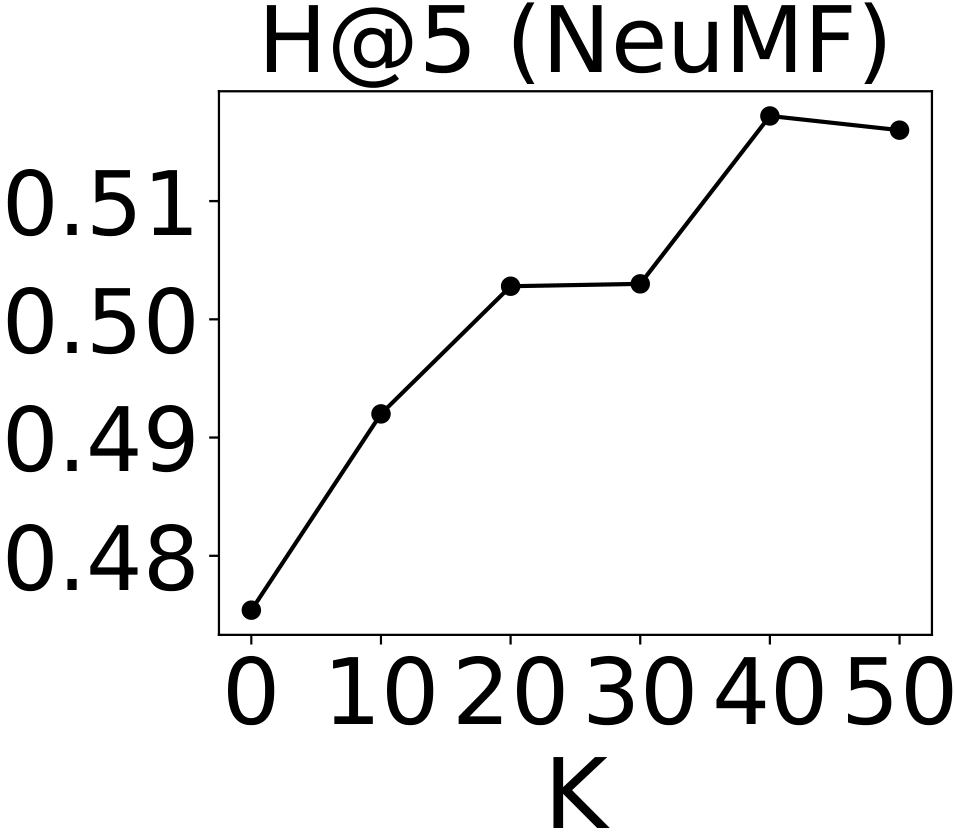}
\end{subfigure} 
\caption*{(b) Effects of $\lambda_{RRD}$ and $K$\quad}
\caption{Effects of the hyperparameters. (a) DE (b) RRD.}
\label{fig:de-rrd_hp}
\end{figure}

\begin{table}[!ht]
\caption{Effects of $\lambda_{DE}$ and $\lambda_{RRD}$ in DE-RRD framework.}
\resizebox{1.01\columnwidth}{!}{%
\RowStretch{0.6}
\setlength\tabcolsep{2.9pt}
\begin{tabular}{|c|c|cccc|cccc|}
\hline
\multicolumn{2}{|c|}{} & \multicolumn{4}{c|}{\textbf{BPR}} & \multicolumn{4}{c|}{\textbf{NeuMF}} \\ \cline{3-10} 
\multicolumn{2}{|c|}{\multirow{-3}{*}{\begin{tabular}[c]{@{}c@{}}\\\\Foursquare\\ (H@5)\\\\\end{tabular}}} & \multicolumn{4}{c|}{{\color[HTML]{000000} $\lambda_{RRD}$}} & \multicolumn{4}{c|}{$\lambda_{RRD}$} \\ \cline{3-10} 
 \multicolumn{2}{|c|}{}& $10^{-4}$ & $10^{-3}$ & $10^{-2}$ & $10^{-1}$ & $10^{-4}$ & $10^{-3}$ & $10^{-2}$ & $10^{-1}$ \\ \hline
\multicolumn{1}{|c|}{} & $10^{-4}$ & 0.5081 & 0.5201 & 0.4590 & 0.3901 & 0.4774 & 0.4896 & 0.5014 & \textbf{0.5193} \\
\multicolumn{1}{|c|}{\multirow{-2}{*}{\rotatebox[origin=c]{90}{$\lambda_{DE}$\quad\,\,}}} & $10^{-3}$ & 0.5186 & 0.5276 & 0.4688 & 0.3906 & 0.4774 & 0.4858 & 0.4942 & 0.5112 \\
\multicolumn{1}{|c|}{} & $10^{-2}$ & 0.5261 & \textbf{0.5308} & 0.4791 & 0.3977 & 0.4846 & 0.4868 & 0.4892 & 0.5110 \\
\multicolumn{1}{|c|}{} & $10^{-1}$ & 0.5269 & \textbf{0.5308} & 0.4928 & 0.4154 & 0.4848 & 0.4881 & 0.4908 & 0.5055 \\ \hline
\end{tabular}%
}
\label{tab:de-rrd_hp}
\end{table}

\subsection{Hyperparameter Analysis}
We provide analyses to offer guidance of hyperparameter selection of DE-RRD.
For the sake of space, we report the results on Foursquare dataset with $\phi=0.1$.
We observe similar tendencies on CiteULike dataset.
For DE, we show the effects of two hyperparameters: $\lambda_{DE}$ that controls the importance of DE and the number of experts in Figure \ref{fig:de-rrd_hp}a.
For RRD, we show the effects of two hyperparameters: $\lambda_{RRD}$ that controls the importance of RRD and the number of interesting items ($K$) in Figure \ref{fig:de-rrd_hp}b. 
In our experiment, the number of uninteresting items is set to the same with $K$.
Note that for all graphs value ‘0’ corresponds to Student (i.e., no distillation).

Because the types of loss function of the proposed methods are different from that of the base models, it is important to properly balance the losses by using $\lambda$. 
For DE, the best performance is achieved when the magnitude of DE loss is approximately 20\% (BPR), 2-5\% (NeuMF) compared to that of the base model’s loss. 
For RRD, the best performance is achieved when the magnitude of RRD loss is approximately 7-10\% (BPR), 1000\% (NeuMF) compared to that of the base model’s loss. 
For the number of experts and $K$, the best performance is achieved near 10-20 and 30-40, respectively. 
Lastly, we show the effects of combinations of $\lambda_{DE}$ and $\lambda_{RRD}$ in DE-RRD framework in Table \ref{tab:de-rrd_hp}.
Generally, the best performance of DE-RRD is observed in the ranges where each method (i.e., DE, RRD) achieves the best performance.

\section{Summary}
\label{sec:DE-RRD_conclusion}
This paper proposes a novel knowledge distillation framework for recommender system, DE-RRD, that enables the student model to learn both from the teacher's predictions and from the latent knowledge stored in a teacher model.
To this end, we propose two novel methods:
(1) DE that directly distills latent knowledge from the representation space of the teacher.
DE adopts the experts and the expert selection strategy to effectively distill the vast CF knowledge to the student. 
(2) RRD that distills knowledge revealed from teacher's predictions with direct considerations of ranking orders among items.
RRD adopts the relaxed ranking approach to better focus on the interesting items.
Extensive experiment results demonstrate that DE-RRD significantly outperforms the state-of-the-art competitors.

\chapter{Personalized Hint Distillation}
\label{chapt:phr}
Nowadays, Knowledge Distillation (KD) has been widely studied for recommender system.
KD is a model-independent strategy that generates a small but powerful student model by transferring knowledge from a pre-trained large teacher model.
Recent work has shown that the knowledge from the teacher's representation space significantly improves the student model.
The state-of-the-art method, named Distillation Experts (DE), adopts cluster-wise distillation that transfers the knowledge of each representation cluster separately to distill the various preference knowledge in a balanced manner.
However, it is challenging to apply DE to a new environment since its performance is highly dependent on several key assumptions and hyperparameters that need to be tuned for each dataset and each base model.
In this work, we propose a novel method, dubbed Personalized Hint Regression (PHR), distilling the preference knowledge in a balanced way without relying on any assumption on the representation space nor any method-specific hyperparameters.
To circumvent the clustering, PHR employs \textit{personalization network} that enables a personalized distillation to the student space for each user/item representation, which can be viewed as a generalization of DE.
Extensive experiments conducted on real-world datasets show that PHR achieves comparable or even better performance to DE tuned by a grid search for all of its hyperparameters.  

\section{Introduction}
In the era of information explosion, Recommender System (RS) has played a key role in helping users' decisions and improving the cooperate profits \cite{NeuMF, BPR}.
Recently, the size of RS is continuously increasing because of a deep and sophisticated model architecture to capture the complex user-item interaction and a growing scale of users and items.
A large model with numerous learning parameters generally has better recommendation performance.
However, it also requires large computational costs and high inference latency, which becomes the major obstacle for model deployment and real-time inference \cite{RD, CD, DERRD}.

To tackle this problem, Knowledge Distillation (KD) has been widely studied for RS \cite{RD, CD, DERRD, BD, TD}.
KD is a model-agnostic strategy that generates a small but powerful model (i.e., student) by distilling knowledge from a previously trained large model (i.e., teacher).
Recent methods \cite{DERRD, zhu2020ensembled, TD} have shown that the knowledge from the teacher's intermediate layer can significantly improve the student model.
In specific, they transfer the knowledge from the \textit{representation space}, where users and items are encoded as vectors, produced by the intermediate layer of a recommendation model\footnote{They can be flexibly applied to any existing RS model \cite{DERRD}.
In the case of the latent factor model (e.g., BPR \cite{BPR}), the space can be thought of as the output of the embedding lookup layer (i.e., embedding matrix) that maps each user/item to an embedding vector.
In the case of the deep model (e.g., NeuMF \cite{NeuMF}, LightGCN \cite{he2020lightgcn}), the space can be thought of as the output of a hidden layer before the prediction layer.
}. 
The representations of users and items in the space contain rich information which is not directly revealed from the final predictions \cite{DERRD, zhu2020ensembled, TD}.
By utilizing the representations from the teacher as additional supervision, the student achieves improved recommendation performance.
Also, the student has low inference latency due to its small size.

The state-of-the-art method, Distillation Experts (DE) \cite{DERRD}, first conducts a clustering on the representations, then distills the knowledge of each cluster separately to the student.
Since there are numerous user/item groups with various preferences, distilling knowledge without considering such group information makes the irrelevant information (e.g., users having dissimilar preferences) mixed. 
This leads to adulterated distillation that hinders the student model from discovering some users’ preferences \cite{DERRD}.
After the clustering, DE transfers the representations in each cluster via a cluster-wise mapping function bridging the teacher space and the student space.
With the cluster-wise distillation, DE effectively mitigates the problem and achieves more balanced distillation \cite{DERRD}.

\begin{figure}[t]
\centering
\begin{subfigure}[t]{0.41\linewidth}
    \includegraphics[width=\linewidth]{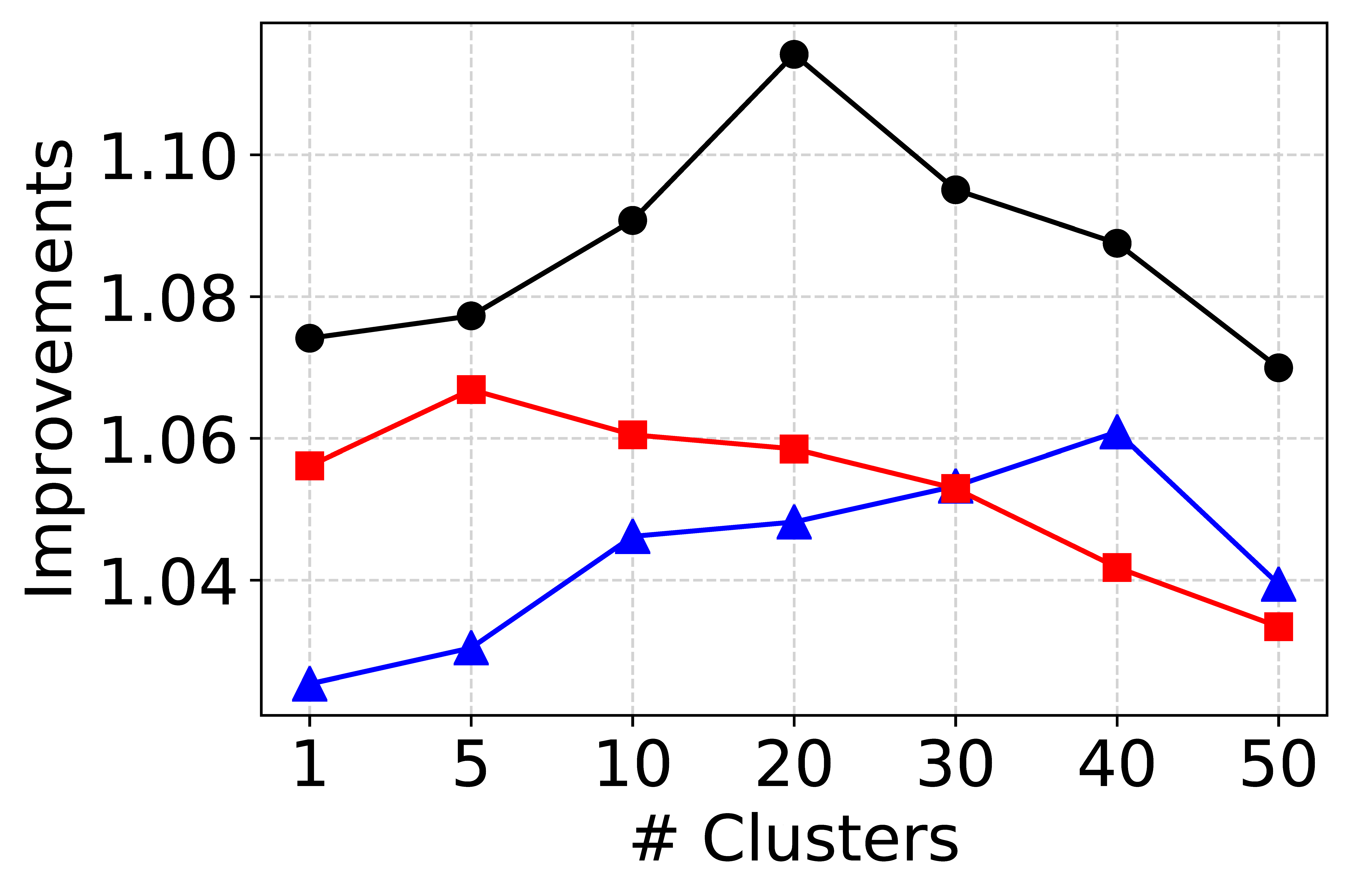}
    \caption{CiteULike dataset}
\end{subfigure}
\begin{subfigure}[t]{0.58\linewidth}
    \includegraphics[width=\linewidth]{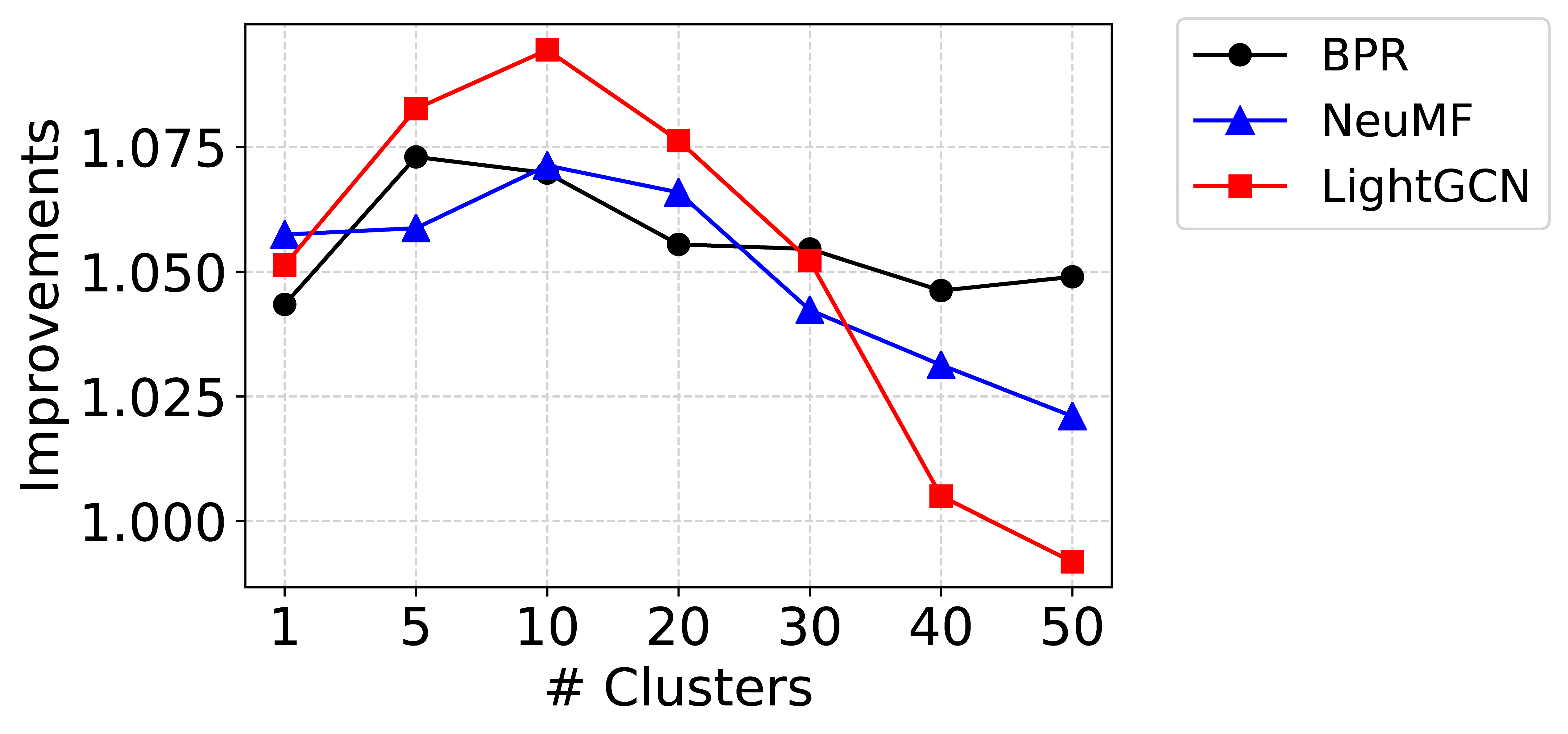}
    \caption{Foursquare dataset \quad \quad \quad \quad}
\end{subfigure} 
\caption{The improvements (R@50) achieved by DE with varying cluster numbers.
The optimal values are different for each dataset (i.e., CiteULike, Foursquare) and each base model (i.e., BPR, NeuMF).}
\label{fig:PHR_intro1}
\end{figure}

\begin{figure}[t]
\centering
\begin{subfigure}[t]{0.32\linewidth}
    \includegraphics[width=\linewidth]{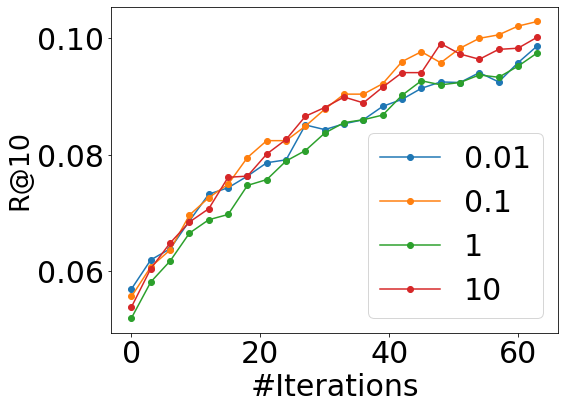}
    \caption{Exponential annealing}
\end{subfigure}
\begin{subfigure}[t]{0.32\linewidth}
    \includegraphics[width=\linewidth]{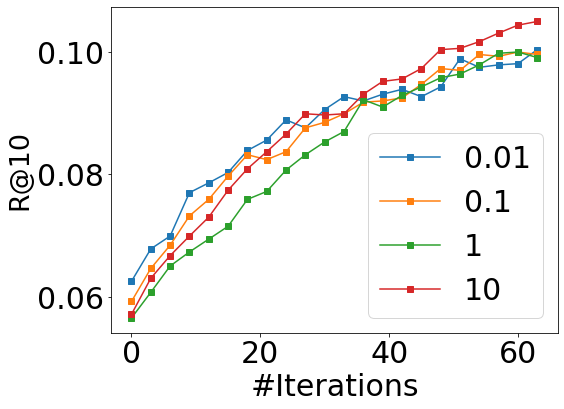}
    \caption{Step annealing (update period: 10)}
\end{subfigure} 
\begin{subfigure}[t]{0.32\linewidth}
    \includegraphics[width=\linewidth]{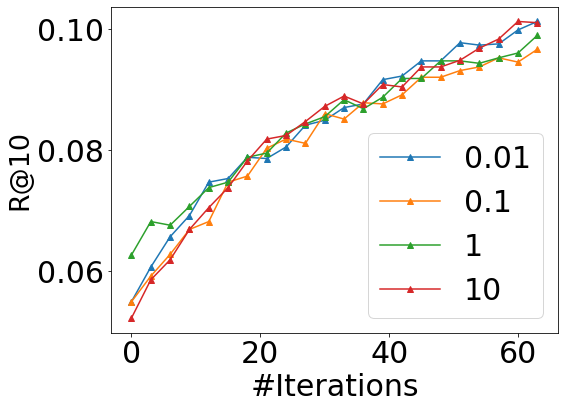}
    \caption{Step annealing (update period: 5)}
\end{subfigure} 
\caption{Training curves with different temperature parameters (0.01, 0.1, 1, and 10). The optimal values are different with each annealing schedule (BPR in CiteULike with $\phi=0.1$). }
\label{fig:PHR_intro2}
\end{figure}

Despite the effectiveness, it is challenging to apply DE to a new environment (e.g., dataset or recommendation model) since its performance is highly dependent on several hyperparameters that need to be carefully tuned for each dataset and each base model.
The most important hyperparameter of DE is \textit{the number of clusters} ($K$) in the representation space.
In other words, DE requires an assumption that there exist $K$ distinct preference groups in the space induced by a model.
Obviously, the optimal number of clusters is different for each dataset and also affected by the base model (Figure \ref{fig:PHR_intro1}).
Thus, for applying DE for a new dataset or a new base model, a thorough search from the scratch is required. 
Moreover, DE adopts Gumbel-Softmax \cite{GumbelSoftmax} to conduct the clustering in an end-to-end manner.
This needs \textit{temperature hyperparameters} controlling the smoothness of the output distribution along with \textit{an annealing schedule} that controls the temperature value during the training (Figure \ref{fig:PHR_intro2}).
These hyperparameters and the complex training scheme\footnote{The hyperparameters and their search ranges are summarized in Table \ref{tbl:PHR_hp_comp}.} remain as the major obstacle for applying DE to a new application.

In this paper, we propose a novel distillation method, named \underline{P}ersonalized \underline{H}int \underline{R}egression (PHR), that effectively distills the knowledge of various preferences without relying on such hyperparameters.
As a workaround for the clustering, PHR employs \textit{personalization network} that provides personalized (or individualized) distillation for each user/item representation.
The personalized network takes each representation along with its neighborhood information as input and generates a mapping function that bridges the teacher space and the student space. 
Then, the knowledge of the representation is distilled to the student via the mapping function.
The mapping functions vary for each user/item representation and provide personalized distillation, which can effectively transfer the knowledge of various preferences.
PHR can be viewed as a generalization of DE in the sense that it reduces to cluster-wise distillation when the personalization network is restricted to generate only $K$ distinct mapping functions.
The main contributions of this work are as follows:
\begin{itemize}
    \item We propose a novel method named PHR, effectively distilling various preferences knowledge in RS, without relying on any assumption on the representation space nor any method-specific hyperparameters.
    
    \item We validate PHR by extensive experiments on real-world datasets.
    PHR achieves comparable or even better performance to DE that is thoroughly tuned by a grid search for all the hyperparameters.
    
    \item We provide an in-depth analysis of PHR. 
    We show that PHR indeed achieves a balanced distillation via the personalization network.
    Also, we provide a detailed ablation study.

\end{itemize}

\section{Preliminary and Related Work}
Training a larger model with numerous learning parameters becomes a common practice to achieve state-of-the-art performances in recommender system (RS) \cite{DERRD, RD, CD}.
However, the large model requires a huge amount of computations and a high inference latency, which becomes the major obstacle for model deployment on real-time platforms.
As such, more and more attention has been paid to reduce the inference latency of RS.
In this section, we briefly review several approaches for reducing the latency and introduce knowledge distillation (KD) that is a model-independent strategy to produce a small but powerful model.
Lastly, we review recent work adopting KD for RS.

\vspace{0.1cm}
\textbf{Reducing inference latency of recommender system}
Several methods have adopted binary representations and hash techniques to reduce the inference cost \cite{hash1, hash2, DCF, candidategeneration}.
Concretely, they first learn binary user/item representations and construct the hash table, which can significantly reduce the search cost.
However, they have a limitation in that their recommendation performance is inferior to a model using real-value representations due to the constrained capability.
Also, several methods have adopted low-rank factorization, which factorizes the original matrix into a product of two low-rank matrices, to compress the model \cite{sun2020generic}.
They use two smaller matrices that represent the original weight matrix to reduce free parameters. 
However, it is only applicable for a few types of layers such as the input layer (i.e., user/item embedding lookup) and fully-connected layer.
Moreover, various model-dependent techniques have been adopted to accelerate the inference.
In specific, tree-based data structures \cite{KDtree}, parameter pruning \cite{pruning_RS2_inner_only}, and approximated nearest neighbor search \cite{LSH, LSH_inner_product} have been applied to RS, and they have effectively decreased the computational costs for the inference phase.
However, they are only applicable to specific models (e.g., kd-tree for metric space-based models \cite{METAS, transCF}) or easily fall into a local optimum because of the local search \cite{DERRD, CD}.

\begin{figure}[t]
\centering
\begin{subfigure}[t]{0.8\linewidth}
    \includegraphics[width=\linewidth]{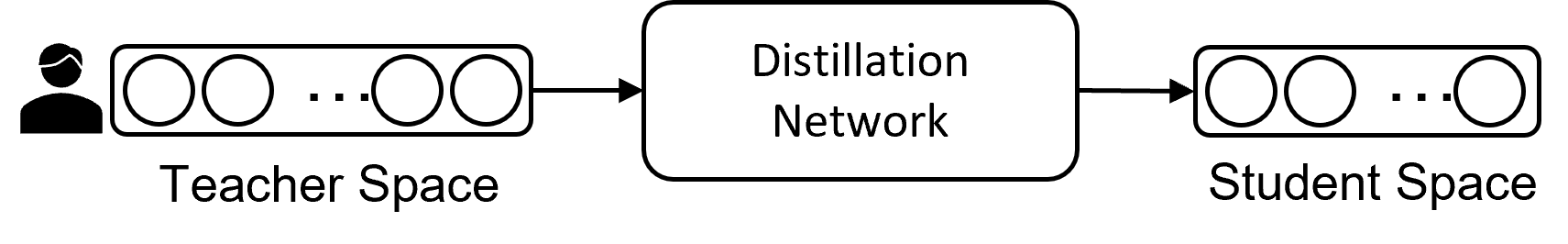}
    \caption{HR (Hint Regression)}
\end{subfigure}
\\
\vspace{0.4cm}
\begin{subfigure}[t]{0.8\linewidth}
    \includegraphics[width=\linewidth]{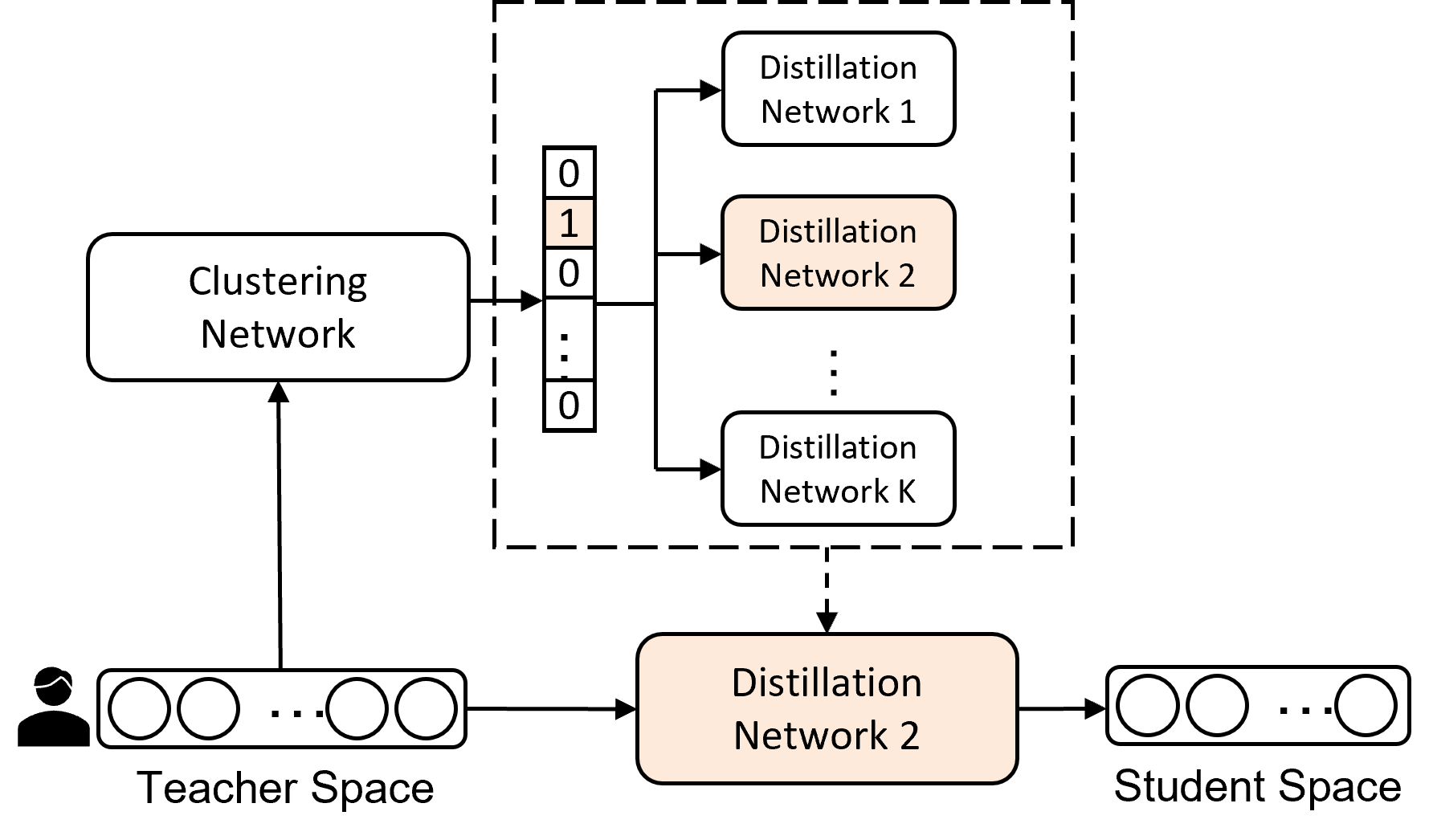}
    \caption{DE (Distillation Experts)}
\end{subfigure} 
\caption{Illustration of hint regression (HR) and distillation experts (DE).}
\label{fig:PHR_HR_DE}
\end{figure}

\vspace{0.1cm}
\textbf{Knowledge Distillation (KD)}
KD is a model-agnostic technique that transfers knowledge from a pre-trained large teacher model to a small student model \cite{KD, FitNet, chen2017learning, RKD, cheng2017survey, chen2020learning}.
During the distillation, the teacher provides additional supervision not explicitly revealed from the training set, which can significantly improve the learning and the performance of the student.
The student model trained with KD has improved performance and also has a low inference latency due to the small size.
A majority of KD methods have focused on the image classification problem in computer vision.
An early work \cite{KD} trains the student with the objective of matching the class distribution (from softmax output) of the teacher.
The softmax distribution shows the inter-class correlations which are not included in the one-hot ground-truth label, providing guidance to the learning process of the student.

Subsequent methods \cite{FitNet, chen2017learning, chen2020learning, Passban2021ALPKDAL, GCN_distill, KD_curri, TAN2021106837} argue that the prediction (i.e., the softmax distribution) incompletely reveals the teacher's knowledge and the student can be further improved by utilizing knowledge from other aspects of the teacher.
\cite{FitNet} proposes \textit{hint regression (HR)} to transfers the knowledge from the intermediate layer of the teacher (Figure \ref{fig:PHR_HR_DE}a).
Concretely, let $h_T: \mathcal{X} \rightarrow \mathbb{R}^{d_T}$ denote the nested function from the input space up to the intermediate layer of the teacher. 
Also, let $h_S: \mathcal{X} \rightarrow \mathbb{R}^{d_S}$ denote the nested function up to the intermediate layer of the student.
$d_T$ and $d_S$ denote the dimensions of the teacher space and the student space, respectively.
The hint regression loss is defined as follows:
\begin{equation}
    \mathcal{L}_{HR} = \frac{1}{2} \lVert h_T(\textbf{x}) - f (h_S(\textbf{x})) \rVert^2_2,
\end{equation}
where $f$ is a small distillation network to bridge the different dimensions.
By minimizing $\mathcal{L}_{HR}$, parameters in the student and $f$ are updated.
The hint regression enables the student to capture core information that can reconstruct the teacher representation, providing guidance to imitate the learning process of the teacher.
Subsequently, \cite{GKD, chen2020learning, RKD} utilizes relations of the intermediate representations as additional supervision for the student,
\cite{Passban2021ALPKDAL} improves the distillation quality by considering layer-wise importance,
\cite{TAN2021106837} utilizes inter-class correlation regularization technique that enables the teacher to best capture explicit correlation among the classes,
\cite{KD_curri} develops a curriculum learning strategy that controls the order of training data to further improves the student model.

\vspace{0.1cm}
\textbf{Knowledge Distillation in Recommender System}
Motivated by the huge success of KD in computer vision, several methods have applied KD to RS \cite{RD, CD, DERRD, zhu2020ensembled, GCN_distill, BD, TD, IRRRD, lee2021dual}\footnote{There are some methods proposed for a specific purpose (e.g., clothing matching \cite{song2018neural}). However, we focus on model-agnostic distillation for the general top-$N$ recommendation task.}.
Most existing work \cite{RD, CD, DERRD, GCN_distill, BD, IRRRD, lee2021dual} have focused on matching the recommendation result of the teacher and that of the student.
\cite{RD, CD, BD} employs a point-wise distillation that makes the student follow a predicted score for a user-item interaction at a time.
Subsequent work \cite{DERRD, GCN_distill, IRRRD, lee2021dual} have shown that distilling ranking information from the recommendation result of the teacher is more effective than the point-wise approach.
In specific, they adopt a listwise approach \cite{xia2008list-wise} and train the student to imitate a relative ranking order among items in the recommendation result. 

Similar to the research line in computer vision, there are a few recent attempts \cite{zhu2020ensembled, DERRD, TD}\footnote{\cite{DERRD} proposes two methods: 1) RRD that distills knowledge from the recommendation result, 2) DE that distills knowledge from the intermediate layer.} that transfers the knowledge from the hidden layer of the teacher.
Pointing out that the recommendation result incompletely reveals the teacher's knowledge, they distill a more detailed preference knowledge from the intermediate representation space.
\cite{zhu2020ensembled} adopts the original hint regression to RS and improves the student model.
The state-of-the-art method DE \cite{DERRD} further elaborates the hint regression for RS (Figure \ref{fig:PHR_HR_DE}b).
To effectively distill the knowledge of various preferences, DE adopts a clustering-based distillation.
Concretely, DE assumes that there exist $K$ distinct preference groups (or clusters) in the representation space, then distills the knowledge of each group separately.
The loss function of DE is defined as follows:
\begin{equation}
    \mathcal{L}_{DE} = \frac{1}{2} \lVert h_T(\textbf{x}) - \sum_{k=1}^K s_k f_k (h_S(\textbf{x})) \rVert^2_2,
\end{equation}
where $\textbf{s}$ is a $K$-dimensional one-hot vector whose element $s_k=1$ if the input representation belongs to $k$-th cluster, otherwise $s_k=0$. $f_k$ is a small  distillation network for each cluster.
The clustering process is evolved via backpropagation in an end-to-end manner by using Gumbel-Softmax \cite{GumbelSoftmax}.
By clearly distinguishing the knowledge transferred by each network $f_k$, DE prevents the distillation process from being biased to a few large preference groups, successfully achieving a more balanced distillation \cite{DERRD}.
The most recent work \cite{TD} distills the knowledge of relations in the teacher representation space, but it also utilizes DE to generate the relational knowledge.
It is worth noting that the methods distilling the intermediate knowledge can be utilized with the methods distilling the recommendation results to fully improve the student \cite{DERRD}.

\noindent
\textbf{Limitation.}
Despite its effectiveness, it is challenging to deploy DE to a new environment (e.g., dataset or base model) since DE is dependent on several hyperparameters that need to be tuned for each dataset and each base model.
First, DE requires the number of clusters ($K$) in the representation space.
As shown in Figure \ref{fig:PHR_intro1}, the optimal value of $K$ differs for each dataset and each base model, thus an exhaustive search from the scratch is inevitable for deploying DE for a new dataset or a new base model. 
Also, DE relies on Gumbel-Softmax to make the clustering process differentiable, which requires a temperature parameter along with an annealing scheduling for the stable training.
These hyperparameters and the complex training scheme remain as the major obstacle for applying DE to a new setting.

\section{Problem Formulation}
In this work, we focus on top-$N$ recommendation task for implicit feedback also known as one-class collaborative filtering \cite{pan2008one, hu2008collaborative, BUIR}.
Let $\mathcal{U}$ denote a set of users, and let $\mathcal{I}$ denote a set of items.
We have user-item interaction history $\boldsymbol { R } \in \{0,1\}^{| \mathcal { U } | \times | \mathcal { I } |}$ whose element $r_{u,i}=1$ if a user $u$ has interacted with an item $i$, otherwise $r_{u,i}=0$.
Also, we denote the set of items that have interacted with a user $u$ as $\mathcal{N}_{u}^{\mathcal { I }}$, and the set of users that have interacted with an item $i$ as $\mathcal{N}_{i}^{\mathcal { U }}$
The goal of top-$N$ recommendation task is to provide a ranked list of $N$ unobserved items for each user.
For each user $u$, we rank all items in $i \in \mathcal{I} \backslash \mathcal{N}_{u}^{\mathcal{I}}$ and provide top-$N$ item list.

The knowledge distillation is conducted as follows:
First, a large teacher model is trained with the training set with the binary labels.
Second, a small student is trained with help from the teacher as well as the binary training set.
For the inference phase, we use the student model which has a low inference latency.
In this work, we aim to design a KD method that effectively distills the vast intermediate knowledge without relying on the method-specific hyperparameters nor complex training schemes.

\section{Proposed Method---PHR}
\begin{figure}[t]
\centering
\begin{subfigure}[t]{1.0\linewidth}
    \includegraphics[width=\linewidth]{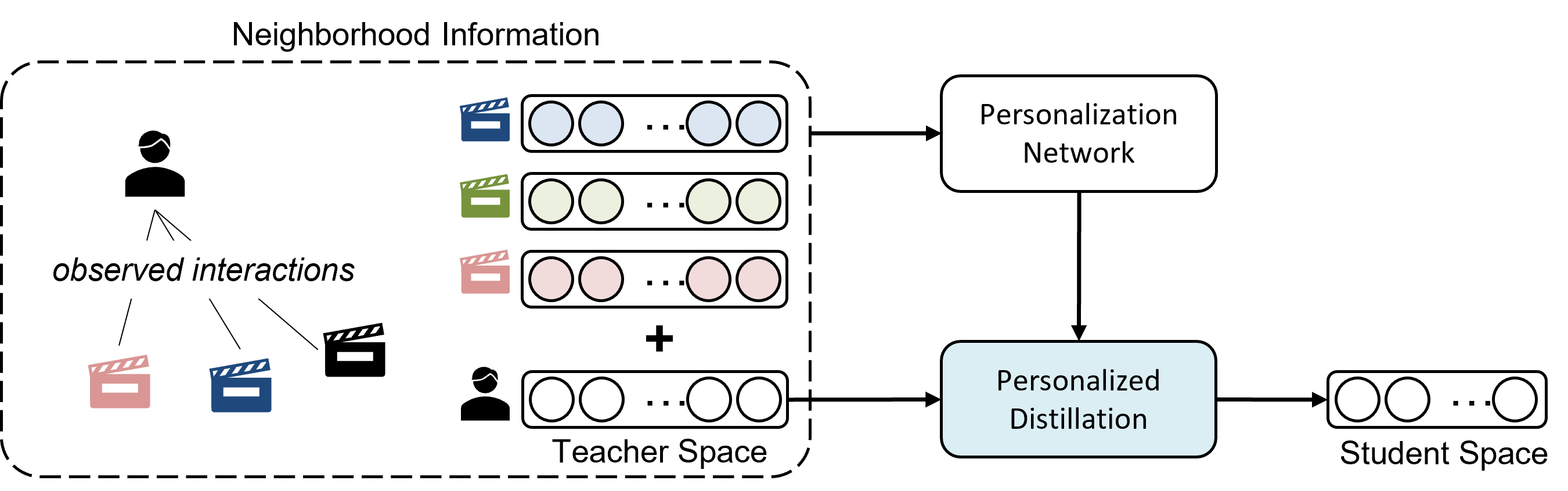}
\end{subfigure}
\caption{A conceptual illustration of personalized hint regression (PHR). PHR utilizes the personalized bridge generated by personalization network based on the neighborhood information.}
\label{fig:PHR_method}
\end{figure}

In this section, we provide the proposed method, named \underline{P}ersonalized \underline{H}int \underline{R}egression (PHR), that effectively distills the various preferences knowledge without any method-specific hyperparameters.
PHR achieves the balanced distillation by using \textit{personalization network}.
For each user/item knowledge, the network provides a personalized mapping function bridging the teacher space and the student space, preventing the distillation from being biased to a few large preference groups.
PHR distills the knowledge to the student model in an end-to-end manner.
Lastly, we discuss the advantages and limitations of PHR, and we also explain the relation of PHR with DE. 
The overview of PHR is provided in Figure \ref{fig:PHR_method} and
its training procedure is provided in Algorithm \ref{algo:PHR}.

\subsection{Personalization Network}
The knowledge of RS contains information of user groups having diverse preferences and item groups having various characteristics.
To effectively transfer the knowledge of each group in a balanced manner, 
we employ the personalization network.
The personalization network generates a mapping network that transfers each user/item representation.
To this end, we need to provide rich and comprehensive information of each representation that is helpful to make the personalized (or individualized) bridge.
Although it is possible to utilize new learning parameters for each representation, we note that this approach is prone to overfitting due to a large number of parameters.

\begin{figure}[t]
\centering
\begin{subfigure}[t]{0.4\linewidth}
    \includegraphics[width=\linewidth]{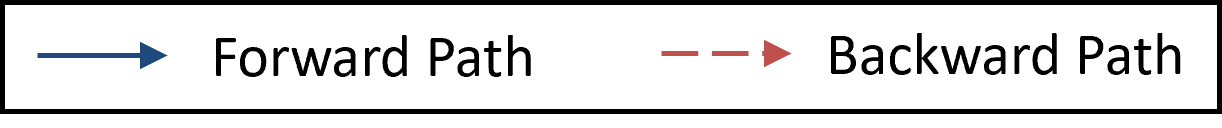}
\end{subfigure}
\vspace{0.1cm}
\\
\begin{subfigure}[t]{0.49\linewidth}
    \includegraphics[width=\linewidth]{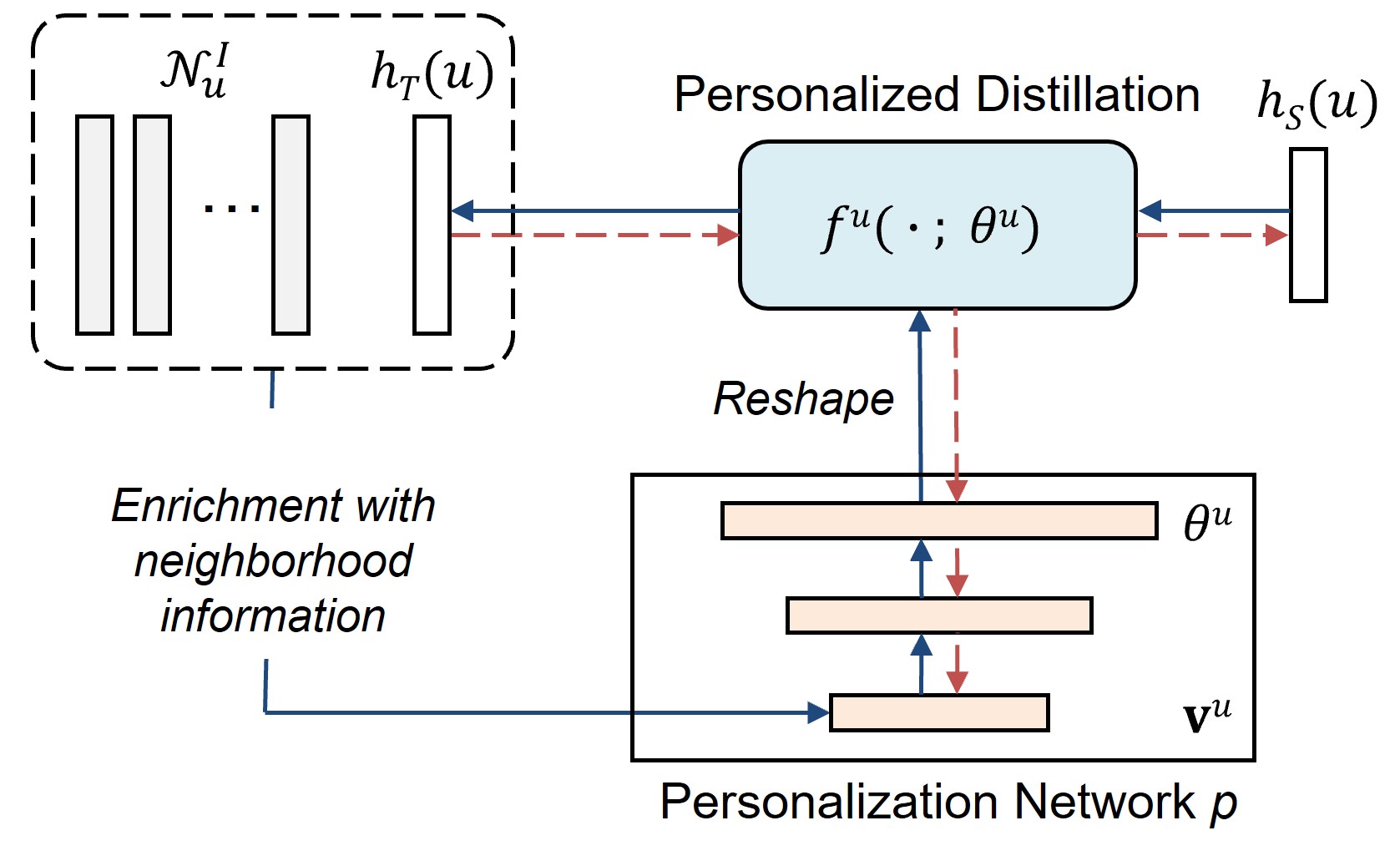}
    \subcaption{Personalized distillation for user $u$.}
\end{subfigure}
\hspace{-0.2cm}
\begin{subfigure}[t]{0.49\linewidth}
    \includegraphics[width=\linewidth]{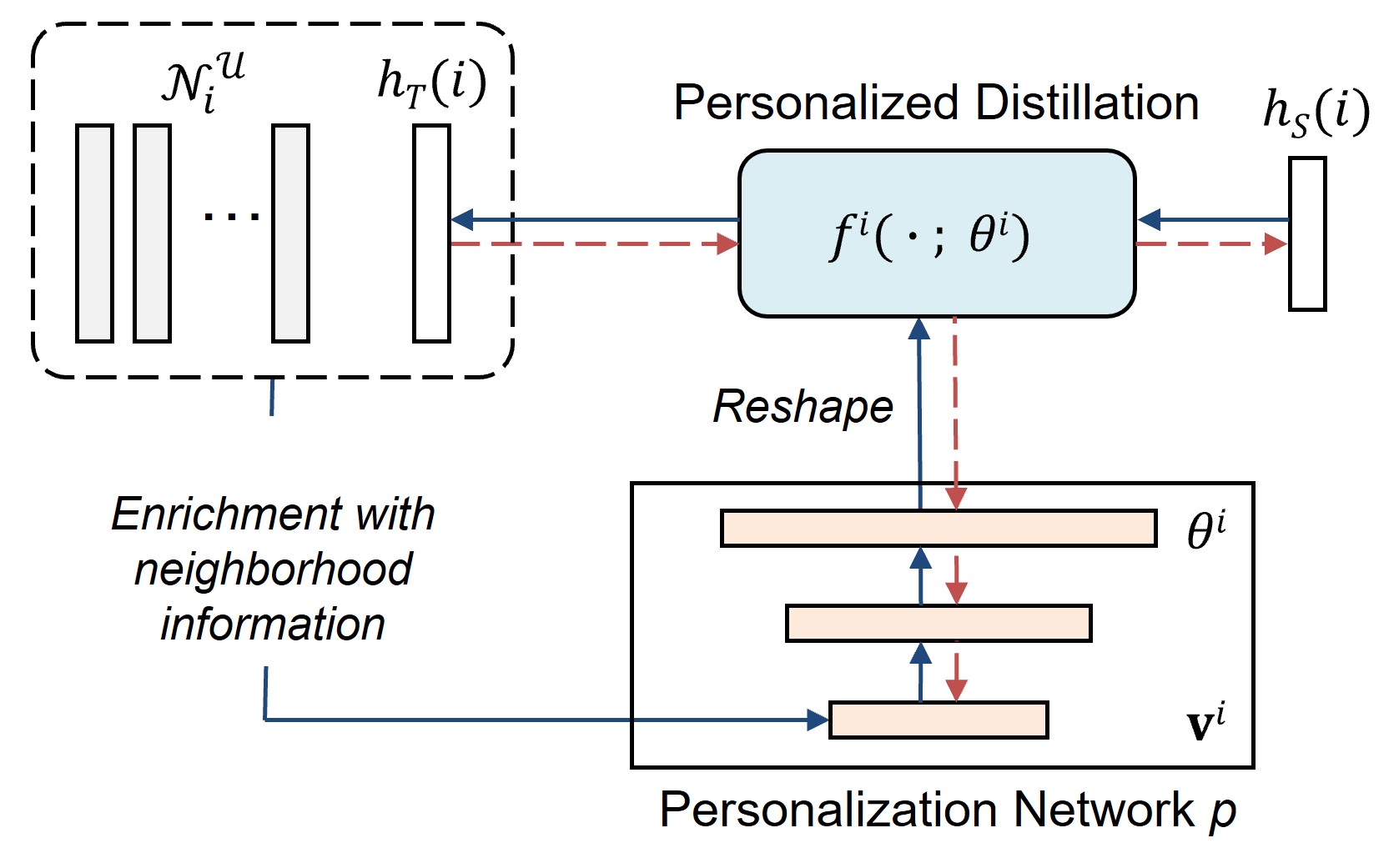}
    \subcaption{Personalized distillation for item $i$.}
\end{subfigure}
\caption{Illustration of personalized distillation process.}
\label{fig:PHR_detailed_method}
\end{figure}

In this work, we utilize \textit{neighborhood information} to enrich each representation for personalization.
The neighborhood information is defined based on the user-item interaction history; 
for each user $u$, the neighbors are the set of items that have interacted with the user (i.e., $\mathcal{N}^{\mathcal{I}}_u$), and for each item $i$, the neighbors are the set of users that have interacted with the item (i.e., $\mathcal{N}^{\mathcal{U}}_i$).
This neighborhood information has been shown to be highly effective for recommendation in modeling prominent features \cite{transCF}. 
For each user representation, we enrich it by using the neighborhood information as follows:
\begin{equation}
  \begin{aligned}
    \textbf{v}^u &= \frac{1}{2} \left( h_T(u) + \frac{1}{|\mathcal{N}^{\mathcal{I}}_u|} \sum_{i \in \mathcal{N}^{\mathcal{I}}_u} h_T(i) \right),
    \end{aligned}
\end{equation}  
Analogously, we produce the enriched item representation as follows: 
\begin{equation}
  \begin{aligned}
    \textbf{v}^i &= \frac{1}{2} \left( h_T(i) + \frac{1}{|\mathcal{N}^{\mathcal{U}}_i|} \sum_{i \in \mathcal{N}^{\mathcal{U}}_i} h_T(i) \right)
    \end{aligned}
\end{equation}  
The personalization network $p: \mathbb{R}^{d_T} \rightarrow \mathbb{R}^{(d_T \times d_S)}$ takes the enriched representation as input and generates a mapping network that transfers each representation.
\begin{equation}
    \begin{aligned}
        \theta^u = p(\textbf{v}^u),\quad \theta^i = p(\textbf{v}^i)
    \end{aligned}
    \label{eq:PHR_theta}
\end{equation}
$\theta^u$ is utilized as the parameters of mapping function $f^u: \mathbb{R}^{d_S} \rightarrow \mathbb{R}^{d_T}$ that distills the knowledge of user $u$, and $\theta^i$ is utilized as the parameters of mapping function $f^i: \mathbb{R}^{d_S} \rightarrow \mathbb{R}^{d_T}$ that distills the knowledge of item $i$.
In this work, we adopt the two-layer multi-layer perceptron for $p$ and reshape $\theta^*$ into a matrix $\mathbb{R}^{d_T \times d_S}$ so that the mapping is conducted by a simple matrix multiplication operation.
The mapping functions vary for each user/item representation and provide personalized distillation, enabling effective distillation of the knowledge with various preferences.
This process is illustrated in Figure \ref{fig:PHR_detailed_method}.
It is possible to use two separate personalization networks for user and item (i.e., $p^u(\cdot)$ and $p^i(\cdot)$), respectively.
However, we consistently obtain similar performance by using a single network shared for both user and item.

\begin{algorithm}[th!]
\SetKwInOut{Input}{Input}
\SetKwInOut{Output}{Output}
\Input{Training set $\mathcal{D}$, A pre-trained Teacher model $T(\cdot; \theta_T)$, Learning rate $\gamma$}
\Output{Student model $S(\cdot; \theta_S)$}
Randomly initialize student model $S(\cdot; \theta_S)$, personalization network $p(\cdot; \theta_p)$\\
\While{not convergence}{

\For{each batch $\mathcal{B} \in \mathcal{D}$}{
\For{each user/item $e \in \mathcal{B}$}{
\BlankLine
\tcc{ \small personalized mapping function}
Compute the enriched representation $\textbf{v}^e = \frac{1}{2} \left( h_T(e) + \frac{1}{|\mathcal{N}_e|} \sum_{e' \in \mathcal{N}_e} h_T(e') \right)$\\
Compute the mapping parameters $\theta^e = p(\textbf{v}^e)$\\
Reshape $\theta^e$ to $\mathbb{R}^{d_T \times d_S}$\\
Generate the personalized mapping function $f^e(\, \cdot \,; \theta^e)$\\
\BlankLine
\tcc{\small personalized hint regression}
Compute the PHR loss $\mathcal{L}(e) = \frac{1}{2} \lVert h_T(e) - f^e\left(h_S\left(e\right);\theta^e\right) \rVert^2_2$\\
}
}

\BlankLine
\tcc{\small optimization}
Compute the loss of the base model $\mathcal{L}_{Base}$\\
Compute the PHR loss for a batch $\mathcal{L}_{PHR} = \sum_{e\in\mathcal{B}}\mathcal{L}(e)$\\
Compute $\mathcal{L} = \mathcal{L}_{Base} + \lambda_{PHR} \mathcal{L}_{PHR}$\\
\BlankLine
Update student model $\theta_S \leftarrow \gamma \frac{\partial \mathcal{L}}{\partial \theta_{S}}$\\
Update personalization network $\theta_p \leftarrow \gamma \frac{\partial \mathcal{L}}{\partial \theta_{p}}$\\
}
\caption{Algorithm of PHR.}
\label{algo:PHR}
\end{algorithm}

\subsection{Personalized Hint Regression}
With the generated mapping functions from the personalization network, we distill the knowledge of each representation.
To this end, we adopt the hint regression that reconstructs the teacher representation from the student representation.
\begin{equation}
    \begin{aligned}
        \widehat{h_T(u)} = f^u(h_S(u); \theta^u),\quad \widehat{h_T(i)} = f^i(h_S(i); \theta^i)
    \end{aligned}
\end{equation}
\begin{equation}
    \mathcal{L}(u) = \frac{1}{2} \lVert h_T(u) - \widehat{h_T(u)} \rVert^2_2, \quad \mathcal{L}(i) = \frac{1}{2} \lVert h_T(i) - \widehat{h_T(i)} \rVert^2_2,
\end{equation}
where $\mathcal{L}(u)$ is the hint regression loss for user $u$, and $\mathcal{L}(u)$ is the corresponding loss for item $i$.
By minimizing the above equation, the parameters in the student model (i.e., $h_S(\cdot)$) and the personalization network (i.e., $p(\cdot)$) are updated.

\subsection{Optimization}
The proposed method is jointly optimized with the student model in the end-to-end manner as follows:
\begin{equation}
    \min \mathcal{L} = \mathcal{L}_{B a s e} + \lambda_{K D} \mathcal{L}_{P H R},
\end{equation}
where $\mathcal{L}_{Base}$ is a loss function of base model, and the base model can be any existing RS.
The PHR loss (i.e., $\mathcal{L}_{PHR}$) is defined for a mini-batch $\mathcal{B}$ used for $\mathcal{L}_{Base}$.
i.e., for each user/item $e$ in $\mathcal{B}$, $\mathcal{L}_{PHR} = \sum_{e \in \mathcal{B}} \mathcal{L}(e)$.
$\lambda_{K D}$ is a hyperparameter that controls the distillation effect, and it is required for all distillation methods.
We compute the gradients for parameters in the student model and the personalization network and update them with mini-batch stochastic gradient descent (SGD).
Note that the personalization network is not used for the inference phase.
It is only used in the offline distillation phase for a more balanced distillation.

\subsection{Discussion}
\subsubsection{Advantages and limitations}\noindent
PHR requires no assumption on the representation space and no method-specific hyperparameters, and this simplicity makes it easier to apply PHR to new datasets and new base models.
DE requires several hyperparameters including the number of clusters ($K$) that need to be tuned for each dataset and each base model.
A \textit{sub-optimal} $K$, which is far from the actual number of clusters in the representation space, injects inaccurate bias about the space to the student.
This may degrade the quality of the distillation.
Also, its training procedure is complicated as it requires temperature annealing and several hyperparameters for the stable clustering process.
Unlike DE, PHR utilizes the personalization network that adaptively generates the personalized mapping function for each knowledge without relying on predefined $K$.
The only hyperparameter of PHR is $\lambda_{K D}$, balancing the base loss and KD loss, which is necessary for all the distillation methods.

Nevertheless, there still exists a limitation in PHR.
That is, PHR utilizes more parameters compared to DE during the distillation process.
We provide an analysis in terms of the number of parameters.
We adopt a multi-layer perceptron with two hidden layers for both the clustering and personalization network.
The clustering network has the shape of $[d_T \rightarrow \frac{1}{2}(d_T + K) \rightarrow K]$, and the personalization network has the shape of $[d_T \rightarrow \frac{1}{2}(d_T + d_T \times d_S) \rightarrow d_T \times d_S]$\footnote{For simplicity, the bias perceptrons are omitted in this parameter analysis. Note that the bias perceptrons are included in our experiments.}.
In sum, DE requires $\frac{1}{2}(d_T + K)^2$ parameters for the clustering network and $(d_T \times d_S) \times K$ parameters for the mapping networks.
On the other hand, PHR requires $\frac{1}{2}d_T^2(1+d_S)^2$ parameters for the personalization network
However, we would like to emphasize that the additional parameters are only utilized in the \textit{offline training phase} for the distillation.
Thus, there are no extra costs for the model inference.
Moreover, PHR achieves the comparable or even better performance compared to DE thoroughly tuned by a grid search for all of its hyperparameters.
Considering these aspects, we believe the proposed method has a great value as a distillation method for the recommender system.

\subsubsection{Relation with DE}\noindent
PHR can be viewed as a generalization of DE in the sense that it reduces to cluster-wise distillation when the personalization network is restricted to generate only $K$ distinct mapping functions.
More specifically, if we limit the output space of $p$ so that it generates $K$ different $\theta$s (i.e., $\theta_1, \theta_2, ..., \theta_K$ from Equation \ref{eq:PHR_theta}), it corresponds to the cluster-wise hint regression adopted in DE.

\section{Experiments}
This section consists of as follows:
First, we introduce the experiment setup.
Second, we provide extensive experiment results supporting the superiority of PHR.
Third, we provide detailed analyses on the effects of PHR.
Then, we present hyperparameter analysis.
Lastly, we provide the results of the proposed method on two large-scale web applications.

\subsection{Experimental Setup}
We closely follow the experiment setup of the state-of-the-art KD method, DE \cite{DERRD}.
However, for more rigorous evaluation we make two changes.
First, we additionally adopt LightGCN \cite{he2020lightgcn}, which is the state-of-the-art RS, as a base model.
Second, we adopt full-evaluation instead of sampling-based evaluation of DE.
Although it is time-consuming, it enables a more thorough evaluation \cite{CD}. 

\subsubsection{Dataset}\noindent
Following \cite{DERRD}, we use two real-world datasets: CiteULike\footnote{\url{ https://github.com/changun/CollMetric/tree/master/citeulike-t}} \cite{wang2013collaborative} and Foursquare\footnote{\url{https://sites.google.com/site/yangdingqi/home/foursquare-dataset}} \cite{liu2017experimental}.
We filter out users and items having fewer than 5 interactions for CiteULike, 20 interactions for Foursquare as done in \cite{BPR, NeuMF, SSCDR, DERRD, transCF, BUIR}.
The data statistics are summarized in Table \ref{tbl:PHR_statistic}.
\begin{table}[h]
\centering
  \caption{Data Statistics}
  \begin{tabular}{ccccc}
    \toprule
    Dataset & \#Users & \#Items & \#Interactions & Sparsity \\
    \midrule
    CiteULike & 5,220 & 25,182 & 115,142 & 99.91\% \\
    Foursquare & 19,466 & 28,594 & 609,655 & 99.89\% \\
    \bottomrule
  \end{tabular}
\label{tbl:PHR_statistic}
\end{table}

\subsubsection{Base Models}\noindent
We evaluate the proposed method on three base models having different optimization strategies and model architectures.
Our base models are highly prominent RS models having representative modeling components.
Specifically, they utilize Matrix Factorization (MF), Multi-Layer Perceptron (MLP), and Graph Convolution Network (GCN), and many subsequent methods have been proposed based on the base models.
\begin{itemize}
    \item \textbf{BPR (Bayesian Personalized Ranking) \cite{BPR}}: 
    BPR models the user-item interaction by using MF.
    It is optimized with the pair-wise objective function with an assumption that all observed items are more preferred than unobserved items. 
    Following \cite{DERRD}, we consider its embedding matrix lookup operation as an embedding layer that maps each user/item to an embedding vector, and we distill the knowledge from the output space of the embedding layer.
    Note that BPR is a highly competitive method that gives strong performance \cite{ADBPR, tay2018latent}.
    
    \item \textbf{NeuMF (Neural Matrix Factorization) \cite{NeuMF}}: 
    NeuMF is a deep model that captures the non-linear user-item relationships.
    It adopts MF and MLP, and it is optimized with binary cross-entropy loss.
    NeuMF is a representative deep method for RS, and numerous subsequent methods have been proposed based on NeuMF.
    
   \item \textbf{LightGCN (Light Graph Convolution Network) \cite{he2020lightgcn}}: 
    Graph neural network has shown outstanding performance in modeling user-item relationships \cite{he2020lightgcn}.
    LightGCN is the state-of-the-art model for top-$N$ recommendation from implicit feedback.
    It adopts the light graph convolution layers to capture multi-hop relationships in the user-item bipartite graph.
    As suggested in the original paper, we optimize it by using the pair-wise objective function.
\end{itemize}

\subsubsection{Teacher and Student}\noindent
For each base model on each dataset, we build the teacher model and the student model as follows:
\begin{itemize}
\item \textbf{Teacher}: We increase the number of learning parameters until the performance is no longer improved.
We use the model with the best performance as the teacher model.
\item \textbf{Student}: 
We build three student models by adjusting the number of learning parameters of the teacher model with three ratios $\phi \in \{0.1, 0.5, 1.0\}$.
Following DE, we adjust the number of parameters based on the size of the last hidden layer\footnote{
The last hidden layer refers to the layer where the final predictions are made based on the user/item representations from the layer.
For BPR, it refers to the embedding lookup layer where the final predictions are made by the inner product of the representations.}, and the sizes of other layers are changed accordingly.
$\phi=0.1$ means that the size of the last hidden layer is 10 times larger than that of the student model, and
$\phi=1.0$ corresponds to the self-distillation where the student and the teacher have the identical model size \cite{self_distill1}.
We denote the student model trained without distillation as ``Student''.
\end{itemize}
The performance and inference time with varying model sizes are summarized in Table \ref{tbl:PHR_sizeandtime}.
A large model has higher recommendation performance, however, it also requires a longer inference time.

\begin{table}[t!]
\centering
  \caption{Model size and inference time. \textit{dims.} refers to the dimensions of the last hidden layer of each model. Time denotes the wall time used for generating a recommendation list for every user.}
  \begin{tabular}{ccc cccc}
    \toprule
    \multicolumn{3}{c}{} & \multicolumn{2}{c}{CiteULike} & \multicolumn{2}{c}{Foursquare}\\
    \cmidrule{4-5}\cmidrule{6-7}
    Base Model & Model Size ($\phi$) & \textit{dims.} & Recall@50 & Time & Recall@50 & Time  \\
    \midrule
     & 0.1 & 20 & 0.2347 & 73s & 0.2164 & 317s \\
    BPR & 0.5 & 100 & 0.3050 & 153s & 0.2420 & 655s \\
     & 1.0 & 200 & 0.3253 & 260s & 0.2732 & 1,089s \\
    \midrule
     & 0.1 & 25 & 0.1970 & 126s & 0.1950 & 564s \\
    NeuMF & 0.5 & 125 & 0.2688 & 348s & 0.2477 & 1,740s \\
     & 1.0 & 250 & 0.2993 & 784s & 0.2529 & 3,245s \\
    \midrule
     & 0.1 & 10 & 0.2512 & 81s & 0.2202 & 370s \\
    LightGCN & 0.5 & 100 & 0.3063 & 191s & 0.2677 & 701s \\
     & 1.0 & 200 & 0.3326 & 330s & 0.2806 & 1,255s \\
    \bottomrule
  \end{tabular}
  \label{tbl:PHR_sizeandtime}
\end{table}

\subsubsection{Compared Methods}\noindent
We compare the proposed method with the following competitors:
\begin{itemize}
    \item \textbf{HR (Hint Regression) \cite{FitNet}}: Hint regression is originally proposed in computer vision and has been successfully applied to RS.
    It reconstructs the teacher representation from the student representation via a distillation network.
    
    \item \textbf{Distillation Experts (DE) \cite{DERRD}}: Distillation Experts is the state-of-the-art KD method for RS.
    It adopts cluster-wise hint regression to effectively distill the various preference knowledge.
\end{itemize}
Note that we do not include the KD method that distills the recommendation results (e.g., RRD \cite{DERRD}) in the compared methods.
This is because they belong to another research direction and they are not competing with the methods distilling the intermediate knowledge \cite{DERRD}.

\subsubsection{Evaluation Protocol and Metrics}\noindent
Following \cite{DERRD}, we adopt \textit{leave-one-out} evaluation protocol that holds out two interacted items for testing and validation and uses the rest for training.
\cite{DERRD} employs a sampling-based evaluation scheme that evaluates how well each method can rank the test items higher than 499 sampled unobserved items. 
However, for a more thorough evaluation, we utilize a full-ranking evaluation that evaluates how well each method can rank the test items higher than all the unobserved items. 
Although it is highly time-consuming, it enables more accurate evaluation \cite{krichene2020sampled, TD}.

We adopt two widely used ranking metrics for evaluating the top-$N$ recommendation performance \cite{TD, DERRD}:
(1) Recall@$N$ (R@$N$) that measures whether the test item is present in the top-$N$ ranking list, 
(2) NDCG@$N$ (N@$N$) that gives higher scores to the hits at higher ranks.  
Recall@$N$ and NDCG@$N$ are defined as follows:
\begin{equation}
\text{Recall} @ N = \frac { 1 } { | \mathcal { U }_\text{test} | } \sum _ { u \in \mathcal { U} _\text{test} } \delta \left( p _ { u } \leq \text { top } N \right), 
\end{equation}
\begin{equation}
\text{NDCG} @ N = \frac { 1 } { | \mathcal { U } _\text{test} | } \sum _ { u \in \mathcal { U }_\text{test} } \frac { 1 } { \log \left( p _ { u } + 1 \right) },
\end{equation}
where $\mathcal { U }_{test}$ is the set of the test users, $\delta ( \cdot )$ is the indicator function, and $p_u$ is the hit position of the test item for the user $u$. 
We report the average values of 5 independent runs.

\begin{table}[t]
\caption{Hyperparameters of HR, DE, and PHR.}
\label{tbl:PHR_hp_comp}
\begin{tabular}{cll}
\toprule
\multicolumn{1}{l}{KD method} & Hyperparameters      & Search Range \\
\midrule
\multicolumn{1}{l}{HR, PHR}   & $\lambda_{KD}$       & $\{ 10^{-1}, 10^{-2}, 10^{-3}, 10^{-4}, 10^{-5}\}$                                                       \\
\midrule
\multirow{6}{*}{DE}           & $\lambda_{KD}$       & $\{ 10^{-1}, 10^{-2}, 10^{-3}, 10^{-4}, 10^{-5}\}$  \\
                              & \#Clusters ($K$)     & $\{5, 10, 20, 30, 40, 50\}$  \\
                              & Initial temperature ($\tau_0$)  &  $\{1, 10^1, 10^2, 10^3, 10^4, 10^5\}$ \\
                              & Final temperature ($\tau_{F}$) & \makecell[l]{$\{1, 10^{-1}, 10^{-2}, 10^{-3}, 10^{-4}, 10^{-5}$\\ $10^{-6}, 10^{-7}, 10^{-8}, 10^{-9}, 10^{-10}\}$} \\
                              & Annealing scheduling & \{Exponential, Step\}\\
                              & Annealing update period& $\{1, 5, 10\}$ \\
\bottomrule
\end{tabular}
\end{table}

\subsubsection{Implementation Details}\noindent

\begin{itemize}
\item The proposed method and all the compared methods are implemented by using PyTorch and optimized by using Adam optimizer.
All hyperparameters are tuned by the grid search on the validation set.
The learning rate is selected from \{0.1, 0.05, 0.01, 0.005, 0.001, 0.0005, 0.0001\}, the model regularizer is selected from \{0.1, 0.01, 0.001, 0.0001, 0.00001\}.
We set the number of total epochs to 500 and adopt the early stopping strategy; stopping if Recall@$50$ on the validation set does not increase for 20 successive epochs.
For all base models (i.e., BPR, NeuMF, and LightGCN), the number of negative samples is set to 1.
For NeuMF and LightGCN, the number of the hidden layers is chosen from \{1, 2, 3, 4\}.
For HR, DE, and PHR, $\lambda_{KD}$ is selected from \{0.1, 0.01, 0.001, 0.0001, 0.00001\}.
Following the setup of DE, for all KD methods, we distill the knowledge from the last hidden layer of the teacher model.
For NeuMF that learns joint representations of users and items, we enrich each representation in the embedding level.
We leave the layer selection and simultaneously using multiple layers as the future work.

\item For DE, we thoroughly tuned all of its hyperparameters (Table \ref{tbl:PHR_hp_comp}) for each dataset and each base model.
Specifically, $K$ is chosen from \{5, 10, 20, 30, 40, 50\}. 
The initial temperature ($\tau_0$) is chosen in the range of $[1, 10^5]$ and the final temperature ($\tau_F$) is chosen in the range of $[1, 10^{-10}]$.
We use the annealing schedule that gradually decrease the temperature value during the training as done in the original paper: $\tau(f)=\tau_{0}\left(\tau_{F} / \tau_{0}\right)^{f / F}$.
$\tau(f)$ is the temperature at epoch $f$ and $F$ is the total epochs.
The search ranges are summarized in Table \ref{tbl:PHR_hp_comp}.
\end{itemize}

\begin{table}[t]
  \caption{Recommendation performances ($\phi=0.1$) on CiteULike dataset. We conduct the paired $t$-test for PHR with DE and report the $p$ values.}
  \begin{tabular}{clcccccc}
    \toprule 
    Base Model & KD Method & R@10 & N@10 & R@20 & N@20 & R@50 & N@50\\
    \midrule
    &Teacher&0.1533&0.0883&0.2196&0.1058&0.3253&0.1247\\
    &Student&0.1014&0.0560&0.1506&0.0684&0.2347&0.0864\\
    BPR&HR&0.1097&0.0595&0.1610&0.0738&0.2521&0.0924\\
    &DE&0.1165&0.0645&0.1696&0.0778&0.2615&0.0960\\
    &PHR&\underline{0.1202}&\underline{0.0663}&\underline{0.1753}&\underline{0.0801}&\underline{0.2694}&\underline{0.0988}\\
    \cmidrule{2-8}
    & $p$-value &0.033&0.032&0.036&0.043&0.018&0.030\\
    \midrule
    &Teacher&0.1487&0.0844&0.2048&0.0986&0.2993&0.1155\\
    &Student&0.0856&0.0449&0.1249&0.0553&0.1970&0.0697\\
    NeuMF&HR&0.0856&0.0469&0.1275&0.0576&0.2020&0.0723\\
    &DE&0.0882&0.0475&0.1306&0.0581&0.2090&0.0736\\
    &PHR&\underline{0.0904}&\underline{0.0481}&\underline{0.1335}&\underline{0.0588}&\underline{0.2132}&\underline{0.0745}\\
    \cmidrule{2-8}
    & $p$-value &0.244&0.364&0.158&0.340&0.109&0.302\\
    \midrule
    &Teacher&0.1610&0.0934&0.2274&0.1091&0.3326&0.1299\\
    &Student&0.1125&0.0618&0.1642&0.0748&0.2512&0.0944\\
    LightGCN&HR&0.1151&0.0642&0.1710&0.0783&0.2653&0.0969\\
    &DE&0.1189&0.0664&0.1733&0.0801&0.2680&0.0988\\
    &PHR&\underline{0.1229}&\underline{0.0681}&\underline{0.1769}&\underline{0.0816}&\underline{0.2763}&\underline{0.1011}\\
    \cmidrule{2-8}
    & $p$-value &0.026&0.024&0.063&0.029&0.001&0.010\\
    \bottomrule
  \end{tabular}
  \label{tab:PHR_main1}
\end{table}

\begin{table}[t]
  \caption{Recommendation performances ($\phi=0.1$) on Foursquare dataset. We conduct the paired $t$-test for PHR with DE and report the $p$ values.}
  \begin{tabular}{clcccccc}
    \toprule 
    Base Model & KD Method & R@10 & N@10 & R@20 & N@20 & R@50 & N@50\\
    \midrule
    &Teacher&0.1187&0.0695&0.1700&0.0825&0.2732&0.1028\\
    &Student&0.0911&0.0544&0.1333&0.0648&0.2164&0.0809\\
    BPR&HR&0.0957&0.0564&0.1386&0.0672&0.2258&0.0845\\
    &DE&0.0979&0.0567&0.1434&0.0681&0.2322&0.0856\\
    &PHR&\underline{0.1019}&\underline{0.0595}&\underline{0.1475}&\underline{0.0705}&\underline{0.2350}&\underline{0.0879}\\
    \cmidrule{2-8}
    & $p$-value &0.007&0.014&0.033&0.021&0.024&0.017\\
    \midrule
    &Teacher&0.1060&0.0590&0.1546&0.0716&0.2529&0.0910\\
    &Student&0.0737&0.0393&0.1125&0.0490&0.1950&0.0653\\
    NeuMF&HR&0.0829&0.0462&0.1243&0.0564&0.2062&0.0729\\
    &DE&0.0855&0.0476&0.1255&0.0576&0.2089&0.0741\\
    &PHR&\underline{0.0881}&\underline{0.0495}&\underline{0.1295}&\underline{0.0599}&\underline{0.2149}&\underline{0.0765}\\
    \cmidrule{2-8}
    & $p$-value &0.076&0.081&0.029&0.051&0.016&0.055\\
    \midrule
    &Teacher&0.1259&0.073&0.1779&0.0865&0.2806&0.1067\\
    &Student&0.0951&0.0564&0.1372&0.0670&0.2202&0.0834\\
    LightGCN&HR&0.0993&0.0587&0.1431&0.0697&0.2315&0.0872\\
    &DE&\underline{0.1051}&0.0617&\underline{0.1503}&\underline{0.0731}&\underline{0.2410}&\underline{0.0910}\\
    &PHR&0.1045&\underline{0.0627}&0.1498&0.0726&0.2402&0.0899\\
    \cmidrule{2-8}
    & $p$-value &0.129&0.112&0.103&0.097&0.104&0.111\\
    \bottomrule
  \end{tabular}
  \label{tab:PHR_main2}
\end{table}

\begin{figure}[t]
\centering
\begin{subfigure}[t]{0.40\linewidth}
    \includegraphics[width=\linewidth]{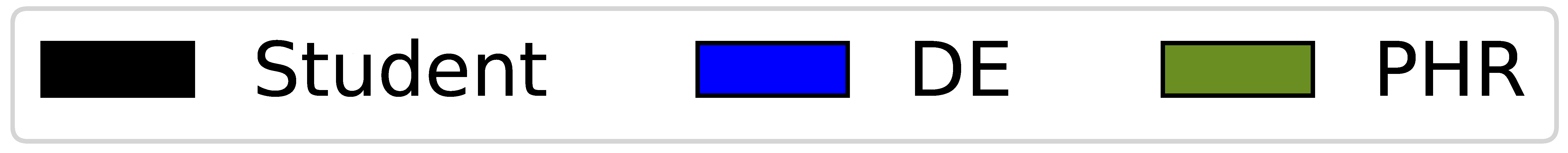}
\end{subfigure}
\\
\begin{subfigure}[t]{0.40\linewidth}
    \includegraphics[width=\linewidth]{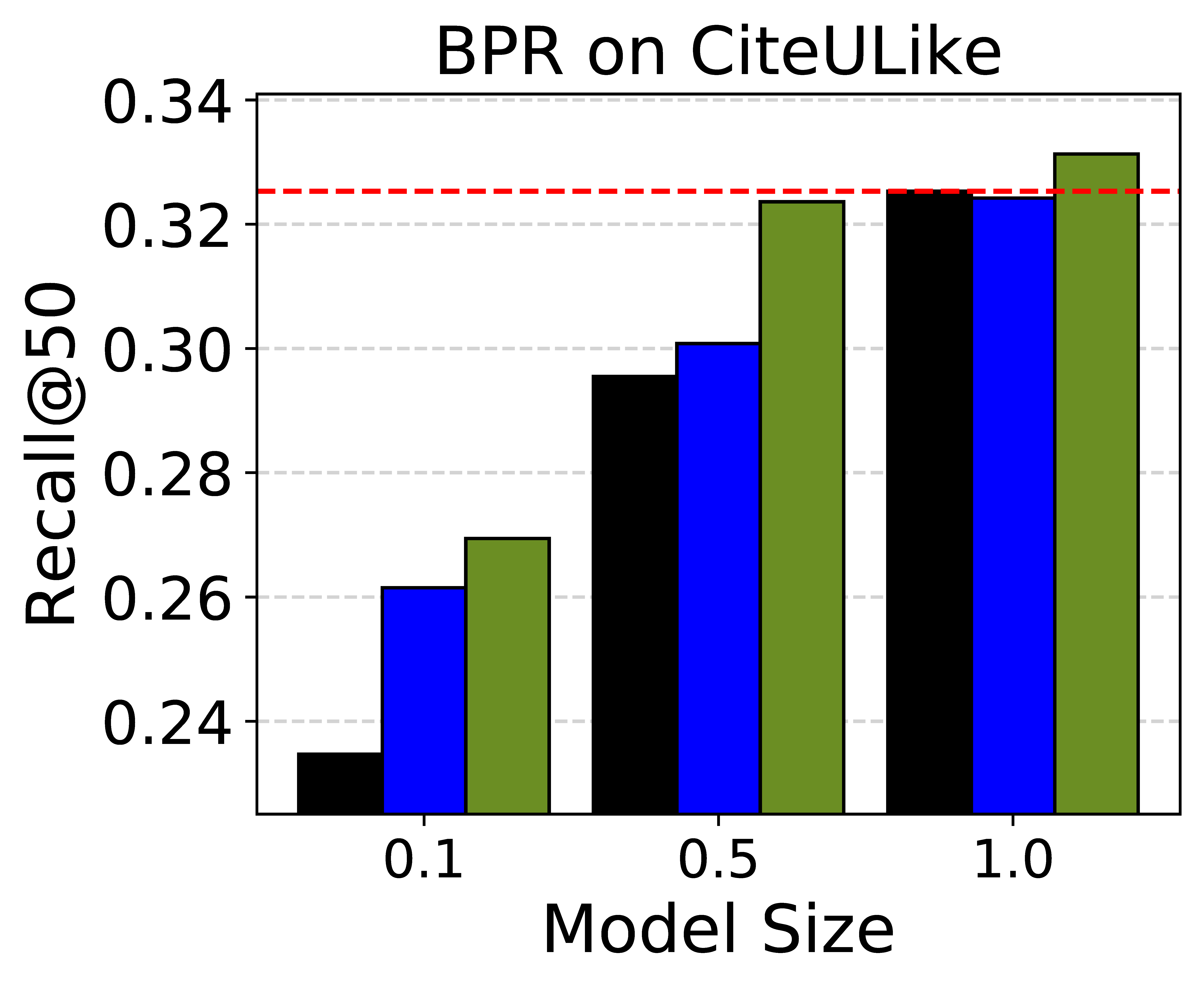}
\end{subfigure}
\begin{subfigure}[t]{0.40\linewidth}
    \includegraphics[width=\linewidth]{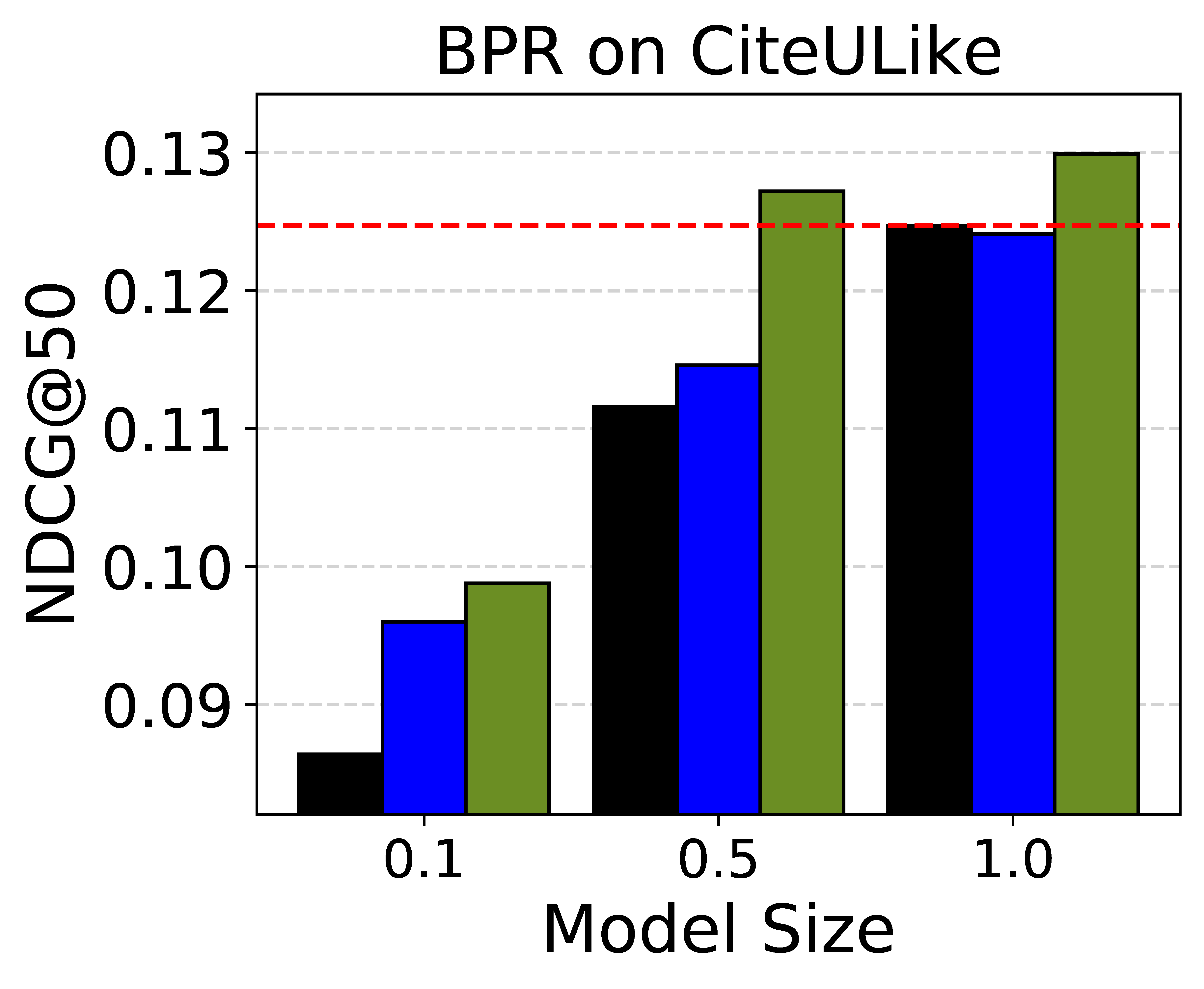}
\end{subfigure} 
\begin{subfigure}[t]{0.40\linewidth}
    \includegraphics[width=\linewidth]{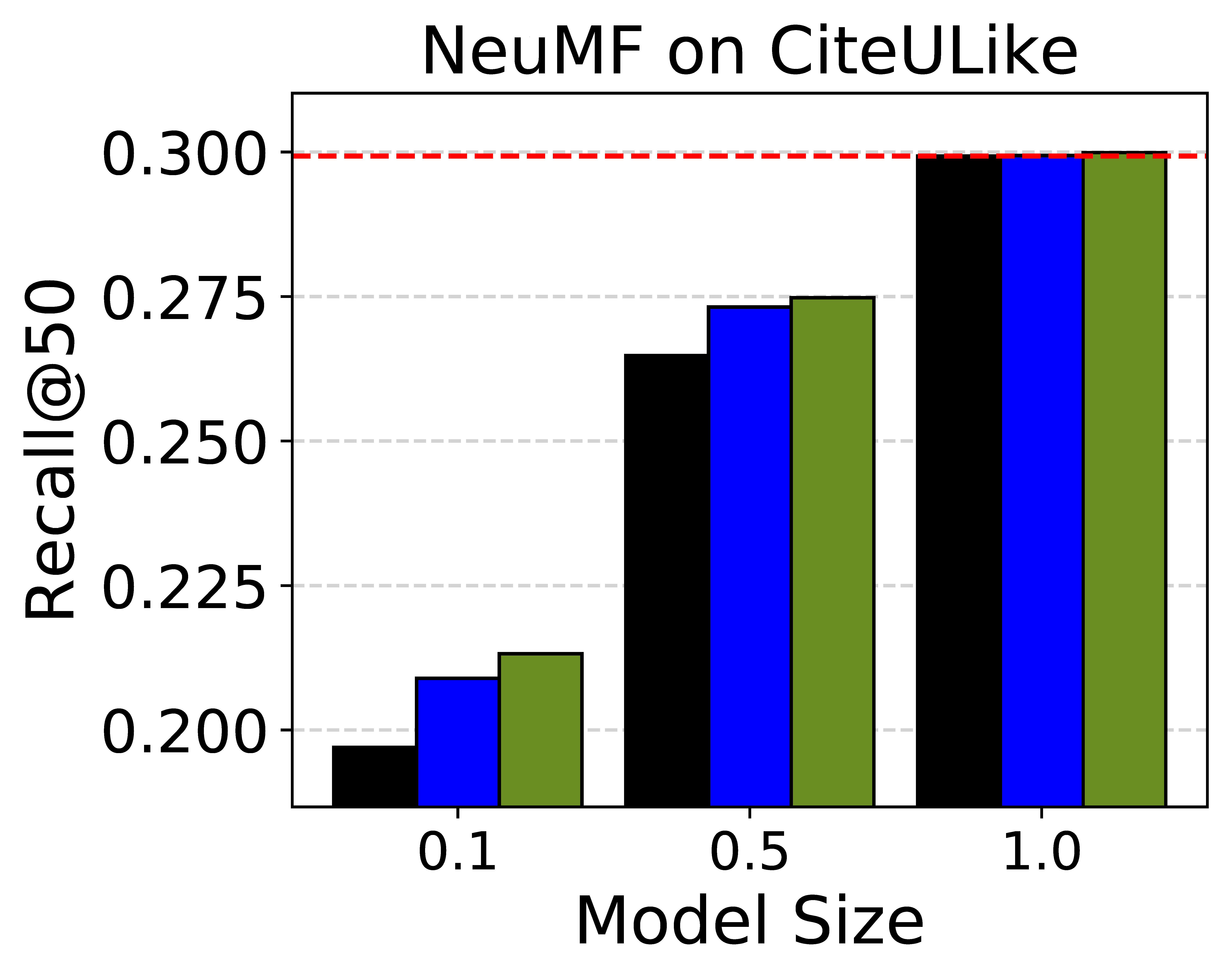}
\end{subfigure} 
\begin{subfigure}[t]{0.40\linewidth}
    \includegraphics[width=\linewidth]{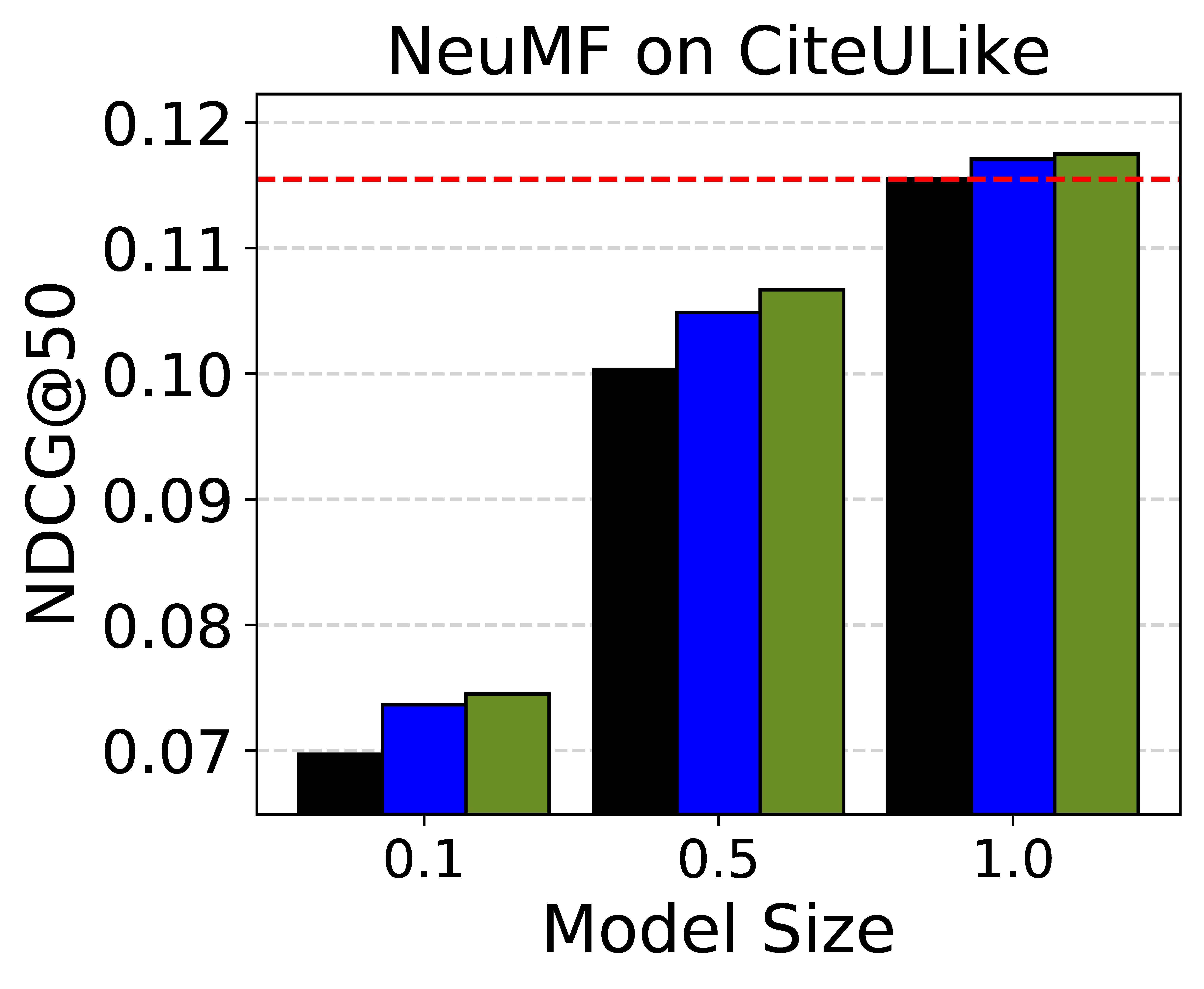}
\end{subfigure}
\begin{subfigure}[t]{0.40\linewidth}
    \includegraphics[width=\linewidth]{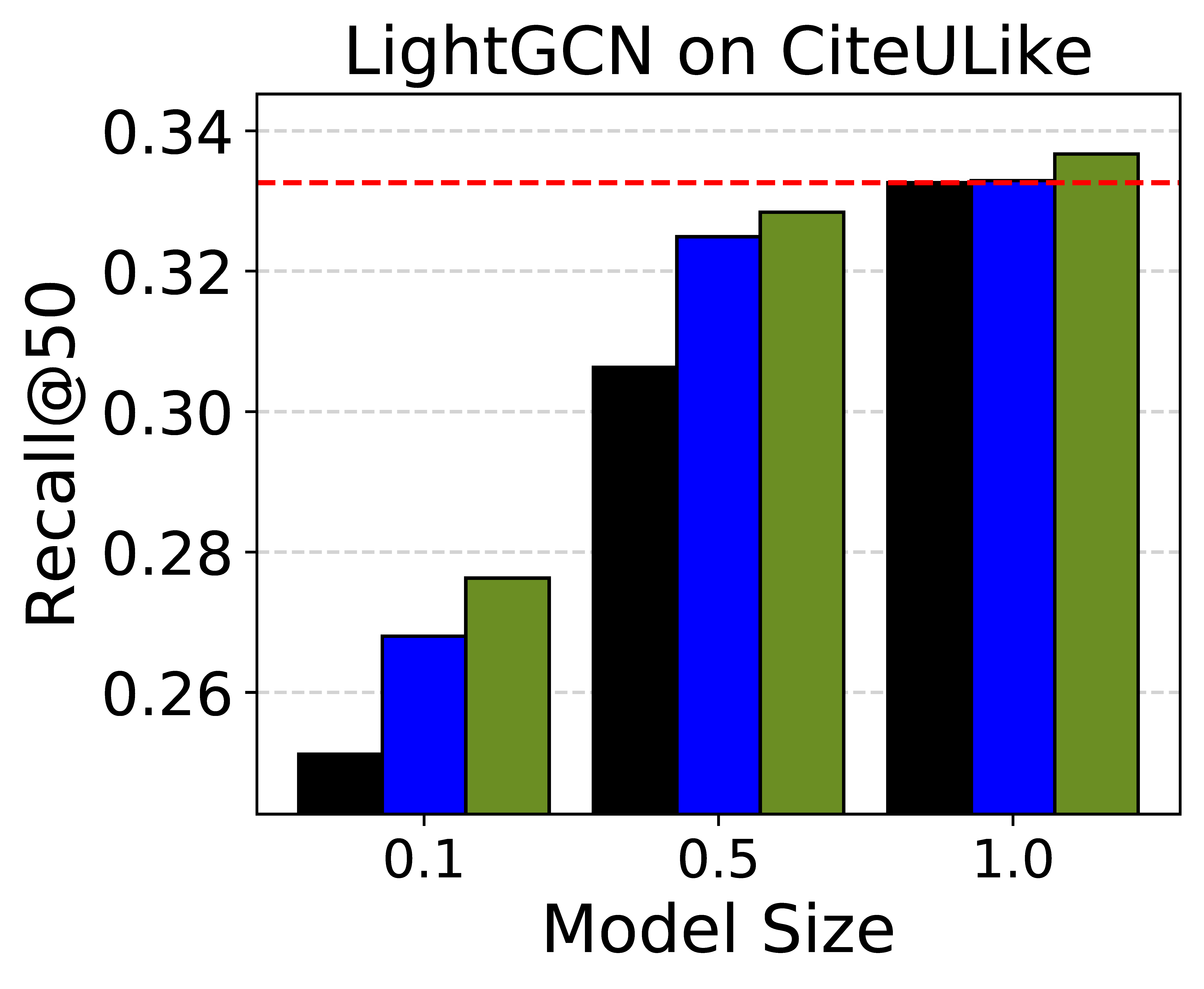}
\end{subfigure} 
\begin{subfigure}[t]{0.40\linewidth}
    \includegraphics[width=\linewidth]{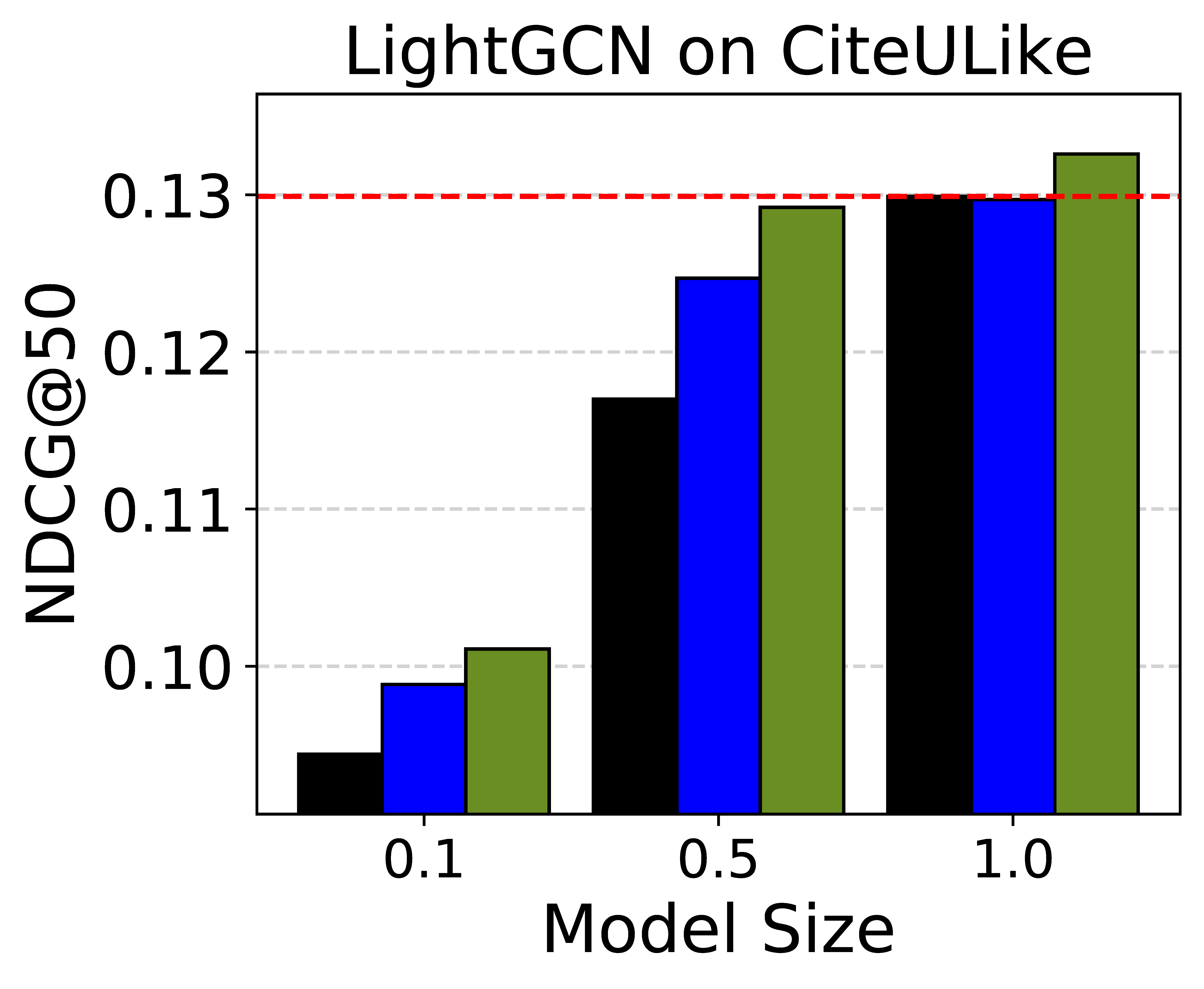}
\end{subfigure}
\caption{Recommendation performance across three different student model sizes. (Red dotted line: Teacher)}
\label{fig:PHR_size1}
\end{figure}

\begin{figure}[t]
\centering
\begin{subfigure}[t]{0.40\linewidth}
    \includegraphics[width=\linewidth]{chapters/phr/images/size_legend2.png}
\end{subfigure}
\\
\begin{subfigure}[t]{0.40\linewidth}
    \includegraphics[width=\linewidth]{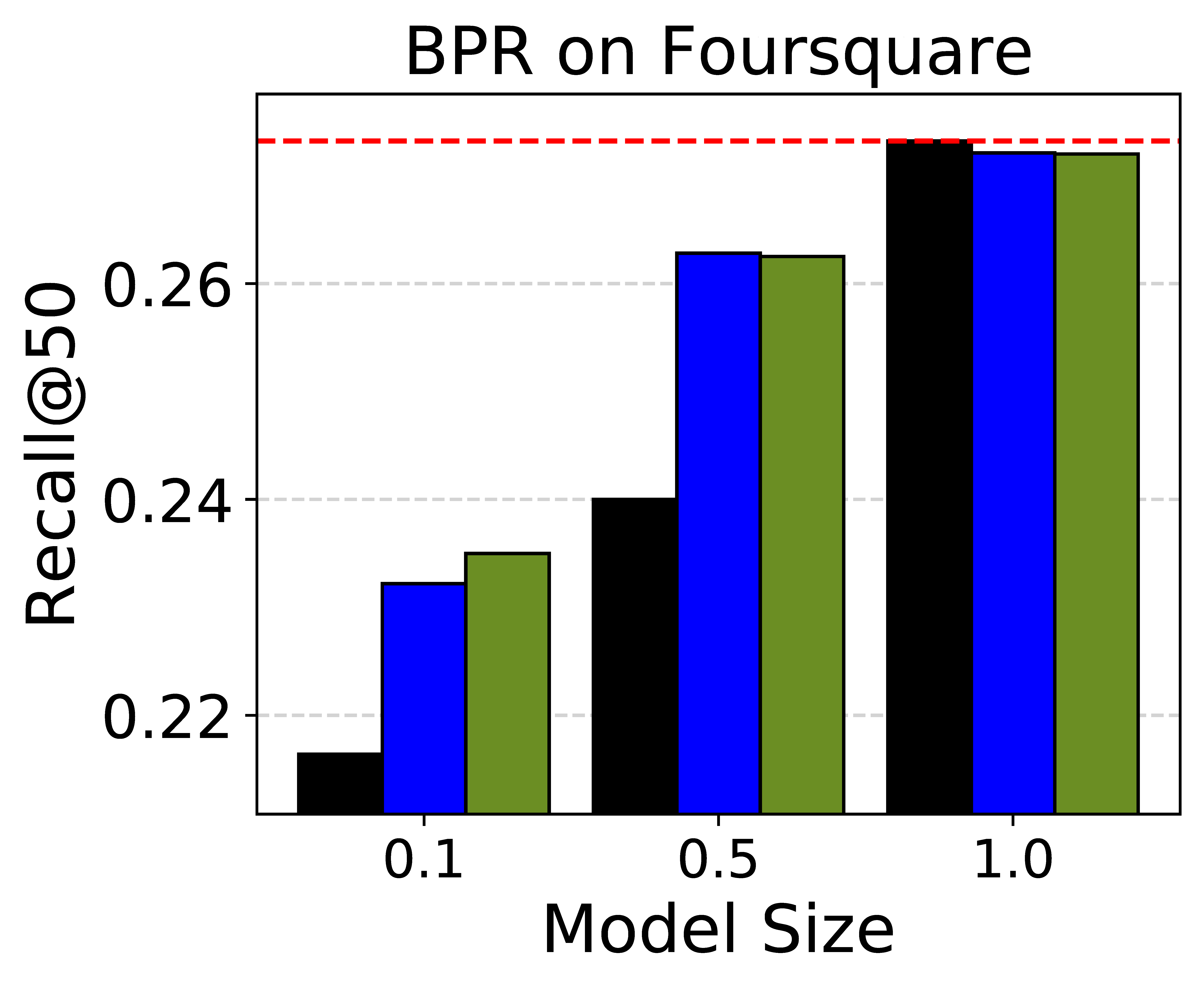}
\end{subfigure}
\begin{subfigure}[t]{0.40\linewidth}
    \includegraphics[width=\linewidth]{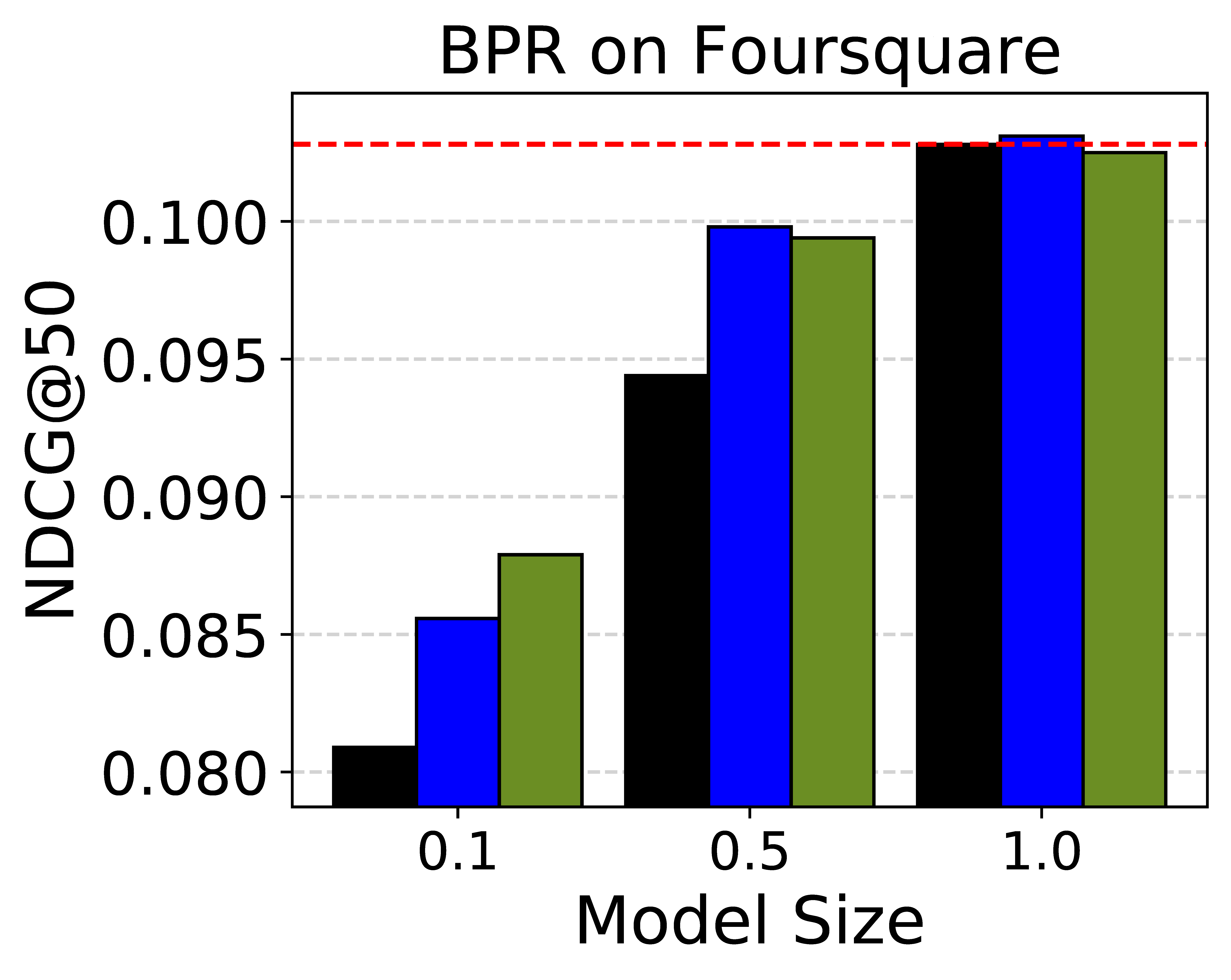}
\end{subfigure} 
\begin{subfigure}[t]{0.40\linewidth}
    \includegraphics[width=\linewidth]{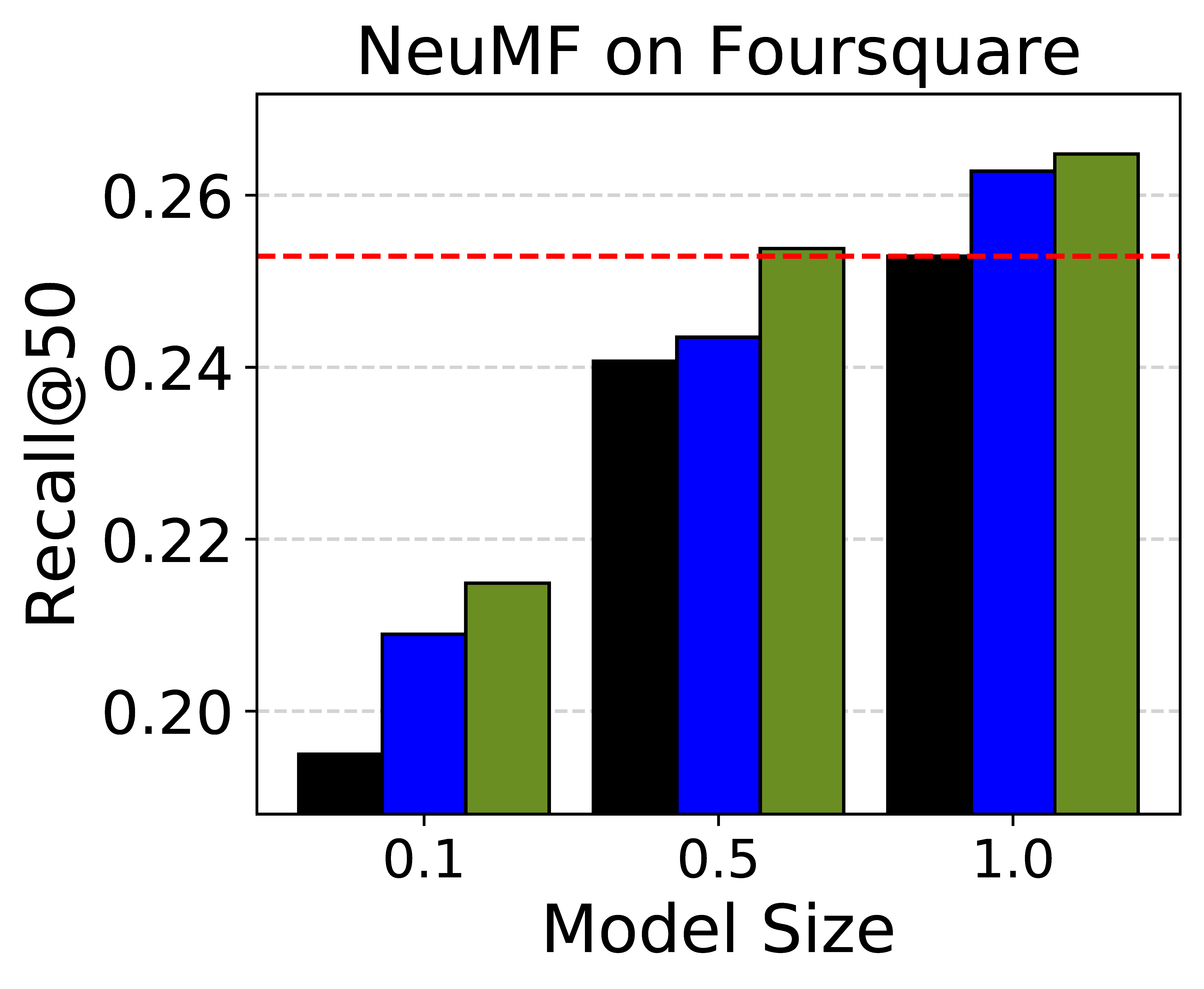}
\end{subfigure} 
\begin{subfigure}[t]{0.40\linewidth}
    \includegraphics[width=\linewidth]{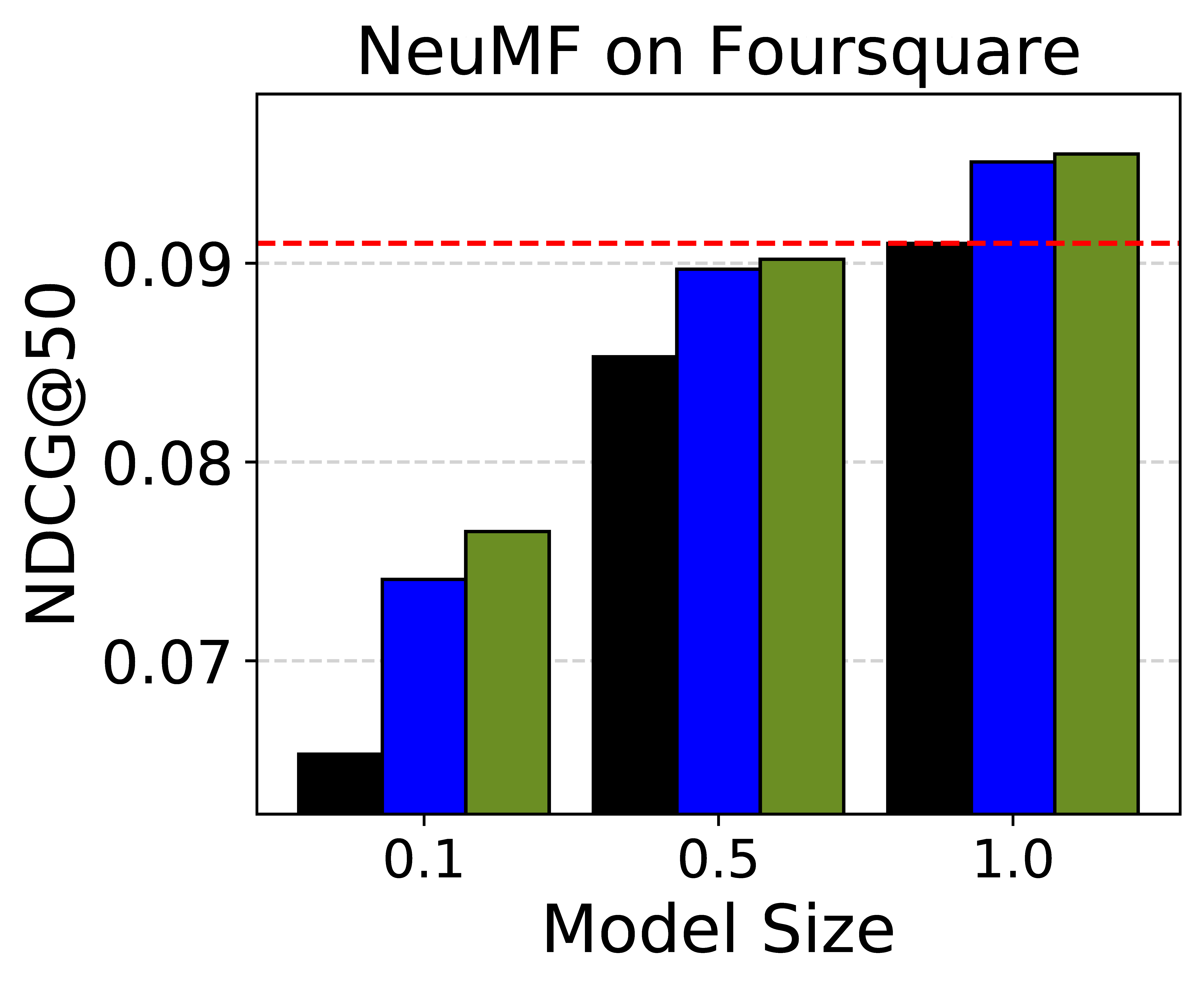}
\end{subfigure}
\begin{subfigure}[t]{0.40\linewidth}
    \includegraphics[width=\linewidth]{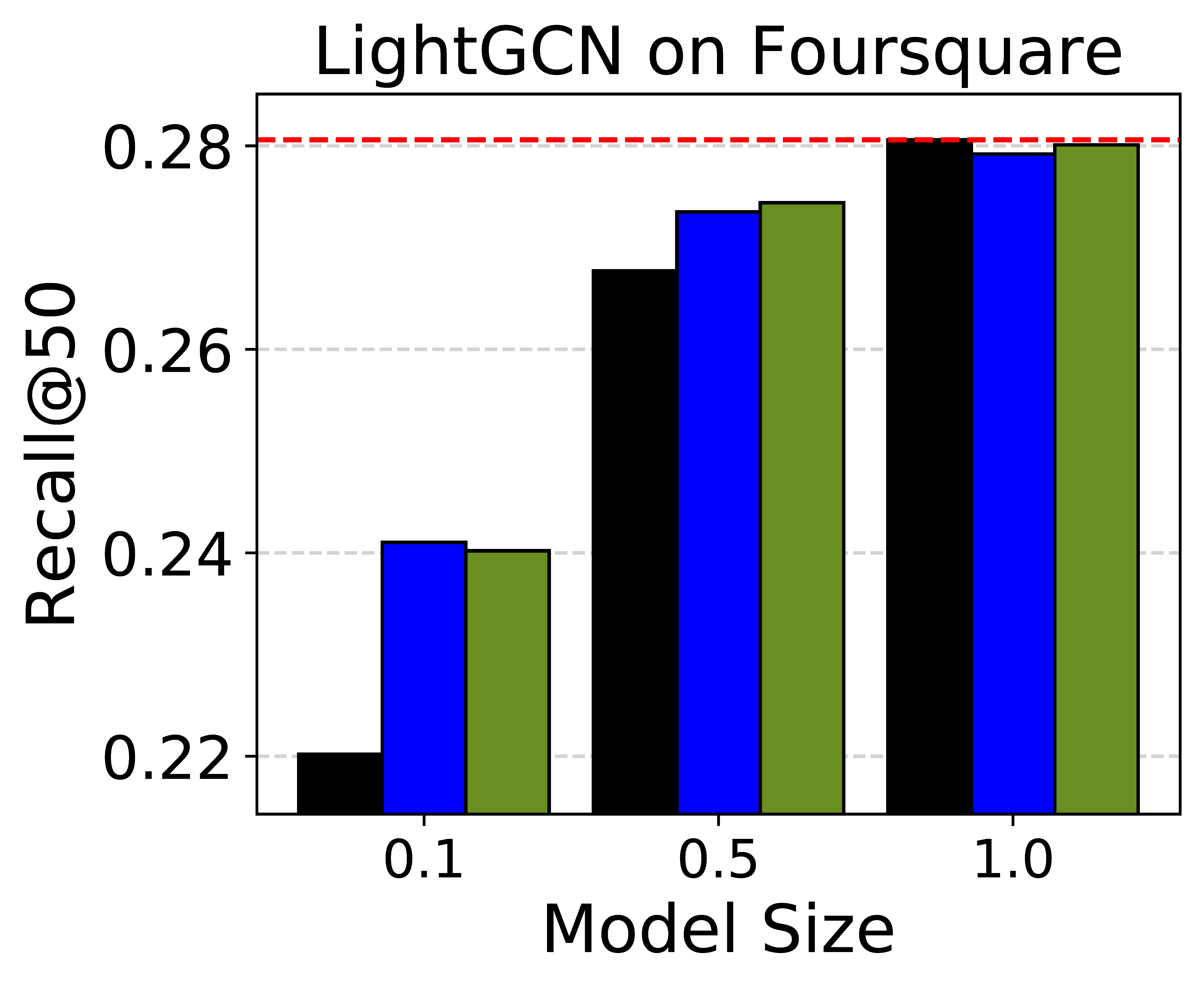}
\end{subfigure} 
\begin{subfigure}[t]{0.40\linewidth}
    \includegraphics[width=\linewidth]{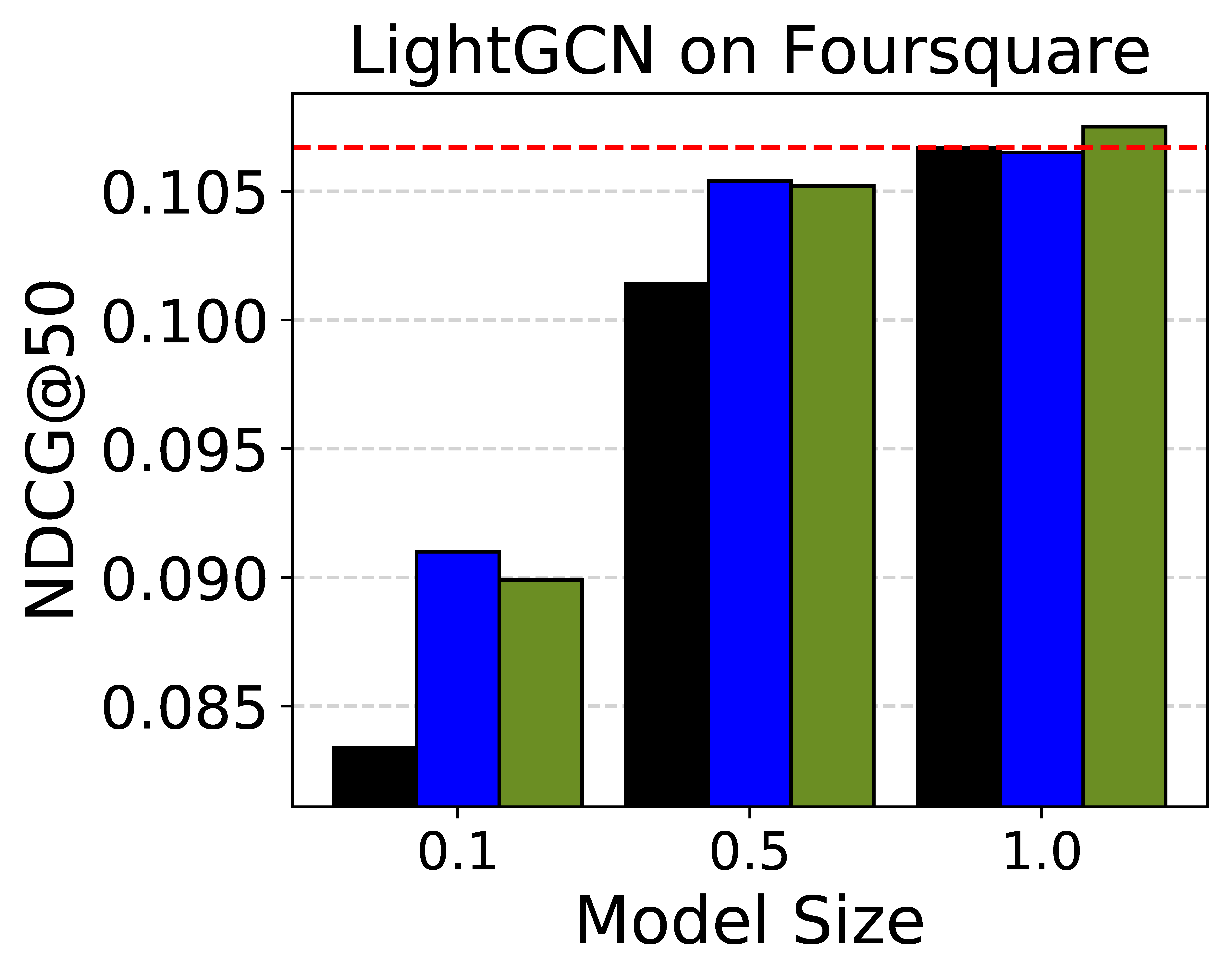}
\end{subfigure}
\caption{Recommendation performance across three different student model sizes. (Red dotted line: Teacher)}
\label{fig:PHR_size2}
\end{figure}

\begin{table}[t]
  \caption{Recommendation performances ($\phi=0.1$) of BPR for ablations.}
  \begin{tabular}{llcccccc}
    \toprule 
    Dataset &KD Method & R@10 & N@10 & R@20 & N@20 & R@50 & N@50\\
    \midrule
    &DE&0.1165&0.0645&0.1696&0.0778&0.2615&0.0960\\
    CiteULike &DE + NB &0.1148&0.0653&0.1663&0.0782&0.2591&0.0967\\
    &PHR - NB&0.1138&0.0613&0.1634&0.0737&0.2556&0.0919\\
    &PHR &0.1202&0.0663&0.1753&0.0801&0.2694&0.0988\\
    \midrule
    &DE&0.0979&0.0567&0.1434&0.0681&0.2322&0.0856\\
    Foursquare &DE + NB &0.1002&0.0582&0.1456&0.0696&0.2323&0.0867\\
    &PHR - NB&0.0999&0.0581&0.1442&0.0694&0.2286&0.0859\\
    &PHR &0.1019&0.0595&0.1475&0.0705&0.2350&0.0879\\
    \bottomrule
  \end{tabular}
  \label{tab:PHR_abl}
\end{table}

\begin{figure}[h!]
\centering
\begin{subfigure}[t]{0.40\linewidth}
    \includegraphics[width=\linewidth]{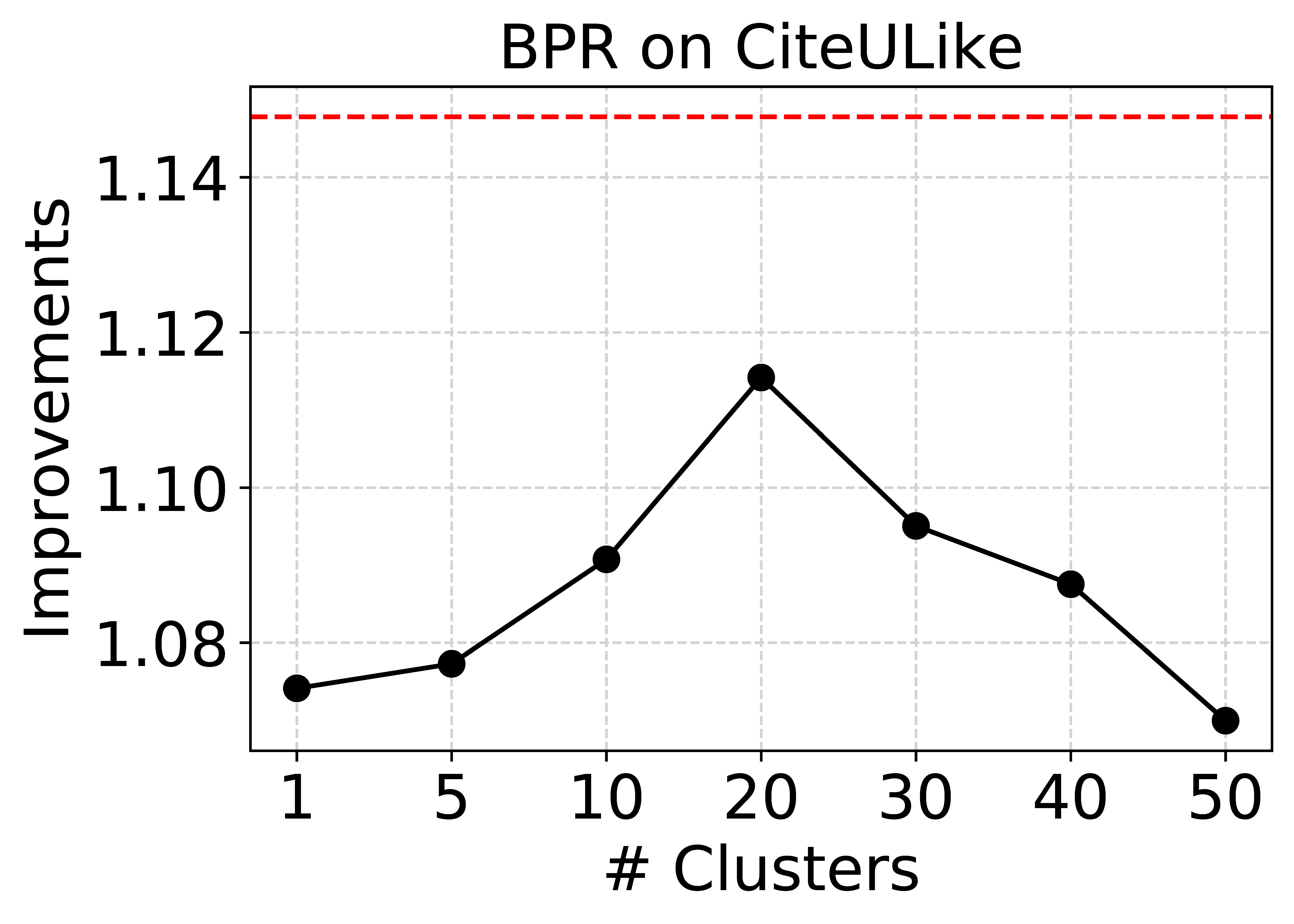}
\end{subfigure}
\begin{subfigure}[t]{0.40\linewidth}
    \includegraphics[width=\linewidth]{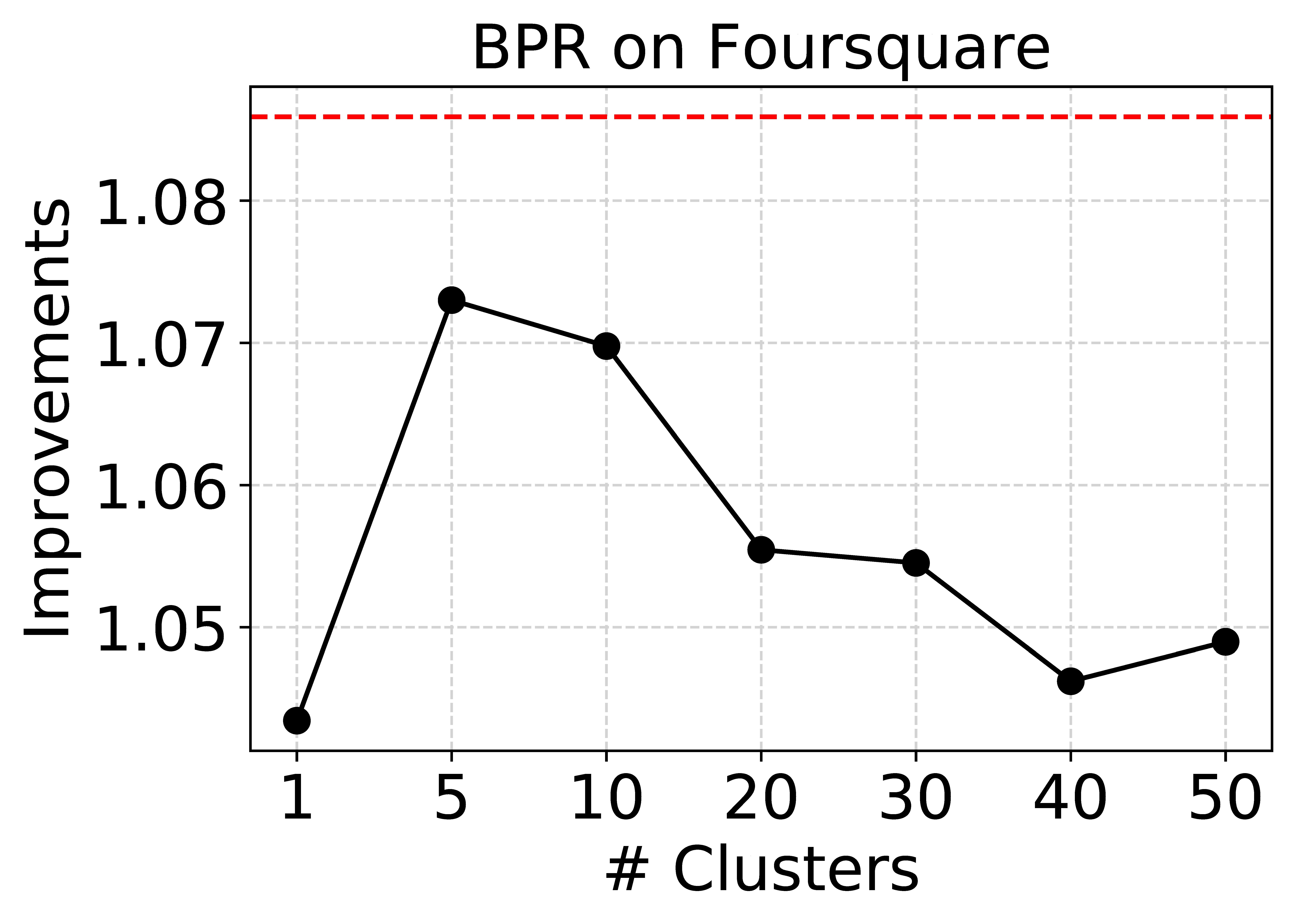}
\end{subfigure} 
\begin{subfigure}[t]{0.40\linewidth}
    \includegraphics[width=\linewidth]{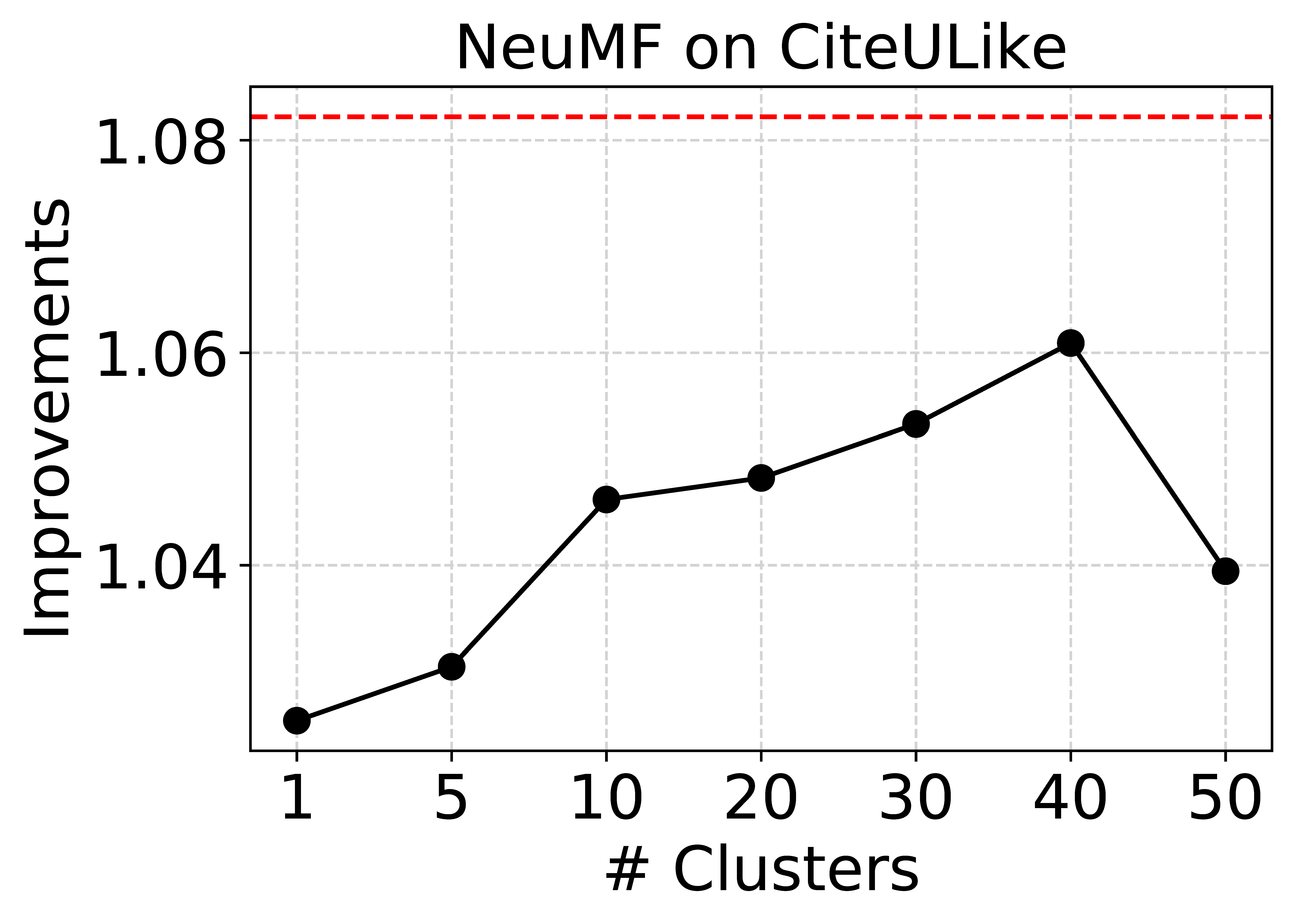}
\end{subfigure} 
\begin{subfigure}[t]{0.40\linewidth}
    \includegraphics[width=\linewidth]{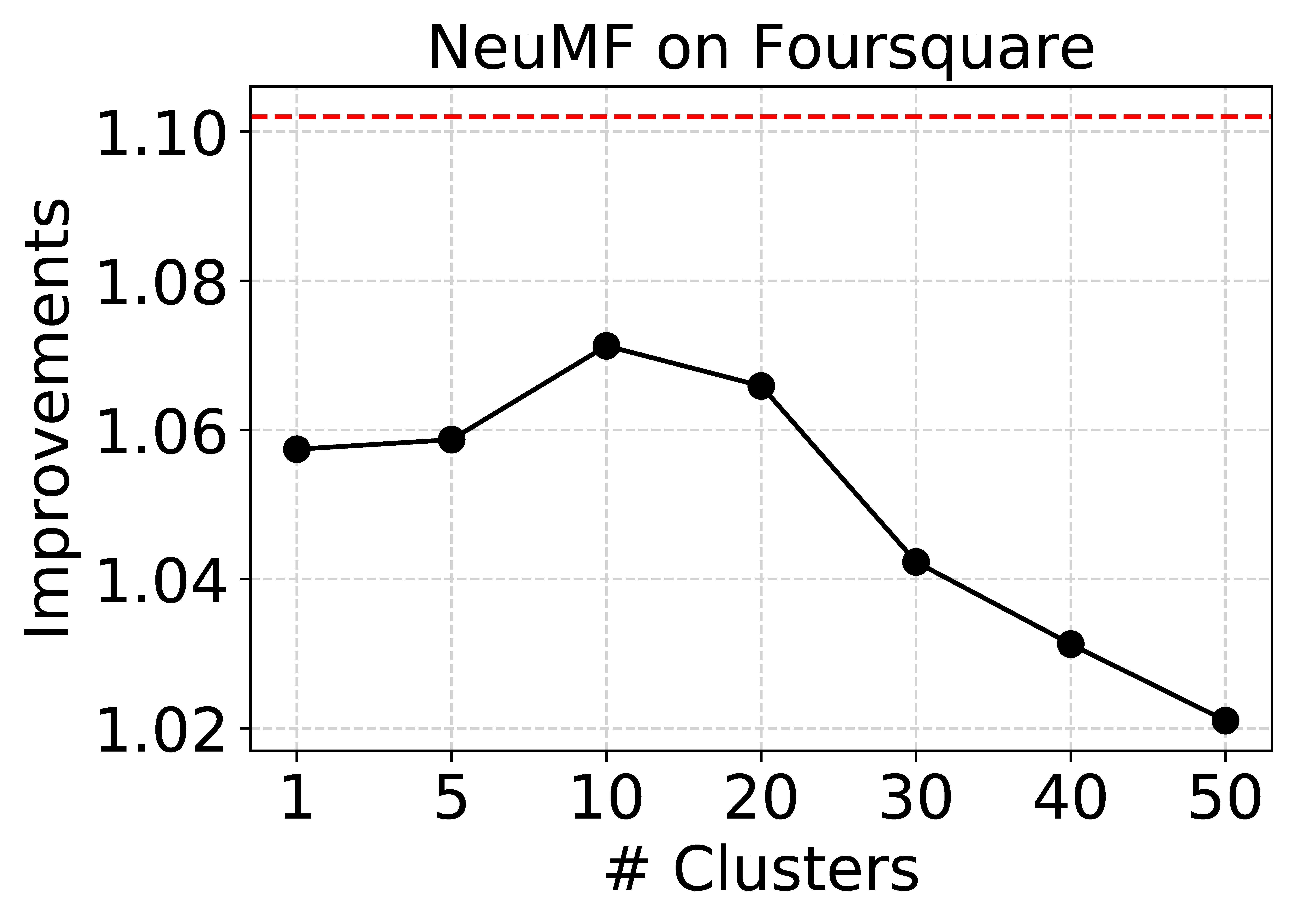}
\end{subfigure}
\begin{subfigure}[t]{0.40\linewidth}
    \includegraphics[width=\linewidth]{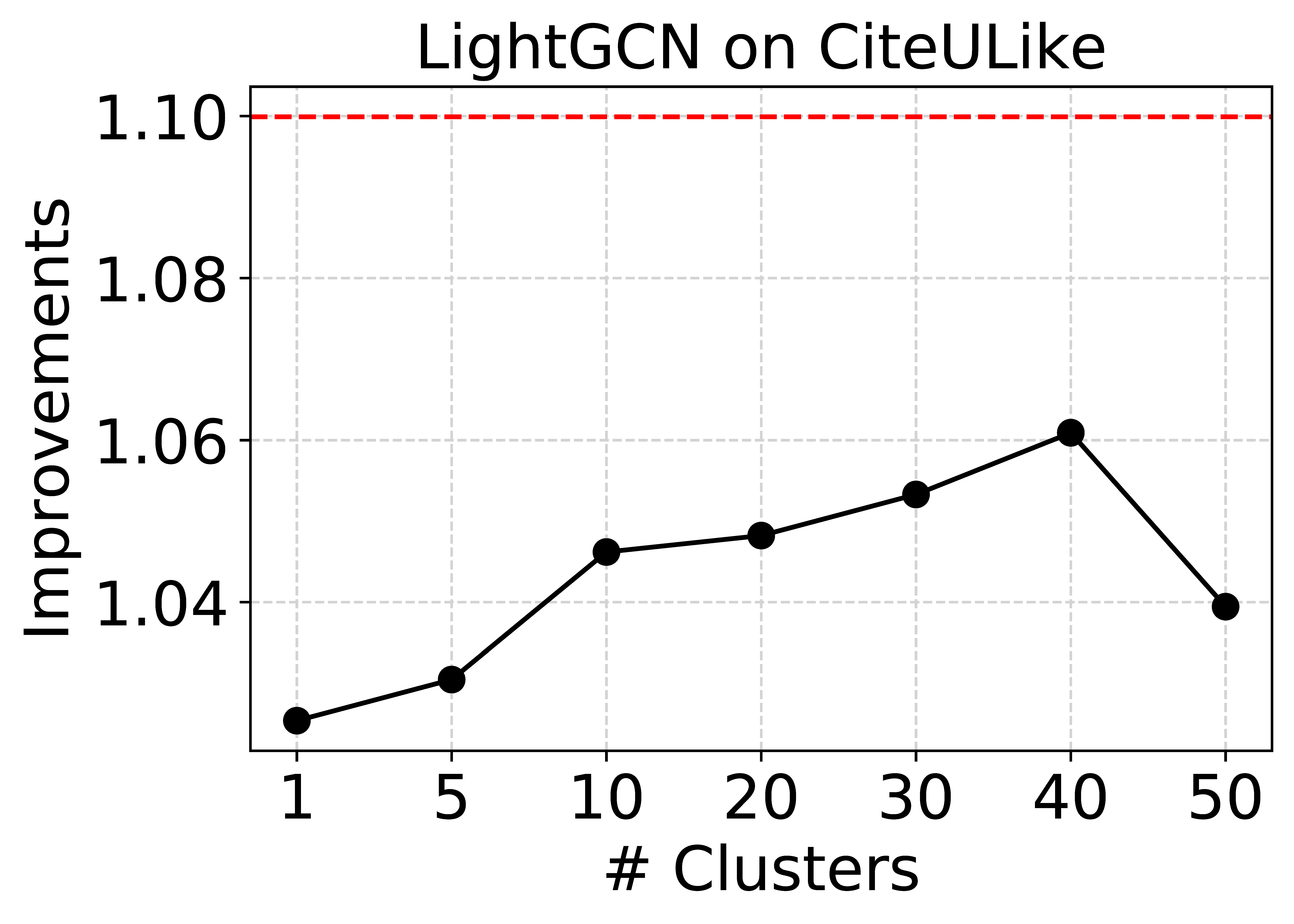}
\end{subfigure}
\begin{subfigure}[t]{0.40\linewidth}
    \includegraphics[width=\linewidth]{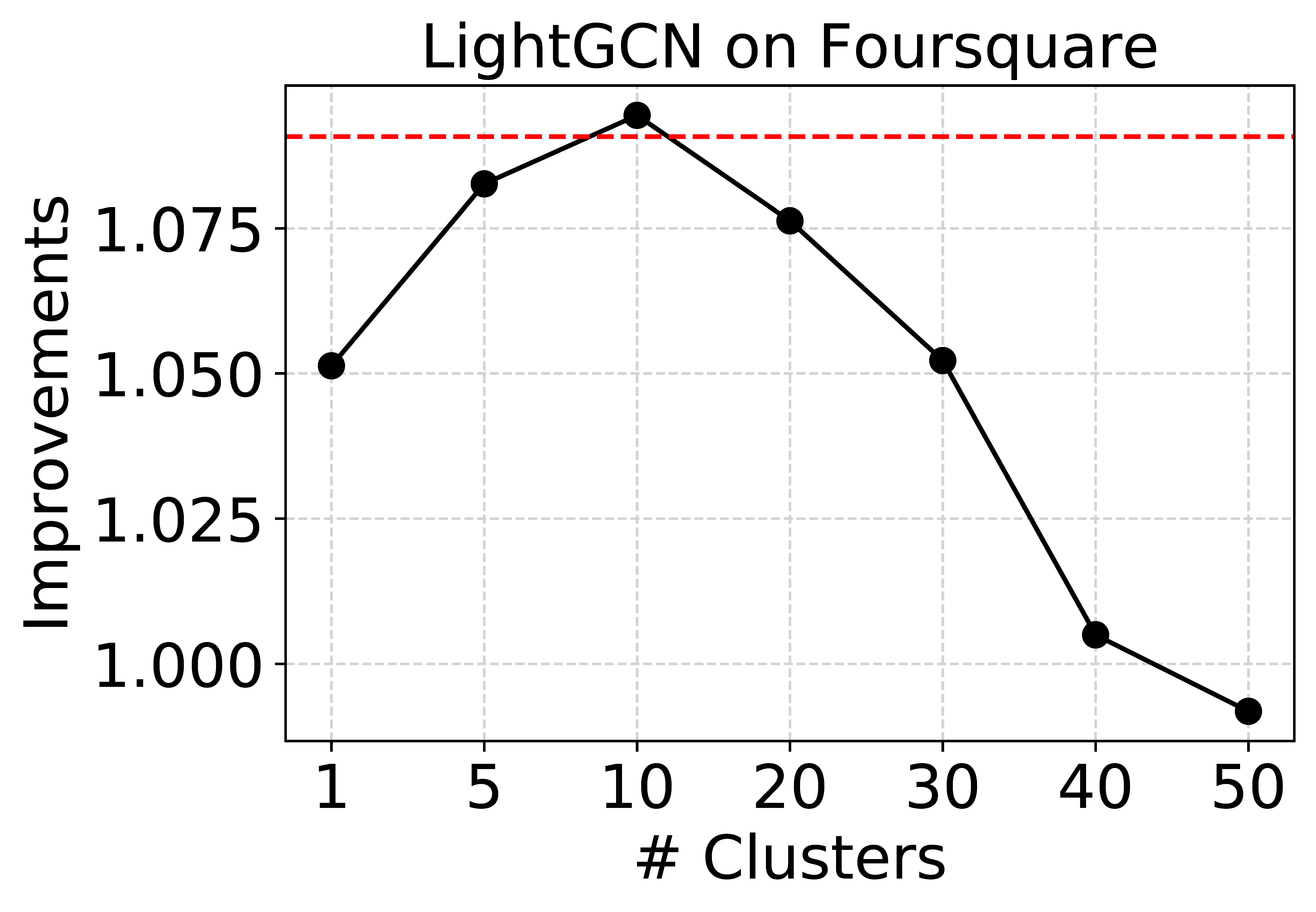}
\end{subfigure} 
\centering
\caption{The improvements (R@50) achieved by DE with varying cluster numbers (Red dotted line: PHR).}
\label{fig:PHR_DE_K}
\end{figure}

\subsection{Performance Comparison}
\label{sec:PHR_percom}
We first measure the top-$N$ recommendation performance of PHR and the baseline methods.
Table \ref{tab:PHR_main1} and Table \ref{tab:PHR_main2} present the results when $\phi=0.1$.
Figure \ref{fig:PHR_size1} and Figure \ref{fig:PHR_size2} show the results with three different student sizes.
To sum up, PHR achieves \textit{comparable or even better} performance compared to DE thoroughly tuned by a grid search.
We analyze the results from various perspectives.

We observe that PHR shows comparable or considerably improved performance compared to DE.
Specifically, PHR achieves statistically significant improvements: BPR and LightGCN on CiteULike dataset, BPR and NeuMF on Foursquare dataset (Table \ref{tab:PHR_main1} and Table \ref{tab:PHR_main2}).
Also, PHR achieves comparable performances to DE in the remaining settings.
We further compare the performance of DE and PHR with varying student sizes.
In Figure \ref{fig:PHR_size1} and Figure \ref{fig:PHR_size2}, we observe the similar tendencies.
PHR effectively improves the student in all settings and achieves comparable performance to DE.
Note that the performance of DE is achieved with its (almost) optimal hyperparameters which are obtained by an exhaustive search on the validation set.

Second, we report 1) the results of DE with the neighborhood information (denoted as DE + NB), 2) the results of PHR without the neighborhood information (denoted as PHR - NB) in Table \ref{tab:PHR_abl}.
For DE+NB, we use the representation enriched by the neighborhood (i.e., Eq.3 and Eq.4) as the input of the clustering network of DE.
For PHR - NB, we use the original representation as the input of the personalization network of PHR.
Note that DE+NB and PHR utilize the same information for enhancing the distillation bridge (i.e., the clustering network and personalization network).
We first observe that the neighborhood information is not always effective in improving the performance of DE.
We conjecture that the neighborhood information may mingle the cluster boundary in the representation space and makes it difficult to distinguish the clusters.
On the other hand, we observe that the neighborhood information consistently improves the performance of PHR.
This result shows that the representation itself provides a limited view insufficient to make the personalized bridge.
Also, this result indicates that our design of the neighborhood-based enrichment is indeed effective for personalization.




Lastly, in Figure \ref{fig:PHR_DE_K}, we report the improvements (over Student) achieved by DE and the improvements achieved by PHR.
We observe that the performance of DE is dependent on the predefined number of clusters.
A sub-optimal value, which is far from the actual number of clusters in the representation space, injects inaccurate bias about the space to the student.
This may degrade the quality of the distillation.
The optimal values are different for each dataset and each base model, which makes it difficult to deploy DE to a new environment.
Unlike DE, PHR adopts the personalization network, effectively distilling the various preference knowledge without relying on such hyperparameters.

Based on these results, we argue that PHR can effectively substitute DE and easing the costs of applying KD in a new dataset and a new recommendation model.

\subsection{Effects of PHR}
\begin{figure}[t!]
\centering
\begin{subfigure}[t]{0.48\linewidth}
    \includegraphics[width=\linewidth]{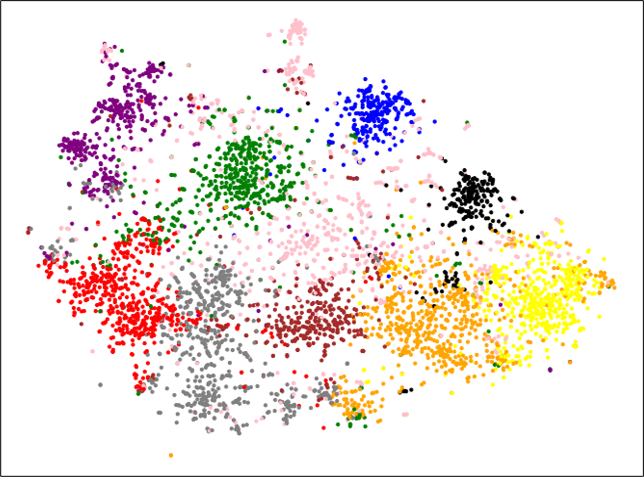}
    \caption*{Teacher representation space}
\end{subfigure}
\begin{subfigure}[t]{0.48\linewidth}
    \includegraphics[width=\linewidth]{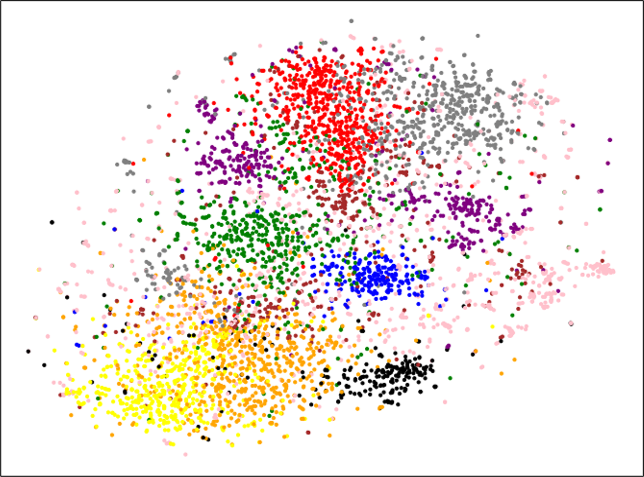}
    \caption*{Output space of personalization network}
\end{subfigure} 
\caption*{(a) CiteULike dataset}
\begin{subfigure}[t]{0.48\linewidth}
    \includegraphics[width=\linewidth]{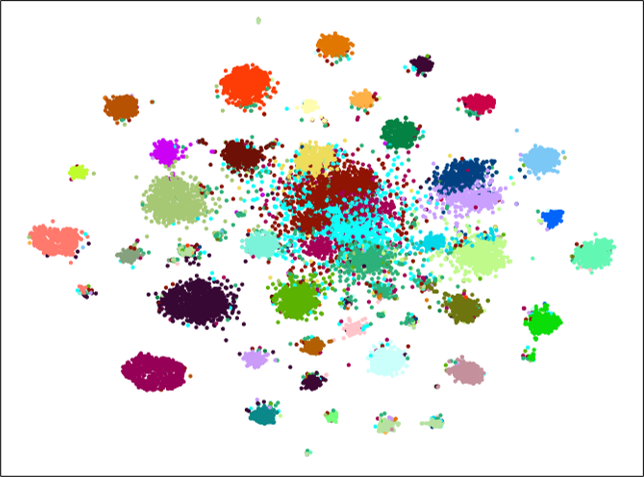}
    \caption*{Teacher representation space}
\end{subfigure} 
\begin{subfigure}[t]{0.48\linewidth}
    \includegraphics[width=\linewidth]{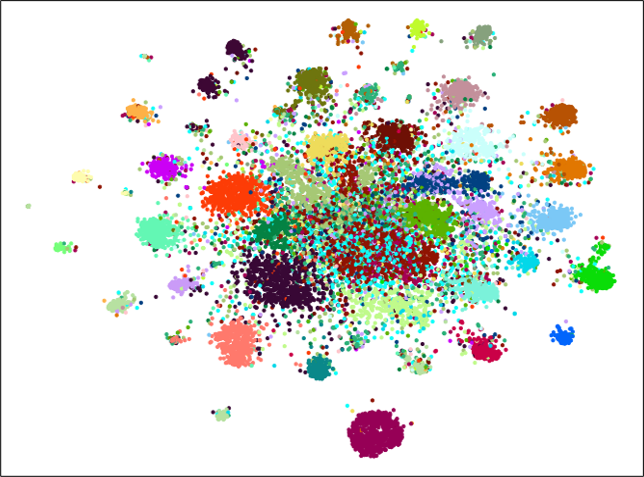}
    \caption*{Output space of personalization network}
\end{subfigure}
\caption*{(b) Foursquare dataset}
\caption{TSNE visualization of (left) teacher representation space and (right) output space of the personalization network.}
\label{fig:PHR_TSNE}
\end{figure}

In this section, we further analyze the effects of PHR to provide a more intuitive understanding of the results in Section \ref{sec:PHR_percom}.
Here, we do not use NeuMF as the base model, because it learns user-item joint representation which is difficult to visualize.
First, we investigate whether the personalization network indeed encodes the personalized information in the mapping function.
We visualize (1) the teacher representation space and (2) the output space of the personalization network ($p$) in Figure \ref{fig:PHR_TSNE}.
We use BPR as the base model and the model size is $0.1$.
We perform $k$-means clustering\footnote{We use Scikit-learn, $k$ is set to $\frac{\text{\#User}}{500}$.} in the teacher representation space, then we use the clustering result to indicate the color.

\begin{figure}[h!]
\centering
\begin{subfigure}[t]{0.40\linewidth}
    \includegraphics[width=\linewidth]{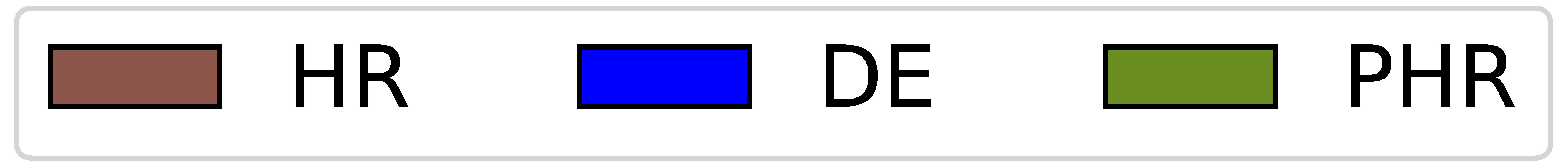}
\end{subfigure}
\\
\begin{subfigure}[t]{0.40\linewidth}
    \includegraphics[width=\linewidth]{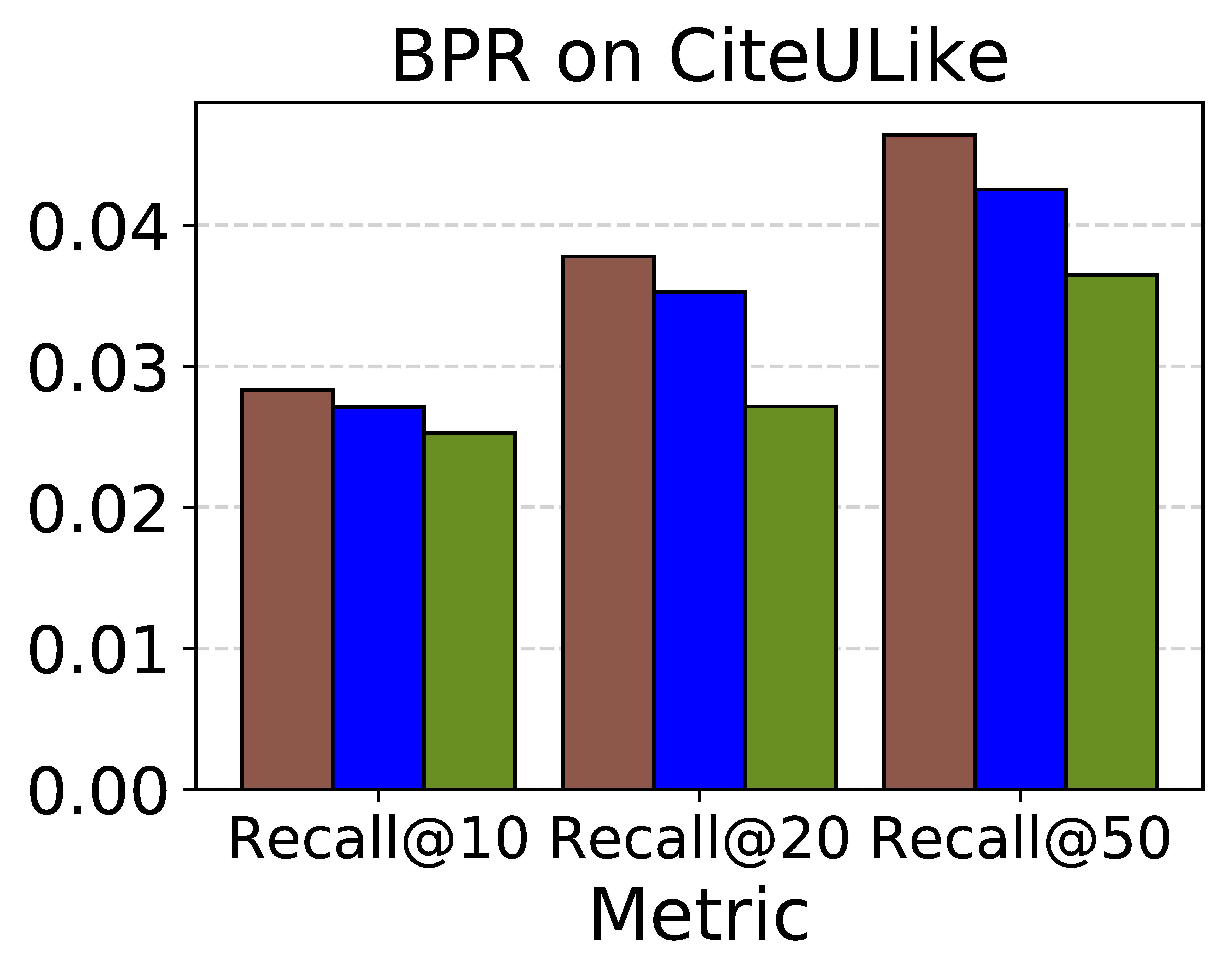}
\end{subfigure}
\begin{subfigure}[t]{0.40\linewidth}
    \includegraphics[width=\linewidth]{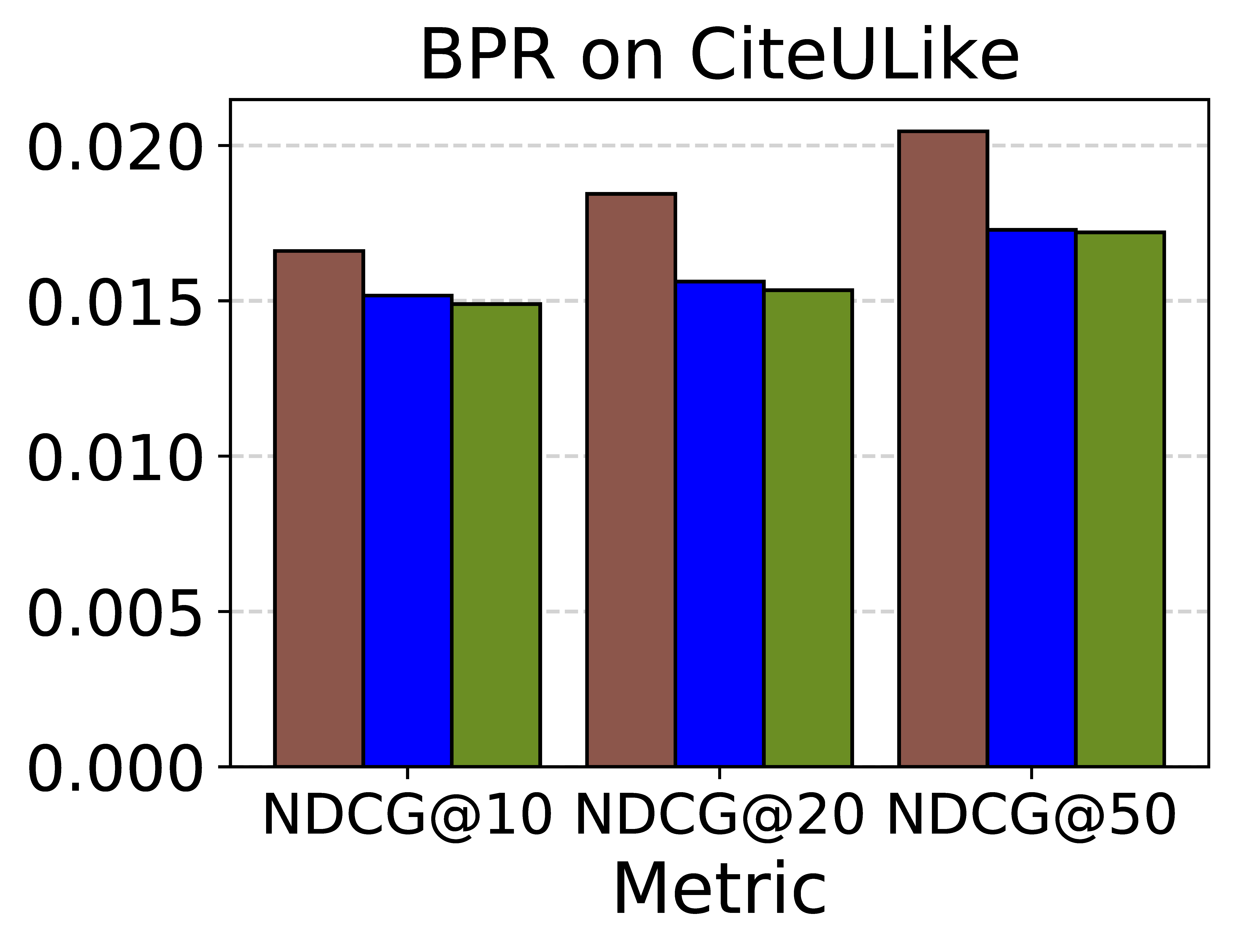}
\end{subfigure} 
\begin{subfigure}[t]{0.40\linewidth}
    \includegraphics[width=\linewidth]{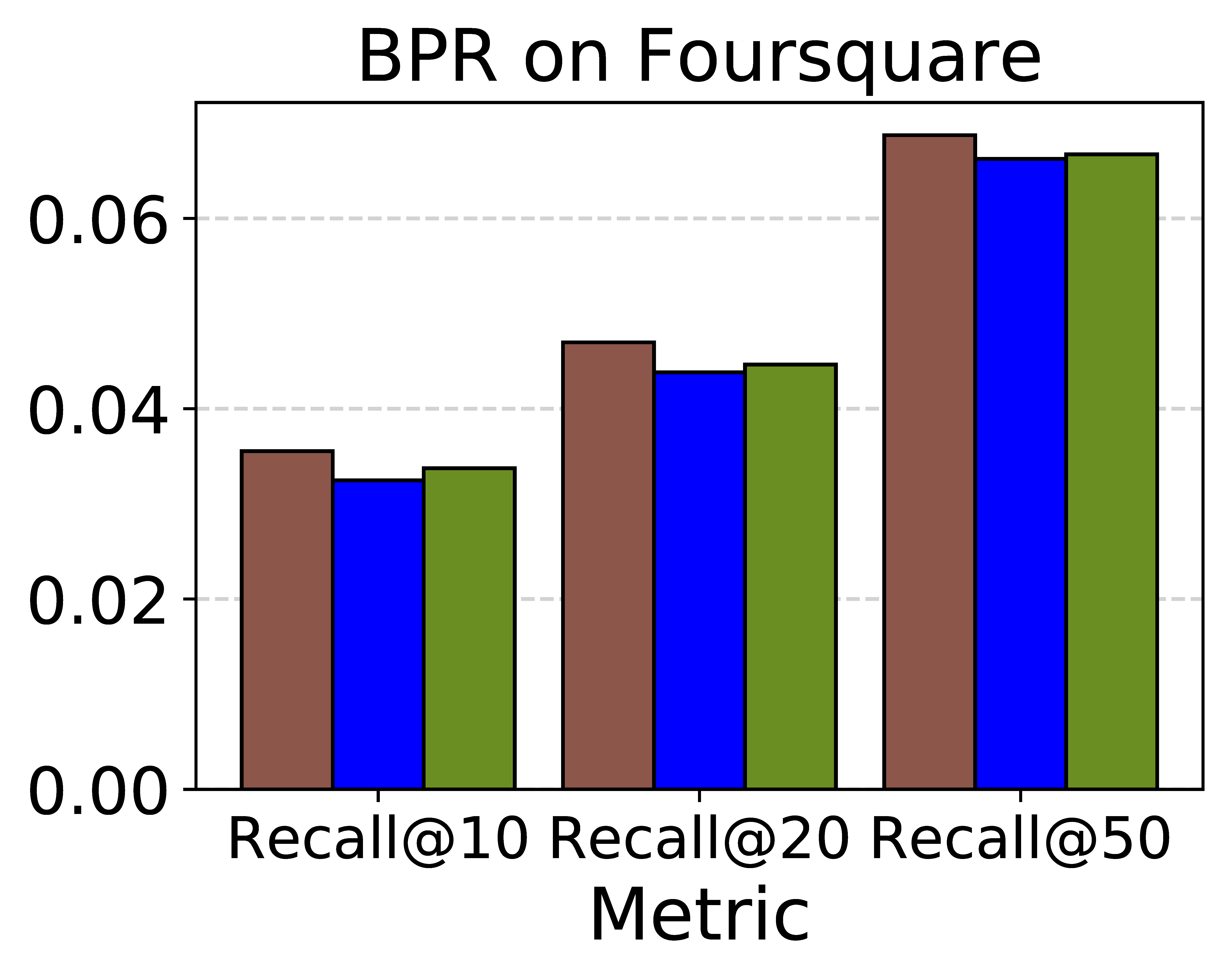}
\end{subfigure} 
\begin{subfigure}[t]{0.40\linewidth}
    \includegraphics[width=\linewidth]{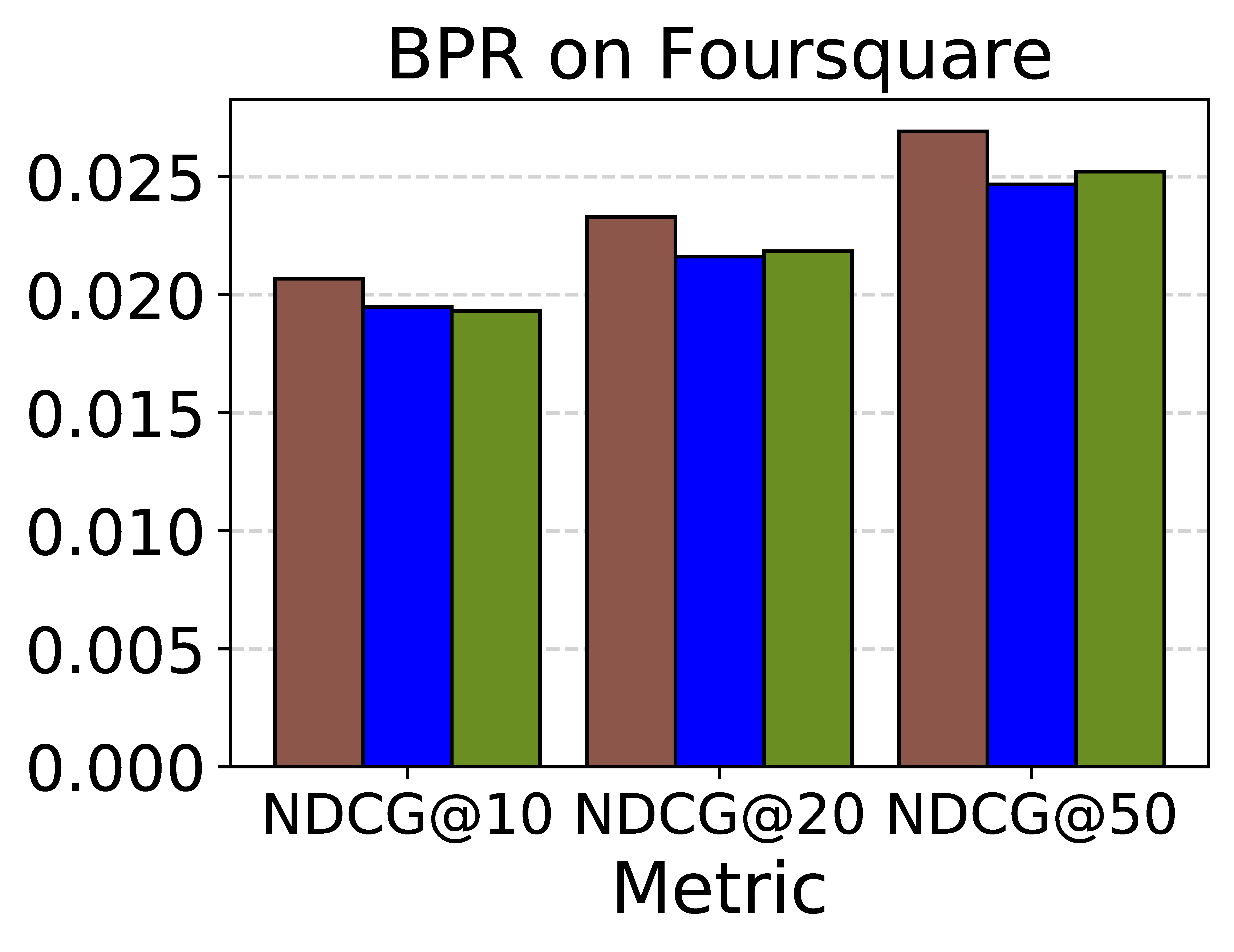}
\end{subfigure}
\begin{subfigure}[t]{0.40\linewidth}
    \includegraphics[width=\linewidth]{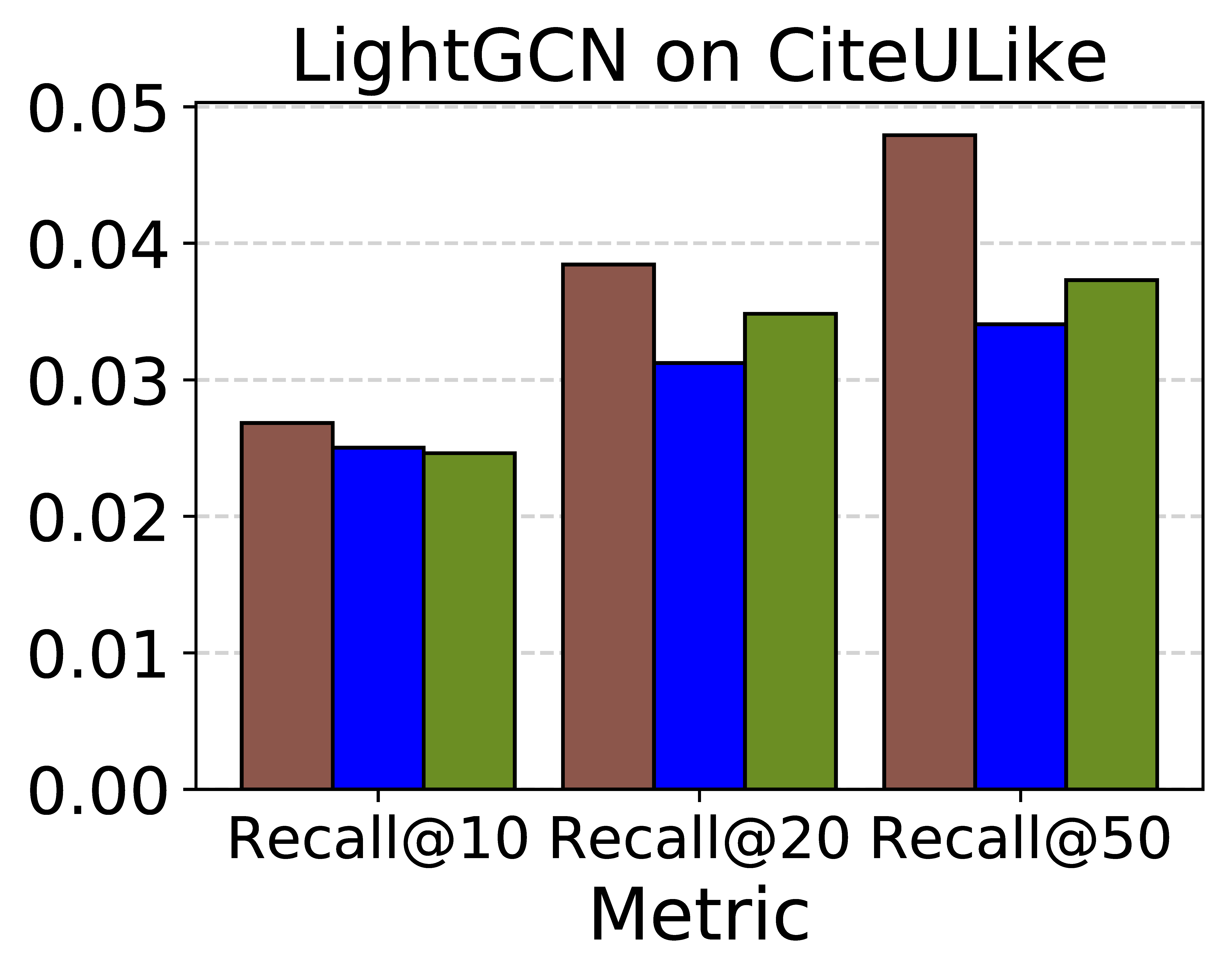}
\end{subfigure}
\begin{subfigure}[t]{0.40\linewidth}
    \includegraphics[width=\linewidth]{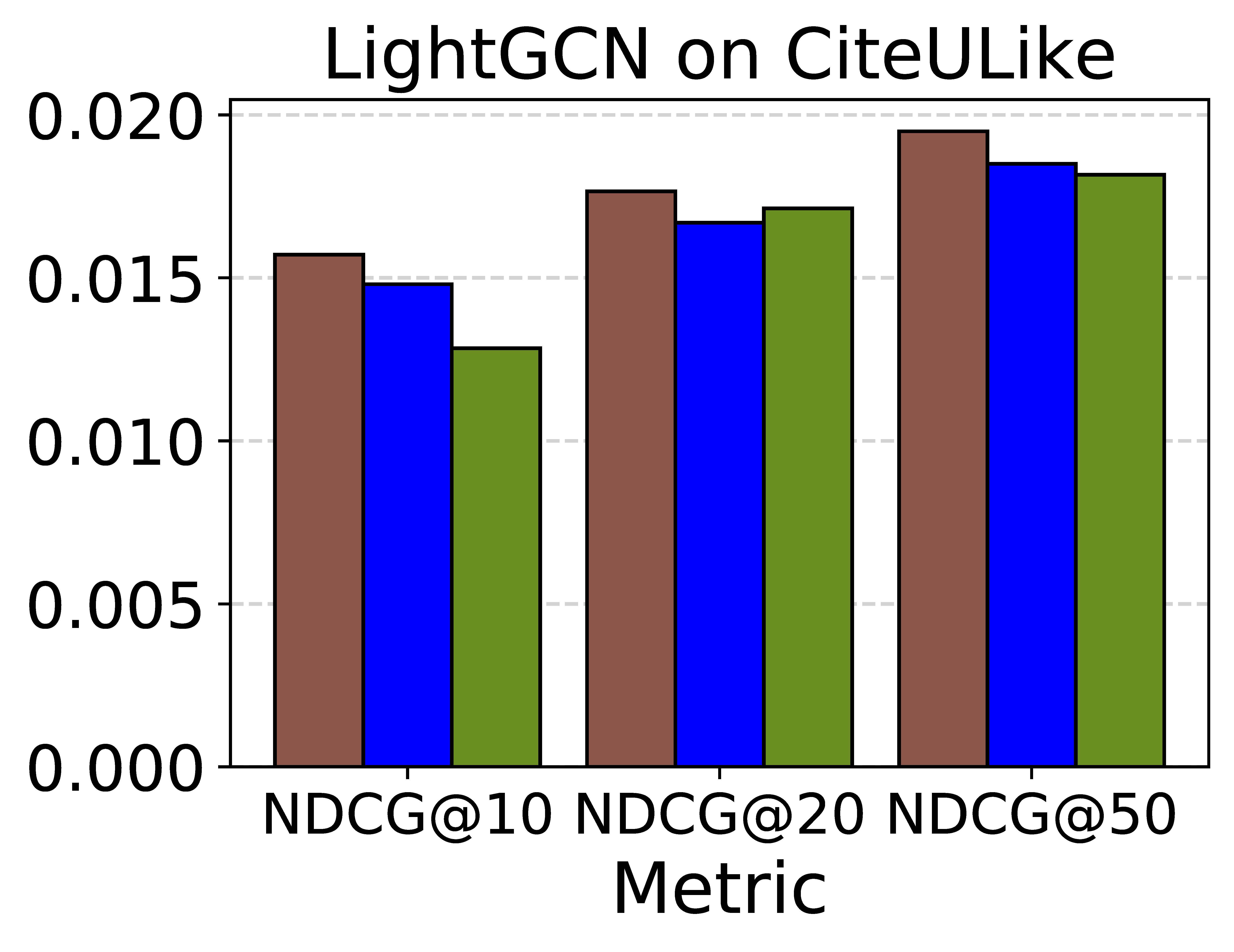}
\end{subfigure} 
\begin{subfigure}[t]{0.40\linewidth}
    \includegraphics[width=\linewidth]{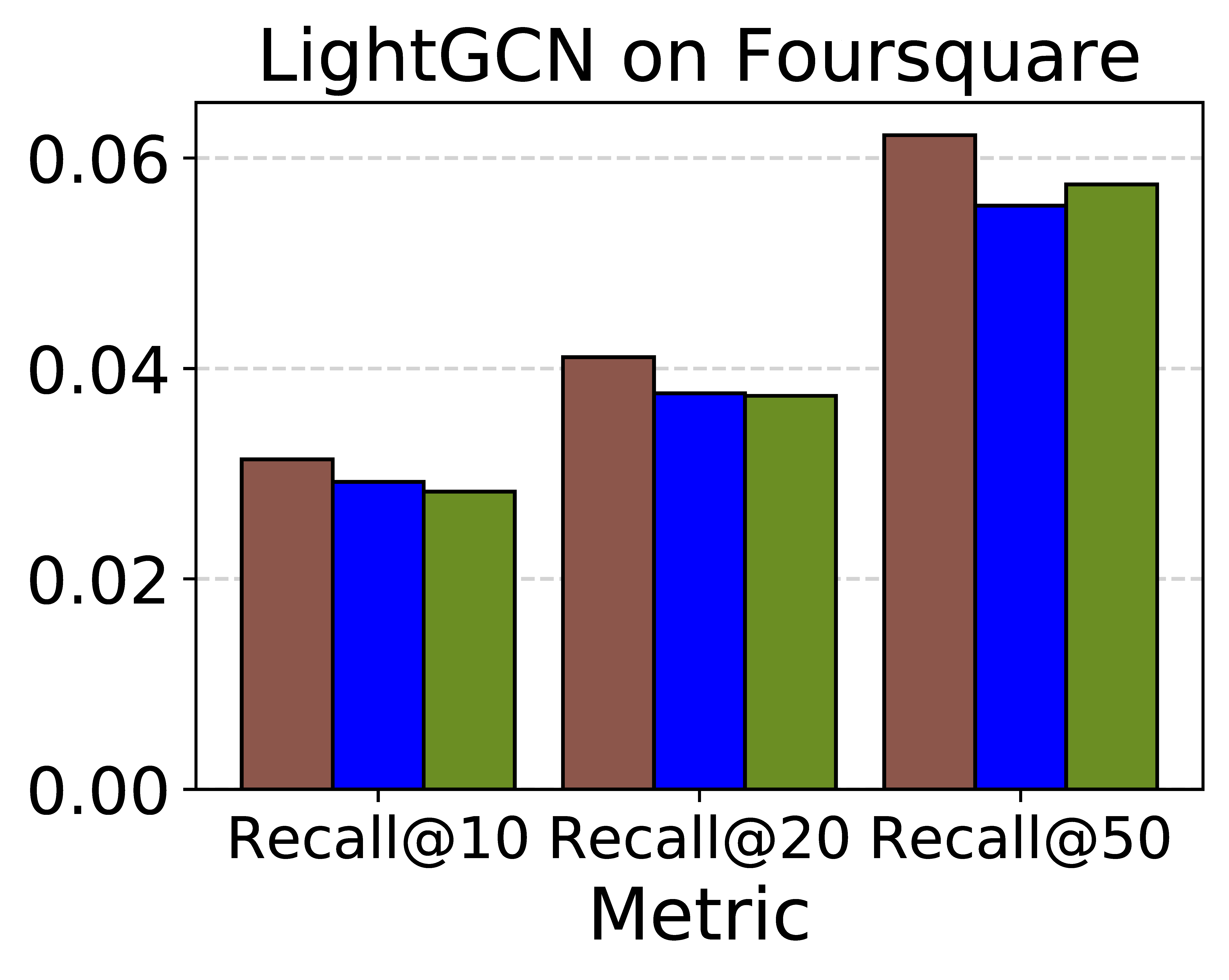}
\end{subfigure} 
\begin{subfigure}[t]{0.40\linewidth}
    \includegraphics[width=\linewidth]{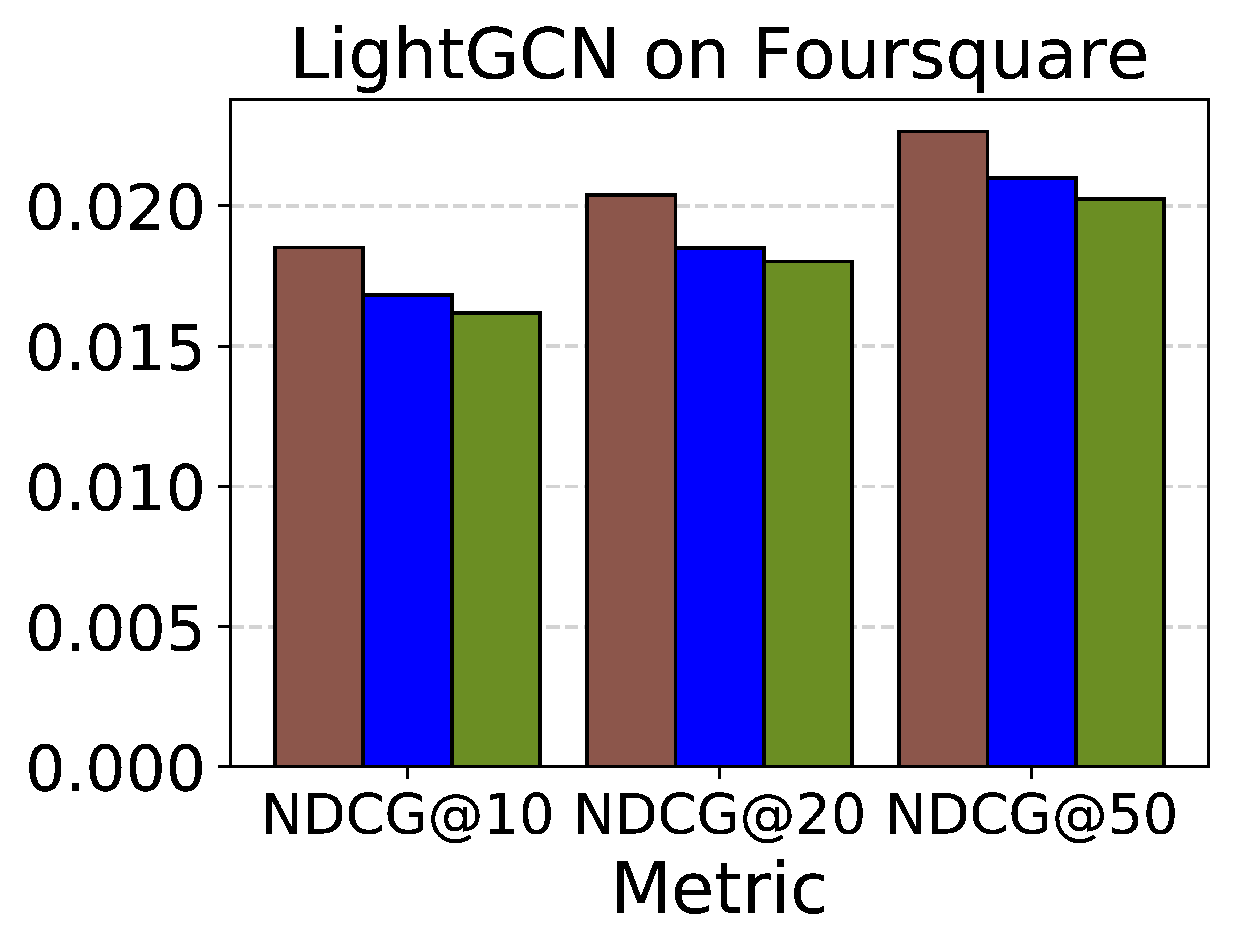}
\end{subfigure}
\centering
\caption{The standard deviation of the average performance of user groups. The lower value means a more balanced recommendation performance for diverse users.}
\label{fig:PHR_std}
\end{figure}

\noindent
\textbf{Observation.}
In Figure \ref{fig:PHR_TSNE}, we observe that the output space of the personalization network well preserves the similarity information in the teacher space;
points generated from similar preference knowledge are located closely in the space. 
Also, we observe that the space allows a certain degree of freedom in each group.
Note that each point corresponds to the parameters of the mapping function ($f$) that bridges the teacher space and the student space.
These results show that PHR 1) effectively encodes the diverse preference knowledge in the distillation and 2) provides a more individualized distillation for each representation than DE.

Second, we investigate whether PHR indeed enables for a more balanced distillation for diverse users.
We first compute the average recommendation performance (R@$50$) for users in each cluster\footnote{We utilize the clustering results obtained earlier.}, then compute the standard deviation.

\noindent
\textbf{Observation.}
In Figure \ref{fig:PHR_std}, we observe that DE and PHR show a lower standard deviation than HR in general.
HR distills all the knowledge with a single distillation network, which leads to adulterated distillation that impedes the student from finding some users’ preference \cite{DERRD}.
On the other hand, DE and PHR (which is a generalization of DE) provide the cluster-wise and personalized distillation respectively, which effectively tackles such problem.
This result supports our claim that PHR indeed enables a more balanced distillation.

\begin{figure}[h!]
\centering
\begin{subfigure}[t]{0.40\linewidth}
    \includegraphics[width=\linewidth]{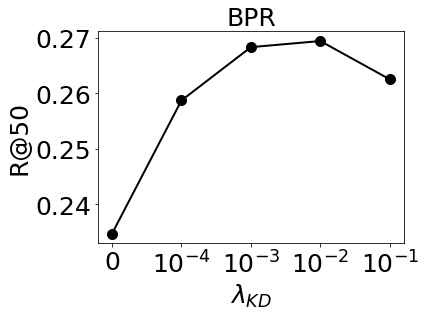}
\end{subfigure}
\begin{subfigure}[t]{0.40\linewidth}
    \includegraphics[width=\linewidth]{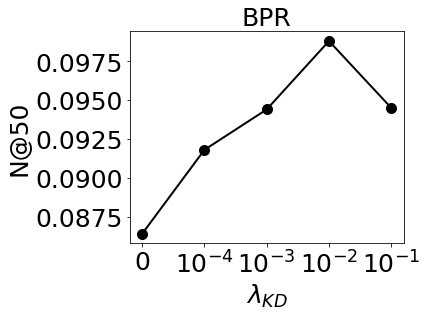}
\end{subfigure} 
\begin{subfigure}[t]{0.40\linewidth}
    \includegraphics[width=\linewidth]{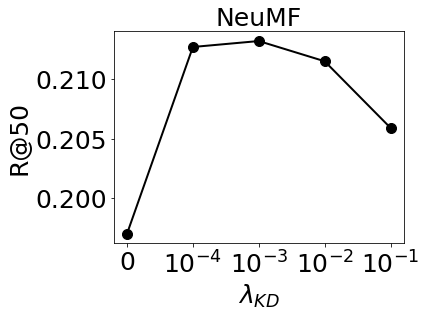}
\end{subfigure} 
\begin{subfigure}[t]{0.40\linewidth}
    \includegraphics[width=\linewidth]{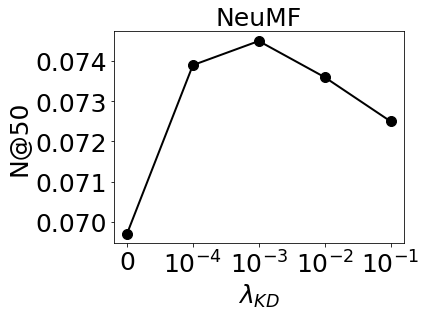}
\end{subfigure}
\begin{subfigure}[t]{0.40\linewidth}
    \includegraphics[width=\linewidth]{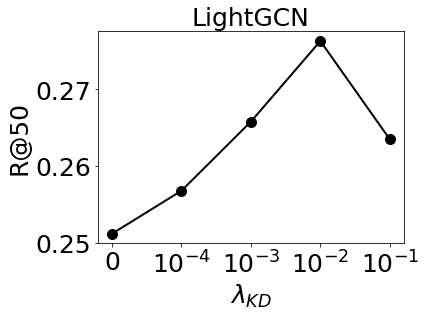}
\end{subfigure}
\begin{subfigure}[t]{0.40\linewidth}
    \includegraphics[width=\linewidth]{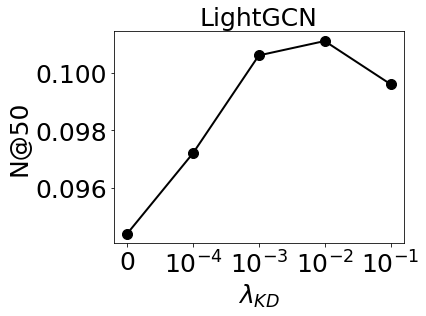}
\end{subfigure} 
\centering
\caption{The effects of $\lambda_{KD}$ on PHR.}
\label{fig:PHR_hp}
\end{figure}

\subsection{Hyperparameter Analysis}
In this section, we provide the performance of PHR with varying $\lambda_{KD}$.
$\lambda_{KD}$, which is necessary for all KD methods, is the only hyperparameter of PHR.
It controls the effects of the distillation with respect to the original loss of the base model.
In Figure \ref{fig:PHR_hp}, we report the results three base models on CiteULike dataset with $\phi=0.1$.
Note that $\lambda_{KD}=0$ corresponds to Student (i.e., no distillation).
For all base models, the best performances are observed when the value of $\lambda_{KD}$ is around $10^{-3}-10^{-2}$.
When the value of $\lambda_{KD}$ is too small ($0-10^{-4}$) or too large (over $10^{-1}$), the effects of distillation are degraded.

Unlike the state-of-the-art method (i.e., DE), PHR does not require any assumption on the teacher representation space.
Also, PHR can be easily optimized via the backpropagation in an end-to-end manner without any complicated training scheme.
Due to this simplicity, it is much easier to apply PHR than DE to a new dataset or base model.

\begin{table}[t]
\caption{Recommendation performances on large-scale application datasets. We conduct the paired $t$-test for PHR with DE and report the $p$ values.}
\begin{tabular}{cccccccc}
\toprule
Dataset                  & KD Method & R@10 & N@10 & R@20 & N@20 & R@50 & N@50 \\
\midrule
& Teacher   & 0.1203                            & 0.0762                            & 0.1630                            & 0.0870                            & 0.2417                            & 0.1024                            \\
& Student   & 0.0817                            & 0.0561                            & 0.1120                            & 0.0637                            & 0.1650                            & 0.0741                            \\
& DE        & 0.0983                            & 0.0675                            & 0.1270                            & 0.0747                            & 0.1910                            & 0.0868                            \\
& PHR       & 0.0977                            & 0.0657                            & 0.1263                            & 0.0744                            & 0.1927                            & 0.0873                            \\
\cmidrule{2-8}
\multirow{-5}{*}{Kakao}  & $p$-value    & 0.195                             & 0.132                             & 0.299                             & 0.297                             & 0.378                             & 0.393                             \\
\midrule
                         & Teacher   & 0.2180                            & 0.1237                            & 0.3170                            & 0.1487                            & 0.4767                            & 0.1804                            \\
                         & Student   & 0.1007                            & 0.0558                            & 0.1453                            & 0.0669                            & 0.2380                            & 0.0853                            \\
                         & DE        & 0.1713                            & 0.0961                            & 0.2660                            & 0.1199                            & 0.4097                            & 0.1484                            \\
                         & PHR       & 0.1710                            & 0.0955                            & 0.2640                            & 0.1188                            & 0.4067                            & 0.1472                            \\
\cmidrule{2-8}
\multirow{-5}{*}{ML-20m} & $p$-value    & 0.325                             & 0.274                             & 0.254                             & 0.212                             & 0.268                             & 0.200          \\
\bottomrule
\end{tabular}
\label{tbl:PHR_large}
\end{table}
\subsection{Results on Large-scale Applications}
Lastly, we compare the performance of DE and PHR on two public large-scale datasets: Kakao Brunch and MovieLens 20M.
Kakao Brunch is one of the largest blog platforms in South Korea, and the dataset contains user historical logs on clicking articles.
Specifically, the dataset has 22,110,706 click logs from October 1, 2018, to March 1, 2019, and there are 310,758 unique users and 505,841 unique articles.
MovieLens 20M is a widely used dataset for evaluating recommendation models, and
the dataset has 20M user-item interactions with 138,493 unique users and 27,278 unique movies.
We adopt LightGCN as the base model because it consistently shows the best performance in the previous experiments.
The latent dimensions of the teacher and the student are set to 100 and 10, respectively. 
The results are reported in Table \ref{tbl:PHR_large}.
We observe that PHR effectively improves the student model and achieves comparable performance to DE on both datasets.
Considering that PHR significantly reduces the hyperparameter search costs of DE, we believe our approach is more suitable for many large-scale applications having numerous users and items.

\section{Summary}
We propose a novel distillation method, named PHR, that effectively distills the knowledge of various preferences without relying on any assumption on the teacher representation space nor any method-specific hyperparameters.
As a workaround for the cluster-wise distillation of DE, PHR trains the personalization network which generates an individualized mapping function for each user/item representation.
PHR can be viewed as the generalization of DE, and it achieves a comparable or even better performance compared to DE with thoroughly tuned hyperparameters.
This shows that applying PHR is much easier than applying DE to a new environment.
We also provide extensive experiment results and analyses showing the effectiveness of PHR. 
In future work, we will further investigate the effects of PHR on more diverse RS models.

\chapter{Topology Distillation}
\label{chapt:TD}
Recommender Systems (RS) have employed knowledge distillation which is a model compression technique training a compact student model with the knowledge transferred from a pre-trained large teacher model. 
Recent work has shown that transferring knowledge from the teacher's intermediate layer significantly improves the recommendation quality of the student.
However, they transfer the knowledge of individual representation point-wise and thus have a limitation in that primary information of RS lies in the relations in the representation space.
This paper proposes a new topology distillation approach that guides the student by transferring the topological structure built upon the relations in the teacher space.
We first observe that simply making the student learn the whole topological structure is not always effective and even degrades the student's performance.
We demonstrate that because the capacity of the student is highly limited compared to that of the teacher, learning the whole topological structure is daunting for the student.
To address this issue, we propose a novel method named Hierarchical Topology Distillation (HTD) which distills the topology hierarchically to cope with the large capacity gap.
Our extensive experiments on real-world datasets show that the proposed method significantly outperforms the state-of-the-art competitors. 
We also provide in-depth analyses to ascertain the benefit of distilling the topology for RS.

\section{Introduction}
\label{sec:TD_introduction}
The size of recommender systems (RS) has kept increasing, as they have employed deep and sophisticated model architectures to better understand the complex nature of user-item interactions \cite{RD, CD, DERRD,  GCN_distill}.
A large model with many learning parameters has a high capacity and therefore generally achieves higher recommendation accuracy.
However, it also requires high computational costs, which results in high inference latency.
For this reason, it is challenging to adopt such a large model to the real-time platform \cite{RD, CD, GCN_distill, DERRD}.

To tackle this problem, \textit{Knowledge Distillation} (KD) has been adopted to RS \cite{RD, CD, DERRD, zhu2020ensembled, BD, GCN_distill}.
KD is a model-independent strategy to improve the performance of a compact model (i.e., student) by transferring the knowledge from a pre-trained large model (i.e., teacher).
The distillation is conducted in two steps.
The teacher is first trained with the training set, and the student is trained with help from the teacher along with the training set.
The student model, which is a compact model, is used in the inference time.
During the distillation, the teacher can provide additional supervision that is not explicitly revealed from the training set.
As a result, the student trained with KD shows better prediction performance than the student trained only with the training set.
Also, it has low inference latency due to its small size.

\begin{figure}[t]
\centering
\captionsetup[subfigure]{justification=centering}
\begin{subfigure}[t]{0.5\linewidth}
    \includegraphics[height=4cm]{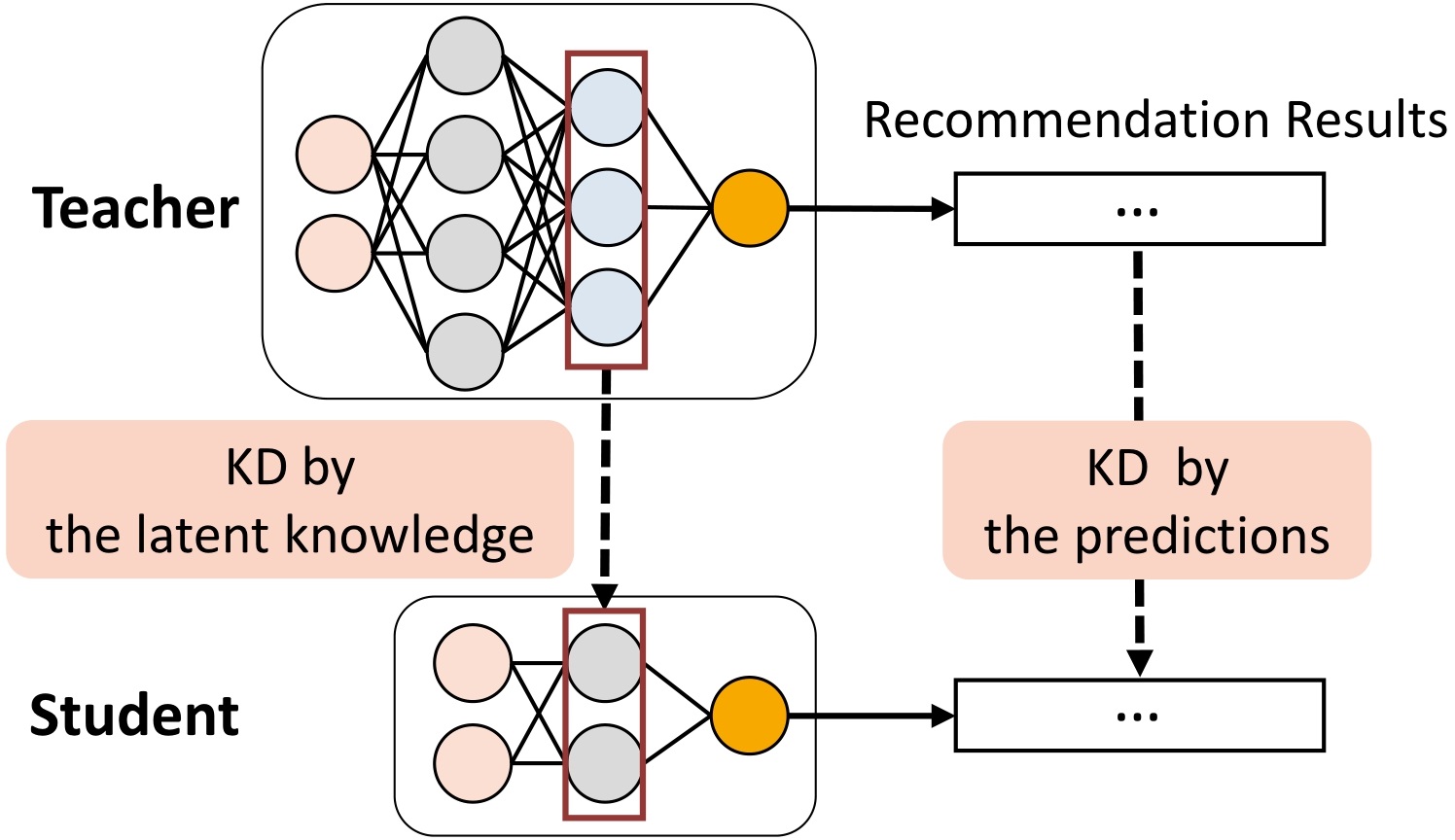}
    \subcaption{}
\end{subfigure}
\\
\begin{subfigure}[t]{0.3\linewidth}
    \includegraphics[height=3.5cm]{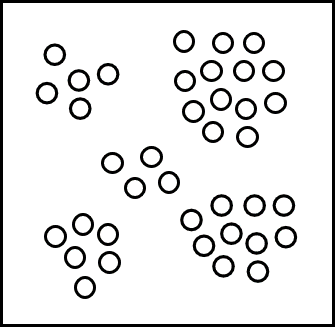}
    \subcaption{Hint Regression}
\end{subfigure}
\begin{subfigure}[t]{0.3\linewidth}
    \includegraphics[height=3.5cm]{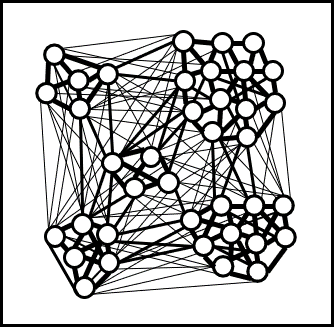}
    \subcaption{FTD}
\end{subfigure}
\begin{subfigure}[t]{0.3\linewidth}
    \includegraphics[height=3.5cm]{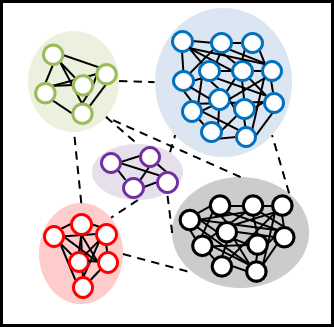}
    \subcaption{ HTD}
\end{subfigure}
\caption{(a) The overview of KD in RS. (b-d) The conceptual illustrations of the latent knowledge that each method transfers from the teacher's representation space. Each point corresponds to a representation of each entity. 
(b) transfers the information of each entity to the student point-wise.
However, our approach (c-d) transfers the relations among the entities.
FTD/HTD refers to Full/Hierarchical topology distillation, and the solid/dotted line denotes the entity/group-level topology, respectively.}
\label{fig:TD_intro}
\end{figure}

Most existing KD methods for RS transfer the knowledge from the teacher's predictions \cite{DERRD, RD, CD, GCN_distill, BD} (Figure \ref{fig:TD_intro}a).
They basically enforce the student to imitate the teacher's recommendation results, providing guidance to the predictions of the student.
There is another recent approach that transfers the \emph{latent} knowledge from the teacher's intermediate layer \cite{DERRD, zhu2020ensembled}, pointing out that the predictions incompletely reveal the teacher’s knowledge and the intermediate representation can additionally provide a detailed explanation on the final prediction of the teacher.
They adopt \textit{hint regression} \cite{FitNet} that makes the student's representation approximate the teacher's representation via a few regression layers.
This enables the student to get compressed information on each entity (e.g., user and item) (Figure \ref{fig:TD_intro}b) that can restore more detailed preference information in the teacher \cite{DERRD}.

However, the existing hint regression-based methods focus on distilling the individual representation of each entity, disregarding the \emph{relations} of the representations.
In RS, each entity is better understood by its relations to the other entities rather than by its individual representation.
For instance, a user's preference is represented in relation to (and in contrast with) other users and items.
Also, the student can take advantage of the space, where the relations found by the teacher are well preserved, in finding more accurate ranking orders among the entities and thereby improving the recommendation performance.

This paper proposes a new distillation approach that effectively transfers the relational knowledge existing in the teacher's representation space.
A natural question is how to define the relational knowledge and distill it to the student.
We build a \emph{topological structure} that represents the relations in the teacher space based on the similarity information, then utilize it to guide the learning of the student via distillation.
Specifically, we train the student with the distillation loss that preserves the teacher's topological structure in its representation space along with the original loss function.
Trained with the topology distillation, the student can better preserve the relations in the teacher space, which not only improves the recommendation performance but also better captures the semantic of entities.

However, we observe that simply making the student learn all the topology information (Figure \ref{fig:TD_intro}c) is not always effective and sometimes even degrades the student's recommendation performance.
This phenomenon is explained by the huge capacity gap between the student and the teacher;
the capacity of the student is highly limited compared to that of the teacher, and learning all the topological structure in the teacher space is often daunting for the student.
To address this issue, we propose a method named Hierarchical Topology Distillation (HTD) which effectively transfers the vast teacher's knowledge to the student with limited capacity.
HTD represents the topology hierarchically and transfers the knowledge in multi-levels using the hierarchy (Figure \ref{fig:TD_intro}d).

Specifically, HTD adaptively finds \textit{preference groups} of entities such that the entities within each group share similar preferences.
Then, the topology is hierarchically structured in group-level and entity-level.
The \emph{group-level topology} represents the summarized relations across the groups, providing an overview of the whole topology.
The \emph{entity-level topology} represents the relations of entities belonging to the same group.
This provides a fine-grained view on important relations among the entities having similar preferences, which directly affects the top-$N$ recommendation performance.
By compressing the complex individual relations across the groups, HTD relaxes the daunting supervision and enables the student to better focus on the important relations.
In summary, the key contributions of our work are as follows:
\begin{itemize}[leftmargin=*]
    \item We address the necessity of transferring the relational knowledge from the teacher representation space and develop a general topology distillation approach for RS.
    
    \item We develop a new topology distillation method, named FTD, designed to guide the student by transferring the full topological structure built upon the relations in the teacher space.  
    
    \item We propose a novel topology distillation method, named~HTD, designed to effectively transfer the vast relational knowledge to the student considering the huge capacity gap.
    
    \item We validate the superiority of the proposed approach by extensive experiments. 
    We also provide in-depth analyses to verify the benefit of distilling the topological structure.
\end{itemize}

\section{Related Work}
\label{sec:TD_relatedwork}
\noindent
\textbf{Knowledge Distillation.}
Knowledge distillation (KD) is a model-independent strategy that accelerates the training of a student model with the knowledge transferred from a pre-trained teacher model.
Most KD methods have mainly focused on the image classification task.
An early work \cite{KD} matches the class distributions (i.e., the softmax output) of the teacher and the student.
The class distribution has richer information (e.g., inter-class correlation) than the one-hot class label, which improves learning of the student model.
Pointing out that utilizing the predictions alone is insufficient because meaningful intermediate information is ignored, subsequent methods \cite{FitNet, yim2017gift, AT, liu2019structured, tung2019similarity} have distilled knowledge from the teacher's intermediate layer along with the predictions.
\cite{FitNet} proposes ``hint regression'' that matches the intermediate representations.
Subsequently, \cite{yim2017gift} matches the gram matrices of the representations, \cite{AT} matches the attention maps from the networks, and \cite{tung2019similarity, li2020local} match the similarities on activation maps of the convolutional layer.

\vspace{0.1cm}
\noindent
\textbf{Reducing inference latency of RS.}
As the size of RS is continuously increasing, various approaches have been proposed for reducing the model size and inference latency.
Several methods \cite{hash1, hash2, DCF} have utilized the binary representations of users and items.
With the discretized representations, the search costs can be considerably reduced via the hash technique.
However, due to their restricted capability, the loss of recommendation accuracy is inevitable \cite{CD, DERRD, GCN_distill}.
Also, various computational acceleration techniques have been successfully adopted to reduce the search costs.
In specific, order-preserving transformations \cite{tree_RS}, pruning and compression techniques \cite{pruning_RS2_inner_only, pruning_RS, compression1}, tree-based data structures \cite{KDtree, tree_RS}, and approximated nearest-neighbor search \cite{LSH_inner_product} have been employed to reduce the inference latency.
However, they have limitations in that the techniques are only applicable to specific models or easy to fall into a local optimum because of the local search~\cite{DERRD, CD}.

\vspace{0.1cm}
\noindent
\textbf{Knowledge Distillation for RS.}
KD, which is the model-agnostic strategy, has been widely adopted in RS.
Similar to the progress on computer vision, the existing methods are categorized into two groups (Figure 1a):
(1) the methods distilling knowledge from the predictions, (2) the methods distilling the latent knowledge from the intermediate layer.
Note that the two groups of methods can be utilized together to fully improve the student \cite{DERRD}.

\noindent
{\textbf{(1) KD by the predictions.}}
Motivated by \cite{KD} that matches the class distributions, most existing methods \cite{RD, CD, DERRD, GCN_distill, BD} have focused on matching the predictions (i.e., recommendation results) from the teacher and the student.
The teacher's predictions convey additional information about the subtle difference among the items, helping the student generalize better than directly learning from binary labels \cite{CD}.
This research direction focuses on designing a method effectively utilizing the teacher's predictions.
First, \cite{RD, CD} distill the knowledge of the items with high scores in the teacher’s predictions.
Since a user is interested in only a few items, distilling knowledge of a few top-ranked items is effective to discover the user's preferable items \cite{RD}.
Most recently, \cite{BD} utilizes rank-discrepancy information between the predictions from the teacher and the student.
Specifically, \cite{BD} focuses on distilling the knowledge of the items ranked highly by the teacher but ranked lowly by the student.
On the one hand, \cite{DERRD, GCN_distill} focus on distilling ranking order information from the teacher's predictions.
Concretely, they adopt listwise learning \cite{xia2008list-wise} and train the student to follow the items' ranking orders predicted by the teacher.

\noindent
{\textbf{(2) KD by the latent knowledge.}}
Pointing out that the predictions incompletely reveal the teacher's knowledge, a few methods \cite{DERRD, zhu2020ensembled}\footnote{\cite{DERRD} proposes two KD methods: one by prediction and the other by latent knowledge.} have focused on distilling latent knowledge from the teacher's intermediate layer.
The existing methods are based on \textit{hint regression} \cite{FitNet} proposed in computer vision.
Let $h^t: \mathcal{X} \rightarrow \mathbb{R}^{d^t}$ denote a mapping function from the input feature space to the representation space of the teacher (i.e., the teacher nested function up to the intermediate layer).
Similarly, let $h^s: \mathcal{X} \rightarrow \mathbb{R}^{d^s}$ denote a mapping function to the representation space of the student.
Also, let $\mathbf{e}_i^t = h^t(\mathbf{x}_i)$ and $\mathbf{e}_i^s = h^s(\mathbf{x}_i)$ denote the representations of entity $i$ from the two spaces\footnote{We use the term `entity' to denote the subject of each representation.}, where $\mathbf{x}_i$ is entity $i$'s input feature.
The hint regression makes $\mathbf{e}_i^s$ approximate $\mathbf{e}_i^t$ as follows:
\begin{equation}
    \mathcal{L}_{Hint} = \lVert \mathbf{e}_i^t -  f(\mathbf{e}_i^s) \rVert^2_2
\end{equation}
where $f:\mathbb{R}^{d^s} \rightarrow \mathbb{R}^{d^t}$ is a small network to bridge the different dimensions ($d^s << d^t$).
By minimizing $\mathcal{L}_{Hint}$, parameters in the student (i.e., $h^s$) and $f$ are updated.
Also, it is jointly minimized with the base model (i.e., $\mathcal{L}_{B a s e} + \lambda \mathcal{L}_{H i n t}$) which can be any existing recommender.
The hint regression enables $\mathbf{e}^s$ to capture compressed information that can restore detailed information in $\mathbf{e}^t$ \cite{zhu2020ensembled, DERRD}.
\cite{zhu2020ensembled} adopts this original hint regression~to~improve~the~student.

The most recent work DE \cite{DERRD} further elaborates this approach for RS.
DE argues that using a single network ($f$) makes the knowledge of entities having dissimilar preferences get mixed, and this degrades the quality of distillation.
Its main idea is that the knowledge of entities having similar preferences should be distilled without being mixed with that of entities having dissimilar preferences. 
To this end, DE clusters the representations into $K$ groups based on the teacher’s knowledge and distills the representations in each group via a separate network $f_k$.
Let $\mathbf{z}_{i}$ be a $K$-dimensional one-hot vector whose element $z_{ik}=1$ if entity $i$ belongs to the corresponding $k$-th group.
For each entity $i$, DE loss is defined as follows:
\begin{equation}
  \begin{aligned}
    \mathcal{L}_{DE} = \lVert \mathbf{e}_i^t -  \sum_{k=1}^K z_{ik} f_k(\mathbf{e}_i^s) \rVert^2_2
\end{aligned}
\end{equation} 
The one-hot vector is sampled from a categorical distribution with class probabilities $\boldsymbol{\alpha}_i = v(\mathbf{e}^{t}_i)$, i.e., $\mathbf{z}_{i} \sim \text{Categorical}_{K} (\boldsymbol{\alpha}_i)$, where $v: \mathbb{R}^{d^t} \rightarrow \mathbb{R}^{K}$ is a small network with Softmax output.
The sampling process is approximated by Gumbel-Softmax \cite{GumbelSoftmax} and trained via backpropagation in an end-to-end manner.
In sum, the representations belonging to the same group share similar preferences and are distilled via the same network without being mixed with the representations belonging to the different groups \cite{DERRD}.

The existing methods \cite{DERRD, zhu2020ensembled} based on the hint regression distill the knowledge of individual entity without consideration of how the entities are related in the representation space. 
Considering a user’s preference is represented in relation to (and in contrast with) other users and items, each entity is better understood by its relations to the other entities rather than by its individual representation.
Also, the student can take advantage of the space, where the relations found by the teacher are well preserved, in finding more accurate ranking orders among the entities and thereby improving the recommendation performance.
In this work, we propose a new distillation approach for RS that directly distills the relational knowledge from the teacher's representation space.

\section{Proposed Approach---Topology Distillation}
\label{sec:TD_method}
We first provide an overview of the proposed approach (Section \ref{sec:TD_overview}).
Before we describe the final solution, we explain a naive method for incorporating the relational knowledge in the distillation process (Section \ref{sec:TD_FTD}).
Then, we shed light on the drawbacks of the method when applying it for KD.
Motivated by the analysis, we present a new method, named HTD, which distills the relational knowledge in multi-levels to effectively cope with a large capacity gap between the teacher and the student (Section \ref{sec:TD_HTD}).
The pseudocodes of the proposed methods are provided in the supplementary material.

\subsection{Overview of Topology Distillation}
\label{sec:TD_overview}
The proposed topology distillation approach guides the learning of the student by the topological structure built upon the relational knowledge in the teacher representation space.
The relational knowledge refers to all the information on how the representations are correlated in the space;
those sharing similar preferences are strongly correlated, whereas those with different preferences are weakly correlated.
We build a (weighted) topology of a graph where the nodes are the representations and the edges encode the relatedness of the representations.
Then, we distill the relational knowledge by making the student preserve the teacher's topological structure in its representation space.
With the proposed approach, the student is trained by minimizing the following loss:
\begin{equation}
    \mathcal{L} = \mathcal{L}_{Base} + \lambda_{TD} \mathcal{L}_{TD}
    \label{eq:TD_total}
\end{equation}
where $\lambda_{TD}$ is a hyperparameter controlling the effects of topology distillation.
The base model can be any existing recommender, and $\mathcal{L}_{Base}$ corresponds to its loss function.
$\mathcal{L}_{TD}$ is defined on the topology of the representations in the same batch used for the base model training.

\subsection{Full Topology Distillation (FTD)}
\label{sec:TD_FTD}
As a straightforward method, we distill the knowledge of the entire relations in the teacher space.
Given a batch, we first generate a fully connected graph in the teacher representation space.
The graph is characterized by the adjacency matrix $\mathbf{A}^t \in \mathbb{R}^{b \times b}$ where $b$ is the number of representations in the batch. 
Each element $a^t_{ij}$ is the weight of an edge between entities $i$ and $j$ representing their similarity and is parameterized as follows:
\begin{align}
    a^t_{ij}=\rho(\mathbf{e}^t_i,\mathbf{e}^t_j)
\end{align}
where $\rho(\cdot,\cdot)$ is a similarity score such as the cosine similarity or negative Euclidean distance, in this work we use the former.
Analogously, we generate a graph, characterized by the adjacency matrix $\mathbf{A}^s\in \mathbb{R}^{b \times b}$, in the student representation space, i.e., $a^s_{ij}=\rho(\mathbf{e}^s_i,\mathbf{e}^s_j)$.

After obtaining the topological structures $\mathbf{A}^t$ and $\mathbf{A}^s$ from the representation space of the teacher and student, respectively, we train the student to preserve the topology discovered by the teacher by the topology-preserving distillation loss as follows:
\begin{align}
    \mathcal{L}_{F T D} = \text{Dist}(\mathbf{A}^t, \mathbf{A}^s) =  \lVert \mathbf{A}^t - \mathbf{A}^s \rVert^2_F,
\end{align}
where $\text{Dist}(\cdot, \cdot)$ is the distance between the topological structures, in this work, we compute it with the Frobenius norm.
By minimizing $\mathcal{L}_{FTD}$, parameters in the student are updated.
As this method utilizes the full topology as supervision to guide the student, we call it Full Topology Distillation (FTD).
Substituting the distillation loss $\mathcal{L}_{TD}$ in Equation \ref{eq:TD_total} derives the final loss for training the student.

\noindent
\textbf{Issues:}
Although FTD directly transfers the relational knowledge which is ignored in the previous work, it still has some clear drawbacks.
Because the student has a very limited capacity compared to the teacher, it is often daunting for the student to learn all the relational knowledge in the teacher.
Indeed, we observe that sometimes FTD even hinders the learning of the student and degrades the recommendation performance.
Therefore, the relational knowledge should be distilled with consideration of the huge capacity gap.

\subsection{Hierarchical Topology Distillation (HTD)}
\label{sec:TD_HTD}
Our key idea to tackle the issues is to decompose the whole topology hierarchically so as to be effectively transferred to the student.
We argue that the student should focus on learning the relations among the strongly correlated entities that share similar preferences and accordingly have a direct impact on top-$N$ recommendation performance.
To this end, we summarize the numerous relations among the weakly correlated entities, enabling the student to better focus on the important relations.

During the training, HTD adaptively finds \emph{preference groups} of strongly correlated entities.
Then, the topology is hierarchically structured in group-level and entity-level:
(1) \emph{group-level topology} includes the summarized relations across the groups, providing the overview of the entire topology.
(2) \emph{entity-level topology} includes the relations of entities belonging to the same group.
This provides fine-grained supervision on important relations of the entities having similar preferences.
By compressing the complex individual relations across the groups, HTD relaxes the daunting supervision, effectively distills the relational knowledge to the student.

\subsubsection{\textbf{Preference Group Assignment}}\noindent
To find the groups of entities having a similar preference in an end-to-end manner considering both the teacher and the student, we borrow the idea of DE \cite{DERRD}.
Formally, let there exist $K$ preference groups in the teacher space.
We use a small network $v: \mathbb{R}^{d^t} \rightarrow \mathbb{R}^{K}$ with Softmax output to compute the assignment probability vector $\boldsymbol{\alpha}_i \in \mathbb{R}^{K}$ for each entity $i$ as follows:
\begin{equation}
    \boldsymbol{\alpha}_{i} = v( \mathbf{e}^{t}_i),
\end{equation}
where each element $\alpha_{ik}$ encodes the probability of the entity $i$ to be assigned to $k$-th preference group.
Let $\mathbf{z}_{i}$ be a $K$-dimensional one-hot assignment vector whose element $z_{ik}=1$ if entity $i$ belongs to the corresponding  $k$-th group.
We assign a group for each entity by sampling the assignment vector from a categorical distribution parameterized by $\{\alpha_{ik}\}$ i.e., $p(z_{ik}=1 \mid v, \mathbf{e}^t_i)=\alpha_{ik}$.
To make the sampling process differentiable, we adopt Gumbel-Softmax \cite{GumbelSoftmax} which is a continuous distribution on the simplex that can approximate samples from a categorical distribution.
\begin{equation}
\begin{aligned}
z_{ik} =\frac{\exp\left(\left( \alpha_{ik} + g_k \right) / \tau\right) }{\sum_{j=1}^K \exp\left(\left( \alpha_{ij} + g_j \right) / \tau\right)} \quad \text{for} \quad k=1,...,K,
\end{aligned}
\end{equation}
where $g_j$ is the gumbel noise drawn from $\text{Gumbel}(0,1)$ distribution \cite{GumbelSoftmax} and $\tau$ is the temperature parameter.
We set a small value on $\tau$ so that samples from the Gumbel-Softmax distribution become one-hot vector.
This group assignment is evolved during the training via backpropagation \cite{DERRD}.

With the assignment process, for a given batch, HTD obtains the grouping information summarized by a $b \times K$ assignment matrix $\mathbf{Z}$ where each row corresponds to the one-hot assignment vector for each entity.
We also denote the set of entities belonging to each group by $G_{k}=\{i|z_{ik}=1\}$.
Note that the group assignment is based on the teacher's knowledge.
Based on the assignment, we decompose the topology hierarchically.

\subsubsection{\textbf{Group-level topology.}}\noindent
HTD introduces a \textit{prototype} representing the entities in each preference group, then use it to summarize the relations across the groups.
Let $\mathbf{E}^t \in \mathbb{R}^{b \times d^t}$ and $\mathbf{E}^s \in \mathbb{R}^{b \times d^s}$ denote the representation matrix in the teacher space and the student space, respectively.
The prototypes $\mathbf{P}^t \in \mathbb{R}^{K \times d^t}$ and $\mathbf{P}^s \in \mathbb{R}^{K \times d^s}$ are defined as follows:
\begin{equation}
\begin{aligned}
    \mathbf{P}^t = \tilde{\mathbf{Z}}^\top \mathbf{E}^t \quad\text{and}\quad \mathbf{P}^s = \tilde{\mathbf{Z}}^\top \mathbf{E}^s,
\end{aligned}
\end{equation}
where $\tilde{\mathbf{Z}}$ is normalized assignment matrix by the number of entities in each group (i.e., $\tilde{\mathbf{Z}}_{[:,i]} = \mathbf{Z}_{[:,i]} / \sum_i \mathbf{Z}_{[:,i]}$).
For $\mathbf{P}$, each row $\mathbf{P}_{[k,:]}$ corresponds to the average representation for the entities belonging to each group $k$, and we use it as a prototype representing the group.

With the prototypes, we consider two design choices with different degrees of relaxation.
In the first choice, we distill the relations between the prototypes.
We build the topology characterized by the $K \times K$ matrix $\mathbf{H}^t$ which contains the relations~as:
\begin{equation}
h^t_{km} = \rho(\mathbf{P}^t_{[k,:]}, \mathbf{P}^t_{[m,:]}),
\end{equation}
where $k, m \in \{1, ..., K\}$.

In the second choice, we distill the relations between each prototype and entities belonging to the other groups.
We build the topology characterized by the $K \times b$ matrix $\mathbf{H}^t$ which contains the relations~as:
\begin{equation}
h^t_{kj} = \rho(\mathbf{P}^t_{[k,:]}, \mathbf{e}^t_j),
\end{equation}
where $k \in \{1, ..., K\}, j \in \{1, ..., b\}$.
It is worth noting that we only distill the relations across the groups (i.e., $j \notin G_k$).
Using one of the choices, we build the group-level topological structure $\mathbf{H}^t$ in the teacher space, and analogously, we build $\mathbf{H}^s$ in the student space.

The first choice puts a much higher degree of relaxation compared to the second choice.
For instance, assume that there are two groups of entities (i.e., $G_1$ and $G_2$).
Without the hierarchical approach (as done in FTD), there exists $|G_1|\times|G_2|$ relations across the groups.
With the first choice, they are summarized to a single relation between two prototypes, and with the second choice, they are summarized to $|G_1|+|G_2|$ relations.
We call the first choice as Group(P,P) and the second choice as Group(P,e) and provide results with each choice in Section \ref{sec:TD_result}.

\subsubsection{\textbf{Entity-level topology.}}\noindent
HTD distills the full relations among the strongly correlated entities in the same group.
In the teacher space, the entity-level topology contains the following~relations:
\begin{equation}
\{ \rho(\mathbf{e}^t_i,\mathbf{e}^t_j) \mid (i, j) \in G_{k} \times G_{k} \}, \;\;\text{for} \;k \in \{1, ..., K\},
\end{equation}
and analogously, we build the entity-level topology in the student space. 
For an efficient computation on matrix form, we introduce the $b \times b$ binary indicator matrix $\mathbf{M} = \mathbf{Z} \mathbf{Z}^\top$ indicating whether each relation is contained in the topology or not.
Intuitively, each element $m_{ij}=1$ if entity $i$ and $j$ are assigned to the same group, otherwise $m_{ij}=0$.
Then, the entity-level topology is defined by $\mathbf{A}^t$ with $\mathbf{M}$ in the teacher space and also defined by $\mathbf{A}^s$ with $\mathbf{M}$ in the student space.
The distance between two topological structures is simply computed by $\lVert \mathbf{M} \odot (\mathbf{A}^t - \mathbf{A}^s) \rVert^2_F$ where  $\odot$ is the Hadamard product.

\subsubsection{\textbf{Optimization}}\noindent
HTD guides the student with the decomposed topological structures.
The loss function is defined as follows:
\begin{equation}
\begin{aligned}
\mathcal{L}_{HTD} &= 
\gamma  \left( \lVert \mathbf{H}^t - \mathbf{H}^s \rVert^2_F + \lVert \mathbf{M} \odot (\mathbf{A}^t - \mathbf{A}^s) \rVert^2_F \right)\\
&+ (1-\gamma) \left(\sum_{i=1}^b \lVert \mathbf{e}^t_i - \sum^K_{k=1} z_{ik}  f_k(\mathbf{e}^s_i) \rVert^2_2\right),
\end{aligned}
\end{equation}
where the first term corresponds to the topology-preserving loss,
the second term corresponds to the hint regression loss adopted in DE \cite{DERRD} that makes the group assignment process differentiable.
We put a network $f_k$ for each $k$-th group, then train each network to reconstruct the representations belonging to the corresponding group, which makes the entities having strong correlations get distilled by the same network \cite{DERRD}.
$\gamma$ is a hyperparameter balancing the two terms. 
In this work, we set 0.5 to $\gamma$.
By minimizing $\mathcal{L}_{HTD}$, parameters in the student, $v$ and $f_*$ are updated.
Note that $v$ and $f_*$ are not used in the inference phase, they are only utilized for the distillation in the offline training phase.
Substituting the distillation loss $\mathcal{L}_{TD}$ in Equation \ref{eq:TD_total} derives the final loss for training the student.

\begin{figure}[t]
\centering
\begin{subfigure}[t]{0.75\linewidth}
    \includegraphics[width=1.\linewidth]{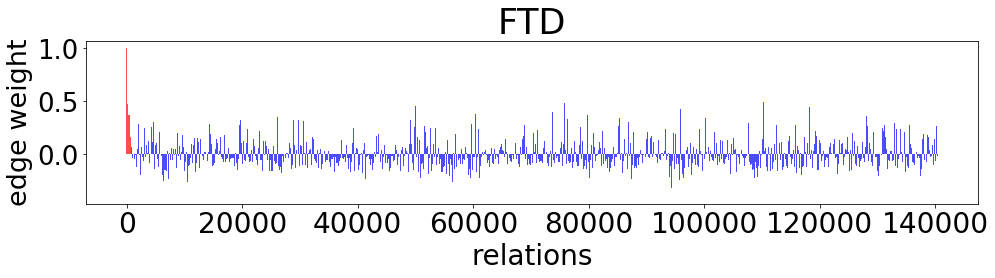}
\end{subfigure}\\
\begin{subfigure}[t]{0.75\linewidth}
    \includegraphics[width=1.\linewidth]{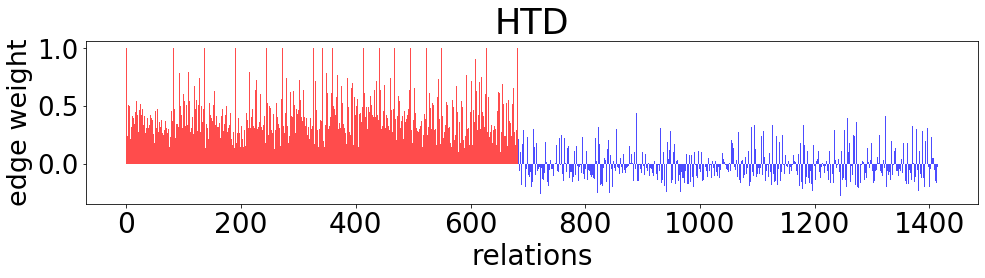}
\end{subfigure}
\caption{The relational knowledge distilled from teacher to student by FTD and HTD (with Group(P,e)). Red/Blue corresponds to relations of the entities belonging to the same/different preference group(s) (BPR on CiteULike).}
\label{fig:TD_FTDvsHTD}
\end{figure}

\vspace{0.1cm}
\noindent
\textbf{Effects of HTD.}
For more intuitive understanding, we provide a visualization of the relational knowledge distilled to the student in Figure \ref{fig:TD_FTDvsHTD}.
We randomly choose a preference group and visualize the relations \textit{w.r.t.} the entities belonging to the group.
Note that there exist the same number of intra-group relations (red) in both figures.
Without the hierarchical approach (as done in FTD), the student is forced to learn a huge number of relations with the entities belonging to the other groups (blue).
On the other hand, HTD summarizes the numerous relations, enabling the student to better focus on learning the detailed intra-group relations, which directly affects in improving the recommendation performance.

\vspace{0.1cm}
\noindent
\textbf{Discussions on Group Assignment.}
Note that HTD is not limited to a specific group assignment method, i.e., DE.
Any method that clusters the teacher representation space or prior knowledge of user/item groups (e.g., item category, user demographic features) can be utilized for more sophisticated topology decomposition, which further improves the effectiveness of HTD. 
The comparison with another assignment method is provided in the supplementary material.

\section{Experiments}
\label{sec:TD_experimentsetup}
We validate the proposed approach on \textbf{18} experiment settings: 2 real-world datasets × 3 base models × 3 different student model sizes.
We first present comparison results with the state-of-the-art competitor and a detailed ablation study.
We also provide in-depth analyses to verify the benefit of distilling the topological structure.
Lastly, we provide a hyperparameter study.

\subsection{Experimental Setup}
We closely follow the experiment setup of DE \cite{DERRD}.
However, for a thorough evaluation, we make two changes in the setup.
\textbf{(1)} we add LightGCN \cite{he2020lightgcn}, which is the state-of-the-art top-$N$ recommendation method, as a base model.
\textbf{(2)} unlike \cite{DERRD} that samples negative items for evaluation, we adopt the full-ranking evaluation which enables more rigorous evaluation.
Refer to the appendix for more detail.

\subsubsection{Datasets}\noindent
We use two real-world datasets: CiteULike and Foursquare.
CiteULike contains tag information for each item, and Foursquare contains GPS coordinates for each item.
We use the side information to evaluate the quality of representations induced by each KD method.
More details of the datasets are provided in the appendix.

\subsubsection{Base Models}\noindent
We evaluate the proposed approach on base models having different architectures and learning strategies, which are widely used for top-$N$ recommendation task.
\begin{itemize}[leftmargin=*]
    \item \textbf{BPR \cite{BPR}}: 
    A learning-to-rank model that models user-item interaction with Matrix Factorization (MF).
    \item \textbf{NeuMF \cite{NeuMF}}: A deep model that combines MF and Multi-Layer Perceptron (MLP) to learn the user-item interaction.
    \item \textbf{LightGCN \cite{he2020lightgcn}}: The state-of-the-art model which adopts simplified Graph Convolution Network (GCN) to capture the information of multi-hop neighbors.
\end{itemize}

\subsubsection{Teacher/Student}
\begin{figure}[t]
\centering
\begin{subfigure}[t]{0.95\linewidth}
    \includegraphics[width=\linewidth]{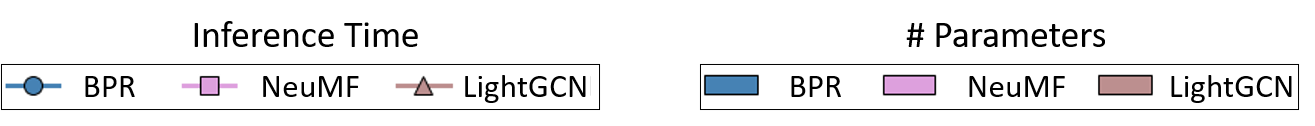}
\end{subfigure}\\
\begin{subfigure}[t]{0.45\linewidth}
    \includegraphics[width=\linewidth]{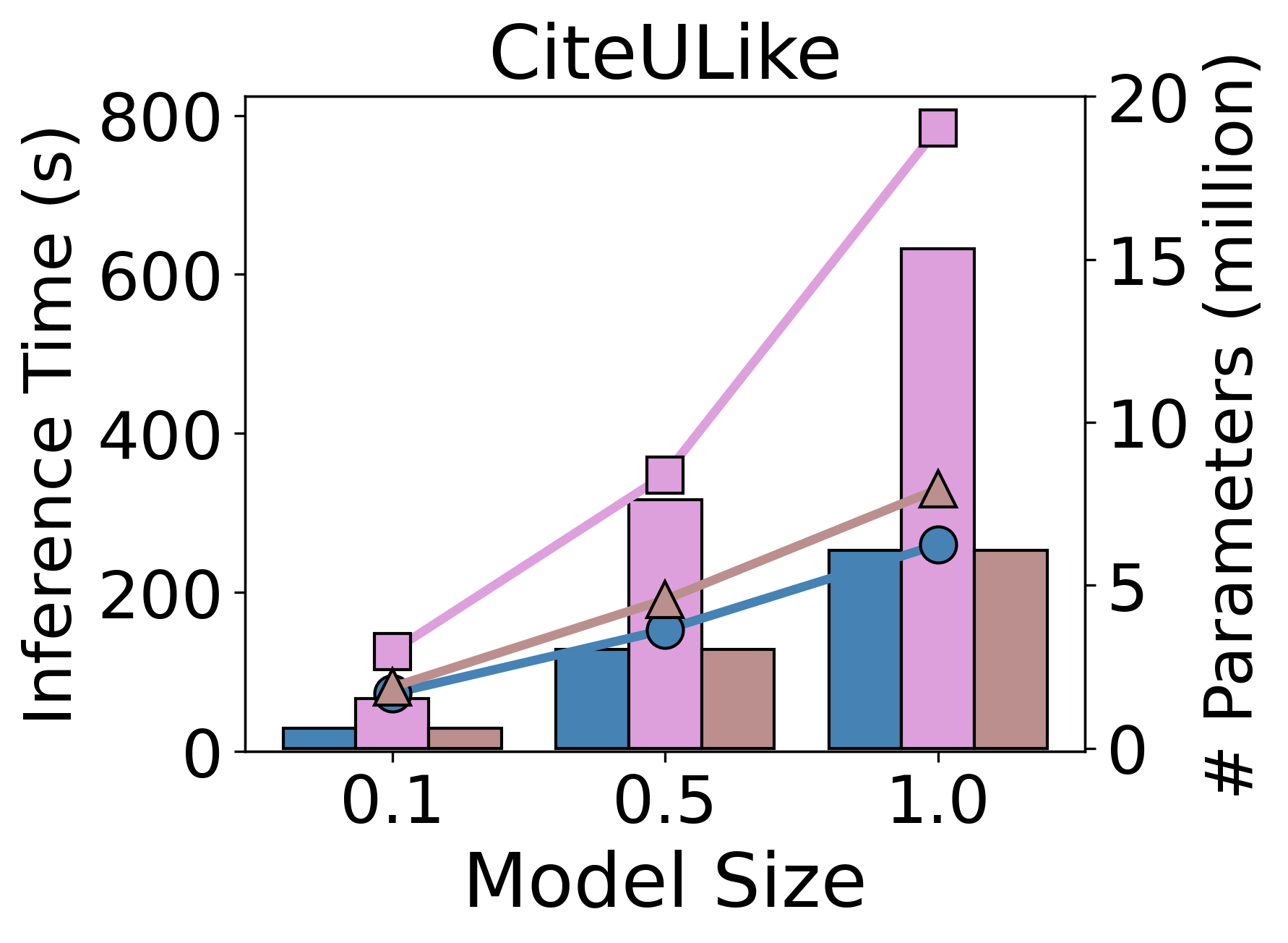}
\end{subfigure}
\hspace{0.1cm}
\begin{subfigure}[t]{0.47\linewidth}
    \includegraphics[width=\linewidth]{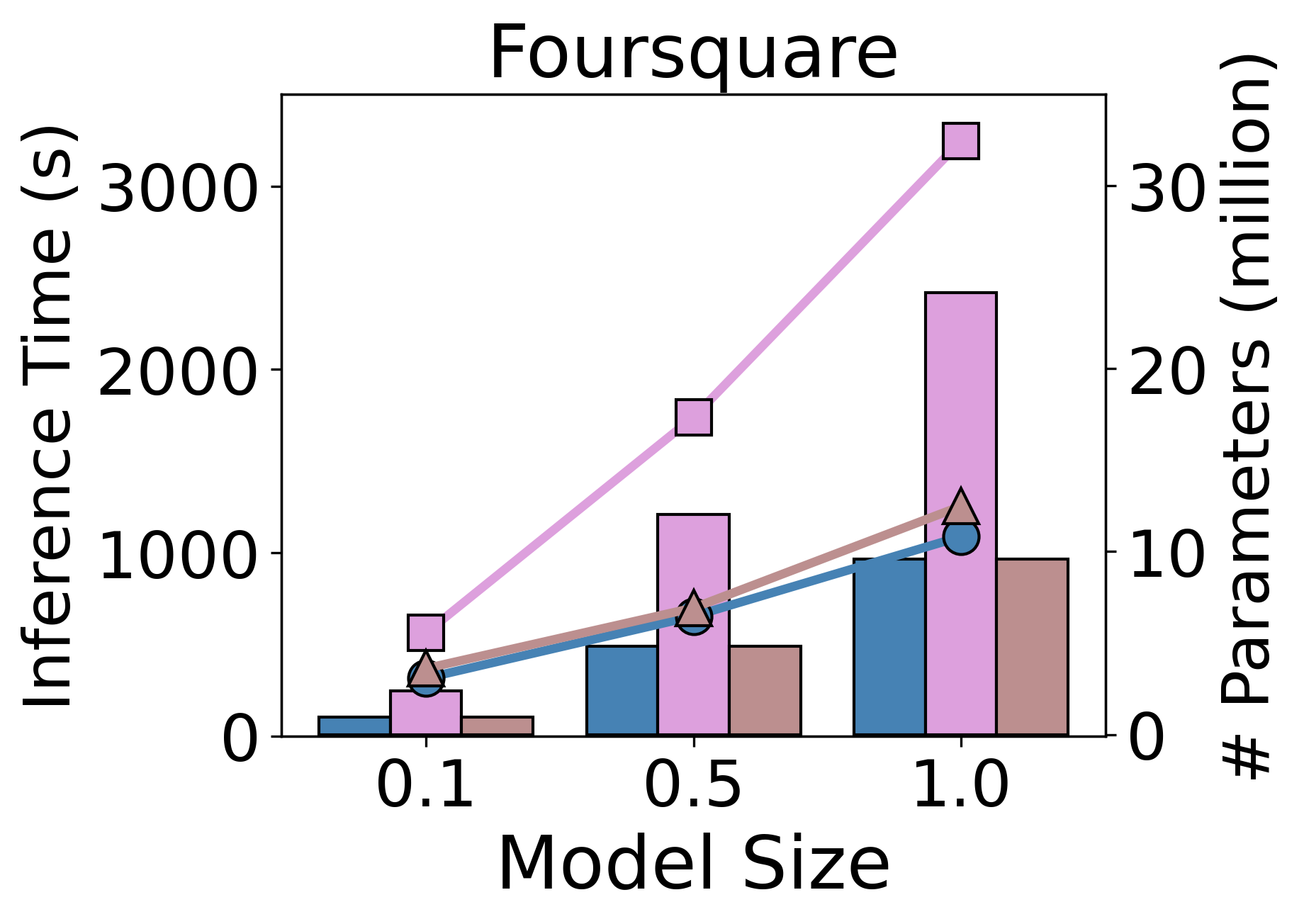}
\end{subfigure}
\caption{Inference time (s) and model size ($\phi$). Inference time denotes the wall time used for generating recommendation list for every user.}
\label{fig:TD_latency}
\end{figure}

\noindent
For each setting, we increase the model size until the recommendation performance is no longer improved and adopt the model with the best performance as \textbf{Teacher} model.
Then, we build three student models by limiting the model size, i.e.,  $\phi \in \{0.1, 0.5, 1.0\}$.
We call the student model trained without distillation as \textbf{Student} in this section.
Figure \ref{fig:TD_latency} summarizes the model size and inference time.
The inferences are made using PyTorch with CUDA from TITAN Xp GPU and Xeon on Gold 6130 CPU.
It shows that the smaller model has lower inference latency.

\subsubsection{Compared Methods}\noindent
We compare the following KD methods distilling the latent knowledge from the intermediate layer of the teacher recommender.
\begin{itemize}[leftmargin=*]
    \item \textbf{FitNet \cite{FitNet}}: A KD method utilizing the original hint regression.
    \item \textbf{Distillation Experts (DE) \cite{DERRD}}: The state-of-the-art KD method distilling the latent knowledge.
    DE elaborates the hint regression.
    \item \textbf{Full Topology Distillation (FTD)}: A proposed method that distills the full topology (Section \ref{sec:TD_FTD}).  
    \item \textbf{Hierarchical Topology Distillation (HTD)}: A proposed method that distills the hierarchical topology (Section \ref{sec:TD_HTD}).  
\end{itemize}
Note that we do not include the methods distilling the predictions (e.g., RD \cite{RD}, CD \cite{CD}, and RRD \cite{DERRD}) in the competitors, because they are not competing with the methods distilling the latent knowledge \cite{DERRD}.
Instead, we provide experiment results when they are combined with the proposed approach in the next section.

\label{sec:TD_result}

\subsection{Performance Analysis}
\begin{figure}[t]
\centering
\begin{subfigure}[t]{0.6\linewidth}
    \includegraphics[width=\linewidth]{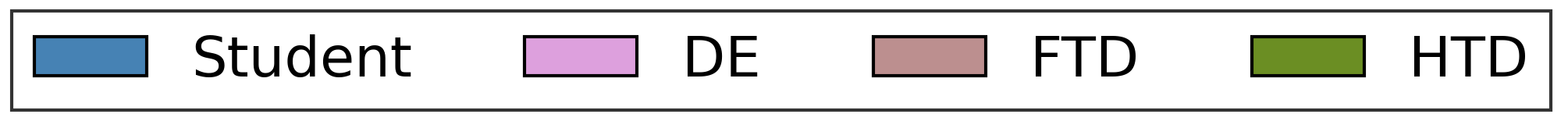}
\end{subfigure}\\
\hspace{-0.3cm}
\begin{subfigure}[t]{0.34\linewidth}
    \includegraphics[height=4cm]{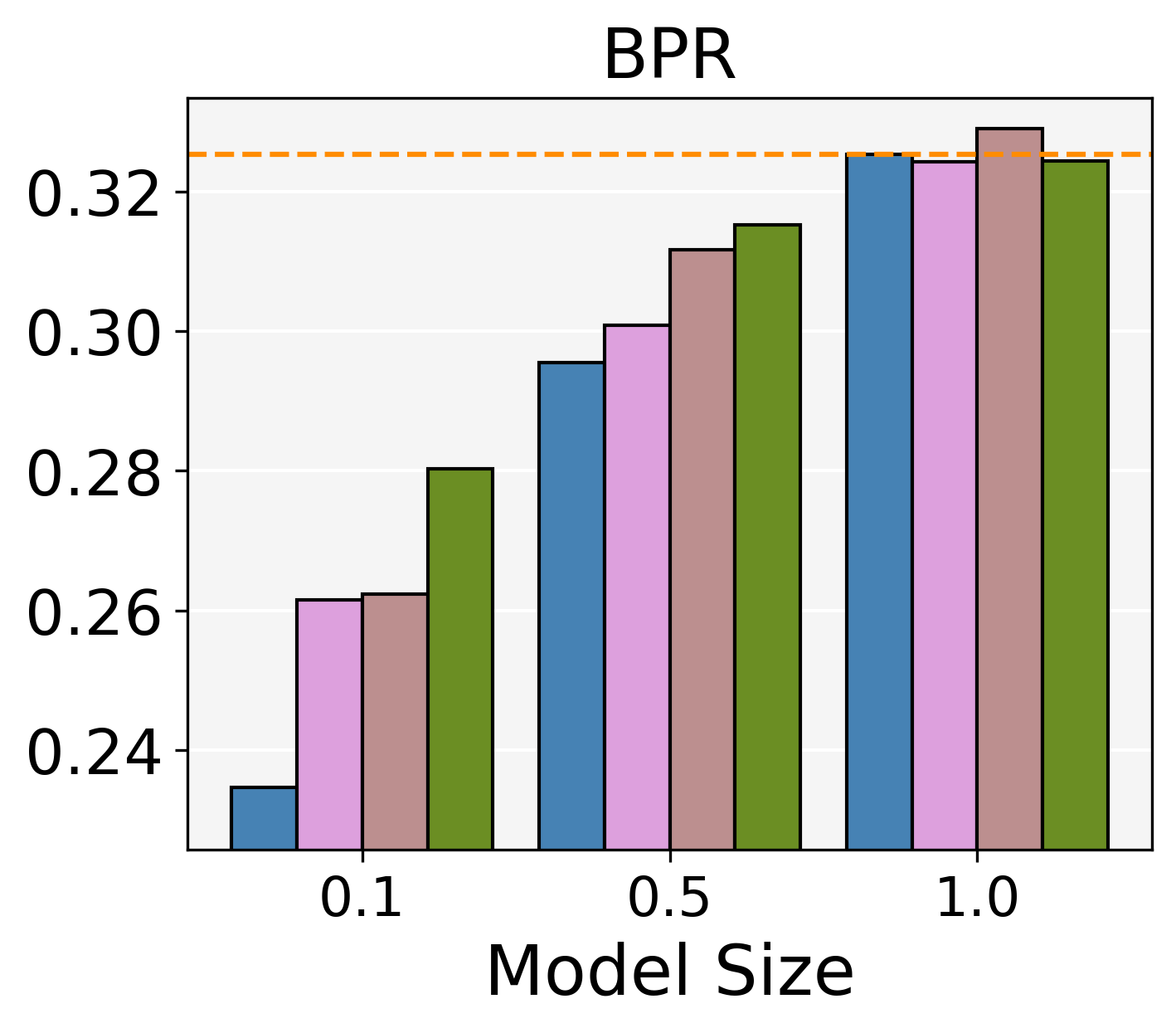}
\end{subfigure}
\hspace{-0.2cm}
\begin{subfigure}[t]{0.34\linewidth}
    \includegraphics[height=4cm]{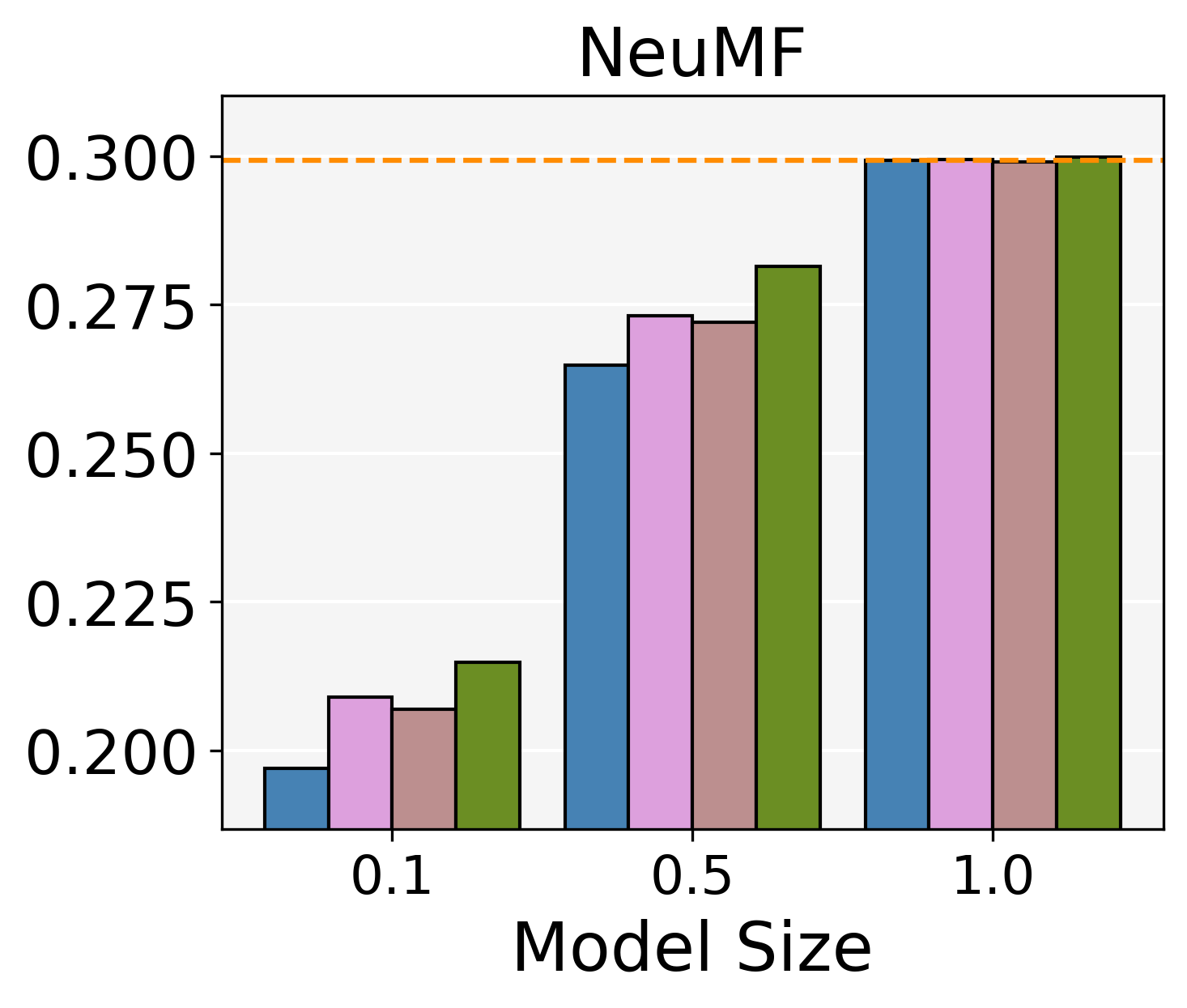}
\end{subfigure}
\hspace{-0.2cm}
\begin{subfigure}[t]{0.34\linewidth}
    \includegraphics[height=4cm]{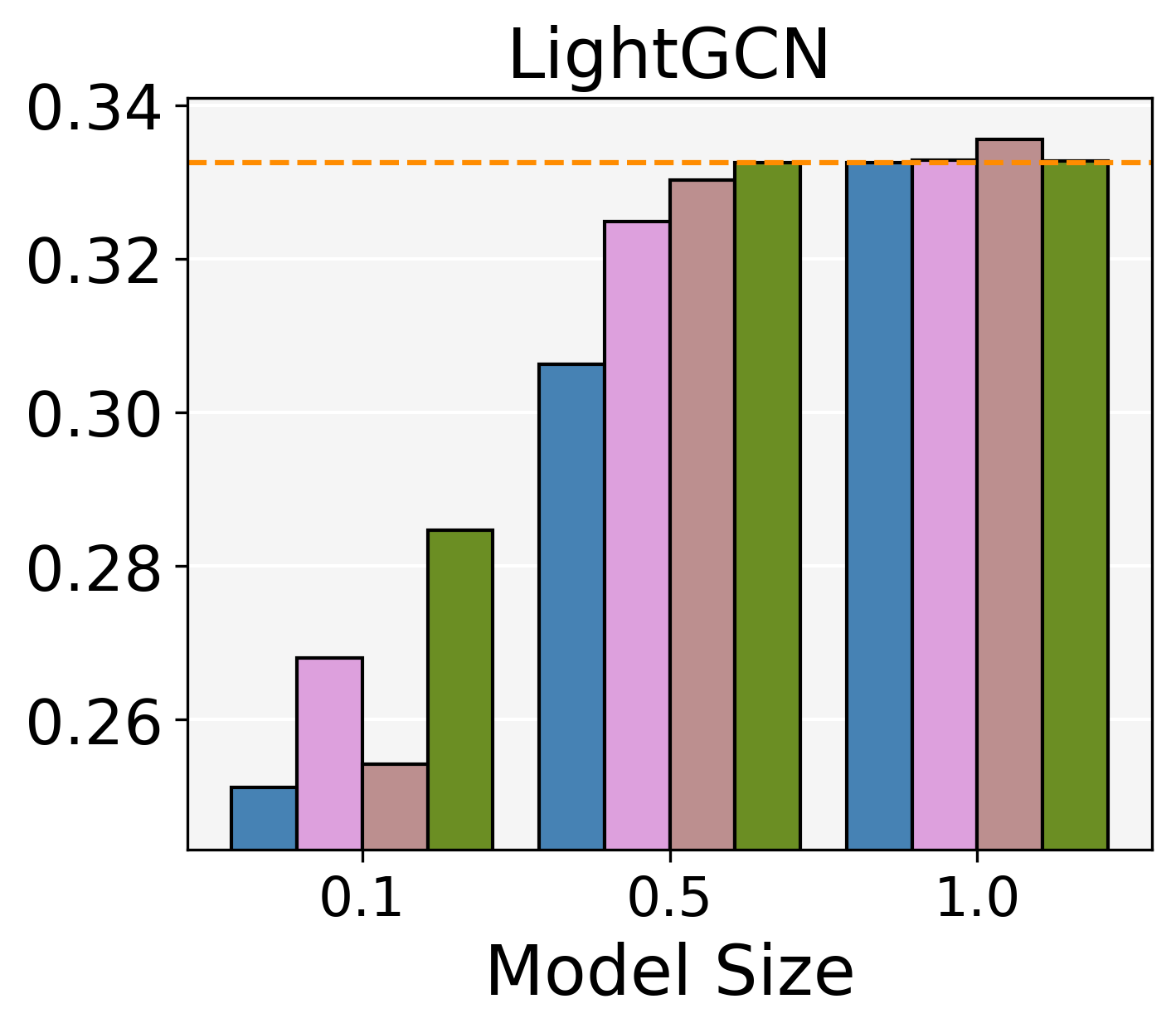}
\end{subfigure}
\hspace{-0.3cm}
\caption*{(a) CiteULike}
\hspace{-0.3cm}
\begin{subfigure}[t]{0.34\linewidth}
    \includegraphics[height=4cm]{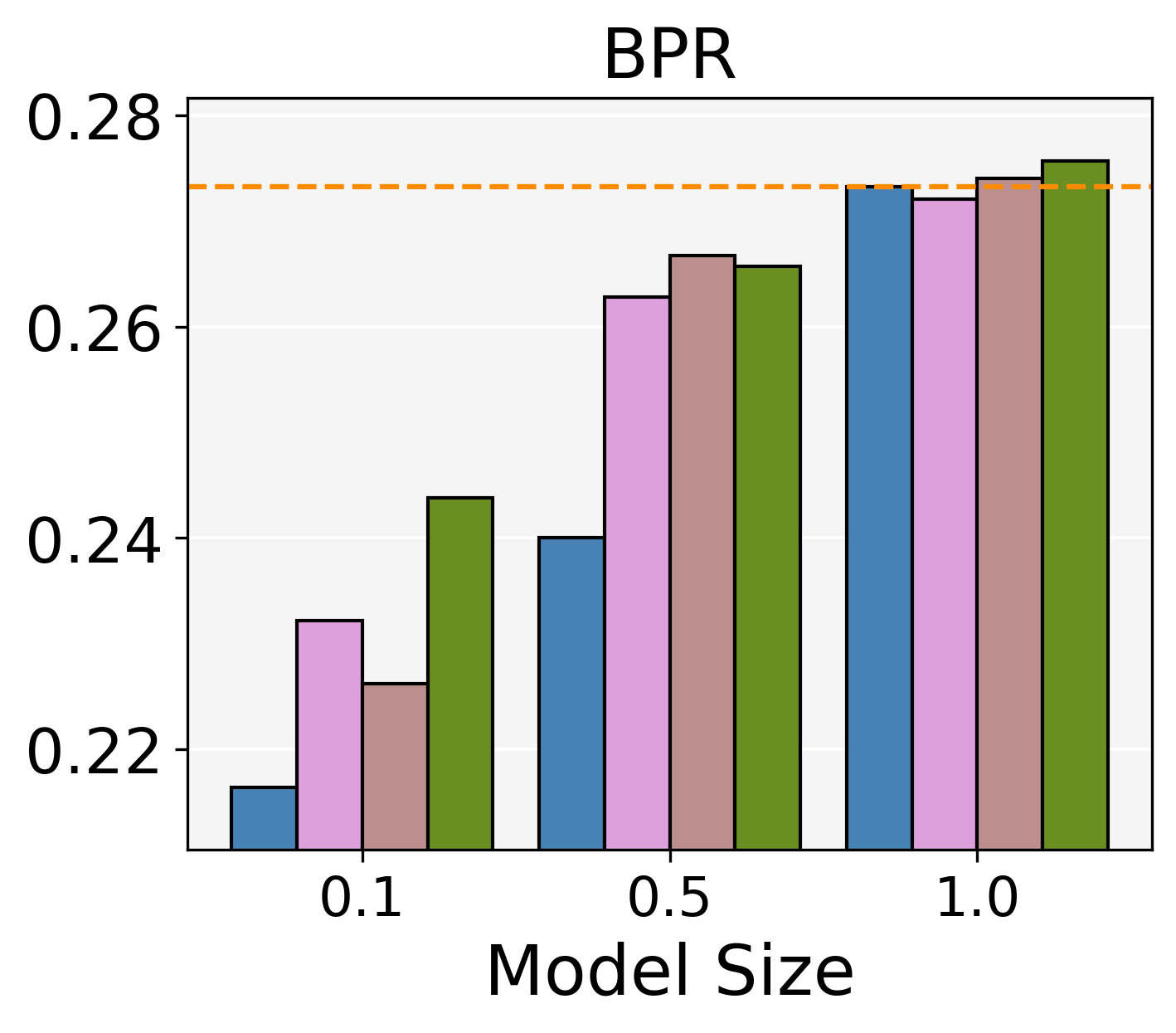}
\end{subfigure}
\hspace{-0.2cm}
\begin{subfigure}[t]{0.34\linewidth}
    \includegraphics[height=4cm]{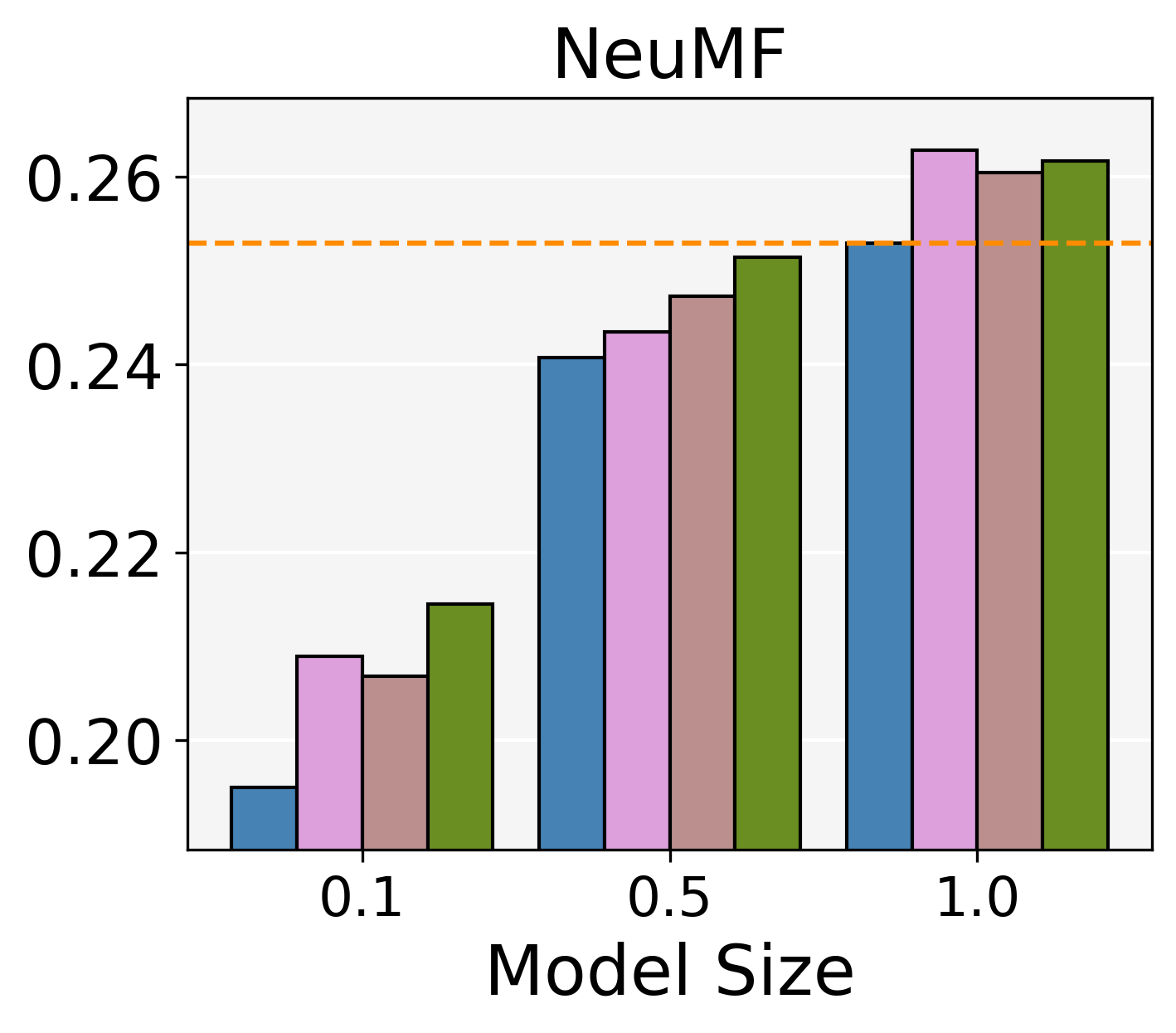}
\end{subfigure}
\hspace{-0.2cm}
\begin{subfigure}[t]{0.34\linewidth}
    \includegraphics[height=4cm]{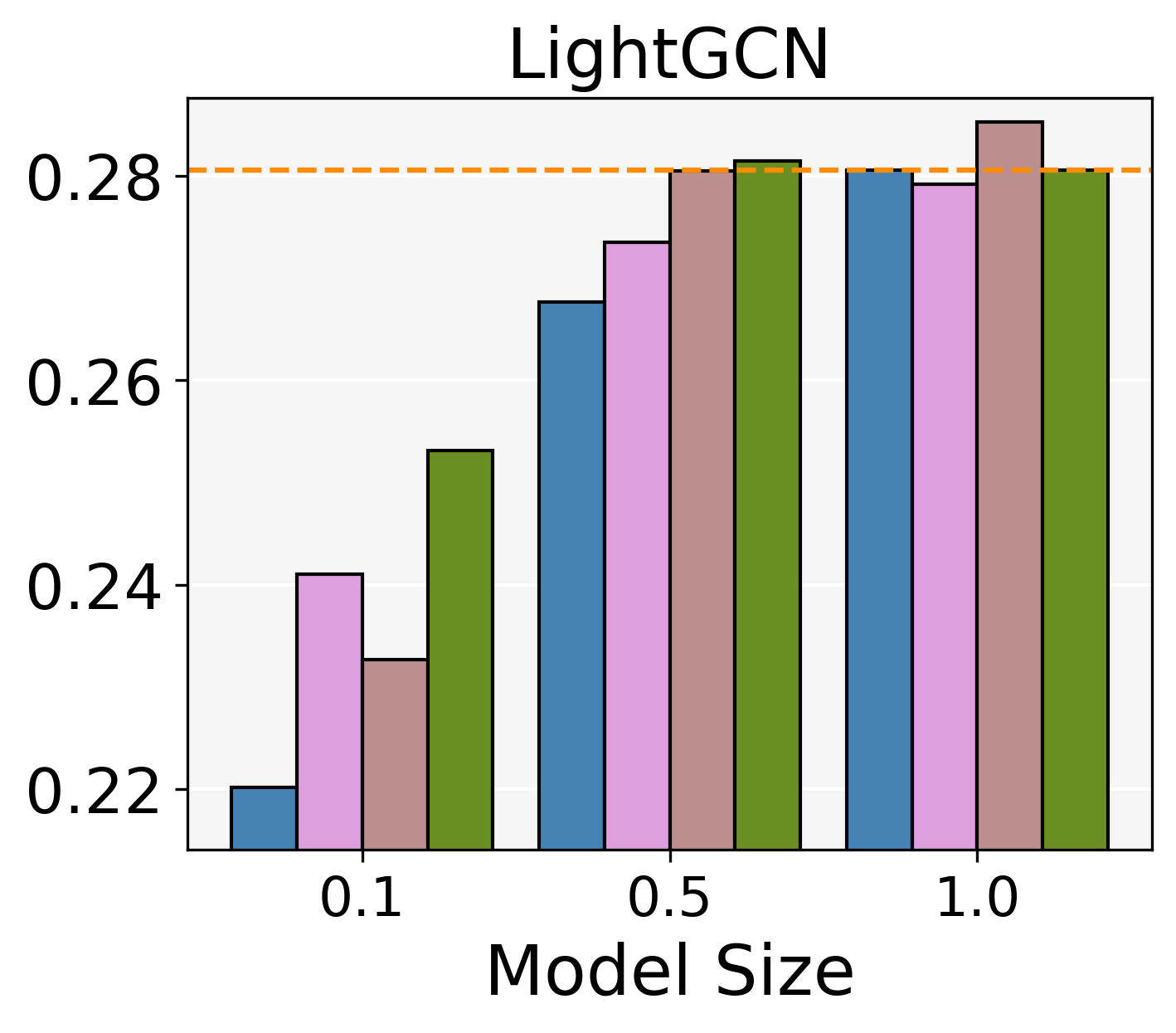}
\end{subfigure}
\hspace{-0.3cm}
\caption*{(b) Foursquare}
\caption{Recall@50 across three different student sizes (Dotted line: Teacher)}
\label{fig:TD_sizes}
\end{figure}

\begin{figure}[t]
\centering
\includegraphics[width=0.7\linewidth]{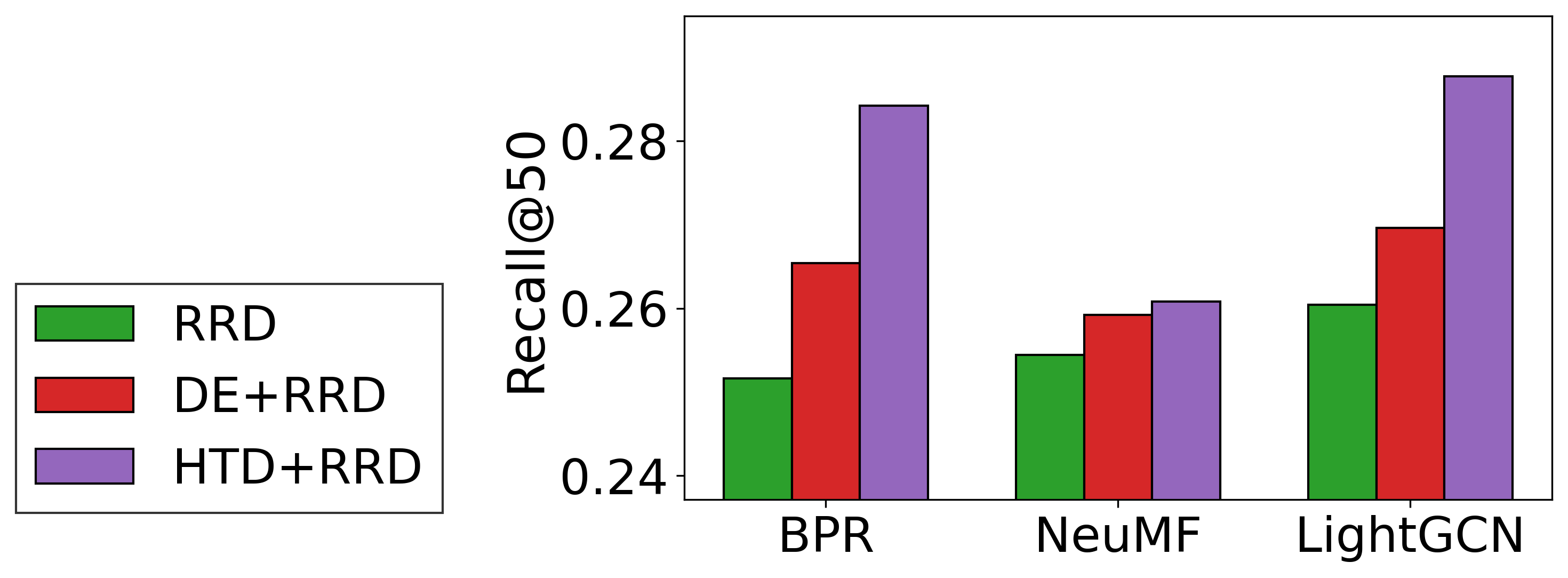}
\caption{Performance comparison with RRD.}
\label{fig:TD_RRD}
\end{figure}

\begin{figure}[t]
\centering
\begin{subfigure}[t]{0.495\linewidth}
    \includegraphics[width=\linewidth]{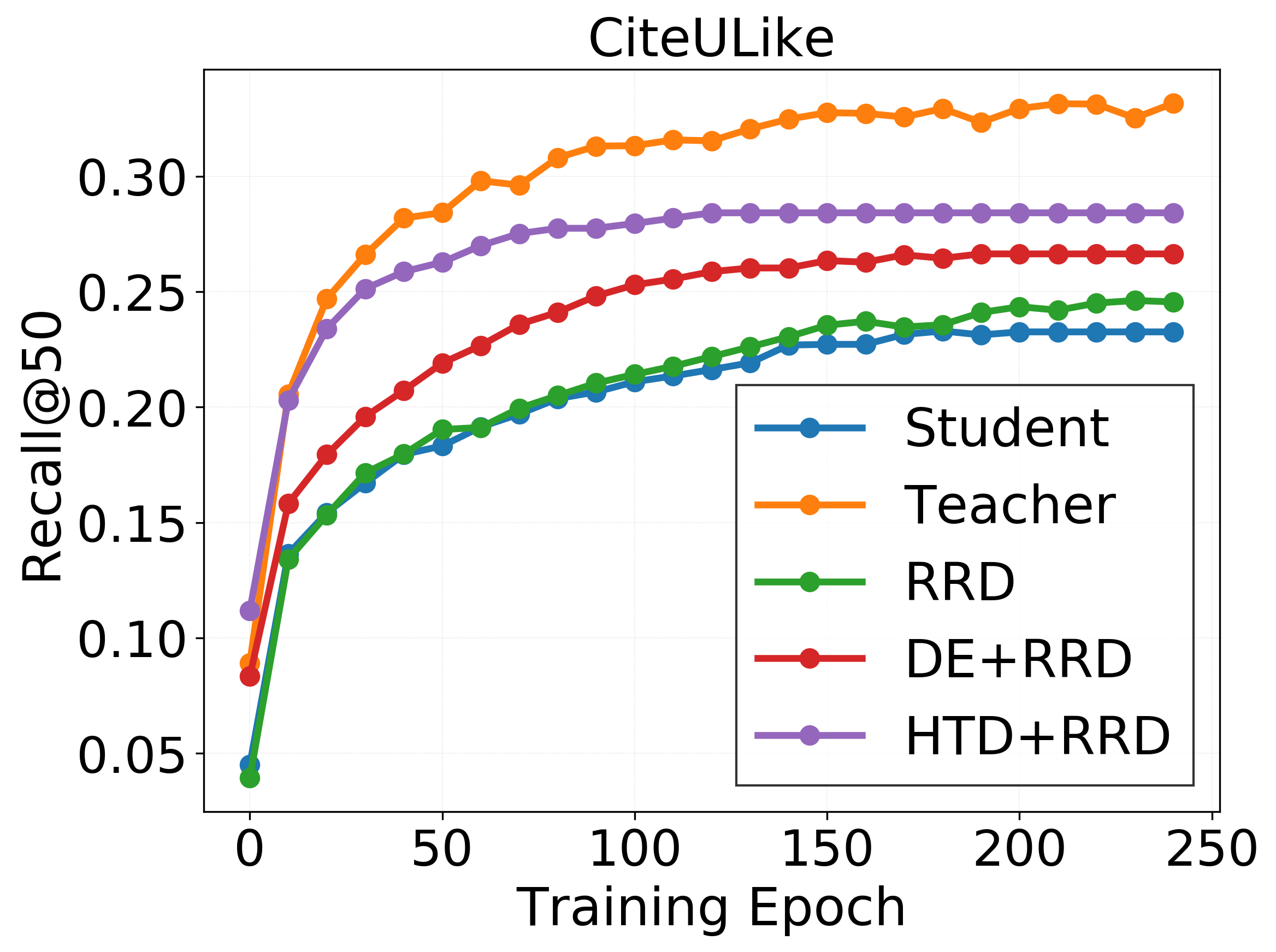}
\end{subfigure}
\begin{subfigure}[t]{0.495\linewidth}
    \includegraphics[width=\linewidth]{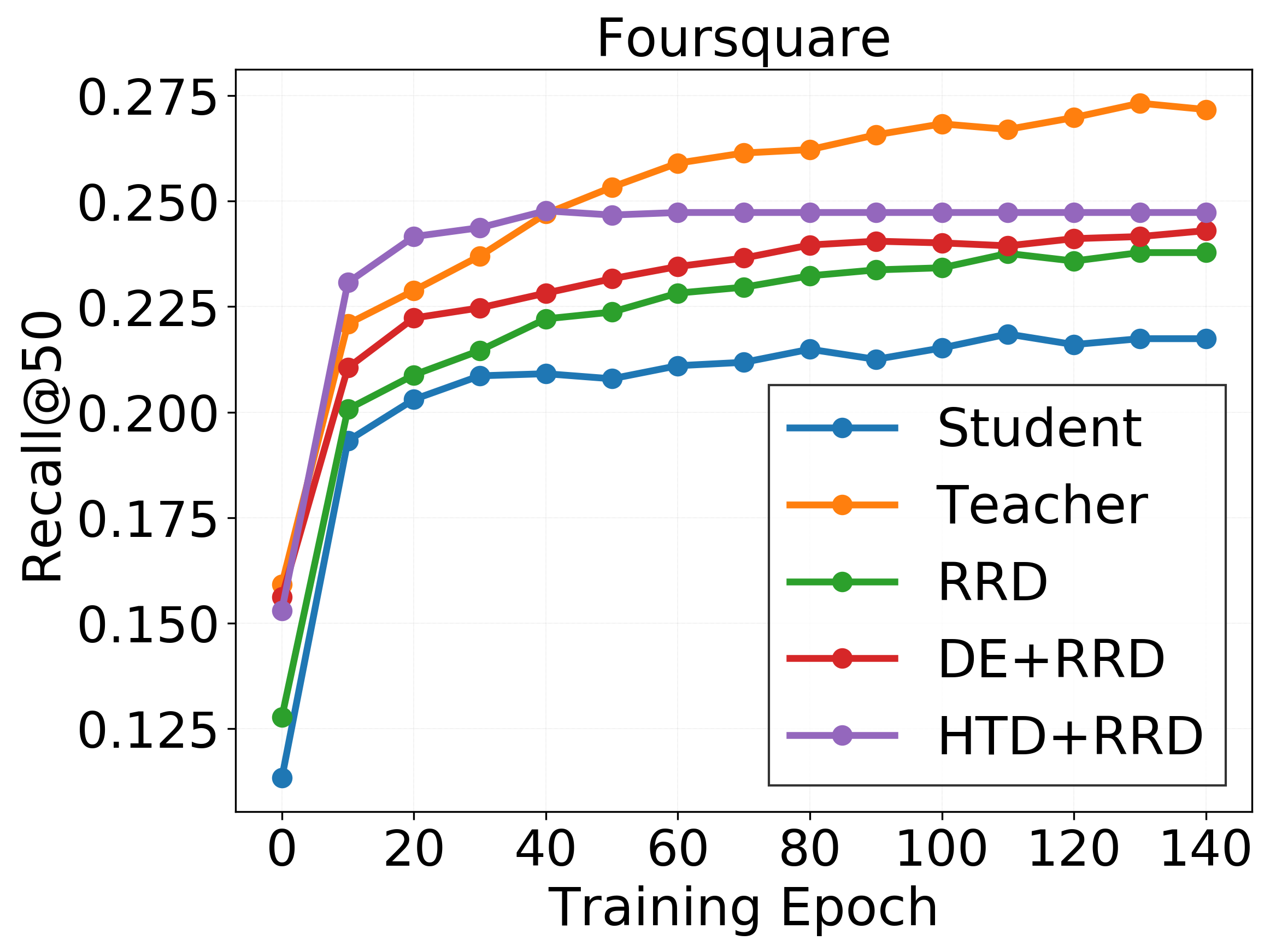}
\end{subfigure}
\caption{Training curves with RRD.}
\label{fig:TD_curve}
\end{figure}

Table \ref{tbl:TD_maintable1} and Table \ref{tbl:TD_maintable2} present top-$N$ recommendation performance of the methods compared ($\phi=0.1$), and Figure \ref{fig:TD_sizes} presents results with three different students sizes. 
For the group-level topology of HTD, we choose Group(P,e), since it consistently shows better results than Group(P,P).
The detailed comparisons along with other various ablations are reported in Table \ref{tbl:TD_ablation}.
Lastly, results with prediction-based KD method are presented in Figure \ref{fig:TD_RRD} and Figure \ref{fig:TD_curve}.

\definecolor{Gray}{gray}{0.95}
\newcolumntype{g}{>{\columncolor{Gray}}c}
\newcolumntype{L}{>{\columncolor{Gray}}l}

\begin{sidewaystable}[!thbp]
\centering
\caption{Performance comparison ($\phi=0.1$). \textit{Gain.DE} denotes the improvement of HTD over DE, \textit{Gain.S} denotes the improvement of HTD over Student.
HTD achieves statistically significant improvements over the best baseline. We use the paired t-test with significance level at 0.05 on Recall@50.}
\renewcommand{\arraystretch}{0.5}
\renewcommand{\tabcolsep}{2.7mm}
  \begin{minipage}[t]{1\linewidth}
  \centering
  \begin{tabular}{ccLgggggg}
    \toprule[.15em]
    \rowcolor{white}
    Dataset &Base Model & Method \quad & Recall@10 & NDCG@10 & Recall@20 & NDCG@20 & Recall@50 & NDCG@50 \\
    \midrule[.15em]
    \rowcolor{white}
    & & Teacher & 0.1533 & 0.0883 & 0.2196 & 0.1058 & 0.3253 & 0.1247 \\
    \rowcolor{white}
    &&Student & 0.1014 & 0.0560 & 0.1506 & 0.0684 & 0.2347 & 0.0864 \\
    \rowcolor{white}
    &&FitNet & 0.1097 & 0.0595 & 0.1610 & 0.0738 & 0.2521 & 0.0924 \\
    \rowcolor{white}
    &\multirow{1}{*}{BPR}&DE & 0.1165 & 0.0645 & 0.1696 & 0.0778 & 0.2615 & 0.0960 \\
    \cmidrule{3-9}
    &&FTD & 0.1131 & 0.0630 & 0.1660 & 0.0763 & 0.2624 & 0.0953 \\
    &&HTD &  0.1247 &  0.0691 &  0.1820 &  0.0836 &  0.2803 &  0.1031 \\
    \cmidrule{3-9}
    \rowcolor{white}
    &&\textit{Gain.DE} & 7.0\% & 7.3\% & 7.3\% & 7.5\% & 7.2\% & 7.4\% \\
    \rowcolor{white}
    &&\textit{Gain.S} & 23.0\% & 23.5\% & 20.9\% & 22.3\% & 19.4\% & 19.3\% \\
   \cmidrule[.1em]{2-9}
   \rowcolor{white}
    & & Teacher & 0.1487 & 0.0844 & 0.2048 & 0.0986 & 0.2993 & 0.1155 \\
    \rowcolor{white}
    &&Student & 0.0856 & 0.0449 & 0.1249 & 0.0553 & 0.1970 & 0.0697 \\
    \rowcolor{white}
    &&FitNet & 0.0856 & 0.0469 & 0.1275 & 0.0576 & 0.2020 & 0.0723 \\
    \rowcolor{white}
    \multirow{1}{*}{CiteULike} &\multirow{1}{*}{NeuMF} &DE & 0.0882 & 0.0475 & 0.1306 & 0.0581 & 0.2090 & 0.0736 \\
    \cmidrule{3-9}
    &&FTD & 0.0875 & 0.0474 & 0.1291 & 0.0579 & 0.2069 & 0.0733 \\
    &&HTD &  0.0914 &  0.0504 &  0.1416 &  0.0618 &  0.2154 &  0.0772 \\
    \cmidrule{3-9}
    \rowcolor{white}
    &&\textit{Gain.DE} & 3.6\% & 6.2\% & 8.4\% & 6.4\% & 3.1\% & 4.8\% \\
    \rowcolor{white}
    &&\textit{Gain.S} & 6.8\% & 12.2\% & 13.4\% & 11.8\% & 9.0\% & 10.8\% \\
    \cmidrule[.1em]{2-9}
    \rowcolor{white}
    & & Teacher & 0.1610 & 0.0934 & 0.2274 & 0.1091 & 0.3326 & 0.1299 \\
    \rowcolor{white}
    &&Student & 0.1125 & 0.0618 & 0.1642 & 0.0748 & 0.2512 & 0.0944 \\
    \rowcolor{white}
    &&FitNet & 0.1151 & 0.0642 & 0.1710 & 0.0783 & 0.2653 & 0.0969 \\
    \rowcolor{white}
    &\multirow{1}{*}{LightGCN} &DE & 0.1189 & 0.0664 & 0.1733 & 0.0801 & 0.2680 & 0.0988 \\
    \cmidrule{3-9}
    &&FTD & 0.1112 & 0.0615 & 0.1635 & 0.0747 & 0.2542 & 0.0926 \\
    &&HTD &  0.1322 &  0.0742 &  0.1902 &  0.0888 &  0.2847 &  0.1075 \\
    \cmidrule{3-9}
    \rowcolor{white}
   & &\textit{Gain.DE} & 11.2\% & 11.8\% & 9.7\% & 10.9\% & 6.2\% & 8.8\% \\
   \rowcolor{white}
    &&\textit{Gain.S} & 17.5\% & 20.1\% & 15.8\% & 18.7\% & 13.3\% & 13.9\% \\    
    \bottomrule[.15em]
  \end{tabular}
  \end{minipage}
    \label{tbl:TD_maintable1}
\end{sidewaystable}

\begin{sidewaystable}[!thbp]
\centering
\caption{Performance comparison ($\phi=0.1$). \textit{Gain.DE} denotes the improvement of HTD over DE, \textit{Gain.S} denotes the improvement of HTD over Student.
HTD achieves statistically significant improvements over the best baseline. We use the paired t-test with significance level at 0.05 on Recall@50.}
\renewcommand{\arraystretch}{0.5}
\renewcommand{\tabcolsep}{2.7mm}
  \begin{minipage}[t]{1\linewidth}
  \centering
  \begin{tabular}{ccLgggggg}
    \toprule[.15em]
    \rowcolor{white}
    Dataset &Base Model & Method \quad & Recall@10 & NDCG@10 & Recall@20 & NDCG@20 & Recall@50 & NDCG@50 \\
    \midrule[.15em]
    \rowcolor{white}
    & & Teacher & 0.1187 & 0.0695 & 0.1700 & 0.0825 & 0.2732 & 0.1028 \\
    \rowcolor{white}
    &&Student & 0.0911 & 0.0544 & 0.1333 & 0.0648 & 0.2164 & 0.0809 \\
    \rowcolor{white}
    &&FitNet & 0.0957 & 0.0564 & 0.1386 & 0.0672 & 0.2258 & 0.0845 \\
    \rowcolor{white}
    &\multirow{1}{*}{BPR} &DE & 0.0979 & 0.0567 & 0.1434 & 0.0681 & 0.2322 & 0.0856 \\
   \cmidrule{3-9}
    &&FTD & 0.0987 & 0.0582 & 0.1417 & 0.0690 & 0.2262 & 0.0857 \\
    &&HTD &  0.1037 &  0.0622 &  0.1505 &  0.0740 &  0.2438 &  0.0921 \\
    \cmidrule{3-9}
    \rowcolor{white}
    &&\textit{Gain.DE} & 5.9\% & 9.7\% & 5.0\% & 8.7\% & 5.0\% & 7.6\% \\
    \rowcolor{white}
   & &\textit{Gain.S} & 13.8\% & 14.3\% & 12.9\% & 14.2\% & 12.7\% & 13.8\% \\
    \cmidrule[.1em]{2-9}
    \rowcolor{white}
    & & Teacher & 0.1060 & 0.0590 & 0.1546 & 0.0716 & 0.2529 & 0.0910 \\
    \rowcolor{white}
    &&Student & 0.0737 & 0.0393 & 0.1125 & 0.0490 & 0.1950 & 0.0653 \\
    \rowcolor{white}
    &&FitNet & 0.0829 & 0.0462 & 0.1243 & 0.0564 & 0.2062 & 0.0729 \\
    \rowcolor{white}
    \multirow{1}{*}{Foursquare} &\multirow{1}{*}{NeuMF} &DE & 0.0855 & 0.0476 & 0.1255 & 0.0576 & 0.2089 & 0.0741 \\
    \cmidrule{3-9}
    &&FTD & 0.0823 & 0.0451 & 0.1233 & 0.0554 & 0.2068 & 0.0719 \\
    &&HTD &  0.0891 &  0.0501 &  0.1294 &  0.0601 &  0.2152 &  0.0770 \\
    \cmidrule{3-9}
    \rowcolor{white}
   & &\textit{Gain.DE} & 4.3\% & 5.3\% & 3.1\% & 4.3\% & 3.0\% & 3.9\% \\
   \rowcolor{white}
   & &\textit{Gain.S} & 20.9\% & 27.2\% & 15.0\% & 22.7\% & 10.0\% & 17.9\% \\
    \cmidrule[.1em]{2-9}
    \rowcolor{white}
   &  & Teacher & 0.1259 & 0.0730 & 0.1779 & 0.0865 & 0.2806 & 0.1067 \\
   \rowcolor{white}
   & &Student & 0.0951 & 0.0564 & 0.1372 & 0.0670 & 0.2202 & 0.0834 \\
   \rowcolor{white}
   & &FitNet & 0.0993 & 0.0587 & 0.1431 & 0.0697 & 0.2315 & 0.0872 \\
   \rowcolor{white}
   & \multirow{1}{*}{LightGCN} &DE & 0.1051 & 0.0617 & 0.1503 & 0.0731 & 0.2410 & 0.0910 \\
    \cmidrule{3-9}
   & &FTD & 0.1018 & 0.0602 & 0.1466 & 0.0714 & 0.2327 & 0.0884 \\
   & &HTD &  0.1119 &  0.0652 &  0.1597 &  0.0772 &  0.2531 &  0.0956 \\
    \cmidrule{3-9}
    \rowcolor{white}
   & &\textit{Gain.DE} & 6.4\% & 5.6\% & 6.3\% & 5.6\% & 5.0\% & 5.1\% \\
   \rowcolor{white}
   & &\textit{Gain.S} & 17.7\% & 15.6\% & 16.4\% & 15.2\% & 14.9\% & 14.6\% \\
    \bottomrule[.15em]
  \end{tabular}
  \end{minipage}
    \label{tbl:TD_maintable2}
\end{sidewaystable}

\begin{table}[t]
\small
\centering
\caption{Performance comparison with ablations ($\phi=0.1$).}
\renewcommand{\arraystretch}{0.8}
\renewcommand{\tabcolsep}{0.3cm}
  \begin{minipage}[t]{1\linewidth}
  \centering
  \begin{tabular}{clcc cc}
    \toprule[.15em]
     \multirow{2}{*}{\makecell{Base\\Model}} & \multirow{2}{*}{Method}  &  \multicolumn{2}{c}{CiteULike}& \multicolumn{2}{c}{Foursquare}\\
    \cmidrule(lr){3-4}\cmidrule(lr){5-6}
     &  &Recall@50 & NDCG@50 & Recall@50 & NDCG@50 \\
    \midrule[.15em]
     &FTD &  0.2624 &  0.0953 &  0.2262 & 0.0857\\
    &FTD+DE  & 0.2650 & 0.0969  & 0.2339 & 0.0867 \\
    \cmidrule{2-6}
    \multirow{1}{*}{BPR}& HTD  &  0.2803 &  0.1031  &  0.2438 &  0.0921 \\
    &Group only  & 0.2619 & 0.0948  & 0.2349 & 0.0874 \\
    &Entity only  & 0.2608 & 0.0968  & 0.2361 & 0.0871 \\
    &Group (P,P) & 0.2648 & 0.0982  & 0.2399 & 0.0887 \\
    \cmidrule{1-6}
     &FTD & 0.2542 &  0.0926 &  0.2327 & 0.0884\\
    &FTD+DE & 0.2572 & 0.0931  & 0.2335 & 0.0872 \\
    \cmidrule{2-6}
    \multirow{1}{*}{LightGCN}& HTD  &  0.2847 &  0.1075 &  0.2531 &  0.0956 \\
    &Group only & 0.2596  & 0.0976   & 0.2432  & 0.0910 \\
    &Entity only  & 0.2709  & 0.1010   & 0.2453  & 0.0910 \\
    &Group (P,P)  & 0.2683  & 0.0987    & 0.2459  & 0.0909 \\
    \bottomrule[.15em]
  \end{tabular}
  \end{minipage}
    \label{tbl:TD_ablation}
\end{table}

\vspace{0.1cm}
\noindent
\textbf{Overall Evaluation.} 
In Table \ref{tbl:TD_maintable1} and Table \ref{tbl:TD_maintable2}, we observe that HTD achieves significant performance gains compared to the main competitor, i.e., DE.
This result shows that distilling the relational knowledge provides better guidance than only distilling the knowledge of individual representation.
We also observe that FTD is not always effective and sometimes even degrades the student’s recommendation performance (e.g., LightGCN on CiteULike).
As the student's size is highly limited compared to the teacher in the KD scenario, learning all the relational knowledge is daunting for the student, which leads to degrade the effects of distillation.
This result supports our claim that the relational knowledge should be distilled considering the huge capacity gap.
Also, the results show that HTD successfully copes with the issue, enabling the student to effectively learn the relational knowledge.

In Figure \ref{fig:TD_sizes}, we observe that as the model size increases, the performance gap between FTD and HTD decreases. 
FTD achieves comparable and even higher performance than HTD when the student has enough capacity (e.g., LightGCN on Foursquare with $\phi=1.0$).
This result again verifies that the effectiveness of the proposed topology distillation approach.
It also shows that FTD can be applied to maximize the performance of recommender in the scenario where there is no constraint on the model size by self-distillation.

\vspace{0.1cm}
\noindent
\textbf{Comparison with ablations.} 
In Table \ref{tbl:TD_ablation}, we provide the comparison with diverse ablations.
For FTD, we report the results when it is used with the state-of-the-art hint regression method (denoted as \textbf{FTD+DE}).
As HTD includes DE for the group assignment, comparison with FTD+DE shows the direct impacts of topology relaxation.
We observe that FTD+DE is not as effective as HTD and sometimes even achieves worse performance than~DE.
This shows that the power of HTD comes from the distillation strategy transferring the relational knowledge in multi-levels.

For HTD, we compare three ablations: \textbf{1) Group only} that considers only group-level topology, \textbf{2) Entity only} that considers only entity-level topology, and \textbf{3) Group (P,P)} that considers the relations of the prototypes only for the group-level topology.
We first observe that both group-level and entity-level topology are indeed necessary.
Without either of them, the performance considerably drops.
The group-level topology includes the summarized relations across the groups, providing the overview of the entire topology.
On the other hand, the entity-level topology includes the full relations in each group, providing fine-grained supervision of how the entities should be correlated.
Based on both two-level topology, HTD effectively transfers the relational knowledge.
Lastly, we observe that \text{Group (P,P)} is not as effective as \text{Group(P,e)} adopted in HTD.
We conjecture that summarizing numerous relations across the groups into a single relation may lose too much information and cannot effectively boost the student.

\vspace{0.1cm}
\noindent
\textbf{With prediction-based KD method.} 
We report the results with the state-of-the-art prediction KD method (i.e., RRD \cite{DERRD}) on CiteULike with $\phi=0.1$ in Figure \ref{fig:TD_RRD}.
Also, we provide the training curves of BPR with $\phi=0.1$ in Figure \ref{fig:TD_curve}\footnote{After the early stopping on the validation set, we plot the final performance.}.
First, we observe that the effectiveness of RRD is considerably improved when it is applied with the KD method distilling the latent knowledge (i.e., DE and HTD).
This result aligns with the results reported in \cite{DERRD} and shows the importance of distilling the latent knowledge.
Second, we observe that the student recommender achieves the best performance with HTD.
Unlike RRD, which makes the student imitate the ranking order of items in each user's recommendation list, HTD distills relations existing in the representation space via the topology matching.
The topology includes much rich supervision including not only user-item relations but also \textit{user-user} relations and \textit{item-item} relations.
By obtaining this additional information in a proper manner considering the capacity gap, the student can be fully improved with HTD.

\subsection{Benefit of Topology Distillation}
\label{sec:TD_benefit}
To further ascertain the benefit of the topology distillation, we provide in-depth analysis on representations obtained by each KD~method ($\phi=0.1$).

\begin{figure}[t]
\centering
\begin{subfigure}[t]{0.4\linewidth}
    \includegraphics[width=\linewidth]{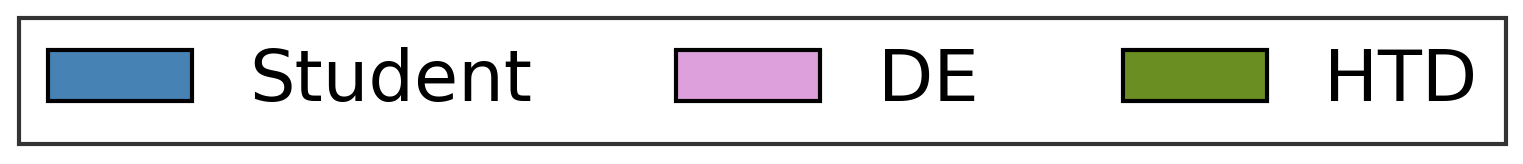}
\end{subfigure}\\
\hspace{-0.2cm}
\begin{subfigure}[t]{0.45\linewidth}
    \includegraphics[width=\linewidth]{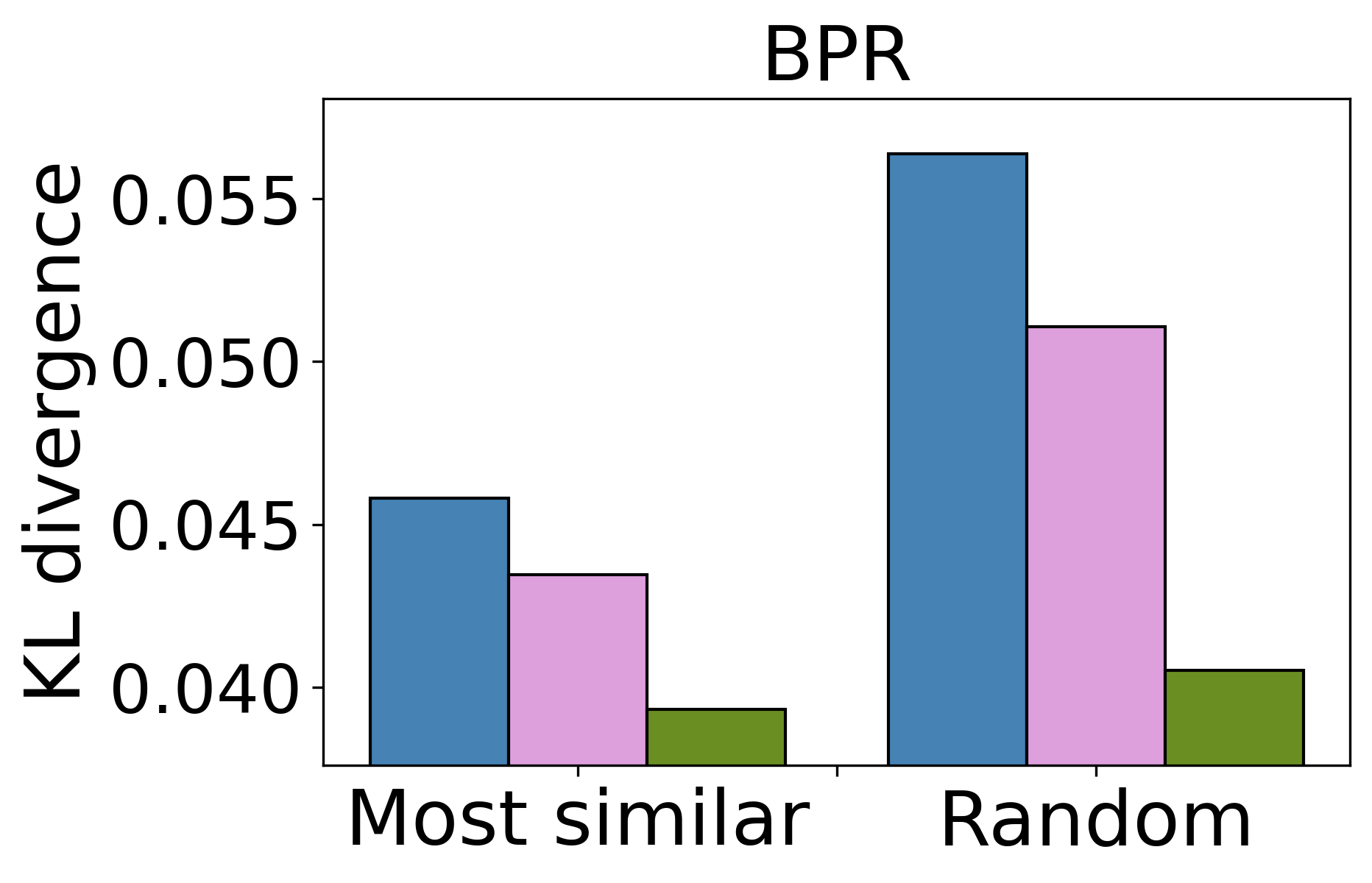}
\end{subfigure}
\hspace{-0.2cm}
\begin{subfigure}[t]{0.45\linewidth}
    \includegraphics[width=\linewidth]{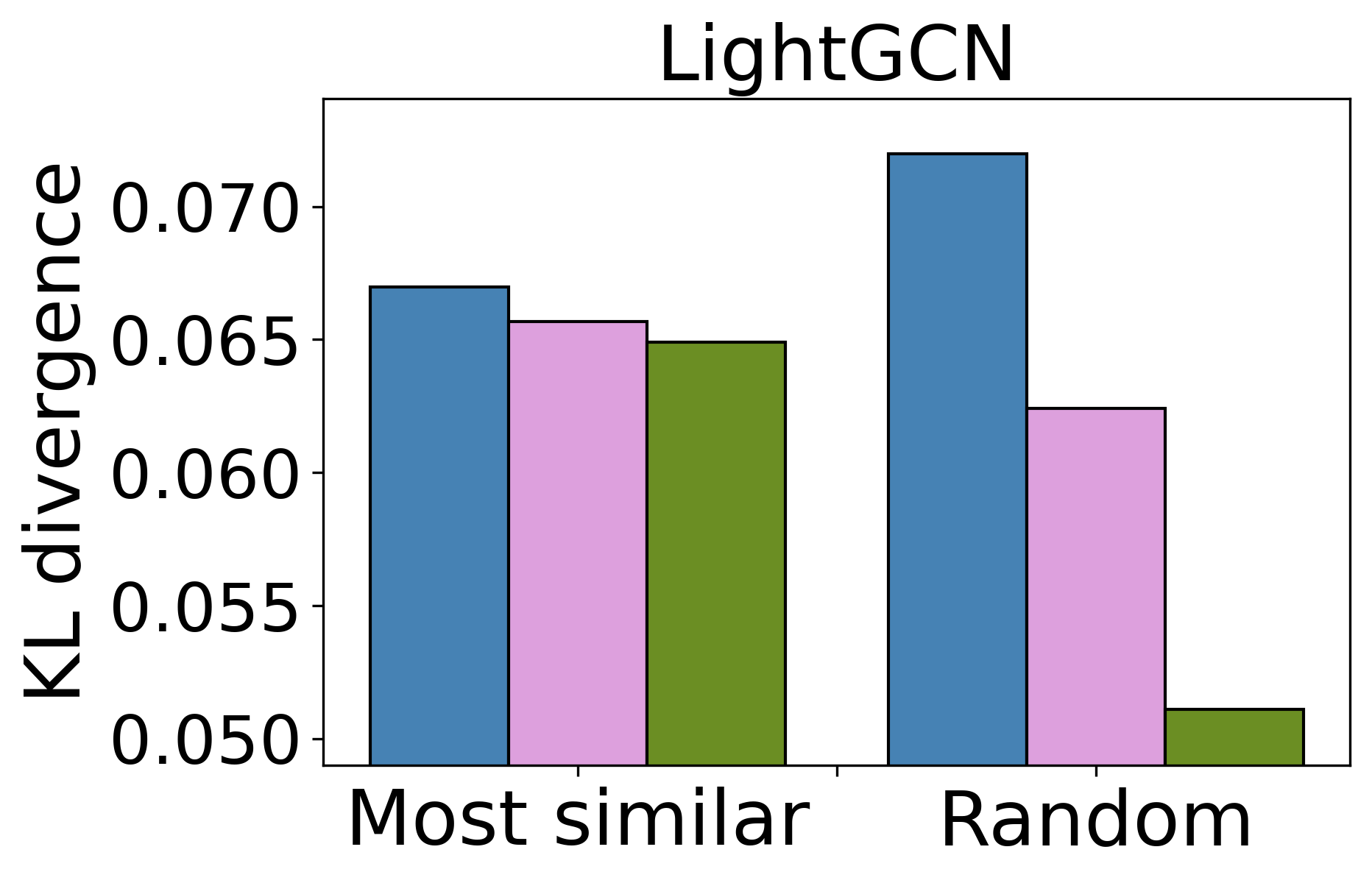}
\end{subfigure}
\hspace{-0.2cm}
\caption{The average KL divergence from similarity distributions obtained in the teacher representation~spaces.}
\label{fig:TD_kld}
\end{figure}

First, we evaluate whether the topology distillation indeed makes the student better preserve the relations in the teacher representation space than the existing method.
For quantitative evaluation, we conduct the following steps:
\textbf{(1)} In the teacher space, for each representation, we compute the similarity distributions with 100 \textit{most similar} representations and 100 \textit{randomly selected} representations, respectively.
\textbf{(2)} In the student space, for each representation, we compute the similarity distributions with the representations chosen in the teacher space.
\textbf{(3)} We compute KL divergence for each distribution and report the average value in Figure \ref{fig:TD_kld}.
KL divergence of `Most similar' indicates how well the detailed relations among the strongly correlated representations are preserved, and that of `Random' indicates how well the overall relations in the space are preserved.
We observe that HTD achieves the lowest KL divergence for both Most similar and Random, which shows that HTD indeed enables the student to better preserve the relations in the teacher~space.

Second, we compare the performance of two downstream tasks that evaluate how well each method encodes the items’ characteristics (or semantics) into the representations.
We perform the tag retrieval task for CiteULike and the region classification task for Foursquare.
We train a linear and a non-linear model to predict the tag/region of each item by using the fixed item representation as the input.
The detailed setup is provided in the appendix. 
In Table \ref{tbl:TD_downstream_tasks}, we observe that HTD achieves consistently higher performance than DE on both of two downstream tasks.
This strongly indicates that the representation space induced by HTD more accurately captures the item’s semantics compared to the space induced by~DE.

In sum, with the topology distillation approach, the student can indeed better preserve the relations in the teacher space.
This not only improves the recommendation performance but also allows it to better capture the semantic of entities.

\begin{table}[t]
\centering
\small
\caption{Performance comparison on downstream tasks.}
\renewcommand{\arraystretch}{0.8}
\renewcommand{\tabcolsep}{1.0mm}
  \centering
  \begin{tabular}{cclcc}
    \toprule[.15em]
    \makecell{Base\\Model} & Type & Method & \makecell{Tag Retrieval \\(Recall@10)}& \makecell{Region Classification \\(Accuracy)} \\
    \midrule[.15em]
    \multirow{8}{*}{\small BPR}&\multirow{4}{*}{linear} & Teacher & 0.3121$\pm$2.7e-3& 0.6531$\pm$2.5e-3 \\
    &&Student & 0.2421$\pm$1.7e-3& 0.4222$\pm$7.4e-3\\
    &&DE & 0.2542$\pm$1.5e-3& 0.5448$\pm$3.2e-3 \\
    &&HTD & \textbf{0.2635}$\pm$1.2e-3& \textbf{0.5653}$\pm$5.0e-3  \\
    \cmidrule{2-5}
    &\multirow{4}{*}{\makecell{non-linear}} & Teacher & 0.3123$\pm$1.4e-3& 0.6357$\pm$8.8e-3  \\
    &&Student & 0.2701$\pm$2.1e-3&0.4224$\pm$7.9e-3  \\
    &&DE & 0.2807$\pm$2.5e-3& 0.5123$\pm$6.6e-3\\
    &&HTD & \textbf{0.2909}$\pm$2.2e-3&\textbf{0.5318}$\pm$3.5e-3  \\
    \midrule
    \multirow{8}{*}{\small LightGCN}&\multirow{4}{*}{linear} & Teacher & 0.3489$\pm$0.6e-3& 0.6787$\pm$6.1e-3 \\
    &&Student & 0.2523$\pm$0.7e-3& 0.4635$\pm$3.1e-3\\
    &&DE & 0.2565$\pm$1.2e-3& 0.5654$\pm$6.5e-3 \\
    &&HTD & \textbf{0.2650}$\pm$0.3e-3&\textbf{0.5854}$\pm$7.7e-3  \\
    \cmidrule{2-5}
    &\multirow{4}{*}{non-linear} & Teacher & 0.3360$\pm$1.2e-3& 0.6504$\pm$5.9e-3\\
    &&Student & 0.2971$\pm$0.3e-3& 0.4354$\pm$6.5e-3 \\
    &&DE & 0.3053$\pm$1.3e-3& 0.5250$\pm$3.4e-3 \\
    &&HTD & \textbf{0.3138}$\pm$1.1e-3& \textbf{0.5485}$\pm$8.1e-3\\
    \bottomrule[.15em]
  \end{tabular}
    \label{tbl:TD_downstream_tasks}
\end{table}
\begin{figure}[t]
\centering
\hspace{-0.3cm}
\begin{subfigure}[t]{0.245\linewidth}
    \includegraphics[height=3cm]{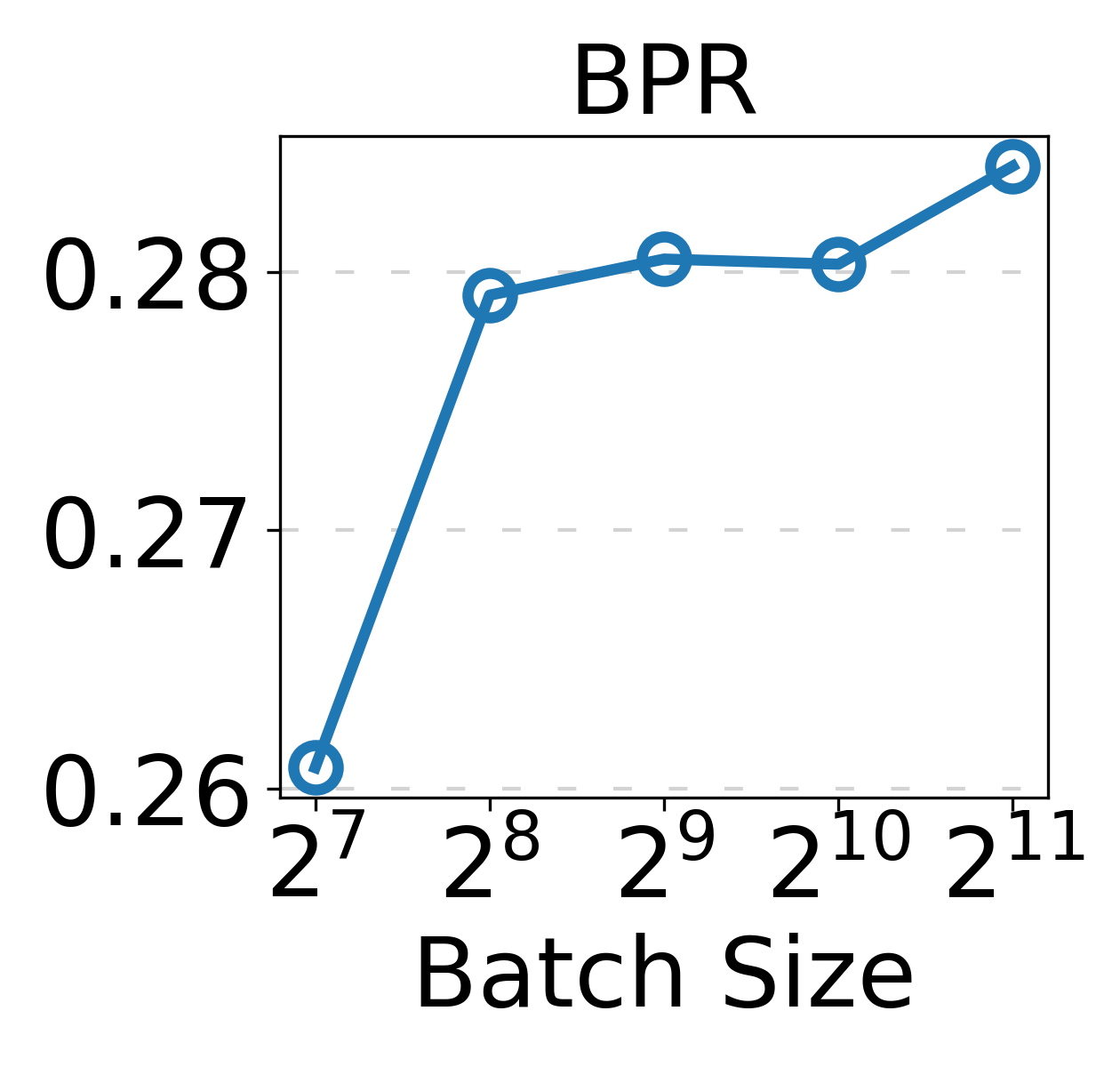}
\end{subfigure}
\hspace{-0.22cm}
\begin{subfigure}[t]{0.245\linewidth}
    \includegraphics[height=3cm]{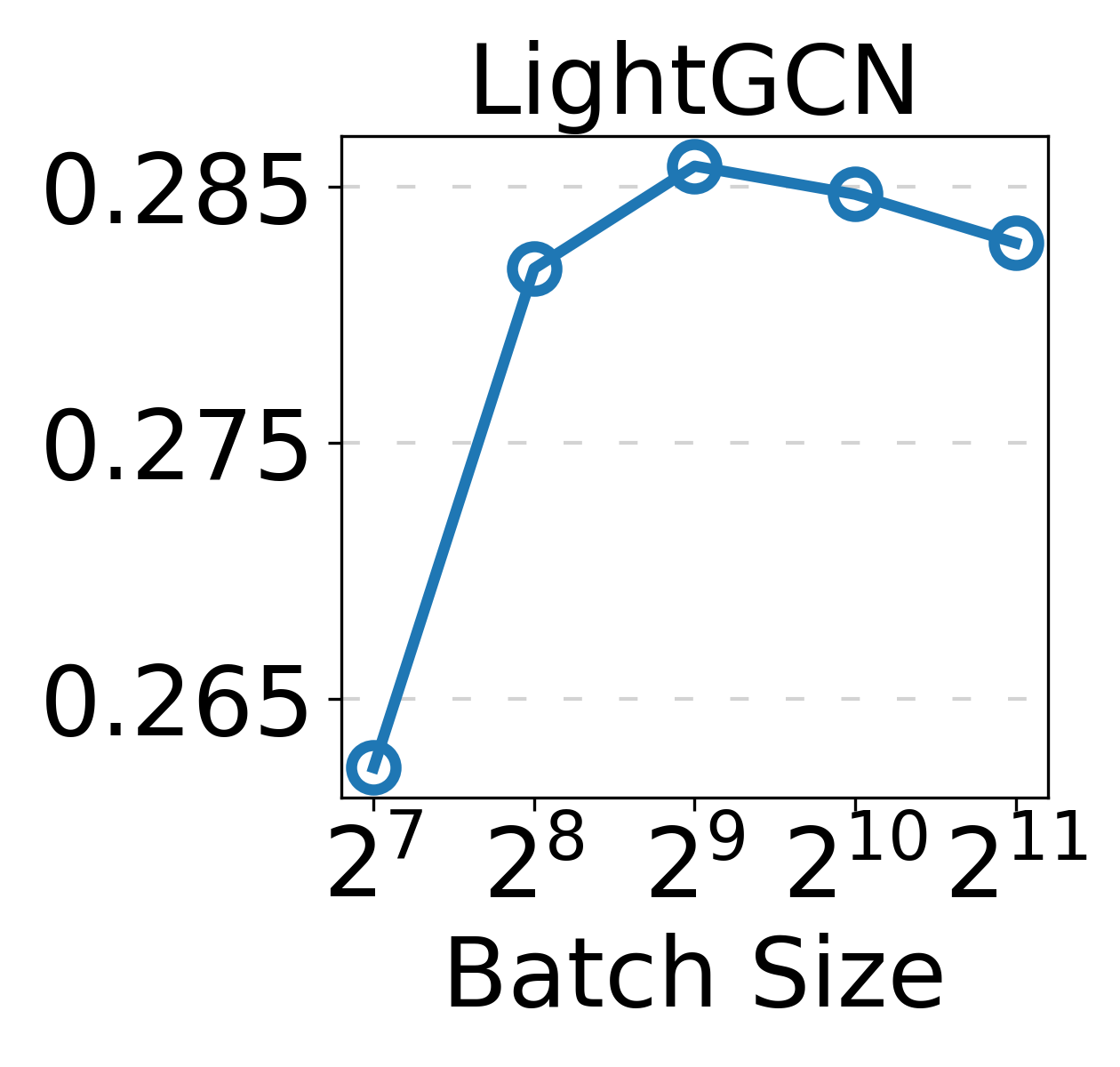}
\end{subfigure}
\hspace{-0.22cm}
\begin{subfigure}[t]{0.245\linewidth}
    \includegraphics[height=3cm]{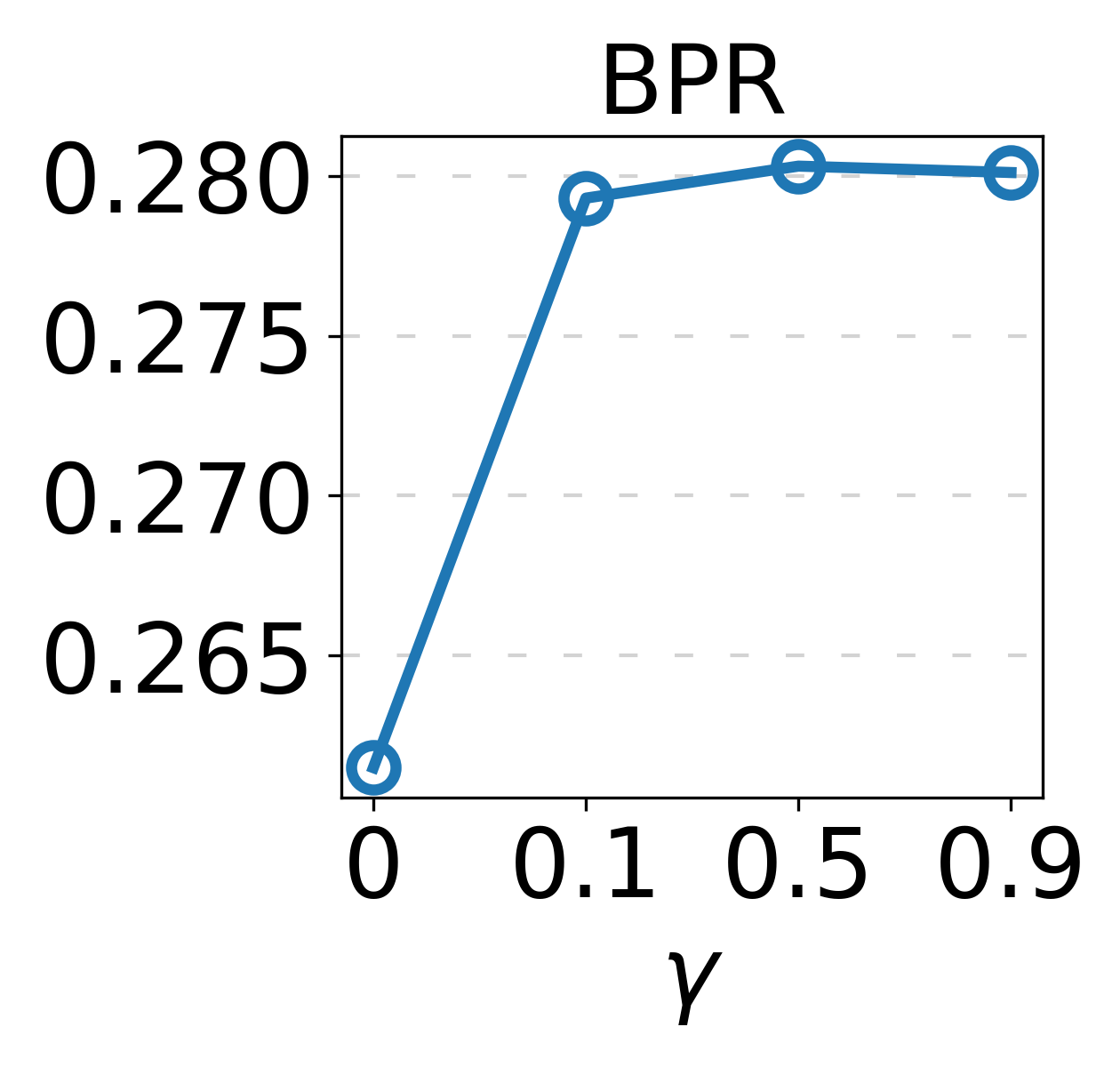}
\end{subfigure}
\hspace{-0.22cm}
\begin{subfigure}[t]{0.245\linewidth}
    \includegraphics[height=3cm]{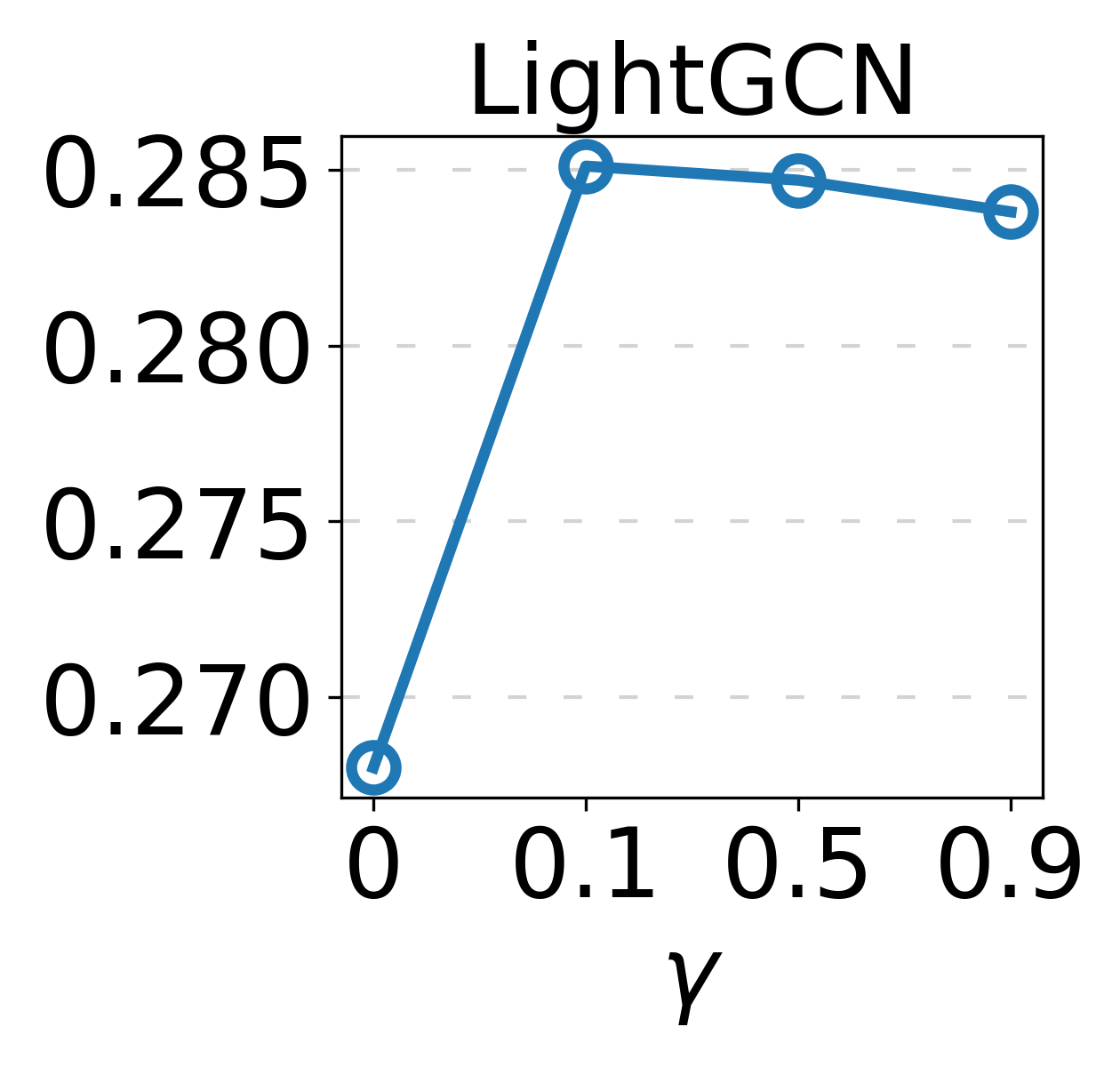}
\end{subfigure}
\hspace{-0.3cm}
\caption*{(a) Effects of batch size and $\gamma$}
\hspace{1cm}
\begin{subfigure}[t]{0.4\linewidth}
    \includegraphics[height=4cm]{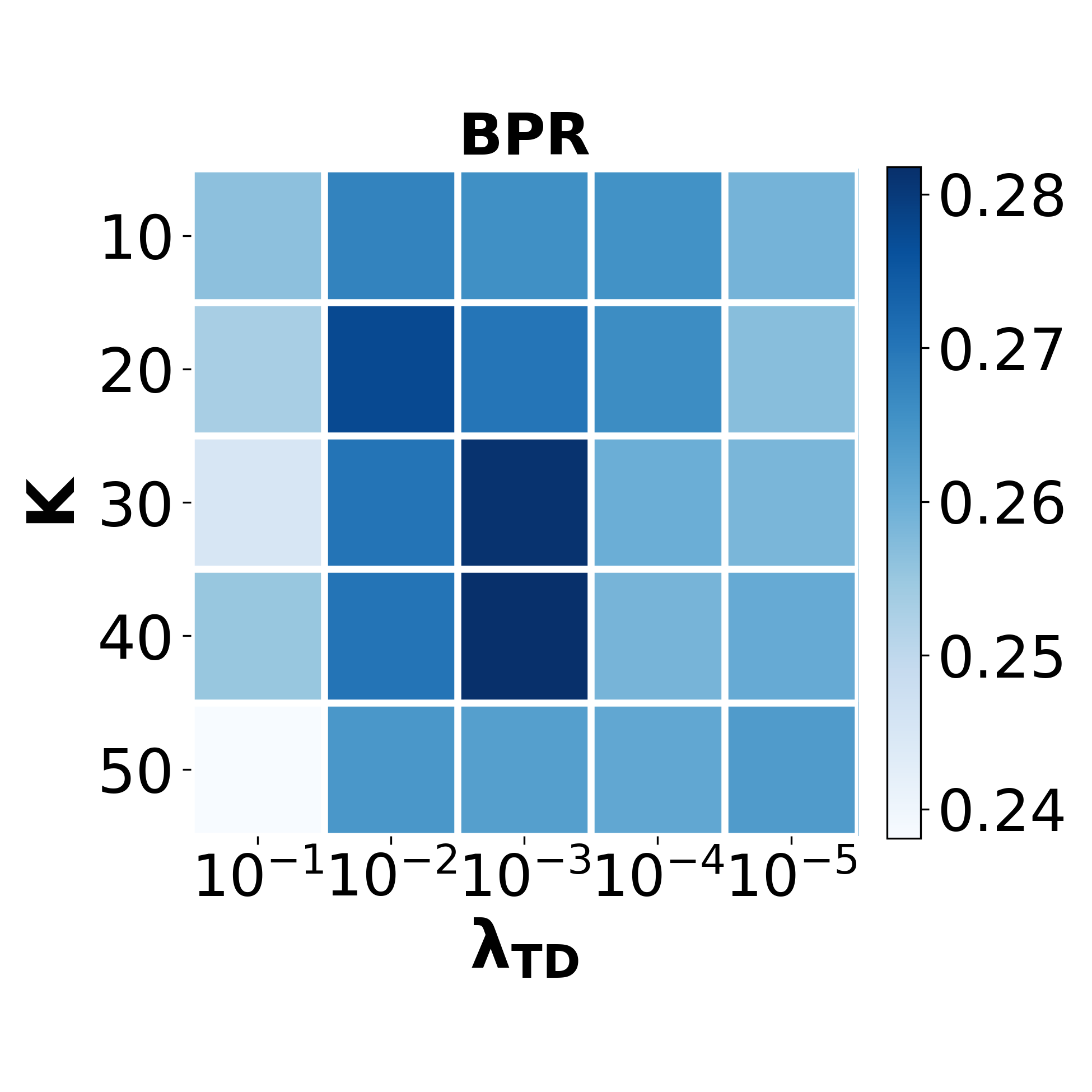}
\end{subfigure}
\begin{subfigure}[t]{0.4\linewidth}
    \includegraphics[height=4cm]{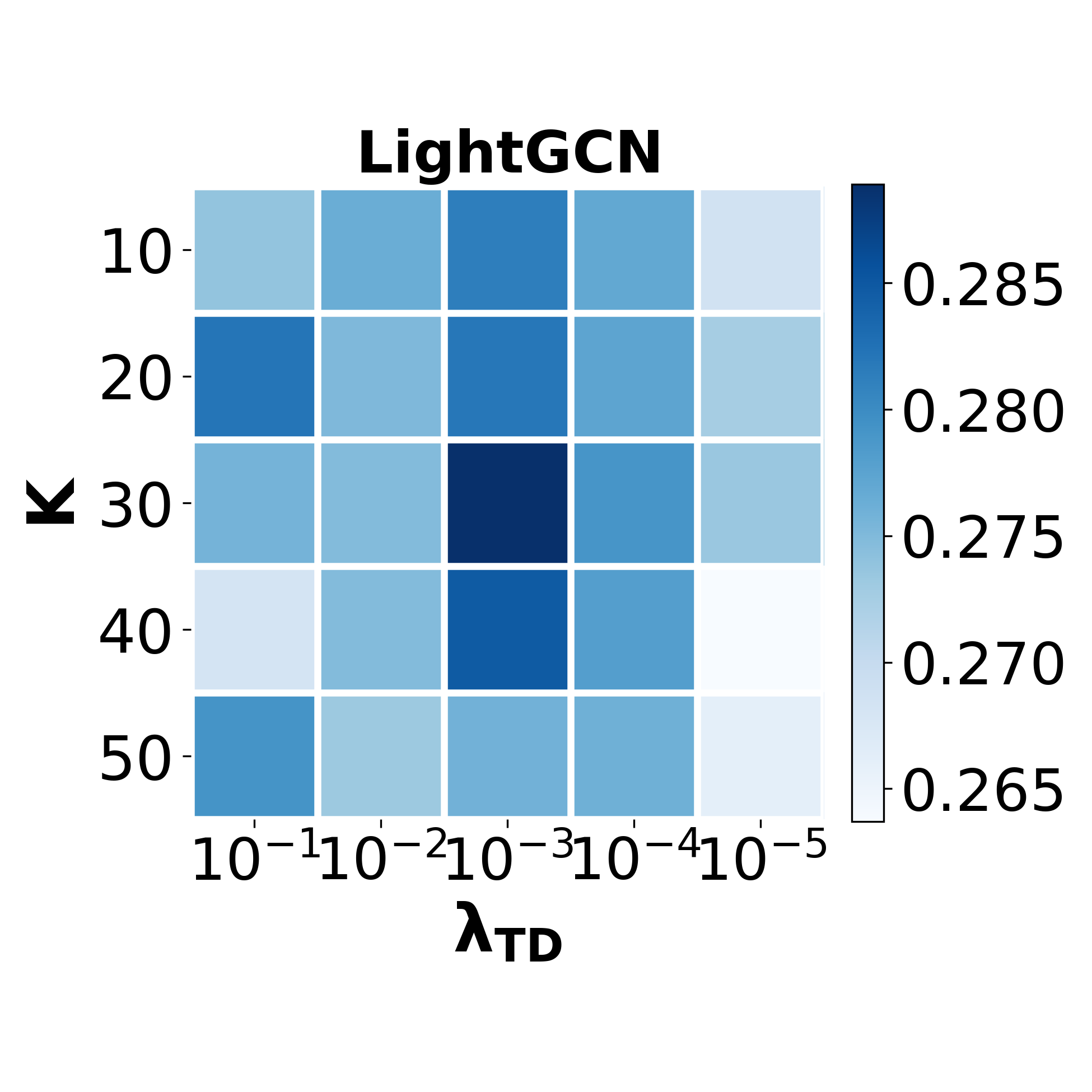}
\end{subfigure}
\caption*{(b) Effects of $\lambda_{TD}$ and $K$}
\caption{Effects of the hyperparameters (Recall@50).}
\label{fig:TD_hp}
\end{figure}

\subsection{Hyperparameter Analysis}
We provide analyses to guide the hyperparameter selection of the topology distillation approach.
For the sake of the space, we report the results of BPR and LightGCN on CiteULike dataset with $\phi=0.1$ (Figure \ref{fig:TD_hp}).
\textbf{(1)} The batch size is an important factor affecting the performance of the topology distillation.
When the batch size is too small, the topology cannot include the overall relational knowledge in the representation space, leading to limited performance.
For CiteULike, we observe that HTD achieves the stable performance around $2^{8}$-$2^{11}$.
In this work, we set the batch size to $2^{10}$.
\textbf{(2)} $\gamma$ is a hyperparameter for balancing the topology-preserving loss and the regression loss.
We observe the stable performance with a value larger than $0.1$.
In this work, we set $\gamma$ to $0.5$.
Note that $\gamma=0$ equals DE.
\textbf{(3)} The number of preference groups ($K$) is an important hyperparameter of HTD. 
It needs to be determined considering the dataset, the capacity gap, and the selected layer for the distillation. 
For this setup, the best performance is achieved when $K$ is around 30-40 in both base models.
\textbf{(4)} $\lambda_{TD}$ is a hyperparameter for controlling the effects of topology distillation.
For this setup, the best performance is achieved when $\lambda_{TD}$ is around $10^{-3}$ in both base models.

\section{Summary}
\label{sec:TD_conclusion}
We develop a general topology distillation approach for RS, which guides the learning of the student by the topological structure built upon the relational knowledge in the teacher representation space.
Concretely, we propose two topology distillation methods: 
(1) FTD that transfers the full topology. 
FTD is used in the scenario where the student has enough capacity to learn all the teacher's knowledge.
(2) HTD that transfers the decomposed topology hierarchically.
HTD is adopted in the conventional KD scenario where the student has a very limited capacity compared to the teacher.
We conduct extensive experiments on real-world datasets and show that the proposed approach consistently outperforms the state-of-the-art competitor. 
We also provide in-depth analyses to ascertain the benefit of distilling the topology.

We believe the topology distillation approach can be advanced and extended in several directions.
First, layer selection and simultaneous distillation from multiple layers are not investigated in this work.
We especially expect that this can further improve the limited improvements by the topology distillation in the deep model (i.e., NeuMF).
Second, topology distillation across different base models (e.g., from LightGCN to BPR) can be also considered to further improve the performance.
Lastly, prior knowledge of user/item groups (e.g., item category, user demographic features) can be utilized for more sophisticated topology decomposition.
We expect that this can further improve the effectiveness of the proposed method by providing better supervision on relational knowledge.

\section{Supplementary Material}

\subsection{Pseudocode of the proposed methods}
The pseudocode of FTD and HTD are provided in Algorithm \ref{algo:TD_FTD} and Algorithm \ref{algo:TD_HTD}, respectively.
The base model can be any existing recommender and $\mathcal{L}_{Base}$ is its loss function.
Note that the distillation is conducted in the offline training phase.
At the online inference phase, the student model is used only.

All the computations of the topology distillation are efficiently computed on matrix form by parallel execution through GPU processor.
In Algorithm \ref{algo:TD_FTD} (FTD), the topological structures (line 5) are computed as follows: 
$$\mathbf{A}^{t} = \text{Cos}(\mathbf{E}^{t}, \mathbf{E}^{t})\;\;\text{and}\;\;\mathbf{A}^{s} = \text{Cos}(\mathbf{E}^{s}, \mathbf{E}^{s}),$$
where $\text{Cos}$ is the operation computing the cosine similarities by $\text{Cos}(\mathbf{B}, \mathbf{D}) = \hat{\mathbf{B}}\hat{\mathbf{D}}^{\top}$, $\hat{\mathbf{B}}_{[i,:]} = \mathbf{B}_{[i,:]} / \lVert \mathbf{B}_{[i,:]} \rVert_2$.
In Algorithm \ref{algo:TD_HTD} (HTD with Group(P,e)), the two-level topological structures (line 5-8) are computed as follows:
\begin{equation*}
\begin{aligned}
\mathbf{Z} = \text{Gumbel}&\text{-Softmax}(v(\mathbf{E}^t))\\
\mathbf{P}^{t} = \Tilde{\mathbf{Z}}^{\top}\mathbf{E}^{t} \;\;&\text{and}\;\; \mathbf{P}^{s} = \Tilde{\mathbf{Z}}^{\top}\mathbf{E}^{s}\\
\mathbf{H}^{t} = \text{Cos}(\mathbf{P}^{t}, \mathbf{E}^{t}) \;\;&\text{and}\;\; \mathbf{H}^{s} = \text{Cos}(\mathbf{P}^{s}, \mathbf{E}^{s})\\
\mathbf{A}^{t} = \text{Cos}(\mathbf{E}^{t}, \mathbf{E}^{t}) \;\;&\text{and}\;\; \mathbf{A}^{s} = \text{Cos}(\mathbf{E}^{s}, \mathbf{E}^{s})\\
\mathbf{M} &= \mathbf{Z} \mathbf{Z}^{\top}\\
\end{aligned}
\end{equation*}
The power of topology distillation comes from distilling the additional supervision in a proper manner considering the capacity gap.

\begin{algorithm}[h]
\SetKwInOut{Input}{Input}
\SetKwInOut{Output}{Output}
\Input{Training data $\mathcal{D}$, Trained Teacher model}
\Output{Student model}
Initialize Student model\\
\While {not convergence}{
\For{each batch $\mathcal{B} \in \mathcal{D}$}{
Compute $\mathcal{L}_{Base}$ \\
Build full topology $\mathbf{A}^t$ and $\mathbf{A}^s$\\
Compute $\mathcal{L}_{FTD}$ \\
Compute $\mathcal{L} = \mathcal{L}_{Base} + \lambda \mathcal{L}_{FTD}$ \\
Update Student model by minimizing $\mathcal{L}$\\
}
}
\caption{Full Topology Distillation.}
\label{algo:TD_FTD}
\end{algorithm}

\begin{algorithm}[h]
\SetKwInOut{Input}{Input}
\SetKwInOut{Output}{Output}
\Input{Training data $\mathcal{D}$, Trained Teacher model}
\Output{Student model}
Initialize Student model, $v$, and $f_*$\\
\While {not convergence}{
\For{each batch $\mathcal{B} \in \mathcal{D}$}{
Compute $\mathcal{L}_{Base}$ \\
Assign preference groups $\mathbf{Z}$ for $\mathcal{B}$\\
Compute prototypes $\mathbf{P}^t$ and $\mathbf{P}^s$ \\
Build group-level topology $\mathbf{H}^t$ and $\mathbf{H}^s$\\
Build entity-level topology by $\mathbf{A}^t$, $\mathbf{A}^s$, and $\mathbf{M}$\\
Compute $\mathcal{L}_{HTD}$ \\
Compute $\mathcal{L} = \mathcal{L}_{Base} + \lambda \mathcal{L}_{HTD}$ \\
Update Student model, $v$, and $f_*$ by minimizing $\mathcal{L}$
}
}
\caption{Hierarchical Topology Distillation.}
\label{algo:TD_HTD}
\end{algorithm}

\subsection{Group Assignment.}
Instead of DE, any clustering method or prior knowledge of user/item groups can be utilized for more sophisticated topology decomposition.
The simplest method is $K$-means clustering. 
Specifically, we first conduct the clustering in the teacher space, then use the results for the group assignment.
The results are summarized in Table \ref{tab:TD_fixed}.
For both DE and $K$-means, the number of preference group ($K$) is set to 30, and the teacher space dimensions ($d^t$)~is~200.

We get consistently better results with the adaptive assignment by DE. 
We conjecture the possible reasons as follows:
(1) Accurate clustering in high-dimensional space is very challenging,
(2) With the adaptive approach, the assignment process gets gradually sophisticated along with the student, and thereby it provides guidance considering the student's learning.
For these reasons, we use the adaptive assignment in HTD.
The performance of HTD can be further improved by adopting a more advanced assignment method and prior knowledge.
We leave exploring better ways of the assignment for future study.

\begin{table}[h]
\centering
\caption{Performance comparison with different group assignment methods.}
\renewcommand{\tabcolsep}{1.8mm}
  \begin{minipage}[t]{1\linewidth}
  \centering
  \begin{tabular}{l cc cc}
    \toprule[.15em]
        &  \multicolumn{2}{c}{CiteULike}& \multicolumn{2}{c}{Foursquare}\\
    \cmidrule(lr){2-3}\cmidrule(lr){4-5}
     &Recall@50 & NDCG@50 & Recall@50 & NDCG@50 \\
    \midrule[.15em]
    $K$-means &0.2661&0.0980&0.2346&0.0871\\
    DE & 0.2803& 0.1031& 0.2438& 0.0921\\
    \bottomrule[.15em]
  \end{tabular}
  \end{minipage}
  \label{tab:TD_fixed}
\end{table}

\subsection{Datasets.}
We use two public real-world datasets: CiteULike and Foursquare.
We remove users having fewer than 5 (CiteULike) and 20 interactions (FourSquare) and remove items having fewer than 10 interactions (FourSquare) as done in \cite{DERRD, BUIR}.
Table \ref{tbl:TD_statistic} summarizes the statistics of the datasets.
In the case of CiteULike, each item corresponds to an article, and each article has multiple tags.
In the case of Foursquare, each item corresponds to a POI (points-of-interest) such as museums and restaurants, and each POI has GPS coordinates (i.e., the latitude and longitude).
We use this side information in Section \ref{sec:TD_benefit}.
Table \ref{tbl:TD_url} shows the URLs from which the datasets can be downloaded.

\begin{table}[!h]
\centering
\renewcommand{\tabcolsep}{2.6mm}
  \caption{Statistics of the datasets.}
  \begin{tabular}{ccccc}
    \toprule[.15em]
    Dataset & \#Users & \#Items & \#Interactions & Density \\
    \midrule[.15em]
    CiteULike & 5,220 & 25,182 & 115,142 & 0.09\% \\
    Foursquare & 19,466 & 28,594 & 609,655 & 0.11\% \\
    \bottomrule[.15em]
  \end{tabular}
    \label{tbl:TD_statistic}
    \vspace{-0.4cm}
\end{table}

\begin{table}[h]
\centering
  \caption{URL links to the datasets.}
  \begin{tabular}{c l}
    \toprule[.15em]
    Dataset & URL link to the dataset \\
    \midrule[.15em]
    CiteULike & \url{https://github.com/changun/CollMetric} \\
    Foursquare & \url{http://spatialkeyword.sce.ntu.edu.sg/eval-vldb17/} \\
    \bottomrule[.15em]
  \end{tabular}
    \label{tbl:TD_url}
\end{table}

\subsection{Evaluation Protocol and Metrics}
We adopt the widely used \textit{leave-one-out} evaluation protocol, whereby two interacted items for each user are held out
for testing/validation, and the rest are used for training.
However, unlike \cite{DERRD} that samples a predefined number (e.g., 499) of unobserved items for evaluation, we adopt the full-ranking evaluation scheme
that evaluates how well each method can rank the test item higher than all the unobserved items.
Although it is time-consuming, it enables a more thorough evaluation compared to the sampling-based evaluation \cite{CD, krichene2020sampled}.
We evaluate all methods by two widely used ranking metrics: 
Recall@$N$ \cite{recall} and Normalized Discounted Cumulative Gain (NDCG@$N$) \cite{jarvelin2002cumulated}.
Recall@$N$ measures whether the test item is included in the top-$N$ list and NDCG@$N$ assigns higher scores on the upper ranked test items.
We compute the metrics for each user, then compute the average score.
Lastly, we report the average value of five independent runs for all methods.

\subsection{Implementation Details.}
We use PyTorch to implement the proposed methods and all the competing methods.
We optimize all methods with Adam optimizer.
For DE and RRD, we use the public implementation provided by the authors.
For each setting, hyperparameters are tuned by using grid searches on the validation set.
The learning rate is searched in the range of \{0.01, 0.005, 0.001, 0.0005, 0.0001\}.
The model regularizer is searched in the range of $\{10^{-2}, 10^{-3}, 10^{-4}, 10^{-5}, 10^{-6}\}$.
We set the total number of epochs to 500 and adopt the early stopping strategy; it terminates when Recall@50 on the validation set does not increase for 20 successive epochs.

For all base models, the number of negative samples is set to 1.
For NeuMF and LightGCN, the number of layers is searched in the range of \{1, 2, 3, 4\}.
For all the distillation methods, weight for the distillation loss ($\lambda$) searched in the range of $\{1, 10^{-1}, 10^{-2}, 10^{-3}, 10^{-4}, 10^{-5}\}$.
For the hint regression-related setup, we closely follow the setup reported in DE paper \cite{DERRD}.
Specifically, two-layer MLP with [$d^s \rightarrow (d^s + d^t)/2 \rightarrow d^t$] is employed for $f$ in FitNet, DE and HTD.
Also, one-layer perceptron with [$d^t \rightarrow K $] is employed for assigning group ($v$) in DE and HTD.
For DE and HTD, the number of preference groups ($K$) is chosen from \{5, 10, 20, 30, 40, 50\}.

\subsection{Experiment Setup for Downstream Tasks.}
We evaluate how well each method encodes the items’ characteristics (or semantics) into the representations.
We train a small network to predict the side information of items by using the \textit{fixed} item representations as the input.
Specifically, we use a linear and a non-linear model (i.e., a single-layer perceptron and three-layer perceptron, respectively) with Softmax output.
The linear model has the shape of [$d^s \rightarrow C$], and the non-linear model has the shape of [$d^s \rightarrow (d^s+C)/2 \rightarrow (d^s+C)/2 \rightarrow C$] with relu, where $C$ is the number of tags/classes.
Let $\mathbf{q}$ denote the output of the model whose element $q_i$ is a prediction score for each tag/class.
Also, let $\mathbf{p}$ denote the ground-truth vector whose element $p_i=1$ if $i$-th tag/class is the answer, otherwise $p_i=0$. 
We train the model by minimizing the negative log-likelihood: $-\sum_{i} p_i \log q_i$.
Note that the side-information is not utilized for training of the base model.

For CiteULike dataset, we perform \textbf{item-tag retrieval task};
by using each item representation as a query, we find a ranking list of tags that are relevant to the item.
We first remove tags used less than 10 times.
Then, there exist 4,153 tags and an item has 6.4 tags on average.
After training, we make a ranking list of tags by sorting the prediction scores.
We evaluate how many the ground-truth tags are included in the top-$10$ list by Recall$@10$.
For Foursquare dataset, we perform \textbf{item-region classification task};
given each item representation, we predict the region class to which the item belongs.
We first perform $k$-means clustering on the coordinates with $k=200$ and use the clustering results as the class labels. 
After training, we evaluate the performance by Accuracy.
Finally, we perform 5-fold cross-validation and report the average result and standard deviation in Table \ref{tbl:TD_downstream_tasks}.

\label{sec:TD_sup}

\chapter{Item-side Ranking Regularization for Ranking Distillation}
\label{chapt:IR-RRD}
Recent recommender system (RS) have adopted large and sophisticated model architecture to better understand the complex user-item relationships, and accordingly, the size of the recommender is continuously increasing.
To reduce the high inference costs of the large recommender, knowledge distillation (KD), which is a model compression technique from a large pre-trained model (teacher) to a small model (student), has been actively studied for RS.
The state-of-the-art method is based on the ranking distillation approach, which makes the student preserve the ranking orders among items predicted by the teacher.
In this work, we propose a new regularization method designed to maximize the effect of the ranking distillation in RS.
We first point out an important limitation and a room for improvement of the state-of-the-art ranking distillation method based on our in-depth analysis.
Then, we introduce the item-side ranking regularization, which can effectively prevent the student with limited capacity from being overfitted and enables the student to more accurately learn the teacher’s prediction results.
We validate the superiority of the proposed method by extensive experiments on real-world datasets.  

\section{Introduction}
Recent Recommender System (RS) has employed Knowledge Distillation (KD) to make a small but powerful recommender.
KD is a model compression technique that distills knowledge from a large pre-trained model (teacher) to a small model (student) \cite{KD}.
During the distillation process, the teacher provides additional supervisions, which are not explicitly revealed in the training set, to the student, and this can significantly accelerate the learning and the performance of the student model \cite{KD, FitNet, chen2017learning, self_distill1, cheng2017survey}.
The student model has much lower inference latency and computational costs compared to the teacher model so that the student model is much suitable for real-time applications \cite{CD, RD, DERRD}.

A few early work \cite{RD, CD} focus on making the student follow the teacher's predictions with particular emphasis on the high-ranked items.
In specific, in \cite{RD}, the student is trained to imitate top-ranked items in the teacher's predictions.
Such top-ranked items reveal the hidden patterns among users and items.
For example, the top-ranked items would have strong correlations to the user and the items that the user has interacted with.
The student can be enhanced by utilizing such additional information for its training without struggling to find it.
However, they have a clear limitation in that the distillation relies on the point-wise approach which considers an item at a time.
In other words, their distillation strategies distill the knowledge of each item separately without considering multiple items simultaneously.
As a result, they cannot directly handle the ranking violations among the items \cite{NCR}, and this leads to the limited ranking performance \cite{DERRD}.

In our previous work \cite{DERRD}, pointing out such limitation, we propose a new \textit{ranking distillation} method named RRD.
Unlike the existing methods, RRD formulates the distillation process as a ranking matching problem between the recommendation list of the teacher and that of the student.
Formally, RRD regards the recommendation list for each user as ground-truth data and enforces the student to maximize the log-likelihood of the ground-truth for all users.
With the list-wise approach, RRD distills the knowledge of each item with direct consideration of its ranking relationships with other items.
This enables the student to better imitate the teacher's predictions compared to the point-wise approach \cite{RD, CD}.

However, from our in-depth analysis, we observe a critical limitation of the ranking distillation method and discover room for further improvement.
Specifically, we observe that distilling the user-side ranking information (i.e., ranking orders among items) can destroy the item-side ranking information (i.e., ranking orders among users) in the student, and this further hinders the student from accurately learning the overall predictions of the teacher.
Also, we discover that directly learning the item-side ranking information can give supplementary knowledge that is complementary to that from the user-side ranking information.

Based on these findings, we propose a new regularization method that complements the limitation of the ranking distillation method and makes the student be further improved.
The proposed regularization is based on the item-side ranking information which is ignored in the previous work.
It effectively prevents the student with limited generalization power from being overfitted to the user-side rankings and enables the student to more accurately learn the teacher’s prediction results.
The main contributions of this work are as follows:
\begin{itemize}
    \item Through the exhaustive analyses, we demonstrate the existing ranking distillation method can hinder the student to learn the teacher's prediction results.
    Based on the results, we also point out that the student can be further improved by directly learning the item-side ranking information.
    
    \item We propose a new regularization method applicable for any ranking distillation method in RS.
    The proposed regularization is designed based on the item-side ranking information predicted by the teacher, considering the capacity gap between the teacher and the student.

    \item We validate the superiority of the proposed method by extensive experiments on real-world datasets. 
    The proposed method considerably improves the performance of the state-of-the-art ranking distillation method. 
    We also provide both qualitative and quantitative analyses to further investigate the effects of the proposed method. 
\end{itemize}

\section{Preliminary and Related Work}
An important requirement for recommender system (RS) is to balance between \textit{effectiveness} and \textit{efficiency}.
That is, the system should provide accurate recommendations with fast inference time.
However, the size of RS is continuously increasing, and accordingly, the computational time and memory costs for the inference are also increasing \cite{RD, CD, GCN_distill, DCF}.
Thus, it is difficult to apply such large RS to the real-time platform.
There have been several approaches to achieve a satisfying balance between the effectiveness and efficiency.
In this section, we first review several approaches to reduce the inference latency of RS.
Then, we introduce knowledge distillation (KD) that is a model-agnostic strategy to compress a large model.
Finally, we review recent methods that adopt KD for RS.

\vspace{0.1cm}
\textbf{Reducing inference latency of recommender system}
Firstly, a number of methods have adopted hash techniques to reduce the inference cost \cite{hash1, hash2, DCF, candidategeneration}.
To construct the hash table, they first restrict the capability of user and item representations (e.g., discrete representations).
By using the hash technique on the discrete representations, they have significantly reduced the inference latency.
However, their recommendation performance is highly limited due to the restricted capability of the representations.
Thus, their performance is limited compared to models that use real-values representations \cite{CD}.

Second, several methods have adopted model-dependent techniques to accelerate the inference phase \cite{pruning_RS2_inner_only, tree_RS, compression1, CML}.
For example, order-preserving transformations \cite{tree_RS}, tree-based data structures \cite{KDtree}, pruning
techniques \cite{pruning_RS2_inner_only}, data compression techniques \cite{compression1}, and approximated nearest neighbor search techniques \cite{LSH, LSH_inner_product} have been adopted, and they have successfully reduced the computational costs during the inference phase.
However, they have some intrinsic limitation is that they are applicable only to specific models (e.g., k-d tree for metric learning-based models \cite{METAS, CML}, the techniques \cite{LSH_inner_product, triangle_inequality_2012} for inner product-based models \cite{BPR}), or easily falling into a local optimum due to the local search \cite{DERRD}.

\vspace{0.1cm}
\textbf{Knowledge Distillation in Computer Vision}
KD is a model-independent strategy that improves the learning and the performance of a small model (student) by transferring knowledge from a pre-trained cumbersome model (teacher) \cite{KD, FitNet, chen2017learning, self_distill1, RKD, cheng2017survey}.
The core idea of KD is that the previously trained teacher model can provide additional supervision not included in the training set so that this can accelerate the learning of the student model.
The student model trained with distillation from the teacher model has much improved performance compared to when being trained separately, and it also has lower inference latency because of its small size.

Most KD methods in the computer vision field have focused on the image classification task.
A pioneer work \cite{KD} matches the class distribution predicted by the teacher classifier and the student classifier.
The predicted class distribution reveals much rich information such as the inter-class correlations compared to the one-hot class label.
It turns out that this additional information can considerably enhance the student model.
Subsequent methods \cite{FitNet, chen2017learning} have focused on transferring knowledge from the middle layers of the teacher model.
They have shown that the intermediate representations from the teacher model can provide hints to improve the training process of the student model.
Because the size of the teacher's middle layers is bigger than that of the student in general, they employ some additional layers to handle the dimension gap.
Recent work \cite{self_distill1, zhang2019your} shows that the teacher model can be further improved by distilling knowledge from itself.

\vspace{0.1cm}
\textbf{Knowledge Distillation in Recommender System}
Recently, motivated by the significant success of Knowledge Distillation in the field of computer vision, a few work \cite{RD, CD, DERRD} have applied KD to the field of RS\footnote{There are a few methods proposed for specific purpose or model (e.g., clothing matching \cite{song2018neural}, binarized RS \cite{GCN_distill}), but we focus on model-agnostic distillation for general top-$N$ recommendation task.}.
Here, we briefly review the existing work. 
They can be categorized into two groups:
1) the methods that distill from the teacher's prediction results and 2) the method that distills the teacher's latent knowledge.

\subsubsection{Distilling from teacher's prediction results}\noindent
Most existing work \cite{RD, CD, DERRD} has focused on making the student recommender to imitate the recommendation list\footnote{In this work, we use the term `recommendation list' to refer a ranked list of unobserved items for each user.} from the teacher recommender.
Ranking Distillation (RD) \cite{RD} and Collaborative Distillation (CD) \cite{CD} adopt the point-wise distillation approach which distills the knowledge of a single item at a time.
Concretely, RD makes the student follow the teacher’s predictions on top-$K$ unobserved items.
By the nature of the RS, only the high-ranked items in the ranking list are important. 
Also, such high-ranked items reveal hidden patterns among users and items;
the high-ranked items would have strong correlations to the user. 
By using such supplementary supervision not explicitly included in the training set, RD successfully improves the effectiveness of the student model.
A subsequent work CD first samples items with ranking-based sampling technique which puts more weights on items with higher rankings.
Then, CD makes the student imitate the teacher’s predictions on the sampled items.
By doing so, CD considers both the high-ranked and low-ranked items in a probabilistic way and thus the student learns both the positive and negative correlations among items to the user.

From our previous work \cite{DERRD}, we propose a new ranking distillation method named RRD.
RRD first points out that RD and CD rely on the point-wise distillation which transfers the knowledge of an item at a time.
Since they cannot consider relations among multiple items simultaneously, they have a limitation in accurately learning the ranking orders predicted by the teacher model.
To tackle this problem, RRD formulates the distillation task as a user-side ranking matching problem between the ranking list of each user predicted from the teacher and that from the student.
RRD also adopts the relaxed ranking approach which makes the student focus on learning the detailed ranking orders among the high-ranked items while ignoring the ranking orders among the low-ranked items.
However, from our in-depth analyses, we find a critical limitation of the ranking distillation method, and we discover room for further improvement.

\vspace{0.1cm}
\textbf{Distilling teacher's latent knowledge}\noindent
Since the teacher recommender makes predictions based on its latent knowledge (e.g., user and item representations), the student recommender can also learn by distilling such latent knowledge.
Such latent knowledge is valuable for the student, as it provides detailed explanations for the teacher's prediction results.
Distillation Experts (DE) \cite{DERRD} is the first method that directly distills the latent knowledge encoded in the teacher recommender.
DE first introduce ``expert'', which is the small feed-forward network, to capture only the core information from the vast teacher's knowledge.
Then, DE adopts the expert selection strategy to effectively distill the knowledge of various users and items.
By directly learning the latent knowledge, the student can have detailed supervision supporting the teacher's predictions, which considerably improves the effectiveness of the student recommender.
Note that DE is not a competing method with the KD methods in the first group; DE can be combined with any KD methods that distill knowledge from teacher's prediction results (e.g., RD, CD, RRD).

\section{Problem Formulation}
\vspace{3pt}
\noindent
\textbf{Top-$N$ Recommender System.}
We focus on top-$N$ recommendations for implicit feedback, which is also known as one-class collaborative filtering problem \cite{pan2008one, hu2008collaborative}.
Let $\mathcal{U}$ and $\mathcal{I}$ denote the set of users and items, respectively.
Given collaborative filtering information, which is the implicit interactions between users and items, we build a binary matrix $\boldsymbol { R } \in \{0,1\}^{| \mathcal { U } | \times | \mathcal { I } |}$,
where each element $r_{u,i}$ indicates whether a user $u$ has interacted with an item $i$ or not.
We also denote the set of items that have not interacted with a user $u$ as $\mathcal{I}_{u}^{-}$, and the set of users that have not interacted with an item $i$ as $\mathcal{U}_{i}^{-}$.
The goal of top-$N$ RS is to provide a recommendation list for each user;
For each user, RS ranks all items in $\mathcal{I}_{u}^{-}$, then provides a ranked list of top-$N$ unobserved items.

\vspace{3pt}
\noindent
\textbf{Knowledge Distillation for Recommender System.}
KD is a model compression strategy that transfers the knowledge of a large model (i.e., teacher) to a small model (i.e., student).
In the \textit{offline training phase}, the distillation process for RS is conducted in two steps.
First, a teacher recommender with numerous learning parameters is trained with the training set having binary labels.
Then, a student recommender with a smaller number of learning parameters is trained with help from the teacher recommender along with the binary labels.
In the \textit{inference phase}, we utilize the student recommender.
The student recommender has much faster inference time compared to the teacher due to its small size.
The goal of our work is to enhance the quality of the distillation so that the recommendation performance of the student model can be fully improved.


\section{Proposed Method---IR-RRD}
\section{Item-side Ranking Regularized Knowledge Distillation}

In this section, we propose \textbf{I}tem-side ranking \textbf{R}egularization (IR) method designed to maximize the effects of the ranking distillation method in RS.
We first point out the limitations of the existing ranking distillation method and clarify the necessity of the proposed method.
Then, we describe the proposed method that enables the student to accurately learn the ranking information from each user's recommendation list with consideration of the ranking orders among the users.
Finally, we explain the end-to-end optimization process of the student model. 
The detailed algorithm of the proposed method is provided in Algorithm \ref{alg:IRRRD_framework}, and its key concept is illustrated in Figure \ref{fig:IRRRD_4.2_key}.

\subsection{Motivation}
\label{sec:IRRRD_motiv}
\begin{figure}[h]
\centering
    \includegraphics[width=0.7\linewidth]{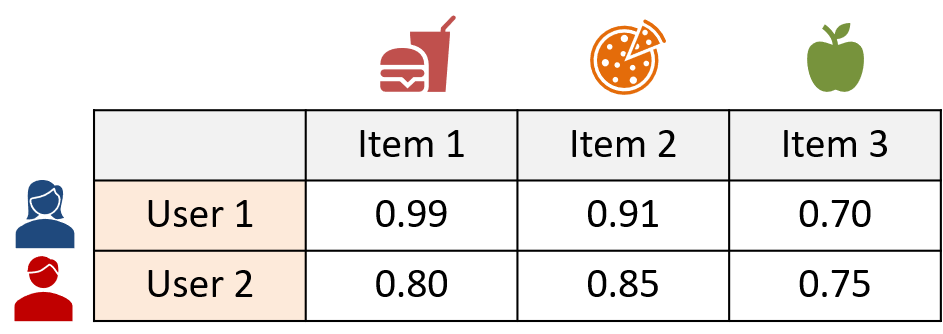}
\caption{A toy example: the teacher's prediction results on two users for three items.}
\label{fig:IRRRD_4.1_toy}
\end{figure}

\noindent
\textbf{User-side ranking distillation may destroy ranking orders among users.}
We first point out that the user-side ranking distillation may destroy ranking orders among users.
Figure \ref{fig:IRRRD_4.1_toy} shows ranking scores predicted by the teacher recommender of two users for three items.
It shows that $u_1$’s recommendation list is $[I_1, I_2, I_3]$ and $u_2$’s recommendation list is $[I_2, I_1, I_3]$.
By the user-side ranking distillation method (e.g., RRD), the student recommender is trained to imitate each user's recommendation list.
Now, assume that the ranking order for $u_2$ predicted by the student during the training is $[I_3, I_1, I_2]$. 
The user-side ranking distillation method, which learns to maximize the likelihood of ranking orders among items for each user, will make the student to be optimized to decrease the ranking score of $I_3$ and to increase the ranking score of $I_2$.
However, it cannot directly consider ranking relationships among users (i.e., item-side ranking orders) during the distillation process.
Specifically, the teacher's predictions show that $u_2$ more prefers $I_3$ than $u_1$ does, but such information is ignored during the distillation process.
Likewise, the score of $u_2$ on $I_2$ needs to be adjusted while considering that $u_1$ more prefers $I_2$ than $u_2$ does.
The distillation process without the direct consideration of item-side ranking information destroys the ranking relationship among users, and further hinders the student from accurately preserving the overall rankings predicted by the teacher. 

In KD, the student has a smaller capacity compared to the teacher and thus has restricted generalization power.
Therefore, the student has limited capability in finding information that is \textit{not explicitly given} during the training.
In other words, unlike the large teacher model, the student has a limitation in accurately inferring the ranking relationships among users from the ranking relationship among items.
Thus, the student can be easily overfitted to the given user-side ranking information, which leads to degraded performance.
We conclude that the ranking information in the teacher's recommendation list should be distilled with proper regularization considering the item-side ranking relationships.

\vspace{3pt}
\noindent
\textbf{Learning ranking orders among users give complementary information.}
To further investigate the effects of distilling the item-side ranking relationships, we conduct in-depth analyses and observe some interesting results.
We first modify RRD, which the state-of-the-art user-side ranking distillation method, to distill the item-side ranking information (denoted as \textbf{ItemRRD}).
Note that ItemRRD only considers the item-side ranking relationships, without consideration of the user-side ranking relationships. 
The recommendation performance of RRD and ItemRRD is provided in Table \ref{tab:IRRRD_4.1_rp}.

\begin{table}[h]
\setlength\tabcolsep{2.8pt}
\footnotesize
\RowStretch{1.}
  \caption{Recommendation performances on CiteULike dataset ($\phi=0.1$).}
  \begin{tabular}{clccc ccc ccc}
    \toprule 
     Base Model & KD Method & H@5 & M@5 & N@5 & H@10 & M@10 & N@10 & H@20 & M@20 & N@20 \\
    \midrule
     &Teacher&0.5135&0.3583&0.3970&0.6185&0.3724&0.4310&0.7099&0.3788&0.4541\\
     \multirow{3}{*}{BPR}&Student&0.4441&0.2949&0.3319&0.5541&0.3102&0.3691&0.6557&0.3133&0.3906\\
     \cmidrule{2-11}
     &RRD&0.4622&0.3076&0.3461&0.5703&0.3220&0.3809&0.6746&0.3293&0.4074\\
     &ItemRRD&0.4909&0.3364&0.3748&0.5909&0.3499&0.4073&0.6929&0.3570&0.4331\\
    \midrule
    \cmidrule{2-11}
    &Teacher&0.4790&0.3318&0.3684&0.5827&0.3457&0.4020&0.6748&0.3521&0.4254\\
    \multirow{3}{*}{NeuMF}&Student&0.3867&0.2531&0.2865&0.4909&0.2670&0.3202&0.5833&0.2738&0.3436\\
    \cmidrule{2-11}
    &RRD &0.4737&0.3086&0.3497&0.5800&0.3236&0.3847&0.6765&0.3305&0.4094\\
    &ItemRRD &0.4397&0.2892&0.3266&0.5426&0.3028&0.3597&0.6354&0.3092&0.3832\\
    \bottomrule
  \end{tabular}
  \label{tab:IRRRD_4.1_rp}
\end{table}

\begin{figure}[h]
\centering
\begin{subfigure}[t]{0.40\linewidth}
    \includegraphics[width=\linewidth]{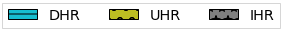}
\end{subfigure}\\
\begin{subfigure}[t]{0.48\linewidth}
    \includegraphics[width=\linewidth]{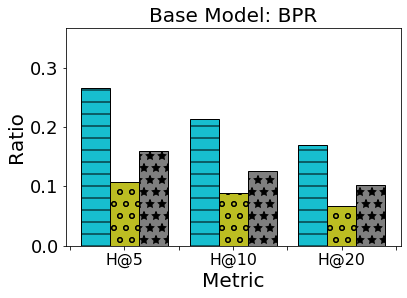}
\end{subfigure}
\begin{subfigure}[t]{0.48\linewidth}
    \includegraphics[width=\linewidth]{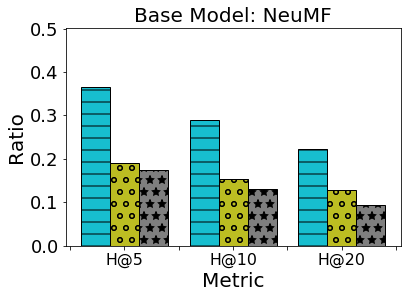}
\end{subfigure} 
\centering
\caption{Comparison between the item set hit by RRD and that by ItemRRD on CiteULike dataset.}
\label{fig:IRRRD_4.1_dhr}
\end{figure}

We first observe that both the user-side ranking distillation (i.e., RRD) and the item-side ranking distillation (i.e., ItemRRD) are effective in improving the ranking performance of the student. 
In specific, the item-side distillation achieves higher performance gain on BPR and user-side distillation achieves higher performance gain on NeuMF.
Based on the results, we inspect the items \textit{hit} by each method.
To this end, we introduce three statistics, which show the difference between the set of items hit by RRD and that by ItemRRD as follows:
\begin{equation}
\text {User-side only Hits Ratio (UHR)}=\frac{|\mathcal{H}_{RRD} - \mathcal{H}_{ItemRRD}|}{|\mathcal{H}_{RRD} \cup \mathcal{H}_{ItemRRD}|}
\end{equation}
\begin{equation}
\text {Item-side only Hits Ratio (IHR)}=\frac{|\mathcal{H}_{ItemRRD} - \mathcal{H}_{RRD}|}{|\mathcal{H}_{RRD} \cup \mathcal{H}_{ItemRRD}|}
\end{equation}
\begin{equation}
\text {Different Hits Ratio (DHR)}=\frac{|\mathcal{H}_{RRD} - \mathcal{H}_{ItemRRD}| + |\mathcal{H}_{ItemRRD} - \mathcal{H}_{RRD}|}{|\mathcal{H}_{RRD} \cup \mathcal{H}_{ItemRRD}|}
\end{equation}
where $\mathcal{H}_{RRD}$ and $\mathcal{H}_{ItemRRD}$ are the set of items hit by RRD and ItemRRD respectively.
The Different Hits Ratio of RRD and ItemRRD are summarized in Figure \ref{fig:IRRRD_4.1_dhr}.
Interestingly, despite the significant difference in recommendation performance between the two methods shown in Table \ref{tab:IRRRD_4.1_rp}, we observe that both UHR and IHR ratios are considerably large on both base models.
For example, on NeuMF where RRD outperforms ItemRRD, ItemRRD shows 17.41\% of IHR in terms of H@5. 
If ItemRRD is completely surpassed by RRD, the value of IHR should be 0.
This strongly indicates that distilling ranking relationships among users and distilling ranking relationships among items give the student information which is considerably different but reciprocal information.
We conclude that the student can be further improved by distilling both the user-side and item-side ranking information from the teacher model.

The findings presented in this subsection motivate us to design a regularization method that enables the student recommender to learn the user-side ranking relationships with consideration of item-side ranking information simultaneously.

\subsection{IR: Item-side Ranking Regularization}
We propose a new item-side ranking regularization method designed for maximizing the effects of a user-side ranking distillation method, named \textbf{IR}.
The proposed regularizer prevents the student with limited generalization power from being overfitted to the user-side rankings and enables the student to more accurately learn the teacher's prediction results.
IR can be flexibly combined with any prediction-based distillation method that considers the user-side ranking.
A user-side ranking distillation method (denoted as \textbf{UKD}) regards ranking orders among items predicted by the teacher as the ground-truth,
and enforces the student to maintain the ranking orders.
For a user $u$, let $\mathcal{I}^u_{UKD}$ denote the set of items for which the student learns the ranking relationship by UKD.
The proposed item-side ranking regularization method is applied to the items belongs to $\mathcal{I}^u_{UKD}$, enabling the student to learn the relative potential preferences of users on the items.
The algorithm of IR is provided in Algorithm 1.

\begin{algorithm}[t]
\SetAlgoLined
\nonl \textbf{Require:} Student recommender $S(\cdot;{\theta_s})$, Teacher recommender $T(\cdot;{\theta_t})$, learning rate $\gamma$\\
generate item-side ranking lists from Teacher recommender\\
randomly initialize $\theta_s$\\
\textbf{while} not converge \textbf{do}\\
\nonl \quad \quad // \textit{Student recommender} with \textit{UKD}\\
\quad \quad compute $\mathcal{L}_{\text{UKD}}$ and  get $\mathcal{I}_{\text{UKD}}$ \\
\nonl \quad \quad // \textit{Item-side ranking regularization}\\
\quad \quad  draw $\boldsymbol{\pi}^{i \in \mathcal{I}_{\text{UKD}}}$ by potential preference-based sampling  (Eq.4, Eq.5)\\
\quad \quad  compute $\mathcal{L}_{\text{IR}}$ on $\boldsymbol{\pi}^{i \in \mathcal{I}_{\text{UKD}}}$ (Eq.7)\\
\nonl \quad \quad // \textit{Optimization}\\
\quad \quad compute $\mathcal{L}$ (Eq.8)\\
\quad \quad  update $\theta_s \leftarrow \theta_s - \gamma \frac{\partial \mathcal{L}}{\partial \theta_s}$\\
\textbf{end while}\\
\caption{The item-side ranking regularized distillation}
\label{alg:IRRRD_framework}
\end{algorithm}

\subsubsection{Generating Item-side ranking list}\noindent
For regularization based on the item-side ranking information, we formally define the item-side ranking list.
For each item $i$, the teacher recommender calculate ranking scores for all users in $\mathcal{U}_{i}^{-}$.
Then, it sorts the users according to the predicted scores in the decreasing order, making the ranking list for each item.
The item-side ranking list reveals the teacher's knowledge about the relative potential preferences that users would have on each item.
Specifically, a few users that would prefer an item would be located near the top of the list, whereas numerous users that would not prefer the item would be located far from the top.
This information, which is not explicitly distilled by UKD, can give complementary knowledge to the student recommender.
It is worth noting that this generating process is conducted only once before the distillation process.

\subsubsection{Item-side ranking regularization}\noindent
Since the capacity of the student is highly limited, the student cannot learn all the ranking orders among users.
Further, enforcing the student to learn all the ranking orders is not even effective, considering only top-ranked ranking orders are important in RS.
Thus, an effective regularization method should enable the student to focus on leaning the ranking orders among the top-ranked users for each item.
To this end, we adopt the list-wise approach \cite{xia2008list-wise} with relaxed permutation probability which has shown remarkable performance in distilling the user-side knowledge \cite{DERRD}.
More specifically, for each item in $\mathcal{I}^u_{UKD}$, we distill the knowledge of users' potential preference orders to the item predicted by the teacher.
To overcome the capacity gap between the teacher and the student, we define (potential) positive users and negative users, and make the student focus on learning the detailed relationships among the positive users.

\vspace{3pt} \noindent
\textbf{Potential preference-based sampling.}
Formally, from the item-side ranking list of item $i$, we sample $P$ positive users and $N$ negative users.
The positive users are those located near the top of the list, and very likely to prefer the item in the future.
To sample the positive users, we adopt the ranking position importance scheme used in the previous work \cite{RD, CD, DERRD}, 
The probability of a user $u$ to be sampled is computed as follows:
\begin{equation}
p_{\text{pos}}(u) \propto \exp\left(\frac{-r a n k(u)}{T}\right),
\end{equation}
where $r a n k(u)$ denotes the ranking position of $u$, the highest ranking is 1, and $T$ is the hyperparameter that controls the sharpness of the exponential function.
With the sampling probability $p_{\text{pos}}$, we sample $P$ positive users according to their potential preference on the item (i.e., users' ranking).
On the other hand, the negative users are those located in the wide rest of the ranking list.
Thus, to sample the negative users, we adopt simple uniform sampling in which the probability of a user $u$ to be sampled is computed as follows:
\begin{equation}
p_{\text{neg}}(u) \propto \frac{1}{|\mathcal{U}_{i}^{-}|}.
\end{equation}
With the sampling probability $p_{\text{neg}}$, we sample $N$ negative users that have lower potential preferences than the previously sampled positive users.

\begin{figure}[t]
\centering
    \includegraphics[width=0.9\linewidth]{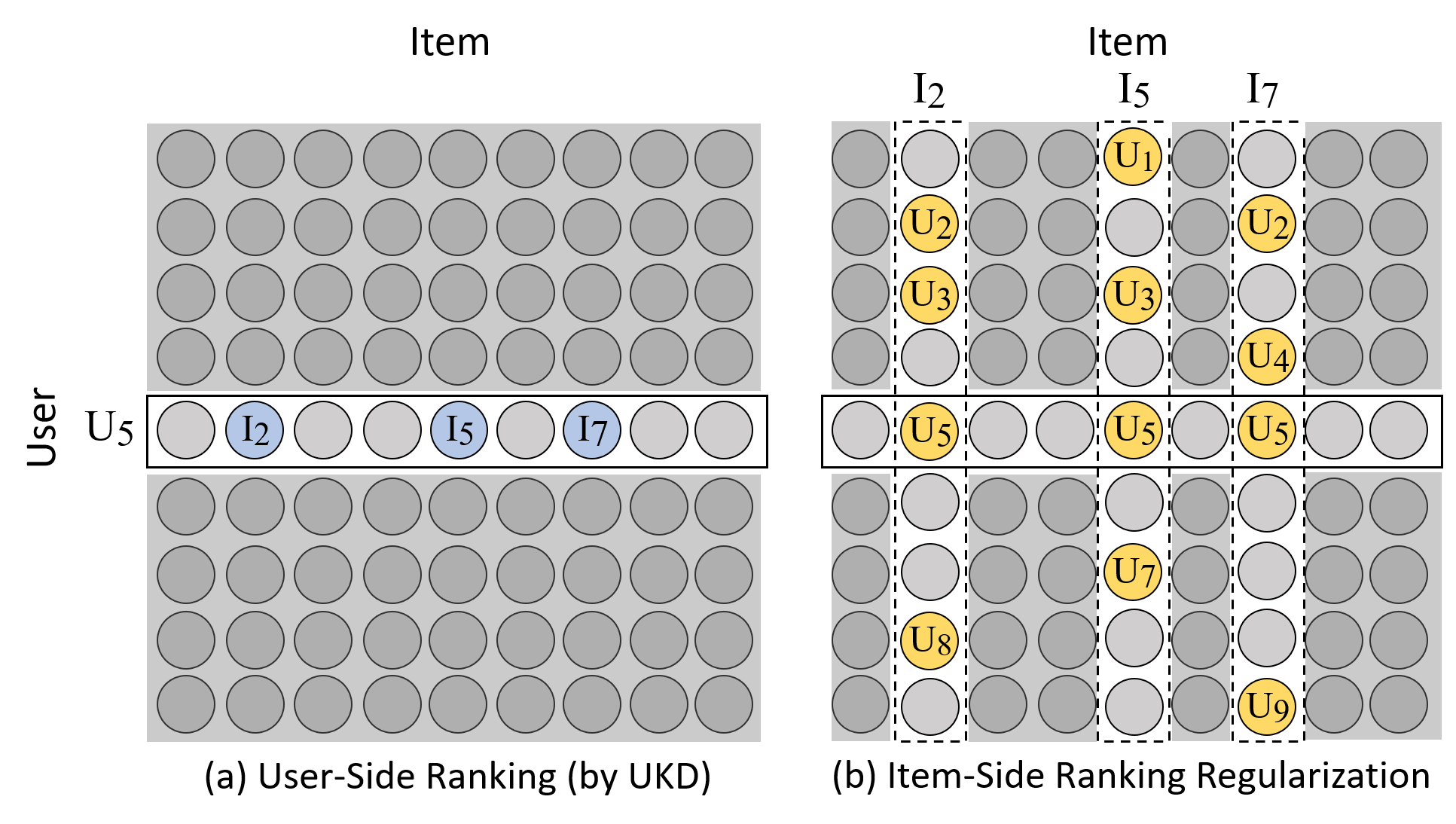}
\caption{An illustration of the key concept of the item-side ranking regularization.}
\label{fig:IRRRD_4.2_key}
\end{figure}

\vspace{3pt} \noindent
\textbf{IR loss.}
For item $i$, let $\boldsymbol{\pi}^i$ denote the list of the sampled users sorted by the original order in the teacher's item-side ranking list.
We are interested in making the student learn the precise ranking orders among the positive users in $\boldsymbol{\pi}^i$, ignoring the detailed ranking orders among the negative users.
The relaxed permutation probability is defined as follows:
\begin{equation}
    p\left(\boldsymbol{\pi}^i_{1:P} | S\right)=\prod_{k=1}^{P} \frac{\exp \left[S(\pi^i_{k}, i)\right]}{\sum_{j=k}^{P} \exp\left[ S(\pi^i_{j}, i)\right]+\sum_{m=P}^{|\pi^i|} \exp \left[ S(\pi^i_{m}, i)\right]},
\end{equation}
where $\pi_k$ is the $k$-th user in $\pi$, $S(u,i) \in \mathbb{R}$ is the ranking score of user-item interaction $(u,i)$ predicted by the student model $S$. 
The final loss function of item-side ranking regularizer is defined as follows:
\begin{equation}
\mathcal{L}_{IR} = - \frac{1}{|\mathcal{I}_{\text{UKD}}|} \sum_{i \in \mathcal{I}_{\text{UKD}}}\log p\left(\boldsymbol{\pi}^{i}_{1:P} | S\right)
\end{equation}
By minimizing the above loss function, the model is optimized to push the positive users higher than all the negative users in the item-side ranking list, while preserving the accurate ranking orders among the positive users.
The item-side ranking regularizer enables the student to directly handle the ranking violations among the users, which can be exacerbated by the UKD, and thus can lead to better ranking performance. 
It is worth noting that the IR loss can be efficiently computed in a parallel manner for all items in $\mathcal{I}_{\text{UKD}}$.
Here, we provide explanations in terms of the iterative manner for better understanding. 

Also, we provide an illustration of the key concept of IR in Figure \ref{fig:IRRRD_4.2_key}.
To distill the knowledge of user $U_5$, UKD (Fig. \ref{fig:IRRRD_4.2_key}a) transfers the ranking orders among the items (i.e., $I_2, I_5, I_7$).
By doing so, the ranking relationships among the users are ignored, which leads to degraded performance. 
The proposed IR method (Fig. \ref{fig:IRRRD_4.2_key}b) makes the student consider the item-side ranking information during the distillation process.
Namely, with IR, the student learns the ranking orders among the items as well as the ranking orders among the users, enabling the student to better imitate the teacher's prediction results.

\subsection{Optimization with IR}
Any RS model adopting the user-side ranking distillation (UKD) can be jointly optimized with the proposed item-side ranking regularization method (IR) in an end-to-end manner.
Let $\mathcal{L}_{U K D}$ denote the base model's loss function trained with UKD.
The optimization with IR is conducted as follows:
\begin{equation}
\min_{\theta_{S}} \mathcal{L} = \mathcal{L}_{U K D} + \lambda_{I R} \cdot \mathcal{L}_{I R}
\end{equation}
where $\theta_{S}$ is the learning parameters of the student model, and $\lambda_{I R}$ is the hyperparameters that control the effects of item-side ranking regularization.
The student model can be any existing recommender such as BPR \cite{BPR}, NeuMF \cite{NeuMF}.
We minimize $\mathcal{L}$ using gradient descent with respect to the parameters in the student model ($\theta_{S}$) so that the ranking scores from the student are accordant with the ranking order $\pi$ predicted by the teacher model.
The algorithm of IR is provided in Algorithm 1.

\subsection{Computational Complexity with IR}
Note that IR does not require any additional computation at the \textit{inference phase} of the student model.
IR is designed to enhance the distillation quality of UKD method at the \textit{offline training phase} and requires no additional learning parameters.
Thus, IR improves the recommendation accuracy of the student model without affecting the inference efficiency of the student model.

The forward computations of IR method in the offline training phase are summarized as follows:
1) sampling users from item-side ranking lists,
2) computing the IR loss ($\mathcal{L}_{IR}$).
Note that the item-side ranking lists are generated before the distillation process.
The cost for the sampling part is related to the implementation, but it increases as the population size (i.e., the number of users) increases.
The second part can be parallelly computed through GPU processor.
Firstly, the student model predicts the ranking scores for the sampled users (i.e., S($\cdot, \cdot$) in Eq.6), which requires a constant cost depending on the architecture of the base model.
Then, IR loss is computed.
For each item, its complexity can be viewed as $O(\max(P, N))$, where $P$ and $N$ are the numbers of positive and negative users respectively. 
Specifically, $\log p\left(\boldsymbol{\pi}^i_{1:P} | S\right)$ is as follows:
\begin{equation}
\begin{aligned}
    \log p\left(\boldsymbol{\pi}^i_{1:P} | S\right)&=\sum_{k=1}^{P}  S(\pi^i_{k}, i) - \sum_{k=1}^{P} \log \left( \sum_{j=k}^{P} \exp\left[ S(\pi^i_{j}, i)\right]+\sum_{m=P}^{|\pi^i|} \exp \left[ S(\pi^i_{m}, i)\right] \right),
\end{aligned}
\end{equation}
Here, the first term can be computed with $O(P)$ cost.
The second term can be computed with the cumulative sum operations (for positive users) and sum operations (for negative users) which require $O(P)$ and $O(N)$ cost respectively.
These operations are parallelly computed for the items used by UKD in a batch (i.e., $\mathcal{I}_{\text{UKD}})$ through GPU processor.
The power of the proposed IR method comes from the additional supervisions based on these extra computations.
Note that the proposed IR method is only applied in the offline training phase.

\section{Experiments}
By following the experiment setup of \cite{DERRD}, we validate the proposed IR method on 12 different experiment settings (2 real-world datasets $\times$ 2 base models $\times$ 3 different student model sizes).
We first introduce the experiment setup.
Then, we provide a performance comparison supporting the superiority of the proposed method.
We also provide in-depth analyses on ranking consistency. 
Moreover, we show that the proposed method significantly accelerates the learning of the student model.
We then provide experiment results showing that the proposed method enables the student to achieve a good balance between effectiveness and efficiency. 
Lastly, we provide hyperparameter study.

\subsection{Experimental Setup}
\subsubsection{Dataset}\noindent
We use two publicly available real-world datasets: CiteULike\footnote{\url{https://github.com/changun/CollMetric/tree/master/citeulike-t}} \cite{wang2013collaborative} and Foursquare\footnote{\url{https://sites.google.com/site/yangdingqi/home/foursquare-dataset}} \cite{liu2017experimental}.
We filter out users and items having fewer than five ratings for CiteULike, twenty ratings for Foursquare as done in \cite{BPR, NeuMF, SSCDR, DERRD, transCF}.
The data statistics after preprocessing are provided in Table \ref{tbl:IRRRD_statistic}.
\begin{table}[t]
\centering
  \caption{Data Statistics}
  \begin{tabular}{ccccc}
    \toprule
    Dataset & \#Users & \#Items & \#Interactions & Sparsity \\
    \midrule
    CiteULike & 5,220 & 25,182 & 115,142 & 99.91\% \\
    Foursquare & 19,466 & 28,594 & 609,655 & 99.89\% \\
    \bottomrule
  \end{tabular}
    \label{tbl:IRRRD_statistic}
\end{table}

\subsubsection{Base Models}\noindent
We verify the proposed method on base models that have different optimization strategies and architectures.
Specifically, we choose a latent factor model and a deep learning model that are very widely used for top-$N$ recommendation task with implicit feedback.
\begin{itemize}
    \item \textbf{BPR \cite{BPR}}: 
    A learning-to-rank recommendation method that adopts Matrix Factorization (MF) to model the user-item interaction. 
    BPR adopts the pair-wise objective function for the optimization with the assumption that observed items are more preferred than unobserved items.
    Note that it is a highly competitive method for item recommendation \cite{ADBPR, tay2018latent}.
    
    \item \textbf{NeuMF \cite{NeuMF}}: A deep recommendation method that adopts MF and Multi-Layer Perceptron (MLP) to model the complex and non-linear user-item relationships. 
    NeuMF adopts the point-wise objective function (i.e., binary cross-entropy) for the optimization.
\end{itemize}

\subsubsection{Teacher/Student}\noindent
To make the teacher model, we adopt as many learning parameters as possible until the ranking performance is no longer increased for each base model on each dataset.
Then, we build three student models by limiting the number of learning parameters\footnote{We adjust the number of parameters based on the size of the last hidden layer.}.
The limiting ratios ($\phi$) are \{0.1, 0.5, 1.0\}.
Following the notation of the previous work \cite{RD, CD, DERRD}, we denote the student model trained without the help of the teacher model (i.e., no distillation) as ``Student'' in this experiment sections.

\subsubsection{Comparison Methods}\noindent
The proposed method is compared with the following competitors:
\begin{itemize}
    \item \textbf{MOSTPOP}: A method that simply ranks items by the number of their interactions (i.e., popularity). 
    It provides benchmark performances of the non-personalized recommendation.
    \item \textbf{Ranking Distillation (RD) \cite{RD}}: A prediction-based knowledge distillation method for recommender system.
    RD uses items with the highest ranking from the teacher's predictions for distilling the knowledge.
    \item \textbf{Collaborative Distillation (CD) \cite{CD}}: A prediction-based knowledge distillation method for recommender system.
    CD samples items from teacher's predictions based on their ranking and uses them for distillation.
    As suggested in the paper, we use unobserved items only for distilling the knowledge.
    \item \textbf{Relaxed Ranking Distillation (RRD) \cite{DERRD}}:
    The state-of-the-art ranking knowledge distillation method for recommender system which is our preliminary model.
    RRD formulates the distillation as a ranking matching problem between each user's ranking list from the teacher and that from the student.
\end{itemize}
Finally, we apply the proposed item-side ranking regularization method on the state-of-the-art ranking distillation KD method (i.e., RRD), and denote it as \textbf{IR-RRD}.
Note that the other KD methods (i.e., RD, CD) are not the user-side ranking distillation method (i.e., UKD).
They are the point-wise distillation methods that do not directly consider the ranking orders among the items.
Since they are not enforcing the student to learn the ranking orders predicted by the teacher, they don't suffer from the problems that we mentioned in Section \ref{sec:IRRRD_motiv}.
However, for the very same reason, they cannot effectively transfer the teacher's knowledge to the student, and thus already been surpassed by RRD with a huge margin \cite{DERRD}.
Thus, we do not apply IR on them.
It is worth noting that DE, which distills the teacher's latent knowledge, is not a competitor of the above prediction-based distillation methods (i.e., RD, CD, RRD, IR-RRD).
Rather, DE can be combined with them to further enhance the student model, but that is beyond the scope of this work.

\subsubsection{Evaluation Protocol}\noindent
We follow the broadly used \textit{leave-one-out} evaluation protocol ~\cite{NeuMF, transCF, SSCDR, DERRD}.
Specifically, for each user, we hold out an interacted item for testing and use the rest for training.
In our experiments, we hold out an additional interacted item for validation purpose.
Following the setup of our previous work \cite{DERRD}, we randomly sample 499 items from a set of unobserved items of the user, then evaluate how well each method can rank the test item higher than these sampled unobserved items. 
We repeat this process of sampling a test item, validation item, and unobserved items five times and report the average results.
On top of that, we provide the performance comparison on the full-ranking evaluation setup that uses all the unobserved items as candidates. 

For evaluating the top-$N$ recommendation performance based on implicit feedback, we adopt broadly used three ranking metrics \cite{NeuMF, SSCDR, DERRD} (i.e., hit ratio (H@$N$), normalized discounted cumulative gain (N@$N$), and mean reciprocal rank (M@$N$)) that measures how well a method pushes the test items to the top of the ranking list.
The hit ratio simply measures whether the test item is present in the top-$N$ list, whereas the normalized discounted cumulative gain and the mean reciprocal rank are ranking position-aware metrics that put higher scores to the hits at upper ranks.
The three ranking metrics are defined as follows:
\begin{equation}
\text{H} @ N = \frac { 1 } { | \mathcal { U }_\text{test} | } \sum _ { u \in \mathcal { U} _\text{test} } \delta \left( p _ { u } \leq \text { top } N \right)   
\end{equation}
\begin{equation}
\text{N} @ N = \frac { 1 } { | \mathcal { U } _\text{test} | } \sum _ { u \in \mathcal { U }_\text{test} } \frac { 1 } { \log \left( p _ { u } + 1 \right) }
\end{equation}
\begin{equation}
\text{M} @ N = \frac { 1 } { | \mathcal { U }_\text{test} | } \sum _ { u \in \mathcal { U } _\text{test} } \frac { 1 } { p _ { u } }
\end{equation}
where $\delta ( \cdot )$ is the indicator function, $\mathcal { U }_{test}$ is the set of the test users, $p_u$ is the hit ranking position of the test item for the user $u$.

\subsubsection{Implementation Details}\noindent
We use PyTorch to implement the proposed method and all the competitors and use Adam optimizer to train all the methods.
For RD, we use the public implementation provided by the authors\footnote{\url{https://github.com/graytowne/rank\_distill}}.
For each base model and distillation method (i.e., RD, CD, RRD), hyperparameters are tuned by using the grid searches on the validation set. 
The learning rate for the Adam optimizer is chosen from \{0.1, 0.05, 0.01, 0.005, 0.001, 0.0005, 0.0001\}, the model regularizer is chosen from \{0.1, 0.01, 0.001, 0.0001, 0.00001\}.
The number of maximum epochs is set to 1000, and the early stopping strategy is adopted.
Specifically, we stop the training process if H@$5$ on the validation set does not increase for 30 successive epochs.
For all base models (i.e., BPR, NeuMF), the number of negative sample is set to 1, and pre-trained technique is not used.
For NeuMF, the number of the hidden layers is selected from \{1, 2, 3, 4\}.
For distillation methods, all the hyperparmeters are tuned in the range reported in \cite{DERRD}.
For the proposed item-side ranking regularization, the weight for regularization loss ($\lambda_{I R}$) is selected from \{0.1, 0.01, 0.001, 0.0001\},
the number of positive users ($P$) is selected from \{10, 20, 30, 40, 50\} and the number of negative users ($N$) is set to the same with $P$.
The weight for controlling the sharpness of the exponential function ($T$) is set to 20.

\subsection{Performance Comparison}
\label{sec:IRRRD_PC}
We evaluate the top-$N$ recommendation performance of IR-RRD and the competitors with three ranking metrics.
Table \ref{tab:IRRRD_main_BPR} and Table \ref{tab:IRRRD_main_NeuMF} show the recommendation accuracy.
In summary, the proposed method considerably improves the performance of the state-of-the-art UKD method (i.e., RRD) on two base models that have different optimization strategies and architectures.
In addition, IR-RRD continuously outperforms the existing methods on the student model with three different sizes in Figure 4 and Figure \ref{fig:IRRRD_model_size_NeuMF}.
Moreover, we further verify the superiority of the proposed IR method on the full evaluation setup in Table \ref{tab:IRRRD_main_Full} and Figure \ref{fig:IRRRD_model_size_F}.
We analyze the results with various perspectives.

\begin{table}[t]
\setlength\tabcolsep{1.8pt}
\footnotesize
\RowStretch{1.2}
  \caption{Recommendation performances of BPR ($\phi=0.1$). 
  IR-RRD outperforms the best baseline at 0.05 level on the paired t-test.}
  
  \begin{tabular}{clccc ccc ccc}
    \toprule 
     Dataset & KD Method & H@5 & M@5 & N@5 & H@10 & M@10 & N@10 & H@20 & M@20 & N@20 \\
    \midrule
    &MOSTPOP&0.1389&0.0820&0.0960&0.1920&0.0890&0.1131&0.2610&0.0938&0.1305\\
    \cmidrule{2-11}
     &Teacher&0.5135&0.3583&0.3970&0.6185&0.3724&0.4310&0.7099&0.3788&0.4541\\
     \multirow{3}{*}{\rotatebox{90}{CiteULike}}&Student&0.4441&0.2949&0.3319&0.5541&0.3102&0.3691&0.6557&0.3133&0.3906\\
     &RD&0.4533&0.3019&0.3395&0.5601&0.3161&0.3740&0.6633&0.3232&0.3993\\
     &CD&0.4550&0.3025&0.3404&0.5607&0.3167&0.3746&0.6650&0.3240&0.4011\\
     &RRD&0.4622&0.3076&0.3461&0.5703&0.3220&0.3809&0.6746&0.3293&0.4074\\
     \cmidrule{2-11}
     &IR-RRD &\textbf{0.5028}&\textbf{0.3395}&\textbf{0.3801}&\textbf{0.6052}&\textbf{0.3531}&\textbf{0.4132}&\textbf{0.6971}&\textbf{0.3562}&\textbf{0.4338}\\
    \midrule
    &MOSTPOP&0.1370&0.0909&0.1023&0.1811&0.0967&0.1165&0.2390&0.1007&0.1312\\
    \cmidrule{2-11}
    &Teacher&0.5598&0.3607&0.4101&0.7046&0.3802&0.4571&0.8175&0.3882&0.4859\\
    \multirow{3}{*}{\rotatebox{90}{Foursquare}}&Student&0.4869&0.3033&0.3489&0.6397&0.3239&0.3984&0.7746&0.3338&0.4333\\
    &RD&0.4932&0.3102&0.3555&0.6453&0.3302&0.4045&0.7771&0.3391&0.4377\\
    &CD&0.5006&0.3147&0.3608&0.6519&0.3354&0.3237&0.7789&0.3440&0.4421\\
    &RRD&0.5132&0.3258&0.3722&0.6616&0.3455&0.4202&0.7862&0.3540&0.4516\\
    \cmidrule{2-11}
    &IR-RRD &\textbf{0.5325}&\textbf{0.3377}&\textbf{0.3859}&\textbf{0.6770}&\textbf{0.3567}&\textbf{0.4325}&\textbf{0.7949}&\textbf{0.3650}&\textbf{0.4621}\\
    \bottomrule
  \end{tabular}
  \label{tab:IRRRD_main_BPR}
\end{table}

\begin{table}[h]
\setlength\tabcolsep{1.8pt}
\footnotesize
\RowStretch{1.2}
  \caption{Recommendation performances of NeuMF ($\phi=0.1$). 
  IR-RRD outperforms the best baseline on 0.05 level for the paired t-test.}
  
  \begin{tabular}{clccc ccc ccc}
    \toprule 
     Dataset & KD Method & H@5 & M@5 & N@5 & H@10 & M@10 & N@10 & H@20 & M@20 & N@20 \\
    \midrule
    &MOSTPOP&0.1389&0.0820&0.0960&0.1920&0.0890&0.1131&0.2610&0.0938&0.1305\\
    \cmidrule{2-11}
    &Teacher&0.4790&0.3318&0.3684&0.5827&0.3457&0.4020&0.6748&0.3521&0.4254\\
    \multirow{3}{*}{\rotatebox{90}{CiteULike}}&Student&0.3867&0.2531&0.2865&0.4909&0.2670&0.3202&0.5833&0.2738&0.3436\\
    &RD&0.4179&0.2760&0.3113&0.5211&0.2896&0.3444&0.6227&0.2958&0.3696\\
    &CD&0.4025&0.2633&0.2979&0.5030&0.2769&0.3306&0.6053&0.2822&0.3550\\
    &RRD &0.4737&0.3086&0.3497&0.5800&0.3236&0.3847&0.6765&0.3305&0.4094\\
    \cmidrule{2-11}
    &IR-RRD &\textbf{0.4862}&\textbf{0.3165}&\textbf{0.3587}&\textbf{0.5936}&\textbf{0.3304}&\textbf{0.3930}&\textbf{0.6908}&\textbf{0.3372}&\textbf{0.4174}\\
    \midrule
    &MOSTPOP&0.1370&0.0909&0.1023&0.1811&0.0967&0.1165&0.2390&0.1007&0.1312\\
    \cmidrule{2-11}
    &Teacher&0.5436&0.3464&0.3954&0.6906&0.3662&0.4430&0.8085&0.3746&0.4731\\
    \multirow{3}{*}{\rotatebox{90}{Foursquare}}&Student&0.4754&0.2847&0.3319&0.6343&0.3060&0.3833&0.7724&0.3157&0.4185\\
    &RD&0.4789&0.2918&0.3380&0.6368&0.3110&0.3878&0.7761&0.3173&0.4205\\
    &CD&0.4904&0.2979&0.3456&0.6477&0.3156&0.3940&0.7845&0.3260&0.4293\\
    &RRD &0.5172&0.3110&0.3621&0.6739&0.3321&0.4132&0.7982&0.3409&0.4450\\
    \cmidrule{2-11}
    &IR-RRD &\textbf{0.5212}&\textbf{0.3157}&\textbf{0.3666}&\textbf{0.6752}&\textbf{0.3363}&\textbf{0.4164}&\textbf{0.7985}&\textbf{0.3450}&\textbf{0.4476}\\
    \bottomrule
  \end{tabular}
  \label{tab:IRRRD_main_NeuMF}
\end{table}

\begin{figure}[h!]
\centering
\begin{subfigure}[t]{0.80\linewidth}
    \includegraphics[width=\linewidth]{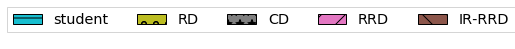}
\end{subfigure}

\begin{subfigure}[t]{0.48\linewidth}
    \includegraphics[width=\linewidth]{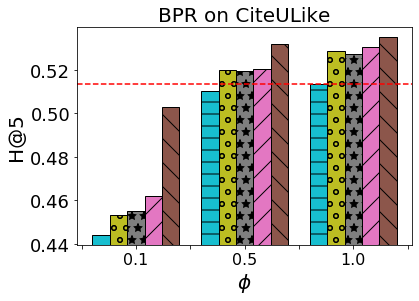}
\end{subfigure}
\begin{subfigure}[t]{0.48\linewidth}
    \includegraphics[width=\linewidth]{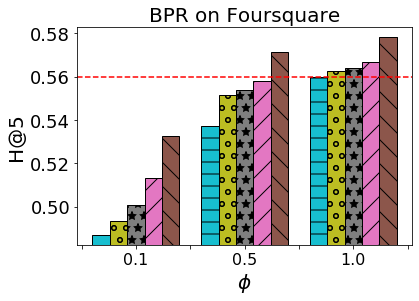}
\end{subfigure} 
\\
\begin{subfigure}[t]{0.48\linewidth}
    \includegraphics[width=\linewidth]{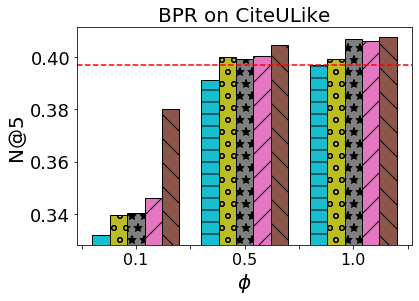}
\end{subfigure} 
\begin{subfigure}[t]{0.48\linewidth}
    \includegraphics[width=\linewidth]{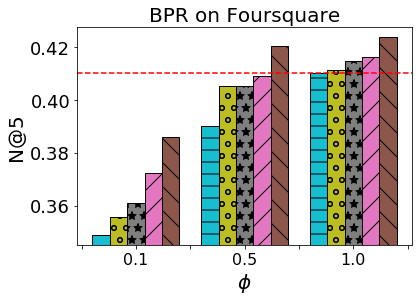}
\end{subfigure}
\\
\begin{subfigure}[t]{0.48\linewidth}
    \includegraphics[width=\linewidth]{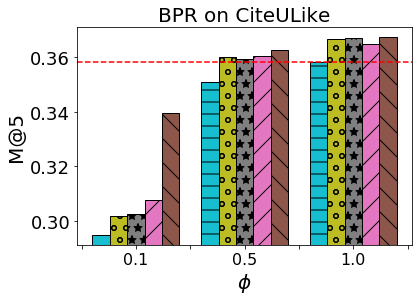}
\end{subfigure} 
\begin{subfigure}[t]{0.48\linewidth}
    \includegraphics[width=\linewidth]{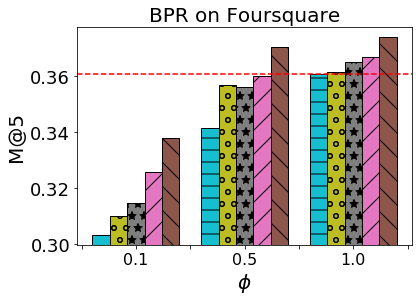}
\end{subfigure} 
\centering
\label{fig:IRRRD_model_size_BPR}
\caption{Recommendation Performance of BPR across three different student model sizes. (Red dotted line: Teacher)}
\end{figure}

\begin{figure}[h!]
\centering
\begin{subfigure}[t]{0.80\linewidth}
    \includegraphics[width=\linewidth]{chapters/irrrd/images/legend.png}
\end{subfigure}

\begin{subfigure}[t]{0.48\linewidth}
    \includegraphics[width=\linewidth]{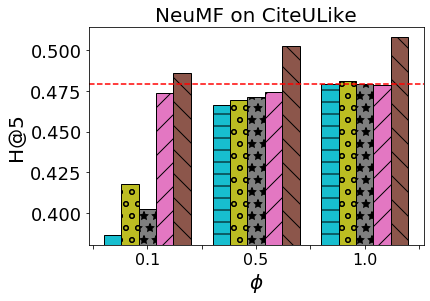}
\end{subfigure}
\begin{subfigure}[t]{0.48\linewidth}
    \includegraphics[width=\linewidth]{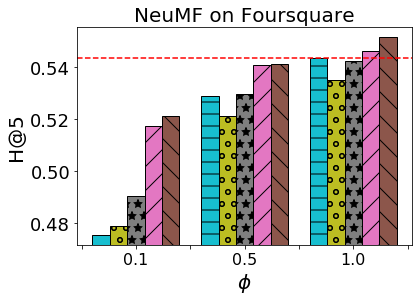}
\end{subfigure} 
\\
\begin{subfigure}[t]{0.48\linewidth}
    \includegraphics[width=\linewidth]{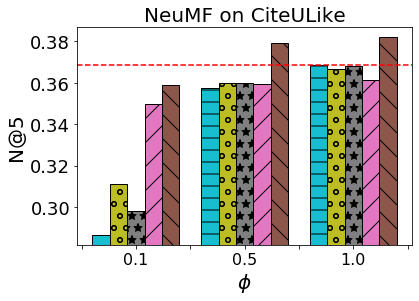}
\end{subfigure} 
\begin{subfigure}[t]{0.48\linewidth}
    \includegraphics[width=\linewidth]{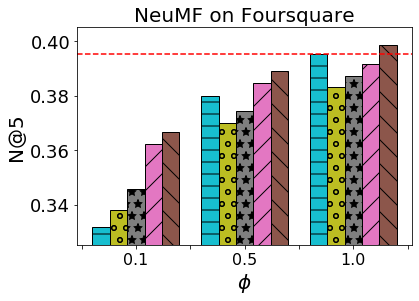}
\end{subfigure}
\\
\begin{subfigure}[t]{0.48\linewidth}
    \includegraphics[width=\linewidth]{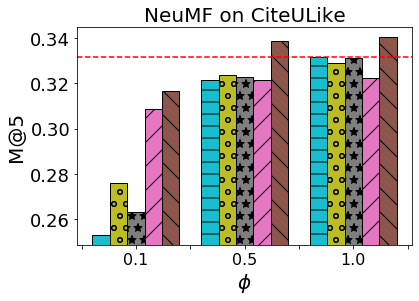}
\end{subfigure} 
\begin{subfigure}[t]{0.48\linewidth}
    \includegraphics[width=\linewidth]{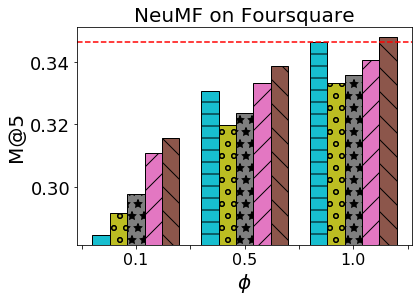}
\end{subfigure} 
\centering
\caption{Recommendation Performance of NeuMF across three different student model sizes. (Red dotted line: Teacher)}
\label{fig:IRRRD_model_size_NeuMF}
\end{figure}

\vspace{3pt}
\noindent
\textbf{Effects of item-side ranking regularization.}
We observe that the state-of-the-art UKD method is considerably improved with the proposed item-side ranking regularization.
In specific, IR-RRD achieves approximately 9\% for BPR and 3\% for NeuMF improvements respectively on CiteULike dataset ($\phi=0.1$).
Also, we observe that the proposed IR method consistently improves the performance of RRD in the three different sizes of the student recommender.
RRD enforces the student to accurately follow the teacher's user-side ranking orders.
However, this approach can destroy the item-side ranking relationships in the student, and further limits the effectiveness of the distillation.
The proposed IR method prevents the student from being overfitted to the user-side ranking by considering both the user-side and item-side information in the teacher's predictions, which leads to improved ranking performance.
With the regularization, IR-RRD achieves the best performance among all the methods.

\begin{table}[t!]
\setlength\tabcolsep{1.8pt}
\RowStretch{1.2}
\footnotesize
  \caption{Recommendation performances on full evaluation setup ($\phi=0.1$). IR-RRD outperforms RRD on 0.05 level for the paired t-test.}
   \begin{minipage}[t]{1\linewidth}
        \centering
  \caption*{\textbf{BPR}}
  \begin{tabular}{clccc ccc ccc}
    \toprule 
     Dataset & KD Method & H@5 & M@5 & N@5 & H@10 & M@10 & N@10 & H@20 & M@20 & N@20 \\
    \midrule
     &Teacher&0.1016&0.0574&0.0682&0.1470&0.0633&0.0828&0.2073&0.0673&0.0978\\
     \multirow{2}{*}{\rotatebox{90}{CiteULike}}&Student&0.0630&0.0348&0.0417&0.0973&0.0396&0.0534&0.1460&0.0428&0.0652\\
     &RRD&0.0730&0.0410&0.0489&0.1109&0.0460&0.0611&0.1670&0.0498&0.0752\\
     \cmidrule{2-11}
     &IR-RRD &\textbf{0.0854}&\textbf{0.0478}&\textbf{0.0571}&\textbf{0.1266}&\textbf{0.0532}&\textbf{0.0704}&\textbf{0.1832}&\textbf{0.0571}&\textbf{0.0846}\\
    \midrule
    &Teacher&0.0809&0.0488&0.0566&0.1137&0.0532&0.0670&0.1615&0.0564&0.0789\\
    \multirow{2}{*}{\rotatebox{90}{Foursquare}}&Student&0.0645&0.0388&0.0454&0.0911&0.0425&0.0544&0.1333&0.0453&0.0648\\
    &RRD&0.0711&0.0426&0.0498&0.1015&0.0466&0.0594&0.1464&0.0496&0.0708\\
    \cmidrule{2-11}
    &IR-RRD &\textbf{0.0761}&\textbf{0.0466}&\textbf{0.0539}&\textbf{0.1086}&\textbf{0.0507}&\textbf{0.0640}&\textbf{0.1557}&\textbf{0.0537}&\textbf{0.0749}\\
    \bottomrule
  \end{tabular}
  \end{minipage}
  \vspace{0.1cm}
  \vfill
     \begin{minipage}[t]{1\linewidth}
     \centering
  \caption*{\textbf{NeuMF}}
  \begin{tabular}{clccc ccc ccc}
    \toprule 
     Dataset & KD Method & H@5 & M@5 & N@5 & H@10 & M@10 & N@10 & H@20 & M@20 & N@20 \\
    \midrule
    &Teacher&0.0985&0.0584&0.0682&0.1420&0.0640&0.0818&0.2010&0.0677&0.0955\\
    \multirow{2}{*}{\rotatebox{90}{CiteULike}}&Student&0.0537&0.0275&0.034&0.0839&0.0319&0.0446&0.1219&0.0349&0.0549\\
    &RRD &0.0660&0.0329&0.0410&0.1078&0.0384&0.0544&0.1696&0.0426&0.0701\\
    \cmidrule{2-11}
    &IR-RRD &\textbf{0.0704}&\textbf{0.0380}&\textbf{0.0458}&\textbf{0.1170}&\textbf{0.0438}&\textbf{0.0601}&\textbf{0.1838}&\textbf{0.0480}&\textbf{0.0762}\\
    \midrule
    &Teacher&0.0704&0.0406&0.0479&0.1065&0.0453&0.0595&0.1541&0.0485&0.0714\\
    \multirow{2}{*}{\rotatebox{90}{Foursquare}}&Student&0.0436&0.0225&0.0277&0.0716&0.0262&0.0367&0.1151&0.0291&0.0476\\
    &RRD &0.0568&0.0307&0.0371&0.0894&0.0351&0.0477&0.1365&0.0382&0.0595\\
    \cmidrule{2-11}
    &IR-RRD &\textbf{0.0589}&\textbf{0.0318}&\textbf{0.0385}&\textbf{0.0902}&\textbf{0.0360}&\textbf{0.0485}&\textbf{0.1403}&\textbf{0.0391}&\textbf{0.0605}\\
    \bottomrule
  \end{tabular}
  \end{minipage}
  \label{tab:IRRRD_main_Full}
\end{table}

\begin{figure}[h!]
\centering
\begin{subfigure}[t]{0.50\linewidth}
    \includegraphics[width=\linewidth]{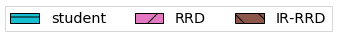}
\end{subfigure}

\begin{subfigure}[t]{0.40\linewidth}
    \includegraphics[width=\linewidth]{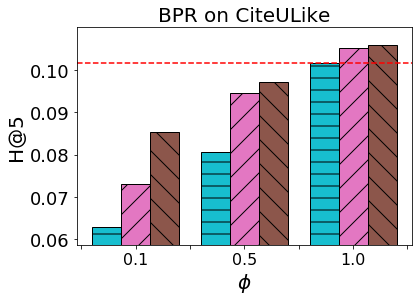}
\end{subfigure}
\begin{subfigure}[t]{0.40\linewidth}
    \includegraphics[width=\linewidth]{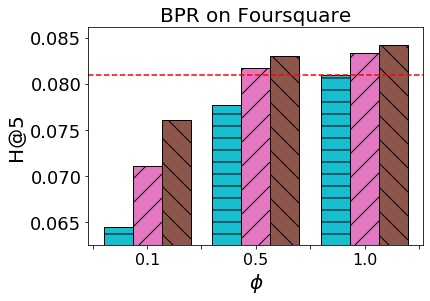}
\end{subfigure} 
\begin{subfigure}[t]{0.40\linewidth}
    \includegraphics[width=\linewidth]{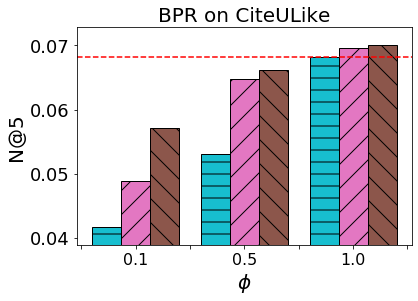}
\end{subfigure} 
\begin{subfigure}[t]{0.40\linewidth}
    \includegraphics[width=\linewidth]{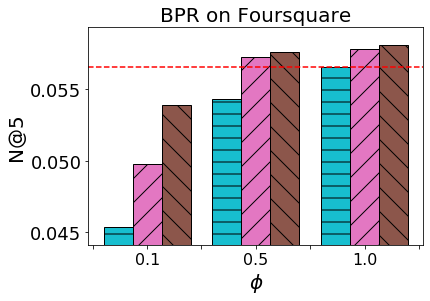}
\end{subfigure}
\\
\begin{subfigure}[t]{0.40\linewidth}
    \includegraphics[width=\linewidth]{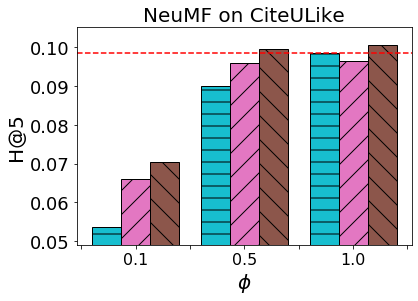}
\end{subfigure}
\begin{subfigure}[t]{0.40\linewidth}
    \includegraphics[width=\linewidth]{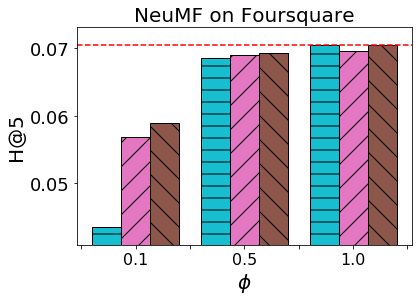}
\end{subfigure} 
\\
\begin{subfigure}[t]{0.40\linewidth}
    \includegraphics[width=\linewidth]{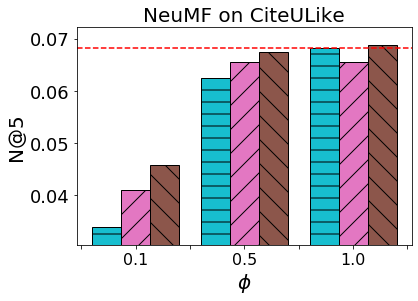}
\end{subfigure} 
\begin{subfigure}[t]{0.40\linewidth}
    \includegraphics[width=\linewidth]{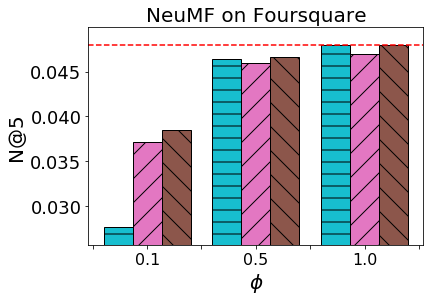}
\end{subfigure}
\caption{Recommendation Performance on full evaluation setup across three different student model sizes. (Red dotted line: Teacher)}
\label{fig:IRRRD_model_size_F}
\end{figure}

\vspace{3pt}
\noindent
\textbf{Analysis: IR in BPR.}
We observe that the proposed item-side ranking regularization achieves particularly large performance improvement in BPR.
One possible reason is that BPR is trained with user-side ranking loss function defined as:
\begin{equation}
\sum_{(u, i, j)} \ln \sigma\left( S(u,i) - S(u,j)\right)
\end{equation}
where $i \in \mathcal{I}/\mathcal{I}_{u}^{-}, j \in \mathcal{I}_{u}^{-}$ for each user $u$, $S$ is the student model.
Formally, BPR learns to assign a higher score on the observed item than the unobserved item for each user.
Since BPR is already trained with the user-side ranking loss function, distilling only the user-side ranking information from the teacher (i.e., RRD) can make BPR be easily biased and overfitted.
The proposed regularization method can alleviate this problem by enabling the student to consider the item-side ranking relations in the teacher's prediction, and thus it can achieve a large performance gain.

\vspace{3pt}
\noindent
\textbf{Analysis: IR in NeuMF.}
We observe that when the model has enough number of learning parameters (NeuMF with $\phi$= 0.5, 1.0 in Figure \ref{fig:IRRRD_model_size_NeuMF}), the existing prediction-based KD methods (i.e., RD, CD, RRD) have adverse effects on the student model.
We conjecture that this is because when the student has sufficient capacity to achieve comparable performance to the teacher, forcing it to accurately imitate the teacher’s predictions can act as a strong constraint for the student.
However, interestingly, we also observe that IR-RRD consistently improves the performance of the student in all settings.
As shown in Section \ref{sec:IRRRD_motiv}, the proposed method can provide additional information that is complementary to the knowledge transferred by UKD.
We believe the student model with a large capacity can take huge benefits from such additional information and can achieve better performance.

\vspace{3pt}
\noindent
\textbf{Analysis: RRD vs. IR-RRD on full evaluation setup.}
We further compare the performance of the best competitor (i.e., RRD) and IR-RRD on the full-ranking evaluation setup that uses all the unobserved items as the candidates.
In summary, we observe similar tendencies to previous results.
The proposed regularization method considerably improves the recommendation performance of RRD on the student model with three different sizes.
The results are provided in Table \ref{tab:IRRRD_main_Full} and Figure 6.

\begin{table}[h!]
\small
\centering
\setlength\tabcolsep{4pt}
\renewcommand{\arraystretch}{1.}
  \caption{Ranking consistency test of BPR ($\phi=0.1$). C@$K$ denotes Consistency@$K$.}
  
  \begin{tabular}{cclccc cc}
    \toprule 
     Dataset & Ranking Side & KD Method & C@5 & C@10 & C@20 & C@50 & C@100 \\
    \midrule
  &&Teacher&1.0 &1.0 &1.0 &1.0 &1.0\\
  &&Student&0.6788&0.7032&0.7294&0.7665&0.7980\\
  &Item Side&RRD&0.7285&0.7486&0.7704&0.8002&0.8258\\
  &&ItemRRD&0.9035&0.8977&0.8866&0.8814&0.8884\\
  &&IR-RRD&0.9074&0.9002&0.8885&0.8833&0.8905\\

    \cmidrule{2-8}
  &&Teacher&1.0 &1.0 &1.0 &1.0 &1.0\\
  \multirow{3}{*}{\rotatebox{90}{CiteULike}}&&Student&0.7337&0.7507&0.7695&0.7985&0.8253\\
  &User Side&RRD&0.8815&0.8883&0.8920&0.8925&0.8980\\
  &&ItemRRD&0.7414&0.7593&0.7789&0.8068&0.8321\\
  &&IR-RRD&0.8949&0.8991&0.9010&0.8999&0.9054\\
  
\cmidrule[1pt]{2-8}
  &&Teacher&1.0 &1.0 &1.0 &1.0 &1.0\\
  &\multirow{2}{*}{Harmonic}&Student&0.7052&0.7262&0.7489&0.7822&0.8114\\
  &\multirow{2}{*}{Mean}&RRD&0.7977&0.8125&0.8268&0.8438&0.8604\\
  &&ItemRRD&0.8145&0.8227&0.8293&0.8425&0.8593\\
  &&IR-RRD&\textbf{0.9011}&\textbf{0.8996}&\textbf{0.8947}&\textbf{0.8915}&\textbf{0.8979}\\
 \bottomrule
  &&Teacher&1.0 &1.0 &1.0 &1.0 &1.0\\
  &&Student&0.5209&0.5563&0.5967&0.6600&0.7190\\
  &Item Side&RRD&0.6481&0.6723&0.6999&0.7447&0.7863\\
  &&ItemRRD&0.8009&0.8077&0.8126&0.8199&0.8341\\
  &&IR-RRD&0.8010&0.8098&0.8179&0.8299&0.8462\\

    \cmidrule{2-8}
  &&Teacher&1.0 &1.0 &1.0 &1.0 &1.0\\
  \multirow{3}{*}{\rotatebox{90}{Foursquare}}&&Student&0.6350&0.6587&0.6823&0.6823&0.7496\\
  &User Side&RRD&0.8286&0.8394&0.8456&0.8454&0.8487\\
  &&ItemRRD&0.6542&0.6812&0.7068&0.7068&0.7729\\
  &&IR-RRD&0.8251&0.8396&0.8487&0.8487&0.8556\\
  
\cmidrule[1pt]{2-8}
  &&Teacher&1.0 &1.0 &1.0 &1.0 &1.0\\
  &\multirow{2}{*}{Harmonic}&Student&0.5723&0.6032&0.6366&0.6710&0.7340\\
  &\multirow{2}{*}{Mean}&RRD&0.7273&0.7466&0.7659&0.7919&0.8163\\
  &&ItemRRD&0.7202&0.7391&0.7560&0.7592&0.8023\\
  &&IR-RRD&\textbf{0.8129}&\textbf{0.8244}&\textbf{0.8330}&\textbf{0.8392}&\textbf{0.8509}\\
 \bottomrule
  \end{tabular}
  \label{tab:IRRRD_ranking_cons_1}
\end{table}    

\begin{table}[h!]
\centering
\small
\setlength\tabcolsep{4pt}
\renewcommand{\arraystretch}{1.}
  \caption{Ranking consistency test of NeuMF ($\phi=0.1$). C@$K$ denotes Consistency@$K$.}
  \begin{tabular}{cclccc cc}
    \toprule 
     Dataset & Ranking Side & KD Method & C@5 & C@10 & C@20 & C@50 & C@100 \\
    \midrule
  &&Teacher&1.0 &1.0 &1.0 &1.0 &1.0\\
  &&Student&0.2823&0.2982&0.3174&0.3408&0.3591\\
  &Item Side&RRD&0.4150&0.4403&0.4630&0.4871&0.5010\\
  &&ItemRRD&0.4450&0.4836&0.5087&0.5284&0.5382\\
  &&IR-RRD&0.4793&0.5262&0.5552&0.5746&0.5824\\

    \cmidrule{2-8}
  &&Teacher&1.0 &1.0 &1.0 &1.0 &1.0\\
  \multirow{3}{*}{\rotatebox{90}{CiteULike}}&&Student&0.3230&0.3226&0.3438&0.4042&0.4577\\
  &User Side&RRD&0.4250&0.4330&0.4676&0.5475&0.5967\\
  &&ItemRRD&0.3173&0.3182&0.3412&0.4070&0.4633\\
  &&IR-RRD&0.4466&0.4529&0.4856&0.5596&0.6068\\
  
\cmidrule[1pt]{2-8}
  &&Teacher&1.0 &1.0 &1.0 &1.0 &1.0\\
  &\multirow{2}{*}{Harmonic}&Student&0.3013&0.3099&0.3301&0.3698&0.4024\\
  &\multirow{2}{*}{Mean}&RRD&0.4199&0.4366&0.4653&0.5155&0.5447\\
  &&ItemRRD&0.3705&0.3838&0.4084&0.4598&0.4979\\
  &&IR-RRD&\textbf{0.4624}&\textbf{0.4868}&\textbf{0.5181}&\textbf{0.5670}&\textbf{0.5943}\\
 \bottomrule
  &&Teacher&1.0 &1.0 &1.0 &1.0 &1.0\\
  &&Student&0.6944&0.7032&0.7209&0.7652&0.8207\\
  &Item Side&RRD&0.7332&0.7392&0.7540&0.7942&0.8456\\
  &&ItemRRD&0.7876&0.7890&0.7975&0.8260&0.8664\\
  &&IR-RRD&0.8065&0.8072&0.8147&0.8401&0.8775\\

    \cmidrule{2-8}
  &&Teacher&1.0 &1.0 &1.0 &1.0 &1.0\\
  \multirow{3}{*}{\rotatebox{90}{Foursquare}}&&Student&0.8511&0.8225&0.7840&0.7251&0.6976\\
  &User Side&RRD&0.9245&0.8975&0.8573&0.7900&0.7532\\
  &&ItemRRD&0.8975&0.8668&0.8238&0.7564&0.7233\\
  &&IR-RRD&0.9056&0.8814&0.8463&0.7863&0.7556\\
  
\cmidrule[1pt]{2-8}
  &&Teacher&1.0 &1.0 &1.0 &1.0 &1.0\\
  &\multirow{2}{*}{Harmonic}&Student&0.7648&0.7582&0.7511&0.7446&0.7542\\
  &\multirow{2}{*}{Mean}&RRD&0.8178&0.8107&0.8023&0.7921&0.7967\\
  &&ItemRRD&0.8390&0.8261&0.8104&0.7897&0.7884\\
  &&IR-RRD&\textbf{0.8532}&\textbf{0.8427}&\textbf{0.8302}&\textbf{0.8123}&\textbf{0.8120}\\
 \bottomrule
  \end{tabular}
  \label{tab:IRRRD_ranking_cons_2}
\end{table}

\subsection{Ranking Consistency Analysis}
\begin{figure}[t]
\centering
    \includegraphics[width=1.0\linewidth]{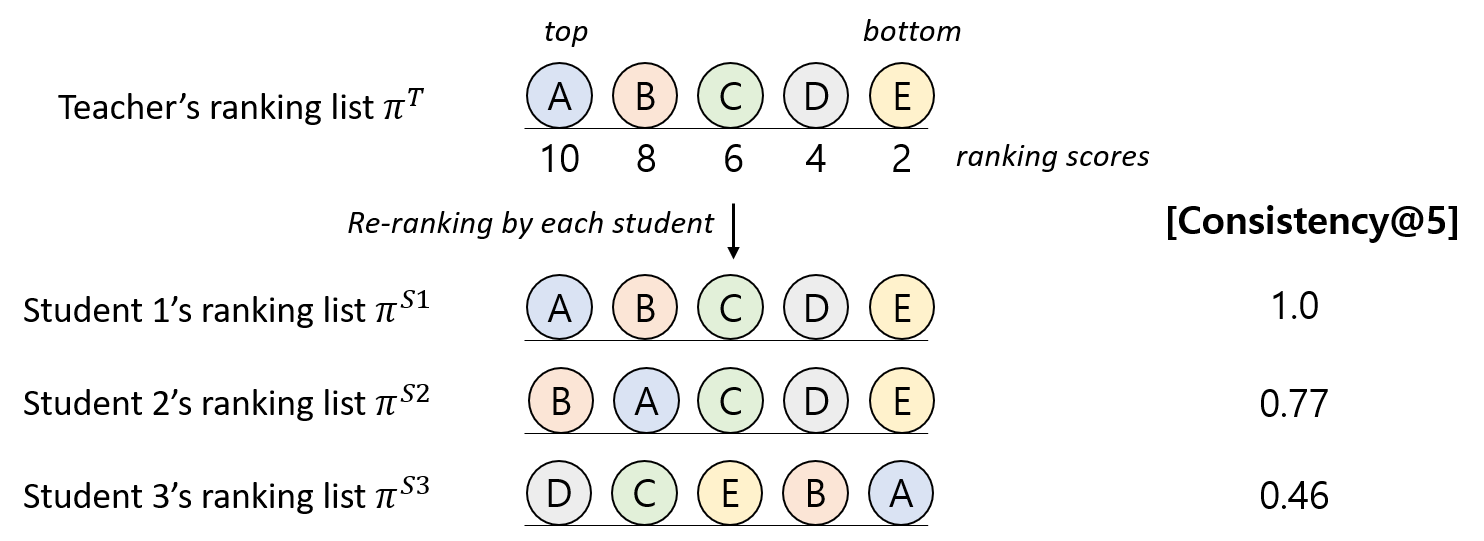}
\caption{An illustration of Consistency@K which measures the degree of deviation from $\pi^T$.}
\label{fig:IRRRD_consistency}
\end{figure}

To further investigate why the proposed regularization method considerably improves UKD method, we analyze the ranking consistency between the student and the teacher.
In other words, we quantitatively analyze how accurately the ranking orders in the teacher's predictions are preserved in the student's predictions.

\vspace{3pt}
\noindent
\textbf{Setup.}
We first make user-side and item-side ranking lists for all users and all items from the teacher recommender.
Then, we evaluate how accurately the student recommender preserves the ranking orders \textit{predicted by the teacher recommender}.
For quantitative evaluation, we introduce a new metric \textbf{Consistency@\textbf{K}} (C@K) which quantify the degree of deviation from the ground-truth permutation (i.e., a ranking order from the teacher).
An illustration of the metric with a toy example is provided in Figure \ref{fig:IRRRD_consistency}.
Let $\pi^T$ denote the ranking list predicted by the teacher, and let $\pi^S$ denote the re-ranked list of $\pi^T$ by the student.
If the student perfectly learns the ranking information from the teacher, $\pi^T$ should be equal to $\pi^S$.
Consistency@$K$ that measures how much $\pi^S$ deviates from $\pi^T$ is defined as follows:
\begin{equation}
\text{DCG}@K(\pi^{T}, \pi^{S}; T)=\sum_{k=1}^{K} \frac{2^{T({\pi^{S}_k})}-1}{\log (k+1)}
\end{equation}
\begin{equation}
\text{Consistency}@K(\pi^{T}, \pi^{S}; T)=\frac{\text{DCG}@K(\pi^{T}, \pi^{S}; T)}{\text{DCG}@K(\pi^{T}, \pi^{T}; T)}
\end{equation}
where $T$ is the teacher, $\pi^{S}_k$ is the $k$-th element in $\pi^{S}$, $T(\pi^{S}_k)$ is the ranking score on $\pi^{S}_k$ predicted by the teacher.
If $\pi^S[:K] = \pi^T[:K]$, Consistency@$K$ has a value of 1.
Obviously, the value of Consistency@$K$ for the teacher is always 1, as we compute the degree of deviation from the teacher's ranking order (i.e., Consistency@$K$($\pi^{T}, \pi^{T}$)=1).
Note that Consistency@$K$ are identical to NDCG@$K$ \cite{weimer2008cofi} where the ground-truth permutation is the teacher's ranking list $\pi^T$.
The above notations (i.e., $\pi^{T}, \pi^{S}, \text{DCG}@K,$ \text{Consistency}@$K$) are only used in this section for explaining the ranking consistency evaluation setup.
We compute Consistency@$K$ for the user-side and item-side ranking lists respectively.
Also, we compute the harmonic mean of the user-side and item-side results to measure how well the student imitates the teacher's predictions overall.
The results are presented in Table \ref{tab:IRRRD_ranking_cons_1} and Table \ref{tab:IRRRD_ranking_cons_2}.

\vspace{3pt}
\noindent
\textbf{Analysis: Effects of single-side ranking distillation.}
We observe that both the user-side ranking distillation (i.e., RRD) and the item-side ranking distillation (i.e., ItemRRD) are making the student better preserve the corresponding side ranking orders predicted by the teacher. 
Specifically, RRD achieves high consistency in terms of the user-side ranking, and Item-RRD achieves high consistency in terms of the item-side ranking.
We also observe that both RRD and ItemRRD increases the harmonic mean of the consistencies of two sides.
These results show that the ranking distillation method can help the student to follow the teacher's predictions.

\vspace{3pt}
\noindent
\textbf{Analysis: Effects of item-side ranking regularization.}
We first observe that the proposed regularization method (i.e., IR-RRD) achieves the best consistency with good balancing between the user-side and the item-side.
In specific, IR-RRD outperforms all the single-side ranking distillation methods (i.e., RRD, ItemRRD) in terms of harmonic mean.
Interestingly, we also observe that IR-RRD achieves comparable or even better single-side consistency compared to the single-side ranking distillation methods.
For example, on CiteULike dataset in Table \ref{tab:IRRRD_ranking_cons_2}, IR-RRD achieves considerably higher item-side consistency than ItemRRD, and similarly, IR-RRD achieves higher user-side consistency than RRD.
In other words, the student can better learn the single-side ranking orders with the proper consideration of the other side ranking information.
These results support our claim that the proposed regularization method indeed prevents the student to be overfitted to the user-side ranking information, and further enables the student to better preserve the teacher's prediction overall.

\begin{figure}[h]
\centering
\begin{subfigure}[t]{0.48\linewidth}
    \includegraphics[width=6.5cm,height=4.4cm]{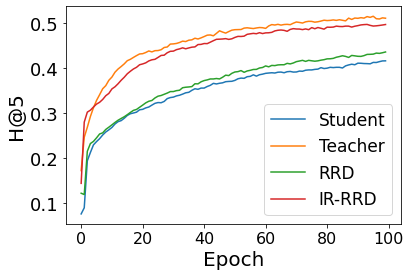}
\end{subfigure}
\begin{subfigure}[t]{0.48\linewidth}
    \includegraphics[width=6.5cm,height=4.4cm]{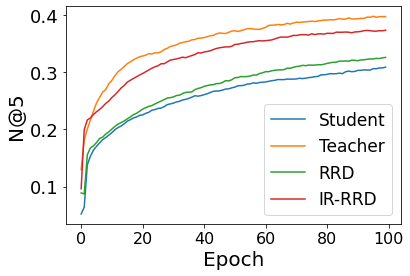}
\end{subfigure} 
\begin{subfigure}[t]{\linewidth}
    \subcaption{Base Model: BPR}
\end{subfigure}
\begin{subfigure}[t]{0.48\linewidth}
    \includegraphics[width=6.5cm,height=4.4cm]{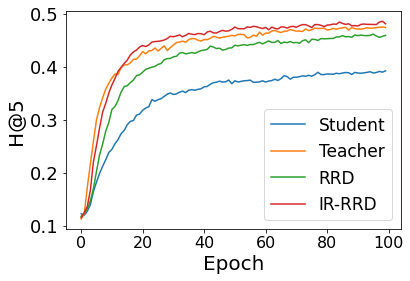}
\end{subfigure}
\begin{subfigure}[t]{0.48\linewidth}
    \includegraphics[width=6.5cm,height=4.4cm]{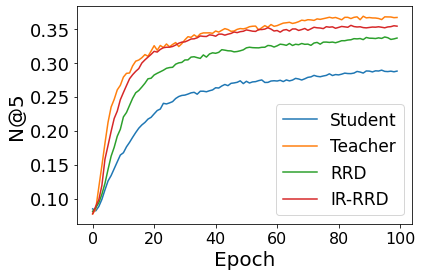}
\end{subfigure}
\begin{subfigure}[t]{\linewidth}
    \subcaption{Base Model: NeuMF}
\end{subfigure}
\centering
\caption{Recommendation performance vs. epoch on CiteULike dataset in the offline training phase.}
\label{fig:IRRRD_learning1}
\end{figure}

\begin{figure}[h]
\centering
\begin{subfigure}[t]{0.48\linewidth}
    \includegraphics[width=6.5cm,height=4.4cm]{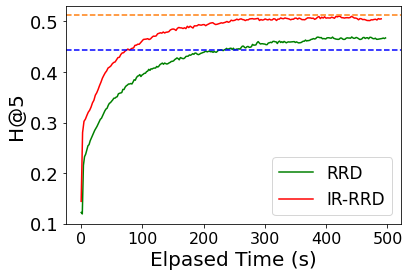}
\end{subfigure}
\begin{subfigure}[t]{0.48\linewidth}
    \includegraphics[width=6.5cm,height=4.4cm]{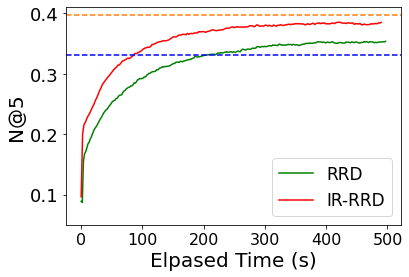}
\end{subfigure} 
\begin{subfigure}[t]{\linewidth}
    \subcaption{Base Model: BPR}
\end{subfigure}
\begin{subfigure}[t]{0.48\linewidth}
    \includegraphics[width=6.5cm,height=4.4cm]{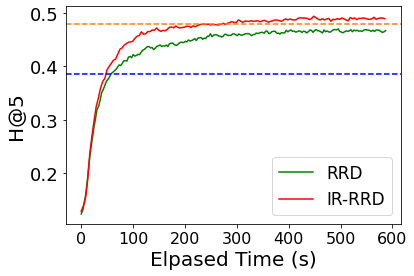}
\end{subfigure}
\begin{subfigure}[t]{0.48\linewidth}
    \includegraphics[width=6.5cm,height=4.4cm]{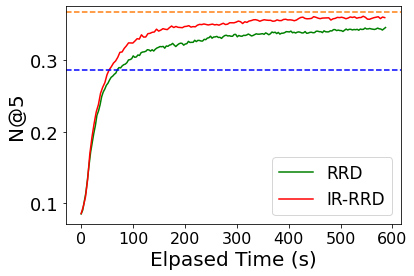}
\end{subfigure}
\begin{subfigure}[t]{\linewidth}
    \subcaption{Base Model: NeuMF}
\end{subfigure}
\centering
\caption{Recommendation performance vs. elapsed time of IR-RRD and RRD on CiteULike dataset in the offline training phase. Orange and blue dotted lines denote the performance of the teacher and the student, respectively.}
\label{fig:IRRRD_learning2}
\end{figure}

\subsection{Effects of IR in Learning}
We provide the learning curves of the base models on CiteULike ($\phi=0.1$) dataset in Figure \ref{fig:IRRRD_learning1} and Figure \ref{fig:IRRRD_learning2}.
We first observe that the proposed IR method considerably accelerates the learning of the student model.
Especially, in BPR, the performance is significantly improved from the very beginning of the training.
This result is consistent with the results from Section \ref{sec:IRRRD_PC}.
The IR method can effectively prevent the student from being overfitted to the user-side ranking, leading to better recommendation performance.
We also observe that the student model converges faster and achieves better performance with the item-side regularization.
IR method can provide additional supervisions that are complementary to the knowledge distilled by UKD in a properly regularized way considering the capacity gap between the teacher and the student.
These results again verify the superiority of the proposed method.

\begin{table}[h]
\centering
\renewcommand{\arraystretch}{1.}
  \caption{Model compactness and online inference efficiency. Time (seconds) indicates the wall time used for generating recommendation list for every user.}
  \begin{tabular}{ccccc}
  \toprule
    Dataset & Base Model & $\phi$ & Time (s) & The number of parameters\\
   \midrule
   &\multirow{3}*{BPR}&1.0&59.27s&6.08M\\
   &&0.5&57.53s&3.04M\\
   \multirow{3}*{\rotatebox[origin=c]{0}{CiteULike}}&&0.1&55.39s&0.61M\\
   \cmidrule{2-5}
   &\multirow{3}*{NeuMF}&1.0&79.27s&15.33M\\
   &&0.5&68.37s&7.63M\\
   &&0.1&58.27s&1.52M\\
   \midrule
&\multirow{3}*{BPR}&1.0&257.28s&9.61M\\
   &&0.5&249.19s&4.81M\\
   \multirow{3}*{\rotatebox[origin=c]{0}{Foursquare}}&&0.1&244.23s&0.96M\\
   \cmidrule{2-5}
   &\multirow{3}*{NeuMF}&1.0&342.84s&24.16M\\
   &&0.5&297.34s&12.05M\\
   &&0.1&255.24s&2.40M\\
   \bottomrule
  \end{tabular}
  \label{tab:IRRRD_effi}
\end{table}


\subsection{Online Inference Efficiency}
We also provide the online inference efficiency test results at the \textit{inference phase} in Table \ref{tab:IRRRD_effi}. 
Note that the inference efficiency of the base model is not affected by KD methods.
The inference efficiency is only affected by its size (i.e., the number of learning parameters).
We use PyTorch with CUDA from Tesla P40 GPU and Xeon on Gold 6148 CPU. 
A small model requires fewer computations and memory costs, so it can achieve lower inference latency.
In particular, neural recommender (i.e., NeuMF) which has a large number of learning parameters and complex structures takes more benefits from the smaller model size.

In our experiments, all the model parameters including the user/item embeddings can be loaded on a single GPU, and we can make the inferences based on the parallel computations with a large batch size.   
However, on a real-time RS application that has numerous numbers of users/items and requires a much deep and complex model architecture, making the inferences could be much difficult.
Specifically, due to the large model size, the model may not be uploaded on a single GPU, and also the batch size should be reduced.
In such scenario, the proposed method can lead to a larger improvement in inference efficiency.

\begin{figure}[t]
\centering

\begin{subfigure}[t]{0.48\linewidth}
    \includegraphics[width=6.5cm,height=5.1cm]{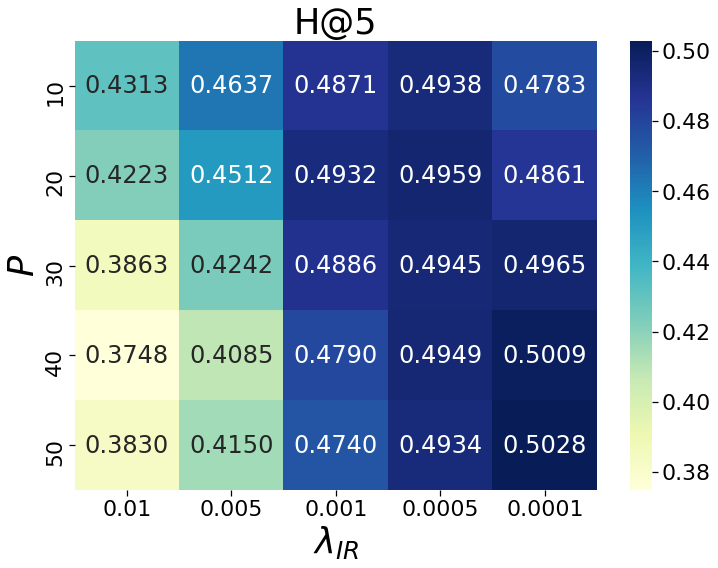}
\end{subfigure}
\begin{subfigure}[t]{0.48\linewidth}
    \includegraphics[width=6.5cm,height=5.1cm]{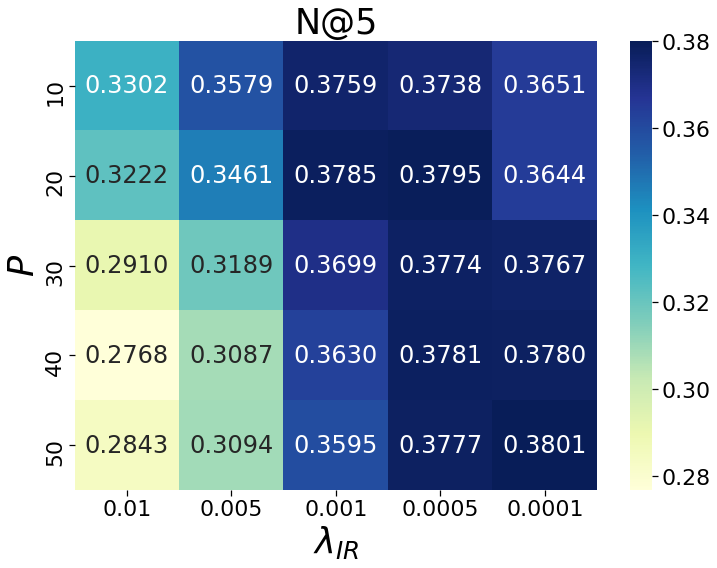}
\end{subfigure} 
\begin{subfigure}[t]{\linewidth}
    \subcaption{Base Model: BPR}
\end{subfigure}

\begin{subfigure}[t]{0.48\linewidth}
    \includegraphics[width=6.5cm,height=5.1cm]{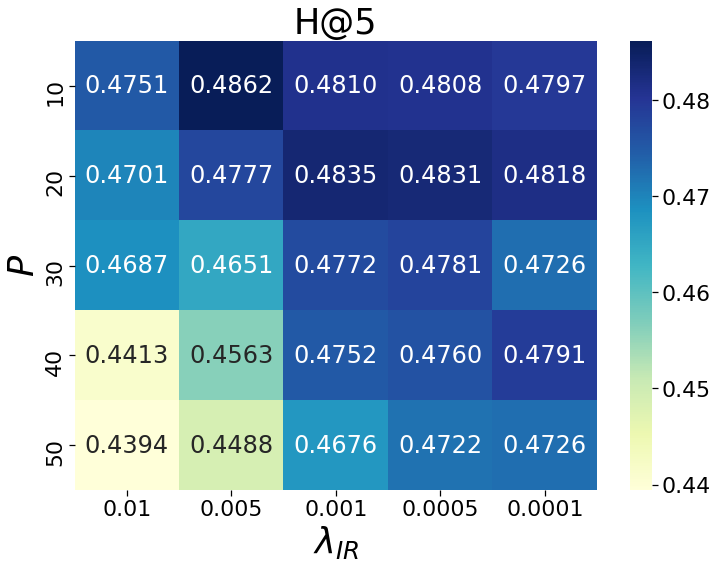}
\end{subfigure}
\begin{subfigure}[t]{0.48\linewidth}
    \includegraphics[width=6.5cm,height=5.1cm]{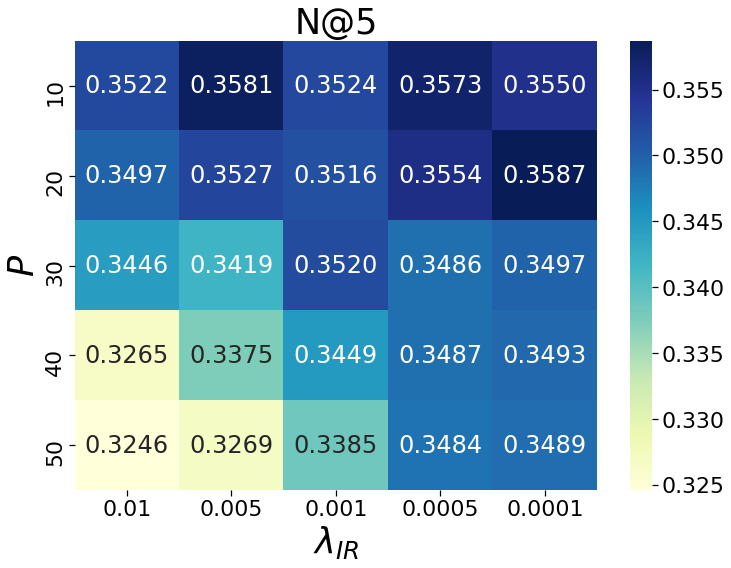}
\end{subfigure}
\begin{subfigure}[t]{\linewidth}
    \subcaption{Base Model: NeuMF}
\end{subfigure}
\centering
\caption{Effects of $P$ and $\lambda_{IR}$ in IR-RRD on citeULike dataset ($\phi=0.1$).}
\label{fig:IRRRD_hp}
\end{figure}

\subsection{Hyperparameter Study}
In this section, we provide hyperparameter study to provide guidance for hyperparameter selection of the proposed method.
We show the effects of two hyperparameters: $\lambda_{IR}$ that controls the importance of the item-side regularization and the number of positive users ($P$).
Note that in our experiment, the number of negative users ($N$) is set to the same with $P$.
IR shows very stable performance with varying $N$.
The results are summarized in Figure \ref{fig:IRRRD_hp}.

Since the type of base model's loss function is very different from that of the proposed IR method, it is important to properly balance the losses by using $\lambda_{IR}$.
For BPR, which uses the pair-wise ranking loss function (i.e., BPR Loss), the best performance is achieved when the magnitude of IR loss is approximately 75-85\% compared to that of the BPR loss with UKD.
For NeuMF, which uses the point-wise loss function (i.e., binary cross-entropy), the best performance is achieved when the magnitude of IR loss is approximately 20-25\% compared to that of the BCE loss with UKD.
Also, the capacity of the student recommender is very limited compared to that of the teacher recommender, it is important to properly set the amount of item-side ranking knowledge to be distilled to the student.
We observe that if $P$ is too large (e.g., more than 40 in Fig. \ref{fig:IRRRD_hp} b), the performance of the student is degraded.
The best performance is achieved when $P$ is near 40-50 for BPR and 10-20 for NeuMF.

\begin{table}[t!]
\setlength\tabcolsep{3.5pt}
\footnotesize
\RowStretch{1.}
  \caption{Recommendation performances on Kakao Brunch dataset. IR-RRD outperforms RRD on 0.05 level for the paired t-test.}
\centering
  \begin{tabular}{lccc ccc ccc}
    \toprule 
     KD Method & H@5 & M@5 & N@5 & H@10 & M@10 & N@10 & H@20 & M@20 & N@20 \\
    \midrule
     Teacher&0.6577&0.4563&0.5072&0.7313&0.4663&0.5311&0.7801&0.4697&0.5435\\
     Student&0.5160&0.3245&0.3726&0.6207&0.3386&0.4063&0.7107&0.3454&0.4307\\
     RRD&0.5300&0.3392&0.3880&0.6420&0.3540&0.4237&0.7200&0.3596&0.4437\\
     \cmidrule{1-10}
     IR-RRD &\textbf{0.5557}&\textbf{0.3553}&\textbf{0.4054}&\textbf{0.6617}&\textbf{0.3696}&\textbf{0.4397}&\textbf{0.7448}&\textbf{0.3754}&\textbf{0.4608}\\
    \bottomrule
  \end{tabular}
  \label{tab:IRRRD_main_kakao}
\end{table}

\subsection{Results on Large-scale Application}
Lastly, we provide the results of the proposed IR method on a large-scale web application:
Kakao Brunch\footnote{\url{https://brunch.co.kr}} which is one of the largest commercial blog platforms in South Korea.
The public dataset\footnote{\url{https://arena.kakao.com/c/6/data}} contains users’ historical logs on clicking blog articles from Kakao Brunch.
In specific, the dataset has 22,110,706 active users’ click logs from October 1, 2018 to March 1, 2019.
There are 310,758 unique users and 505,841 unique articles.
Here, we use BPR as the base model as it shows consistently better performance than NeuMF.
The latent dimensions of the teacher and the student are set to 100 and 10, respectively.
The results are reported in Table \ref{tab:IRRRD_main_kakao}.

We observe that the proposed item-side ranking regularization method effectively improves the effectiveness of RRD.
The student model trained with IR achieves higher recommendation performance compared to the student model trained with UKD method, RRD.
We believe the proposed IR method can be successfully applied to enhance the distillation quality of the existing UKD method on many real-world applications where low online inference latency is required.

\section{Summary}
In this paper, we propose a novel item-side ranking regularization method designed for maximizing the effects of the existing ranking distillation in RS.
We first provide in-depth analyses of the state-of-the-art ranking distillation method. 
Based on the analyses, we point out its limitation and room for further improvement.
Then, we introduce IR, Item-side ranking Regularization, that prevents the student to be overfitted to the user-side ranking information.
With IR, it turns out that the learning and the performance of the student can be significantly improved compared to when being trained with the existing distillation method.
We provide extensive experiment results supporting the superiority of the proposed method.
Also, we provide ranking consistency analysis verifying that the proposed method enables the student to better preserving the ranking orders predicted by the teacher.
In future work, we will investigate its effects (1) on more diverse base models, (2) when being combined with the existing KD method that distilling the latent knowledge from the teacher.

\chapter{Consensus Learning from Heterogeneous Learning Objectives for One-Class Collaborative Filtering}
\label{chapt:ConCF}

Over the past decades, for One-Class Collaborative Filtering (OCCF), many learning objectives have been researched based on a variety of underlying probabilistic models.
From our analysis, we observe that models trained with different OCCF objectives capture distinct aspects of user-item relationships, which in turn produces complementary recommendations.
This paper proposes a novel OCCF framework, named as ConCF, that exploits the complementarity from heterogeneous objectives throughout the training process, generating a more generalizable model.
ConCF constructs a multi-branch variant of a given target model by adding auxiliary heads, each of which is trained with heterogeneous objectives.
Then, it generates \textit{consensus} by consolidating the various views from the heads, and guides the heads based on the consensus.
The heads are collaboratively evolved based on their complementarity throughout the training, which again results in generating more accurate consensus iteratively.
After training, we convert the multi-branch architecture back to the original target model by removing the auxiliary heads, thus there is no extra inference cost for the deployment.
Our extensive experiments on real-world datasets demonstrate that ConCF significantly improves the generalization of the model by exploiting the complementarity from heterogeneous objectives.

\section{Introduction}

One-Class Collaborative Filtering (OCCF) aims to discover users' preferences and recommend the items that they might like in the future, given a set of observed user-item interactions (e.g., clicks or purchases) \cite{pan2008one, BUIR}.
To effectively learn users' preferences from such implicit feedback, many learning objectives (or objective functions) have been researched based on a variety of underlying probabilistic models.
In particular, pair-wise ranking objective \cite{BPR}, metric-learning objective \cite{CML}, and binary cross-entropy \cite{NeuMF} have shown good performance.
However, there is no absolute winner among them that can always achieve the best performance, because their superiority varies depending on datasets, model architectures, and evaluation metrics \cite{sun2020we}.
Since empirical comparison of all the choices is exceedingly costly in terms of both computing power and time consumption, most existing work simply leverages a \textit{generally good} one for their model optimization.


In this paper, we analyze OCCF models optimized by various learning objectives and observe that \emph{models trained with different objectives capture distinct aspects of user-item relationships.}
We observe that the test interactions (i.e., the ground truth of the users' preference) correctly predicted\footnote{We consider the test interactions included in the top-$N$ ranking list as correct predictions, also known as top-$N$ recommendation \cite{BUIR}.}
by each model are significantly different, regardless of their quantitative recommendation performance;
a model with low performance still correctly predicts considerable portions of the test interactions that the other models predict incorrectly.
Further, we demonstrate that the different and complementary knowledge induced by heterogeneous objectives can be exploited to provide more accurate recommendations to a larger number of users, compared to the case of considering each single-faceted knowledge.
The observations lead us to exploit the complementary knowledge for training a model to have a more complete understanding of user-item relationships.

\begin{figure}[t]
\centering
\begin{subfigure}[t]{0.7\linewidth}
    \includegraphics[width=\linewidth]{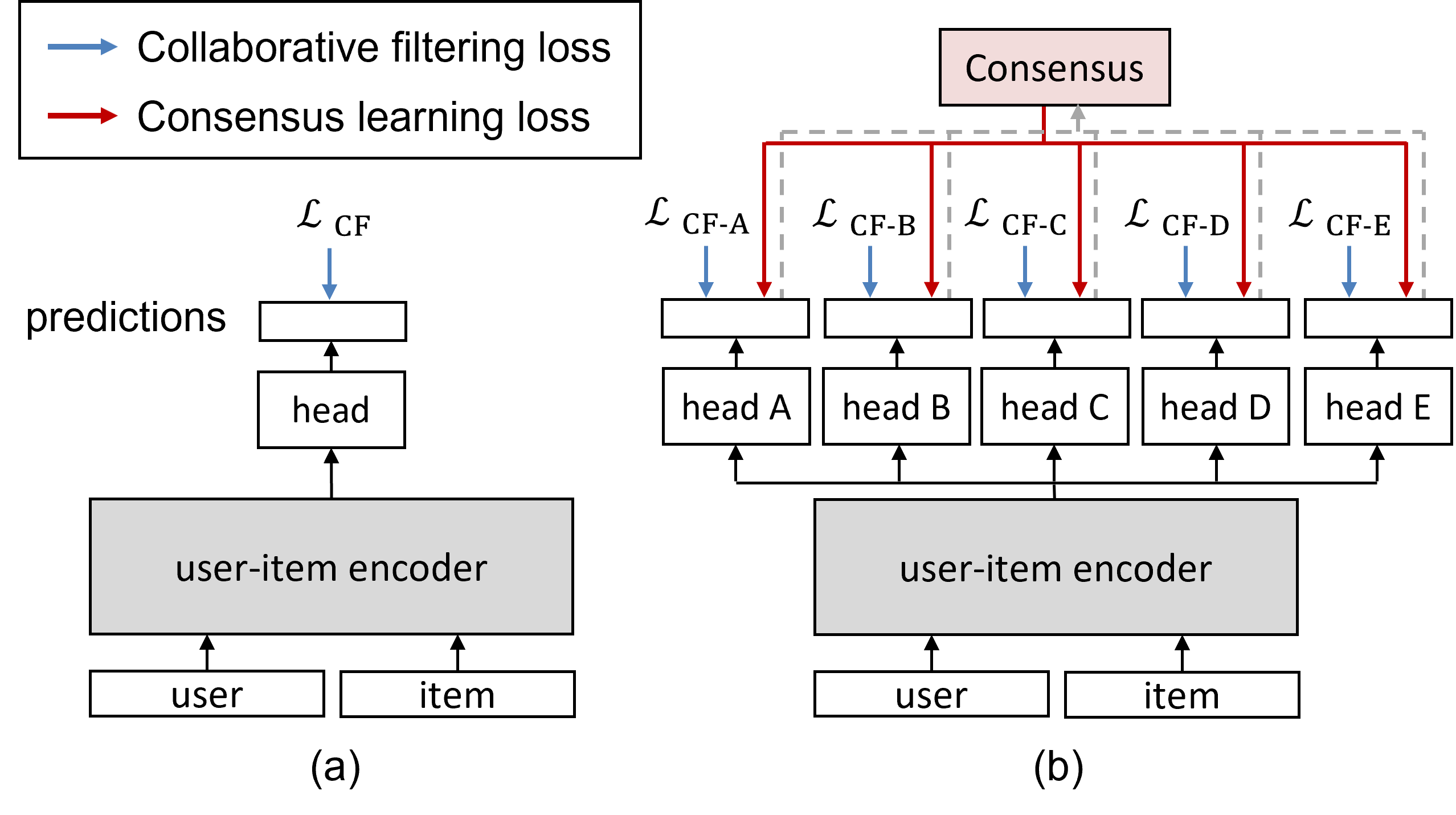}
\end{subfigure}
\caption{The conceptual illustrations of (a) the existing learning scheme and (b) the proposed framework. 
Unlike (a) that uses a given objective function, (b) exploits multi-faceted knowledge induced by the heterogeneous learning objectives, enabling to train a more generalizable model.}
\label{fig:ConCF_overview}
\end{figure}

We propose a new end-to-end OCCF framework, named as ConCF (\underline{Con}sensus learning for OC\underline{CF}), that exploits the complementarity from heterogeneous objectives throughout the training process, generating a more generalizable model (Figure \ref{fig:ConCF_overview}).
ConCF constructs a multi-branch variant of a given target model by adding auxiliary heads, each of which is trained with different objective functions. 
Then, it generates \textit{consensus} of multiple views from different heads, and guides the heads based on the consensus.
Concretely, each head is trained with two loss terms -- (1) the original collaborative filtering loss, and (2) a consensus learning loss that matches its prediction to the consensus.
The consensus, which consolidates the predictions from differently optimized heads, contains rich and accurate information not sufficiently discovered by a single head.
With the guidance, the heads are collaboratively evolved based on their complementarity during training, which again results in generating more accurate consensus iteratively.

The consensus learning, exploiting the complementarity from heterogeneous learning objectives, poses three unique challenges:

\vspace{0.05cm} \noindent
\textbf{C1}:
Since each objective is based on different probabilistic models, the distributions and semantics of their prediction scores are different.
Thus, the predictions from different heads need to be represented as a unified form so that they can learn interchangeably from each other.
For example, BPR objective \cite{BPR} encodes user-item relations as unbounded scores where larger values indicate higher relevance, whereas CML objective \cite{CML} learns the unit-ball metric space where user-item relations are encoded as their Euclidean distances in the range of [$0, 2$].
So, a score `0' for a user-item pair implies a weak relation in the former, but a very strong relation in the latter.
These discrepancies make the naive score-based learning \cite{DML, BD, ONE} among the heads inapplicable.

\vspace{0.1cm} \noindent
\textbf{C2}:
To generate informative consensus beneficial for all heads, it is essential to identify \textit{reliable} predictions from each head and selectively reflect them into the consensus.
As shown in our analysis, the user-item relationships correctly captured by each objective are significantly different.
Thus, only the reliable predictions correctly captured by each head should be included in the consensus.
Without considering the reliability, the consensus will contain incorrect predictions, providing inaccurate supervision for the heads.

\vspace{0.1cm} \noindent    
\textbf{C3}:
To exploit the complementarity, heterogeneous OCCF objectives having different convergence behaviors need to be properly balanced so that model parameters converge to useful features across all heads.
In case that some of the heads are too dominant during training, it incurs the imbalance that impedes the remaining heads from being fully optimized. 
This results in the degradation of the quality of the consensus and the overall effectiveness~of~ConCF.

\vspace{0.1cm}
ConCF introduces solutions to the aforementioned challenges:
\textbf{First}, to cope with the discrepancy among the outputs of the heads, ConCF utilizes \textit{ranking information}, a universal form of knowledge that can be derived from all heads.
Concretely, ConCF uses the information of relative preference priorities (e.g., user $u$ prefers item $i$ over item $j$.) revealed from the predictions to exchange the knowledge among the heads.
Utilizing the ranking information also has strengths in the top-$N$ recommendation, which provides a short item ranking list \cite{DERRD}.
\textbf{Second}, to generate informative consensus, ConCF takes into account not only the predicted rankings but also their reliability revealed from \textit{temporal consistency} of the predictions.
The temporal consistency measures how consistent the predicted ranking is during training, and according to our findings, this is an important indicator to identify reliable predictions.
By selectively reflecting the temporal-consistent predictions from each head, ConCF generates accurate consensus that every head can agree on, providing informative supervision for all heads.
\textbf{Third}, for balancing the heads optimized in a different way, we enforce all the heads to be trained at similar rates by dynamically adjusting the gradient scale of each head on the shared parameters.
We apply the gradient normalization technique \cite{gradnorm}, enabling the shared parameters to converge to beneficial features across all heads without additional hyperparameters for controlling~the~effects~of~each~heads.

In test time, we convert the multi-branch architecture back to the original single-branch model by removing the auxiliary heads, thus there is no additional inference cost for deployment.
Further, the consensus from multiple heads can be also used as a high capacity model in the scenario where there is less constraint on the inference cost.
The key contributions of our work are as follows:
\vspace{-0.1cm}
\begin{itemize}[leftmargin=*]
    \item Through our extensive analyses, we address the necessity of consolidating multiple views from heterogeneous OCCF objectives, which has not been studied well in the previous literature.
    
    \item We propose ConCF that exploits the complementarity from heterogeneous objectives throughout the training process.
    ConCF effectively deals with the challenges of the consensus learning from the heterogeneous objectives having distinct nature.
    
    \item We validate the superiority of ConCF by extensive experiments on real-world datasets. 
    ConCF achieves superior performance compared to the model optimized by a single objective.
    Also, ConCF generates a more generalizable model in a single-stage process than the alternative strategies of the conventional two-stage knowledge distillation or ensemble models.    
    
\end{itemize}

\section{Preliminaries}
\label{sec:ConCF_preliminary}

\subsection{Problem Formulation}
Let $\mathcal{U}$ and $\mathcal{I}$ denote a set of users and a set of items, respectively.
Given implicit user–item interactions (e.g., click, purchase), we build a binary matrix $\mathbf{R} \in \{0,1\}^{|\mathcal{U}| \times |\mathcal{I}|}$, where $r_{ui}=1$ if the user $u$ has interacted with the item $i$, otherwise $r_{ui}=0$.
Let $\mathcal{O}^{+}=\left\{(u, i) \mid r_{u i}=1\right\}$ denote a set of observed user-item interactions, and let $\mathcal{O}^{-}=\left\{(u, i) \mid r_{u i}=0\right\}$ denote a set of unobserved user-item pairs.
Also, let $\widetilde{\mathcal{O}}= \{ (u,i,j) \mid r_{u i}=1 \wedge r_{u j}=0 \}$ denote a set of triples $(u,i,j)$ such that user $u$ has interacted with $i$ but has not interacted with $j$.  
The goal of one-class collaborative filtering (OCCF) is to obtain the relevance score $\hat{r}_{ui} \in \mathbb{R}$ indicating how likely the user would interact with (or prefers to) the item $i$.
Based on the predicted scores, we recommend top-$N$ unobserved items with the highest scores for each user, called as~top-$N$~recommendation.

\subsection{Learning Objectives for OCCF}
\label{sec:ConCF_learning}
We select five learning objectives for OCCF that have been widely adopted in recent work.
We briefly review their key concepts and underlying probabilistic models.
In case that there are lots of variants, we focus on the representative basic form.
Note that any OCCF objective can be flexibly adopted in the proposed framework.
In this regard, we use alphabets to denote the objectives throughout the paper, i.e., CF-$x$, $x \in \{A, B, C, D, E\}$.

Let $\hat{r}_{ui}$ denote a relevance score of $(u, i)$ pair predicted by a model (Figure \ref{fig:ConCF_overview}a), and let $\theta$ denote the model parameters.

\vspace{0.1cm}
\noindent
\textbf{CF-A.} 
Bayesian personalized ranking (BPR) \cite{BPR} has been broadly used for model optimization \cite{BPR2021a, BPR2020a, BPR2021b, ADBPR, he2020lightgcn, NGCF}.
It maximizes the posterior probability $p(\theta \mid \widetilde{\mathcal{O}}) \propto p(\widetilde{\mathcal{O}} \mid \theta) p(\theta)$.
The probability that user $u$ prefers item $i$ to item $j$ is defined as $\sigma (\hat{r}_{ui}-\hat{r}_{uj})$, where $\sigma(\cdot)$ is the sigmoid function and $\hat{r}_{ui}$ has unbounded range. 
By taking the negative log-posterior, the loss function is formulated as follows:
\begin{equation}
    \begin{aligned}
        \mathcal{L}_{CF-A} = -\sum_{(u,i,j) \in \widetilde{\mathcal{O}}} \log \sigma (\hat{r}_{ui}-\hat{r}_{uj}). 
    \end{aligned}
\end{equation}
The prior $p(\theta)$ is computed as $\lVert \theta \rVert^2$ by adopting a normal distribution.
We omit the prior from Eq.1 for simplicity.
It is also known that optimizing BPR loss has a meaning of maximizing AUC (area under the ROC curve) \cite{BPR}.

\vspace{0.1cm} \noindent
\textbf{CF-B.}
Many recent studies have adopted metric learning \cite{CML, feng2015personalized, zhou2019collaborative, tay2018latent, li2020dynamic, li2020symmetric}.
The strength of the user’s preference on the item is considered inversely (or negatively) proportional to their Euclidean distance in the representation space, i.e., $\hat{r}_{ui}= -\lVert \mathbf{u}-\mathbf{i}\rVert_2$.
One of the most representative work \cite{CML} adopts triplet loss as follows:
\begin{equation}
    \begin{aligned}
        \mathcal{L}_{CF-B} = \sum_{(u,i,j) \in \widetilde{\mathcal{O}}} [- \hat{r}_{ui} + \hat{r}_{uj} + m]_+,
    \end{aligned}
\end{equation}
where $[x]_+ = max(x, 0)$ is the hinge loss, and $m$ is the margin.
The space has a unit-ball constraint that enforces the size of all user and item representations less than or equal to 1.

\vspace{0.1cm} \noindent
\textbf{CF-C.}
Binary cross-entropy is a highly popular choice for model optimization \cite{BCE2020a, NeuMF, BCE2020b, BCE2020c, BCE2020d, NCR}.
It assumes the observations are sampled from the Bernoulli distribution, i.e., 
$p(\mathcal{O}^+, \mathcal{O}^- \mid \theta) = \prod_{(u,i) \in \mathcal{O}^+ \cup \mathcal{O}^-} {\hat{r}_{ui}}^{r_{ui}}{(1-\hat{r}_{ui})}^{(1-r_{ui})}$ where $\hat{r}_{ui} \in [0,1]$.
By taking the negative log-likelihood, the loss function is defined as follows:
\begin{equation}
    \begin{aligned}
        \mathcal{L}_{CF-C} = -\sum_{(u,i) \in \mathcal{O}^+ \cup  \mathcal{O}^-} r_{ui} \log \hat{r}_{ui} + (1-r_{ui}) \log(1- \hat{r}_{ui})
    \end{aligned}
\end{equation}

\vspace{0.1cm} \noindent
\textbf{CF-D.} 
Mean squared error is another popular loss function \cite{PMF, SQR2020a, hu2008collaborative, CDAE, VAE}. 
It utilizes Gaussian distribution for the observation sampling distribution, i.e., $p(\mathcal{O}^+, \mathcal{O}^- \mid \theta) = \prod_{(u,i) \in \mathcal{O}^+ \cup \mathcal{O}^-} \mathcal{N}(r_{ui} \mid \hat{r}_{ui}, \sigma^2)$, with a fixed variance $\sigma^2$. The score $\hat{r}_{ui}$ has unbounded range.
By taking the negative log-likelihood, the loss function is defined as follows:
\begin{equation}
    \begin{aligned}
        \mathcal{L}_{CF-D} = \frac{1}{2} \sum_{(u,i) \in \mathcal{O}^+ \cup  \mathcal{O}^-}  \left( r_{ui} -\hat{r}_{ui}\right)^2
    \end{aligned}
\end{equation}


\vspace{0.1cm} \noindent
\textbf{CF-E.} The multinomial distribution, which is widely used in~language models \cite{blei2003latent}, has been adopted in many recent work \cite{VAE, MUL2020a, MUL2020b, MUL2020c}. 
It assumes each user's observations are sampled as
$p(\mathcal{O}^+_u, \mathcal{O}^-_u \mid \theta) = \frac{\Gamma(\sum_{i\in \mathcal{I}} r_{ui}+1)}{\prod_{i\in \mathcal{I}}\Gamma(r_{ui}+1)} \prod_{i\in \mathcal{I}}{\hat{r}_{ui}}^{r_{ui}} $ where $\sum_{i \in \mathcal{I}} \hat{r}_{ui} =1$ and $\hat{r}_{ui} \in [0,1]$.
By taking the negative log-likelihood, the loss function is defined as follows:
\begin{equation}
    \begin{aligned}
        \mathcal{L}_{CF-E} = -\sum_{(u,i) \in \mathcal{O}^+ \cup  \mathcal{O}^-} r_{ui}  \log \hat{r}_{ui}
    \end{aligned}
\end{equation}

\noindent
\textbf{Remarks.}
Most previous studies use a single OCCF objective for model optimization.
However, the heterogeneous objectives, which are based on different assumptions and probabilistic models, inject different inductive biases into the model \cite{inductive2}.
These inductive biases make the model prefer some hypotheses over other hypotheses, and thus better at solving certain problems and worse at others \cite{inductive2}.
In the next section, we investigate the complementarity of knowledge induced by heterogeneous learning~objectives.

\begin{figure}[t]
\centering
    \begin{minipage}[r]{1.0\linewidth}
        \centering
        \begin{subfigure}{0.4\textwidth}
            \includegraphics[height=4cm]{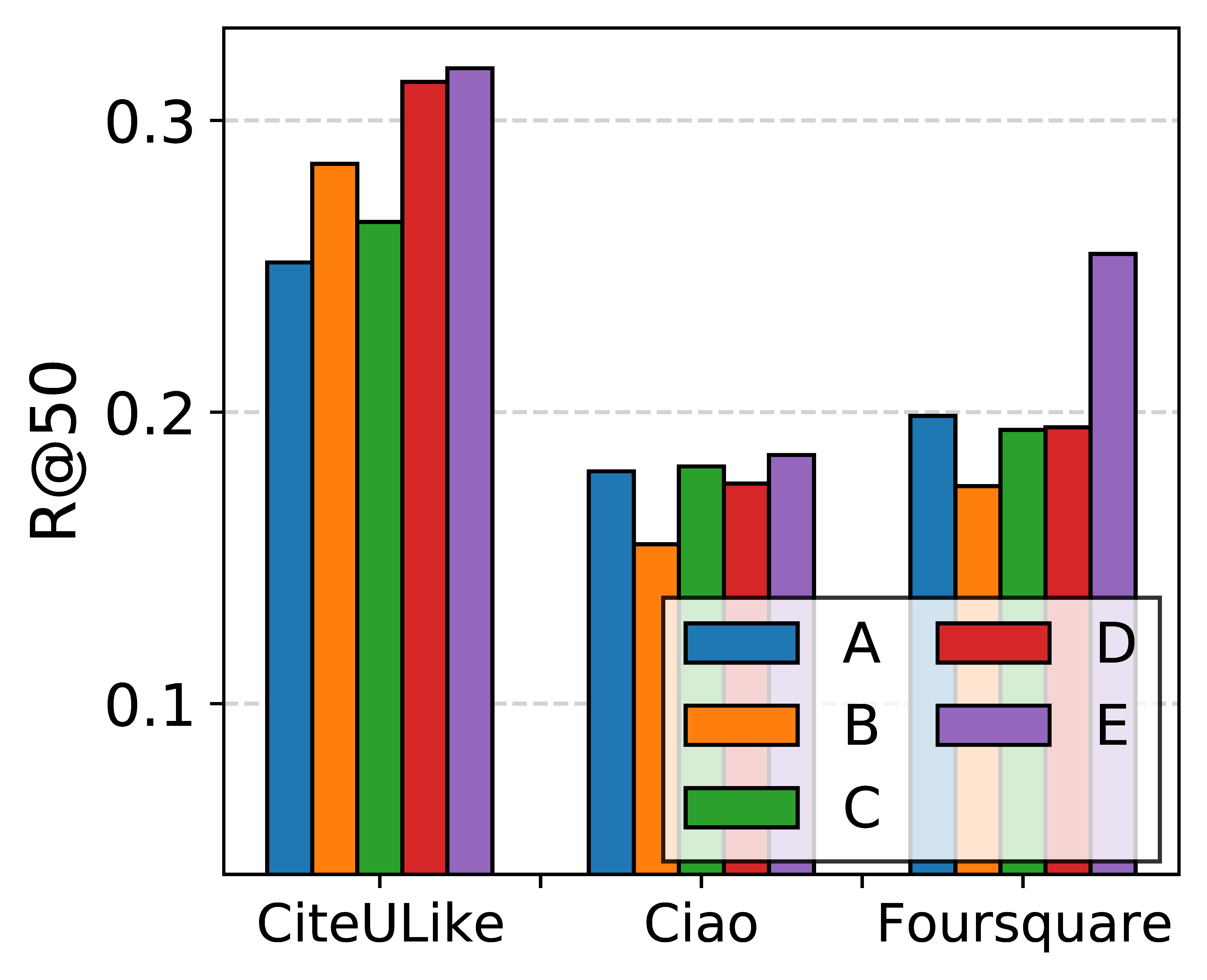}
        \end{subfigure}\\
        \begin{subfigure}{0.3\textwidth}
            \includegraphics[height=4cm]{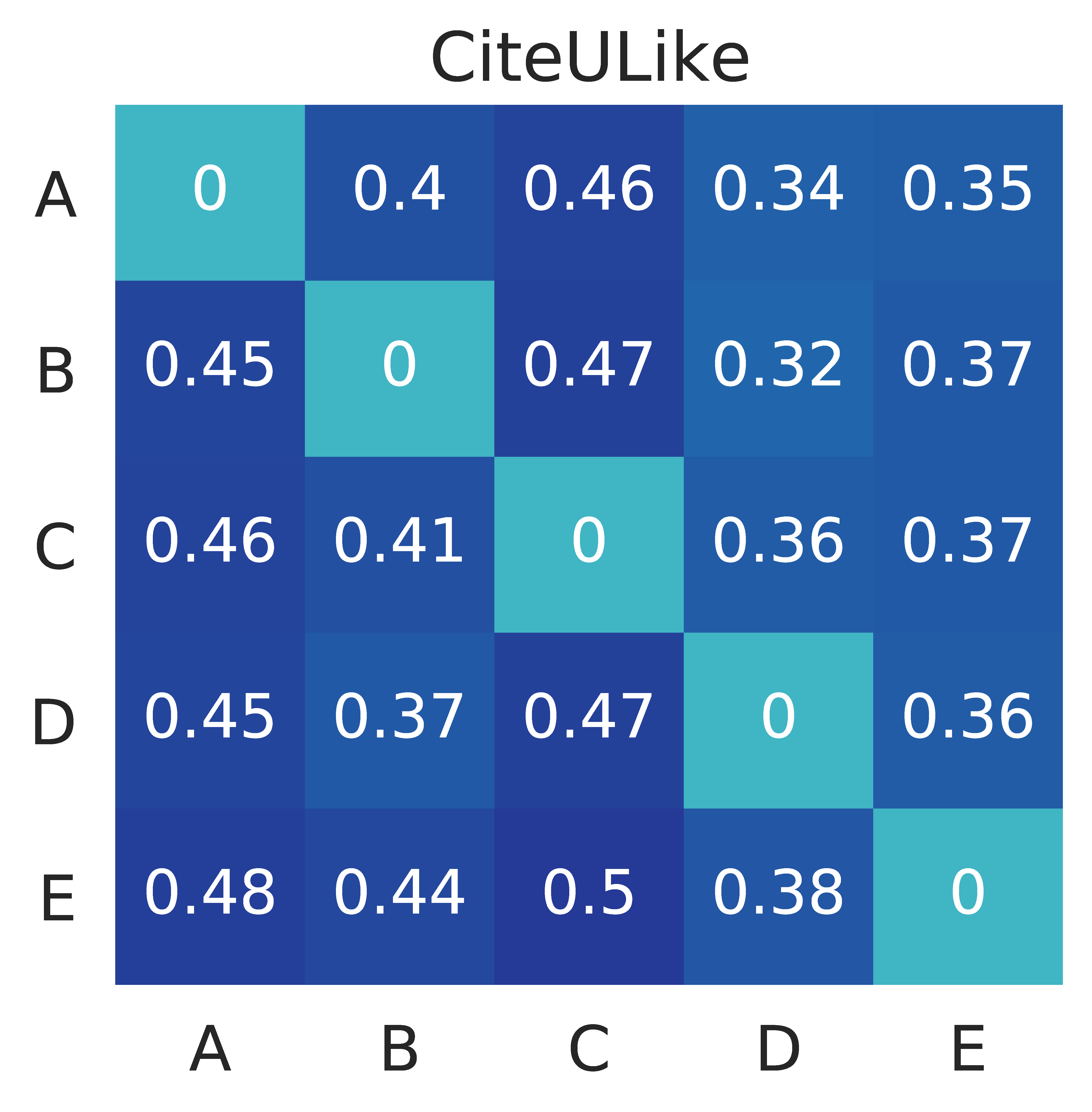}
        \end{subfigure}
        \hspace{0.1cm}
        \begin{subfigure}{0.3\textwidth}
         \includegraphics[height=4cm]{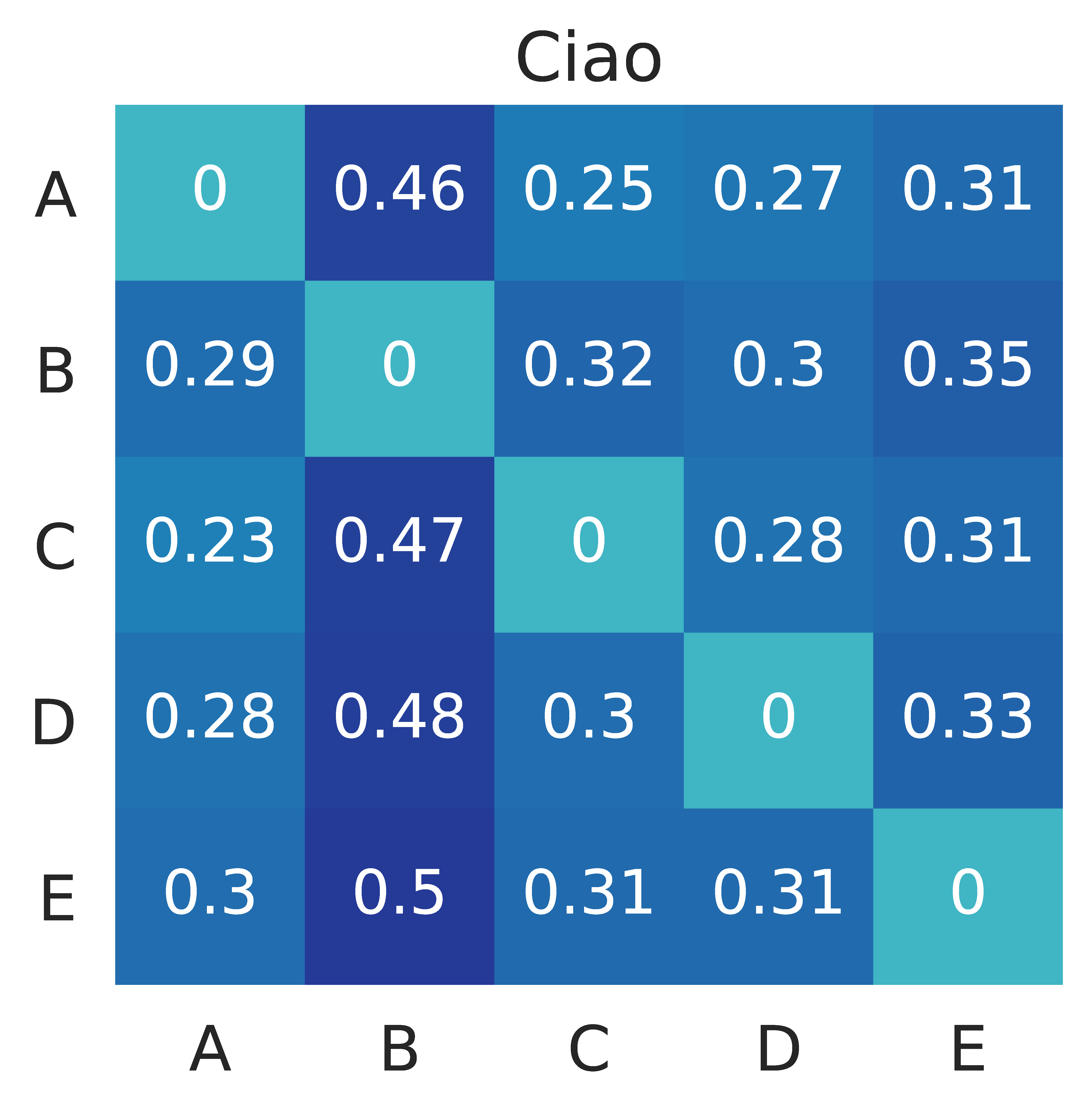}
        \end{subfigure}
        \hspace{0.1cm}
        \begin{subfigure}{0.3\textwidth}
          \includegraphics[height=4cm]{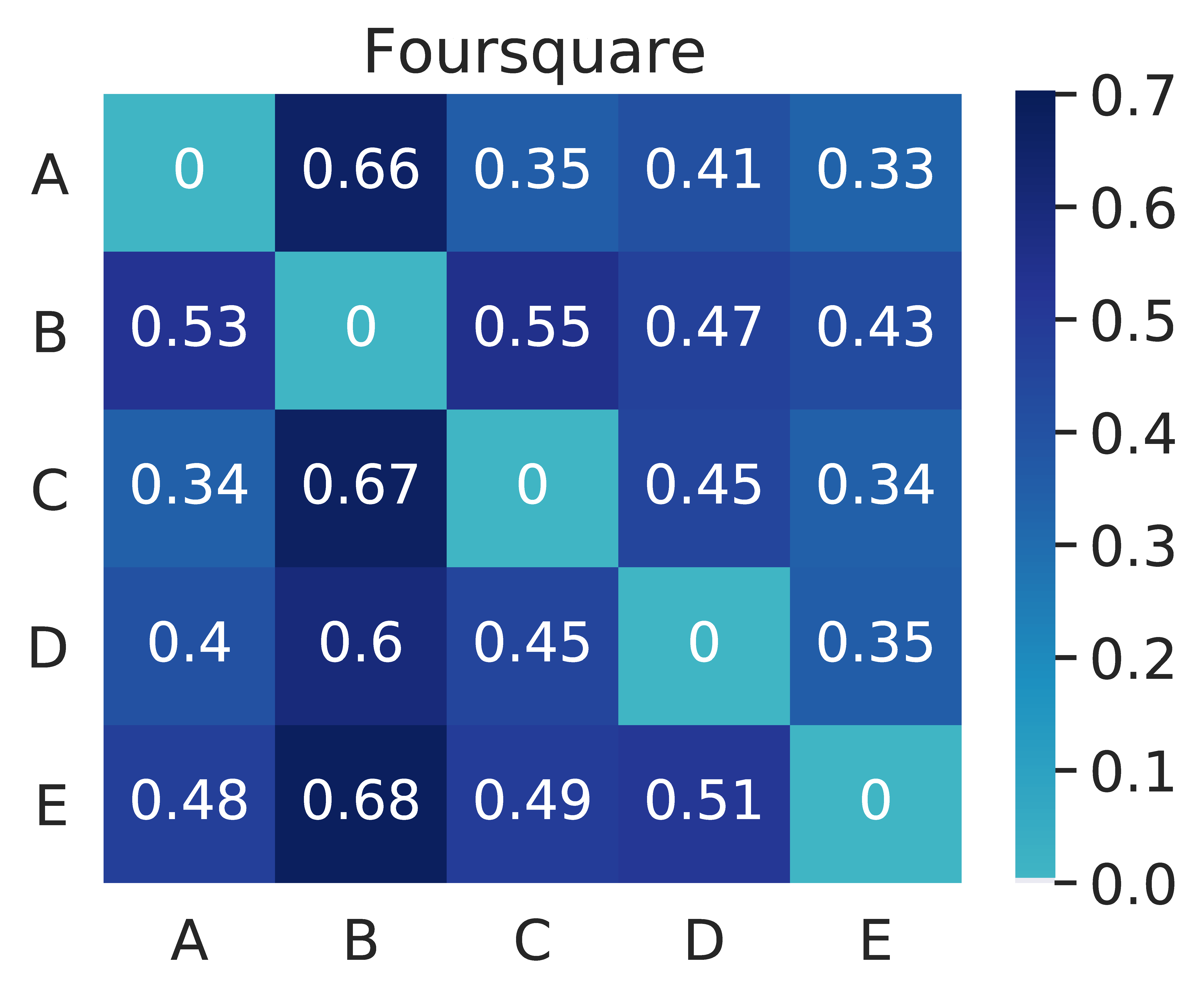}
        \end{subfigure}
        \subcaption{Recommendation performance (Recall@50) and PER($x$: row  ;  $y$: column) maps.}
    \end{minipage}
    \begin{minipage}[r]{0.4\linewidth}
        \centering
        \begin{subfigure}{1\textwidth}
          \includegraphics[height=4cm]{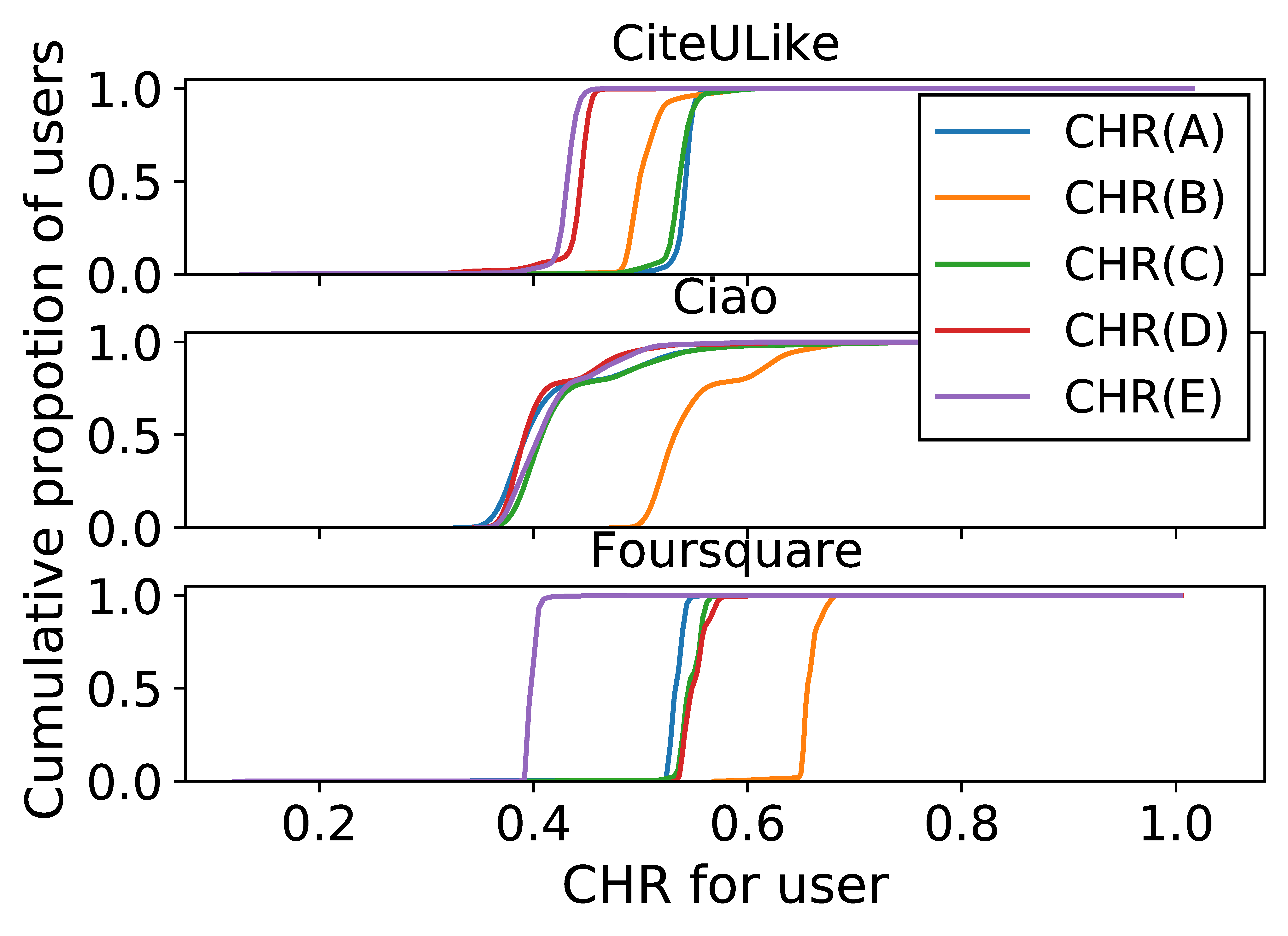}
        \end{subfigure}
        \subcaption{The CDF of CHR($y$; $\mathcal{M}_{O}$)}
    \end{minipage}
    \caption{Analyses of complementarity from different learning objectives. A model $x$ is trained with learning objective CF-$x$.}
    \label{fig:ConCF_prelim_analysis}
\end{figure}

\subsection{Analyses of Complementarity induced by Heterogeneous Learning Objectives}
\label{sec:ConCF_analysis}
We provide in-depth analyses of the knowledge induced by heterogeneous objectives on three real-world datasets.
We independently train five models (Fig.\ref{fig:ConCF_overview}a) with each CF function in Section \ref{sec:ConCF_learning},
and their performances are summarized in the leftmost chart in Fig.\ref{fig:ConCF_prelim_analysis}a.
All other controllable factors except for the loss function are fixed.
See Section \ref{app:ConCF_setup} for details of experiment setup.

\subsubsection{Pairwise Exclusive-hit Ratio (PER)}
 \noindent
$\text{PER}(x;y)$ quantifies the knowledge of user-item relationship captured by the model $x$ but failed to be effectively captured by the model $y$.
Let $\mathcal{H}_{x}$ denote the set of test interactions correctly predicted\footnote{We assume that the knowledge captured by a model is revealed by its predictions. Here, we consider the test items (i.e., the ground truth of the user’s preference) that are in the top-50 ranking list as the correct predictions.} by the model $x$.
We define PER as follows:
\begin{equation}
    \begin{aligned}
        \text {PER} (x; y) =\frac{| \mathcal{H}_{x}  - \mathcal{H}_{y}|}{|\mathcal{H}_{x}|}.
    \end{aligned}
\end{equation}
A high PER value means that the two models capture significantly different aspects from the same dataset.
We compute PER for all pairs of models trained with each learning objective, and the PER maps are summarized in Figure \ref{fig:ConCF_prelim_analysis}a.

\vspace{0.02cm}
\noindent
\textbf{Observations.} The models trained with different objectives have considerably high PER overall.
Even the model with low performance still better predicts considerable portions of the test interactions than the other models with much higher performance, 
e.g., on CiteULike, the model trained by CF-E achieves 26\% higher performance than that by CF-A, but it fails to correctly predict 35\% of the test interactions correctly predicted by the latter, i.e.,~PER(A;E)=0.35.

\subsubsection{Complementary Hit Ratio (CHR)}
 \noindent
We investigate the complementarity that a model can learn from the other models.
Given a set of trained models $\mathcal{M}$, we define CHR for a model $y$ as follows:
\begin{equation}
    \begin{aligned}
        \text {CHR} (y; \mathcal{M}) = \frac{|\bigcup_{x \in \mathcal{M}}\mathcal{H}_x - \mathcal{H}_y|}{|\bigcup_{x \in \mathcal{M}} \mathcal{H}_x|}.
    \end{aligned}
    \label{eq:ConCF_chr}
\end{equation}
$\bigcup_{x \in \mathcal{M}} \mathcal{H}_x$ is the total test interaction set correctly predicted by~$\mathcal{M}$.
CHR($y; \mathcal{M}$) quantifies the complementary knowledge that cannot be fully captured by the model $y$ but can be better understood by the other models.
Intuitively, a high CHR value means that the model can take huge benefits from the complementarity of the other models, whereas CHR$(\cdot)=0$ means that the model already captures all the knowledge of the other models.

We provide two statistical analyses using CHR.
\textbf{(1)} We compare CHR values induced by learning objective with values induced by two other factors, which influence the information that a model learns, investigated in recent work:
initialization \cite{DML} and model size (i.e., capacity) \cite{BD}.
Given a model $y$, we generate a set of models (including $y$) with five different conditions for each factor\footnote{
For the initial condition, we use the default initialization schemes provided in PyTorch and control it by setting different random seeds.
For the model size, we control the user/item embedding dimension $\in \{32, 48, 64, 80, 96\}$, and the size of subsequent layers are changed accordingly.
It is worth noting that different model sizes necessarily entail different initial conditions.}, i.e., $\mathcal{M}_{I}$ with different initial conditions, $\mathcal{M}_{C}$ with different capacities, and $\mathcal{M}_{O}$ with different learning objectives.
Then, we compute CHR for each factor, i.e., CHR($y;\mathcal{M}_{I}$), CHR($y;\mathcal{M}_{C}$), and CHR($y;\mathcal{M}_{O}$), and report the results in Table \ref{tbl:ConCF_CHR}. 
\textbf{(2)} Given $\mathcal{M}_{O}$, we compute \textit{user-level} CHR, then we present CDF of CHR for the users in Figure \ref{fig:ConCF_prelim_analysis}b.


\begin{table}[t]
\centering
\caption{CHR induced by three factors.
All values are computed for the same model $y$ having the identical initial condition, embedding dimension 64, and optimized by~CF-A.}
\renewcommand{\arraystretch}{0.7}
\renewcommand{\tabcolsep}{0.35cm}
\begin{tabular}{c ccc}
\toprule
Factors & CiteULike   & Ciao  & Foursquare   \\
\midrule
Initial condition & 0.30  & 0.23 & 0.29 \\
Model size & 0.32 & 0.28 & 0.38 \\
Learning objective & 0.56 & 0.43 & 0.54 \\
\toprule
\end{tabular} 
\label{tbl:ConCF_CHR}
\end{table}

\vspace{0.1cm}
\noindent
\textbf{Observations.}
\textbf{(1)} Learning objective incurs significantly high CHR (up to 75 \% higher than that from model size on CiteULike).
This indicates the learning objective is indeed an important factor affecting the knowledge captured by a model.
The other factors can be also exploited in the proposed framework to enhance the complementarity.
In this work, however, we focus on the learning objective which has not been studied well in the previous literature.
\textbf{(2)} Considerable complementarity exists for the majority of users, e.g., on CiteULike, most users have CHR larger than 0.4, which indicates that only less than 60\% of the total correct prediction set (i.e., $\bigcup_{x \in \mathcal{M}_o} \mathcal{H}_x$) is reached with a single learning~objective.

In sum, we conclude as follows:
(1) The learning objective is a key factor that determines the knowledge captured by the model.
(2) Models trained with different objectives learn distinct aspects of user-item relationships, which in turn produces complementary recommendations.
(3) The complementarity can be exploited to provide more accurate recommendations to more users, compared to the case of considering each single-faceted knowledge.

\section{Proposed Framework---ConCF}
\label{sec:ConCF_method}
We present our framework, named as ConCF (\underline{Con}sensus learning for OC\underline{CF}),
which exploits the complementarity from heterogeneous objectives throughout the training process, generating a more generalizable model (Fig.\ref{fig:ConCF_overview}b).
We first provide an overview of ConCF (Section \ref{sec:ConCF_overview}). 
Then, we introduce a consensus learning strategy from heterogeneous objectives (Section \ref{sec:ConCF_CE}).
Lastly, we present our solution to balance the heterogeneous objectives without additional hyperparameters (Section \ref{sec:ConCF_blance}).
The training details and Algorithm of ConCF are provided in Appendix~\ref{app:ConCF_training}.

\subsection{Overview}
\label{sec:ConCF_overview}
In training, ConCF constructs a multi-branch variant of a given target model by adding auxiliary heads, each of which is trained with heterogeneous learning objectives. 
Then, it generates \textit{consensus} of multiple views from different heads, and it guides the heads based on the consensus.
The superscript $t$ denotes $t$-th training step (or epoch).
The overall loss function is formulated as:
\begin{equation}
    \begin{aligned}
        \mathcal{L}^t = \sum_{x \in \mathcal{F}} \lambda_{x}^t \mathcal{L}^t_{x},
    \end{aligned}
    \label{eq:ConCF_ConCF_L}
\end{equation}
where $\mathcal{F}$ is the set of the heads, $\mathcal{L}_x^t$ is the loss function for head $x$ at training step $t$, and $\lambda_{x}^t$ is the trainable parameter for dynamically balancing the heads during training.
Each head $x$ is trained with two loss terms: the original CF loss ($\mathcal{L}^t_{CF\text{-}x}$) and the consensus learning loss which aligns its prediction to the consensus ($\mathcal{L}^t_{CL\text{-}x}$). 
\begin{equation}
    \begin{aligned}
        \mathcal{L}^t_x =  \mathcal{L}^t_{CF\text{-}x} + \alpha \mathcal{L}^t_{CL\text{-}x}
    \end{aligned}
    \label{eq:ConCF_Lx}
\end{equation}
where $\alpha$ is the hyperparameter controlling the effect of the consensus learning.
After training, we convert the multi-branch architecture back to the original single-branch model by removing the auxiliary heads, thus there is no additional inference cost for the deployment.
Further, the consensus from multiple heads can be also used as a high capacity model in the scenario where there is less constraint on the inference cost.

\subsection{Collaborative Evolution with Consensus}
\label{sec:ConCF_CE}
The consensus learning from the heterogeneous objectives raises several challenges.
First, the distributions and semantics of their prediction scores are different, thus the predictions from the heads need to be represented as a unified form so that they can learn interchangeably from each other.
For example, $\hat{r}_{ui}=0$ means a weak relation with CF-A, but a very strong relation with CF-B.
These discrepancies make the naive score-based knowledge exchange \cite{DML, BD, ONE} inapplicable.
Second, to generate informative consensus beneficial for all heads, it is essential to identify \textit{reliable} predictions from each head and selectively reflect them into the consensus.
As shown in Section \ref{sec:ConCF_analysis}, the user-item relationships correctly predicted by each objective are significantly different.
Without considering the reliability, the consensus will contain incorrect predictions, providing inaccurate supervision~for~the~heads.

We propose an effective strategy to resolve the challenges.
Instead of the score (i.e., $\hat{r}_{ui}$) itself, we utilize \textit{ranking information} (e.g., $\hat{r}_{ui} > \hat{r}_{uj} > \hat{r}_{uk} $).
These relative preference priorities are the universal form of knowledge that can be derived from all heads, enabling the knowledge exchange among the heads with different output distributions.
Further, we devise a new way of generating the consensus, considering both the predicted rankings and their reliability~revealed~from~\textit{temporal consistency}~of~the~predictions.

\subsubsection{Generating consensus}
 \noindent
For each user, we generate the item ranking list by sorting the prediction scores from each head.
Let $\pi^t_x$ denote the ranking list from head $x$ at training epoch $t$ ($u$ is omitted for simplicity).
We assess the importance of each item in terms of top-$N$ recommendation, then consolidate the importance from each head to generate the consensus.
Concretely, we generate the ranking consensus $\pi^t$ from $\{\pi^t_x \mid x \in \mathcal{F}\}$ by reranking items according to their overall importance.
The importance is assessed by two factors: \textit{ranking position} and \textit{consistency}.
The ranking position indicates the user's potential preference on each item, naturally, high-ranked items need to be considered important for the top-$N$ recommendation.
The consistency indicates how consistent the predicted ranking is during training, and according to our analysis, this is an important indicator to identify reliable~predictions.

\vspace{0.1cm}
\noindent
\textbf{Ranking position.}
Let $\text{rank}^{t}(x,i)$ denote the rank of item $i$ in $\pi^t_x$ where a lower value means a higher ranking position, i.e., $\text{rank}^{t}(x,i)=0$ is the highest rank.
The importance of item $i$ by ranking position is defined as follows:
\begin{equation}
\begin{aligned}
R^t_{x,i} = f(\text{rank}^{t}(x,i)),
\end{aligned}
\end{equation}
where $f$ is the monotonically decreasing function.
In this work, we use $f(k) = e^{-k/T}$ to put more emphasis on top positions where $T$ is a hyperparameter controlling the emphasis.

\vspace{0.1cm}
\noindent
\textbf{Consistency.}
To identify reliable predictions (i.e., correct top-ranked predictions) from each ranking list $\pi^t_x$, we define consistency of predicted ranking position as follows:
\begin{equation}
\begin{aligned}
C^t_{x,i} &= f(\text{std}[\text{rank}^{t-W}(x,i), \ldots, \text{rank}^{t}(x,i)])
\end{aligned}
\label{eq:ConCF_c}
\end{equation}
where $\text{std}[\cdot]$ is the standard deviation.
We compute the consistency of recent predictions from $t-W$ to $t$, where $W$ is the window size.
This metric gives higher importance to more consistent predictions.

\begin{figure}[t]
\centering
\begin{subfigure}[t]{0.7\linewidth}
    \includegraphics[width=\linewidth]{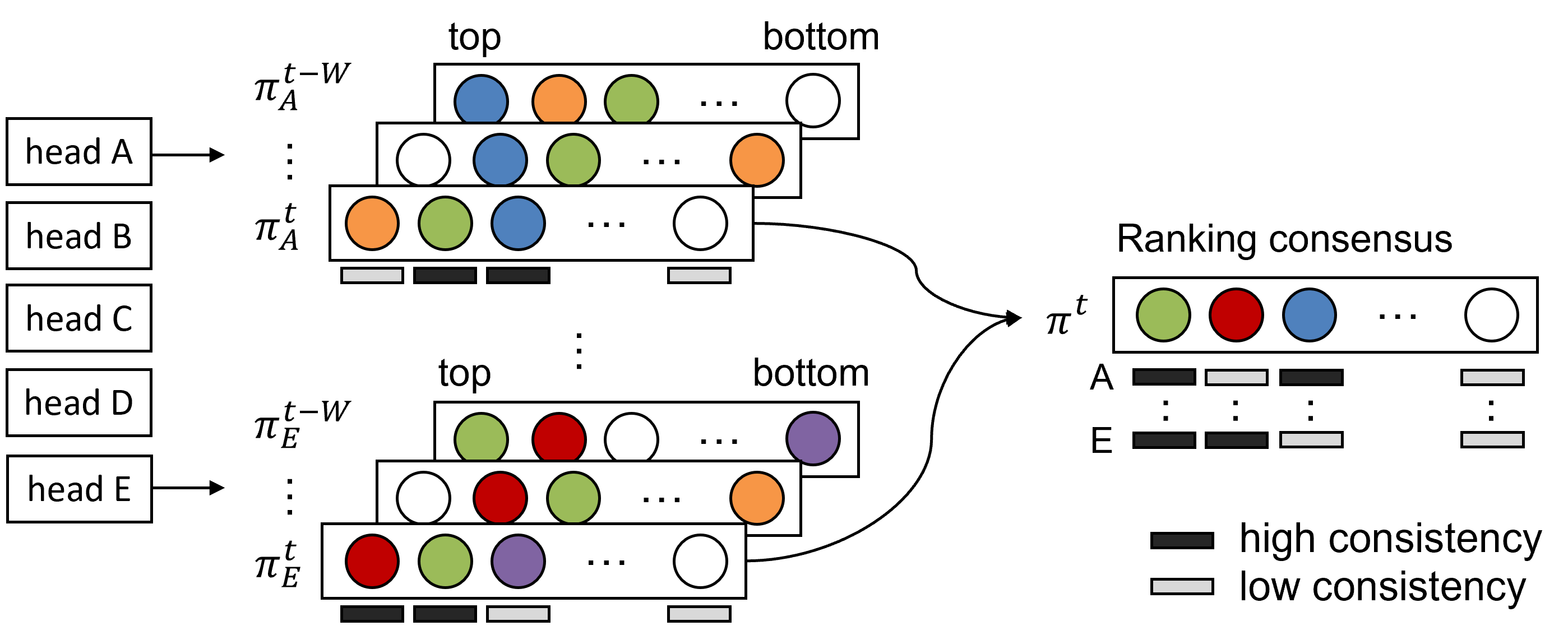}
\end{subfigure}
\caption{The conceptual illustration of the consensus generation process. Best viewed in color.}
\label{fig:ConCF_consensus1}
\end{figure}

\vspace{0.05cm}
Finally, the ranking consensus is generated by reranking items based on the consolidated importance.
For each item $i$, the importance $I^t_{i}$ is computed as follows:
\begin{equation}
\begin{aligned}
I^t_{i} = \mathbb{E}_{x \in \mathcal{F}}[RC^t_{x,i}], \,\,\, \text{where} \,\,\, RC^t_{x,i} = combine(R^t_{x,i}, C^t_{x,i}).
\end{aligned}
\end{equation}
$combine(\cdot,\cdot)$ is the function to consider both factors simultaneously. 
In this work, we use a simple addition.
The consensus generation process is illustrated in Figure \ref{fig:ConCF_consensus1}.
The proposed strategy penalizes the frequently changed predictions, and pushes the items having not only high rank but also high consistency (green and blue in head A, red and green in head E) across the heads to the top of the ranking consensus $\mathcal{\pi}^t$.

\subsubsection{Consensus learning}\noindent
After generating the consensus, we enhance the performance of each head using the consensus.
The consensus can be thought of as the teacher model in the general two-stage knowledge distillation (KD) \cite{KD}.
However, unlike the general KD relying on the pretrained teacher that makes static predictions, the consensus collaborative evolves with the heads based on their complementarity, which can generate more accurate supervision beyond the static teacher.
That is, the improved heads generate more accurate consensus again, interactively boosting the recommendation quality throughout the training process.


We train each head $x$ to match its ranking orders with the consensus $\pi^t$.
To this end, we use the listwise learning-to-rank approach \cite{xia2008list-wise}. 
It defines a likelihood of the ranking order as a permutation probability based on the Plackett-Luce model \cite{marden2019analyzing} and trains the model to maximize the likelihood of the ground-truth ranking order.
We define the top-$N$ version of the permutation probability, which ignores the detailed ranking orders below the $N$-th rank, enabling each head to better focus on learning the top-$N$ ranking~results~\cite{DERRD}.
\begin{equation}
\begin{aligned}
p(\pi^t_{0:N}|\, \theta ) =  \prod_{{k}={1}}^{N} 
 \frac{{\exp} ( \hat{r}_{u, \pi^t(k)} ) }
 {\sum_{{i}= {k}}^{|\pi^t|}  {\exp} ( \hat{r}_{u, \pi^t(i)} )},
\end{aligned}
\label{eq:ConCF_perm}
\end{equation}
where $\pi^t_{0:N}$ is the top-$N$ partial list, $\theta$ is the model parameters.
$\pi^t(k)$ is the $k$-th item of $\pi^t$, and $\hat{r}_{u, \pi^t(k)}$ is the relevance score for $(u, \pi^t(k))$ predicted by head $x$.
Then, we train each head to derive scores that agree with the consensus ranking by maximizing its log-likelihood.
The consensus learning loss is defined for users in the same batch $B$ used for the original CF loss as follows:
\begin{equation}
\begin{aligned}
 \mathcal{L}^t_{CL-x} = - \sum_{u \in B} \log p\left(\pi^{t, u}_{0:N} \mid \theta \right),
\end{aligned}
\label{eq:ConCF_LCDx}
\end{equation}
where $\pi^{t, u}_{0:N}$ is the top-$N$ ranking consensus for user $u$.

\subsubsection{Empirical evidence}
\begin{figure}[t]
\centering
\begin{subfigure}[t]{0.44\linewidth}
    \includegraphics[height=4cm]{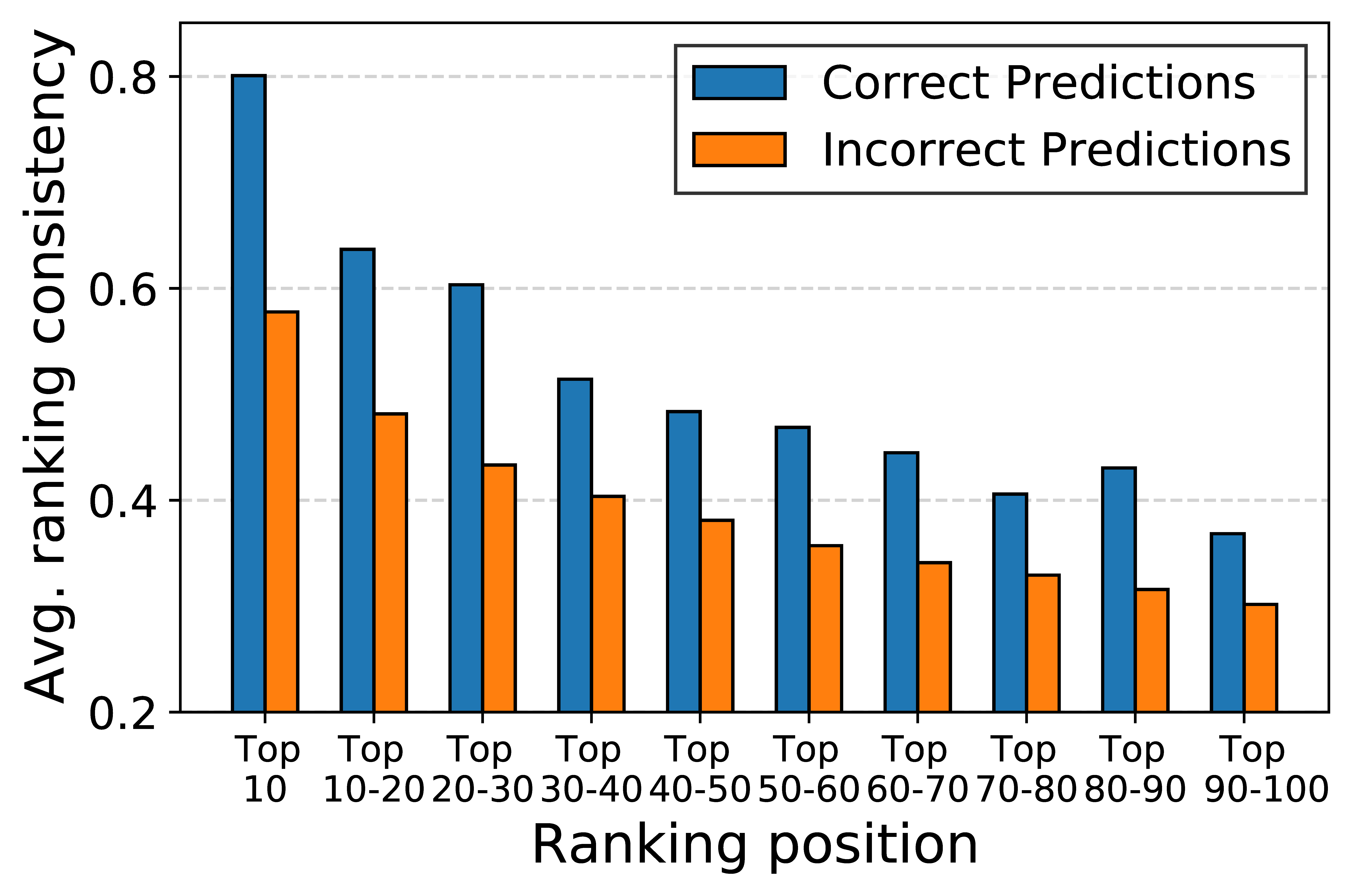}
\end{subfigure}
\hspace{0.1cm}
\begin{subfigure}[t]{0.44\linewidth}
    \includegraphics[height=4cm]{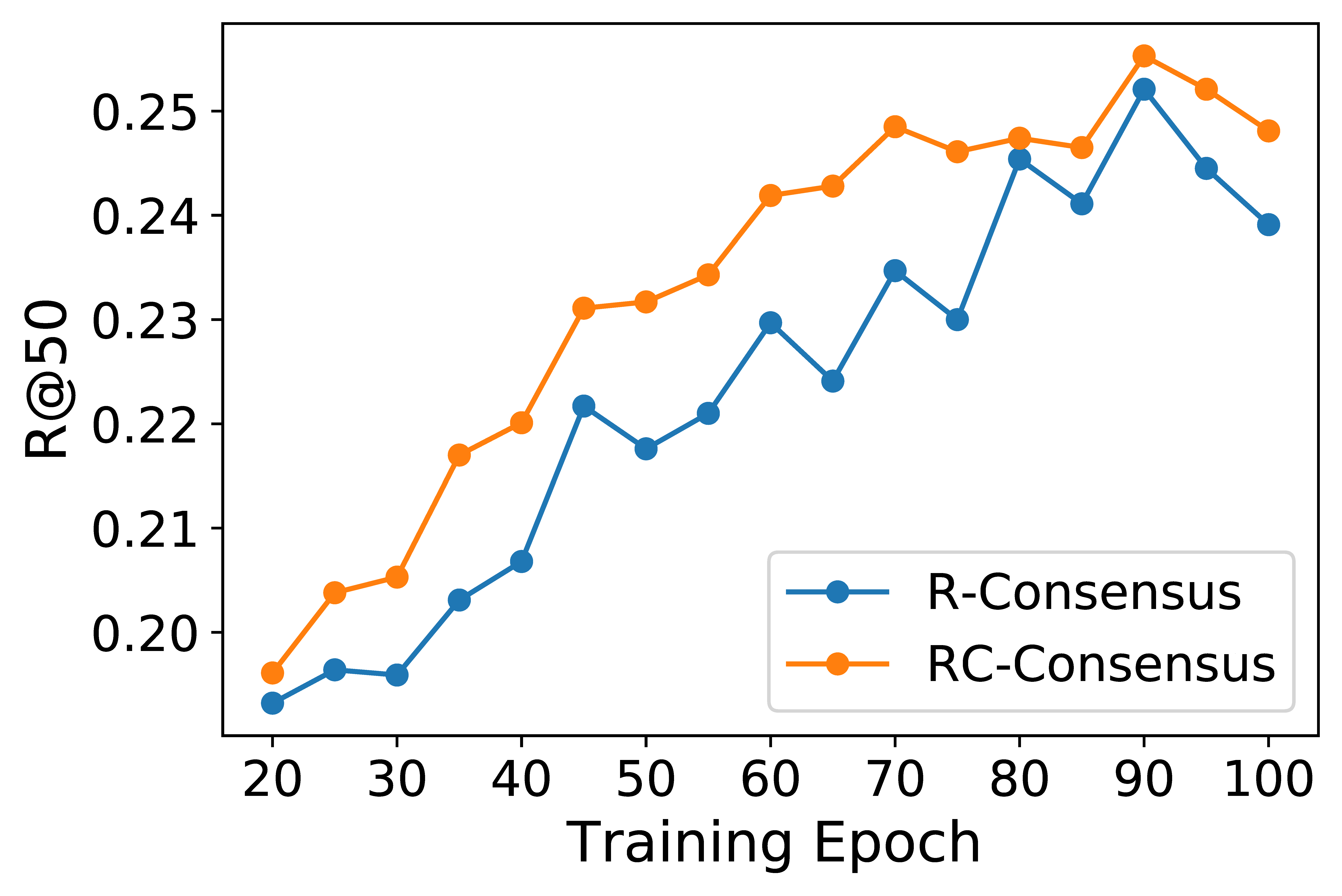}
\end{subfigure}
\caption*{\small (a) a single learning objective: $x \in \{A\}$}
\begin{subfigure}[t]{0.44\linewidth}
    \includegraphics[height=4cm]{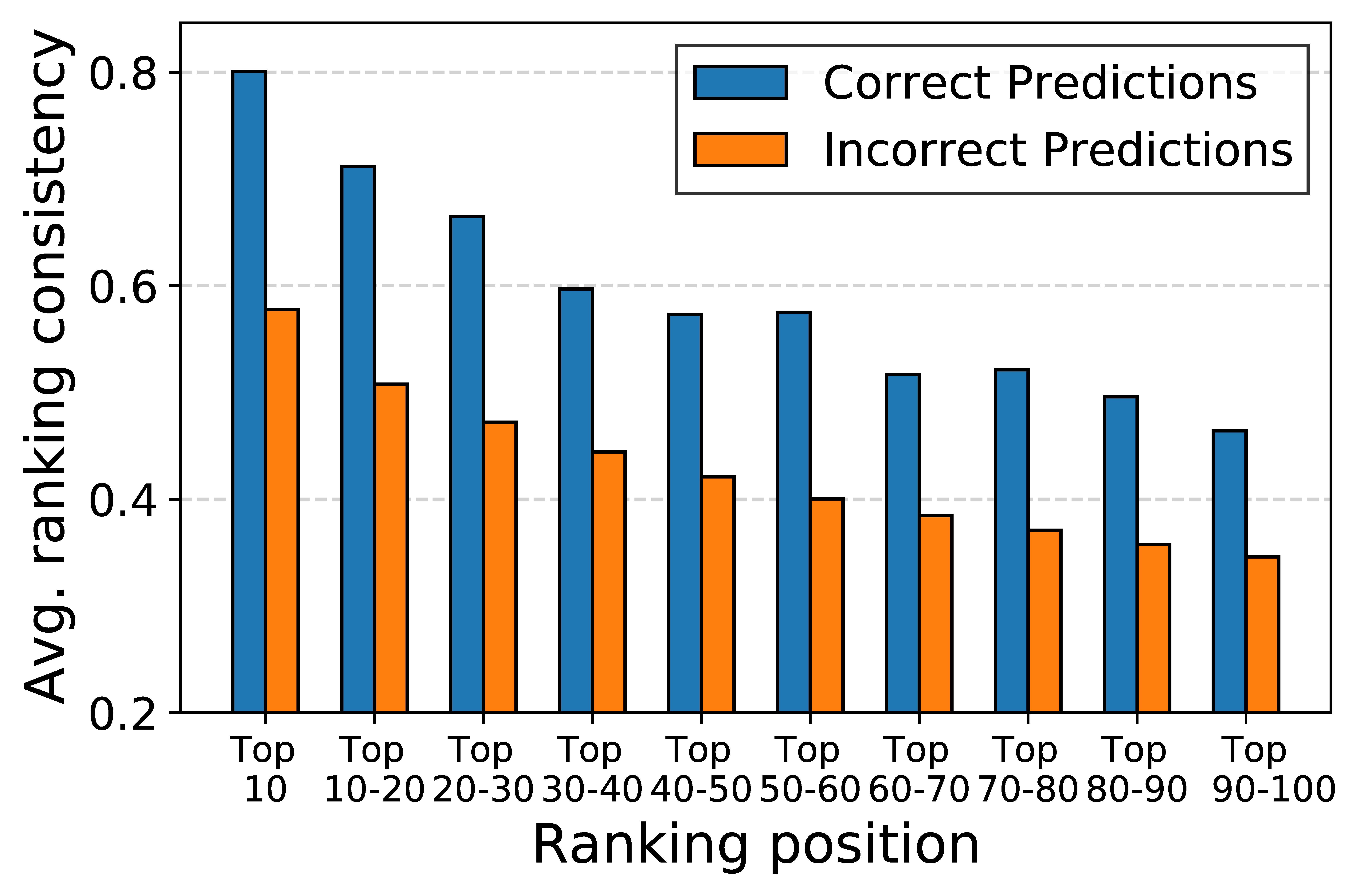}
\end{subfigure}
\hspace{0.1cm}
\begin{subfigure}[t]{0.44\linewidth}
    \includegraphics[height=4cm]{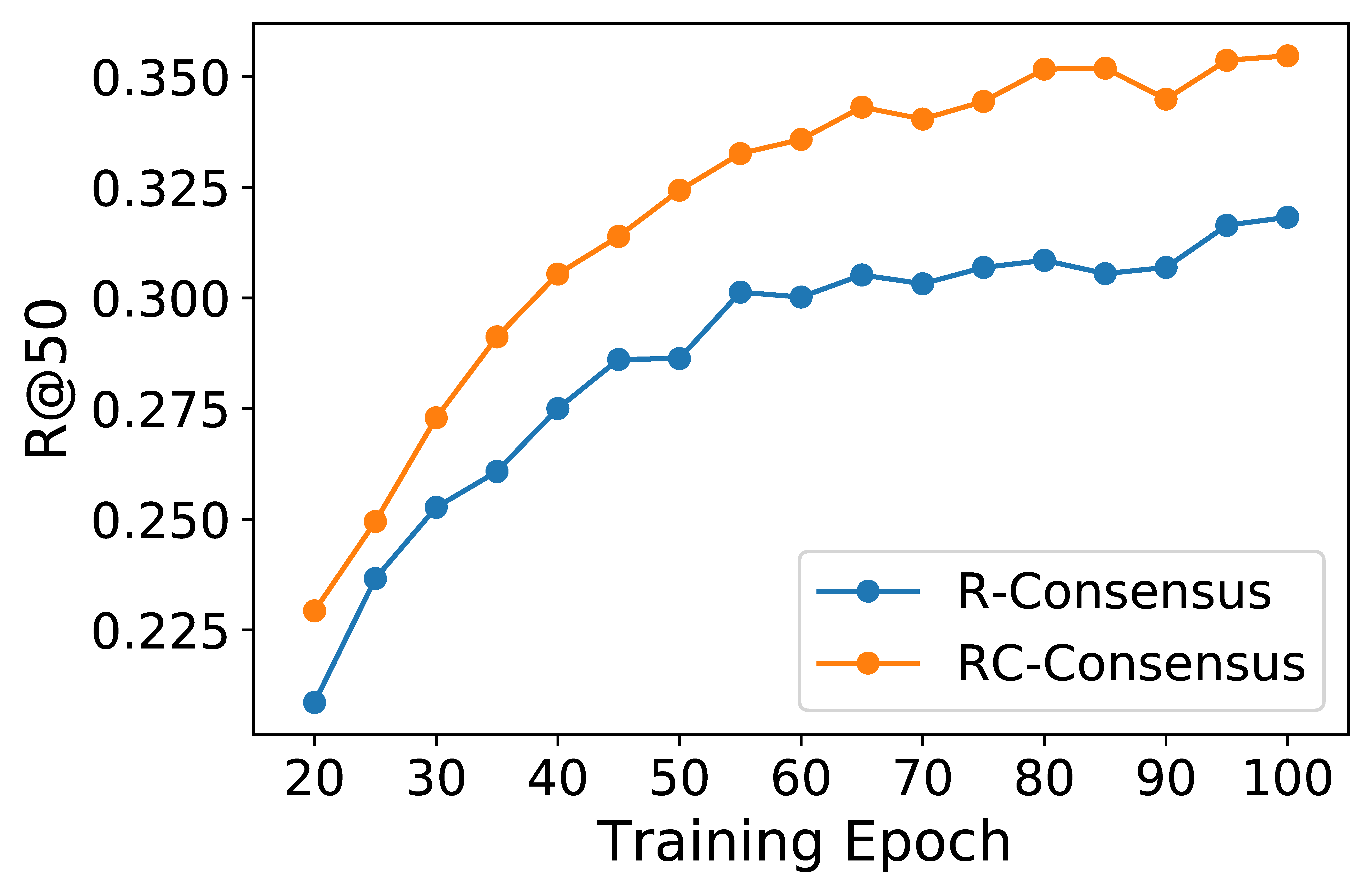}
\end{subfigure}
\caption*{\small (b) multiple learning objectives: $x \in \{A, B, C, D, E\}$}
\caption{(left) Computed consistency, (right) Recall@50 of the generated consensus.}
\label{fig:ConCF_consistency}
\end{figure}

\noindent
We investigate the reliability of the prediction measured by the consistency metric.
We analyze the dynamics of model predictions during training on CiteULike\footnote{The experimental setup is identical to Section \ref{sec:ConCF_analysis}.
The window size is set to 20.}.
First, we compare the computed consistency of correct predictions (i.e., test interactions) and that of incorrect predictions in the top-100 ranking list at epoch 50.
For each user's ranking list, we compute consistency for the item in each ranking position.
We then compute the average consistency of all users for each ranking position and summarize the results for 10 equispaced bins for better legibility. 
We report the results from CF-A (i.e., $C^{50}_{A,i}$) in Fig.\ref{fig:ConCF_consistency}a (left) and the results from all objectives (i.e., $\mathbb{E}_{x}[C^{50}_{x,i}]$) in Fig.\ref{fig:ConCF_consistency}b (left).

We observe that regardless of the learning objectives, for each rank bin
(1) the model tends to make consistent predictions on some user-item interactions but changes its prediction more frequently on others,
(2) the correct predictions have higher average consistency than the incorrect predictions. 
These observations indicate that the consistency reveals the reliability of each prediction and also show that even the items with the same rank need to have different importance based on the consistency in forming consensus.
Also, the model tends to make more consistent predictions for high-ranked items, indicating that the ranking consistency is a suitable indicator for assessing the reliability of the top-$N$ recommendation.

Further, we investigate whether consistency is indeed effective for generating accurate consensus.
In Figure \ref{fig:ConCF_consistency} (right), we compare the accuracy of consensus generated by two importance criteria:
\textbf{(1) R-consensus} by $\mathbb{E}_{x}[R^t_{x,i}]$,
\textbf{(2) RC-consensus} by $\mathbb{E}_{x}[RC^t_{x,i}]$.
R-consensus can be thought of as a kind of majority voting scheme of ensemble;
items with the same rank are considered to have the same importance, and items with an overall high rank across the heads would be ranked near the top in the consensus.
As discussed earlier, this is insufficient to create an informative consensus.
It even generates the consensus more incorrect than the predictions of some models.
Note that in Fig.\ref{fig:ConCF_consistency}a, R-consensus is equivalent to the original predictions by the model.

We observe that RC-consensus produces more accurate consensus than R-consensus (up to 16\% in Fig.\ref{fig:ConCF_consistency}b).
Moreover, considering the consistency from all heads produces larger improvements compared to the case of a single head (Fig.\ref{fig:ConCF_consistency}a vs. Fig.\ref{fig:ConCF_consistency}b).
This again shows the efficacy of the proposed consensus generation strategy, which identifies and consolidates reliable predictions from~each~head.

\vspace{0.05cm}
\noindent
\textbf{Remarks.}
In sum, ConCF uses ranking information along with the consistency to create consensus, and guides the heads by using the ranking consensus.
Our strategy is inspired by the consistency regularization of semi-supervised learning (SSL) in image classification \cite{temporal_ensemble, tc-ssl} that deals with the uncertainty of unlabeled data by penalizing inconsistent predictions that change frequently during training.
Also, our observations are consistent with recent findings of SSL that frequently changed~predictions~are~unreliable~\cite{tc-ssl}.

\subsection{Balancing Heterogeneous Objectives}
\label{sec:ConCF_blance}
In ConCF, the heads are optimized by heterogeneous objectives having different convergence behaviors.
Thus, proper balancing of the heads is essential to effectively exploiting the complementarity.
If a few heads are too dominant during training, the shared parameters would be biased in favor of the heads, which incurs imbalances that impede the remaining heads from being fully optimized.
This degrades the overall effectiveness of the framework.

We enforce all the heads to be trained at similar rates by dynamically adjusting the gradient scale of each head on the shared parameters ($\theta_s$).
We apply gradient normalization technique \cite{gradnorm}, enabling the shared parameters to converge to beneficial features across all heads without additional hyperparameters for balancing the heads.
At each training step $t$, we define relative training ratio for each head $x$ as follows:
\begin{equation}
    \begin{aligned}
        \gamma^t_x = \frac{\mathcal{L}^t_x \, / \mathcal{L}^0_x}{\mathbb{E}_{x \in \mathcal{F}}[\mathcal{L}^t_x \,/ \mathcal{L}^0_x ]}, 
    \end{aligned}
\end{equation}
where $\mathcal{L}^0_x$ is the initial loss of the training. 
For $\gamma^t_x$, a lower value indicates head $x$ is optimized faster compared to the other heads. 
Based on the computed ratio, we adjust the gradient scale by each head.
For each head $x$, let $G^t_x = ||\nabla_{\theta_s} \lambda^t_x \mathcal{L}^t_x||_2$ denote the gradient scale by the loss of head $x$.
The target gradient scale is defined as $\mathbb{E}_{x\in \mathcal{F}}[G^t_x] \times \gamma^t_x$.
The gradient normalization is conducted by minimizing the distance between the gradient scales.
\begin{equation}
    \begin{aligned}
        \mathcal{L}^t_{b} = \sum_{x\in \mathcal{F}} \Big\vert G^t_x  - \mathbb{E}_{x\in \mathcal{F}}[G^t_x] \times \gamma^t_x \Big\vert.
    \end{aligned}
    \label{eq:ConCF_LB}
\end{equation}
The target scale is treated as a constant, and $\mathcal{L}_{b}$ is differentiated only with respect to $\lambda^t_x$.
The computed gradient $\nabla_{\lambda^t_x} \mathcal{L}_{b}$ is then applied via standard gradient descent update rules to update $\lambda^{t+1}_x$.
Finally, $\lambda^{t+1}_x$ is normalized so that $\sum_{x\in \mathcal{F}} \lambda^{t+1}_x = 1$.

\section{Experiments}
\label{sec:ConCF_experimentsetup}
\subsection{Experiment Setup}
\noindent
\textbf{Dataset.}
We use three real-world datasets: CiteULike \cite{wang2013collaborative}, Ciao \cite{tang2012mtrust}, and Foursquare \cite{liu2017experimental}.
These datasets are publicly available and widely used in recent work \cite{BUIR, DERRD, CML, BD}.

\noindent
\textbf{Evaluation protocol.}
We randomly split each user’s interaction history into train/valid/test sets in a 60\%/20\%/20\% split \cite{CML}.
We evaluate all models by two top-$N$ ranking metrics \cite{VAE, CML}: Recall@$N$ (R@$N$) and NDCG@$N$ (N@$N$). 
We report the average value of five independent runs, each of which uses differently split data.

\noindent
\textbf{Compared training schemes.}
\textbf{(a)} SingleCF uses a single objective for optimization as done in the most previous studies (Fig.\ref{fig:ConCF_overview}a).
We train five independent models with each CF objective described in Section \ref{sec:ConCF_learning}.
\textbf{(b)} ConCF uses \textit{multi-branch architecture} trained by \textit{heterogeneous objectives} along with~\textit{consensus learning} (Fig.\ref{fig:ConCF_overview}b).
After training, we can make recommendations by using either 
1) a preferred head (e.g., based on the system requirement for further processing)
or 2) the consensus generated by all heads.
The former adds no extra inference cost compared to SingleCF and the latter can be adopted in the scenarios where computational cost is less constrained.
We report the results of both~cases~in~Table~\ref{tbl:ConCF_main}.

\begin{sidewaystable}
\centering
\caption{Recommendation performances. \textit{Gain.Best} denotes the improvement of the best objective, which generates the best SingleCF model, in ConCF.
\textit{Gain.Con.} denotes the improvement of consensus over the best SingleCF model. We conduct the paired t-test with 0.05 level and all \textit{Gain}s are statistically significant.}
\small
\renewcommand{\arraystretch}{0.9}
\renewcommand{\tabcolsep}{1.4mm}
\begin{tabular}{cc cccc cccc cccc}
\toprule
\multicolumn{2}{c}{Dataset}                      & \multicolumn{4}{c}{CiteULike}             & \multicolumn{4}{c}{Ciao}                  & \multicolumn{4}{c}{Foursquare}            \\
\cmidrule{1-2} \cmidrule(lr){3-6} \cmidrule(lr){7-10} \cmidrule(lr){11-14}
Training scheme & Objective  & R@20 & N@20 & R@50 & N@50 & R@20 & N@20 & R@50 & N@50 & R@20 & N@20 & R@50 & N@50 \\
\midrule
\multirow{5}{*}{\makecell{SingleCF}} 
& CF-A   & 0.1411    & 0.0870  & 0.2513    & 0.1160  & 0.1151    & 0.0766  & 0.1797    & 0.0952  & 0.1207    & 0.0897  & 0.1987    & 0.1151  \\
& CF-B  & 0.1551    & 0.0892  & 0.2851    & 0.1231  & 0.0645    & 0.0348  & 0.1547    & 0.0615  & 0.0897    & 0.0573  & 0.1746    & 0.0847  \\
& CF-C   & 0.1483    & 0.0774  & 0.2652    & 0.1086  & 0.1182    & 0.0784  & 0.1814    & 0.0964  & 0.1123    & 0.0786  & 0.1939    & 0.1052  \\
& CF-D    & 0.1739    & 0.1011  & 0.3132    & 0.1375  & 0.1042    & 0.0676  & 0.1755    & 0.0879  & 0.1026    & 0.0655  & 0.1948    & 0.0956  \\
& CF-E    & 0.1938    & 0.1125  & 0.3179    & 0.1451  & 0.0975    & 0.0539  & 0.1853    & 0.0797  & 0.1548    & 0.1123  & 0.2542    & 0.1457  \\
\midrule
\multirow{8}{*}{ConCF} 
& CF-A  & 0.2371          & 0.1378          & 0.3701          & 0.1734          & 0.1347          & 0.0845          & 0.2237          & 0.1134          & 0.1605          & 0.1115          & 0.2638          & 0.1451          \\
& CF-B   & 0.2350          & 0.1339          & 0.3733          & 0.1698          & 0.1350          & 0.0878          & 0.2213          & 0.1123          & 0.1720          & 0.1219          & 0.2773          & 0.1562          \\
& CF-C   & 0.2412          & 0.1395          & 0.3763          & 0.1756          & 0.1334          & 0.0880          & 0.2221          & 0.1131          & 0.1639          & 0.1125          & 0.2782          & 0.1495          \\
& CF-D   & 0.2300          & 0.1259          & 0.3716          & 0.1634          & 0.1349          & 0.0884          & 0.2266          & 0.1140          & \textbf{0.1757}          & 0.1235          & 0.2876          & 0.1579          \\
& CF-E  & 0.2418          & 0.1407          & 0.3721          & 0.1750          & 0.1367          & 0.0881          & 0.2248          & 0.1137          & 0.1714          & \textbf{0.1243}          & 0.2759          & \textbf{0.1584}          \\
& Consensus & \textbf{0.2533} & \textbf{0.1474} & \textbf{0.3896} & \textbf{0.1836} & \textbf{0.1420} & \textbf{0.0897} & \textbf{0.2303} & \textbf{0.1144} & 0.1749 & 0.1218 & \textbf{0.2877} & 0.1583 \\
\cmidrule{2-14}
& \textit{Gain.Best}  &	24.77\%	& 25.07\%	& 17.05\%	& 20.61\%	& 12.86\%	& 12.24\%	& 21.32\%	& 17.32\%	& 10.72\%	& 10.69\%	& 8.54\%& 8.72\% \\
& \textit{Gain.Con.}  & 30.70\%         & 31.02\%         & 22.55\%         & 26.53\%         & 20.14\%         & 14.41\%         & 24.28\%         & 18.67\%         & 12.98\%         & 8.46\%          & 13.18\%         & 8.65\%           \\
                                  \bottomrule
\end{tabular}
\label{tbl:ConCF_main}
\end{sidewaystable}

\begin{figure}[t]
\centering
\hspace{0.5cm}
        \begin{subfigure}{0.7\linewidth}
          \includegraphics[height=4cm]{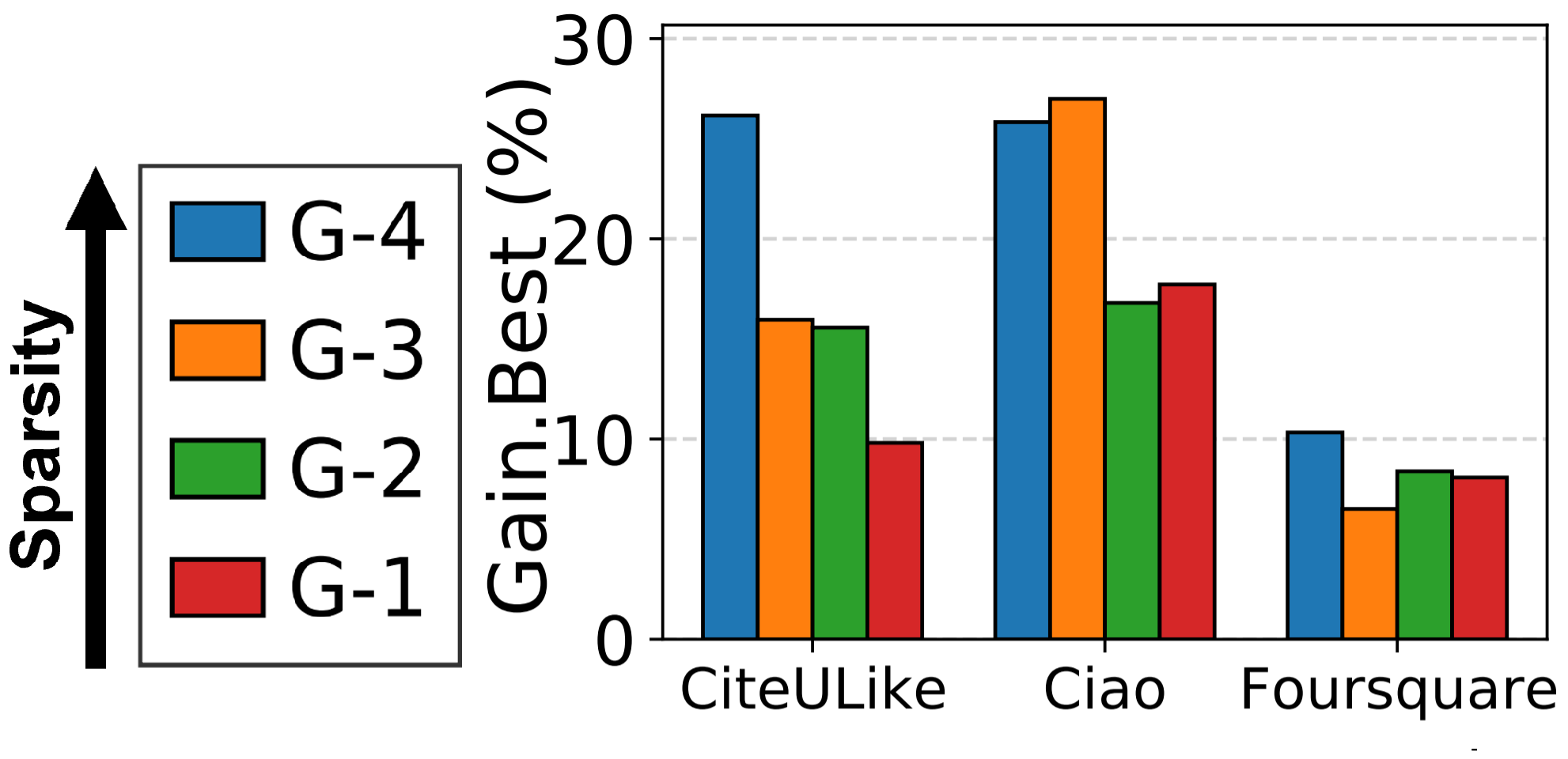}
        \end{subfigure}
        \caption*{(a) \textit{Gain.Best} for user groups.}
		\hspace{0.3cm}
        \begin{subfigure}{0.4\textwidth}
        \hspace{0.9cm}
	\vspace{0.8cm}
          \includegraphics[height=1.5cm]{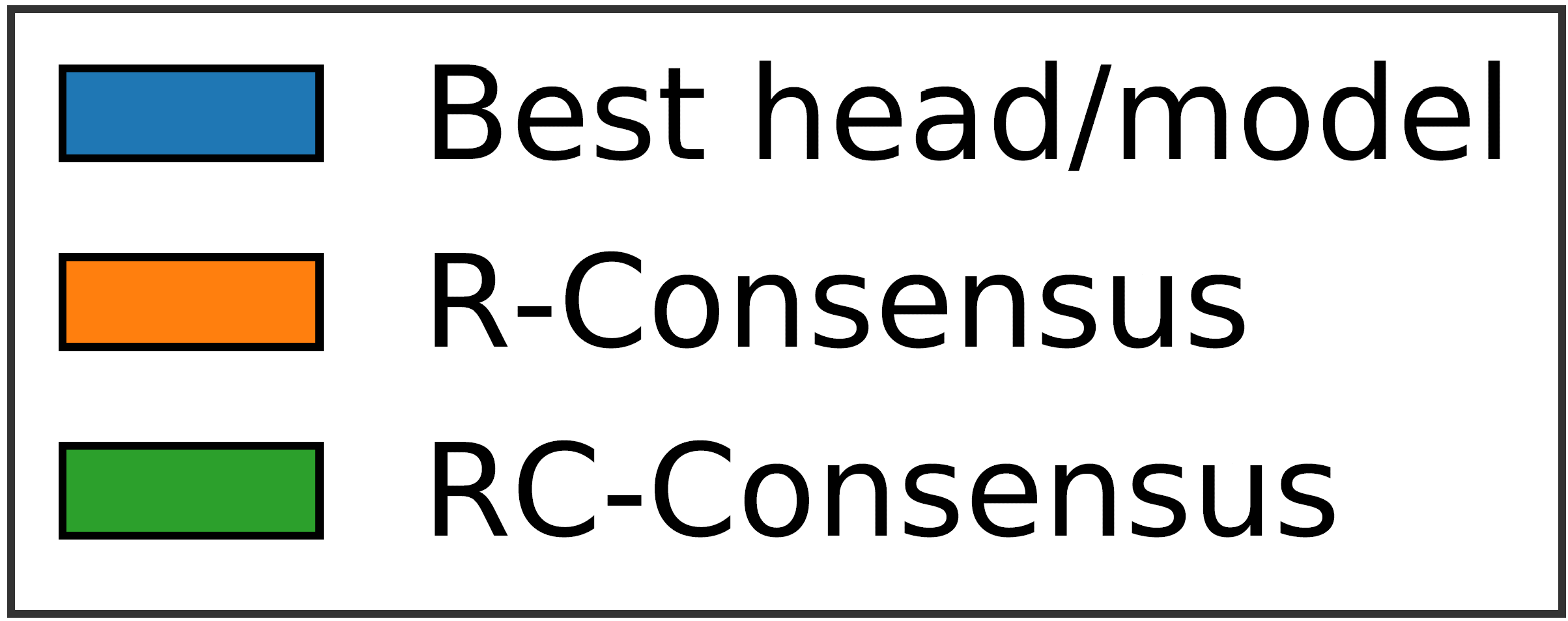}
        \end{subfigure}
\hspace{-0.32cm}
        \begin{subfigure}{0.4\textwidth}
          \includegraphics[height=4cm]{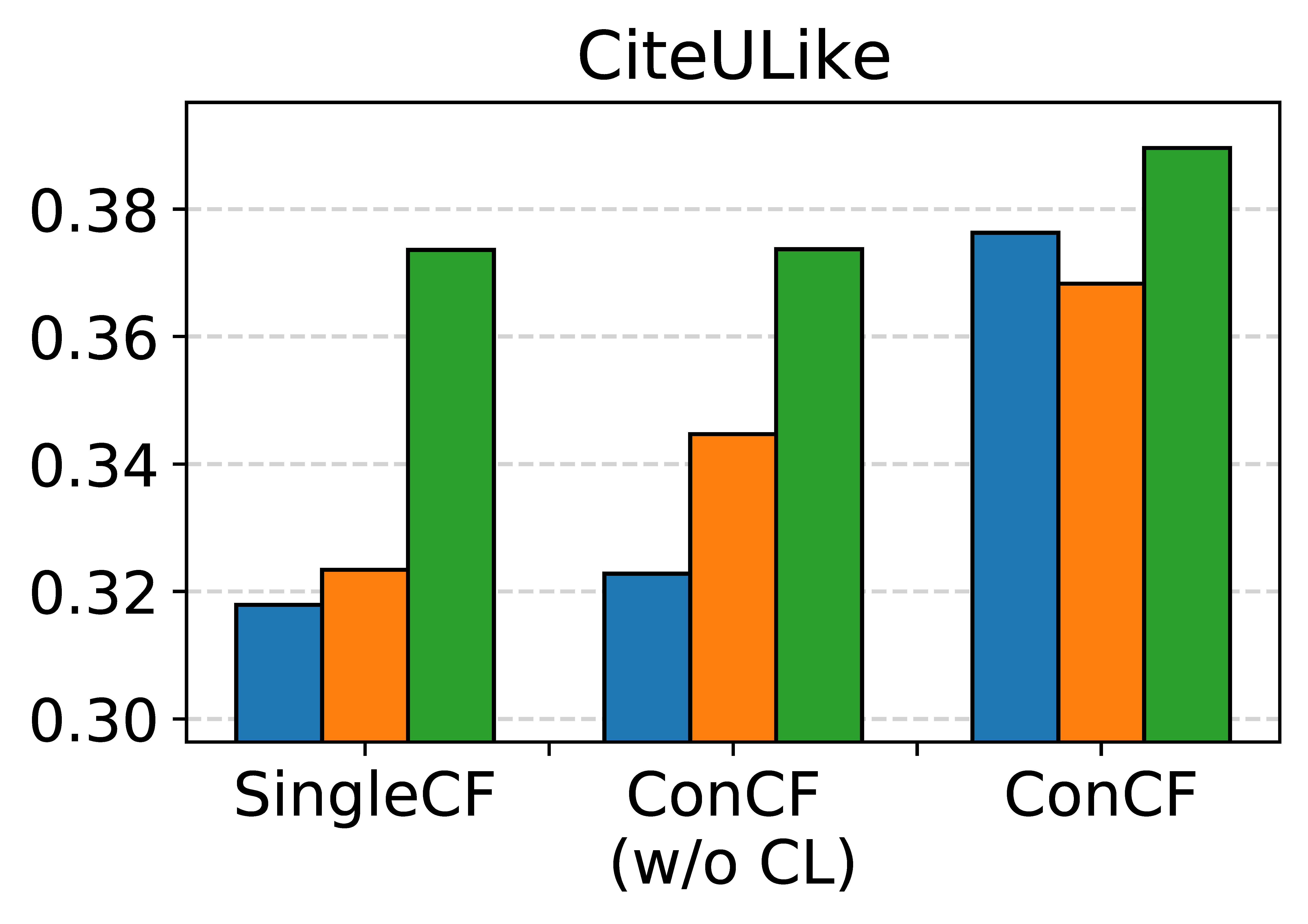}
        \end{subfigure}
        \\
        \begin{subfigure}{0.4\textwidth}
         \includegraphics[height=4cm]{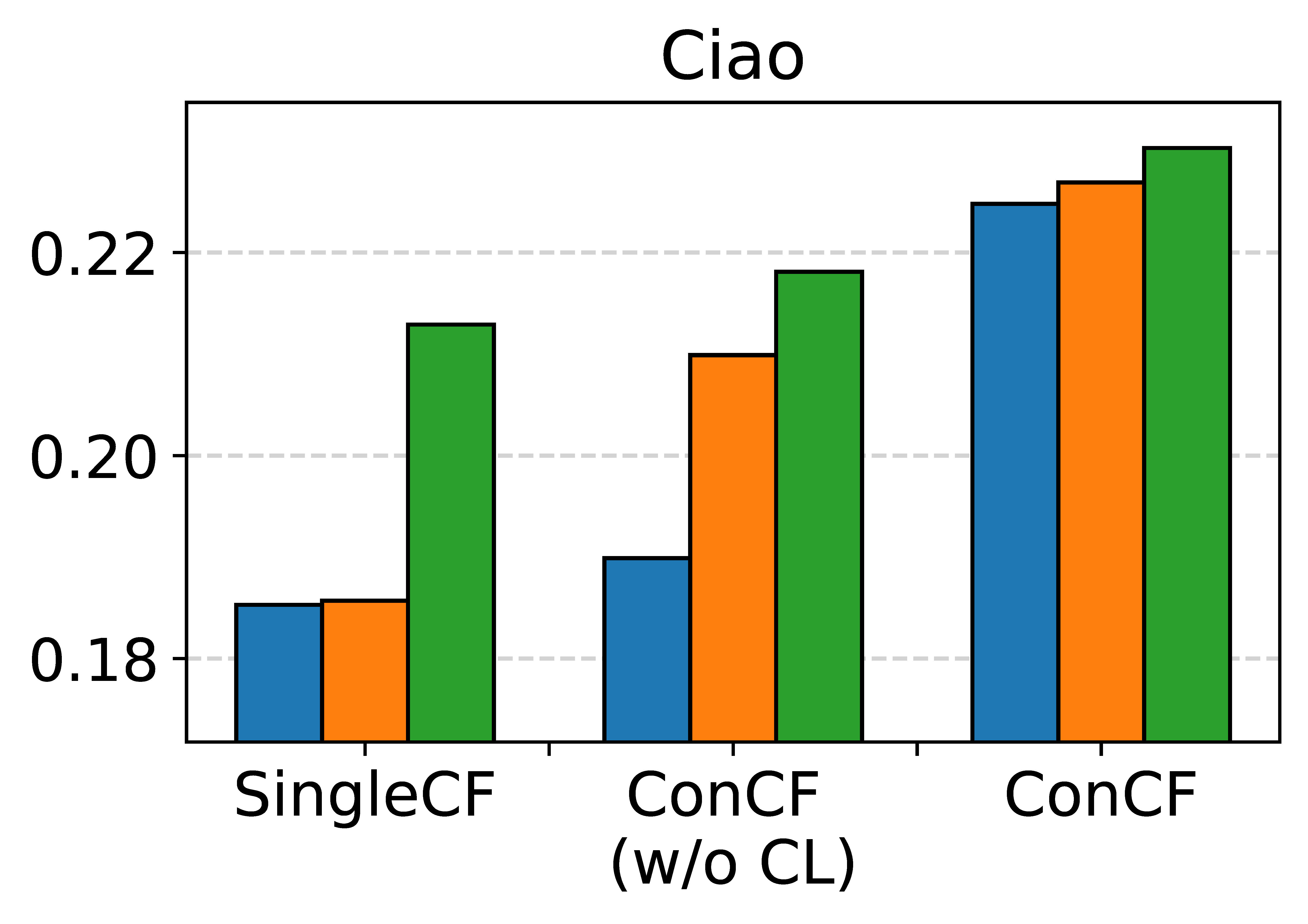}
        \end{subfigure}
        \hspace{0.1cm}
        \begin{subfigure}{0.4\textwidth}
          \includegraphics[height=4cm]{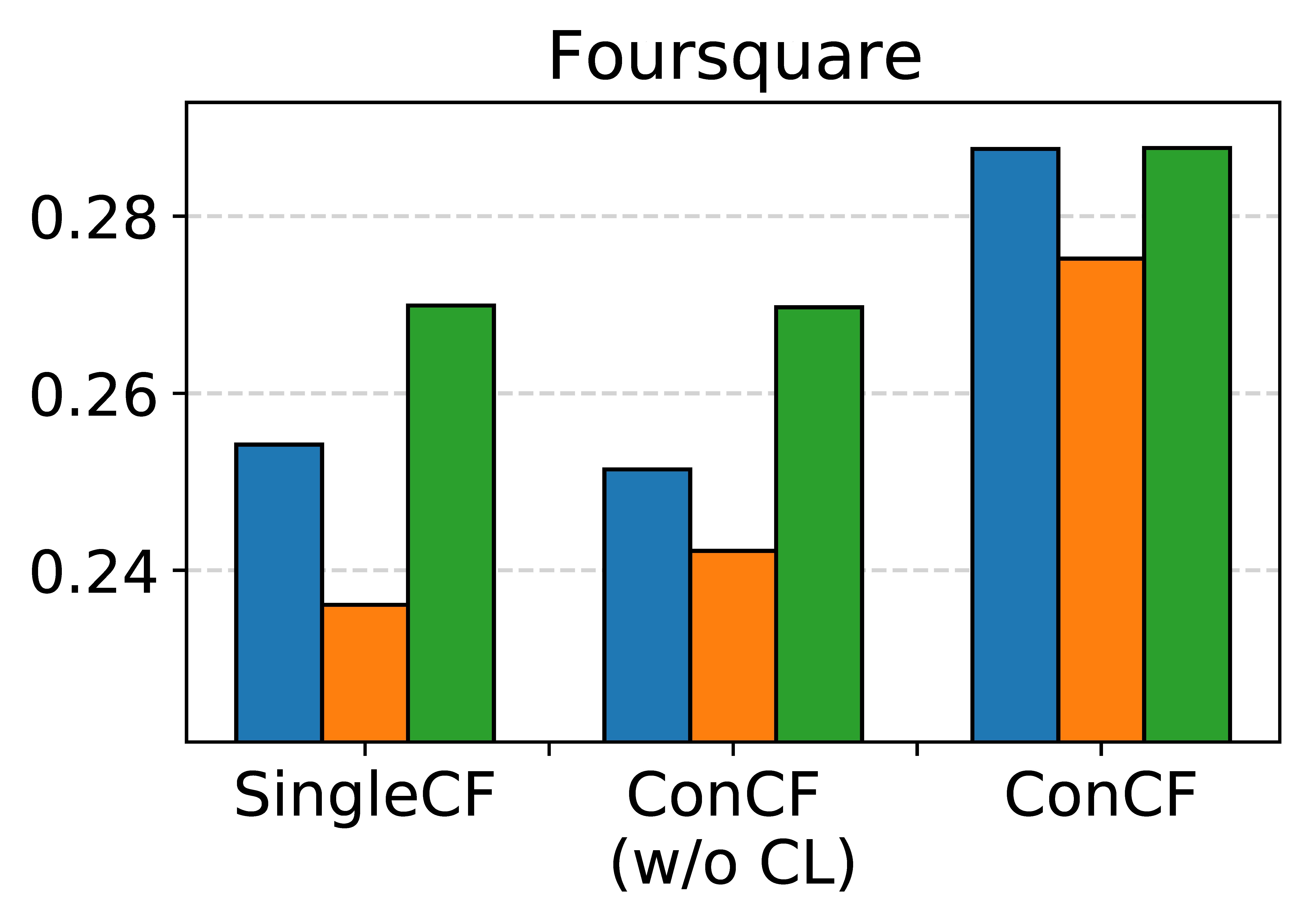}
        \end{subfigure}
        \caption*{(b) Comparison of three schemes. `ConCF (w/o CL)' is ConCF without consensus learning~loss.}
    \hspace{-0.5cm}
    \caption{Further analyses on the consensus learning. All metrics are R@50.}
    \label{fig:ConCF_further}
\end{figure}




\subsection{Effectiveness of ConCF}\noindent
\label{sec:ConCF_result}
We provide various experiment results supporting the superiority of ConCF.
Supplementary experiments including hyperparameter study can be found in Appendix.
In Table \ref{tbl:ConCF_main}, ConCF achieves significantly higher performance than all SingleCF models for all the datasets.
In particular, the performances achieved by each learning objective are considerably improved in ConCF beyond the best SingleCF~model.

We further analyze the impacts of ConCF on users with different sparsity levels.
We sort users in decreasing order according to their number of interactions (i.e., sparsity) and equally divide them into four groups; G-1 is the group of top-25\% users having many interactions, whereas G-4 is the group of users having few interactions.
We compute user-level $Gain.Best$ and report the average value for each user group in Fig.\ref{fig:ConCF_further}a.
We observe that ConCF provides more accurate recommendations to all user groups, but particularly, the users with high sparsity (e.g., G-4) take more benefits from ConCF.
In ConCF, various aspects captured by the heterogeneous objectives provide a more complete understanding of the users, which can be particularly effective for the users with~very~limited~information.


\subsection{Consensus Learning Analysis}\noindent
To validate the effectiveness of the consensus learning, we compare three training schemes: SingleCF, ConCF (w/o CL), and ConCF.
Since all these schemes can generate the consensus by consolidating the predictions from all models/heads, we report the performance of their best model (for SingleCF) or best head (for ConCF) as well as that of their consensus in Fig.\ref{fig:ConCF_further}b.
Table \ref{tbl:ConCF_chrvalue} presents CHR induced by each scheme.
In sum, ConCF achieves the best performance for both the best head and the final consensus (i.e., RC-con).
We analyze the results with various perspectives:

\begin{itemize}[leftmargin=*]
    
\item The consensus of the SingleCF models is not as effective as that of ConCF.
It can be seen as the \textit{ensemble model}; the predictions from multiple models are consolidated to make the final prediction.
Because the knowledge captured by a single objective is limited, each SingleCF model has limited performance.
This leads to unsatisfactory ensemble results despite the effectiveness of RC-con.

\item ConCF (w/o CL) cannot effectively improve the performance of each head, and also its consensus shows limited performance.
ConCF (w/o CL), which ablates the consensus learning (i.e., $\mathcal{L}_{CL}$) from ConCF, can be seen as conventional \textit{multi-task learning}.
Here, the knowledge of the objectives is exchanged only through the shared parameters without guidance on the predictions.
Its limited performance shows the importance of providing accurate and direct guidance on the predictions. 
Unlike the other two schemes, in ConCF, the heads learn reliable predictions from each other by negotiating their differences and reaching consensus throughout the training. 

\item RC-con is highly effective in generating consensus.
R-con sometimes even generates more inaccurate consensus than the best head/model (e.g., Foursquare), whereas RC-con always generates informative consensus beneficial for all heads.
This shows our strategy considering the reliability effectively consolidates the knowledge from heterogeneous objectives.

\item In Table \ref{tbl:ConCF_chrvalue}, ConCF shows the lowest CHR values, which indicates the head effectively learns the complementary knowledge from the other objectives.
For ConCF (w/o CL) and ConCF, CHR is computed in terms of the head optimized by CF-A, i.e., CHR$(A;\mathcal{F})$, as done in the previous analysis (Table \ref{tbl:ConCF_CHR}).
This result supports the superior performance of ConCF on Table \ref{tbl:ConCF_main} and Fig.\ref{fig:ConCF_further}b.
\end{itemize}

\begin{table}[t]
\centering
\caption{CHR by three training schemes.}
\small
\renewcommand{\arraystretch}{0.9}
\renewcommand{\tabcolsep}{3.5mm}
\begin{tabular}{cccc}
\toprule
Training schemes  & CiteULike & Ciao   & Foursquare \\
\midrule
\begin{tabular}[c]{@{}l@{}}SingleCF\end{tabular} & 0.56    & 0.43 & 0.54     \\
\begin{tabular}[c]{@{}l@{}}ConCF (w/o CL) \end{tabular} & 0.42    & 0.39 & 0.51     \\
ConCF & 0.18    & 0.10 & 0.17    \\
\bottomrule
\end{tabular}
\label{tbl:ConCF_chrvalue}
\end{table}

\subsection{Comparison with Homogeneous Objectives}\noindent
In Table \ref{fig:ConCF_homo}, we analyze the importance of heterogeneous objectives in ConCF on CiteULike.
ConCF (w/o Het) is a variant of ConCF that uses the homogeneous objectives. 
We also present the results of PCL \cite{PCL} which is the state-of-the-art \textit{online KD} method for image classification.
It trains multiple classifier heads simultaneously by (multiclass) cross-entropy (i.e., CF-E) and guides the classifiers to minimize KL divergence from the ensemble Softmax logits that integrate the heads.
Here, the complementarity of the heads is mainly originated from the different initializations.

We observe that ConCF (w/o Het) still achieves considerably higher performance than the SingleCF model, however, it cannot achieve comparable performance to ConCF. 
These results are consistent with the findings of Table \ref{tbl:ConCF_CHR} and again show the importance of exploiting the heterogeneous objectives.
Also, ConCF (w/o Het) performs better than PCL, which shows the effectiveness of utilizing ranking information for top-$N$ recommendation~\cite{RD, DERRD}.

\subsection{Comparison with two-stage KD}\noindent
In Table \ref{fig:ConCF_kd}, we compare ConCF with conventional KD approach \cite{KD} on CiteULike.
Unlike ConCF where the consensus is dynamically updated during training, the teacher in two-stage KD makes static predictions.
We apply KD to SingleCF and ConCF (w/o CL).
Here, we use CF-E as it shows the best performance in the previous experiments.
ConCF (w/o CL) is guided by the static teacher via distillation.
All distillations are conducted in the same way using the listwise learning.

We observe that the two-stage KD cannot generate the target model which significantly exceeds the teacher's performance.
Also, ConCF achieves the best performance among all training schemes.
In ConCF, the consensus collaboratively evolves with the heads based on their complementarity, which can generate more accurate supervision beyond the static teacher.


\begin{table}[t]
\centering
\small
\renewcommand{\arraystretch}{0.9}
\renewcommand{\tabcolsep}{2.2mm}
\caption{Comparison with homogeneous learning~objectives.}
\begin{tabular}{cccccc}
\toprule
Training scheme & Objective & R@20   & N@20   & R@50   & N@50   \\
\midrule
\multicolumn{1}{c}{\begin{tabular}[c]{@{}l@{}}SingleCF\end{tabular}} & CF$\text{-}$E          & 0.1938 & 0.1125 & 0.3179 & 0.1451 \\
\midrule
 & CF$\text{-}$E 1       & 0.2196 & 0.1350 & 0.3363 & 0.1662 \\
\multirow{4}{*}{PCL \cite{PCL}}  & CF$\text{-}$E 2        &0.2101 & 0.1291 & 0.3340 & 0.1621 \\
& CF$\text{-}$E 3        & 0.2133 & 0.1295 & 0.3471 & 0.1644 \\
& CF$\text{-}$E 4        &0.2139 & 0.1298 & 0.3344 & 0.1615 \\
& CF$\text{-}$E 5        & 0.2164 & 0.1318 & 0.3383 & 0.1635 \\
& Ensemble        & 0.2103 & 0.1331 & 0.3422 & 0.1674\\
\midrule
 & CF$\text{-}$E 1        & 0.2236 & 0.1369 & 0.3479 & 0.1689 \\
\multirow{4}{*}{ConCF (w/o Het)} & CF$\text{-}$E 2        & 0.2227 & 0.1400 & 0.3458 & 0.1718 \\
& CF$\text{-}$E 3        & 0.2282 & 0.1389 & 0.3495 & 0.1702 \\
& CF$\text{-}$E 4        & 0.2266 & 0.1400 & 0.3486 & 0.1712 \\
& CF$\text{-}$E 5        & 0.2256 & 0.1378 & 0.3441 & 0.1682 \\
& Consensus        & 0.2321 & 0.1407 & 0.3521 & 0.1731\\
\bottomrule
\end{tabular}
\label{fig:ConCF_homo}
\end{table}

\begin{table}[t]
\centering
\caption{R@50 comparison with two-stage KD. T1: the best SingleCF model (0.3179), T2: RC-con of all SingleCF models (0.3736).}
\small
\renewcommand{\arraystretch}{0.9}
\renewcommand{\tabcolsep}{1mm}
\begin{tabular}{ccccc}
\toprule
\multicolumn{1}{l}{}       & Teacher               &    Training scheme  & Best model/head & Consensus \\
\midrule
 & T1                                 & SingleCF (CF-E) + KD &    0.3207  &   -  \\
    two-stage  & T2               & SingleCF (CF-E) + KD &     0.3672  &-     \\
                           &      T2                              & ConCF (w/o CL)  + KD                   &    0.3704    &  0.3750   \\
\midrule
\multicolumn{1}{c}{one-stage} &  & ConCF                   &     0.3763  &  0.3896  \\
\bottomrule
\end{tabular}
\label{fig:ConCF_kd}
\end{table}


\section{Related Work}
\label{sec:ConCF_relatedwork}
We briefly introduce the work related to our study.
Recent OCCF methods are summarized in terms of their learning objectives in Section \ref{sec:ConCF_learning}, and other related work including AutoML and multi-objective optimization are provided in Appendix \ref{app:ConCF_related_work}.

\noindent
\textbf{Multi-task Learning.}
Multi-task learning (MTL) trains a single model that handles multiple tasks \cite{crawshaw2020multi, mmoe}.
MTL model learns commonalities and differences across different tasks, which can result in both improved efficiency and accuracy for each task \cite{mmoe}.
Extensive research has been conducted to improve the quality of MTL.
\cite{mmoe, tang2020progressive} have focused on designing the architecture for MTL, such as how to model task-specific and shared information.
\cite{gradnorm, liu2019loss} have focused on balancing the losses from multiple tasks.
ConCF also takes advantage of MTL, achieving better generalization and significantly reducing the parameters compared to training a separate model for each objective.
We also tried the various MTL architectures \cite{mmoe, tang2020progressive}, but no significant improvement is observed.

\vspace{0.05cm}
\noindent
\textbf{Knowledge Distillation.} 
Most KD methods \cite{KD, RD, DCD, DERRD, TD, IRRRD, PHR} have focused on model compression that transfers knowledge from a large pre-trained teacher model to improve a small target model.
Based on the extra supervision from the teacher, they have successfully improved the performance of the target model.
Despite their effectiveness on model compression, they require at least a two-stage training process and are also heavily dependent on the pre-trained teacher model \cite{DML, ONE}.
In computer vision, many studies \cite{DML, song2018collaborative, ONE, PCL, guo2020online} have focused on online KD which optimizes a target model by distilling knowledge among multiple models (or branches) in a single-stage training.
They consolidate the output from the multiple classifiers and guide the classifiers to follow the ensemble Softmax logits.
The ensemble logits integrate varying predictions of the classifiers that start from different initial conditions, which can help to improve the learning of the target model \cite{ONE}. 
For recommender system, \cite{BD} proposes an online KD method that simultaneously trains two models having different capacities, and \cite{zhang2020distilling} jointly trains an embedding-based model along with a meta-path based model for better explainability.

Although the aforementioned methods have improved performance by training multiple models simultaneously, they rely on a single objective to train the models.
Thus, they cannot exploit the complementarity from the heterogeneous objectives.
Also, they do not utilize ranking information for consolidating and guiding the models, which leads to sub-optimal ranking performance.

\vspace{-0.1cm}

\section{Summary}
\label{sec:ConCF_conclusion}

\label{sec:ConCF_appendix}

\section{Supplementary Material}

\subsection{Experiment Setup}
The source code of ConCF is publicly available through the author's GitHub repository\footnote{\url{https://github.com/SeongKu-Kang/ConCF_WWW22}}.

\vspace{0.2cm}
\label{app:ConCF_setup}
\noindent
\textbf{Dataset.}
We use three real-world datasets: CiteULike \cite{wang2013collaborative}, Ciao \cite{tang2012mtrust}, and Foursquare \cite{liu2017experimental}.
These datasets are publicly available and widely used in recent studies \cite{BUIR, DERRD, CML}.
We follow the preprocessing of \cite{BUIR}.
Table~\ref{tbl:ConCF_datastats} summarizes the statistics of the datasets.
\begin{table}[h]
\caption{The statistics of the datasets.}
\small
\renewcommand{\arraystretch}{0.9}
\renewcommand{\tabcolsep}{2.5mm}
\centering
\begin{tabular}{ccccc}
\toprule
Dataset & $\#$ Users  & $\#$ Items  & $\#$ Interactions & Density \\
\midrule
CiteULike  & 5,219  & 25,181  & 125,580       & 0.096\% \\
Ciao & 7,265  & 11,211  & 149,141        & 0.183$\%$ \\
Foursquare  & 19,465 & 28,593 & 1,115,108      & 0.200$\%$ \\
\bottomrule
\end{tabular}
\label{tbl:ConCF_datastats}
\end{table}

\vspace{0.2cm}
\noindent
\textbf{Evaluation methodology.}
We randomly split each user’s interaction history into training/validation/test sets in a 60\%/20\%/20\% split \cite{CML}.
We set the maximum number of epochs to 500 and adopt the early stopping strategy;
it terminates when R@50 on the validation set does not increase for 20 successive epochs.
We keep the model with the best validation R@50 and report test set metrics with it.
Users and items that have less than 10 interactions are only included in the training set as done in \cite{CML}. 
As we focus on the top-$N$ recommendation task for implicit feedback, we evaluate all methods by using two widely used ranking metrics \cite{VAE, CML, SSCDR}: Recall@$N$ (R@$N$) and Normalized Discounted Cumulative Gain@$N$ (N@$N$). 
R@$N$ measures how many test items are included in the top-$N$ list and N@$N$ assigns higher scores on the upper-ranked test items.
We report the average value of five independent runs, each of which uses different random seeds for the data splits.

\vspace{0.2cm}
\noindent
\textbf{Model architecture.}
An overview of SingleCF and ConCF is illustrated in Figure \ref{fig:ConCF_overview}.
In this work, we focus on the pure OCCF setting where the identity of each user (and item) is available \cite{NeuMF}.
The user-item encoder first maps the user/item id into embedding vector, and the embeddings are fed into subsequent layers \cite{NeuMF}.
Then, the head predicts a relevance score for each user-item pair by using the output representations of the user-item encoder.
We employ a two-layer perceptron with \textit{relu} activation for the user-item encoder and a single-layer perceptron for the head.
Note that a variety of modeling architectures can be flexibly adopted in the proposed framework.
We leave trying more diverse architectures including feature-based models and sequential models for future study.

\vspace{0.2cm}
\noindent
\textbf{Implementation details.}
We implement ConCF and all the baselines by using PyTorch and use the Adam optimizer to train all models.
The learning rate is set to 0.01 as it generally achieves the best performance for the separate networks.
The batch size is set to 1024.
For all models, we set the dimension size ($d$) of user/item embedding to 64 and [$d \rightarrow d/2 \rightarrow d/4$] for the encoder.
We uniformly sample a negative pair for each observed interaction as it shows good performance in recent work \cite{sun2020we, BUIR}.
We also notice that advanced sampling techniques \cite{NS_std} can be used, but it is not the focus of this work.
For ConCF, the importance of top positions in generating consensus ($T$) is set to 10, the importance of consensus learning loss ($\alpha$) is set to 0.01, and the target length of the recommendation list ($N$ in Eq.\ref{eq:ConCF_perm}) is set to 50.
We apply CF-E on both row-wise and column-wise of the binary matrix $\mathbf{R}$.
The gradient normalization for balancing is applied to the shared user/item embeddings.
The details of the consistency computation are provided in Appendix.

\begin{algorithm}[t]
\SetKwInOut{Input}{Input}
\SetKwInOut{Output}{Output}
\Input{Training data $\mathcal{D}$, \# total epochs $T_E$, Ranking prediction queue $Q$ with updating period $p$}
\Output{A trained target model $\{\theta_s, \theta_x\}$,\\ or a multi-branch model $\{\theta_s, \theta_x \text{ } \forall x\}$}
Initialize model parameters $\theta_{s}, \theta_{x}$ $\text{ } \forall x$\\
Initialize balancing parameters $\lambda^0_x = 1$ $\text{ } \forall x$\\

\For{$t=0,1,..., T_E$}{
\For{each batch $\mathcal{B} \in \mathcal{D}$}{
Compute $\mathcal{L}^t_{CF\text{-}x}$ $\text{ } \forall x$\\
\If{Warm-up ($t < \lvert Q \rvert \times p$)}{
    Compute $\mathcal{L}_x^t = \mathcal{L}^t_{CF\text{-}x}$ $\text{ } \forall x$\\
}
\Else{
    Compute $\mathcal{L}_x^t = \mathcal{L}^t_{CF\text{-}x} + \alpha \mathcal{L}^t_{CL\text{-}x}$ $\text{ } \forall x$ \\
}
Compute $\mathcal{L}^t = \sum_{x \in \mathcal{F}} \lambda^t_x \mathcal{L}^t_x$ \\
Compute $\mathcal{L}^t_b$ and Update $\lambda_x$~$\text{ } \forall x$\\
Update model parameters $\theta_{s}, \theta_{x}$ $\text{ } \forall x$
}
\If{$t \text{ } \% \text{ } p == 0$}{
    Compute ranking predictions and Update $Q$\\
    Generate consensus $\pi^{t}$ for all users
}
}

Deploy with a target model $\{\theta_s, \theta_x\}$\\
Deploy with a multi-branch model $\{\theta_s, \theta_x \text{ } \forall x\}$
\caption{Algorithm of ConCF.}
\label{algo:ConCF_algo}
\end{algorithm}

\subsection{Training Details}
\label{app:ConCF_training}
The training procedure of ConCF is provided in Algorithm 1.
Also, we provide an analysis of hyperparameters for training in Table~\ref{tbl:ConCF_hp}.

\vspace{0.2cm}
\noindent
\textbf{Efficient consistency computation.}
For generating the consensus, we compute the consistency of recent ranking predictions during training.
For an efficient implementation, we utilize a queue $Q$ to store the ranking predictions of $|Q|$ different epochs while gradually reflecting the latest predictions, then compute the consistency by using $Q$.
We update the queue and the consensus every $p$ epochs, so the window size $W$ can be thought of as $W=p\times|Q|$.
In this work, we set $p$ to 20, $|Q|$ to 5.
We provide the performance with different values in Table \ref{tbl:ConCF_hp}.
We observe that ConCF achieves stable performance when the recent predictions are reflected more than a certain level (i.e., when the window size is large enough.).
Thus, we can efficiently compute the consistency by tracking the predictions of a few epochs ($|Q|$).
Interestingly, in the case of $|Q|=total\,\, epochs$, the performance is degraded, indicating that the recent prediction dynamics need to be considered important for the consistency computation.
 
\begin{table}[t]
\centering
\caption{R@50 comparison with different~hyperparameters on CiteULike.}
\small
\renewcommand{\arraystretch}{0.9}
\renewcommand{\tabcolsep}{3.7mm}
\begin{tabular}{llcc}
\toprule
\multicolumn{2}{c}{Hyperparameters} & Best head & Consensus \\
\midrule[.1em]
\multicolumn{2}{l}{Without balancing (Section \ref{sec:ConCF_blance})}           & 0.3453    & 0.3575  \\
\midrule[.1em]
\multicolumn{2}{c}{$\alpha=0$}               & 0.3228    & 0.3747  \\
\multicolumn{2}{c}{$\,\,\,\,\,\,\,\alpha=0.01$}                  & 0.3763    & 0.3896  \\
\multicolumn{2}{c}{$\,\,\,\,\alpha=0.1$}                 & 0.3712    & 0.3836  \\
\multicolumn{2}{c}{$\alpha=1$}                & 0.3719    & 0.3823  \\
\midrule[.1em]
 & $p=1$                  & 0.3632    & 0.3661  \\
       $|Q|=total\,\, epochs$    & $p=10$                 & 0.3648    & 0.3685  \\
                         & $p=20$              & 0.3660     & 0.3664  \\
\cmidrule{1-4}
    & $p=1$                 & 0.3579    & 0.3691  \\
     $|Q|=5$        & $p=10$               & 0.3772    & 0.3872  \\
                         & $p=20 $                & 0.3763    & 0.3896  \\
\cmidrule{1-4}
   & $p=1$                  & 0.3708    & 0.3829  \\
     $|Q|=10$       & $p=10$                 & 0.3781    & 0.3907  \\
                         & $p=20$                 & 0.3779    & 0.3865 \\
\bottomrule[.1em]
\end{tabular}
\label{tbl:ConCF_hp}
\end{table}

\vspace{0.2cm}
\noindent
\textbf{Other details.}
For the model parameters, $\theta_s$ denotes the shared parameters for all heads (i.e., the user/item embedding, the user-item-encoder), and $\theta_x$ denotes the parameters of each head $x$ (line 1).
In the early stages of the training, we warm up the heads only with the original CF losses, ensuring enough prediction dynamics for generating consensus be collected and making each head sufficiently specialized for each CF objective (line 6-7).
For $\alpha$ which controls the effects of $\mathcal{L}^t_{CL}$, we observe the stable performance with $\alpha >= 0.01$ (Table \ref{tbl:ConCF_hp}).
In this work, we set $\alpha=0.01$ (line 9).
Lastly, the balancing parameters $\lambda_*$ can be updated every batch or every epoch, we consistently obtain better results with the former.
As shown in Table \ref{tbl:ConCF_hp}, the performance is degraded without balancing.

\subsection{Other Related Work}
\label{app:ConCF_related_work}

\noindent
\textbf{AutoML and Loss function selection.}
In the field of AutoML, there have been a few attempts to automate the loss function search \cite{li2019lfs, SLF, AutoML}.
Recently, \cite{SLF, AutoML} select the most appropriate loss function among several candidate loss functions for each data instance by using Gumbel-Softmax.
There exists a fundamental difference between our work and the AutoML approach.
Our work is to \textit{consolidate} the complementary aspects from the heterogeneous objectives, whereas the AutoML approach \textit{selects} the most proper loss function \textit{in an exclusive manner.}
However, our analyses in Section \ref{sec:ConCF_analysis} show that each objective captures different aspects of the user-item relationships, thus selecting a single objective provides an incomplete understanding.

To further ascertain our approach, we test the idea of exclusive selection.
ConCF-S is a variant of ConCF where each instance is trained by a single head (instead of all heads) selected by Gumbel-Softmax, and AutoML \cite{AutoML, SLF} trains each instance with a selected loss function without the consensus learning.
We observe that ConCF-S achieves better performance than SingleCF (Table \ref{tbl:ConCF_main}), however, ConCF outperforms ConCF-S by a large margin.
For the AutoML approach, the performance of the best head is slightly degraded compared to SingleCF because it utilizes fewer training instances.
The ensemble result, which consolidates the predictions from all heads by the selection mechanism, shows a slightly better performance than SingleCF.
These results show that the exclusive selection is not effective enough to fully exploit the heterogeneous objectives, and again verify the effectiveness of our approach.

\begin{table}[t]
\small
\centering
\renewcommand{\arraystretch}{0.9}
\renewcommand{\tabcolsep}{2.7mm}
\caption{Performance comparison with loss function selection and AutoML approach on CiteULike.}
\begin{tabular}{ccccccc}
\toprule
                     &           & R@20    & N@20    & R@50    & N@50    \\
\midrule
\multirow{2}{*}{ConCF}          & Best head & 0.2412 & 0.1395 & 0.3763 & 0.1756 \\
                     & Consensus    & 0.2533 & 0.1474 & 0.3896 & 0.1836 \\
\midrule
\multirow{2}{*}{ConCF-S} & Best Head & 0.2158 & 0.1255 & 0.3487 & 0.1601 \\
                     & Consensus    & 0.2309 & 0.1351 & 0.3638 & 0.1750  \\
\midrule
\multirow{2}{*}{AutoML}  & Best Head & 0.1758 & 0.1032 & 0.3039 & 0.1371 \\
                     & Ensemble  & 0.1993 & 0.1204 & 0.3216 & 0.1529\\
\bottomrule
\end{tabular}
\end{table}

\vspace{0.2cm}
\noindent
\textbf{Multi-Objective Optimization.}
Multi-Objective Optimization (MOO) \cite{MOO, MOR2} aims to optimize more than one criterion (or desired goals) that may have trade-offs \cite{MOO}. 
Although the term `objective' serves as the main keyword in both MOO and our work, it is a distinct research direction from our work.
Unlike MOO which aims at generating a single model simultaneously satisfying multiple conflicting goals (e.g., accuracy and diversity), our work aims at consolidating the knowledge from the differently optimized models for the same criterion (i.e., ranking accuracy).
To this end, we utilize online consensus of separate heads (or models) having independent learning parameters, which raises novel challenges different from MOO.

\subsection{Supplementary Results}
\label{app:ConCF_sup_result}
In Table \ref{tbl:ConCF_sharing}, we compare different levels of parameter sharing among the objectives in ConCF on CiteULike.
`Full sharing' shares all parameters except for the heads, whereas `No sharing' shares no parameters, and each objective occupies a separate model.
Overall, parameter sharing between `full sharing' and `no sharing' shows good results, and in our experiment, sharing the user/item embeddings shows the best results. 
It is worth noting that the embeddings typically account for most parameters of OCCF model (more than 99\% on CiteULike). 


\begin{table}[h]
\centering
\caption{R@50 comparison of different parameter sharing.}
\small
\renewcommand{\arraystretch}{0.9}
\renewcommand{\tabcolsep}{2.8mm}
\begin{tabular}{ccccc}
\toprule
\multicolumn{1}{l}{}         &   Parameter sharing & Best head & R-con & RC-con \\
\midrule
\multirow{4}{*}{ConCF}    & Full sharing        & 0.3515            & 0.3512      & 0.3688       \\
                             & Embedding + 1 layer & 0.3702            & 0.3618      & 0.3805       \\
                             & Embedding           & 0.3763            & 0.3683      & 0.3896       \\
                             & No sharing      & 0.3662            & 0.3617      & 0.3836       \\
\bottomrule
\end{tabular}
\label{tbl:ConCF_sharing}
\vspace{-0.55cm}
\end{table}

\chapter{Distillation from Heterogeneous Models for Top-K Recommendation}
\label{chapt:HetComp}
\section{Introduction}
Recommender system (RS) has been deployed in various applications to facilitate decision-making \cite{he2020lightgcn}.
The core of RS is to provide a personalized ranking list of items to each user.
In the past decades, a variety of models with different architectures and loss~functions, from matrix factorization \cite{BPR} to graph neural networks \cite{he2020lightgcn}, have been studied to generate high-quality ranking lists.
It is known that these heterogeneous models possess different inductive biases that make the model prefer some hypotheses over others \cite{inductive3, MT_KD2}, and accordingly, they better capture certain user/item preferences that better fit the bias of each model \cite{zhu2020ensembled, concf, jointAE}.
As a result, utilizing their multi-faceted knowledge via model ensemble often achieves significantly increased accuracy over a single model \cite{MT_KD2, MT_KD4, zhu2020ensembled, concf, rank_aggregation}.
However, the fundamental limitation is that its computational cost for inference can be many times greater than that of a single model, which makes it impracticable to apply to real-time services.

An increasingly common way to reduce inference latency is to compress a large model into a smaller model via knowledge distillation (KD) \cite{KD}.
KD trains a compact model (student) by transferring the knowledge from a well-trained heavy model (teacher), effectively narrowing the performance gap between them.
Inspired by its huge success in computer vision, KD has been actively studied to compress the ranking model.
Recent \textit{ranking matching} approach \cite{darkrank, DERRD, DCD, reddi2021rankdistil, GCN_distill}
adopts listwise learning that trains the student to emulate the permutations of items (i.e., ranking list) from the teacher.
This approach has shown remarkable performance in many ranking-oriented applications such as top-$K$ recommendation \cite{DERRD, DCD, GCN_distill}, document retrieval \cite{reddi2021rankdistil, CL-DRD}, and person identification \cite{darkrank}.
Nevertheless, they mostly focus on distillation from a homogeneous teacher that has the same model type as the student.
Cross-model distillation from heterogeneous teachers, which have distinct architectures and loss functions, has not been studied well.


\begin{figure*}[t]
\centering
\begin{subfigure}[t]{0.5\linewidth}
\includegraphics[height=4cm]{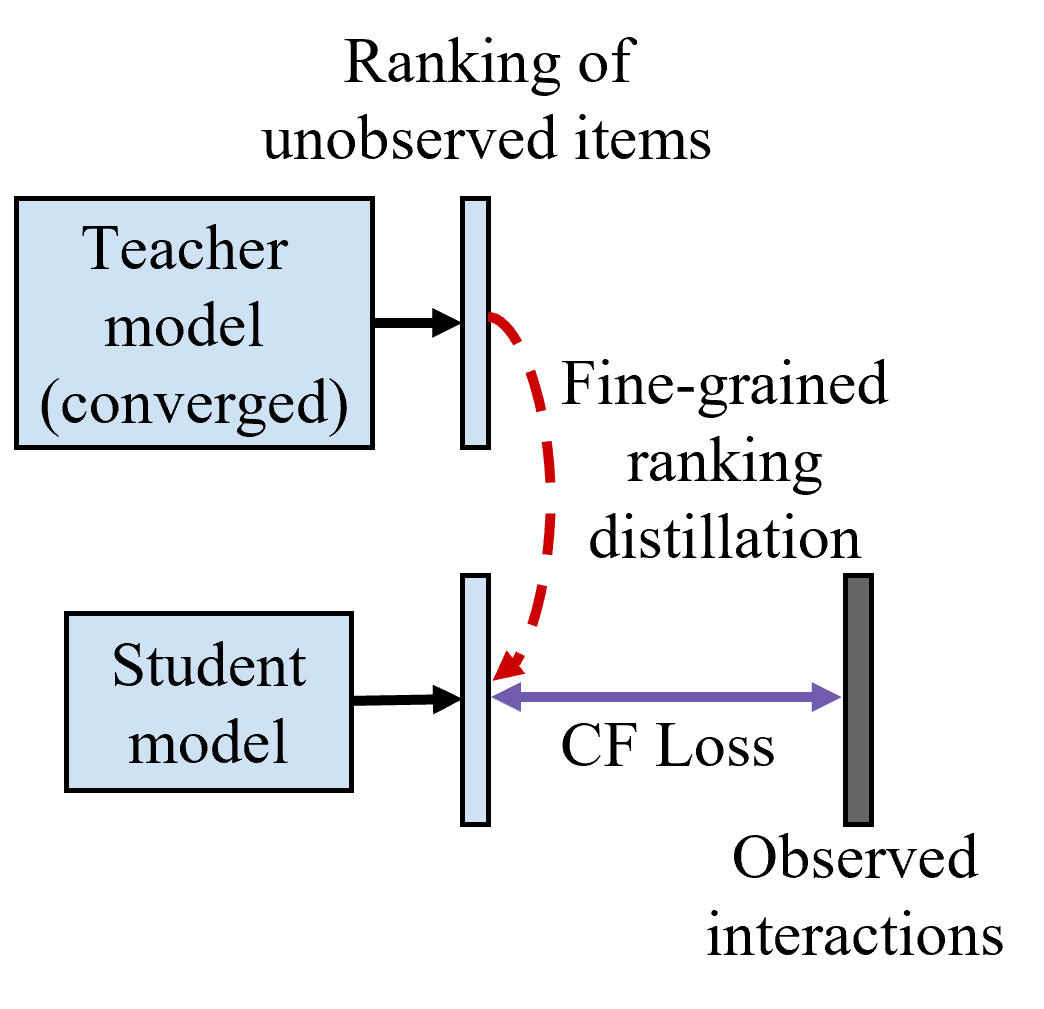}
\end{subfigure}\\
\begin{subfigure}[t]{0.9\linewidth}
\includegraphics[height=4cm]{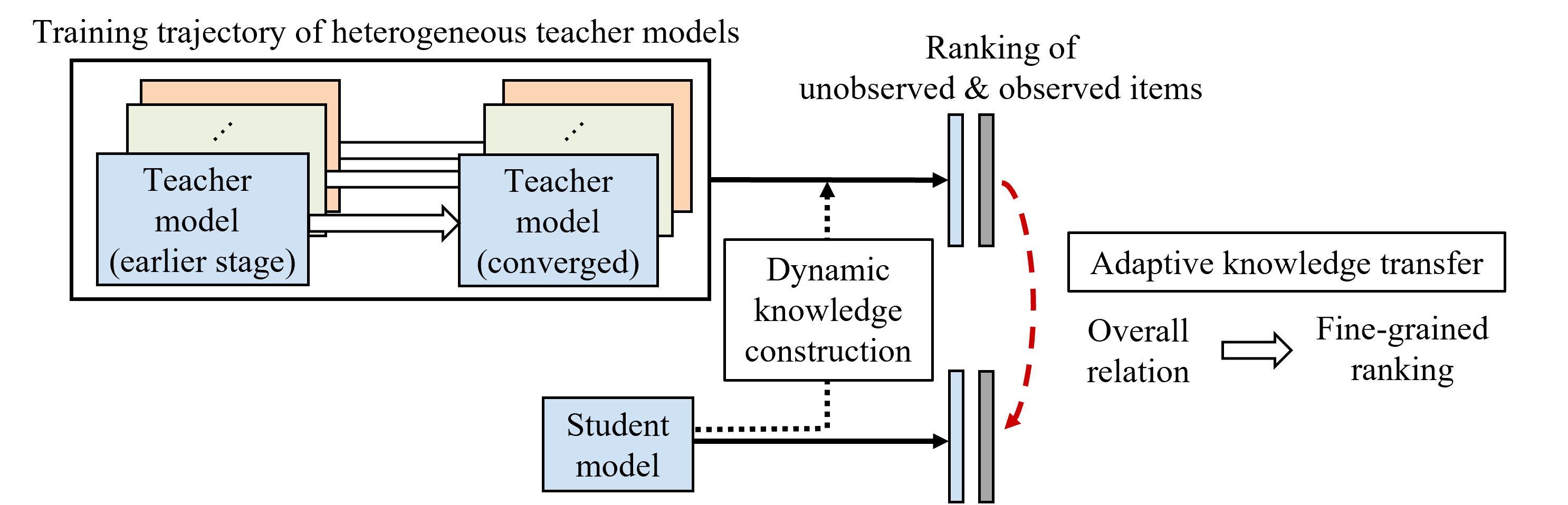}
\end{subfigure}
\caption{
A comparison of (left) the existing KD approach \cite{RD, DERRD, DCD, GCN_distill} for recommender system and (right) our proposed HetComp for heterogeneous teacher models.
HetComp supervises the student model by using an easy-to-hard sequence of ranking knowledge along with the distillation objective adjusted according to the student's learning state.
}
\label{fig:intro}
\end{figure*}

From our analysis, we observe that the efficacy of the prior KD methods significantly drops when transferring knowledge from heterogeneous teacher models, which eventually prevents the student from retaining their ensemble accuracy.
To be specific, we investigate discrepancies between ranking predictions from the student and the teacher, and we observe that the student learning from the heterogeneous teacher has a notably large discrepancy compared to the student learning from the homogeneous teacher.
Fortunately, we find that an important clue that can help to reduce the discrepancy can be obtained from the teacher's intermediate training states;
the students learning from the teacher's earlier training states have considerably lower discrepancies than the student learning from the final converged teacher, regardless of the type of teacher model.
This indicates that during the teacher's training, the teacher's knowledge gets gradually harder for the student to learn.
Indeed, we empirically show that the teacher's predictions become increasingly complex, as the latter predictions contain more diversified and individualized item~rankings.

Motivated by the observation, our key idea for improving the distillation efficacy is to dynamically transfer knowledge in an easy-to-hard sequence by using the teacher models' training trajectories (Figure \ref{fig:intro}).
As the teachers' knowledge is gradually becoming harder during their training, we aim to supervise the student model to gradually follow such sequences for easing the difficulties of learning from the converged teachers.
Our approach is based on the easy-to-hard learning paradigm \cite{jiang2015self, curriculum, Curri_survey} which has been extensively studied in various fields of machine learning.
It trains the model by using easy samples first and progressively more difficult samples so that the model can leverage previously seen concepts to ease the acquisition of more difficult ones \cite{jiang2015self, curriculum}.
Following the idea, we start distillation with relatively easy knowledge from the teachers' early training stage, then gradually increase the difficulty along the teachers' trajectories,  reducing the huge discrepancy incurred when learning from heterogeneous teachers.


In this work, we present a new KD framework, termed as HetComp, that effectively compresses the valuable but difficult ensemble knowledge of heterogeneous models, generating a lightweight model with high recommendation performance.
HetComp concretizes the easy-to-hard distillation in the following aspects:
\begin{itemize}[leftmargin=*] \vspace{-\topsep}
    \item \textbf{What to transfer}: HetComp supervises the student model via \textit{dynamic knowledge construction} which provides the easy-to-hard sequence of permutations considering the student's learning state.
    It first identifies proper knowledge from the trajectory of each teacher, and then dynamically constructs the target knowledge to guide the student model.
    This knowledge construction is applied in a personalized manner to transfer each user's recommendation result considering their different learning difficulties.

    \item \textbf{How to transfer}: HetComp uses \textit{adaptive knowledge transfer} which adjusts the distillation objective according to the student's learning state.
    It trains the student model to first focus on the overall relations in the target permutations, and gradually move on to learning to rank the fine-grained orders of the preferable items.
    Furthermore, we introduce a new transfer strategy to exploit the knowledge of both observed and unobserved user-item interactions, which has been not considered in prior~works.
\end{itemize}
\vspace{-\topsep}

\vspace{0.2cm}
\noindent
Our contributions are summarized as follows:
\begin{itemize}[leftmargin=*] \vspace{-\topsep}
    \item We reveal the difficulty of the ranking knowledge transfer from heterogeneous models and tackle the issue from the perspective of easy-to-hard distillation, which is new for recommendation.
    
    \item We propose HetComp, a new KD framework that effectively compresses the ensemble of heterogeneous models into a compact model. 
    HetComp can significantly ease the huge computational burdens of the model ensemble while retaining its high accuracy.

    \item We validate the superiority of HetComp by extensive experiments on real-world datasets.
    We also provide a comprehensive analysis of our proposed approach.
\end{itemize}
\vspace{-\topsep}

\section{Related Work}
\label{sec:HetComp_relatedwork}

\noindent
\textbf{Knowledge Distillation.}
Knowledge distillation (KD) has been actively studied for model compression in various fields \cite{curri_dialog, darkrank, RCO, KD, FitNet, xia2022device}.
KD transfers the knowledge captured by a teacher model through large capacity into a lightweight student model, significantly lowering the inference cost while maintaining comparable performance.
Pointing out that the knowledge from a single teacher model is insufficient to provide accurate supervision, many recent studies \cite{MT_KD1, MT_KD2, MT_KD3, MT_KD4, MT_KD5, MT_KD6, zhu2020ensembled} employ multiple teacher models and show great effectiveness in further improving a student model.
Notably, the state-of-the-art methods \cite{MT_KD2, MT_KD4, MT_KD6} exploit heterogeneous teacher models varying in configurations, architectures, loss functions, and many other factors to incorporate their complementary knowledge, which can provide more comprehensive guidance than a single view from a single or homogeneous teacher model.

\noindent
\textbf{Knowledge Distillation for Ranking.}
KD has been also studied for ranking problems.
Many studies \cite{RD, CD, BD, zhu2020ensembled, chen2022learning} transfer~point-wise importance on each user-item pair (or query-document pair).
However, the point-wise approach cannot consider the relations of multiple items simultaneously, which leads to the limited ranking performance \cite{darkrank, GCN_distill, DERRD}.
Recent methods \cite{DERRD, DCD, GCN_distill, CL-DRD, reddi2021rankdistil, darkrank} formulate the distillation process as a \textit{ranking matching} task.
They utilize the ranking orders from the teacher as supervision and train the student to preserve the teacher's permutation.
By directly transferring the ranking knowledge, this approach has shown state-of-the-art performance in various ranking-oriented applications such as top-$K$ recommendation \cite{DERRD, DCD, IRRRD, GCN_distill}, document retrieval \cite{reddi2021rankdistil, CL-DRD}, and person identification \cite{darkrank}.
Further, the ranking matching approach can be flexibly applied to knowledge transfer between heterogeneous models having distinct output score distributions to which the point-wise approach cannot be~directly~applied~\cite{concf}.

\noindent
\textbf{Easy-to-hard Learning Paradigm.}
Inspired by the learning process of humans, the easy-to-hard learning paradigm has been extensively studied in various fields of machine learning \cite{jiang2015self, curriculum, kumar2010self, liu2017easy, macavaney2020training, curri_RS, Curri_survey, wu2020curricula}.
It has been widely used when direct optimization of a non-convex objective function may converge to poor local minima and has been proven to play an important role in achieving a better generalization \cite{curriculum}.
Curriculum learning \cite{curriculum, Curri_survey} trains a model by gradually including data samples in ascending order of difficulty defined by prior knowledge.
On the other hand, self-paced learning \cite{kumar2010self} makes the curriculum dynamically adjusted during the training, usually based on training loss \cite{kumar2010self} or performance on the validation set \cite{curri_RS, xiang2020learning}.
The easy-to-hard learning has been applied to KD to improve the distillation efficacy in computer vision \cite{TA_KD, shi2021follow, RCO} and natural language processing \cite{curri_dialog, CL-DRD}.
\cite{TA_KD} employs a middle-sized assistant model to provide more student-friendly knowledge, \cite{RCO, shi2021follow, cazenavette2022dataset} exploits the teacher's optimization route~to~form a curriculum for the student, \cite{CL-DRD} gradually includes an increasing number of fine-grained document pairs during the~training.

\noindent \textbf{Remarks.}
The existing KD methods for RS \cite{DERRD, DCD, RD, CD, BD} mostly focus on knowledge transfer from a homogeneous teacher that has the same model type to the student model.
Distillation from heterogeneous teachers, which have distinct architectures and learning objectives to the student model, has not been studied well.
In this work, we show the necessity and difficulty of distilling the ensemble of heterogeneous teacher models and apply the easy-to-hard learning paradigm to cope with the problem.
Further, the prior KD works with the easy-to-hard learning focus on classification \cite{TA_KD, shi2021follow, RCO} or rely on domain-specific features \cite{curri_dialog}, which makes them hard to apply to the ranking problem and recommender system.
Our work provides a systematic framework tailored to compress diverse ranking models by distilling an easy-to-hard sequence of ranking knowledge considering the student's learning state.

\section{Preliminaries}
\label{sec:HetComp_preliminary}
\subsection{Problem Formulation}
Let $\mathcal{U}$ and $\mathcal{I}$ denote the user and item sets, respectively.
 Given implicit user-item interaction (e.g., click) history, a recommendation model $f: \mathcal{U} \times \mathcal{I} \rightarrow \mathbb{R}$ learns the ranking score of each user-item pair. 
Based on the predicted scores, the recommender system provides a ranked list of top-$K$ unobserved items for each user, called as top-$K$ recommendation.
Given a set of cumbersome teacher models $\mathcal{F} = \{f^1, f^2, ... , f^M\}$, our goal is to effectively compress an ensemble of the teachers into a lightweight student model $f$.
The student model has a significantly reduced computational cost for inference, and thus it is more suitable for real-time services and resource-constrained environments.
We pursue a model-agnostic solution, which enables any kind of recommendation model can be flexibly used for both teacher and student, allowing service providers to use any preferred model according to their environments.

We exploit heterogeneous teacher models with various architectures and loss functions, as diversity is a core of ensemble~\cite{rank_aggregation} and multi-teacher KD \cite{MT_KD2}.
In this work, we choose six representative types of models extensively studied for RS: MF (Matrix Factorization) \cite{BPR}, ML (Metric Learning) \cite{CML}, DNN (Deep Neural Network) \cite{NeuMF}, GNN (Graph Neural Network) \cite{he2020lightgcn}, AE (AutoEncoder) \cite{VAE}, I-AE (Item-based AE) \cite{autorec}.
We leave trying diverse combinations of other types of models for future study.
A detailed analysis of the teacher models and their ensemble is provided in~Appendix~\ref{sec:app_MTS}.

\noindent
\textbf{Notations.}
Given a ranked list (i.e., permutation of items) $\pi$, $\pi_k$ denote $k$-th item in $\pi$, and $r(\pi, i)$ denotes the ranking of item $i$ in $\pi$ where a lower value indicates a higher position, i.e., $r(\pi, i)=0$ is the highest ranking.
Note that $\pi$ is defined for each user $u$.
For notational simplicity, we omit $u$ from $\pi$ throughout the paper.

\subsection{Ranking Matching Distillation}
\label{sec:rankingKD}
Ranking matching distillation \cite{DERRD, DCD, GCN_distill, darkrank, reddi2021rankdistil} trains the student model to emulate the teacher's permutation.
A dominant strategy is to associate a probability with every permutation based on the ranking score, then train the student model to maximize the likelihood of the teacher's permutations \cite{xia2008list-wise}.
Given the teacher's permutation $\pi^t$ on a user $u$, the recent studies \cite{reddi2021rankdistil, DCD, DERRD} match the ranking of top-ranked items ($P$) while \textit{ignoring} the ranking of the remaining items ($N$).
The listwise KD loss is defined as the negative log-likelihood of permutation probability of $[P;N]$ as follows:
\begin{equation}
\mathcal{L}_{F}(P, N) = - \log \prod_{{k}={1}}^{|P|} 
 \frac{\exp \, f(u, P_k)}
  {\sum_{{j}={k}}^{|P|}  {\exp} \,  f(u, P_j) + {\sum_{{l}={1}}^{|N|} {\exp} \, f(u, N_l)}}
\label{Eq:soft_listmle}
\end{equation}
$P$ and $N$ are mostly chosen to be a few top-ranked items and a subset of items randomly drawn from the numerous remaining items, respectively \cite{DERRD, reddi2021rankdistil}.
By minimizing the loss, the student model learns to preserve the fine-grained orders in $P$, while penalizing items in $N$ below the lowest ranking of items in $P$.
It is worth noting that the orders of items within $N$ are not necessarily preserved.

\subsection{Study on Ranking Knowledge Distillation}
We present our analysis showing the difficulty of our task that compresses the ensemble knowledge of heterogeneous teacher models.
Further, we show that a clue that helps to ease the difficulty can be obtained from the teachers' intermediate training states.
Here, we use MF with embedding size 6 as the student model, and all teacher models have embedding size 64. 
Similar results are also observed with other types of students.
We train the student solely with distillation (Eq.\ref{Eq:soft_listmle}).
Please refer to Sec.\ref{sec:HetComp_experimentsetup} for the detailed setup.

\vspace{-0.05cm}
\subsubsection{\textbf{Discrepancy.}}
Since recommendation accuracy only reveals the efficacy of KD in an indirect way, we introduce a new metric for directly assessing how closely the student model learns the teacher's permutation $\pi^t$.
Let $\pi$ denote the permutation predicted by the student model.
We define the discrepancy between~$\pi$~and~$\pi^t$~by
\begin{equation}
\begin{aligned}
    D@K(\pi, \pi^{t}) = 1 - NDCG@K(\pi, \pi^{t}),
\end{aligned}
\end{equation}
where $D@K(\pi, \pi^{t})=0$ indicates the student model perfectly preserves top-$K$ ranking of $\pi^t$.
$NDCG$ is a widely used listwise ranking evaluation metric \cite{CofiRank}.
Here, we consider $\pi^t$ as the optimal ranking. 
\begin{equation}
\begin{aligned}
\hspace{-0.1cm}
\textit{\small{NDCG@K}}(\pi, \pi^t)=\frac{\textit{\small{DCG@K}}(\pi)}{\textit{\small{DCG@K}}\left(\pi^t\right) },\,  \textit{\small{DCG@K}}(\pi)=\sum_{k=1}^{K} \frac{2^{y_{\pi_k}}-1}{\log (k+1)}
\end{aligned}
\end{equation}
The relevance of each item $i$ (i.e., $y_i$) is mostly defined as ratings (for explicit feedback) or binary values (for implicit feedback).
To put a higher emphasis on a top-ranked item in $\pi^t$, we use the parametric geometric distribution \cite{rendle2014improving}, i.e., $y_i = \exp(-r(\pi^{t}, i) / \lambda)$ if $i$ is within the top-$K$ of $\pi^t$, otherwise 0.
$\lambda \in \mathbb{R}^+$ is the hyperparameter that controls the sharpness of the distribution.


\vspace{-0.05cm}
\subsubsection{\textbf{Observations and analysis.}}
We train the student model (MF) by distillation from various ranking supervisions and analyze the discrepancy between the student and the supervision.
In Table~\ref{tab:pre}, (a) denotes a homogeneous teacher, which has the same model type (i.e., MF) to the student, as used in most previous work.
(b) and (c) denote the ensemble of six homogeneous teachers with different initialization and the ensemble of six heterogeneous teachers, respectively.
`NLL' denotes the negative log-likelihood of the student model for the given supervision (Eq.\ref{Eq:soft_listmle}) where a lower value implies the student better emulates the given supervision.
We compute the metrics for each user and report the average value.
In Table \ref{tab:pre}, we observe that \textit{compressing (c) incurs a notably large discrepancy compared to compressing (a) and (b).}
In other words, the efficacy of distillation is severely degraded when we transfer the ensemble knowledge of heterogeneous teachers.
This is an interesting observation showing that items' ranking orders in permutations bring a huge difference to learning~difficulty.

\begin{table}[t]
\caption{Discrepancy to the given supervision after KD.
}
\label{tab:pre}
\small
\centering
\renewcommand{\arraystretch}{0.9}
\renewcommand{\tabcolsep}{1.2mm}
\begin{tabular}{c|lc|ccc}
\hline
\multirow{2}{*}{\textbf{Dataset}} & \multicolumn{2}{c|}{\textbf{Supervision (Teacher)}} & \multicolumn{3}{c}{\textbf{Discrepancy}} \\
\cline{2-6}
 & \multicolumn{1}{c}{\textbf{Type}} & \textbf{Recall@50} & \textbf{D@10} & \textbf{D@50} & \textbf{NLL} \\
\hline
 & (a) Single-teacher (MF) & 0.2202 & 0.6640 & 0.5167 & 0.5805 \\
Amusic & (b) Ensemble (MF)  & 0.2396 & 0.6699 & 0.5162 & 0.6048 \\
 & (c) Ensemble (Het)  & 0.2719 & 0.7417 & 0.5958 & 0.7206\\ \hline
 & (a) Single-teacher (MF) & 0.2604 & 0.5101 & 0.3716 & 0.5962 \\ 
CiteULike  & (b) Ensemble (MF) & 0.2763 & 0.5373 & 0.3910 & 0.5977 \\
 & (c) Ensemble (Het) & 0.3144 & 0.6983 & 0.5269 & 0.6906 \\ 
\hline
\end{tabular}
\end{table}

\begin{figure}[t]
\centering
\hspace{-0.15cm}
    \includegraphics[width=0.51\linewidth]{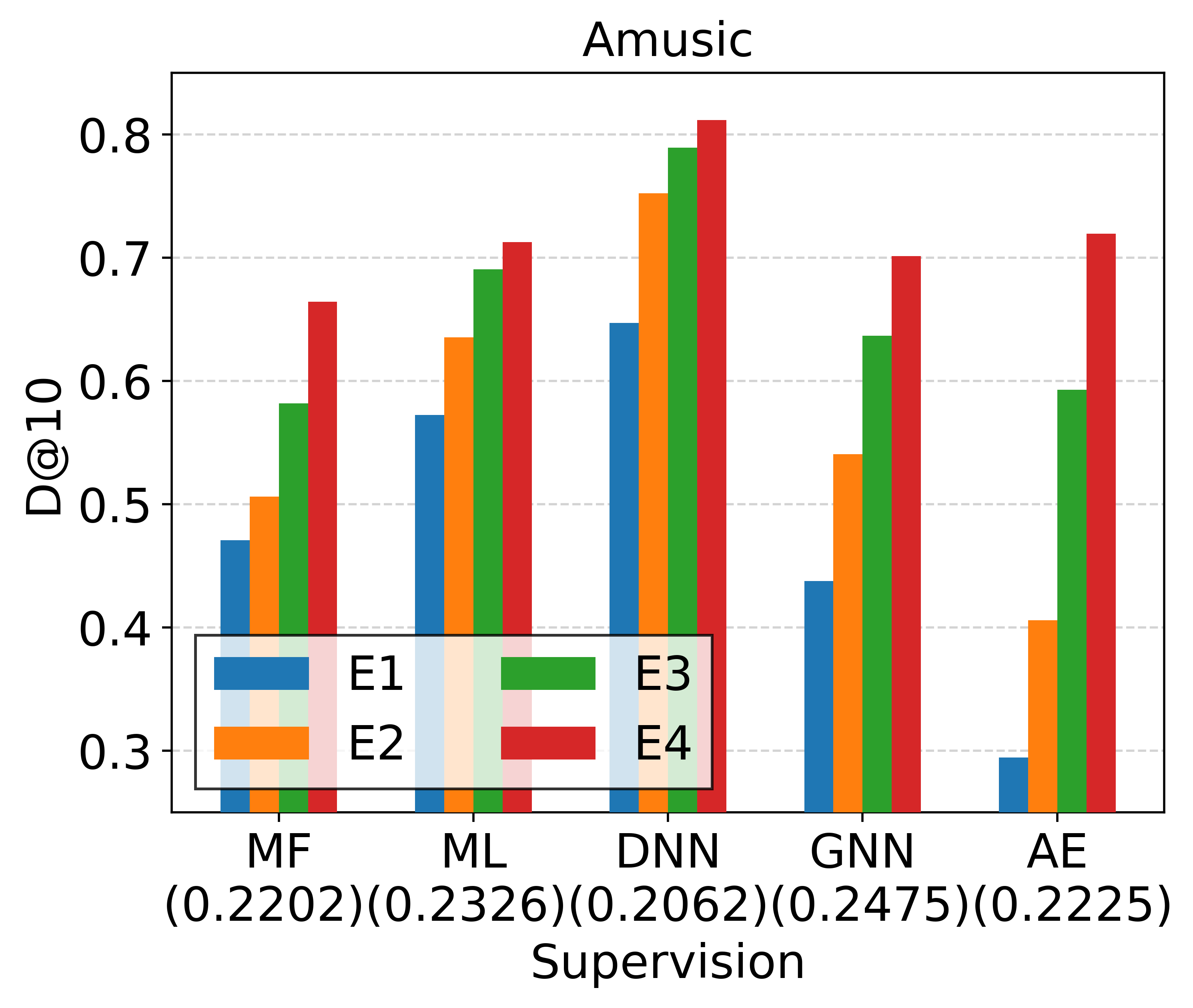}
    \hspace{-0.2cm}
    \includegraphics[width=0.51\linewidth]{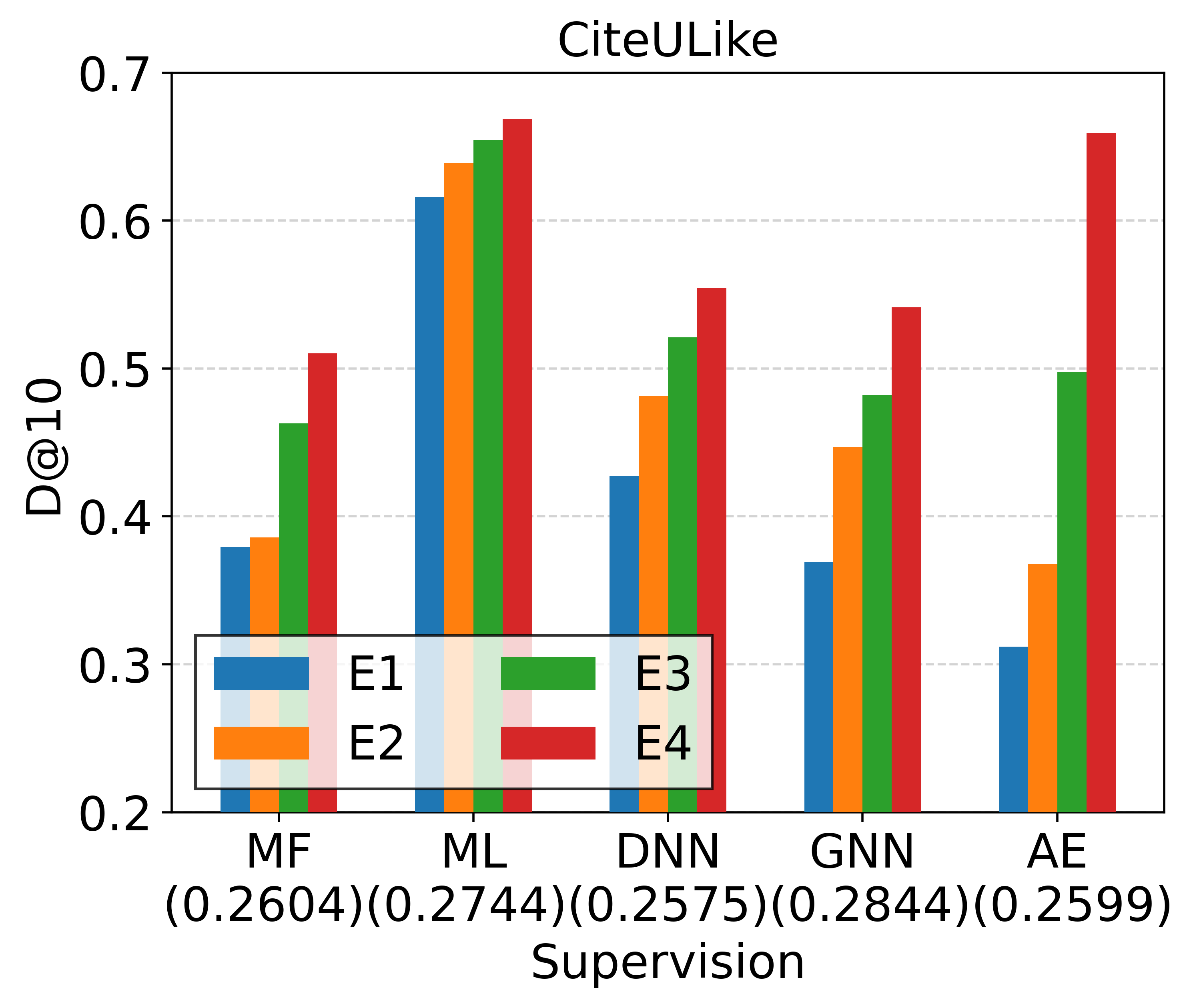}\hspace{-0.15cm}
    \caption{Discrepancy to various supervisions from intermediate training states (E1/E2/E3) and the final converged state (E4) of each teacher model. 
    We also annotate Recall@50 of each converged teacher model below the x-axis.
    }
    \label{fig:pre}
\end{figure}

To investigate where such a large discrepancy originated from, we analyze the cases of learning from each teacher in (c).
We independently train the student model (MF) with distillation from each converged teacher and its intermediate training states\footnote{For each teacher, we use 4 training states (or checkpoints), from E1 to E4, each of which corresponds to the checkpoint at 25\%, 50\%, 75\%, and 100\% of the converged~epochs.
I-AE shows similar results to AE, and its results are omitted due to the limited space.
}.
Then, we present the discrepancy to each supervision in Figure \ref{fig:pre}.
We observe that \textit{compared to the case of the homogeneous model (i.e., MF with E4), distillation from a heterogeneous model (i.e., others with E4) consistently incurs a larger discrepancy.}
It is known that heterogeneous models possess different inductive biases that make the model prefer some hypotheses over others \cite{inductive2, inductive3, MT_KD2}.
In this sense, learning from heterogeneous teachers can be particularly challenging since it needs to learn relationships that do not fit well with the student model. 
Interestingly, higher teacher accuracy does not necessarily result in a larger discrepancy.
For example, on the CiteULike dataset, GNN achieves higher accuracy and AE achieves comparable accuracy compared to MF.
However, the discrepancy is much higher in the case of learning from AE.

On the one hand, \textit{the discrepancy gradually gets larger during the teachers' training (from E1 to E4), regardless of the model type.}
That is, the teachers' knowledge gets more difficult to emulate during their training process.
As will be shown in Sec.\ref{subsec:studyH}, a model tends to learn overall patterns first, then gradually learns personalized~preferences.
We also show that teachers' knowledge becomes increasingly complex during the training, as the latter predictions contain more diverse and individualized item rankings for each user.
These observations motivate us to utilize the teachers’ training trajectories as a natural curriculum for the student model.


\section{Proposed Framework---HetComp}
\label{sec:HetComp_method}
HetComp (\textbf{Het}erogeneous model \textbf{Comp}ression for RS) framework supervises the student model using the teachers' training trajectories, based on the idea of easy-to-hard learning.
HetComp consists of the two major components designed for easy-to-hard distillation:
\begin{itemize}[leftmargin=*] \vspace{-\topsep}
    \item \textbf{(Sec.\ref{subsec:dkc}) Dynamic knowledge construction} dynamically generates knowledge with appropriate difficulty, considering the student's learning state.
    It first identifies \textit{proper} knowledge from the trajectory of each teacher and constructs target knowledge to guide the student model.
    This process is applied in a personalized manner to transfer each user's recommendation result considering their different learning difficulties.
    
    \item \textbf{(Sec.\ref{subsec:KT}) Adaptive knowledge tranfer} adjusts the distillation objective according to the student's learning state.
    It trains the student to first focus on the overall relations in the target permutations and gradually move on to learning to rank the fine-grained orders of the preferable items.
    Furthermore, we propose a new strategy to transfer the knowledge of both observed and unobserved interactions which is ignored in the~previous~works.
\end{itemize}
\vspace{-\topsep}
The overall training process of HetComp is provided in Sec. \ref{subsec:tp}.

\subsection{Dynamic Knowledge Construction}
\label{subsec:dkc}
The goal of our knowledge construction is to generate target knowledge to be transferred from the teachers' training trajectories.
This process has to meet the following desiderata:
(1) it should reflect the different learning difficulties of knowledge from each teacher model and each user;
our analysis shows that the learning difficulty varies across teacher models, 
and further, ranking with respect to each user naturally has different difficulties.
(2) it should work in a model-agnostic manner so that it can handle any kind of teacher/student model.
For this reason, we avoid using statistics that differ for each model type (e.g., training loss) for deciding on knowledge to be transferred.
We also avoid using a performance metric on additional held-out data, as it is difficult to obtain enough interactions of each user due to the~high~sparsity.

Let $\mathcal{T}=\{T^x\}_{x \in \{1,...,M\}}$ denote the teachers' training trajectories.
For each teacher $f^x$, we use its predictions (i.e., rankings of unobserved items) at $E$ different training states\footnote{In this work, we set $E$ as 4 and evenly distribute them across each teacher's trajectory.
We empirically found that $E$ hardly affects the final performance as long as they are well distributed over the teacher training process (Appendix \ref{app:cost}).}, i.e., $T^x = [\pi^{x, 1}, ..., \pi^{x, E}]$.
The last permutation $\pi^{x, E}$ corresponds to the final prediction after convergence.
Our analysis shows that the teacher's knowledge is getting harder during its training, so that $\pi^{x,e}$ is easier to emulate than $\pi^{x,e+1}$.
So, we start from $\pi^{x,1}$ and then gradually move to $\pi^{x, E}$.
We use $v$ to denote the $M$-dimensional selection vector where each element $v[x] \in \{1,...,E\}$ indicates which training state of the teacher $f^x$ is currently selected to guide the student model.
During the training, we identify proper knowledge from each teacher's trajectory $T^x$, then construct the dynamic supervision $\pi^\text{d}$ by consolidating~them~based on $v$.
The overview of the knowledge construction is provided in Algorithm~\ref{algo:dkc}.

\begin{algorithm}[t]
\SetKwInOut{Input}{Input}
\SetKwInOut{Output}{Output}
\Input{Teachers' trajectories $\mathcal{T}$, student model $f$, current selection variable $v$ with initial discrepancies $d$}
\Output{New dynamic target permutation $\pi^\text{d}$}
\BlankLine
Predict $\pi$ by student model $f$  

\ForEach{teacher $\,x \in \{1,...,M\}$}{
    \If{$(v[x] < E)$ \, $\operatorname{\mathbf{and}}$ \, $(\gamma^x > \alpha)$}{
        $v[x] = v[x] + 1$\\ 
        $d[x] = D@K(\pi, \pi^{x,\, \min(v[x]+1,\, E)})$
    }
}
Generate $\pi^\text{d} = g(\{\pi^{x, v[x]}\}_{x \in \{1,...,M\}})$
\caption{Dynamic Knowledge Construction}
\label{algo:dkc}
\end{algorithm}


To control the difficulty of distillation, we use the discrepancy between ranking predictions from the student and teachers.
Our key idea is to use the earlier predictions to ease the learning difficulty of the latter predictions.
For each teacher trajectory $T^x$, we keep tracking the discrepancy to the permutation from the next training state $D@K(\pi, \pi^{x, (v[x]+1)})$, and move to the next state if the discrepancy decreases to a certain degree.
To this end, we define the relative discrepancy ratio as follows:
\begin{equation}
\begin{aligned}
    \gamma^x = \frac{d[x]}{D@K(\pi, \pi^{x, (v[x]+1)})},
\end{aligned}
\end{equation}
where $d[x]$ denotes the \textit{initial} discrepancy to $\pi^{x, (v[x]+1)}$ computed when the student \textit{begins to} learn $\pi^{x, v[x]}$.
Note that $d[x]$ is treated as a constant and $D@K(\pi, \pi^{x, (v[x]+1)})$ decreases as the student evolves during the training.
$\gamma^x$ reveals the degree to which the learning difficulty of $\pi^{x, (v[x]+1)}$ is lowered.
Then, we employ a greedy strategy that moves to the next teacher state if $\gamma^x$ becomes larger than the threshold $\alpha$.
$\alpha \geq 1$ is a hyperparameter controlling the transition speed where a lower value incurs a faster transition.
Our greedy strategy based on the discrepancy provides an efficient curriculum considering both the student's learning state and the varying difficulties of the teachers.
Also, as the discrepancy can be measured for any model, it works in~a~model-agnostic~manner.

After updating the selection variable $v$, we generate the target permutation by 
\begin{equation}
\pi^\text{d} = g(\{\pi^{x, v[x]}\}_{x \in \{1,...,M\}})
\label{eq:pid_ensemble}
\end{equation}
where $g$ is the ranking ensemble function to consolidate the permutations.
Here, various ensemble techniques \cite{rank_aggregation}, from a simple averaging to a more sophisticated one with learnable importance, can be flexibly used in HetComp.
In this work, we use a simple technique, which leverages the consistency of prediction as importance, widely used in recent work \cite{NS_std, concf} (Appendix \ref{app:ensemble_technique}).

Once $v[x]$ equals $E$ for all $x$, the knowledge becomes equivalent to the final ensemble knowledge used in the conventional KD.
However, the key difference is that the student model is now more capable of learning from the more difficult knowledge.

\subsection{Adaptive Knowledge Transfer}
\label{subsec:KT}
We present how we transfer the dynamically-constructed permutation $\pi^\text{d}$ to the student.
We first introduce our distillation objective which is adaptively adjusted according to the student's learning state in Sec \ref{subsubsec:ado}.
Then, we explain our strategy to transfer both observed and unobserved ranking knowledge in Sec \ref{subsubsec:obs}.

\subsubsection{\textbf{Adaptive distillation objective}}
\label{subsubsec:ado}
Considering the target permutation varies during the student's training, learning the detailed ranking in the earlier phase is not only unnecessary but also daunting.
We first train the student model to learn the overall relations in the target permutation by $\mathcal{L}_{O}$.
Then, once the target permutation is constructed from the final converged predictions (i.e., $v[x]$ equals $E$, $\forall x$), we move on to learning to rank the fine-grained ranking orders by $\mathcal{L}_{F}$.
By modifying Eq.\ref{Eq:soft_listmle}, we define the distillation objective to transfer the overall relations~as~follows:
\begin{equation}
    \mathcal{L}_{O}(P, N) = - \log \prod_{{k}={1}}^{|P|} 
    \frac{\exp \, f(u, P_k)}
    {{\exp} \,  f(u, P_k) + {\sum_{{l}={1}}^{|N|} {\exp} \, f(u, N_l) }},
\label{Eq:overall_listmle}
\end{equation}
It simply pushes the items in $P$ to be ranked higher than items in $N$ without imposing any constraint among the items in $P$.

\subsubsection{\textbf{Transferring knowledge of observed/unobserved items}}
\label{subsubsec:obs}
The prior KD methods \cite{RD, CD, DERRD, DCD} mostly transfer the ranking of unobserved items without considering the ranking of observed items.
We argue that the relative priorities among the observed items are also valuable knowledge of user preference.
A straightforward way to exploit the knowledge of observed items is to transfer a permutation of the whole item set constructed from the teachers.
However, because RS models are mostly trained by collaborative filtering (CF) losses (e.g., BCE \cite{NeuMF}, BPR \cite{BPR} loss) penalizing unobserved items to have lower scores than the observed items, ``relevant but not-yet observed'' items are likely to have lower ranks compared to their true relevance in the permutation of whole item set.
We observe that if we directly distill the whole item permutation, this causes some not-yet observed items to be overly penalized in the student's prediction, which hinders the learning of user preferences.

We propose a new strategy to exploit both observed and unobserved ranking knowledge effectively.
We \textit{independently transfer} the ranking of observed items and top-ranked unobserved items to prevent such over-penalizing.
Let $P^-$ denote the ranking of top-ranked unobserved items and $N$ denote the remaining unobserved items obtained from $\pi^\text{d}$.
We additionally define $P^+$, the ranking of observed items\footnote{$P^+$ is obtained by the ensemble of the converged teachers on the observed items.
It is worth noting that $P^+$ needs to be generated only once before the student's training as the set of observed items is fixed and $\mathcal{L}_O$ doesn't transfer their detailed ranking.}.
The distillation loss is defined as~follows:
\begin{equation}
    \mathcal{L} = \mathcal{L}_{KD}(P^+, N) + \mathcal{L}_{KD}(P^-, N),
\label{Eq:final_loss}
\end{equation}
where $\mathcal{L}_{KD}$ is either $\mathcal{L}_{O}$ (Eq.\ref{Eq:overall_listmle}) or $\mathcal{L}_{F}$ (Eq.\ref{Eq:soft_listmle}) depending on the student's learning state.
Note that our strategy does not enforce the top-ranked unobserved items ($P^-$) to be located below the observed items ($P^+$), preventing pushing the not-yet observed items away from the top ranking.
Instead, it enables some unobserved items with high relevance to be naturally mixed with the observed items near the top of the ranking list.
$\mathcal{L}_{KD}(P^+, N)$ is distinguishable from the CF losses in that (1) it transfers the ranking of observed items, and (2) it does not penalize top-ranked unobserved items.

\begin{algorithm}[t]
\SetKwInOut{Input}{Input}
\SetKwInOut{Output}{Output}
\Input{Teachers' trajectories $\mathcal{T}$, student $f$, an update period $p$}
\Output{Trained student model $f$}
\BlankLine
Randomly initialize student model $f$ \\
Initialize selection variables $v_u[x] = 1$ and $d_u[x]  \,\,\,\, \forall x, \forall u$\\
Obtain $P^+_u$ by the ensemble of converged teachers \,\,\,\, $\forall u$\\
\For{$i=1,... ,\text{ }epoch_{max}$  }{
    \ForEach{user $u \in \mathcal{U}$}{
        \If{$(i \text{ } \% \text{ } p == 0)$ \,$\operatorname{\mathbf{and}}$\,
        $\operatorname{\mathbf{not}}$ $(v_u[x] == E\,, \forall x)$}{
            Update $v_u, d_u$, and $\pi^\text{d}_u$ via dynamic knowledge construction (Algorithm \ref{algo:dkc})
        }
        \If{$\operatorname{\mathbf{not}}$ $(v_u[x] == E\,, \forall x)$}{
            $\mathcal{L} = \mathcal{L}_{O}(P^+_u, N_u)$ + $\mathcal{L}_{O}(P^-_u, N_u)$
        }
        \Else{
            $\mathcal{L} = \mathcal{L}_{F}(P^+_u, N_u)$ + $\mathcal{L}_{F}(P^-_u, N_u)$
        }
        Update student model $f$
    }
}
\caption{Training Procedure of HetComp}
\label{algo:tp}
\end{algorithm}

\noindent
\subsection{The Overall Training Process}
\label{subsec:tp}
Algorithm \ref{algo:tp} summarizes the whole training process.
The knowledge construction is applied in a personalized manner for each user $u$.
Also, the knowledge construction and $\mathcal{L}_{O}$ are applied until the student learns from the final converged teachers (i.e., $v[x]$ equals $E$, $\forall x$).
We conduct the knowledge construction every $p$ epoch, since changing the target permutation every epoch is time-consuming and unnecessary.
In this work, we set $p$ as 10.
A detailed analysis of HetComp's offline training costs is provided in Appendix \ref{app:cost}.

\section{Experiments}

\subsection{Experiment Setup}
\label{sec:HetComp_experimentsetup}
Due to space limitation, we provide a summary of the experiment setup.
Please refer to Appendix \ref{sub:setup} for details.

\noindent
\textbf{Experiment settings.}
We use three real-world datasets widely used in recent KD work \cite{CD, DERRD, DCD, BD}: Amazon-music (Amusic), CiteULike, and Foursquare.
We randomly divide each user’s interaction history into train/valid/test sets in a 60\%/20\%/20\% split \cite{CML}.
We use two top-$K$ ranking metrics\footnote{Since a sampled metric (e.g., leave-one-out \cite{DERRD, DCD}) cannot correctly assess the true performance of RS \cite{krichene2020sampled}, we follow the all-ranking protocol \cite{NGCF, BUIR} for evaluation.}: Recall@$K$ (R@$K$) and NDCG@$K$ (N@$K$)\footnote{Here, NDCG is computed for implicit feedback. 
To avoid confusion, we use NDCG to refer to the recommendation accuracy in the experiment section.}.
For the student model, we use MF, ML, and DNN \cite{DERRD, DCD, BD}. 
We set the user/item embedding dimension (or the bottleneck size for autoencoder) as 64 for all teacher models and 6 for~all~student~models so that the student has roughly one-tenth of the learning parameters used by the teacher as done in \cite{DERRD, DCD, CD, BD}.

\vspace{0.1cm}
\noindent
\textbf{Baseline methods.}
We compare HetComp with various methods.
All compared methods except MTD transfer the ensemble ranking of the converged heterogeneous teachers (denoted as \textbf{Ensemble}) which is generated in the same way as HetComp.
The first group of baselines includes a point-wise KD approach.
\vspace{-\topsep}
\begin{itemize}[leftmargin=*]
    \item \textbf{RD \cite{RD}} transfers the importance of top-ranked items. The importance is defined by each item's ranking position.
\end{itemize}
\vspace{-\topsep}
\noindent
The second group includes ranking matching KD methods (Sec.\ref{sec:rankingKD}).
\begin{itemize}[leftmargin=*]
\vspace{-\topsep}
    \item \textbf{RRD \cite{DERRD}} is a ranking matching method proposed for the recommender system. It uses listwise loss focusing on top-ranked items.
    
    \item \textbf{MTD (Multi-Teacher Distillation)}: 
    We note that multi-teacher KD methods in other fields \cite{MT_KD1, ONE, MT_KD6, zhu2020ensembled} commonly use the trainable importance of each teacher on each data instance.
    Borrowing the idea, we develop MTD that transfers the  knowledge with the trainable importance of each teacher on each user's ranking.
\end{itemize}
\vspace{-\topsep}

\noindent
The third group includes the state-of-the-art KD methods for ranking that use advanced schemes to improve the distillation~quality.
\begin{itemize}[leftmargin=*] \vspace{-\topsep}
    \item \textbf{CL-DRD \cite{CL-DRD}} is the state-of-the-art KD method for document retrieval. 
    It applies curriculum learning where the learning difficulty is predefined by the absolute ranking position.

    \item \textbf{DCD \cite{DCD}} is the state-of-the-art KD method for recommender system.
    It uses the dual correction loss, which corrects what the student has failed to accurately predict, along with RRD.
\end{itemize}
\vspace{-\topsep}
\noindent
Note that KD methods \cite{KD, BD, CD, zhu2020ensembled, xia2022device} that directly use predicted scores are not applicable, as the output score distributions of the teachers and the student are very different in our task.

\vspace{-0.15cm}

\begin{sidewaystable}[!thbp]
\caption{The recommendation performance comparison. 
$Imp$ denotes the improvement of HetComp over the best baseline. $*$ denotes significance from the paired t-test (0.05 level) against the best baseline.}
\label{tab:main_b}
\footnotesize
\renewcommand{\arraystretch}{0.9}
\renewcommand{\tabcolsep}{1.2mm}
\begin{minipage}[t]{1\linewidth}
\centering
\begin{tabular}{c|c|cccc|cccc|cccc}
\hline
& \multirow{2}{*}{\textbf{Method}} & \multicolumn{4}{c|}{\textbf{Amusic}} & \multicolumn{4}{c|}{\textbf{CiteULike}} & \multicolumn{4}{c}{\textbf{Foursquare}} \\
\cline{3-14}
  &  & \textbf{R@10} & \textbf{N@10} & \textbf{R@50} & \textbf{N@50} & \textbf{R@10} & \textbf{N@10} & \textbf{R@50} & \textbf{N@50} & \textbf{R@10} & \textbf{N@10} & \textbf{R@50} & \textbf{N@50} \\
\hline
  & Best Teacher & 0.0972 & 0.0706 & 0.2475 & 0.1139  & 0.1337 & 0.0994 & 0.2844 & 0.1392 & 0.1147 & 0.1085 & 0.2723 & 0.1635\\
  & Ensemble& 0.1096 & 0.0820 & 0.2719 & 0.1273 & 0.1550 & 0.1156 & 0.3144 & 0.1571 & 0.1265 & 0.1213 & 0.2910 & 0.1786  \\
  \hline
\multirow{8}{*}{\begin{tabular}[c]{@{}c@{}}Student   \\      Model:\\      MF\end{tabular}} & w/o KD & 0.0449 & 0.0303 & 0.1451 & 0.0594 & 0.0568 & 0.0422 & 0.1372 & 0.0634 & 0.0726 & 0.0666 & 0.1806 & 0.1047 \\
  & RD & 0.0522 & 0.0387 & 0.1602 & 0.0693 & 0.0610 & 0.0472 & 0.1514 & 0.0725 & 0.0778 & 0.0703 & 0.1921 & 0.1153  \\
  & RRD & 0.0890 & 0.0659 & 0.2353 & 0.1077  & 0.0973 & 0.0740 & 0.2422 & 0.1113 & 0.0982 & 0.0905 & 0.2539 & 0.1446\\
  & MTD & 0.0901 & 0.0649 & 0.2279 & 0.1043  & 0.0993 & 0.0749 & 0.2425 & 0.1118 & 0.0955 & 0.0890 & 0.2402 & 0.1394\\
  & CL-DRD & 0.0883 & 0.0648 & 0.2375 & 0.1071  & 0.1033 & 0.0794 & 0.2512 & 0.1175 & 0.1001 & 0.0933 & 0.2528 & 0.1464\\
  & DCD & 0.0956 & 0.0675 & 0.2380 & 0.1079  & 0.1106 & 0.0851 & 0.2640 & 0.1246 & 0.1034 & 0.0965 & 0.2547 & 0.1491\\
  & HetComp & \textbf{0.1036}* & \textbf{0.0747}* & \textbf{0.2469}* & \textbf{0.1157}*  & \textbf{0.1379}* & \textbf{0.1031}* & \textbf{0.2814}* & \textbf{0.1396}* & \textbf{0.1118}* & \textbf{0.1036}* & \textbf{0.2722}* & \textbf{0.1594}*\\
\cline{2-14}
  & \textit{Imp}  & 8.37\% & 10.67\% & 3.74\% & 7.23\%  & 24.68\% & 21.15\% & 6.59\% & 12.04\% & 8.12\% & 7.36\% & 6.87\% & 6.91\%\\
\hline
\multirow{8}{*}{\begin{tabular}[c]{@{}c@{}}Student   \\      Model:\\      ML\end{tabular}} & w/o KD  & 0.0447 & 0.0310 & 0.1522 & 0.0623 & 0.0210 & 0.0148 & 0.0859 & 0.0323 & 0.0184 & 0.0139 & 0.0804 & 0.0356 \\
  & RD & 0.0706 & 0.0507 & 0.1874 & 0.0840 & 0.0835 & 0.0615 & 0.1914 & 0.0890 & 0.0729 & 0.0677 & 0.1811 & 0.1059\\
  & RRD & 0.0903 & 0.0643 & 0.2422 & 0.1074  & 0.0981 & 0.0701 & 0.2529 & 0.1116 & 0.0925 & 0.0813 & 0.2505 & 0.1366\\
  & MTD & 0.0843 & 0.0590 & 0.2293 & 0.1003  & 0.0944 & 0.0690 & 0.2519 & 0.1098 & 0.0909 & 0.0811 & 0.2440 & 0.1347\\
  & CL-DRD & 0.0866 & 0.0621 & 0.2409 & 0.1061  & 0.0989 & 0.0718 & 0.2583 & 0.1131 & 0.0931 & 0.0825 & 0.2541 & 0.1387\\
  & DCD & 0.0928 & 0.0653 & 0.2466 & 0.1086 & 0.1003 & 0.0724 & 0.2592 & 0.1144 & 0.0943 & 0.0845 & 0.2530 & 0.1399 \\
  & HetComp & \textbf{0.1020}* & \textbf{0.0751}* & \textbf{0.2470} & \textbf{0.1156}* & \textbf{0.1251}* & \textbf{0.0916}* & \textbf{0.2686}* & \textbf{0.1287}* & \textbf{0.1039}* & \textbf{0.0962}* & \textbf{0.2645}* & \textbf{0.1521}* \\
\cline{2-14}
  & \textit{Imp} & 9.91\% & 15.01\% & 0.16\% & 6.45\% & 24.73\% & 26.52\% & 3.63\% & 12.50\% & 10.18\% & 13.85\% & 4.09\% & 8.72\% \\
\hline
\multirow{8}{*}{\begin{tabular}[c]{@{}c@{}}Student   \\      Model:\\      DNN\end{tabular}} & w/o KD & 0.0460 & 0.0324 & 0.1396 & 0.0597 & 0.0414 & 0.0339 & 0.1095 & 0.0518 & 0.0693 & 0.0665 & 0.1608 & 0.0987 \\
  & RD & 0.0531 & 0.0378 & 0.1545 & 0.0670  & 0.0584 & 0.0445 & 0.1440 & 0.0671 & 0.0746 & 0.0683 & 0.1820 & 0.1060\\
  & RRD & 0.0851 & 0.0613 & 0.2255 & 0.1016  & 0.1034 & 0.0792 & 0.2552 & 0.1186 & 0.1016 & 0.0939 & 0.2584 & 0.1484\\
  & MTD & 0.0802 & 0.0563 & 0.2210 & 0.0958 & 0.0982 & 0.0710 & 0.2322 & 0.1058 & 0.0888 & 0.0797 & 0.2321 & 0.1305  \\
  & CL-DRD & 0.0889 & 0.0623 & 0.2365 & 0.1047 & 0.1083 & 0.0816 & 0.2575 & 0.1183 & 0.1039 & 0.0983 & 0.2635 & 0.1536 \\
  & DCD & 0.0919 & 0.0646 & 0.2404 & 0.1071  & 0.1114 & 0.0838 & 0.2668 & 0.1240 & 0.1060 & 0.1017 & 0.2671 & 0.1576\\
  & HetComp & \textbf{0.1045}* & \textbf{0.0768}* & \textbf{0.2534}* & \textbf{0.1190}*  & \textbf{0.1381}* & \textbf{0.1050 }*& \textbf{0.2864}* & \textbf{0.1413}* & \textbf{0.1136}* & \textbf{0.1079 }*& \textbf{0.2759}* & \textbf{0.1642}*\\
\cline{2-14}
  & \textit{Imp}& 13.71\% & 18.89\% & 5.41\% & 11.11\% & 23.97\% & 25.30\% & 7.35\% & 13.95\% & 7.17\% & 6.10\% & 3.29\% & 4.19\% \\
\hline
\end{tabular}
\end{minipage}
\end{sidewaystable}

\subsection{Distillation Effects Comparison}
\label{sec:HetComp_result}
Table \ref{tab:main_b} presents the recommendation performance of the student models trained by different KD methods.
Also, Table \ref{tab:D_b} summarizes the discrepancy values of the best baseline (i.e., DCD) and HetComp.
`Best Teacher' denotes the teacher model showing the best performance among all heterogeneous teachers on each dataset.
\vspace{-\topsep}
\begin{itemize}[leftmargin=*]
\item For all datasets and student models, HetComp significantly outperforms all baselines, effectively improving the performance of the student\footnote{We also tried KD from various teachers (e.g., Best Teacher/Homogeneous teacher ensemble). HetComp improves the distillation efficacy in all cases, and the best student performance is achieved by using the heterogeneous teacher ensemble (Appendix \ref{app:sup}).}.
Further, in terms of the number of recommended items ($K$), HetComp shows larger improvements for R@10/N@10 compared to R@50/N@50, which is advantageous for real-world services that aim to provide the most preferred items to their users.
Also, Table \ref{tab:D_b} shows that HetComp indeed achieves considerably lower discrepancy compared to the~best~baseline.

\item Compared to the point-wise KD approach (i.e., RD), the other KD methods that directly transfer the ranking orders consistently show higher performances, which again shows the importance of ranking knowledge in the top-$K$ recommendation.
On the one hand, MTD shows limited performance compared to RRD. 
Due to the high sparsity of interaction data, the user-wise learnable importance can be easily overfitted.

\item Ranking KD methods with advanced schemes (i.e., CL-DRD, DCD) improve the distillation effectiveness to some extent. 
However, for CL-DRD, we observe considerable variations for each dataset and student model.
One possible reason is that it uses predefined rules for defining the difficulty and controlling the difficulty level without considering the student's state.
On the other hand, DCD directly transfers what the student model has failed to predict, achieving consistently higher performance than~RRD.

\item Table \ref{tab:inference_time} presents the number of parameters and inference latency\footnote{We use PyTorch with CUDA from RTX A5000 and Xeon Gold 6226R CPU.} of the ensemble and HetComp.
We increase the size of the student (MF) until it achieves comparable performance to the ensemble.
Compared to the ensemble that incurs high inference costs due to the multiple model forwards, HetComp can significantly reduce the costs by distilling knowledge into the compact student model.
\end{itemize}
\vspace{-\topsep}

\noindent
\textbf{Further comparison on model generalization.}
As the easy-to-hard learning paradigm is known to yield a more generalizable model \cite{jiang2015self}, we further assess how accurately the student model captures the user preferences when the knowledge from the teachers is severely limited.
We randomly split the set of users into two sets with 80\%/20\% ratio (i.e., $\mathcal{U}_{g_1}, \mathcal{U}_{g_2}$).
Then, we train all teacher models by using only 80\% of training interactions of $\mathcal{U}_{g_1}$,
i.e., the teachers have limited knowledge of $\mathcal{U}_{g_1}$ and are not aware of $\mathcal{U}_{g_2}$.
Finally, we train the student model by using all training data with the distillation on $\mathcal{U}_{g_1}$\footnote{For training interactions not used for the teachers' training, we use the original~CF~loss.}.
That is, the student model is required to learn from the data by itself with incomplete guidance from the teachers.
In Table \ref{tab:main_g}, we observe that HetComp achieves significant improvements compared to DCD in all cases.
These results show that the student model trained by HetComp has a better generalization ability and can find more accurate hidden preferences by itself.
We believe this can be advantageous in settings where new interactions are constantly being added, and we leave the study of applying HetComp to the continual learning \cite{Continual} in the future~work.

\begin{table}[t!]
\centering
\caption{Discrepancy comparison of DCD and HetComp.}
\label{tab:D_b}
\renewcommand{\arraystretch}{0.85}
\begin{tabular}{cc|cc|cc|cc}
\hline
\multirow{2}{*}{} & \multirow{2}{*}{\textbf{Method}} & \multicolumn{2}{c|}{\textbf{Amusic}} & \multicolumn{2}{c|}{\textbf{CiteULike}} & \multicolumn{2}{c}{\textbf{Foursquare}} \\ \cline{3-8}
 &  & \textbf{D@10} &\textbf{ D@50} & \textbf{D@10} & \textbf{D@50} & \textbf{D@10} & \textbf{D@50} \\
 \hline
\multirow{4}{*}{\rotatebox{90}{MF}} & w/o KD & 0.8984 & 0.8028 & 0.7668 & 0.6608 & 0.6211 & 0.4625 \\
 & DCD & 0.6610 & 0.5286 & 0.4433 & 0.3184 & 0.2856 & 0.1690 \\
 & HetComp & 0.5929 & 0.4709 & 0.3401 & 0.2359 & 0.2135 & 0.1210 \\ \cline{2-8}
 & \textit{Imp}& 10.30\% & 10.92\% & 23.28\% & 25.91\% & 25.25\% & 28.40\% \\ \hline
\multirow{4}{*}{\rotatebox{90}{ML}} & w/o KD & 0.9072 & 0.8149 & 0.9583 & 0.8936 & 0.9532 & 0.8654 \\
 & DCD & 0.6699 & 0.5283 & 0.5272 & 0.3659 & 0.3429 & 0.1967 \\
 & HetComp & 0.5967 & 0.4756 & 0.3856 & 0.2693 & 0.2653 & 0.1481 \\ \cline{2-8}
 & \textit{Imp} & 10.93\% & 9.98\% & 26.86\% & 26.40\% & 22.63\% & 24.71\% \\ \hline
\multirow{4}{*}{\rotatebox{90}{DNN$\,\,\,$}} & w/o KD & 0.9070 & 0.8232 & 0.8225 & 0.7330 & 0.6438 & 0.5119 \\
 & DCD & 0.6862 & 0.5462 & 0.4836 & 0.3427 & 0.2377 & 0.1315 \\
 & HetComp & 0.5338 & 0.4170 & 0.2942 & 0.2101 & 0.2014 & 0.1110\\ \cline{2-8}
 & \textit{Imp} & 22.21\% & 23.65\% & 39.16\% & 38.69\% & 15.27\% & 15.59\% \\\hline
\end{tabular}
\end{table}

\begin{table}[t!]
	\caption{Accuracy-efficiency trade-off. Time (s) indicates the average wall time for generating each user's recommendation. We use PyTorch with CUDA from RTX A5000 and Xeon Gold 6226R CPU.}
	\label{tab:inference_time}
	\centering
	\renewcommand{\tabcolsep}{0.85mm}
	\renewcommand{\arraystretch}{0.9}
		\begin{tabular}{c|c|cc|cc}
			\hline
			\multirow{2}{*}{\textbf{Dataset}} & \multirow{2}{*}{\textbf{Method}} & \multicolumn{2}{c|}{\textbf{Accuracy}}  & \multicolumn{2}{c}{\textbf{Efficiency}} \\\cline{3-6}
			& & \textbf{R@10} & \textbf{N@10} & \textbf{\#Params (emb.size)} & \textbf{Time} \\
			\hline
			\multirow{2}{*}{Amusic} & Ensemble & 0.1096 & 0.0820 & 5.79M$\,\,\,$ (64) & 10.57s \\ 
			& HetComp & 0.1102 & 0.0817 & 0.27M$\,\,\,$ (18) & 0.82s \\
			\hline
			\multirow{2}{*}{CiteULike} & Ensemble & 0.1550 & 0.1156 & 11.72M (64) & 22.10s \\
			& HetComp & 0.1548 & 0.1150 & 0.45M$\,\,\,$ (15) & 1.10s \\
			\hline
			\multirow{2}{*}{Foursquare} & Ensemble & 0.1265 & 0.1213 & 18.52M (64) & 35.47s \\
			& HetComp & 0.1263 & 0.1214 & 0.96M$\,\,\,$ (20) & 2.12s\\
			\hline
	\end{tabular}
\end{table}

\subsection{Study of HetComp}
\label{subsec:studyH}
We provide in-depth analysis to provide a deeper insight of HetComp.
Supplementary results are provided in Appendix \ref{app:sup}.

\begin{figure}[t]
\centering
\hspace{-0.15cm}
\includegraphics[width=0.47\linewidth]{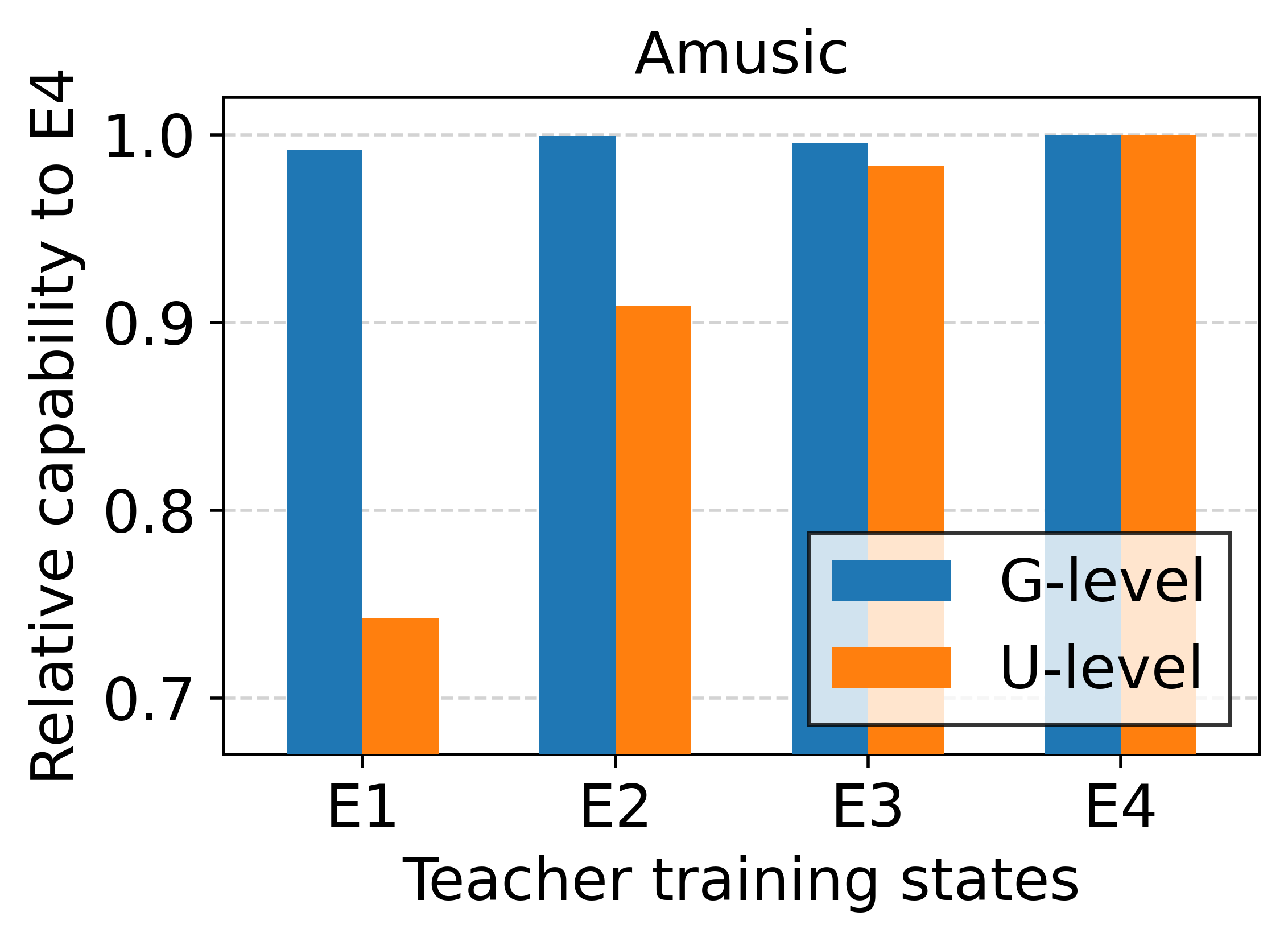}
    \hspace{-0.2cm}
    \includegraphics[width=0.47\linewidth]{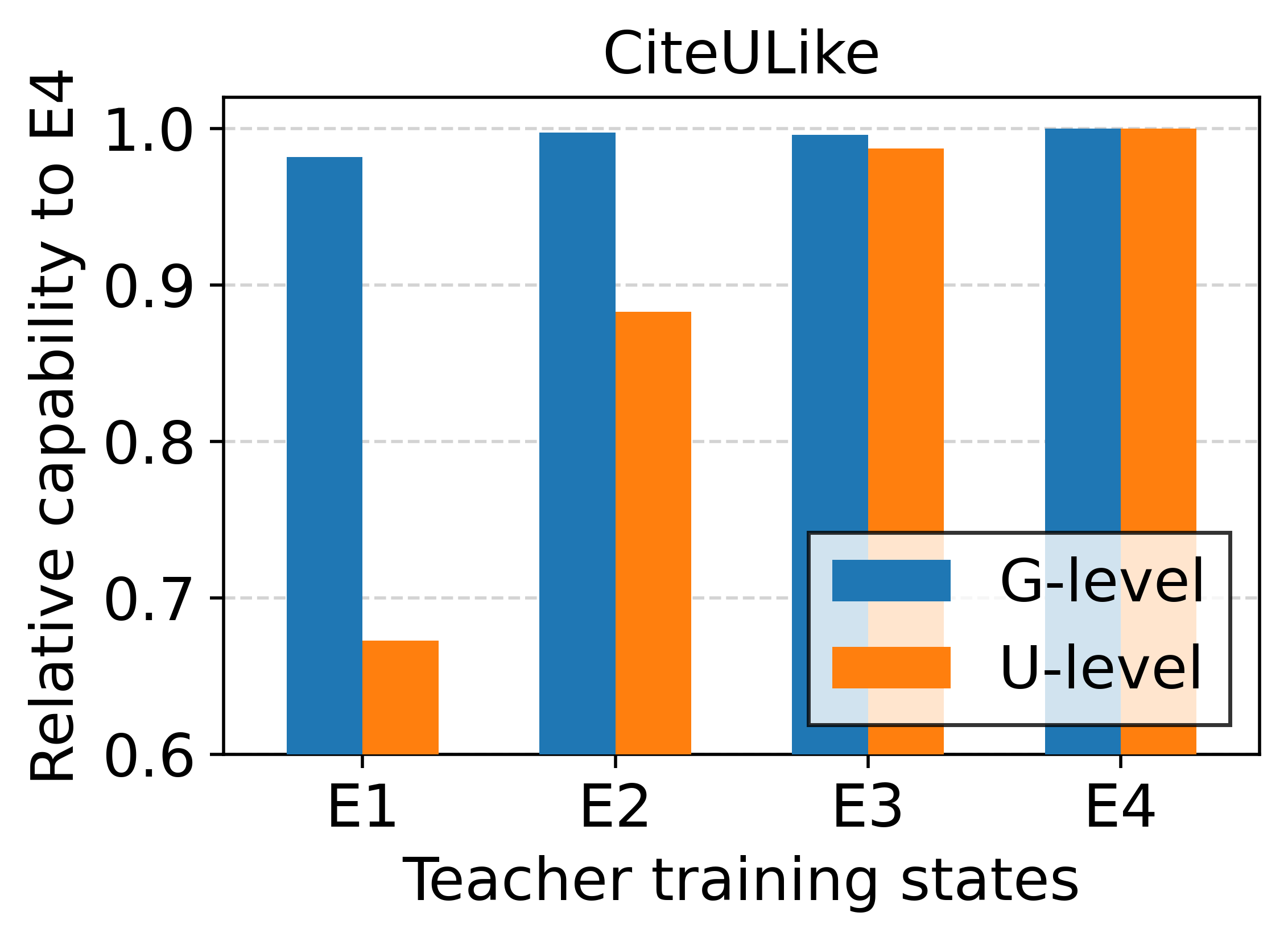}\hspace{-0.1cm}
    \vspace{-0.1cm}
    \caption{The relative ratio of the capability to capture group-level (Accuracy) and user-level (Recall@50) preference.}
    \label{fig:PG_level}
\end{figure}

\vspace{0.05cm}
\noindent
\textbf{Why are teachers' training trajectories helpful?}
To get clues to the question, we analyze what knowledge is revealed from the teachers' intermediate states (i.e., E1-E4, E4: the converged state).
All reported results correspond to the average value from all teachers for each intermediate state.
For intuitive understanding, we report the relative ratio to the value from E4.
We first analyze each state's capability of capturing the Group/User-level preferences\footnote{G-level: We find 50 user groups by $k$-means on interaction history. Then, we perform the user classification task that predicts the group to which each user belongs, using the top-50 predicted items of the user as input. We use a three-layer perception taking the items as a multi-hot vector ($\in \{0,1\}^{\mid \mathcal{I} \mid}$), U-level: the original recommendation~task.}.
In Figure \ref{fig:PG_level}, the early states capture a limited user-level preference, but an accurate group-level preference.
That is, although teachers' earlier predictions include less personalized (fine-grained) preferences, they reveal overall group-level (coarse-grained) patterns.
Further, Table \ref{tab:unique_unpop} presents the ratios of the number of unique items and unpopular items in top-50 ranking lists from the teachers\footnote{In the union of all users' top-50 ranking lists, we count (1) how many unique items exist and (2) how many items belong to unpopular items.
We regard items with the lowest 30\% of interaction numbers as unpopular items.}.
We observe that the items included in the recommendations are gradually diversified during the training.
In addition, the proportions of unpopular items, which reflect more personalized preferences than popular items, are also progressively increasing.

To sum up, in HetComp, by using the teachers' intermediate states, the items making up the knowledge are gradually changed so that it progressively reveals more diverse and personalized preferences.
This aligns well with the idea of curriculum learning that first focus on the overall concept and gradually learn specific patterns \cite{curriculum}.
It is worth noting that all our baselines transfer \textit{fixed knowledge} throughout the student's training.

\begin{sidewaystable}[!thbp]
\caption{The recommendation performance comparison of DCD and HetComp on the generalization setup.  $*$ denotes significance from the paired t-test (0.05 level) against the best baseline.
}
\label{tab:main_g}
\footnotesize
\renewcommand{\arraystretch}{0.9}
\renewcommand{\tabcolsep}{1.2mm}
\begin{minipage}[t]{1\linewidth}
\centering
\begin{tabular}{c|c|c|cccc|cccc|cccc}
\hline
& \textbf{User} & \multirow{2}{*}{\textbf{Method}} & \multicolumn{4}{c|}{\textbf{Amusic}} & \multicolumn{4}{c|}{\textbf{CiteULike}} & \multicolumn{4}{c}{\textbf{Foursquare}}\\
\cline{4-15}
& \textbf{group} &  & \textbf{R@10} & \textbf{N@10} & \textbf{R@50} & \textbf{N@50} & \textbf{R@10} & \textbf{N@10} & \textbf{R@50} & \textbf{N@50} & \textbf{R@10} & \textbf{N@10} & \textbf{R@50} & \textbf{N@50} \\
\hline
 & \multirow{2}{*}{$\mathcal{U}_{g_1}$} & Best Teacher & 0.0741 & 0.0562 & 0.2075 & 0.0948 & 0.1065 & 0.0789 & 0.2276 & 0.1111 & 0.1024 & 0.0977 & 0.2401 & 0.1457 \\
& & Ensemble & 0.0972 & 0.0691 & 0.2344 & 0.1089 & 0.1304 & 0.0955 & 0.2805 & 0.1346 & 0.1141 & 0.1079 & 0.2655 & 0.1601 \\
\hline
\multirow{8}{*}{\begin{tabular}[c]{@{}c@{}}Student \\      Model:\\      MF\end{tabular}} & \multirow{4}{*}{$\mathcal{U}_{g_1}$} & w/o KD & 0.0452 & 0.0307 & 0.1460 & 0.0600 & 0.0551 & 0.0430 & 0.1351 & 0.0642 & 0.0736 & 0.0683 & 0.1811 & 0.1063\\
 &  & DCD & 0.0784 & 0.0548 & 0.2126 & 0.0933 & 0.0955 & 0.0724 & 0.2256 & 0.1056 & 0.0984 & 0.0937 & 0.2317 & 0.1401 \\
 &  & HetComp & \textbf{0.0903}* & \textbf{0.0661}* & \textbf{0.2313}* & \textbf{0.1059}* & \textbf{0.1096}* & \textbf{0.0834}* & \textbf{0.2547}* & \textbf{0.1206}* & \textbf{0.1050 }*& \textbf{0.0993}* & \textbf{0.2483}* & \textbf{0.1494}* \\
 \cline{3-15}
 &  & \textit{Imp} & 15.18\% & 20.62\% & 8.80\% & 13.50\% & 14.76\% & 15.19\% & 12.90\% & 14.20\% & 6.71\% & 5.98\% & 7.16\% & 6.64\%  \\
 \cline{2-15}
 & \multirow{4}{*}{$\mathcal{U}_{g_2}$} & w/o KD & 0.0434 & 0.0286 & 0.1411 & 0.0570 & 0.0555 & 0.0427 & 0.1389 & 0.0649 & 0.0728 & 0.0676 & 0.1808 & 0.1058 \\
 &  & DCD & 0.0567 & 0.0370 & 0.1706 & 0.0699  & 0.0722 & 0.0517 & 0.1785 & 0.0805 & 0.0738 & 0.0717 & 0.1839 & 0.1087\\
 &  & HetComp & \textbf{0.0668}* & \textbf{0.0465}* & \textbf{0.1841}* & \textbf{0.0788}*  & \textbf{0.0801}* & \textbf{0.0558}* & \textbf{0.2008}* & \textbf{0.0876}* & \textbf{0.0880}* & \textbf{0.0829}* & \textbf{0.2088}* & \textbf{0.1227}*\\
 \cline{3-15}
 &  & \textit{Imp} & 17.81\% & 25.68\% & 7.91\% & 12.73\% & 10.94\% & 7.93\% & 12.49\% & 8.82\% & 19.24\% & 15.62\% & 13.54\% & 12.88\% \\
 \hline
\multirow{8}{*}{\begin{tabular}[c]{@{}c@{}}Student \\      Model:\\      ML\end{tabular}} & \multirow{4}{*}{$\mathcal{U}_{g_1}$} & w/o KD & 0.0442 & 0.0308 & 0.1518 & 0.0623 & 0.0209 & 0.0148 & 0.0856 & 0.0322 & 0.0201 & 0.0158 & 0.0860 & 0.0386 \\
 &  & DCD & 0.0817 & 0.0585 & 0.2179 & 0.0970 & 0.0861 & 0.0618 & 0.2214 & 0.0974 & 0.0891 & 0.0799 & 0.2289 & 0.1288\\
 &  & HetComp & \textbf{0.0908}* & \textbf{0.0641}* & \textbf{0.2284}* & \textbf{0.1027}*  & \textbf{0.1110}* & \textbf{0.0831}* & \textbf{0.2445}* & \textbf{0.1173}* & \textbf{0.1025}* & \textbf{0.0960}* & \textbf{0.2499}* & \textbf{0.1475}*\\
 \cline{3-15}
 &  & \textit{Imp} & 11.14\% & 9.57\% & 4.82\% & 5.88\%  & 28.92\% & 34.47\% & 10.43\% & 20.43\% & 15.04\% & 20.15\% & 9.17\% & 14.52\%\\
 \cline{2-15}
 & \multirow{4}{*}{$\mathcal{U}_{g_2}$} & w/o KD & 0.0468 & 0.0318 & 0.1537 & 0.0627 & 0.0208 & 0.0147 & 0.0869 & 0.0324 & 0.0201 & 0.0156 & 0.0861 & 0.0384 \\
 &  & DCD & 0.0532 & 0.0368 & 0.1601 & 0.0680 & 0.0522 & 0.0371 & 0.1644 & 0.0665 & 0.0631 & 0.0562 & 0.1826 & 0.0976 \\
 &  & HetComp & \textbf{0.0651}* & \textbf{0.0453}* & \textbf{0.1906}* & \textbf{0.0805}* & \textbf{0.0806}* & \textbf{0.0560}* & \textbf{0.2012}* & \textbf{0.0868}* & \textbf{0.0709}* & \textbf{0.0626}* & \textbf{0.2052}* & \textbf{0.1099}* \\
 \cline{3-15}
 &  & \textit{Imp} & 22.37\% & 23.10\% & 19.05\% & 18.38\%  & 54.41\% & 50.94\% & 22.38\% & 30.53\% & 12.36\% & 11.39\% & 12.38\% & 12.60\%\\
 \hline
\multirow{8}{*}{\begin{tabular}[c]{@{}c@{}}Student \\      Model:\\      DNN\end{tabular}} & \multirow{4}{*}{$\mathcal{U}_{g_1}$} & w/o KD & 0.0463 & 0.0329 & 0.1398 & 0.0602  & 0.0408 & 0.0335 & 0.1088 & 0.0514 & 0.0691 & 0.0665 & 0.1606 & 0.0986\\
 &  & DCD & 0.0805 & 0.0579 & 0.2118 & 0.0947 & 0.0882 & 0.0675 & 0.2177 & 0.1010 & 0.0914 & 0.0854 & 0.2231 & 0.1314 \\
 &  & HetComp & \textbf{0.0902}* & \textbf{0.0664}* & \textbf{0.2308}* & \textbf{0.1063}* & \textbf{0.1106}* & \textbf{0.0841}* & \textbf{0.2527}* & \textbf{0.1206}* & \textbf{0.1014}* & \textbf{0.0960}* & \textbf{0.2472}* & \textbf{0.1438}* \\
 \cline{3-15}
 &  & \textit{Imp} & 12.05\% & 14.68\% & 8.97\% & 12.25\% & 25.40\% & 24.59\% & 16.08\% & 19.41\% & 10.94\% & 12.41\% & 10.80\% & 9.44\%  \\
 \cline{2-15}
 & \multirow{4}{*}{$\mathcal{U}_{g_2}$} & w/o KD & 0.0450 & 0.03034 & 0.1385 & 0.05795  & 0.0435 & 0.0352 & 0.1121 & 0.0532 & 0.0698 & 0.0671 & 0.1615 & 0.0993\\
 &  & DCD & 0.0501 & 0.0342 & 0.1551 & 0.0642  & 0.0586 & 0.0440 & 0.1516 & 0.0673 & 0.0670 & 0.0621 & 0.1686 & 0.0974\\
 &  & HetComp & \textbf{0.0647}* & \textbf{0.0448}* & \textbf{0.1829}* & \textbf{0.0751}*  & \textbf{0.0780}* & \textbf{0.0572}* & \textbf{0.1901}* & \textbf{0.0833}* & \textbf{0.0820}* & \textbf{0.0778}* & \textbf{0.2001}* & \textbf{0.1188}*\\
 \cline{3-15}
 &  & \textit{Imp}& 29.14\% & 30.99\% & 17.92\% & 16.98\% & 33.11\% & 30.00\% & 25.40\% & 23.77\% & 22.39\% & 25.28\% & 18.68\% & 21.97\% \\
 \hline
\end{tabular}
\end{minipage}
\end{sidewaystable}

\begin{figure}[t]
\centering
\hspace{-0.2cm}
    \includegraphics[width=0.51\linewidth]{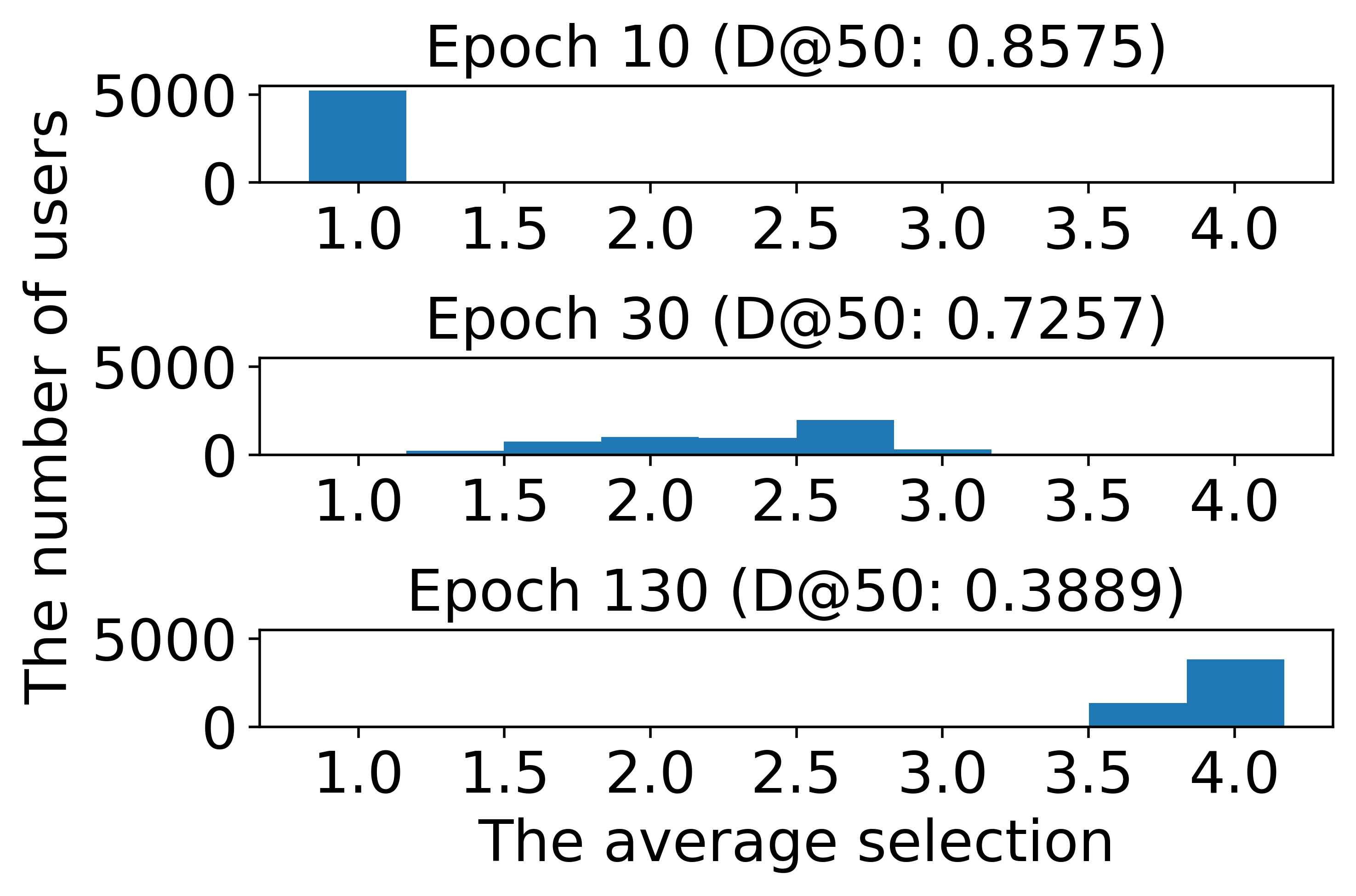}
    \hspace{-0.2cm}
    \includegraphics[width=0.49\linewidth]{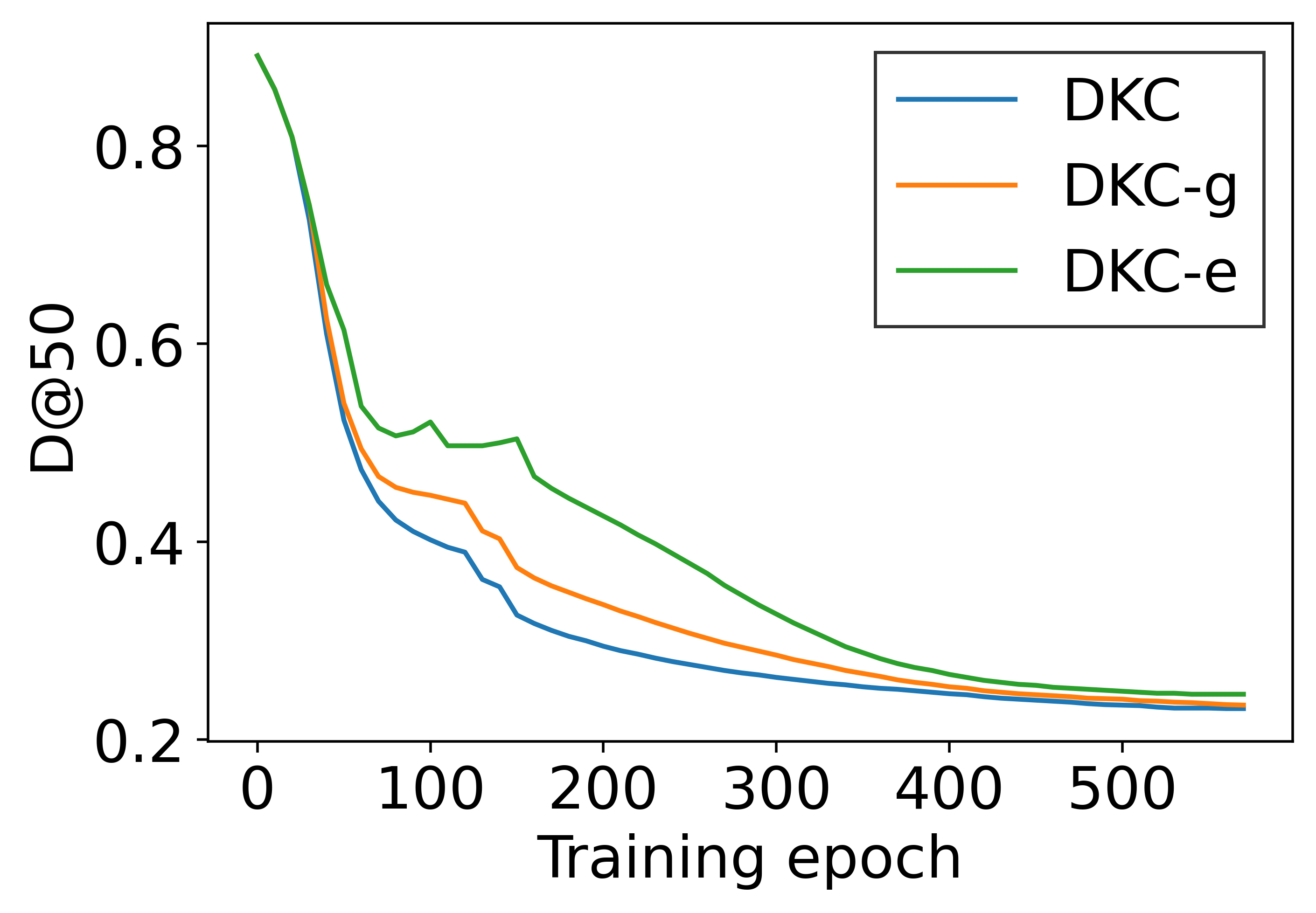}\hspace{-0.1cm}
    \caption{
    (left) the average selection distributions at epoch 10/30/130 of the student.
    (right) $D@50$ curves with variants of dynamic knowledge construction. The discrepancy is computed to the ensemble ranking from converged teachers.}
    \label{fig:DKC_anal}
\end{figure}

\vspace{0.05cm}
\noindent
\textbf{Study of dynamic knowledge construction (DKC).}
We provide a detailed analysis of DKC.
Here, we use MF (student) on CiteULike dataset.
Fig.\ref{fig:DKC_anal}(left) presents how users' average knowledge selections (i.e., $\mathbb{E}_x[v_u[x]]$) change.
During the student's training, the discrepancy to the final ranking is lowered by learning earlier predictions, and HetComp gradually moves toward more difficult knowledge following the teachers' trajectories.
Also, the knowledge selection differs for each user, which reflects different learning difficulties for each user's knowledge.
In Fig.\ref{fig:DKC_anal}(right), we compare two variants of DKC:
`DKC-g' that uses a global selection variable computed by averaging user-wise selection,
`DKC-e' that uses a simple rule that moves to the next state after a certain epoch interval.
The interval is determined by equally dividing the epoch when DKC ends by 3.
Compared to the variants, DKC provides more efficient curricula for the student model.
DKC-e rather increases the discrepancy (around epoch 100) because it cannot reflect the student's learning state.
These results show the benefits of considering both diverse learning difficulties and the student model~in~DKC.


\begin{table}[t]
\centering
\caption{The relative ratios of the number of unique items and unpopular items in top-50 ranking lists from the teachers.}
\label{tab:unique_unpop}
\renewcommand{\arraystretch}{0.85}
\begin{tabular}{c|cc|cc}
\hline
\textbf{Teacher} & \multicolumn{2}{c|}{\textbf{Amusic}} & \multicolumn{2}{c}{\textbf{CiteULike}} \\ \cline{2-5}
\textbf{training states} & \multicolumn{1}{l}{\textbf{unique}} & \multicolumn{1}{l|}{\textbf{unpopular}} & \multicolumn{1}{l}{\textbf{unique}} & \multicolumn{1}{l}{\textbf{unpopular}} \\  \hline
E1  & 0.8386 & 0.7205 & 0.7055 & 0.7190 \\
E2  & 0.8863 & 0.8618 & 0.8207 & 0.8366 \\
E3  & 0.9479 & 0.9254 & 0.9230 & 0.9191 \\
E4  & 1.0 & 1.0 & 1.0 & 1.0 \\
\hline
\end{tabular}
\end{table}

\vspace{0.05cm}
\noindent
\textbf{Ablation study.}
Table \ref{tab:ablation} provides comparison with various ablations.
We report the results of MF (student) on the CiteULike dataset.
First, (a-b) shows that our two components (i.e., DKC and ADO) designed for the easy-to-hard learning effectively improve the student model.
Also, we compare diverse ways of transferring knowledge of observed items ($P^+$);
(c) shows that utilizing $P^+$ is indeed beneficial to improve the student, and (d), which corresponds to the naive approach transferring the whole item permutation, shows that ranking transfer without separating $P^+$ and $P^-$ has adverse effects as discussed in Sec.\ref{subsubsec:obs}.
Lastly, (e) shows that penalizing all unobserved items to have lower ranks than observed items is not effective.
These results support the superiority of our strategy that independently transfers $P^+$ and $P^-$.

\begin{table}[t]
\renewcommand{\arraystretch}{0.85}
\centering
 \caption{Ablation study. DKC: dynamic knowledge construction (Sec.\ref{subsec:dkc}), ADO: adaptive distillation objective (Sec.\ref{subsubsec:ado}).}
\label{tab:ablation}
\renewcommand{\tabcolsep}{0.75mm}
\begin{tabular}{l|cccc|c}
\hline
\textbf{Ablation} & \textbf{R@10} & \textbf{N@10} & \textbf{R@50} & \textbf{N@50} & \footnotesize{\textbf{\textit{Imp.}R@10}}\\
\hline
HetComp & 0.1379 & 0.1031 & 0.2814 & 0.1396 & - \\
\hline
(a) w/o DKC (i.e., KD from Ensemble) & 0.1264 & 0.0951 & 0.2711 & 0.1308 & 9.10\%  \\
(b) w/o ADO (i.e., only $\mathcal{L}_F$) & 0.1311 & 0.0992 & 0.2765 & 0.1360 & 5.19\%\\
(c) w/o $\mathcal{L}_{KD}(P^+,N)$ & 0.1303 & 0.0994 & 0.2754 & 0.1349 & 5.83\%\\
(d) $\mathcal{L}_{KD}(\{P^+, P^-\}, N)$ & 0.1241 & 0.0903 & 0.2796 & 0.1314 & 11.12\%\\
(e) $\mathcal{L}_{KD}(P^+,\{P^-,N\}) + \mathcal{L}_{KD}(P^-,N)$ & 0.1262 & 0.0950 & 0.2665 & 0.1310 & 9.27\% \\
\hline
\end{tabular}
\end{table}

\section{Summary}
\label{sec:HetComp_conclusion}
This paper proposes \proposed, a \cdr framework for cold-start users based on a semi-supervised approach. 
We first provide an analysis showing that only a few users overlaps between two domains in real-world \cdr scenarios, and point out that the supervised approach is not effective due to the lack of labeled data.
\proposed models the users and items in the metric spaces where the similarities among users and items are represented as their distances, and trains a cross-domain mapping function using the distance-based loss defined by all the items (unlabeled data) as well as the overlapping users (labeled data).
Furthermore, we introduce a multi-hop neighborhood inference technique to infer the latent vectors of cold-start users by fully utilizing their neighborhood information.
Through our extensive experiments, we demonstrate that \proposed learns the cross-domain relationship more accurately than the existing supervised approaches, which considerably improves the recommendation accuracy for the cold-start users.

\label{sec:HetComp_appendix}
\section{Supplementary Material}

\subsection{Experiment Setup}
\label{sub:setup}
\subsubsection{\textbf{Dataset}}
We use three real-world datasets: Amazon-music (Amusic), CiteULike, and Foursquare.
These datasets are publicly accessible and also widely used in previous work \cite{BD, DERRD, CD, DCD, BUIR}.
We follow the preprocessing of \cite{BUIR} (CiteULike, Foursquare) and apply 10-core filtering (Amusic).
Table \ref{tbl:datastats} provides the data statistics.

\begin{table}[h]
\centering
\renewcommand{\arraystretch}{0.9}
\caption{Statistics of the datasets.}
\begin{tabular}{c|c|c|c|c}
\hline
\textbf{Dataset} & \textbf{User \#} & \textbf{Item \#} & \textbf{Interaction \#} & \textbf{Density} \\\hline
Amazon-music & 5,729 & 9,267 & 65,344 & 0.001231\\
CiteULike & 5,219 & 25,181 & 125,580 & 0.000956 \\
Foursquare & 19,465 & 28,593 & 1,115,108 & 0.002004 \\
\hline
\end{tabular}
\label{tbl:datastats}
\end{table}

\subsubsection{\textbf{Experiment details}}
For all experiments and inferences, we use PyTorch with CUDA from RTX A5000 and Xeon Gold 6226R CPU.
We report the average value of five independent runs.
For all baselines, we use the public implementations provided by the authors.
However, as done in \cite{CL-DRD}, we found that their sampling processes for top-ranked unobserved items (i.e., $P^-$) are unnecessary, and removing the processes gave considerable performance improvements for the ranking matching KD methods.
For this reason, we remove the sampling process for all ranking matching methods in our experiments.
In D$@K$, $\lambda$ that controls the sharpness of the exponential function is set to 10.
For dynamic knowledge construction, we use D$@50$, $p$ is set to 10, and $\alpha$ is set to $1.05$. 
As the student model gradually converges, we adopt a simple annealing schedule that decreases the value $\alpha=\alpha \times 0.995$ every $p$ epoch.
We set $E$ as 4, each of which corresponds to the checkpoint at 25\%, 50\%, 75\%, and 100\% of the converged epoch for each teacher.
Lastly, $|P^-|$ is set to 50.
For baseline-specific hyperparameters, we tune them in the ranges suggested by~the~original~papers.

\subsection{Study on the Ensemble of Teacher Models}
\label{sec:app_MTS}

\subsubsection{\textbf{Recommendation models for teacher}}
\label{sec:ranking_teacher}
In this work, we use six recommendation models with different architectures and learning objectives.
These models are the representative models for each model type and have shown competitive performance.
\begin{itemize}[leftmargin=*] \vspace{-\topsep}
    \item \textbf{MF} (BPR \cite{BPR}): a matrix factorization-based model trained by a pair-wise ranking loss. The ranking score is defined by the inner product of the user and item latent factors.
    \item \textbf{ML} (CML \cite{CML}): a metric learning-based model trained by a pair-wise hinge loss. The ranking score is defined by the Euclidean distance in the unit-ball metric space.
    \item \textbf{DNN} (NeuMF \cite{NeuMF}): a deep neural network-based model trained by binary cross-entropy. The ranking score is computed by the non-linear function of multi-layer perceptrons.
    \item \textbf{GNN} (LightGCN \cite{he2020lightgcn}): a graph neural network-based model~trained by a pair-wise loss. 
    The ranking score is computed by aggregating the  user and item representations from multiple~GNN~layers.
    \item \textbf{AE} (VAE \cite{VAE}): a variational autoencoder-based model. The ranking score is computed by the generative module (i.e., decoder).
    \item \textbf{I-AE}: a variant of VAE that learns item-side interactions \cite{autorec, jointAE}. It is known that the item-side autoencoder captures complementary aspects to its user-side counterpart \cite{jointAE}.
\end{itemize} \vspace{-\topsep}
\noindent
The best ensemble performance is achieved by using all six models, and we provide empirical evidence supporting our configuration in the next subsection.
As HetComp is a model-agnostic framework, providers can use any model according to their needs.
We leave trying diverse combinations of other types of models for~future~study.

\vspace{-0.1cm}
\subsubsection{\textbf{Necessity of heterogeneous teachers}}
Table \ref{tab:ensemble_study} presents an empirical study on the configuration of the teacher models.
`Best Teacher' denotes the teacher model showing the best performance on each dataset.
`Ensemble' denotes the ensemble results of six heterogeneous models.
`Ensemble-id' denotes the ensemble of six identical models having different initialization\footnote{We use the model showing the best performance on each dataset. We report the best performance among the score averaging and the importance-aware ensemble (\ref{app:ensemble_technique})}. 
`w/o model' denotes the ensemble of five models excluding the model~from~`Ensemble'.

First, Ensemble consistently shows higher performance than Ensemble-id and the best teacher model.
We investigate the correlations of model predictions for Ensemble and Ensemble-id\footnote{We compare test interactions included in the top-50 ranking of each model.
We compute the correlation of all pairs of models for each user and compare the average value.
We use the Matthews correlation coefficient (MCC) provided in sklearn.}.
The former shows 23\% (Amusic), 20\% (CiteULike), and 38\% (Foursquare) lower prediction correlations compared to the latter.
It is well known that models with high diversity boost the overall ensemble accuracy, and the lower correlations of Ensemble support its higher performance to some extent.
This observation also aligns with the recent studies of the ensemble \cite{rank_aggregation} and the multi-teacher KD \cite{MT_KD2, MT_KD4, MT_KD6} showing that the diversity of models is a key factor of performance.
Lastly, the best performance of Ensemble is achieved by consolidating all six models; all cases of five models (i.e., w/o model) show limited performance compared to Ensemble.


\begin{sidewaystable}[!thbp]
\caption{Ensemble Study. The best performance is achieved by using six heterogeneous models.}
\label{tab:ensemble_study}
\footnotesize
\begin{minipage}[t]{1\linewidth}
\centering
\begin{tabular}{c|cccc|cccc|cccc}
\hline
\multirow{2}{*}{Method} & \multicolumn{4}{c|}{\textbf{Amusic}} & \multicolumn{4}{c|}{\textbf{CiteULike}} & \multicolumn{4}{c}{\textbf{Foursquare}} \\
\cline{2-13}
 & \textbf{R@10} & \textbf{N@10} & \textbf{R@50} & \textbf{N@50} & \textbf{R@10} & \textbf{N@10} & \textbf{R@50} & \textbf{N@50} & \textbf{R@10} & \textbf{N@10} & \textbf{R@50} & \textbf{N@50} \\
 \hline
Best Teacher & 0.0972 & 0.0706 & 0.2475 & 0.1139 & 0.1337 & 0.0994 & 0.2844 & 0.1392 & 0.1147 & 0.1085 & 0.2723 & 0.1635 \\
Ensemble & 0.1096 & 0.0820 & 0.2719 & 0.1273 & 0.1550 & 0.1156 & 0.3144 & 0.1571 & 0.1265 & 0.1213 & 0.2910 & 0.1730 \\
\hline
Ensemble-id & 0.1013 & 0.0736 & 0.2569 & 0.1181 & 0.1511 & 0.1130 & 0.2952 & 0.1505 & 0.1215 & 0.1174 & 0.2853 & 0.1709 \\ \hline
w/o MF & 0.1091 & 0.0826 & 0.2634 & 0.1266 & 0.1545 & 0.1161 & 0.3128 & 0.1564 & 0.1206 & 0.1124 & 0.2883 & 0.1708 \\
w/o ML & 0.1050 & 0.0798 & 0.2574 & 0.1232 & 0.1541 & 0.1137 & 0.3085 & 0.1541 & 0.1222 & 0.1181 & 0.2854 & 0.1746 \\
w/o DNN & 0.1074 & 0.0819 & 0.2642 & 0.1263 & 0.1543 & 0.1147 & 0.3134 & 0.1560 & 0.1193 & 0.1099 & 0.2855 & 0.1678 \\
w/o GNN & 0.1085 & 0.0823 & 0.2634 & 0.1259 & 0.1544 & 0.1150 & 0.3117 & 0.1557 & 0.1197 & 0.1124 & 0.2857 & 0.1703 \\
w/o AE & 0.1032 & 0.0776 & 0.2520 & 0.1204 & 0.1482 & 0.1101 & 0.3046 & 0.1508 & 0.1177 & 0.1102 & 0.2775 & 0.1658 \\
w/o I-AE & 0.1040 & 0.0797 & 0.2613 & 0.1243 & 0.1519 & 0.1129 & 0.3046 & 0.1528 & 0.1204 & 0.1150 & 0.2812 & 0.1710 \\ \hline
\end{tabular}
\end{minipage}
\end{sidewaystable}

\begin{figure*}[t]
\centering
\captionsetup[subfigure]{justification=centering}
\begin{minipage}[r]{0.9\linewidth}
\begin{subfigure}[t]{0.495\linewidth}
    \includegraphics[width=\linewidth]{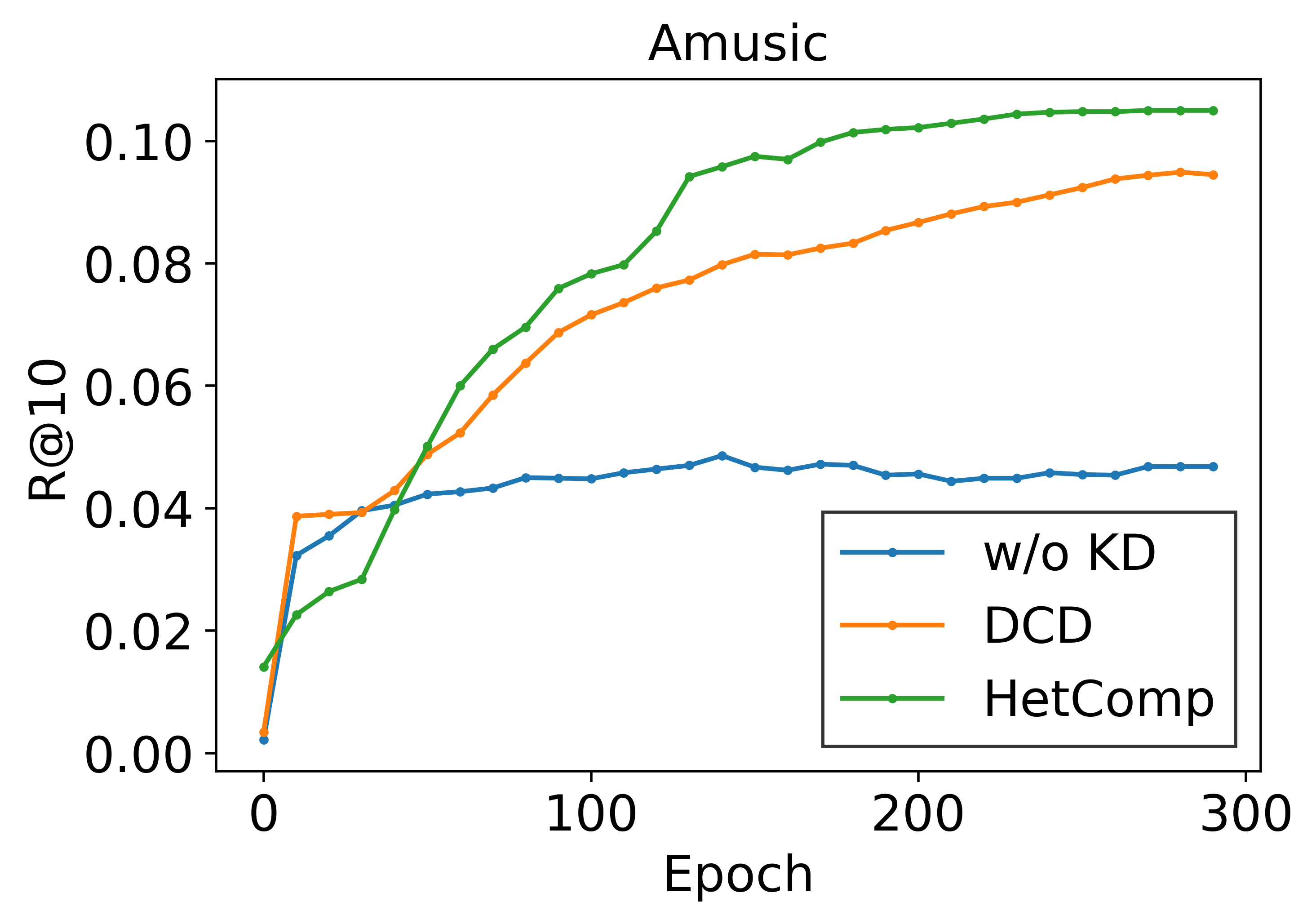}
\end{subfigure}
\begin{subfigure}[t]{0.495\linewidth}
    \includegraphics[width=\linewidth]{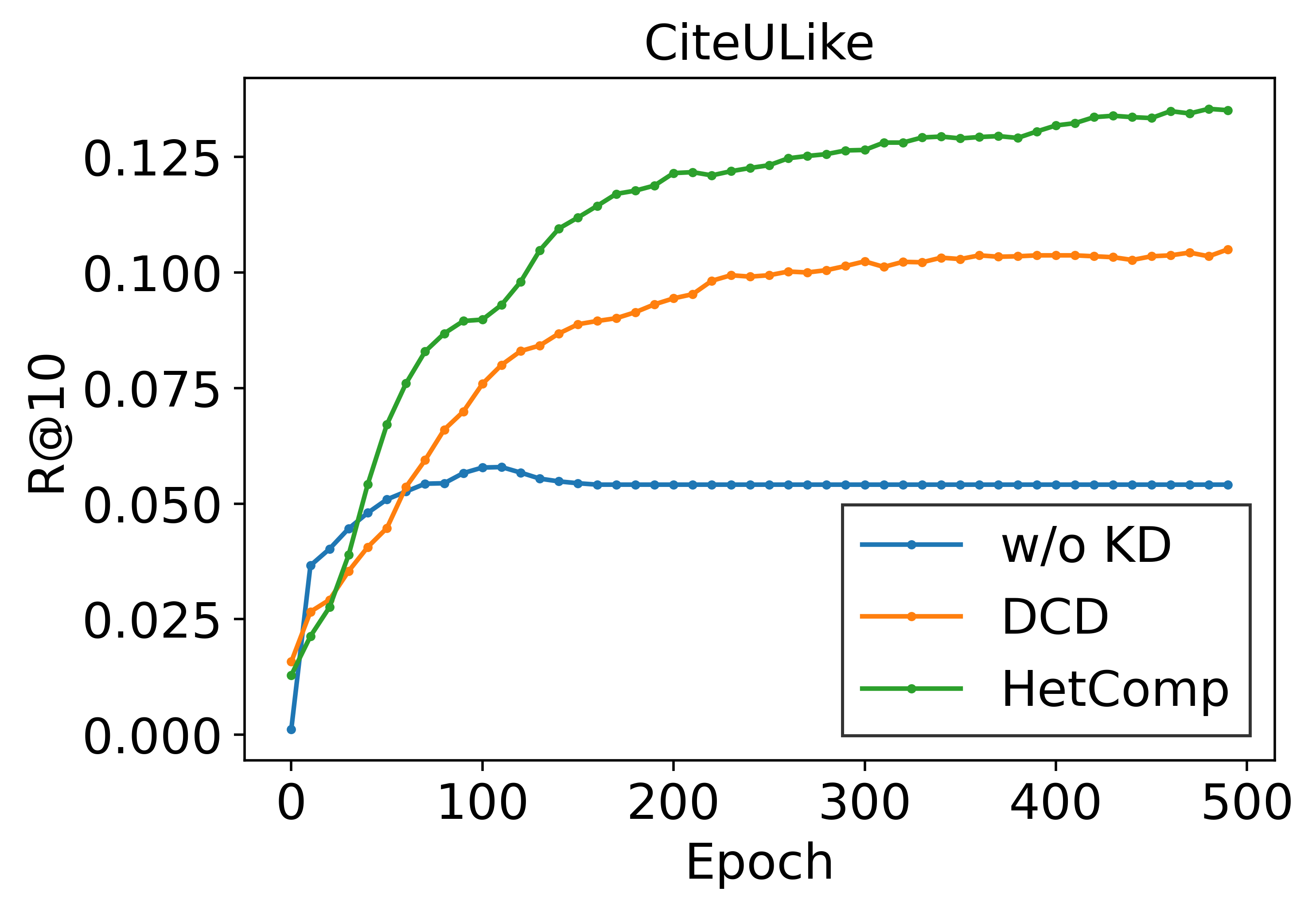}
\end{subfigure}
    \subcaption{The benchmark setup (Table \ref{tab:main_b}).}
\end{minipage}\\
\begin{minipage}[r]{0.9\linewidth}
\begin{subfigure}[t]{0.495\linewidth}
    \includegraphics[width=\linewidth]{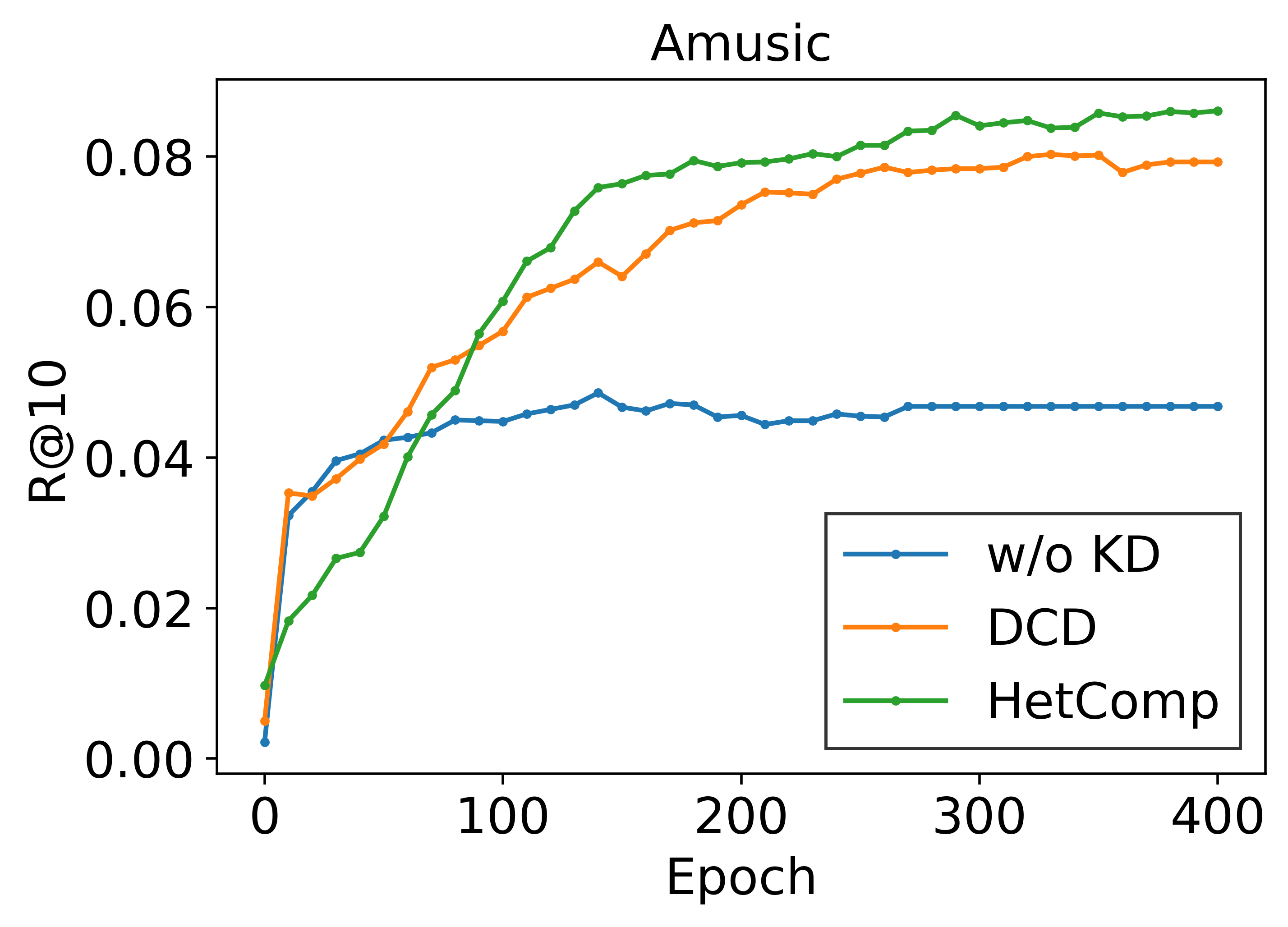}
\end{subfigure}
\begin{subfigure}[t]{0.495\linewidth}    \includegraphics[width=\linewidth]{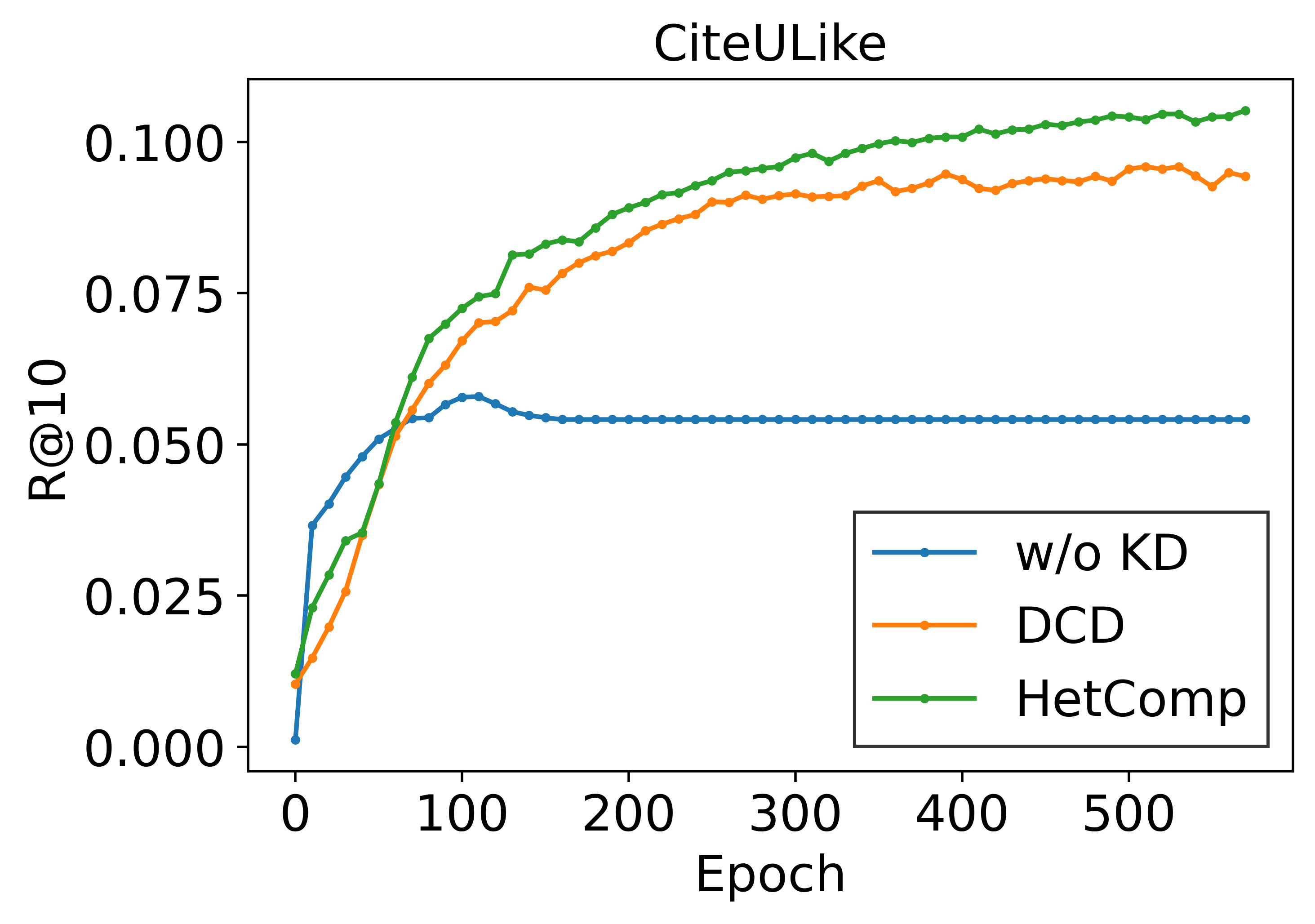}
\end{subfigure}
    \subcaption{The generalization setup (Table \ref{tab:main_g}).}
\end{minipage}
\caption{Training curves of w/o KD, DCD, and HetComp. Testing recall per 10 epochs. After convergence, we plot the last value.}
\label{fig:curve}
\end{figure*}

\subsubsection{\textbf{Ensemble technique}}
\label{app:ensemble_technique}
We now describe our choice of the ensembling function, $g$ for ranking knowledge construction.
Since each teacher model better predicts certain user-item interactions than others, it is vital to selectively reflect their knowledge into the ensemble.
We note that \textit{the consistency of model prediction} is a key factor revealing the reliability of the prediction \cite{tc-ssl}. 
This factor has been successfully employed for RS for obtaining reliable negative samples \cite{NS_std} and for consolidating multiple heads of the multi-task learning model \cite{concf}.
More sophisticated techniques can be considered, but we empirically obtain satisfactory performance with our choice.
We provide comparisons with the technique using trainable importance in the experiments.

$g$ generates an ensemble ranking $\pi^\text{d}$ by consolidating a set of permutations $\Pi=[\pi^1, \pi^2, ..., \pi^M]$.
For top-$K$ of each permutation $\pi^x$, each ranking prediction $r(\pi^x,i)$ has an importance score $c^x_{i}$:
\begin{equation}
\begin{aligned}
c^x_{i} = \exp(-r(\pi^x,i)/\lambda) + \exp(-\operatorname{std}[r(\pi^x,i)]/\lambda)
\end{aligned}
\end{equation}
The first term put the higher score on items with a higher ranking, and the second term uses the variance of predictions to favor items that the model makes consistent predictions.
Following \cite{NS_std}, we define the consistency based on the variance of the latest~5~epochs:
\begin{equation}
\begin{aligned}
\operatorname{std}\left[r(\pi^x,i)\right] =&\sqrt{\sum_{s=t-4}^{t}\left(\left[r(\pi^x,i)\right]_{s}-\operatorname{Mean}\left[r(\pi^x,i)\right]\right)^{2} / 5}, \\
\operatorname{Mean}\left[r(\pi^x,i)\right] =&\sum_{s=t-4}^{t}\left[r(\pi^x,i)\right]_{s} / 5 .
\end{aligned}
\end{equation}
where $\left[r(\pi^x,i)\right]_{s}$ denotes the ranking prediction at epoch $s$.
Finally, the ensemble ranking is generated by reranking items based on the overall importance, i.e., $\mathbb{E}_{x}[c^x_{i}]$.
Note that the consistency is precomputed only once before the distillation and incurs no additional costs for HetComp.

\subsection{Offline Training Cost of HetComp}
\label{app:cost}
HetComp requires additional space and computation costs mostly for the knowledge construction process.
For the teachers' training trajectories $\mathcal{T}$, we store the permutations of top-ranked ($K$) unobserved items.
Note that the rankings of the remaining unobserved items are unnecessary.
Also, we use the permutations of observed items.
In sum, HetComp uses $(K \times E)$ + $|P^+_u|$ space for user $u$ on each teacher.
$K$ and $|P^+_u|$ usually have small values in many recommendation scenarios,
and we empirically obtain satisfactory results with $E$ around 3 as long as they are well distributed (Figure~\ref{fig:alpha}).

\begin{table}[ht]
\centering
\renewcommand{\arraystretch}{0.85}
\caption{The average time cost of each epoch during the offline training.}
\label{tab:epochtime}
\begin{tabular}{c|c|c|c}
\hline
\textbf{Method} & \textbf{Amusic} & \textbf{CiteULike} & \textbf{Foursquare} \\
\hline 
RRD & 0.69s & 2.27s & 8.12s \\
DCD & 1.38s & 3.74s & 12.01s \\
HetComp & 1.15s & 3.08s & 11.81s\\
\hline
\end{tabular}
\end{table}

In Table \ref{tab:epochtime}, we report the average time cost of HetComp and ranking matching KD baselines.
We use MF as the student model, and similar results are also observed with other base models.
Compared to RRD which uses basic listwise learning, DCD and HetComp additional training costs as they require further computations for the distillation.
However, unlike DCD which requires such computation throughout the training, our knowledge construction mostly occurs at the earlier training of the student model.
In our experiments, the average numbers that knowledge construction (Algorithm \ref{algo:tp}, line 7) occurs are 9.06, 8.69, and 5.53 on the Amazon-music, CiteULike, and Foursquare datasets, respectively.
As a result, HetComp shows lower average time costs compared to DCD.
Also, they show similar convergence behavior (Figure \ref{fig:curve}).
As shown in Section 5, HetComp can significantly improve the distillation quality by reducing the discrepancy to the teachers.
In this regard, HetComp can be considered an effective solution to reduce the online inference costs at the expense of some extra computations in offline training.

\subsection{Supplementary Results}
\label{app:sup}
\begin{figure}[h]
\centering
    \includegraphics[width=0.265\linewidth]{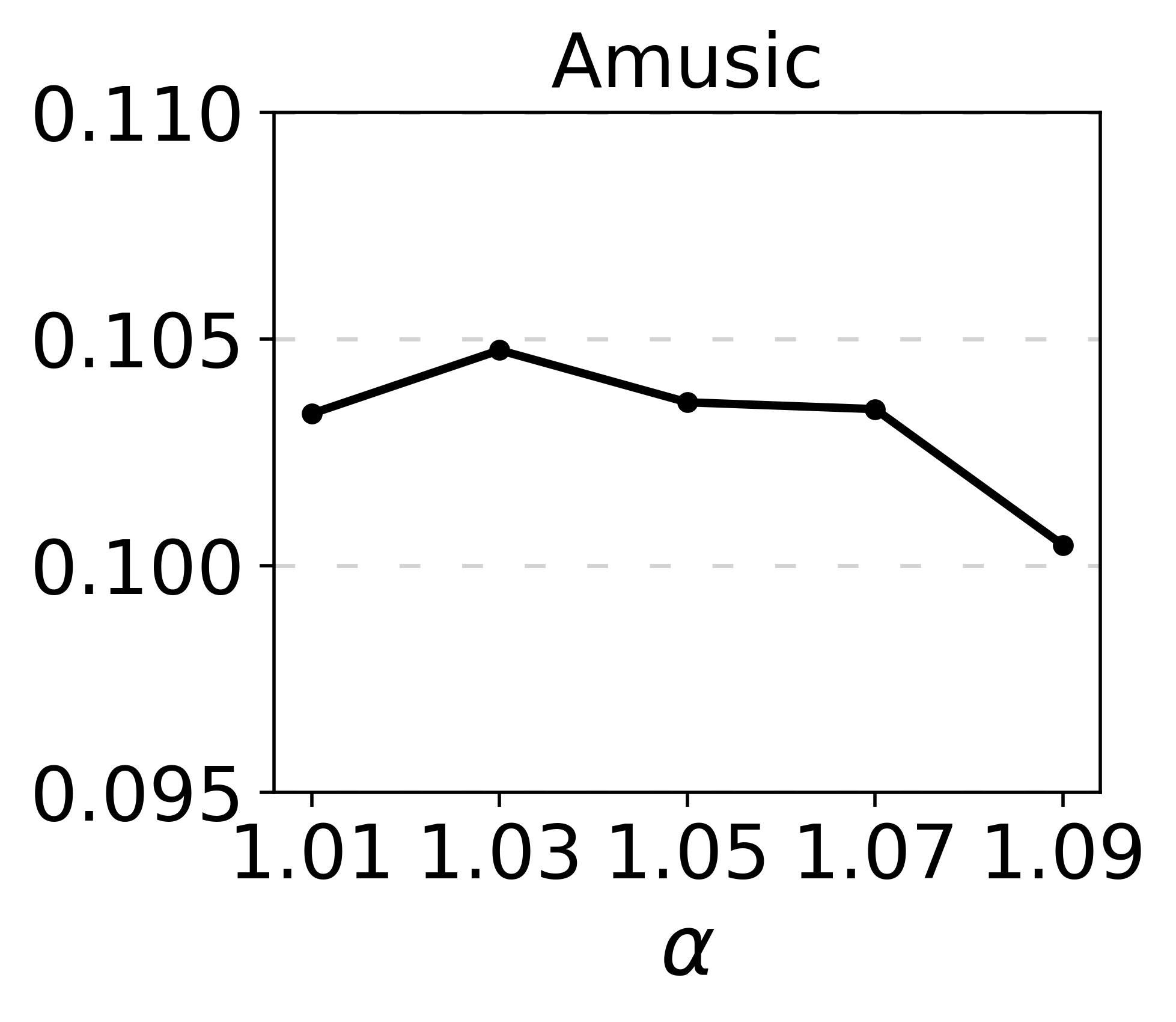}
    \includegraphics[width=0.265\linewidth]{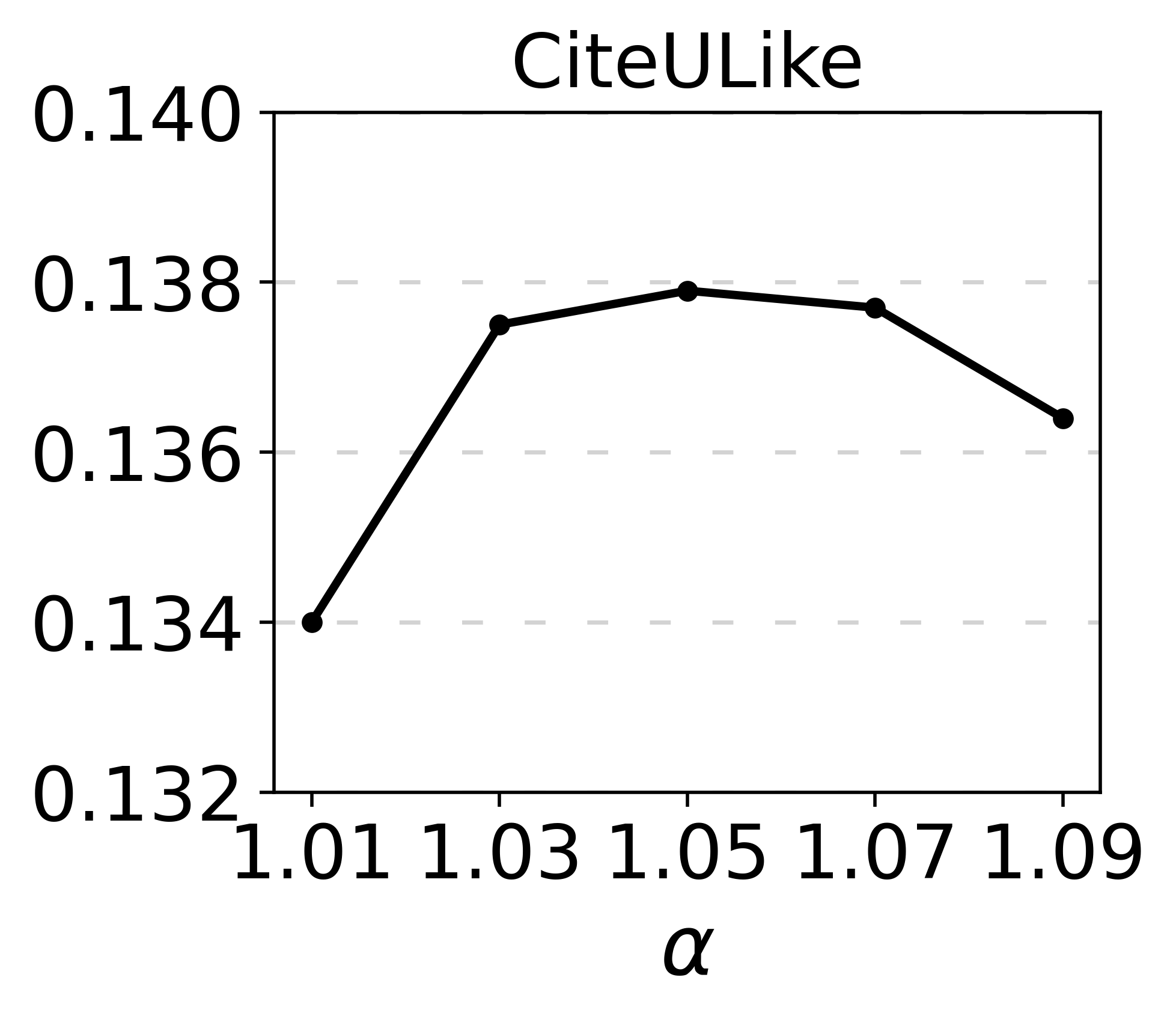}\\
    \includegraphics[width=0.26\linewidth]{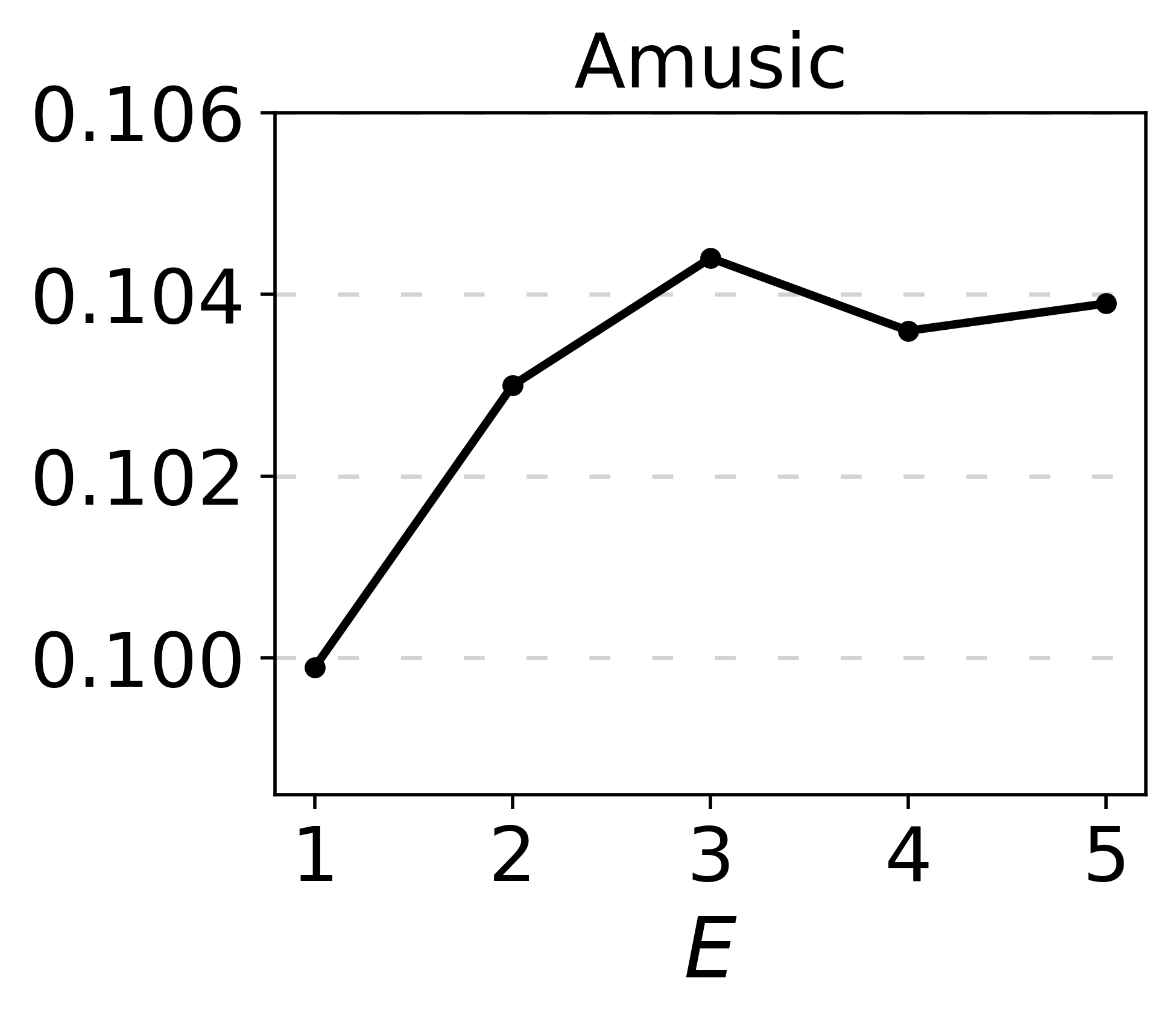}
    \includegraphics[width=0.26\linewidth]{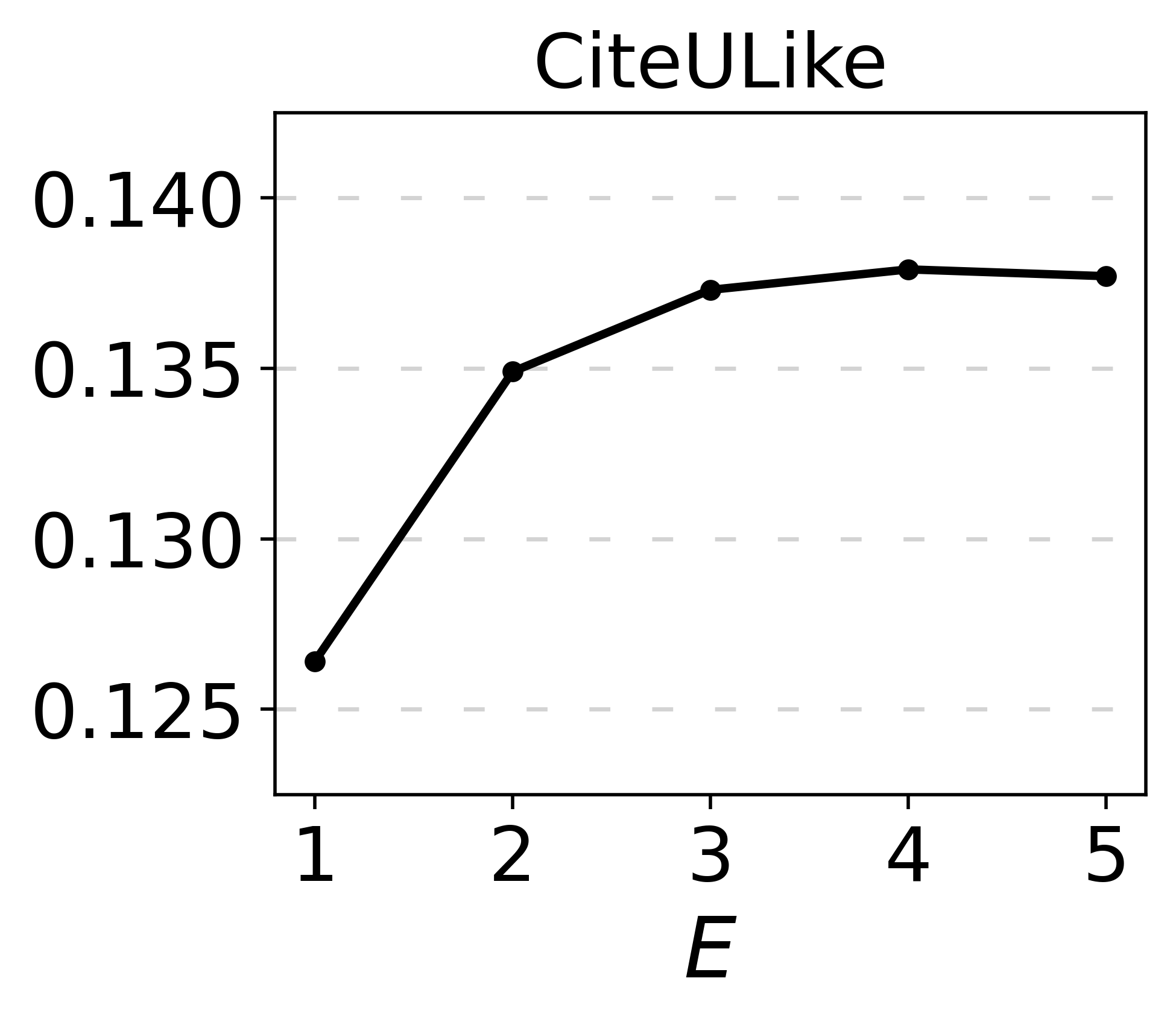}
    \caption{R@10 comparison with varying $\alpha$ and $E$.}
    \label{fig:alpha}
\end{figure}

Figure \ref{fig:alpha} presents the recommendation performance of HetComp (Student: MF) with varying $\alpha$ which controls the transition speed in the dynamic knowledge construction.
The best performance is observed at around $1.03$-$1.05$ on both datasets.
Lastly, Table 11 presents the performance of DCD and HetComp when transferring knowledge of various teacher types.
We use MF as the student model and the CiteULike dataset.
We observe that HetComp effectively improves the distillation quality and achieves the best recommendation performance in all settings, from (a-b) homogeneous model distillation to (c-d) cross-model distillation, and (e) distillation from the ensemble of heterogeneous models. 

\begin{table}[h]
\caption{Performance comparison with various teachers.}
\centering
\label{tab:various}
\renewcommand{\tabcolsep}{1.2mm}
\renewcommand{\arraystretch}{0.85}
\begin{tabular}{c|cccc|c}
\hline
  & \textbf{R@10} & \textbf{N@10} & \textbf{R@50} & \textbf{N@50} & \textbf{D@10} \\
\hline 
(a)  Teacher: MF & 0.1249 & 0.0915 & 0.2604 & 0.1273 & - \\
 DCD & 0.0937 & 0.0698 & 0.2206 & 0.1044 & 0.3831 \\
 HetComp & 0.1092 & 0.0794 & 0.2373 & 0.1128 & 0.3247 \\ \hline
(b)  Teacher: Ensemble (MF) & 0.1395 & 0.1037 & 0.2763 & 0.1395 & - \\
 DCD & 0.1004 & 0.0749 & 0.2307 & 0.1088 & 0.4044 \\
 HetComp & 0.1194 & 0.0878 & 0.2584 & 0.1237 & 0.3286 \\ \hline
(c) Teacher: LightGCN & 0.1337 & 0.0994 & 0.2844 & 0.1392 & - \\
  DCD & 0.1041 & 0.0778 & 0.2367 & 0.1120 & 0.4164 \\
  HetComp & 0.1228 & 0.0896 & 0.2510 & 0.1230 & 0.3590 \\ \hline
(d) Teacher: Ensemble (LightGCN) & 0.1511 & 0.1130 & 0.2952 & 0.1505 & - \\
  DCD & 0.1108 & 0.0838 & 0.2495 & 0.1186 & 0.4180 \\
  HetComp & 0.1305 & 0.0973 & 0.2630 & 0.1255 & 0.3377 \\ \hline
(e)  Teacher: Ensemble & 0.1550 & 0.1156 & 0.3144 & 0.1571 & - \\ 
  DCD & 0.1106 & 0.0851 & 0.2640 & 0.1246 & 0.4433 \\
  HetComp & 0.1379 & 0.1031 & 0.2814 & 0.1396 & 0.3401\\ \hline
\end{tabular}
\end{table}

\chapter{Conclusion and Future Works}
\label{chapt:conclusion}
In this dissertation, we proposed several KD methods for the recommender system having a better accuracy-efficiency trade-off.
We categorized our approaches into two categories based on their knowledge sources, i.e., \textbf{latent knowledge} and \textbf{ranking knowledge}, and introduced several research directions in both approaches.
We expect that the proposed approaches can contribute to a better accuracy-efficiency trade-off of the recommender systems, widening its applications for many resource-constraint environments.

\begin{figure}[h]
	\centering
	\begin{subfigure}[t]{0.99\linewidth}
		\includegraphics[width=\linewidth]{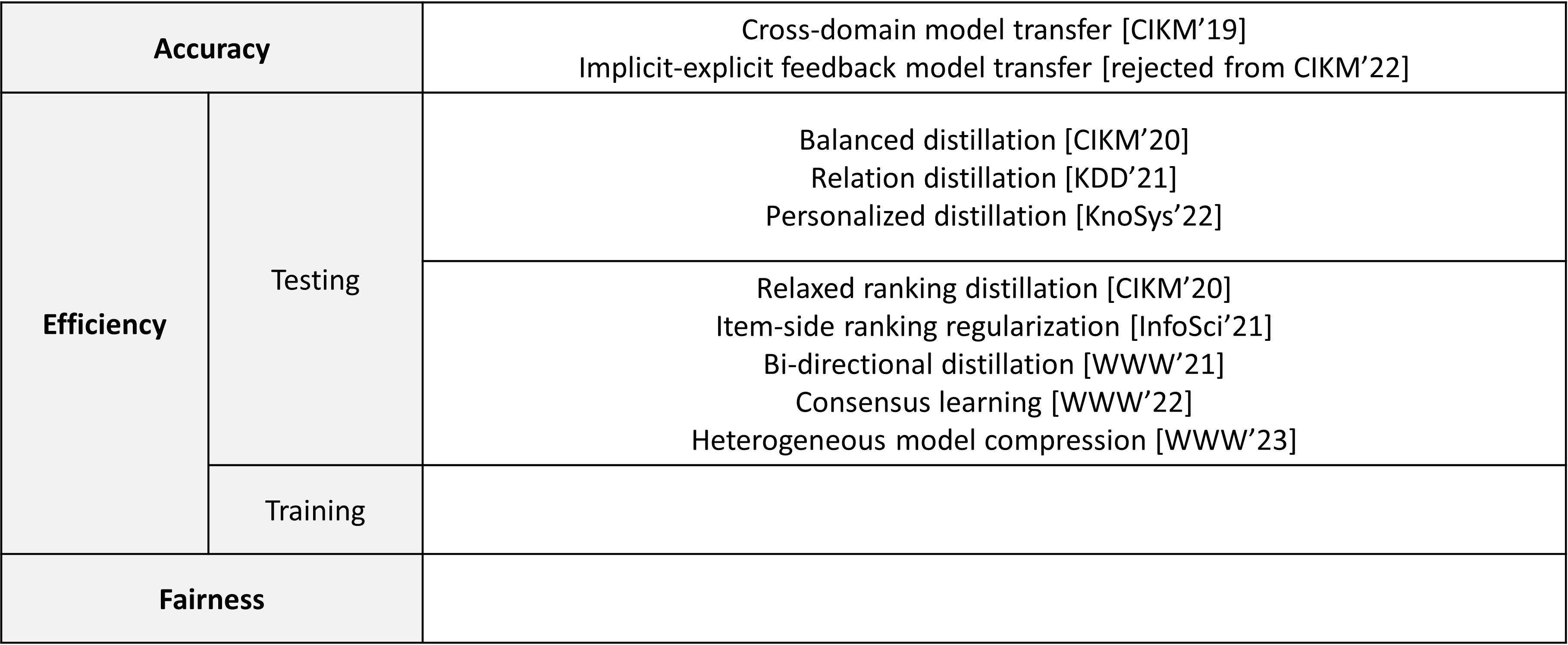}
	\end{subfigure}
	\caption{An overview of future work.}
	\label{fig:thesis_conclusion}
\end{figure}

In the future, we plan to continue our research on distillation approaches, particularly for three key requirements of recommender systems: (1) Accuracy, (2) Efficiency, and (3) Fairness (Figure \ref{fig:thesis_conclusion}).
Although the accuracy-efficiency trade-off at test time is thoroughly investigated in this dissertation, efficiency at training time and fairness aspects have not been covered here.
Fortunately, we have found several clues for the above goals.
In Chapter \ref{chapt:ConCF}, we discovered that the multi-faceted knowledge of heterogeneous models can be used to broaden the recommendation coverage.
Furthermore, in Chapter\ref{chapt:HetComp}, we showed ranking distillation with the easy-to-hard learning paradigm significantly improves the model generalization, which can be particularly beneficial for continual learning.
We will elaborate and concretize those findings towards a long-time goal of our research---an accurate, efficient, and fair recommender system.

\begin{summarykorean}
\begin{spacing}{1.1}
정보의 양이 폭발적으로 증가하고 있는 오늘날, 추천 시스템은 Naver, Google 등의 웹 페이지 검색 서비스나 Amazon, Netflix등의 개인화된 아이템 추천 서비스에서 핵심 기술로 사용되고 있다. 
특히, 추천 시스템은 무수히 많은 아이템 중에서 사용자에게 가장 적합한 항목들을 선별하여 사용자의 결정을 돕고, 기업의 이윤을 극대화한다는 점에서 전자 상거래(E-commerce) 및 미디어 서비스(OTT)의 필수적인 기술로서 자리 잡았다. 

최근 딥러닝 기술의 발전으로 추천 시스템의 정확성(accuracy)은 빠른속도로 발전해온 반면, 효율성(efficiency) 문제는 지속해서 증가하고 있으며 그 중요도에 비해 많은 연구가 이루어지지 않았다. 
방대한 양의 파라미터를 갖는 거대 추천 모델은 우수한 표현 능력을 통해 사용자의 복잡한 선호 정보를 정확하게 포착할 수 있으며, 이로 인해 높은 추천 정확도를 갖는다. 
그러나, 이러한 거대 모델은 아이템 추천을 위해 높은 계산 비용과 긴 지연시간(latency)을 발생시키며, 낮은 효율성으로 인해 실시간 서비스에서 활용되기 어렵다는 한계를 갖는다.
이러한 문제를 해결하고자, 본 연구에서는 성공적인 추천 시스템을 위한 2가지의 필요조건—정확성, 효율성—을 갖춘 모델 생성을 위한 지식 증류 기술들을 제안한다. 

큰 규모의 추천 시스템은 긴 지연시간 뿐 아니라 과 모수화에 의한 과적합 (overfitting) 문제를 수반한다는 점을 고려했을 때, 본 연구를 통해 개발되는 지식 증류 기법들은 높은 효율성과 확장성, 그리고 높은 추천 정확성을 모두 달성할 수 있는 중요한 열쇠가 될 것이다. 
또한, 실제 서비스에 큰 규모의 추천시스템을 적용하는 소모 비용을 줄이고, 아직 추천 시스템이 적용되지 않은 분야들로의 확장을 촉진하는데 기여할 수 있다. 
더 나아가, 지식 증류를 통해 개인화된 모바일 디바이스에서의 추천 모델 활용을 가능하게 하여, 효율성과 개인정보보호 문제를 동시에 해결할 수 있는 디바이스의 발전에 기여할 수 있다.
\end{spacing}
\end{summarykorean}


\bibliographystyle{unsrt}
\bibliography{mybib}

\newpage
\begin{center}
	\textbf{\huge Acknowledgement}
\end{center}

지난 5년 동안 저를 지도해주신 유환조 교수님, 그동안의 지도와 격려에 감사드립니다. 교수님께서 많이 도와주신 덕분에 이렇게 성장할 수 있었습니다. 
그리고 바쁘신 와중에도 논문 심사위원 요청을 흔쾌히 받아주신 조민수, 김상욱, 김동우, 박찬영 교수님 감사드립니다. 프로포절 때 주신 조언들이 디펜스 준비에 큰 도움이 되었습니다. 앞으로도 좋은 연구자가 되도록 노력하겠습니다.

대학원 생활 동안 여러 사람들에게 과분할 정도로 많은 도움을 받았습니다. 이 한 장에 제 감사한 마음을 모두 표현하지 못함을 헤아려주시기 바랍니다.
동현 형, 대학원 진학에 대한 확신이 없었을 때 연구참여 기간 동안 형과 디스커션을 하면서 연구를 좀 더 해보고 싶다는 생각을 갖게 되었습니다, 결혼 축하드립니다.
동하, 너한테 정말 많이 배웠어, 늘 내 고민들 들어줘서 고마워, 앞으로도 도움 구하게 될 것 같아, 부임 축하해.
영혼의 운동 파트너 성제 형, 늘 감사합니다. 계속 같이 득근해나갑시다!
유강 이웃 현준 형, 정말 많이 도움 받았습니다, 감사합니다.
함께 연구하고 토론했던 추천 세미나 그룹 여러분, 현준, 준영, 준수, 원빈, 재현, 창수, 규석, 주영, 다들 많이 고맙고 함께 나누었던 기억들이 자랑스럽습니다.
논문 제출 때마다 함께 했던 준영, 원빈, 너희가 아니었다면 대학원 생활이 더 많이 힘들었을거야. 정말 고마워, 나도 많이 도울게.
수학 문제 분류 과제로 2년 가까이 함께 고생한 정우, 승하 정말 고마워.
그리고, 함께 달렸던 DI-RUN 여러분, 춘천 마라톤 잊지 못할 겁니다.
지난 2년 동안 랩장을 하면서 많은 분들께 도움 받고 신세를 졌습니다. 
특히 많이 도와주신 김지혜 선생님, 상환, 수연 님, 스타랩 업무들을 열심히 도와준 후배 분들을 비롯해, 제 부족함들을 너그러이 이해해주신 모든 연구실 분들께, 이 지면을 빌려 다시 한번 감사드립니다.

마지막으로, 인생의 선택들을 지지해준 부모님, 동생 세영이, 고맙습니다. 늘 건강하고 행복합시다.
고등학생 때부터 이 질긴 인연을 이어가고 있는 고마운 친구들의 서평으로 글을 맺습니다. 얘들아, 이제 포항은 그만 와도 된다.

\begin{center}\textit{뛰어난 논문이지만 한편으론 두렵다. 미래에 인공지능이 우리를 지배하는데 이 연구가 사용되는 것은 아닐까?} \rightline{최원호 (문과, 변호사 지망생)}\end{center}

\begin{center}\textit{추천시스템의 대가가 되었을때도 지금의 마음가짐을 잊지 말것.} \rightline{이영준 (0개 국어 구사자)}\end{center}

\begin{center}\textit{초소형 인공지능... 현실인가?} \rightline{권현성 (던파 닉네임: 내가머더라)}\end{center}

\begin{center}\textit{21세기 최고의 걸작, 부드럽지만 강력하다.. 소름끼치도록 완벽한 서사.. 예측 불가능한 전개.. 나는 이것을 이렇게 부르고싶어요 ``Masterpiece''} \rightline{손영조 (녹번동 된장남)}\end{center}

\newpage

\curriculumvitae[korean]

\begin{personaldata}
	\name       {SeongKu Kang}
\end{personaldata}

\begin{education}
	\item[2012. 3.\ --\ 2018. 2.] B.S. in Computer Science and Engineering, Hanyang University
	\item[2018. 3.\ --\ 2023. 8.] Ph.D. in Computer Science and Engineering, Pohang University of Science and Technology
\end{education}

My proposal and defense slides are available via the below QR code.
\begin{figure}[h]
	\centering
	\includegraphics[width=5cm]{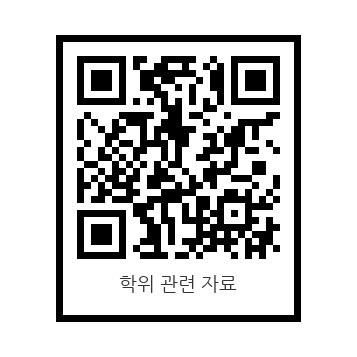}
\end{figure}

\afterpage{\blankpage}  

\end{document}